\pdfoutput=1 
\documentclass[fleqn,usenatbib]{mnras}
\usepackage{amsmath}
\usepackage{txfonts}

\usepackage{lscape}
\usepackage{gensymb}
\usepackage{enumerate}
\usepackage{array} 
\usepackage{graphicx} 
\usepackage{upgreek}
\DeclareGraphicsExtensions{.pdf}
\usepackage{grffile}

\usepackage{savesym} 
\savesymbol{iint}
\usepackage{wasysym}
\restoresymbol{WAS}{iint}

\newcommand\iontwo[2]{[#1\ {\small\rmfamily\@{#2}}\relax]}%

\newcommand{\zgas}{$Z_{\rm gas}$}
\newcommand{\rtwofive}{$R_{25}$}
\newcommand{\frtwofive}{$f_{R_{25}}$}

\newcommand{\OI}{\iontwo{O}{I}}

\newcommand{\SII}{\iontwo{S}{II}}
\newcommand{\NII}{\iontwo{N}{II}}
\newcommand{\OII}{\iontwo{O}{II}}

\newcommand{\OIII}{\iontwo{O}{III}}

\newcommand{\Hb}{H$\beta\,$}

\newcommand{\HII}{{\textrm{H}}\,{\textrm{\sc II}}}

\def\apjl{ApJL}
\def\apj{ApJ}
\def\apjs{ApJS}
\def\mnras{MNRAS}
\def\araa{ARAA}
\def\aj{AJ}
\def\aap{A\&A}
\def\aaps{A\&A Suppl.}
\def\nat{Nature}
\def\pasp{PASP}
\def\fcp{FCPH}
\def\na{NewA}
\def\pasa{PASA}

\title[VENGA: resolved gas metallicity in spirals]{The VIRUS-P Exploration of Nearby Galaxies (VENGA): Spatially Resolved Gas-Phase Metallicity Distributions in  Barred and Unbarred Spirals}
\author[K. F. Kaplan et al.]{\parbox{\textwidth}{Kyle F. Kaplan$^1$\thanks{E-mail: kfkaplan@astro.as.utexas.edu (KFK)}, Shardha Jogee$^1$\thanks{E-mail: sj@astro.as.utexas.edu (SJ)}, Lisa Kewley$^2$, Guillermo A. Blanc$^{3,4,5}$, Tim Weinzirl$^6$,  Mimi Song$^1$, Niv Drory$^1$, Rongxin Luo$^7$, \&  Remco C. E. van den Bosch$^8$}\\
\\
$^1$Department of Astronomy, University of Texas at Austin, 2515 Speedway, Stop C1400, Austin, TX 78712-1205, USA \\
$^2$Research School of Astronomy and Astrophysics, The Australian National University, Cotter Road, Weston Creek, ACT 2611 \\
$^3$Departamento de Astronom\'ia, Universidad de Chile, Camino del Observatorio 1515, Las Condes, Santiago, Chile \\
$^4$Centro de Astrof\'isica y Tecnolog\'ias Afines (CATA), Camino del Observatorio 1515, Las Condes, Santiago, Chile \\
$^5$ Visiting Astronomer, Observatories of the Carnegie Institution for Science, 813 Santa Barbara St, Pasadena, CA, 91101, USA \\
$^6$School of Physics \& Astronomy, University of Nottingham, Nottingham NG7 2RD, UK \\
$^7$Shanghai Astronomical Observatory, 80 Nandan Road, Shanghai 200030, China \\
$^8$Max Planck Institute for Astronomy, K\"onigstuhl 17, D-69117 Heidelberg, Germany \\
}

\pubyear{2016}

\begin{document}
\label{firstpage}
\pagerange{\pageref{firstpage}--\pageref{lastpage}}
\maketitle

\begin{abstract}
We present a study of the excitation conditions and  metallicity of  ionized gas (\zgas{})  
in eight nearby barred and unbarred spiral galaxies from the VIRUS-P Exploration of 
Nearby Galaxies (VENGA) survey, which provides high spatial sampling and 
resolution (median $\sim$~387 pc),  large coverage from the bulge to outer disc, 
broad wavelength range (3600-6800 \AA{}), and medium  spectral resolution 
($\sim$~120 km s$^{-1}$ at 5000 \AA{}). Our results are: 
{\bf (1)}~
We present high resolution gas excitation maps to differentiate between regions 
with excitation typical of  Seyfert, LINER, or recent star formation. 
We find LINER-type  excitation at large distances (3-10 kpc) from the centre,  
and associate this excitation with diffuse ionized gas (DIG).
{\bf (2)}~
After excluding spaxels dominated by Seyfert, LINER, or DIG, 
we produce maps with the best spatial resolution and sampling to date 
of the ionization parameter  $q$, star formation rate, and  \zgas{} using 
common strong line diagnostics.
We find that isolated barred and unbarred spirals exhibit similarly  
shallow \zgas{} profiles from the inner kpc out to large radii (7-10 kpc or 
0.5-1.0 $R_{\rm 25}$). This implies that if profiles had steeper 
gradients at earlier epochs, then the present-day bar is not the primary 
driver flattening gradients over time. This result contradicts earlier claims,
but agrees with recent IFU studies. 
{\bf (3)}~The \zgas{} gradients  in our  $z\sim 0$  massive spirals are
markedly shallower, by $\sim 0.2$ dex kpc$^{-1}$,  than published gradients 
for lensed lower mass galaxies at  $z\sim 1.5-2.0$. Cosmologically-motivated 
hydrodynamical simulations best match this inferred evolution, but the match 
is sensitive to adopted stellar feedback prescriptions.
\end{abstract}


\section{Introduction}

The spatially resolved distribution of a galaxy's gas phase metallicity
is an important signpost 
of its  assembly history and reflects the complex cycle of baryonic 
inflows and outflows, star formation (SF), 
as well as stellar and AGN feedback.
The metallicity of the gas phase in \HII{} regions is traced by the relative abundance
of oxygen to hydrogen, which we quantify as \zgas{} $\equiv$ log(O/H)+12.
The accretion of metal-poor  gas by galaxies  in the form of  halo
accretion and proposed cold mode accretion \citep{keres05, dekel06, keres09, dekel09a,
dekel09b} can 
initially depress \zgas{}  within the galaxy,  but can subsequently fuel
SF that leads to enrichment.
Galaxy mergers and interactions influence \zgas{} 
by inducing stellar  bars, driving gas inflows, triggering SF,  
and possibly fuelling AGN activity
\citep{somerville99, cole2000, springel2005, jogee09, conselice2009, lotz11}.
Secular processes, such as
gas inflows driven by the bar inside its co-rotation resonance 
\citep{kormendy04,  jogee05} transport metal-poor gas from
the outer disc  of a galaxy toward the circumnuclear region,  and  often trigger
powerful  bursts of SF that chemically enrich the circumnuclear gas.
Simulations suggest that  
gas  outflows powered by supernovae, stellar winds and/or  photons
from massive stars,  and variations in the  star formation efficiency 
play important roles in  shaping the gas phase metallicity 
of galaxies
\citep{brooks2007, oppenheimer2010, faucher2011, dave2011, pilkington2012, gibson2013}.
Many observational studies over the last decade  have explored how
the global integrated metallicity of galaxies  correlates with 
other properties,  such as luminosity, stellar mass, and SFR.
The so-called mass metallicity relation  (MZR) between total  
stellar mass and gas phase metallicity is remarkably tight over  a wide
range  in stellar mass
\citep{tremonti2004, lee2006, zhao2010}.
Some studies suggest that the MZR has 
a second-parameter dependence on the SFR  
\citep{ellison2008, mannucci2010, lara2010},
but  this is debated by
other studies 
\citep{sanchez2013, wuyts2014}.

In contrast, our knowledge of the spatially resolved 
distributions of \zgas{}  within galaxies has been very limited. 
The first studies of the distributions of \zgas{}  in nearby galaxies 
were done through multi-slit spectroscopy targeting bright \HII{} regions,
or with long-slit drift scans (e.g. \citealt{moustakas2010}).
Several  studies \citep{vilacostas1992, zaritsky1994, martin1994,
dutil1999, henry1999, considere2000, dors2005, florido2012}
have explored  \zgas{} in galaxies of  different  morphologies,  and
some of them   \citep{vilacostas1992, zaritsky1994, martin1994, dutil1999, henry1999} claim that barred spirals
 exhibit flatter \zgas{} gradients than unbarred spirals.

Studies by \cite{krabbe2008}, \cite{rupke2010}, \cite{kewley2010},  \cite{krabbe2011},
\cite{rosa2014}, and \cite{torres2014} claim that pairs
of interacting galaxies  have flatter \zgas{} gradients than isolated
galaxies due to gas flows driven by the interactions.
However, studies with
slit spectroscopic data often suffer from poor spatial sampling and incomplete coverage. 
Additionally, those studies without adequate spatial information or wavelength coverage cannot identify and
exclude contaminated regions where gas is not primarily excited by photons from young massive stars.

High resolution, high-quality,  integral field unit (IFU) spectroscopic data 
allow us to more accurately  explore the gas phase metallicity distribution  
within barred and unbarred spirals of different Hubble types. 
Two recent  studies by \cite{sanchez2012} and \cite{sanchez2014}  
based on the PINGS \citep{pings2010}  and CALIFA \citep{sanchez11} 
IFU surveys  have explored \zgas{}  in spiral galaxies.
Both studies  find the \zgas{} gradients in their samples are   
independent of Hubble Type or being barred vs. unbarred when scaled   
to a common effective radius ($R_e$).
For interacting spirals,  the IFU study by \cite{rich2012}
finds that \zgas{} gradients in their sample of strongly interacting galaxies
to be flatter than isolated spirals, in agreement with the earlier studies.

In this paper, we present a complementary  IFU-based study of
 the gas phase
metallicity in  a sample of barred and unbarred  spirals  drawn from 
the VIRUS-P  Exploration of Nearby Galaxies (VENGA) survey 
\citep{blanc2013}.  
Our sample of  eight  barred and unbarred spirals was drawn from
 the full VENGA sample of 30 spirals by selecting galaxies 
at  intermediate distances (8 to 33 Mpc) so  that the IFU data  cover a large
fraction  of the galaxy's outer disc,  while  simultaneously
providing a high spatial resolution  (a few hundred pc). 
The combination of spatial coverage and  resolution allows us to resolve
individual galactic components, such as the bulge, primary stellar bar,
outer disc, and separate regions of widely
different
excitation
(e.g., \HII{} regions, spiral arms, starburst or AGN driven outflows,
diffuse ionized gas, etc.).

The IFU-based study in this paper
complements the studies by  \cite{sanchez2012} and \cite{sanchez2014}  
in several respects.
While our study has a smaller  sample than \cite{sanchez2012} and 
\cite{sanchez2014}, it benefits from a 
high spatial resolution (median of 387 pc), 
a high spectral resolution (120 km s$^{-1}$ at 5000 \AA{}), 
a broad blue-to-red wavelength coverage (3600-6800 \AA{}), 
and the use of seven  \zgas{} diagnostics.
The cross-comparisons between these seven \zgas{}  
diagnostics  allows us to break 
degeneracies in values of \zgas{}  ($\S$ \ref{sec:how-zgas}) 
and  can aid other studies 
(e.g., at high redshifts) limited to only a few \zgas{} diagnostics. 

Another strength of this study is that it systematically tackles
the following issues that have plagued many earlier 
long-slit and IFU-based \zgas{} studies:
 {\bf (i)} 
Many \zgas{} diagnostics work well when using emission line ratios 
from gas that is predominantly ionized by photons from local 
massive stars,  but break down when the gas is predominantly ionized by a
hard radiation field from an AGN or is shocked (e.g., in starburst-driven
outflows). 
Studies without  the spatial resolution needed to exclude 
contaminated regions often yield  erroneous \zgas{} values, as emphasized
by \cite{kewley2002}, \cite{kewley2008}, and  \cite{yuan2012b}.
We avoid this pitfall 
by removing regions dominated by  Seyfert or  LINER conditions,
as well as regions dominated by diffuse  
ionized gas (DIG)
before calculating \zgas{}.
{\bf (ii)}
The value of some \zgas{}  diagnostics  
(e.g., \mbox{$R_{23} \equiv$ (\OII{}$\lambda$3727+\OIII{})$\lambda\lambda$4959,5007/\Hb{})}
depend on the
ionization parameter $q$, but not all calibrations of \zgas{} take this into
account. In this work, we calculate {\it spatially resolved} maps of $q$ across
the bulge, bar, and outer disc.

This paper is organized as follows:  $\S$ \ref{sec:obs-venga}
introduces the VENGA IFU survey;  $\S$ \ref{sec:obs-samplesel} describes
the selection of our sub-sample of eight nearby spirals;  and $\S$ \ref{sec:reduc1} covers
the data reduction.  For our methodology, $\S$ \ref{sec:bpt-method} 
discusses the use of excitation diagnostic diagrams to remove Seyfert and LINER
contaminated regions; 
$\S$~\ref{sec:dig} discusses how we identify and remove regions
dominated by  emission from DIG;
$\S$~\ref{sec:sfr} details how we compute SFRs;
$\S$~\ref{sec:how-q} describes our
computation of the ionization parameter $q$; and
 $\S$~\ref{sec:how-zgas} shows how we derive 
the seven different \zgas{} diagnostics.   For our results,
$\S$~\ref{sec:results-components}  presents the
spatially-resolved maps and deprojected radial profiles of \zgas{}, $q$, \& SFR;
$\S$~\ref{sec:results-zgas} compares the absolute value of \zgas{} between the different diagnostics;
$\S$~\ref{sec:results-bars}  presents our results on 
the \zgas{}  distributions  in our barred and unbarred galaxies;
$\S$~\ref{sec:results-redshift} compares 
our \zgas{}  profiles in nearby spirals to published profiles for  high redshift galaxies; 
 and $\S$ \ref{sec:results-models}  compares the  observed evolution in
\zgas{} profiles from $z\sim 2$ to 0 with the evolution
predicted by different suites of simulations. 
$\S$~\ref{sec:summary} presents our summary and conclusions.

\section{Observations and Sample} \label{sec:obs}

\subsection{VENGA} \label{sec:obs-venga}
\begin{figure}
\includegraphics[width=0.51\textwidth]{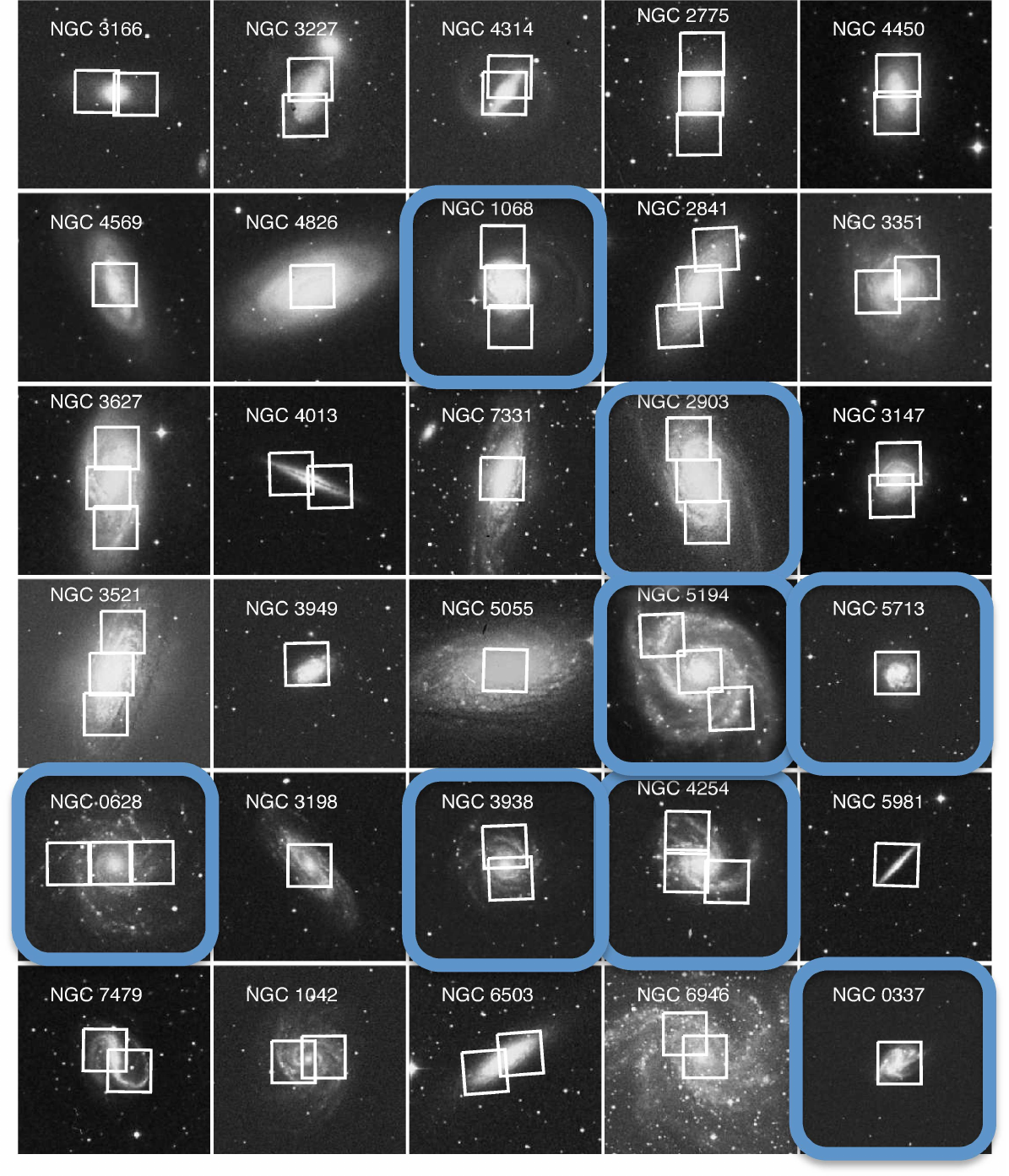}
\caption[For LoT]%
{DSS$^a$
images of the VENGA sample  of 30 nearby spiral galaxies are 
overlaid with squares, which indicate the individual pointings of the Mitchell Spectrograph IFU.
Blue highlights our sub-sample of eight spirals which are selected
to maximize both our spatial resolution 
and spatial coverage of the disc for studying \zgas{} (see $\S$~\ref{sec:obs-samplesel}).
\\
$^a$\ The Digitized Sky Surveys were produced at the Space Telescope Science Institute under U.S. Government grant NAG W-2166. The images of these surveys are based on photographic data obtained using the Oschin Schmidt Telescope on Palomar Mountain and the UK Schmidt Telescope. The plates were processed into the present compressed digital form with the permission of these institutions.
}
\label{fthumb}
\end{figure}

VENGA is an integral field spectroscopic survey 
of the inner and outer regions of the disc
of a sample of 30 
nearby spiral galaxies with the Mitchell Spectrograph (formerly called VIRUS-P) IFU 
on the 2.7 metre telescope at McDonald Observatory \citep{blanc2013}.  
The Mitchell Spectrograph has large 
($5.6 \arcsec$ full width half maximum [FWHM]) 
sensitive fibres and the largest 
FOV ($110\arcsec  \times 110\arcsec$ or  3.36 arcmin$^2$) among
existing IFUs \citep{virusp}. 
Over four years,
this survey has been allocated  $\sim$ 150 
nights of observing time (PIs: G. Blanc and  T. Weinzirl).
Three dithers are performed on each galaxy to compensate for the $1/3$
filling factor of the Mitchell Spectrograph, and multiple pointings are used to 
acquire spectra over a large fraction of the $R_{25}$ radius  
(the galactocentric radius where the B-band surface brightness = 25 mag arcsec$^{-2}$) 
of each galaxy's disc (Figure~\ref{fthumb}). 
Each galaxy is observed with both a blue  (3600-5800 \AA) and red (4600-6800 \AA)
setup to obtain a wide wavelength coverage.
The spectral resolution is $R\approx 1000$ or $\sim$ 5 \AA\ FWHM at 5000 \AA{}, which corresponds to
$\sim 120$ km s$^{-1}$.
An example 1D VENGA spectrum from the galaxy NGC 0628 can be found in \cite{blanc2013}.

The VENGA  sample consists of 30 nearby spirals
at distances out to 50 Mpc,
80\% of which are closer than 20 Mpc.
All the target galaxies are shown in Figure \ref{fthumb}.
The VENGA sample
was chosen to cover
 a range of Hubble types (Sa to Sd), inclinations from face to edge on, include
galaxies with classical bulges and pseudo-bulges (e.g., \citealt{kormendy04, fisher08, weinzirl}),
as well as include barred and unbarred spirals.   
The VENGA galaxies have a global SFR typically in the range of 0.5 to 10 $M_\odot$ yr$^{-1}$,
and stellar masses primarily in the range of $10^9$ to $10^{11}$ $M_\odot$, 
and for those galaxies with
stellar mass above   $10^{10}$ $M_\odot$, they span a representative range of the 
stellar mass-SFR  plane, as  shown  in \cite{blanc2013}.
Most galaxies have ancillary data from a variety of sources including
$HST$, Spitzer, GALEX, CO maps from BIMA SONG \citep{bimasong} and the
CARMA CO survey STING \citep{sting}, archival HI 21 cm maps from
THINGS \citep{walter08} and ALFALFA \citep{giovanelli05}.   
Half of the VENGA sample is covered by the KINGFISH survey \citep{kingfish}.

\begin{table*}
\begin{center}
\caption{Our Sub-sample of  VENGA Spirals.}   
\begin{tabular}{cccccccccccccc}
\label{tab:venga-subsample} \\
\hline \hline
\multicolumn{1}{c}{NGC}&
\multicolumn{1}{c}{D}&
\multicolumn{1}{c}{$a_{\rm 25} \times b_{\rm 25}$}&
\multicolumn{1}{c}{i}&
\multicolumn{1}{c}{Type}&
\multicolumn{1}{c}{$M_\star$}&
\multicolumn{1}{c}{Bar?}&
\multicolumn{1}{c}{Inter.?}&
\multicolumn{1}{c}{$B/T$}&
\multicolumn{1}{c}{$R_{e}$-bulge}&
\multicolumn{1}{c}{Bulge $n$}&
\multicolumn{1}{c}{FPSF}&
\multicolumn{1}{c}{Ratio} &
\multicolumn{1}{c}{\frtwofive{}}\\
\multicolumn{1}{c}{}&
\multicolumn{1}{c}{(Mpc)}&
\multicolumn{1}{c}{($'$)}&
\multicolumn{1}{c}{(\degree)}&
\multicolumn{1}{c}{}&
\multicolumn{1}{c}{}&
\multicolumn{1}{c}{}&
\multicolumn{1}{c}{}&
\multicolumn{1}{c}{}&
\multicolumn{1}{c}{(pc)}&
\multicolumn{1}{c}{}&
\multicolumn{1}{c}{(pc)}&
\multicolumn{1}{c}{}&
\multicolumn{1}{c}{}\\
\multicolumn{1}{c}{(1)}&
\multicolumn{1}{c}{(2)}&
\multicolumn{1}{c}{(3)}&
\multicolumn{1}{c}{(4)}&
\multicolumn{1}{c}{(5)}&
\multicolumn{1}{c}{(6)}&
\multicolumn{1}{c}{(7)}&
\multicolumn{1}{c}{(8)}&
\multicolumn{1}{c}{(9)}&
\multicolumn{1}{c}{(10)}&
\multicolumn{1}{c}{(11)}&
\multicolumn{1}{c}{(12)}&
\multicolumn{1}{c}{(13)}&
\multicolumn{1}{c}{(14)}\\
 \hline
0337 & 19.3$^\alpha$ & $ 1.44\times 0.91$ &     55 & SB(s)d & 1.6E+10 & y &  y$^{a}$  &   &   &   &  523 &   &  0.72 \\
0628 &  8.6$^\beta$ & $ 5.24\times 4.78$ &       9$^b$ & SA(s)c & 2.0E+10 & n &  n &  0.10 &  600 &  1.35 &  268 &  2.24 &  0.47 \\
1068 & 10.1 & $ 3.54\times 3.01$ &       35 & SA(rs)b &  7.9E+10 &  n & n &   &   &   &  273 &   &  0.72 \\
2903 &  8.9 & $ 6.29\times 3.01$ &       66 & SAB(rs)bc & 5.0E+10 & y & n &  0.09 &  360 &  0.42 &  241 &  1.49 &  0.38 \\
3938 & 17.9 & $ 2.69\times 2.45$ &       27 & SA(s)c & 3.2E+10 & n & n &  0.07 &  560 &  1.18 &  485 &  1.15 &  0.60 \\
4254 & 14.3 & $ 2.69\times 2.34$ &       33 & SA(s)c &  4.0E+11 &  n & n &  0.39 &  2110 &  2.68 &  387 &  5.45 &  0.81 \\
5194 &  8.4 & $ 5.61\times 3.46$ &       56 & SA(s)bc & 7.9E+10 & n & y &   &   &   &  228 &   &  0.56 \\
5713 & 32.6 & $ 1.38\times 1.23$ &       30 & SAB(rs)bc & 7.9E+10 & y & n  &  0.33 &  2460 &  1.84 &  883 &  2.79 &  0.66 \\
\hline\hline
\end{tabular}
\end{center}
\raggedright
Columns:
(1)~Galaxy NGC number.  NGC~5194 is also commonly known as M51a.
(2)~Distance D to galaxy in Mpc.  For NGC 03337 \& 4254, we use \cite{springob07}.
For NGC 0628, we adopt the value from  \cite{hermann2008}, also used in \cite{blanc2013}.
For NGC 1068, we use \cite{tully08}. For NGC 2903 we use \cite{drozdovsky00}.  For NGC 3938 \& 5194, we use \cite{poznanski09}.
For NGC 5713, we use  \cite{tully88}, corrected $H_0$ from 75 to 70 km/s/Mpc.
(3)~The major axis $a_{\rm 25}$ and minor axis $b_{\rm 25}$ of
the galaxy's projected radius \rtwofive{} from RC3 \citep{devaucouleurs1991}.  
\rtwofive{}  is defined as the radius where the surface brightness
of the outer disc reaches 25 mag arcsec$^{-2}$ in the $B$-band.
(4)~For all galaxies, except NGC~0628, we adopt  the inclination based on
Tully (1988), calculated from inclination $= \cos^{-1} \left( \sqrt{ \left[(a_{\rm 25}/b_{\rm 25})^2-0.2^2\right] / \left[1 - 0.2^2\right]} \right) + 3^{\circ}$.
For NGC~0628, we adopt the inclination value of $8.7\degree$ published by
\cite{blanc2013}, compared to the value of $27\degree$ from \cite{tully88}.
(5)~Hubble type from RC3 \citep{devaucouleurs1991}.
(6)~Total stellar mass in units of $M_\odot$ from \cite{blanc2013}.
(7)~Does galaxy host a large-scale stellar bar? In assessing the presence or absence of a large-scale stellar bar,
we do not rely only on the RC3 classification based on visual
inspection of optical photographic plates, but instead base the final assessment
on the quantitative analyses of near-infrared images, as detailed in Table \ref{tab:bars}.
(8)~Is galaxy interacting? We use  interaction status from NED$^a$ unless otherwise indicated.
$^a$In the case of NGC 0337, which NED characterizes as non-interacting,
The Carnegie Atlas of Galaxies \citep{sandage1994} classifies NGC 0337 as  Sc(s)II.2 pec due to the patchiness
and asymmetry in the stellar distribution.  Our own data also shows the same patchiness and asymmetry in both the stellar and H$\alpha$
distributions (see Figure \ref{fig:vengamaps1}), prompting us to classify this galaxy as weakly interacting.
(9)~Bulge-to-total ($B/T$) mass ratio. 
Reported ratios for NGC 0628 \& 2938 are from \cite{fisher08}, the ratio for NGC 2903 is from \cite{dong06}, and the ratios
for NGC 4254 \& 5713 are from \cite{weinzirl}.
For galaxies where bulge-plus-disc decomposition has not been performed, no bulge parameters are listed.
(10)~Bulge half-light radius R$_e$-bulge.  
See note for column 9 for references.
(11)~Bulge S\'ersic index. 
See note for column 9 for references.
(12)~Spatial resolution  FPSF  (defined as the  full width half maximum of the point  spread function)  of the Mitchell
    Spectrograph.
(13)~Ratio  ($R_e$-bulge/FPSF) of the bulge half-light radius $R_e$ to the spatial resolution of the Mitchell Spectrograph.
(14)~Fraction of the major axis $a_{\rm 25}$ of the \rtwofive{} radius covered by VENGA.  
This is illustrated in  Figure \ref{fthumb}. \\
$^a$\ The NASA/IPAC Extragalactic Database (NED) is operated by the Jet Propulsion Laboratory, California Institute of Technology, under contract with the National Aeronautics and Space Administration. \url(https://ned.ipac.caltech.edu/)
\end{table*}

\subsection{Sub-sample Selection} \label{sec:obs-samplesel}

We selected a subset of eight spiral galaxies from the 30 VENGA spirals for our study of \zgas{},
based on the four criteria below.
In particular, we simultaneously require IFU data coverage over a large fraction of the galaxy's disc (criterion ii), as well as high spatial resolution (criteria  iii \&  iv; Table \ref{tab:venga-subsample}).

\begin{enumerate}
\item {\bf Inclination:} 
A very high inclination results in large extinction by dust and makes it difficult to spatially resolve the separate galactic components.
Face-on galaxies lack information on the kinematics of the stars and gas, which is important for detecting azimuthal and radial motions.
To balance out these two extremes, our sub-sample covers a range of low to moderate inclinations,
between $9\degree$ to $66\degree$\footnote{In
the initial selection of our sub-sample, we used the inclination from
\cite{tully88} for all galaxies, and our subsample had an inclination
range of $27\degree$ to $66\degree$.  However, at the time of submission
of this paper, we adopted the lower inclination of $8.7\degree$ published
for NGC 0628 \citep{blanc2013}, compared to the value of $27\degree$
from \cite{tully88}.}, with a median of $35\degree$.
\item {\bf Fraction of \rtwofive{}:} 
\rtwofive{} is the galactocentric radius where the B-band surface brightness equals 25 mag arcsec$^{-2}$.
We select galaxies where the fraction  of the disc's  \rtwofive{} radius (\frtwofive{}) covered by
the Mitchell Spectrograph  observations   is  at  least 35\%  (\frtwofive{} $>$ 0.35).  Our sample has
\frtwofive{} ranging from 
0.38 to  0.81  with a median of 0.66.\\
\item {\bf ($R_e$-bulge/FPSF):} 
For galaxies with significant bulges (bulge-to-total ratio  $B/T >0.01$), we ensure
that we resolve the bulge by requiring  that ($R_e$-bulge/FPSF) $>$ 1, 
where $R_e$-bulge denotes the  half-light radius of the bulge
and FPSF  is the FWHM
of the point
spread function (PSF) of the observations ($5.6''$).
Our sub-sample has a median ($R_e$-bulge/FPSF) of 2.2.
\\
\item {\bf Spatial Resolution:}  
In addition to the relative requirement ($R_e$-bulge/FPSF) $>$ 1,
we also require an absolute spatial resolution of FPSF $<$ 900 pc 
in order to resolve the individual galactic components 
such as the bulge, primary stellar bar, outer disc, and to 
separate regions of 
different 
excitations
(e.g., \HII{} regions, starburst or AGN driven outflows, DIG, etc.). 
This condition sets the maximum distance of our sub-sample to be $\sim 33$ Mpc. 
With a combination of criteria (ii) to (iv), we obtain a 
sub-sample of eight  galaxies with a median distance of 14.3 Mpc, giving us a median 
spatial resolution of 387 pc.
\end{enumerate}

\footnotetext[3]{}

\begin{table*}
\caption{Comparison of IFU studies of \zgas{} in nearby spirals}
\begin{center}
\begin{tabular}{cccccccc}
\hline\hline
Study & \# Spirals  & Parent  & Median  & Median  & Wavelength & Spectral  & \# of \zgas{}  \\
&  analysed & IFU Survey  & Dist. (Mpc) & Res. (pc) & Range (\AA{})  & Res. (km s$^{-1}$) & Diagnostics \\
(1) & (2) & (3)  & (4) & (5) & (6) & (7) & (8) \\
\hline
This Study & 8 & VENGA  & 14.3 & 387 & 3600-6800 & 120 @ 5000 \AA{} & 7 \\
\hline
\cite{sanchez2012} & 7 & PINGS  & 10.6 & 103 & 3700-6900 & 600 @ 5400 \AA{}  & 3 \\
 & 31 &  CALIFA  & 66.4 & 644 & 3700-6900 & 600 @ 5400 \AA{} & 3 \\
\hline
\cite{sanchez2014} & 306 & CALIFA  & 68 & 985 & 3745-7500 & 350 @ 5600 \AA{}  & 1 \\
\hline \hline
 \label{tab:survey-comp}
\end{tabular}
\end{center}
\raggedright
Columns: (1) IFU study of \zgas{} radial profiles.
(2) Number of spiral galaxies analysed in study.
(3) Parent IFU study.
(4) Median distance to galaxies in study in Mpc.
(5) Median spatial resolution of study in pc.
(6) Wavelength range from blue to red of IFU used in \AA{}. 
(7) Spectral resolution of study in km s$^{-1}$.
(8) Number of \zgas{} diagnostics used in study.
\end{table*}


Table \ref{tab:venga-subsample} shows our sub-sample of eight
VENGA  spiral galaxies satisfying  the above selection criteria. 
The sample  spans a range of Hubble types from Sab to Sd.
It includes six isolated galaxies, 
of which   two are barred (NGC 2903 \& 5173) and four are unbarred 
(NGC 0628, 1068,  3938, \& 4254),  as well as 
one weakly interacting  barred  galaxy (NGC 0337), and 
one weakly interacting unbarred galaxy (NGC 5194 or M51a).
We describe the interacting systems in our study as weakly
interacting because they are consistent 
with minor interactions/mergers with mass ratios below 1:3.
NGC 5194 or M51a is interacting with  a dwarf galaxy (M51b),  
while NGC 0337 shows subtle  asymmetries in the stellar distribution
of its disc.
When assigning a bar type to the sample,  we do not use only the optically-based 
visual bar classification in RC3, but further verify the bar type using 
quantitative analyses of near-infrared images in $\S$~\ref{sec:results-bars}.

\begin{figure*}
\includegraphics[height=3.3in, angle=0]{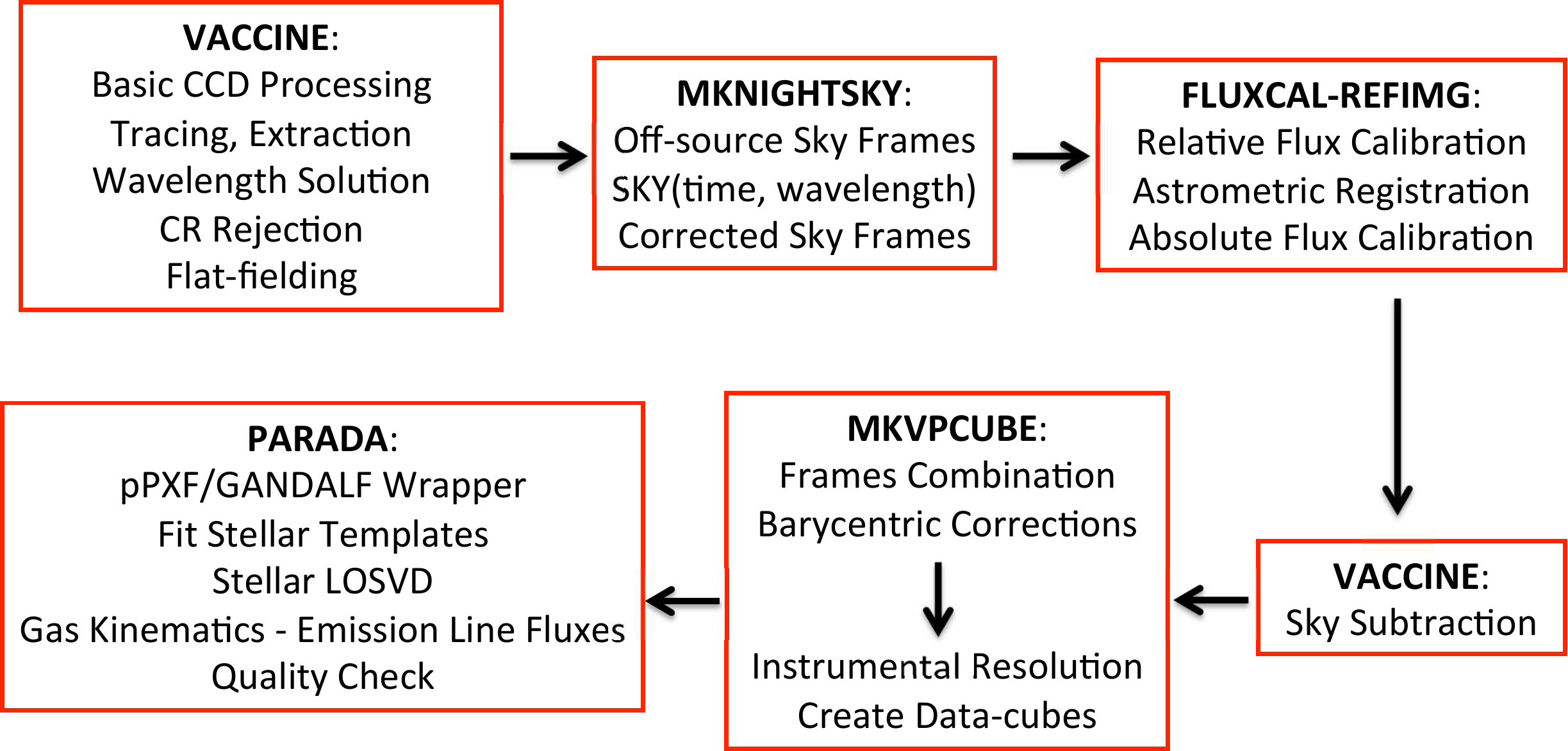}
\caption{
Flow chart of our data reduction and analysis pipeline, as outlined in 
$\S$~\ref{sec:reduc1}.}
\label{freduc}
\end{figure*}

Table \ref{tab:survey-comp}  compares previous IFU
studies of  the distribution of \zgas{} in spiral galaxies to the IFU
study in this paper.
The study by \cite{sanchez2012} analyses data of 7 spirals from the
PINGS \citep{pings2010}  IFU  survey and 31 spirals from the CALIFA
 \citep{sanchez11} IFU survey.  Their median spatial resolution  is 
 644 pc, the spectral resolution is 600 km s$^{-1}$ at 5000
\AA{},   and  only three diagnostics of \zgas{} are used (Table \ref{tab:survey-comp}).
The more recent study by \cite{sanchez2014} uses 306 spiral galaxies 
from CALIFA with a median spatial resolution of 985 pc, a spectral
resolution of 350 km s$^{-1}$ at 5000 \AA{}, and uses only one \zgas{}
diagnostic (Table \ref{tab:survey-comp}).  
In contrast, our IFU study of 8 spirals in this paper 
has a higher median spatial resolution (387 pc), 
a higher spectral resolution (120 km s$^{-1}$ at 5000 \AA{}), 
and uses a full set of seven \zgas{} diagnostics.

\section{Data Reduction}\label{sec:reduc1}

The data reduction pipeline VACCINE \citep{hetdexps}
is used to perform
the basic data reduction, including 
the bias and dark subtraction, flat fielding, 
wavelength calibration, cosmic ray rejection and sky subtraction.
Figure~\ref{freduc} illustrates our data reduction and analysis steps 
(\citealt{hetdexps, blanc2013}).
After running VACCINE, a custom made sky subtraction method based on a spline interpolation of all available sky exposures throughout each night is performed.
Flux calibration is conducted in two steps. The first step uses spectrophotometric
standard star frames for wavelength-dependent flux calibration 
for each observing run. The second step uses 
SDSS broad-band images \citep{alam2015} to perform absolute flux calibration and astrometry correction
to compensate
for changing weather conditions and uncertainty in the pointings.
The uncertainty for the flux in each fibre is a combination of the readout noise, Poisson uncertainty, and systematic error.
Calculation of the uncertainty is detailed in \cite{blanc2013}.
Finally, all science frames are combined into a datacube which is regularly spaced in right ascension, declination, and wavelength. The spaxel size in the datacube is 2'',
which implies a Nyquist sampling of the
PSF FWHM.

We extract the emission lines from the underlying stellar continuum 
by using PARADA, a modified version of GANDALF \citep{sauron5}. 
To fit the stellar continuum,
we use the MILES \citep{miles} empirical stellar library, 
which has over 900 stars with a wide range of spectral types and
metallicities,
and is accurately flux-calibrated (with typical differences between photometric databases and synthetic colours derived from MILES of $<$ 0.1 magnitudes).
Each MILES stellar spectrum is convolved to the wavelength-dependent Mitchell Spectrograph
spectral resolution (median of $\sim 5$ \AA{} FWHM at 5000 \AA{}). 
The observed spectra are fit with the templates in pixel space using the 
Penalized Pixel-Fitting (pPXF; \citealt{ppxf}) method,
from which the stellar continuum and kinematics are extracted.

Galactic extinction is corrected using the extinction
map of \cite{dustmap}. 
Spectra are decomposed by
Gaussian-shaped emission lines being simultaneously fitted with the
underlying stellar continuum. A multiplicative high order Legendre polynomial
is used to account for the internal reddening.
We check the quality of our data using a technique outlined by
\cite{qualitycheck}, comparing the residuals of our fits of the stellar continuum and emission lines to the statistical noise
in the data. 
Regions with residuals greater than three times the noise are flagged and removed.
Typically only a few percent of the regions are flagged by this
quality check and many of these regions are also flagged and removed 
by other means (e.g., signal-to-noise cuts, DIG, excitation diagnostics; $\S$ \ref{sec:bpt-method},
\ref{sec:dig}).

Galactic stars are a source of contamination in the spaxels of our 2D maps and must be masked out.  
They are identified as regions which have a stellar velocity that deviates more than 50 km s$^{-1}$
from the the average stellar velocity over a 10 arcsecond radius.  Identified galactic stars
are then removed from further analysis by masking out spaxels within a 2 arcsecond radius of the identified star.

\begin{figure}
\hspace{-0.55 cm}
\includegraphics[width=0.25\textwidth]{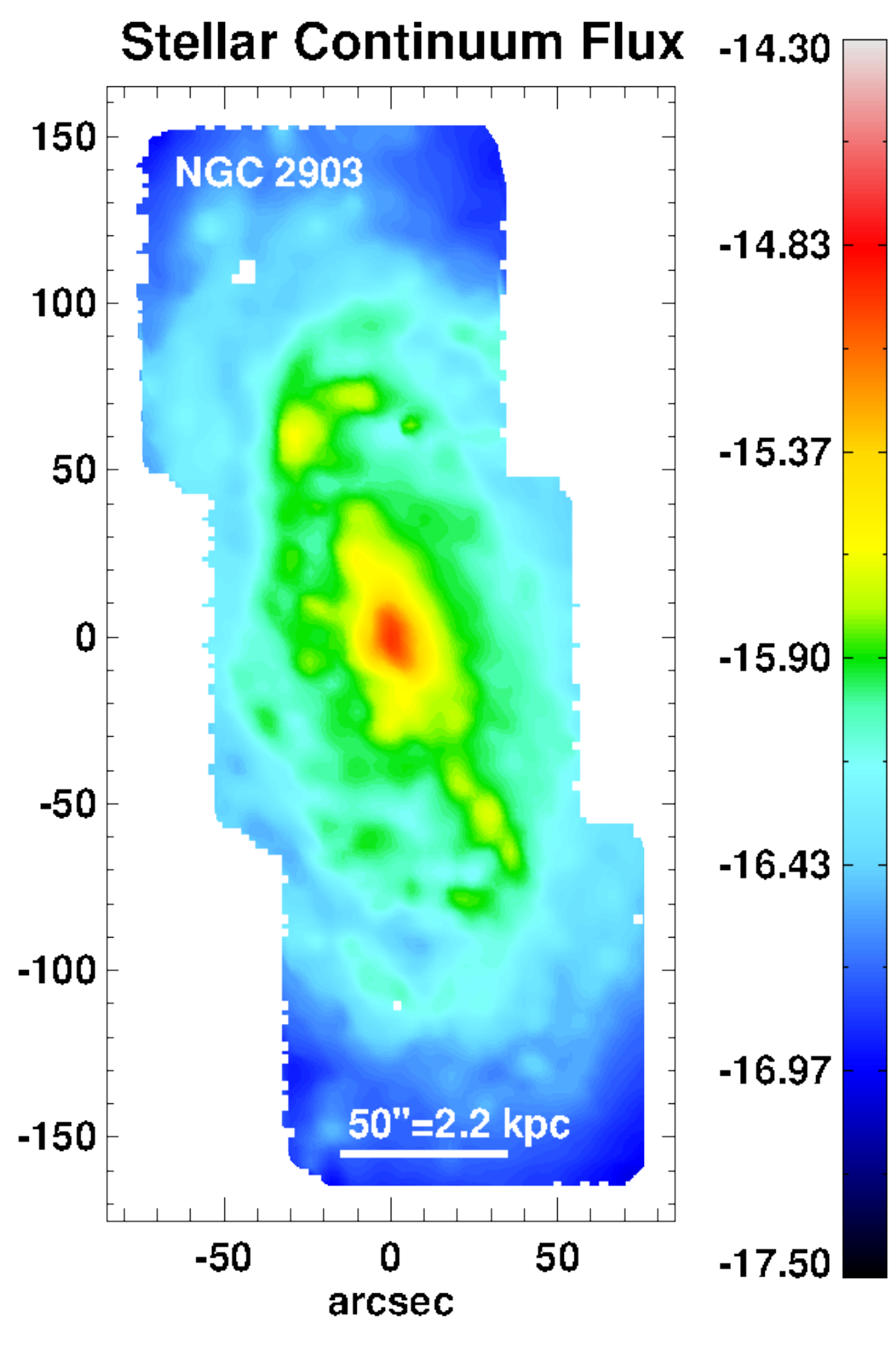}
\includegraphics[width=0.25\textwidth]{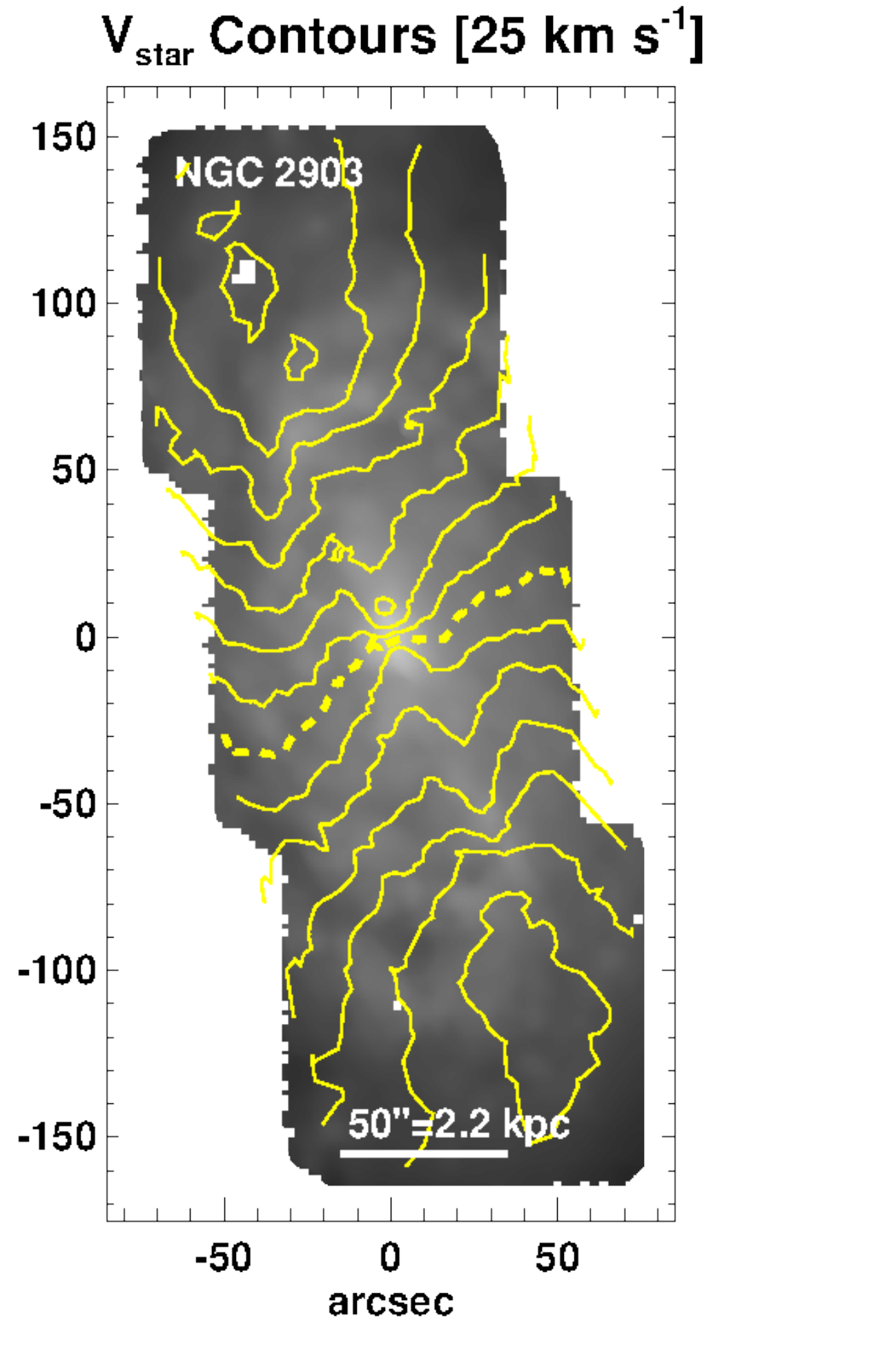} \hspace*{-0.55 cm}
\includegraphics[width=0.25\textwidth]{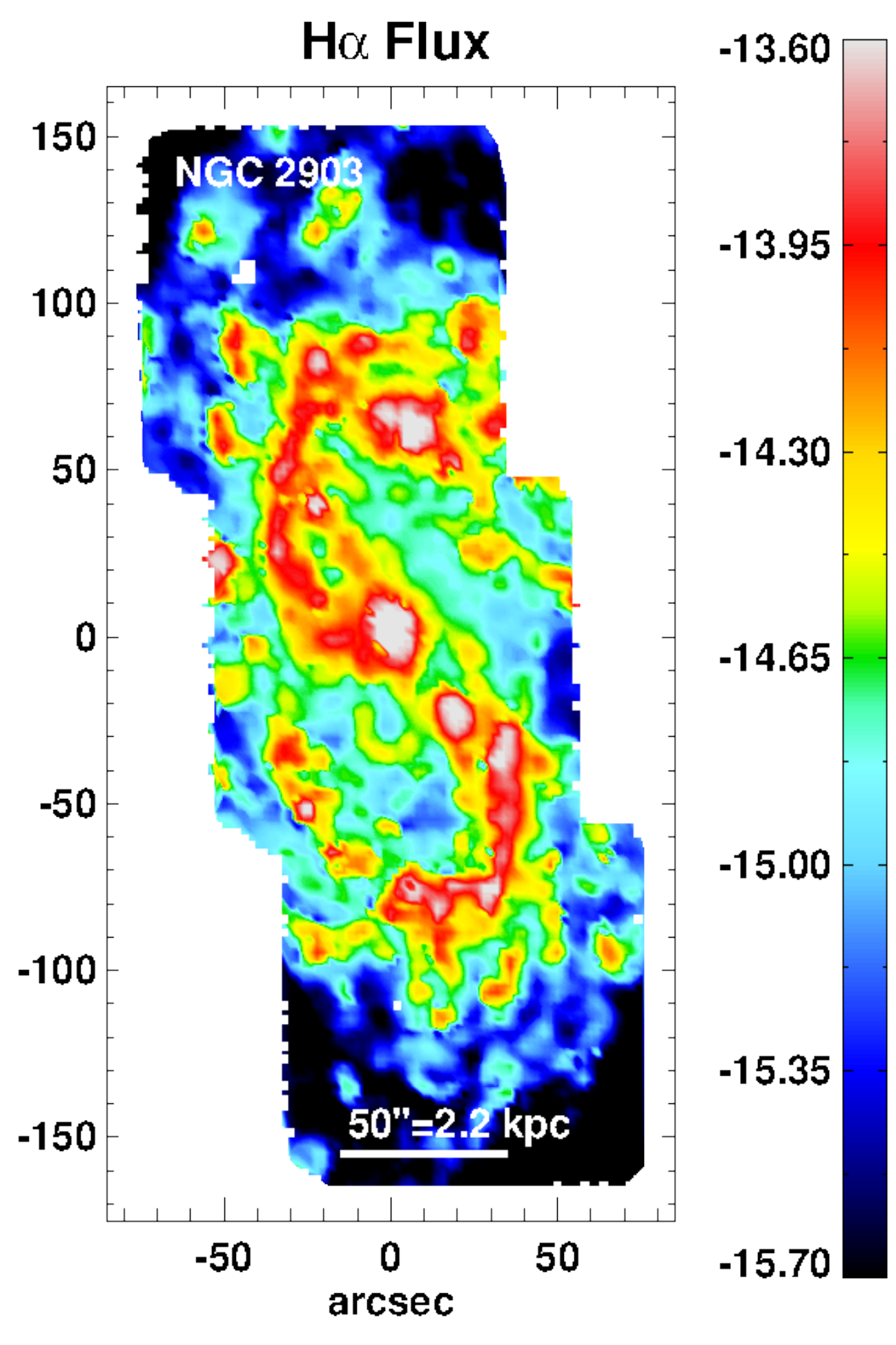}
\includegraphics[width=0.25\textwidth]{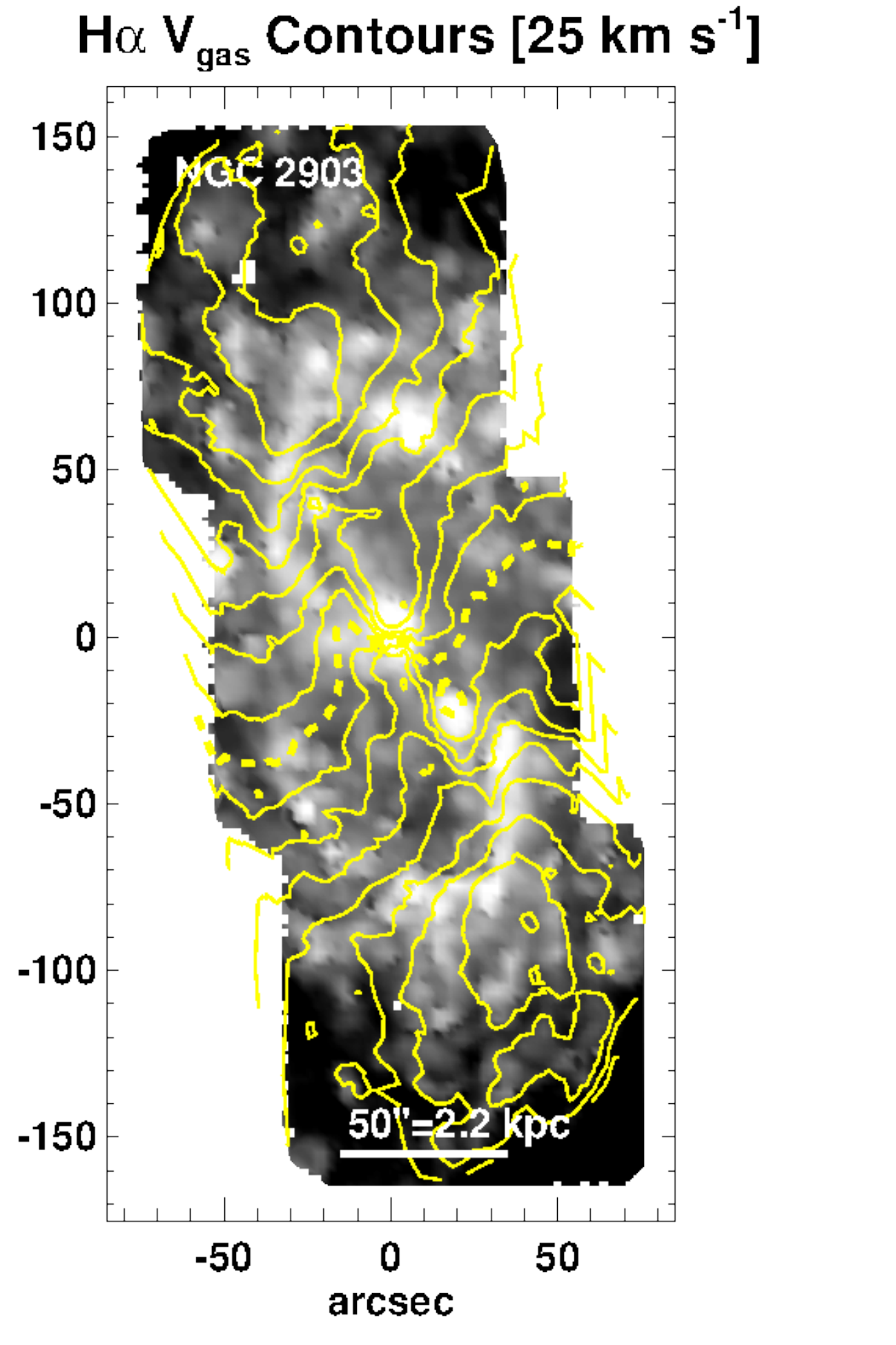} 
\caption{Example IFU-based data
 products for NGC 2903.  
From top to bottom, left to right: the optical stellar continuum [erg s$^{-1}$ cm$^{-2}$ \AA{}$^{-1}$],
stellar velocity field,  extinction corrected H$\alpha$ map  [erg s$^{-1}$ cm$^{-2}$],  and
H$\alpha$ velocity field.
The solid yellow isovelocity contours are in increments of 25 km s$^{-1}$, with the dashed yellow contour showing the systemic velocity where $v_{sys} = 0$ km s$^{-1}$.
For the isovelocity contours, we use Voronoi binning for regions with low $S/N$, with  a minimum of  $S/N$ $\ge$ 3 for H$\alpha$ and a minimum $S/N$ $\ge$ 50 for the stellar continuum.
}
\label{fig:vengamaps1}
\end{figure}

The resulting data cubes store  spatially resolved 2D
information on the gas emission line fluxes, stellar continuum spectra,
stellar and gas velocities, and velocity dispersions from which we
will derive excitation diagrams, \zgas{}, SFRs,  and the ionization parameter $q$.  
All emission line fluxes are corrected for extinction by dust intrinsic to the observed galaxies
using the 
H$\alpha$/H$\beta$ = 2.86
line ratio decrement described in \cite{osterbrock}.
Examples of the data products, such as the stellar flux, 
stellar velocity field,  extinction corrected H$\alpha$ map,  and H$\alpha$ velocity field
are shown for NGC 2903 in Figure~\ref{fig:vengamaps1}.
For the isovelocity contours, we use Voronoi binning for regions with
low  signal-to-noise ($S/N$), with   
a minimum of  $S/N \ge 3$ for H$\alpha$ and a minimum $S/N \ge 50$ for the stellar continuum.

\section{Methodology} \label{sec:method}

For our analysis,  we first identify   in each galaxy 
`star-forming'    regions or spaxels  where the  ionized
 gas  is predominantly photoionized by massive hot young stars, 
and we  exclude contaminating spaxels where the gas is mainly 
excited by the hard radiation field of an AGN or by  shocks 
($\S$~\ref{sec:bpt-method}).
We also exert particular care in dealing  with diffuse ionized gas
($\S$~\ref{sec:dig}). 
We then use only the `star-forming' spaxels for calculating SFRs
($\S$~\ref{sec:sfr}), the ionization parameter $q$ ($\S$~\ref{sec:how-q}), \& \zgas{}
 ($\S$~\ref{sec:how-zgas}). 

\subsection{Identifying Gas in Different Excitation Conditions (SF, 
Seyfert, LINER)} \label{sec:bpt-method}

Gas metallicity (\zgas{}) diagnostics are all based on the assumption
that the  emission  line fluxes we measure stem from gas that is predominantly
being photo-ionized by UV light from massive hot young stars.  
However, if the gas is predominantly ionized by other sources, 
such as AGN   or shocks in the gas,  then  \zgas{} values computed
from the measured emission line ratios will be incorrect
as discussed 
by  \cite{kewley2002}, \cite{kewley2008}, and \cite{yuan2012b}.
In this study, we avoid regions predominantly ionized by sources other than young massive 
stars
by using  high spatial resolution  
IFU  data  to probe regions on scales of a few hundred pc, and by 
using  lines observed at high spectral resolution in each region 
to construct  
standard optical excitation diagnostic diagrams
(e.g.,
\citealt{bpt, veilleux1987,  kewley2001, kewley2002,  kewley2006b,
  rich2010}), which can separate gas in different excitation
conditions.

Common excitation diagnostic diagrams include  plots of
\mbox{\OIII{}$\lambda\lambda$4959,5007/\Hb{}} vs. \mbox{\SII{}$\lambda\lambda$6717,6731/H$\alpha$,}
\mbox{\OIII{}$\lambda\lambda$4959,5007/\Hb{}} vs. \mbox{\NII{}$\lambda$6584/H$\alpha$,}
and  
\mbox{\OIII{}$\lambda\lambda$4959,5007/\Hb{}} vs. \mbox{\OI{}$\lambda$6300/H$\alpha$.}
For this work, we use the \OIII{}/\Hb{} vs. \SII{}/H$\alpha$ diagnostic diagram (hereafter  \SII{} excitation diagnostic diagram) using spaxels where all the lines used have $S/N > 5$ .
In the  \iontwo{S}{II} excitation  diagnostic diagrams seen on the far left panel of Figure \ref{fig:bpt-sliced},  regions to the left of  
the SF threshold curve (middle black curve; \protect\citealt{kewley2001}) 
represent gas predominantly photoionized by massive stars, 
while regions to the right  host gas predominantly  excited by shocks
or AGN. The  latter regions are separated into  Seyfert and LINERs 
via  the blue diagonal line  defined by  \cite{kewley2006b}.
For local galaxies, shocked regions with velocities $< 500$ km s$^{-1}$
tends to have enhanced \iontwo{N}{II}/H$\alpha$ \& \iontwo{S}{II}/H$\alpha$
and low \iontwo{O}{III}/H$\beta$
and these regions are
typically found in the LINER part of the excitation diagnostic diagrams  \citep{farage2010, rich2011},
while shocked gas with velocities $>500$ km s$^{-1}$  and gas 
excited by a hard UV radiation field from an AGN
tends to have high \iontwo{O}{III}/H$\beta$,  \iontwo{N}{II}/H$\alpha$, \& \iontwo{S}{II}/H$\alpha$ and
fall in the Seyfert part of the diagram \citep{allen2008, kewley2013}.

Next we aim to map  the spatial location of  gas in different excitation
phases  in the galaxy. To do this, we first 
make a sliced excitation diagnostic diagram (far left panel in Figure
\ref{fig:bpt-sliced}) by
dividing the excitation diagnostic diagram into six slices defined by five curves, 
based on the method by \cite{davies2014}. 
One of these curves is the SF threshold (middle black curve), while the other four
curves are scaled upward and downward by 0.4 and 0.8 dex along 
the log(\iontwo{O}{III}/H$\beta$) axis.
Spaxels between these curves are colour-coded based on how far they are above or below the SF threshold.
Next, these colour-coded spaxels are over-plotted back onto the galaxy image 
to produce \iontwo{S}{II} excitation maps (centre left panel in Figure
\ref{fig:bpt-sliced}).
The term LINER stands for a low-ionization {\it{nuclear}} emission-line
and it is conventionally assumed low luminosity AGN or central starbursts excite LINER conditions  found near 
the centre of galaxies, while  shocked gas such as in  starburst-driven
outflows can be found outside the nucleus.
We find that in our seven spiral galaxies, LINER 
conditions exist  well outside nuclear regions and are seen out 
to projected distances of several kpc  (centre left panel in Figure \ref{fig:bpt-sliced}). 
As we will show in  $\S$~\ref{sec:dig},   the extra-nuclear gas with LINER
excitation characteristics  often seem to be spatially associated with
diffuse ionized gas.

Two of our sample galaxies, NGC 1068 and NGC 5194 (M 51a) are
classified as having Seyfert 2 nuclei.
In  NGC 1068, which has an AGN-driven outflow \citep{antonucci1985, evans1991, macchetto1994, riffel2014}, the
\iontwo{S}{II} excitation maps show Seyfert type conditions out to
large projected distances
of 7 kpc .
NGC 5194 (M 51a) shows LINER like excitation in its centre (previously
discussed in \citealt{blanc2009}).  
Our data and a search through the literature shows no discernible evidence for AGN activity in
any of the other galaxies.


\begin{landscape}

\begin{figure}
\includegraphics[width=0.25\textwidth]{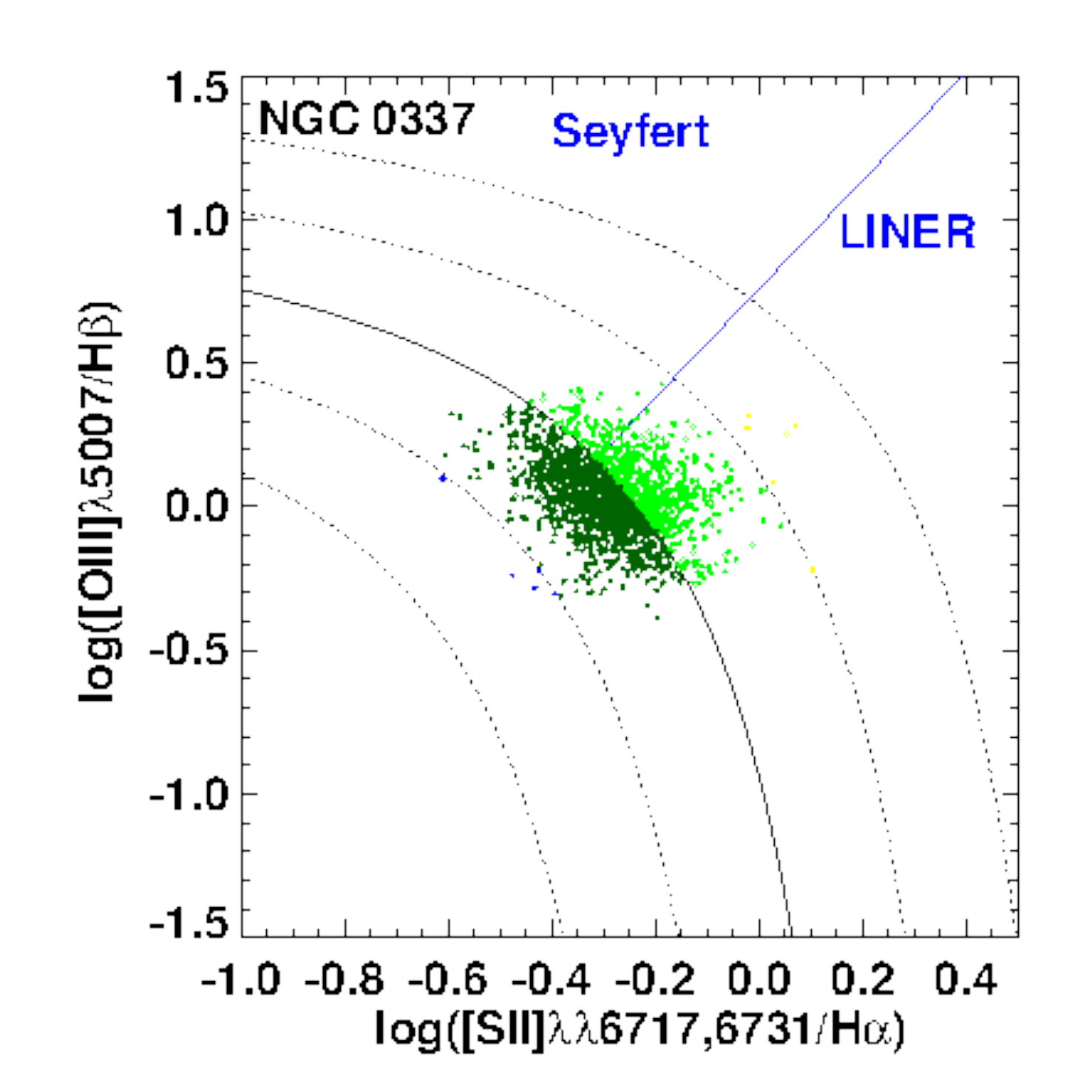}
\includegraphics[width=0.285\textwidth]{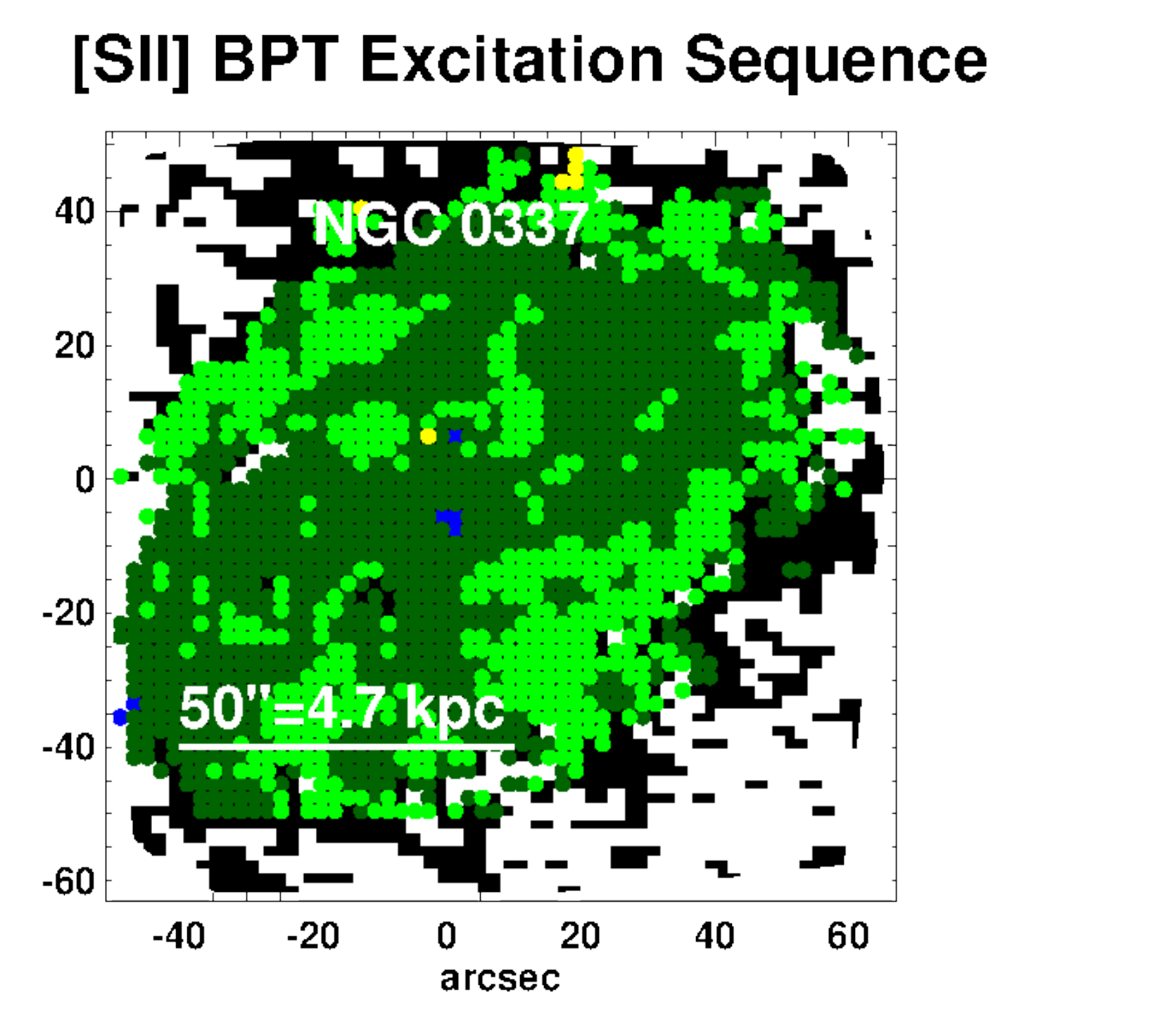}
\hspace{-0.5 cm}
\includegraphics[width=0.285\textwidth]{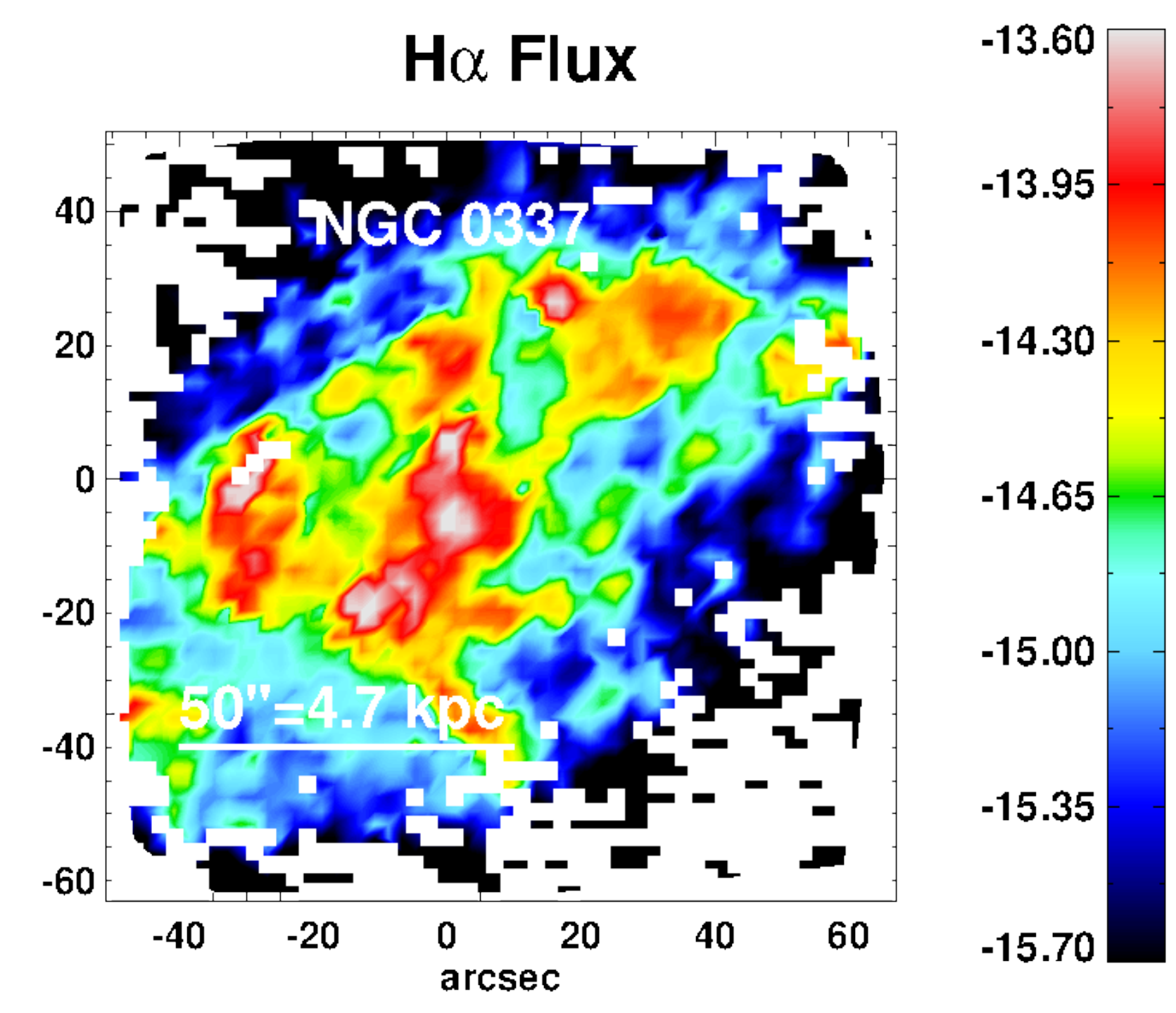} 
\includegraphics[width=0.285\textwidth]{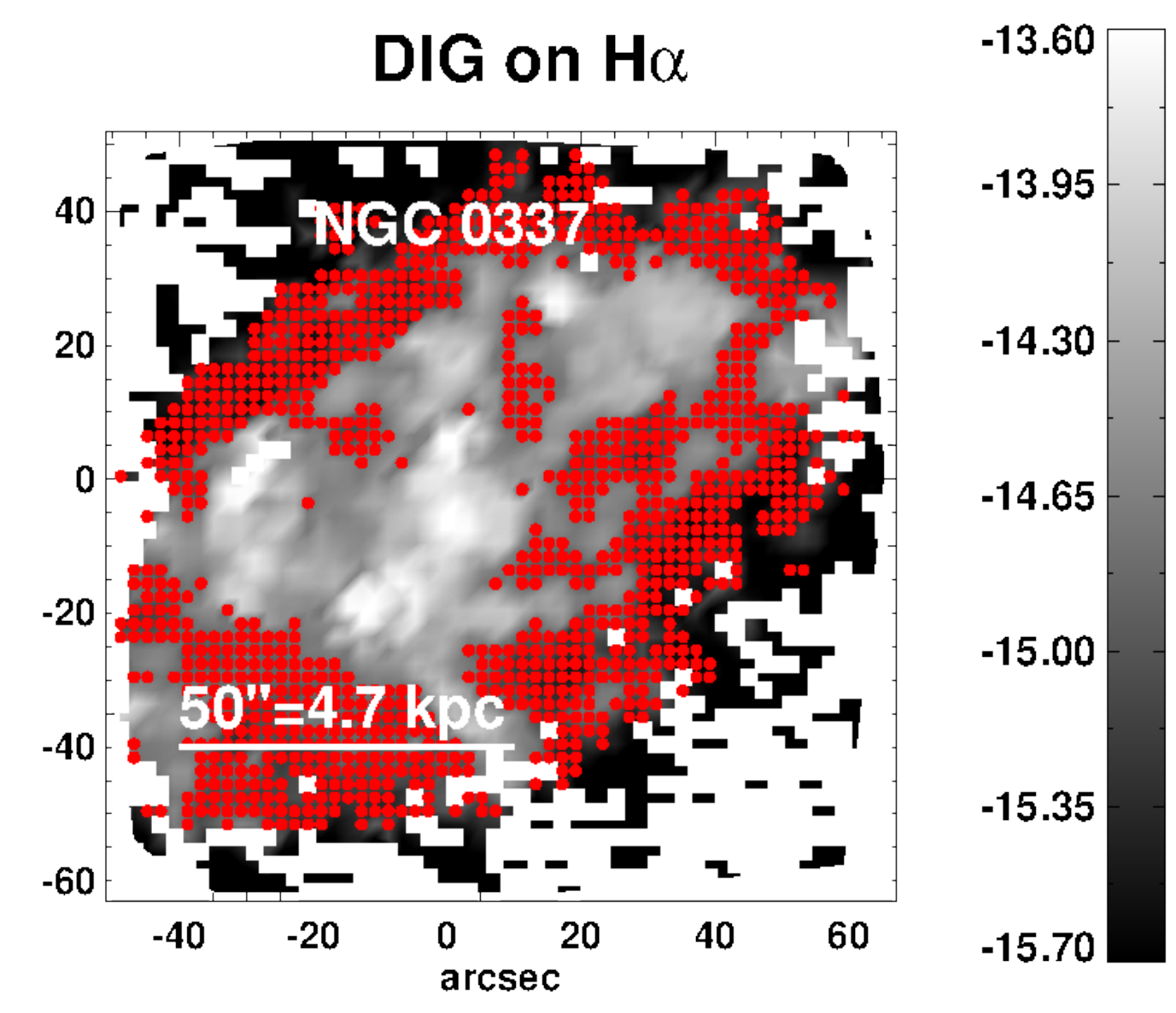}
\includegraphics[width=0.25\textwidth]{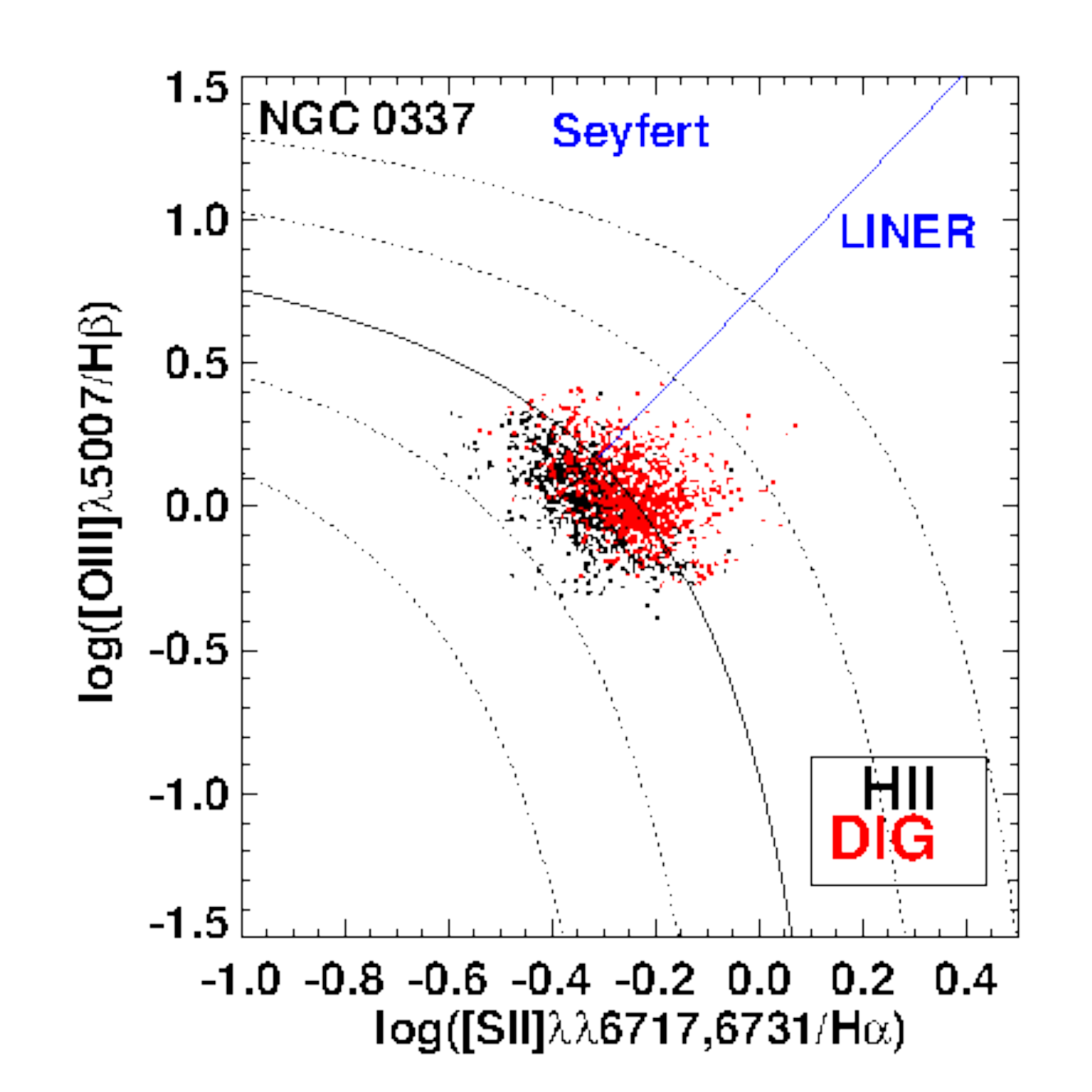}
\caption{
For NGC 0337 we show the following;
{\bf Far Left:}
The {\it sliced}  \iontwo{S}{II} excitation diagnostic diagram where the regions are plotted in different colours, according to their distance 
above or below the SF threshold curve (middle black curve) defined by \protect\cite{kewley2001} to separate regions dominated by massive SF from those dominated by shocks or AGN-excited gas.  The diagram is divided into six slices defined by five curves based on the method by \protect\cite{davies2014}.
One of these curves is the SF threshold (middle black curve), while the other four
curves are scaled upward and downward by 0.4 and 0.8 dex along the log(\iontwo{O}{III}/H$\beta$) axis (dotted curves).
For regions above the SF threshold, the diagonal blue line separates the Seyfert \& LINER regions based on prescriptions in \protect\cite{kewley2006b}.
{\bf Centre Left:}
The \iontwo{S}{II} excitation map shows the spatial distribution of the colour-coded
regions on the galaxy.  Note  that many of the green LINER-dominated regions lie in the outer parts of the galaxy.  This is related to
diffuse ionized gas, as we show in 
the centre right and far right panels.
{\bf Centre:}
For comparison, we show distribution of H$\alpha$ flux which traces the current SF.
{\bf Centre Right:} The diffuse ionized gas (DIG) dominated regions are over-plotted on the 
H$\alpha$ map of the galaxy: they lie mostly in the outer parts of the galaxy, 
away from the spiral arms and regions of high SFR. 
Note that these DIG-dominated regions account for most of the puzzling 
LINER type regions present in the outer parts of the galaxies, as seen above.
{\bf Far Right:}
The \iontwo{S}{II} excitation diagnostic diagram is shown with regions dominated by 100\% diffuse ionized gas (DIG) 
colour-coded as red. These regions tend to fall on the upper right diagram due to their
high \iontwo{S}{II}/H$\alpha$ ratio. 
}
\label{fig:bpt-sliced}
\end{figure}

\addtocounter{figure}{-1}

\begin{figure}
\includegraphics[width=0.25\textwidth]{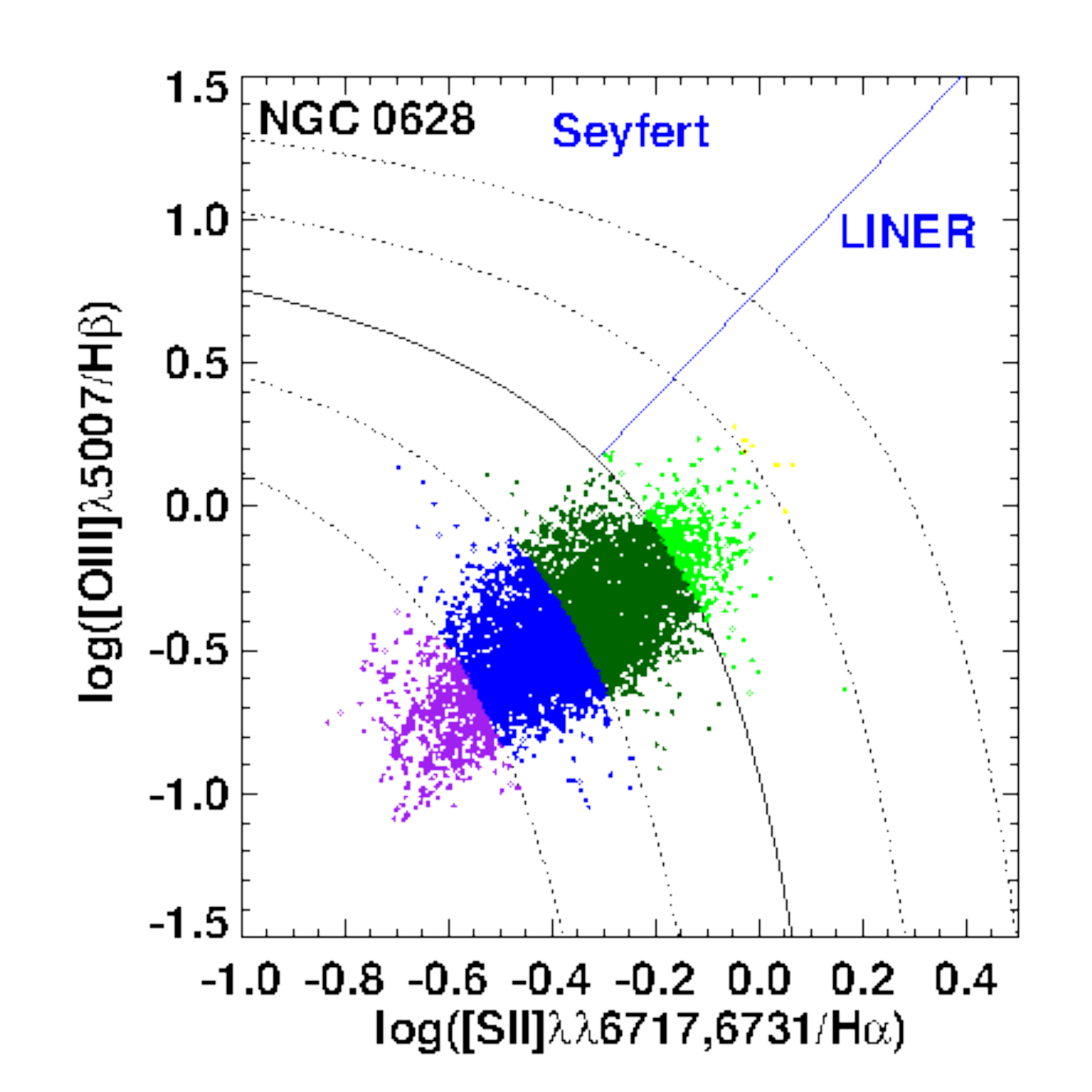}
\includegraphics[width=0.285\textwidth]{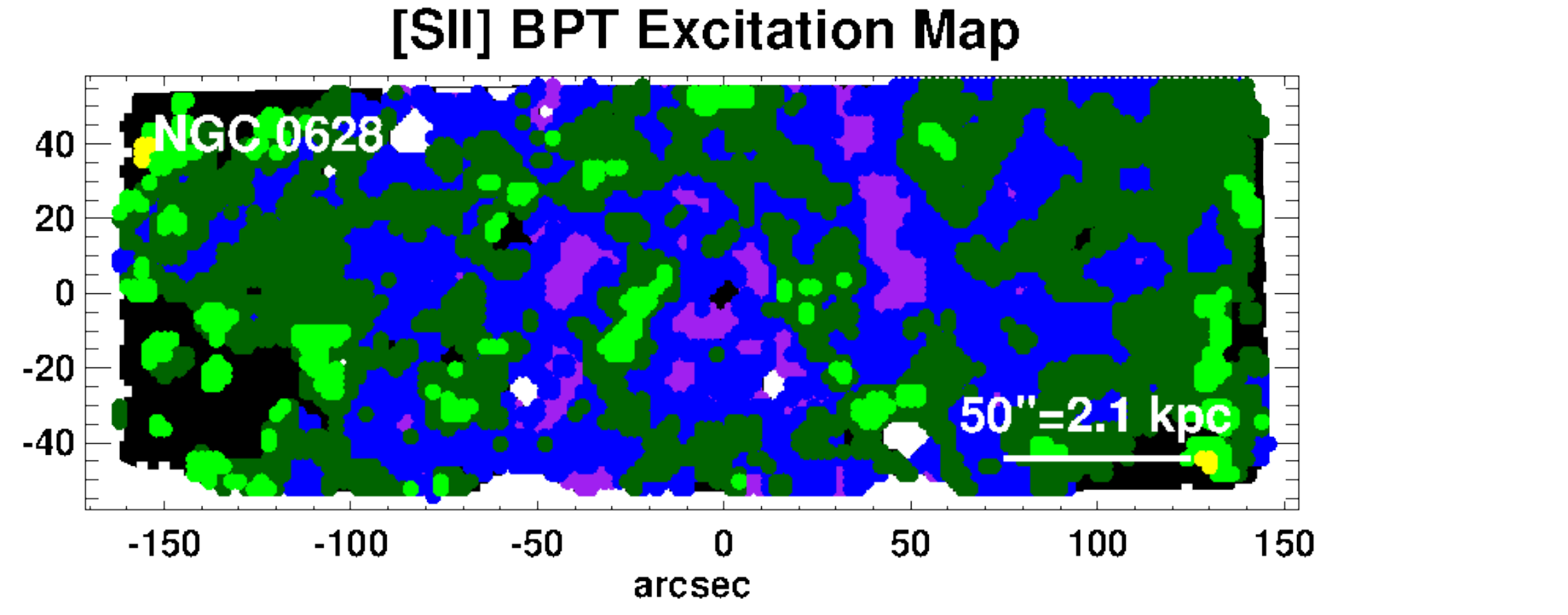} 
\hspace{-0.5 cm}
\includegraphics[width=0.285\textwidth]{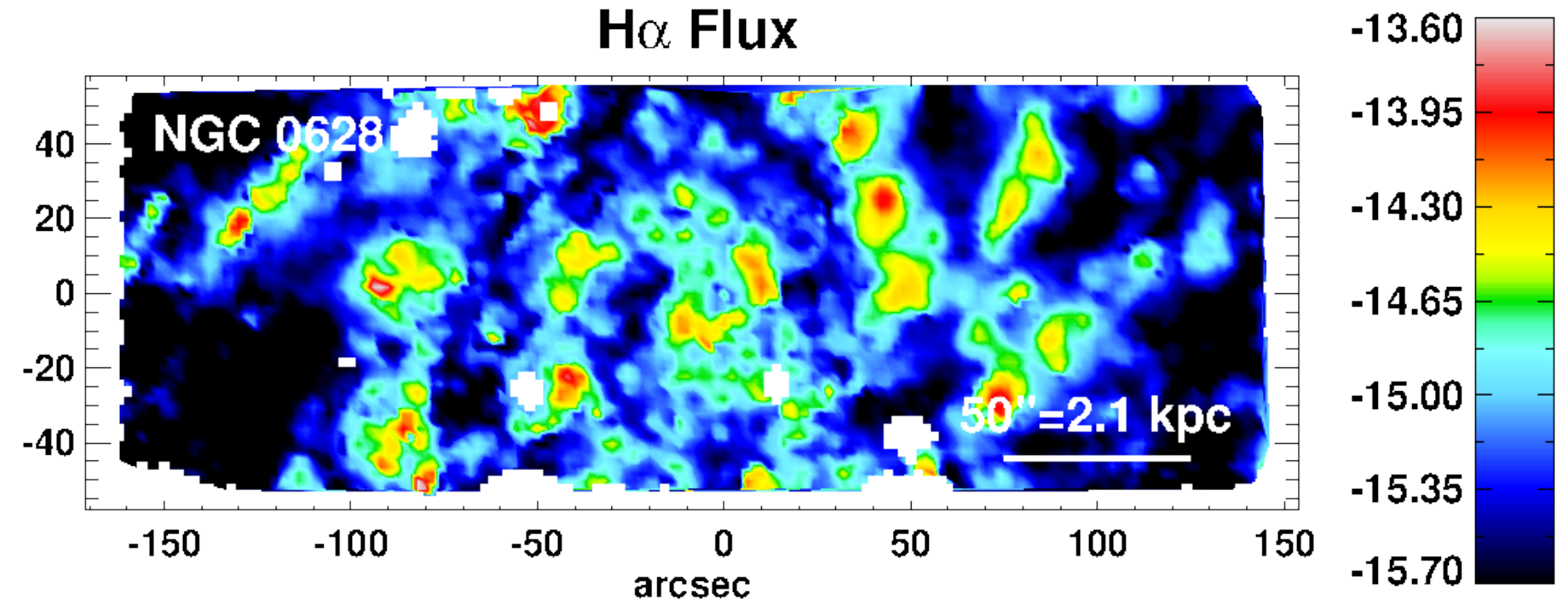} 
\includegraphics[width=0.285\textwidth]{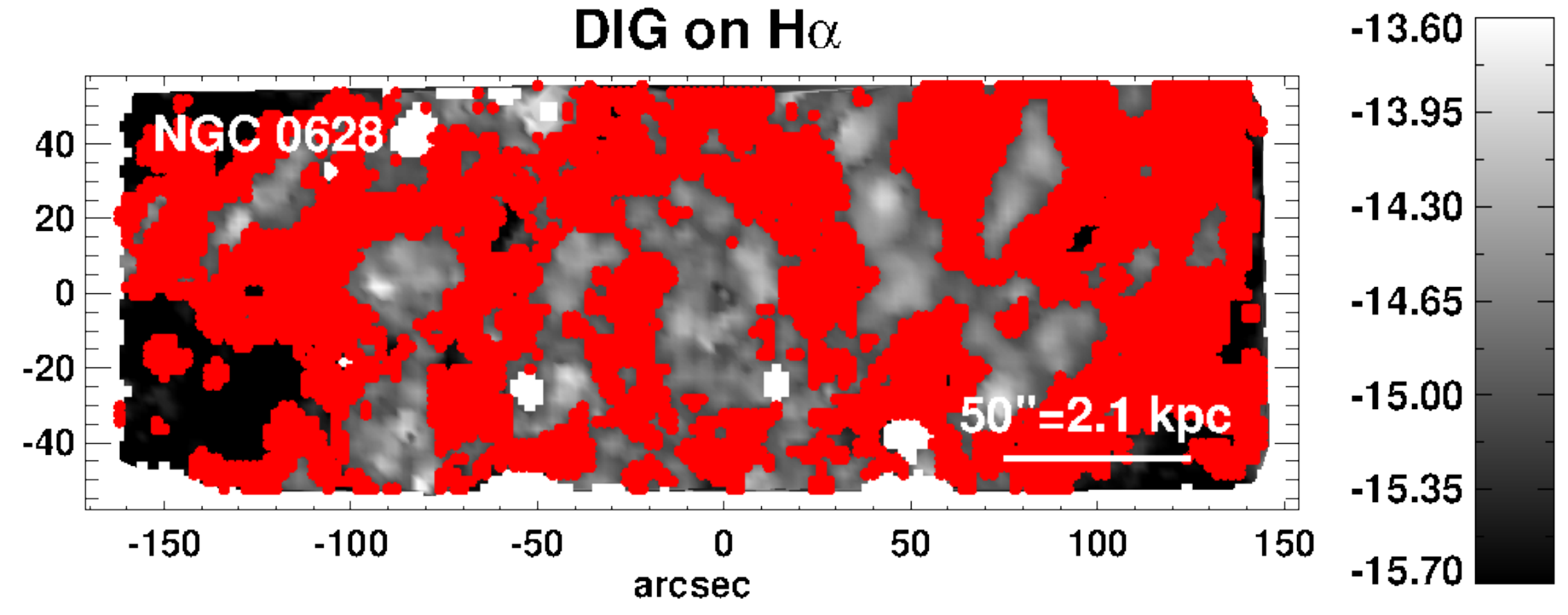}
\includegraphics[width=0.25\textwidth]{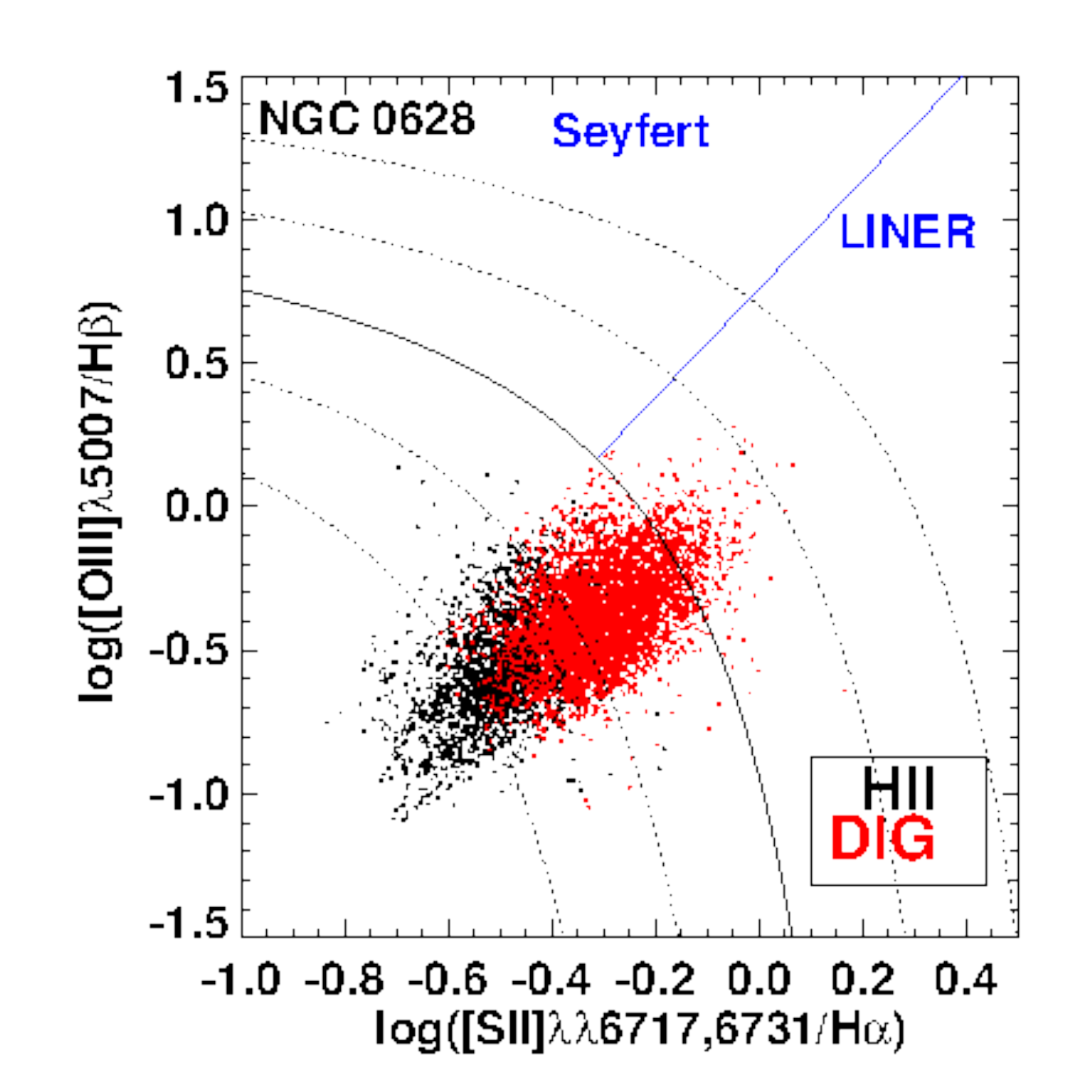}
\caption{Continued:  For NGC 0628}
\end{figure}

\end{landscape}

\addtocounter{figure}{-1}
\clearpage

\begin{landscape}

\begin{figure}
\includegraphics[width=0.25\textwidth]{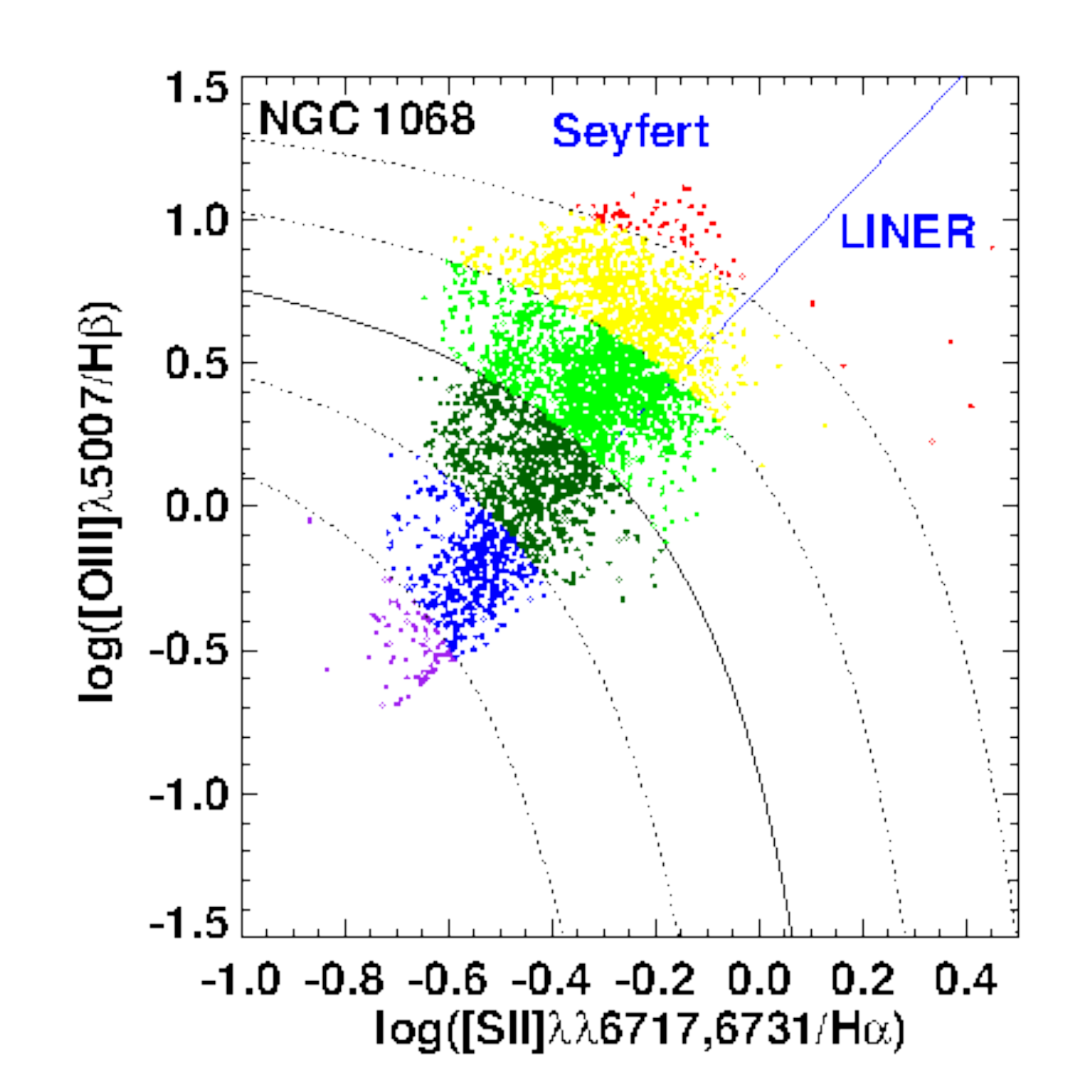}
\includegraphics[width=0.23\textwidth]{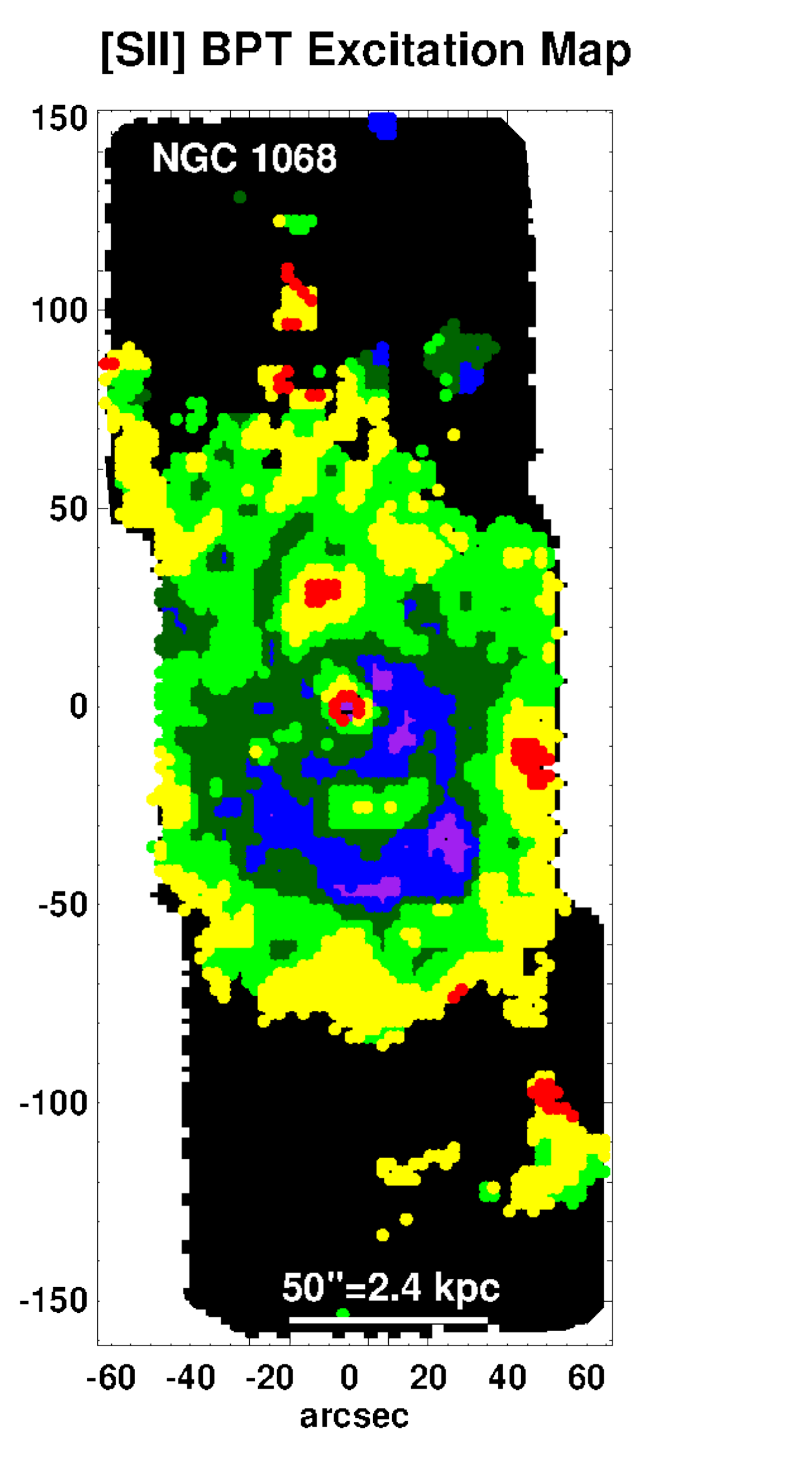} 
\hspace{-0.5 cm}
\includegraphics[width=0.23\textwidth]{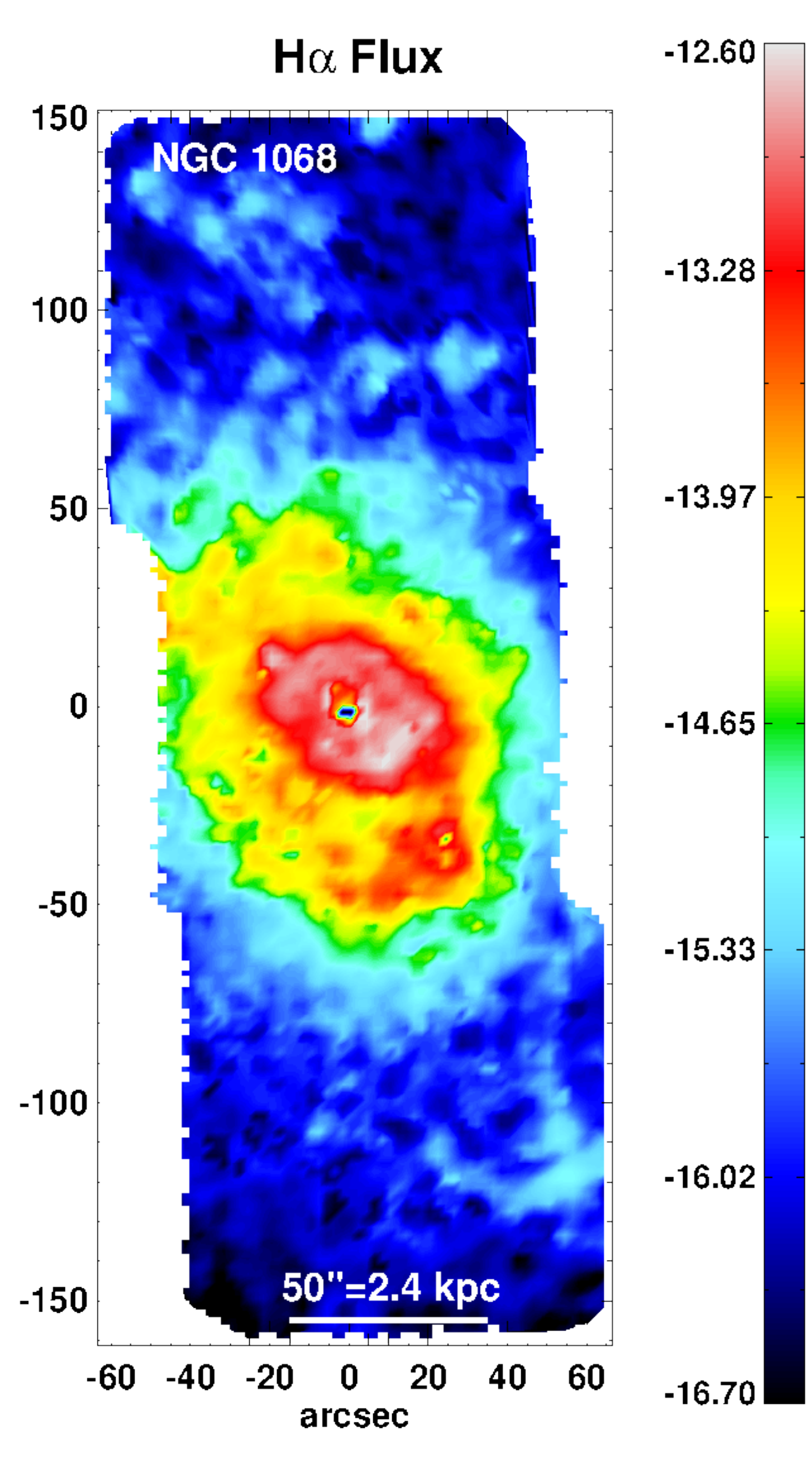} 
\includegraphics[width=0.23\textwidth]{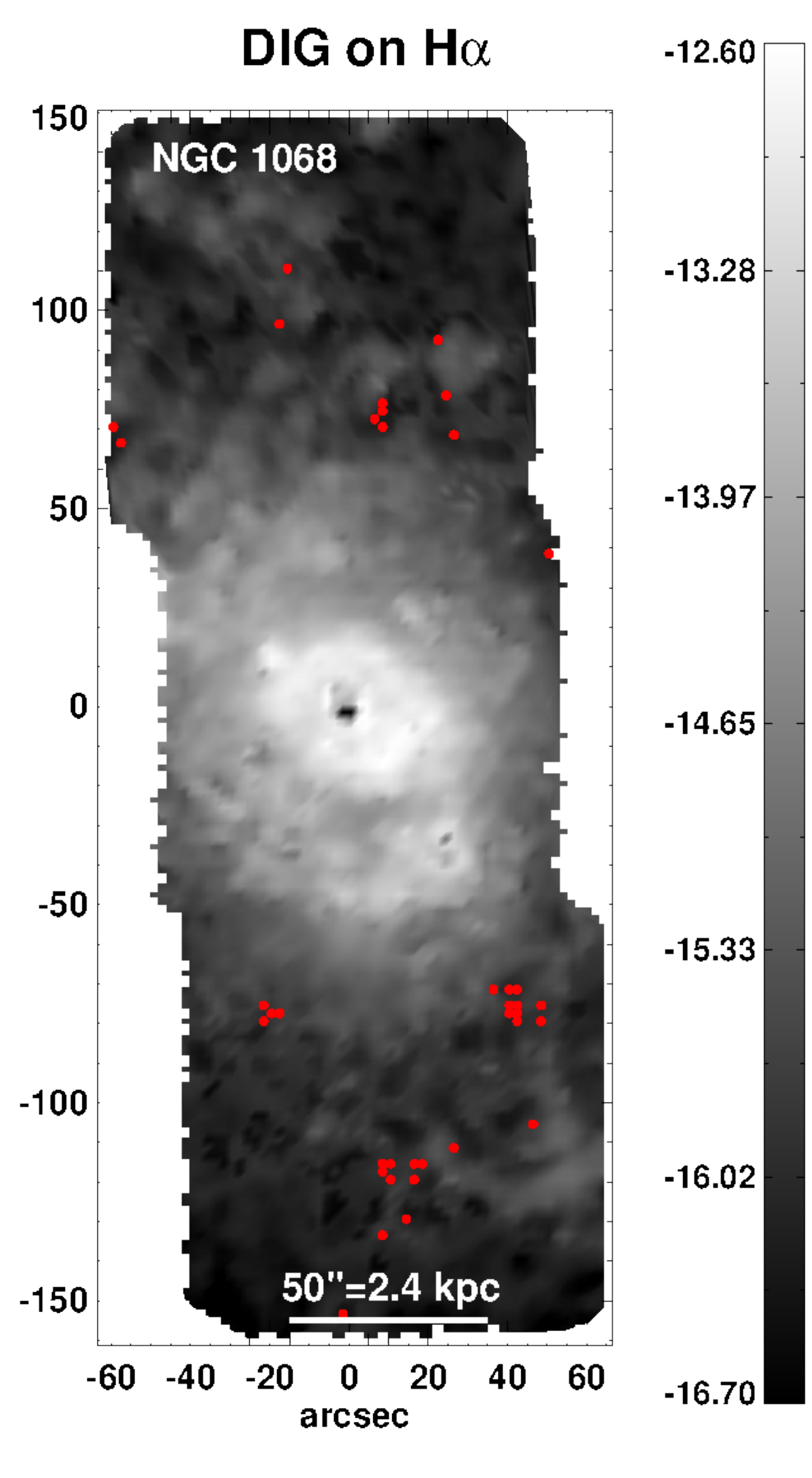}
\includegraphics[width=0.25\textwidth]{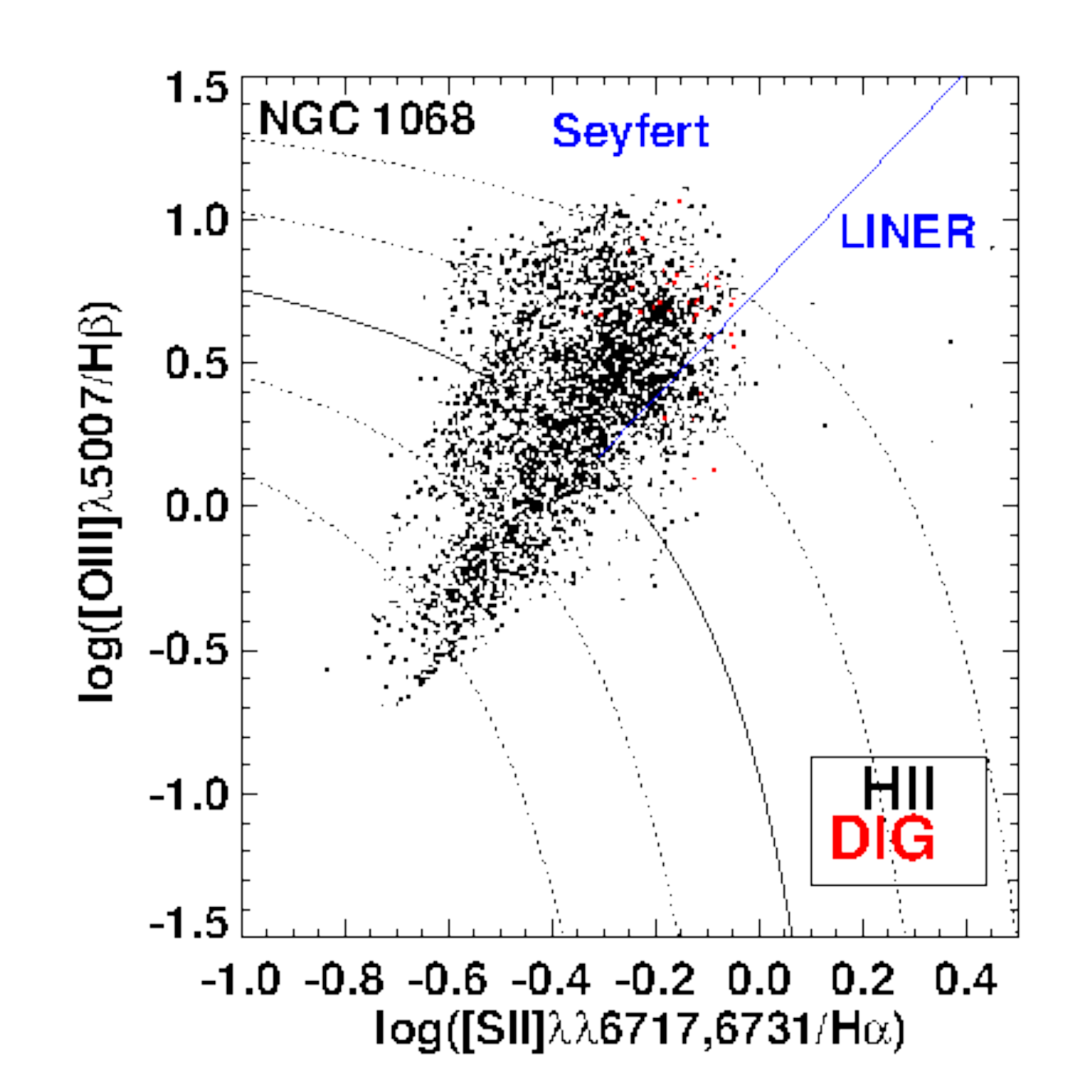}
\caption{Continued:  NGC 1068}
\end{figure}

\addtocounter{figure}{-1}

\begin{figure}
\includegraphics[width=0.25\textwidth]{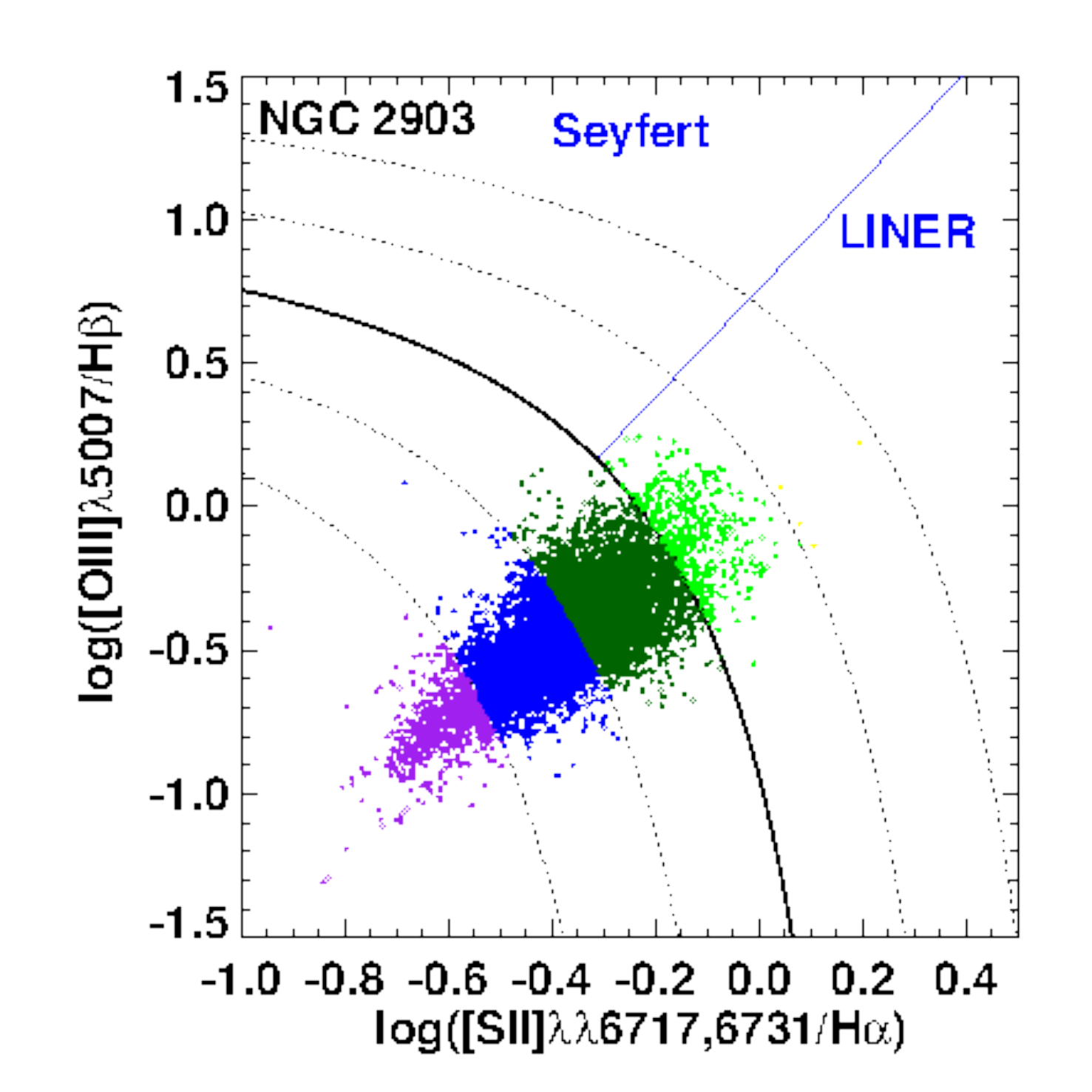}
\includegraphics[width=0.285\textwidth]{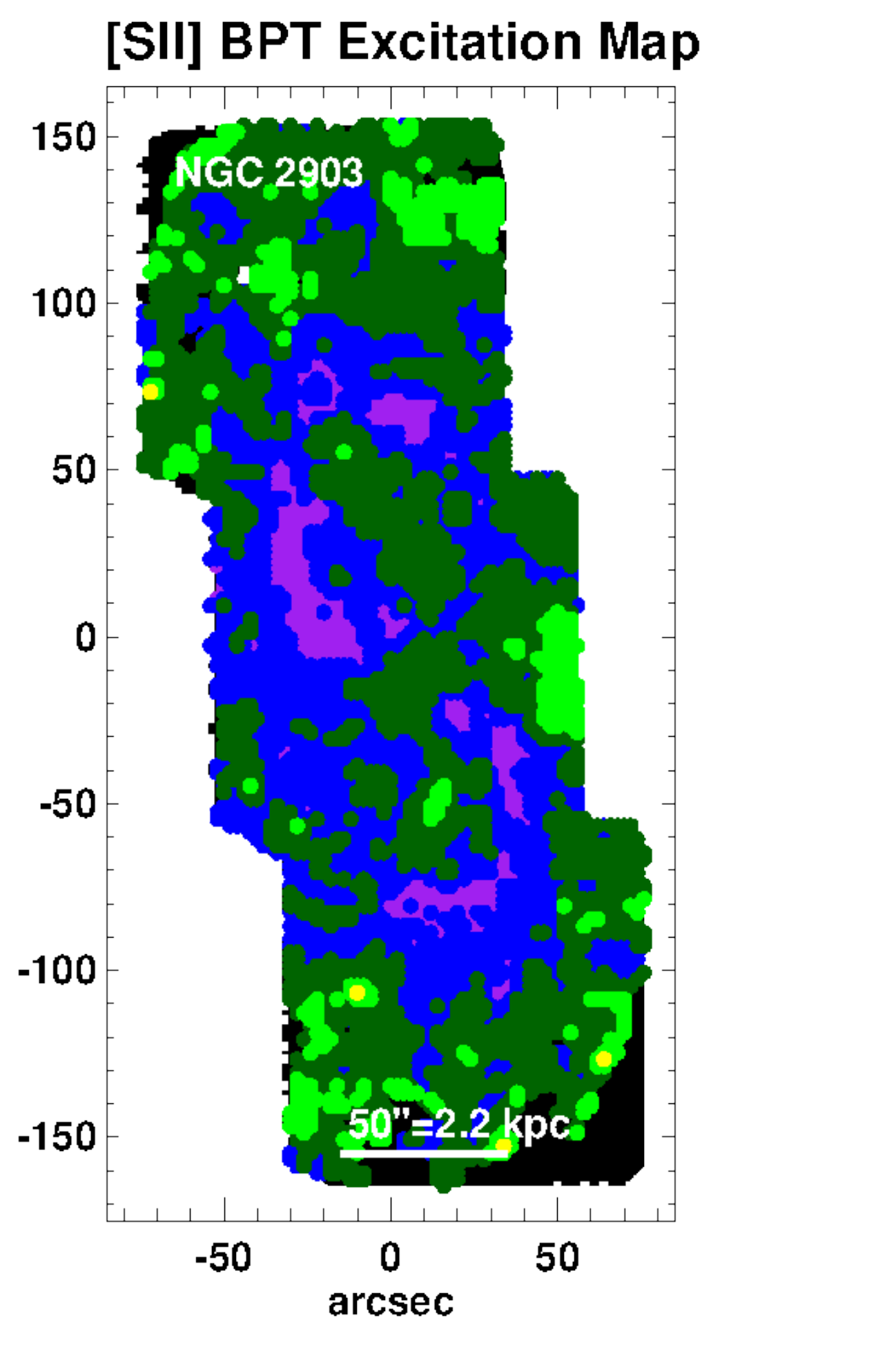} 
\hspace{-0.5 cm}
\includegraphics[width=0.285\textwidth]{figures/spring2013/A3_map_color_ha_flux.ngc2903.pdf} 
\includegraphics[width=0.285\textwidth]{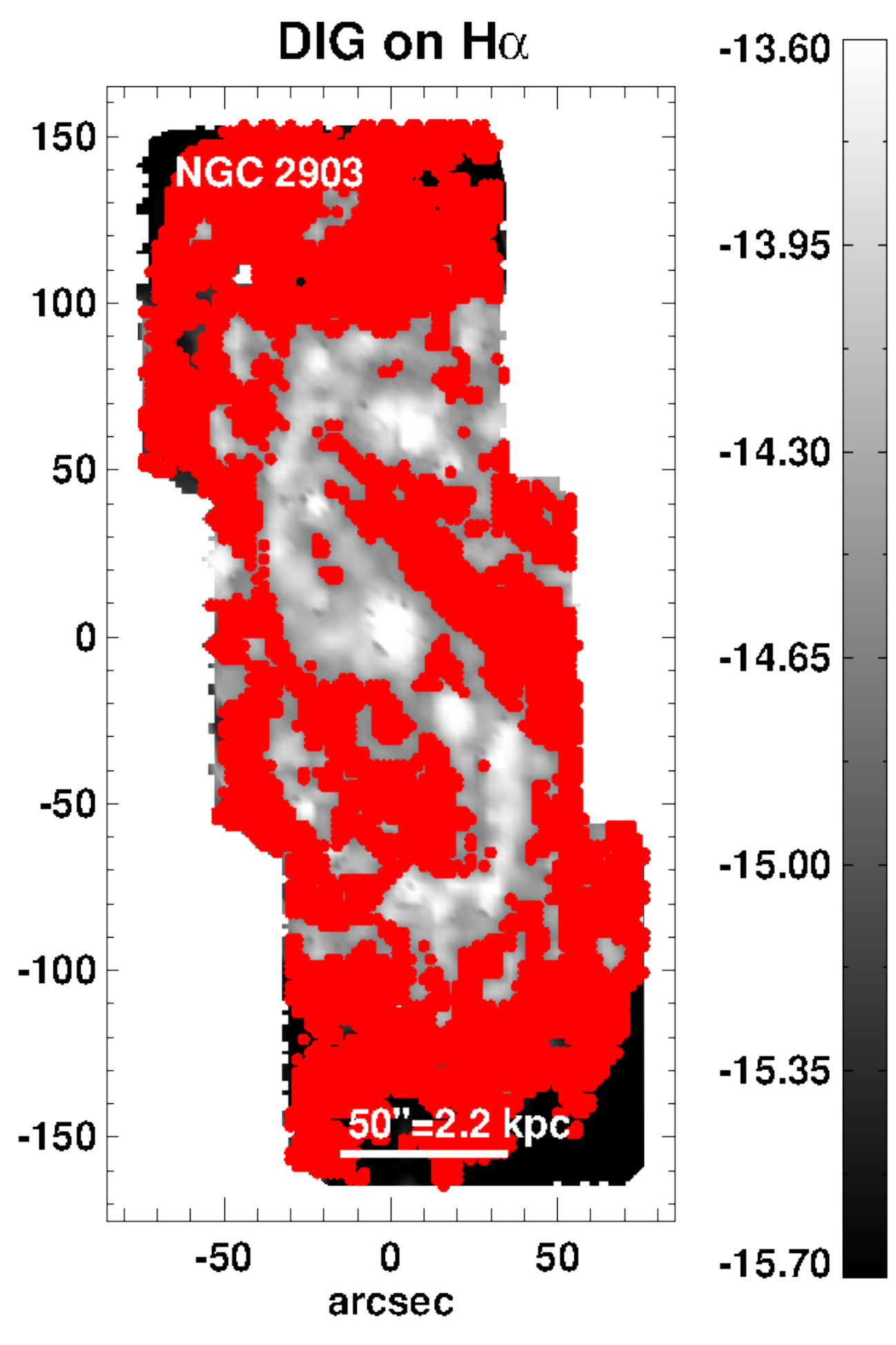}
\includegraphics[width=0.25\textwidth]{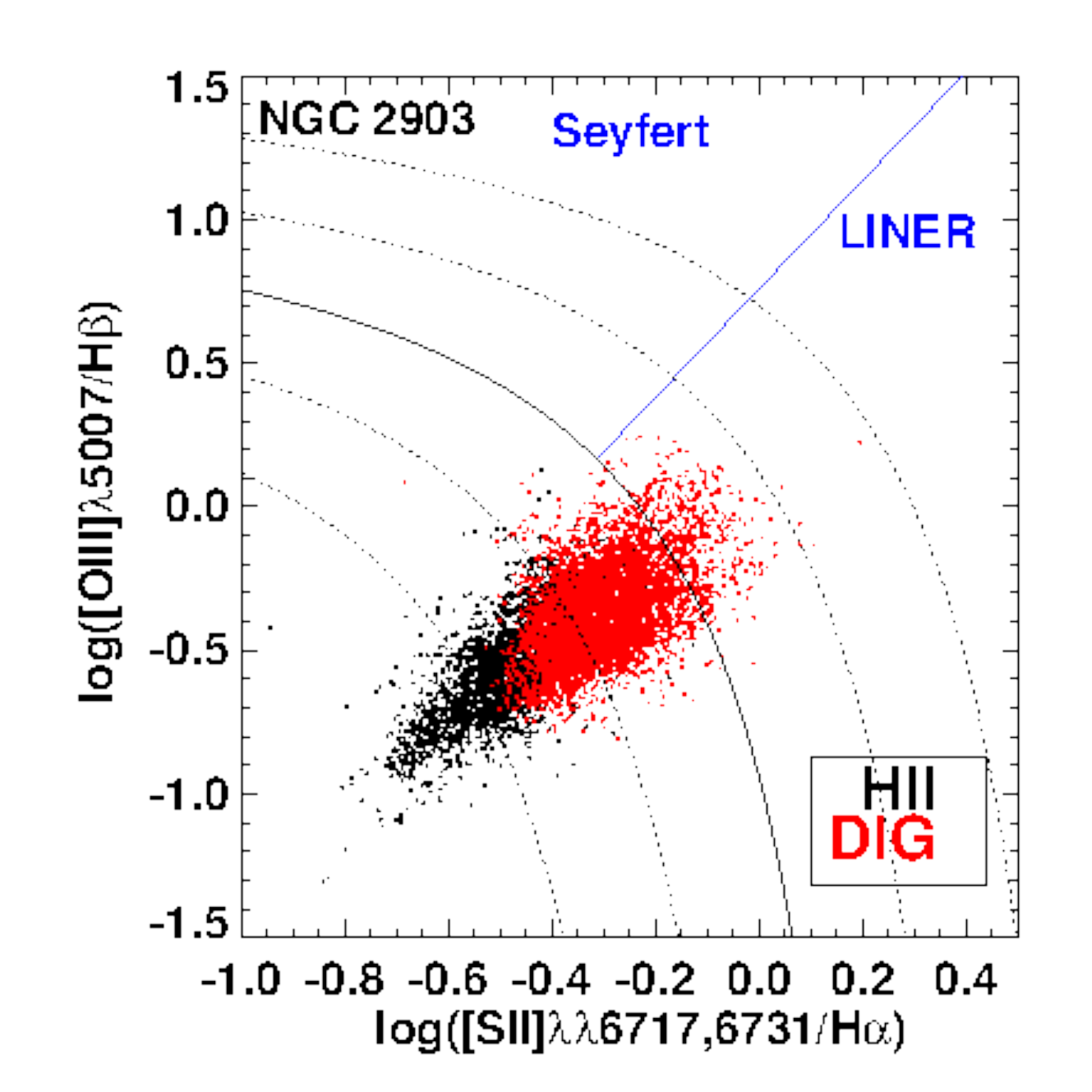}
\caption{Continued: For NGC 2903}
\end{figure}

\addtocounter{figure}{-1}
\clearpage

\end{landscape}

\begin{landscape}

\begin{figure}
\includegraphics[width=0.25\textwidth]{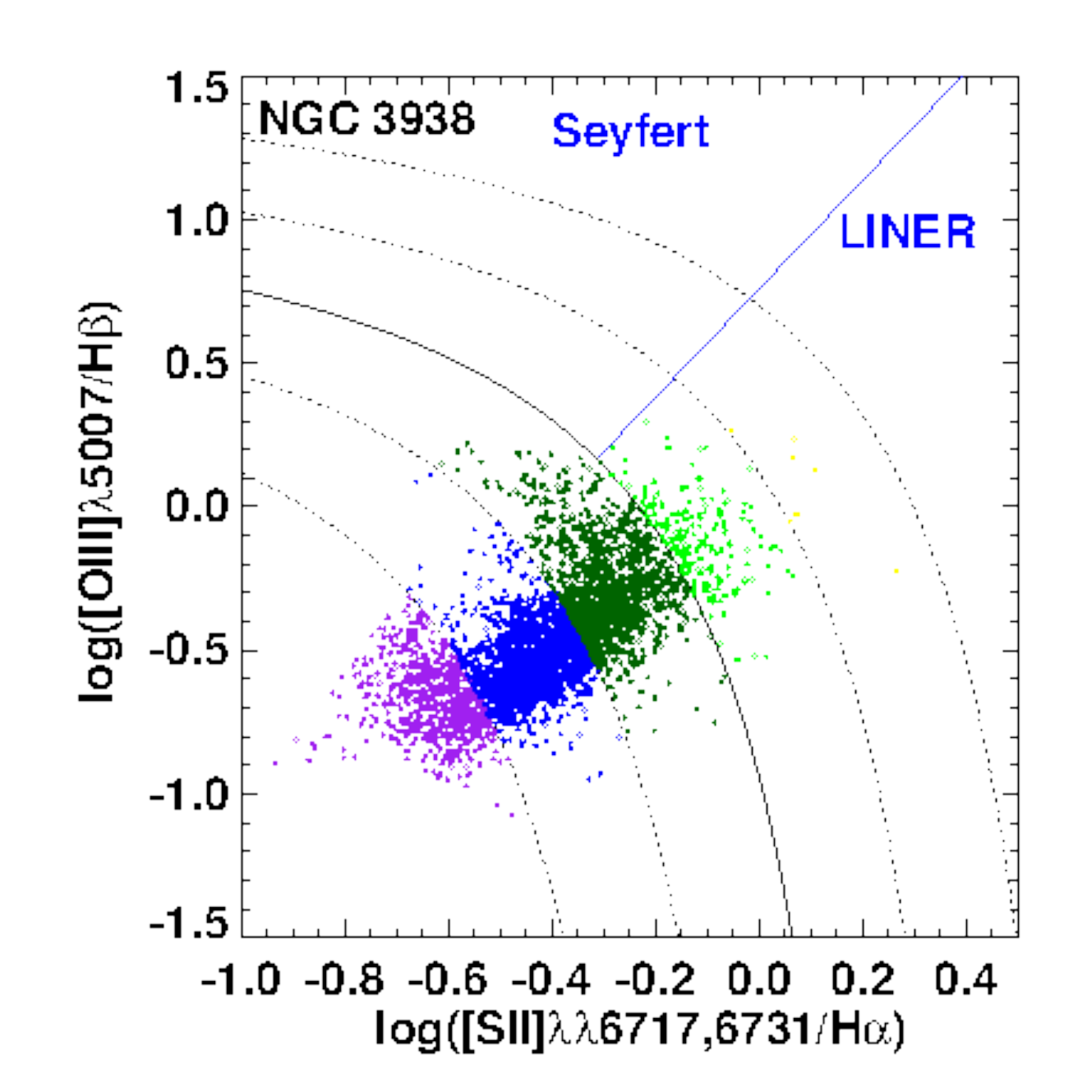}
\includegraphics[width=0.285\textwidth]{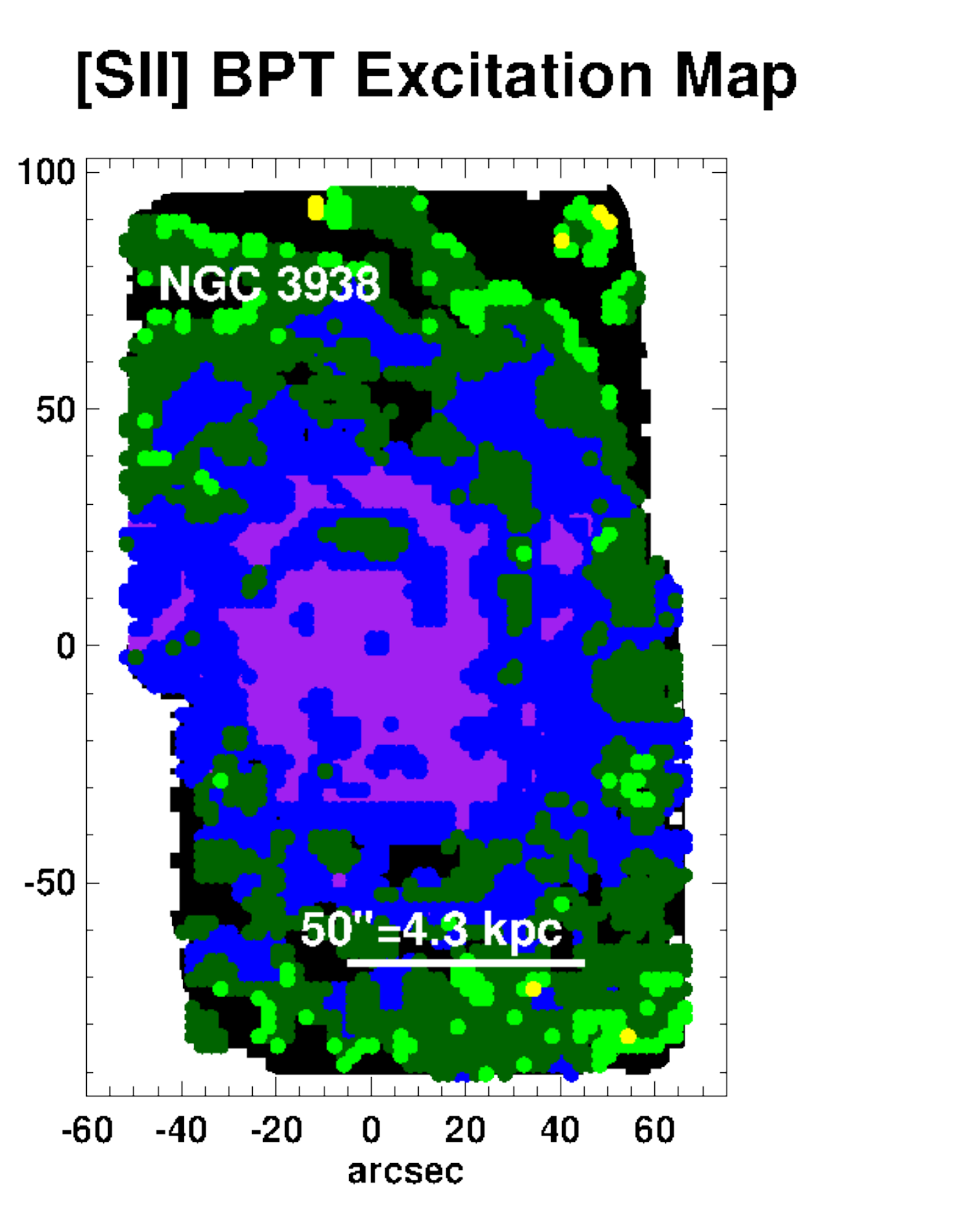} 
\hspace{-0.5 cm}
\includegraphics[width=0.285\textwidth]{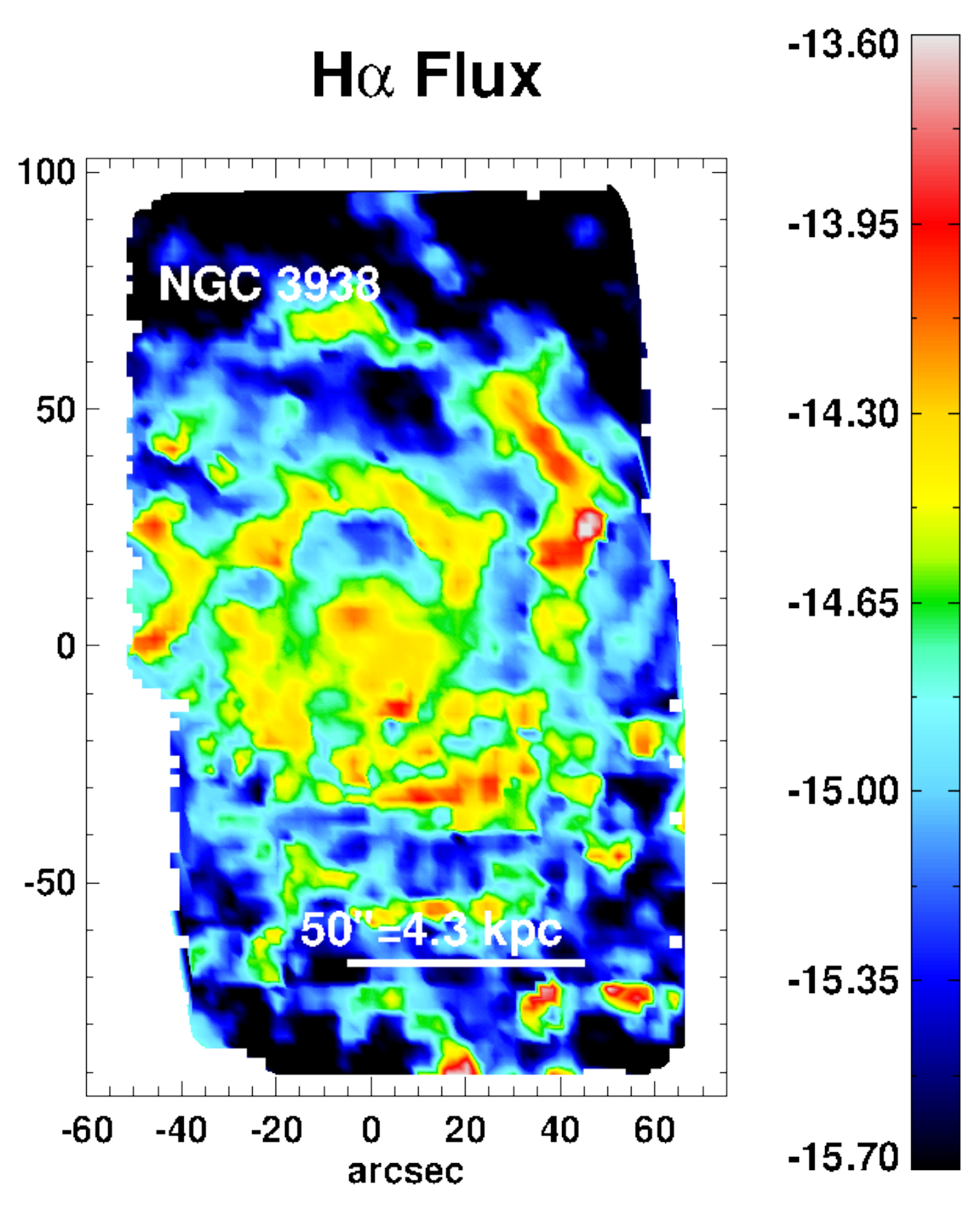} 
\includegraphics[width=0.285\textwidth]{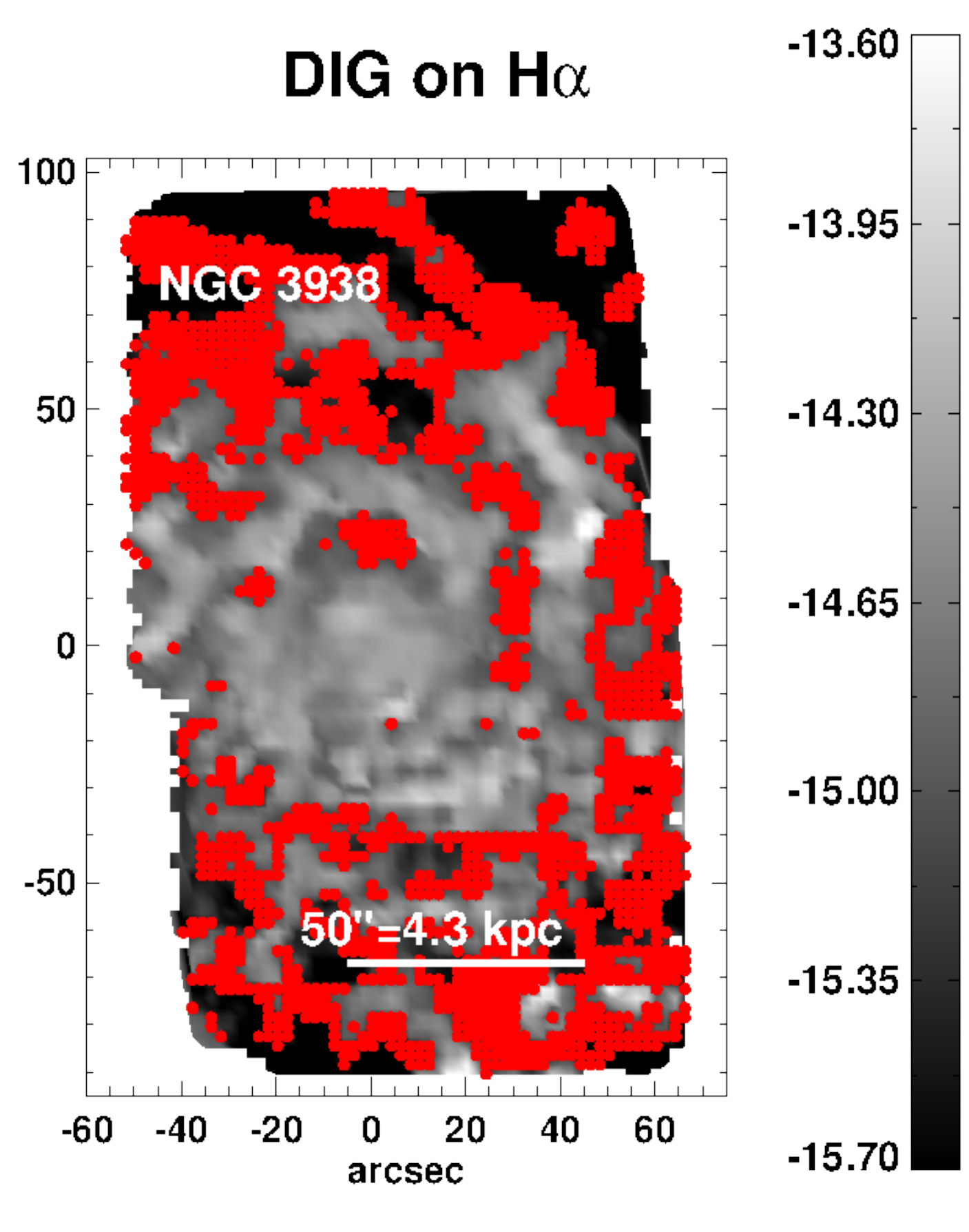}
\includegraphics[width=0.25\textwidth]{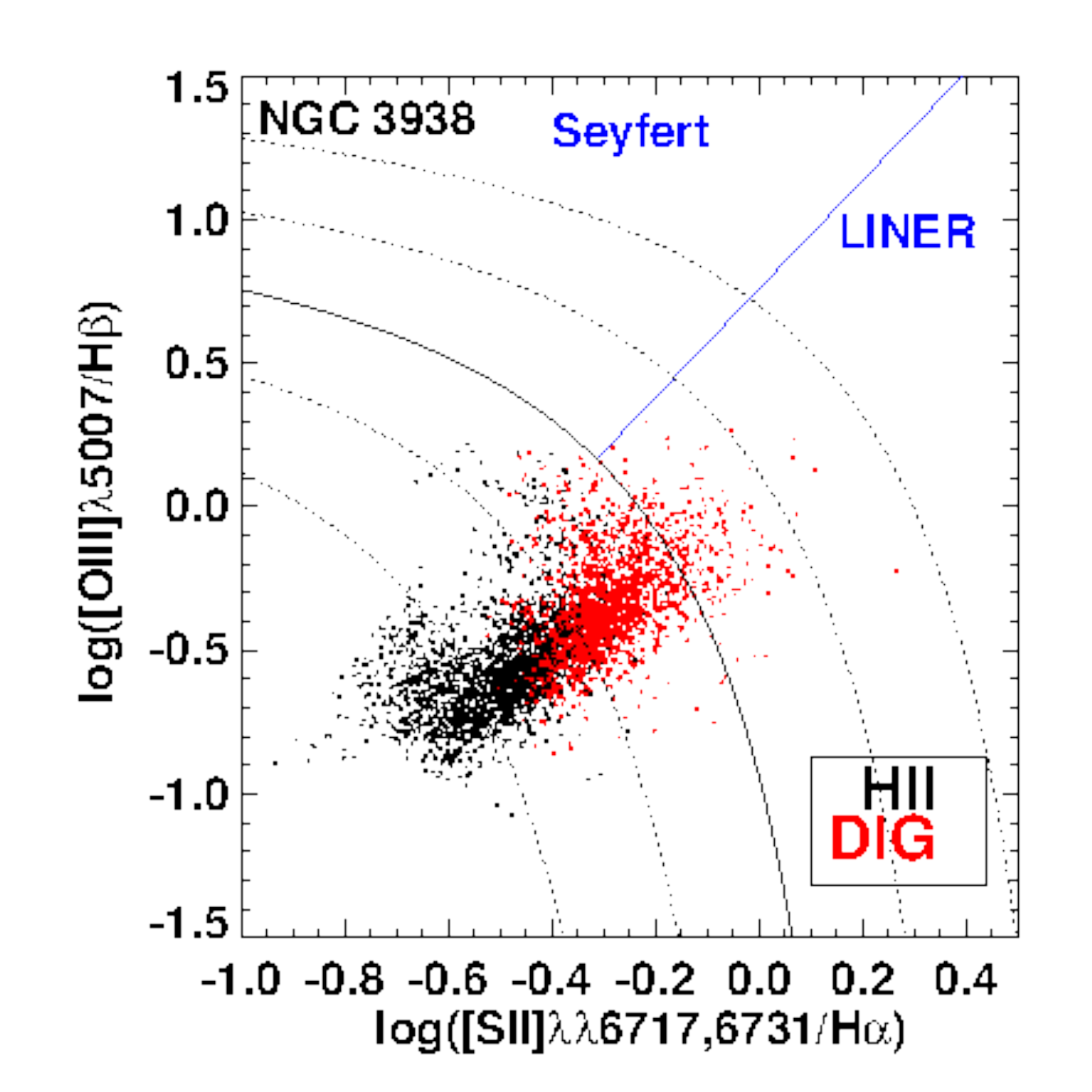}
\caption{Continued:  For NGC 3938}
\end{figure}

\addtocounter{figure}{-1}

\begin{figure}
\includegraphics[width=0.25\textwidth]{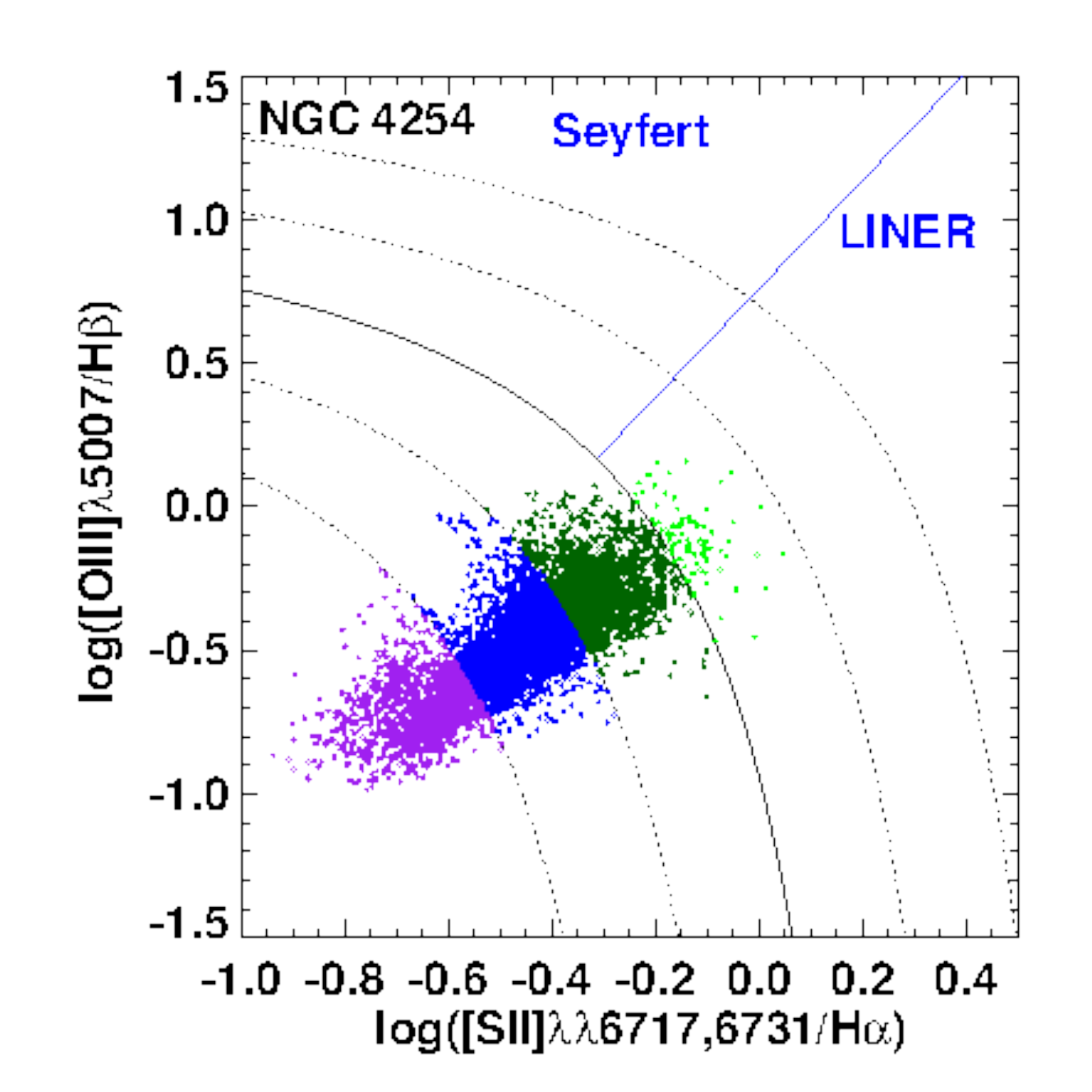}
\includegraphics[width=0.285\textwidth]{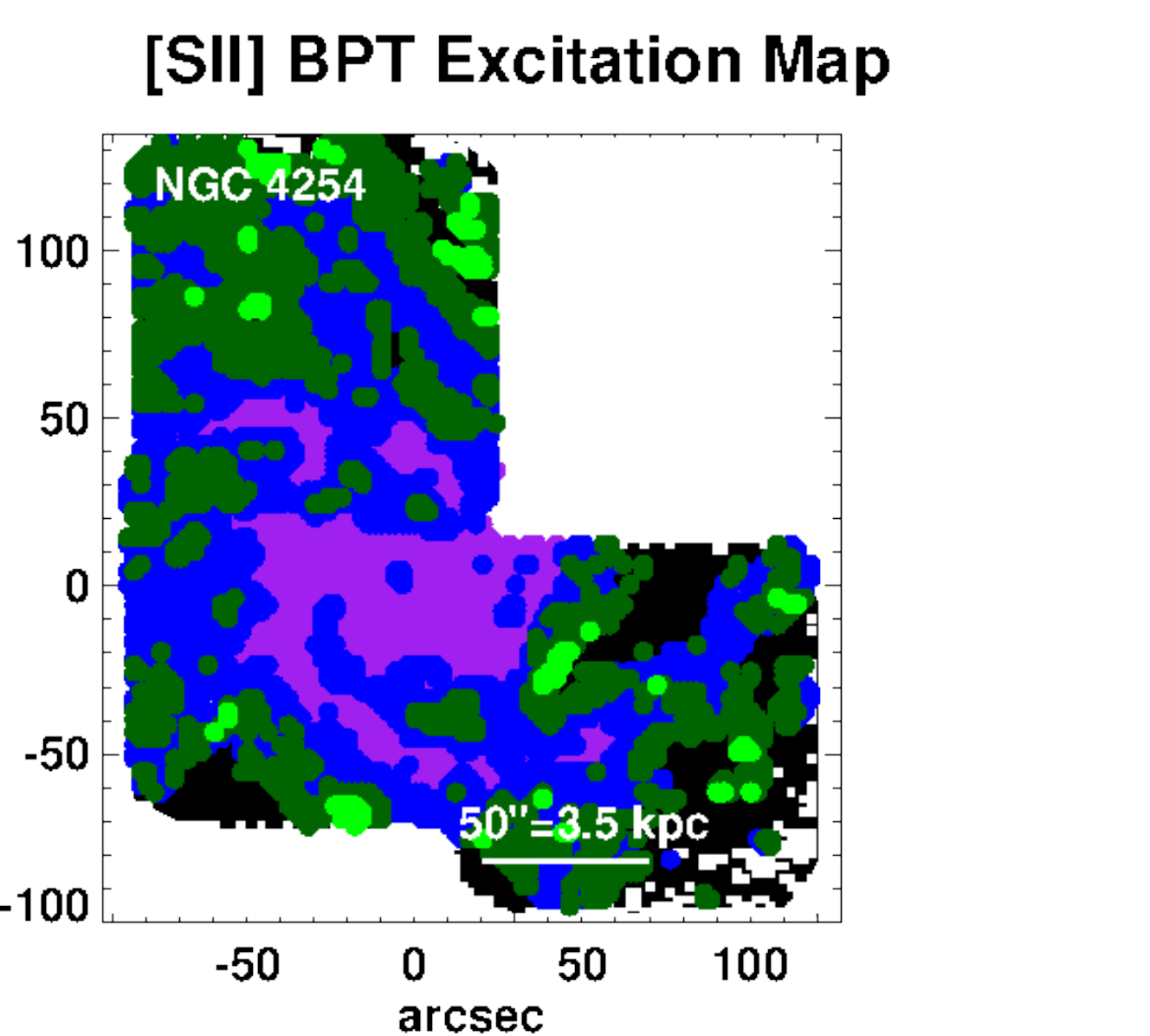} 
\hspace{-0.5 cm}
\includegraphics[width=0.285\textwidth]{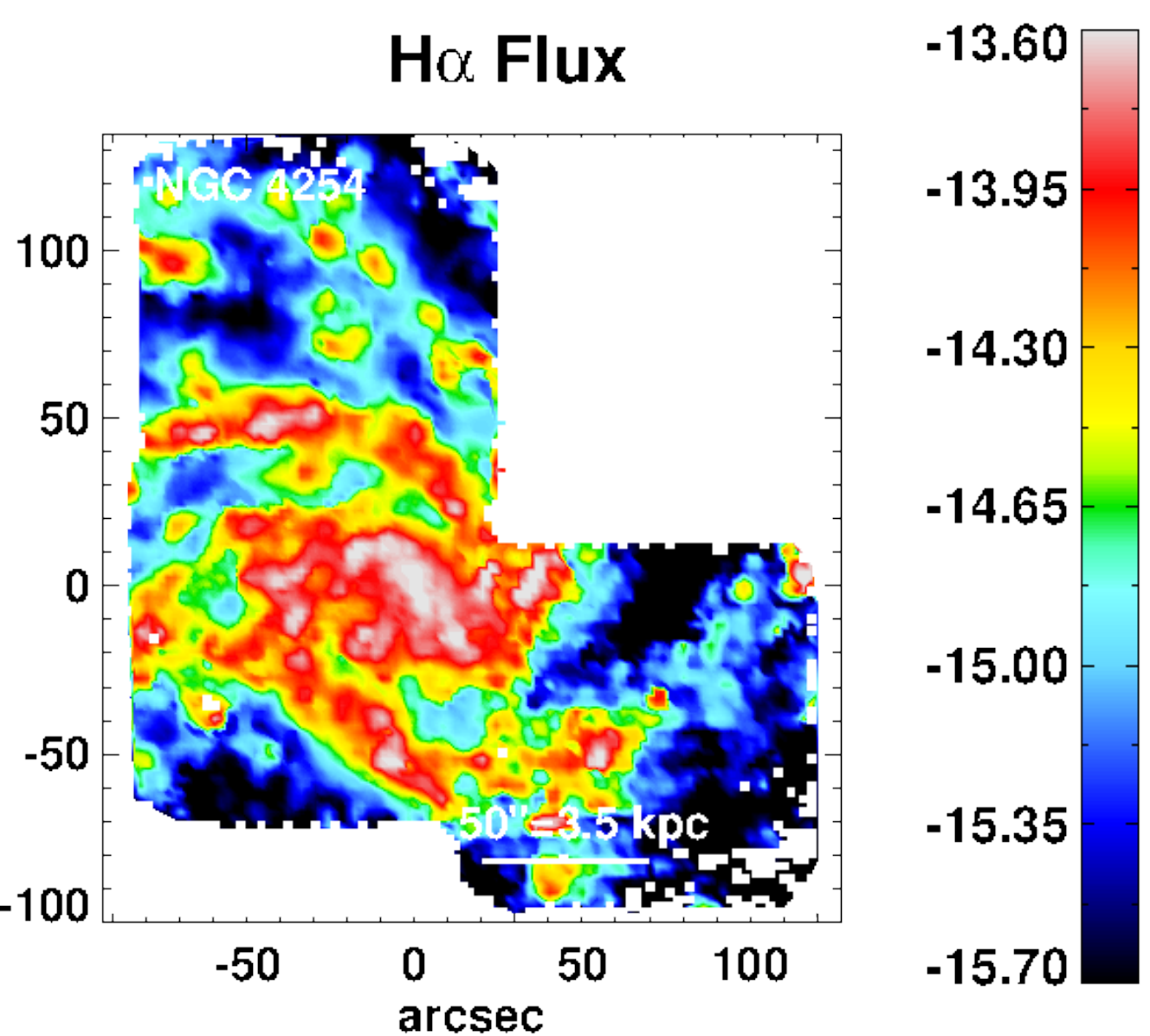} 
\includegraphics[width=0.285\textwidth]{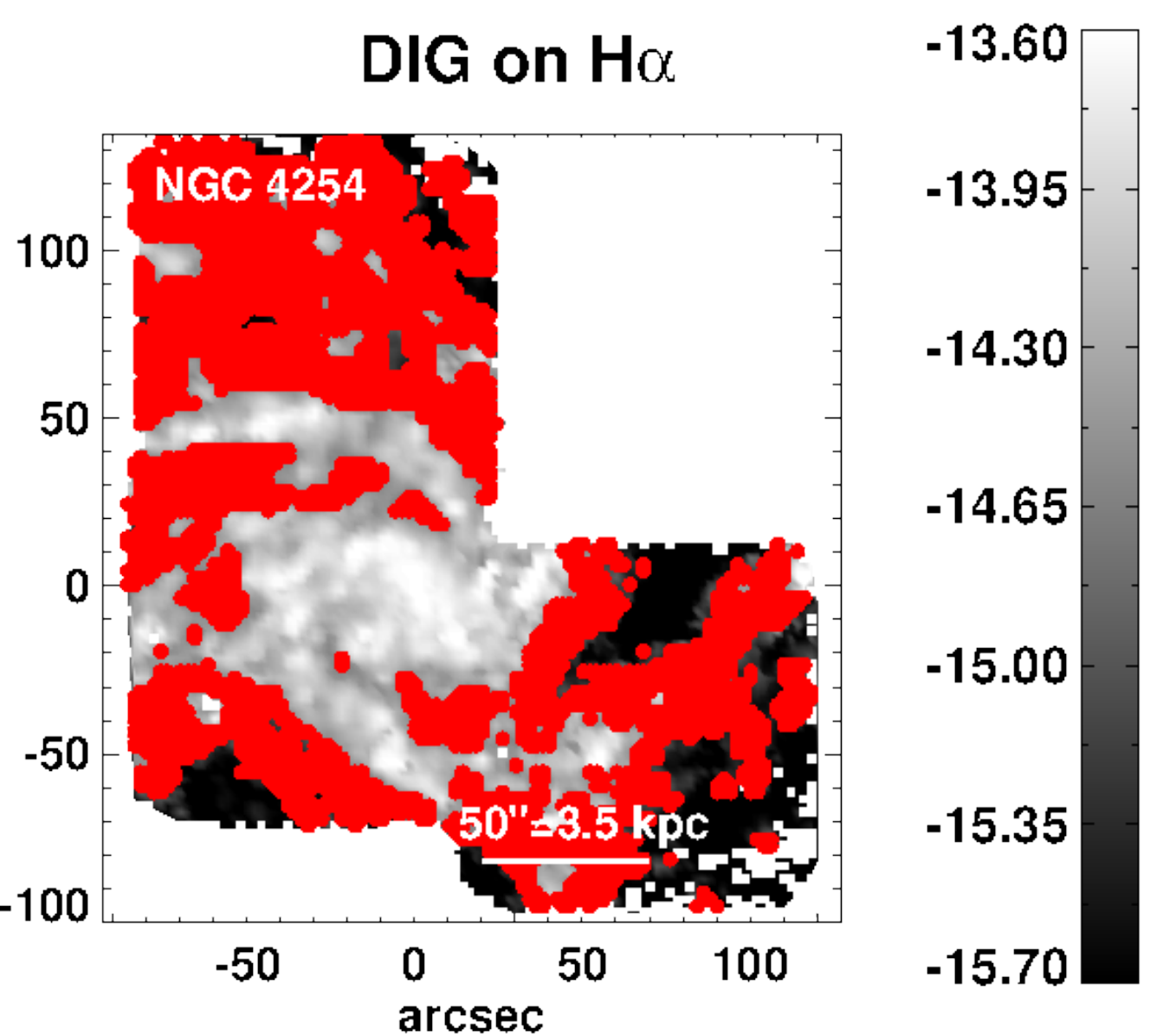}
\includegraphics[width=0.25\textwidth]{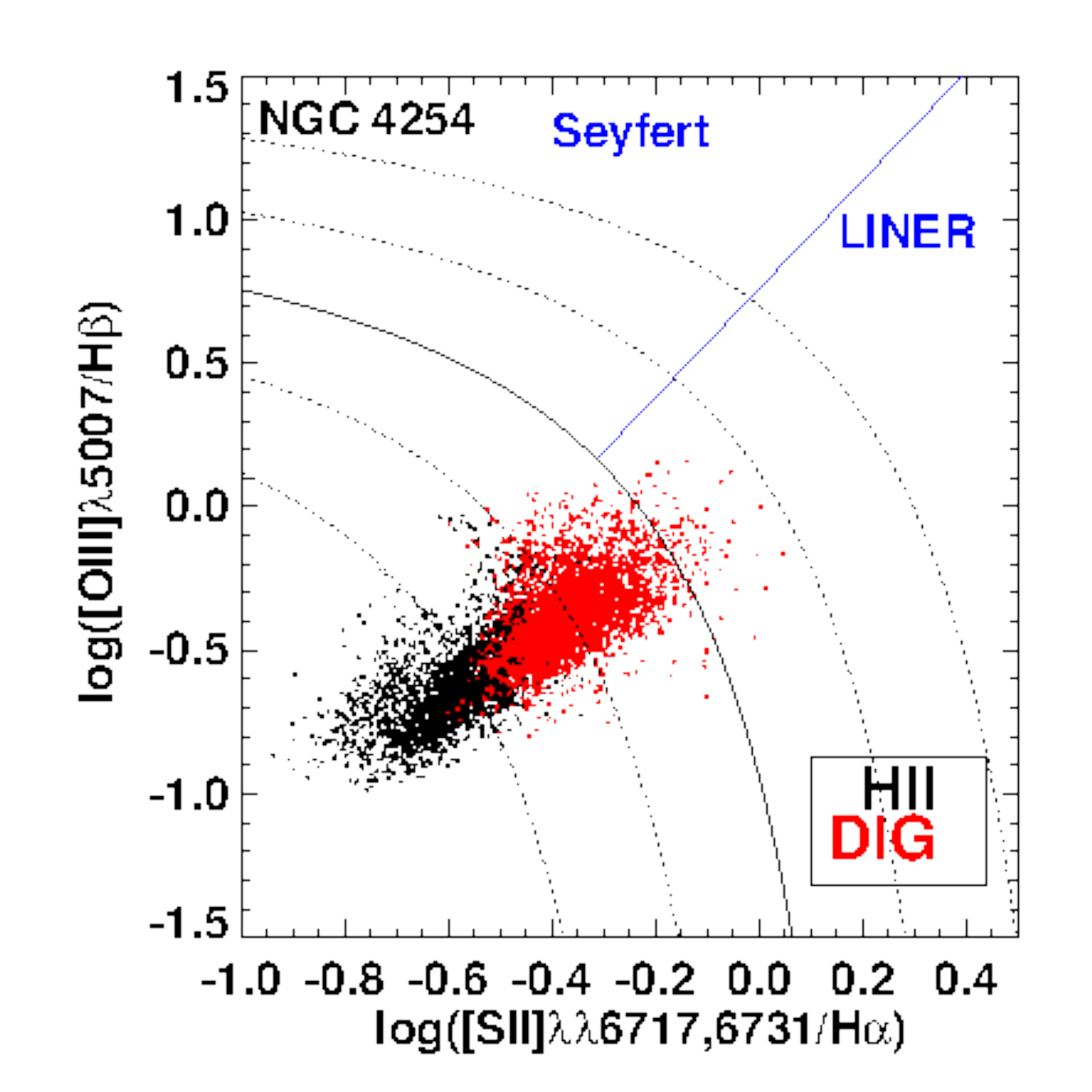}
\caption{Continued:  For NGC 4254}
\end{figure}

\addtocounter{figure}{-1}
\clearpage

\end{landscape}

\begin{landscape}

\begin{figure}
\includegraphics[width=0.25\textwidth]{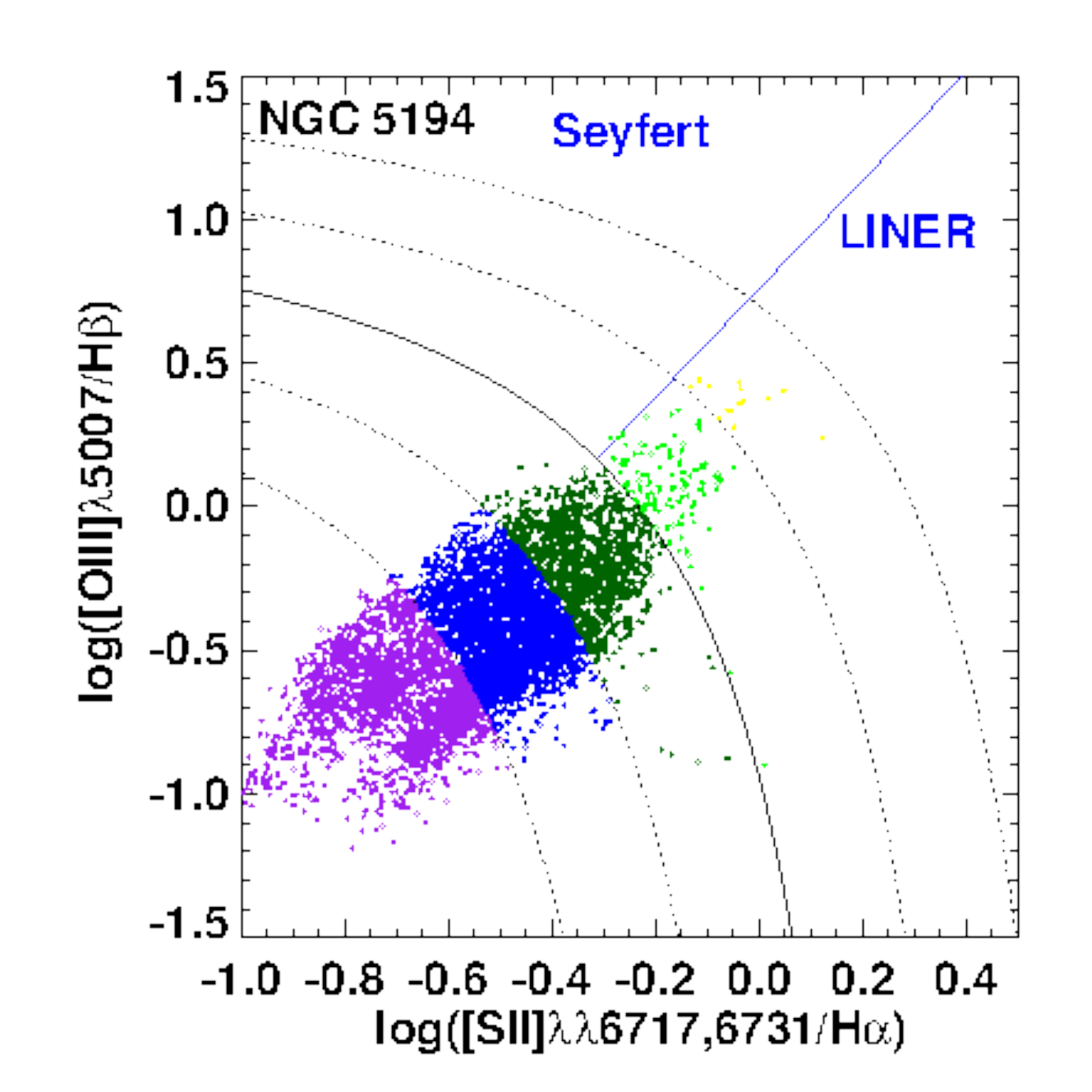}
\includegraphics[width=0.285\textwidth]{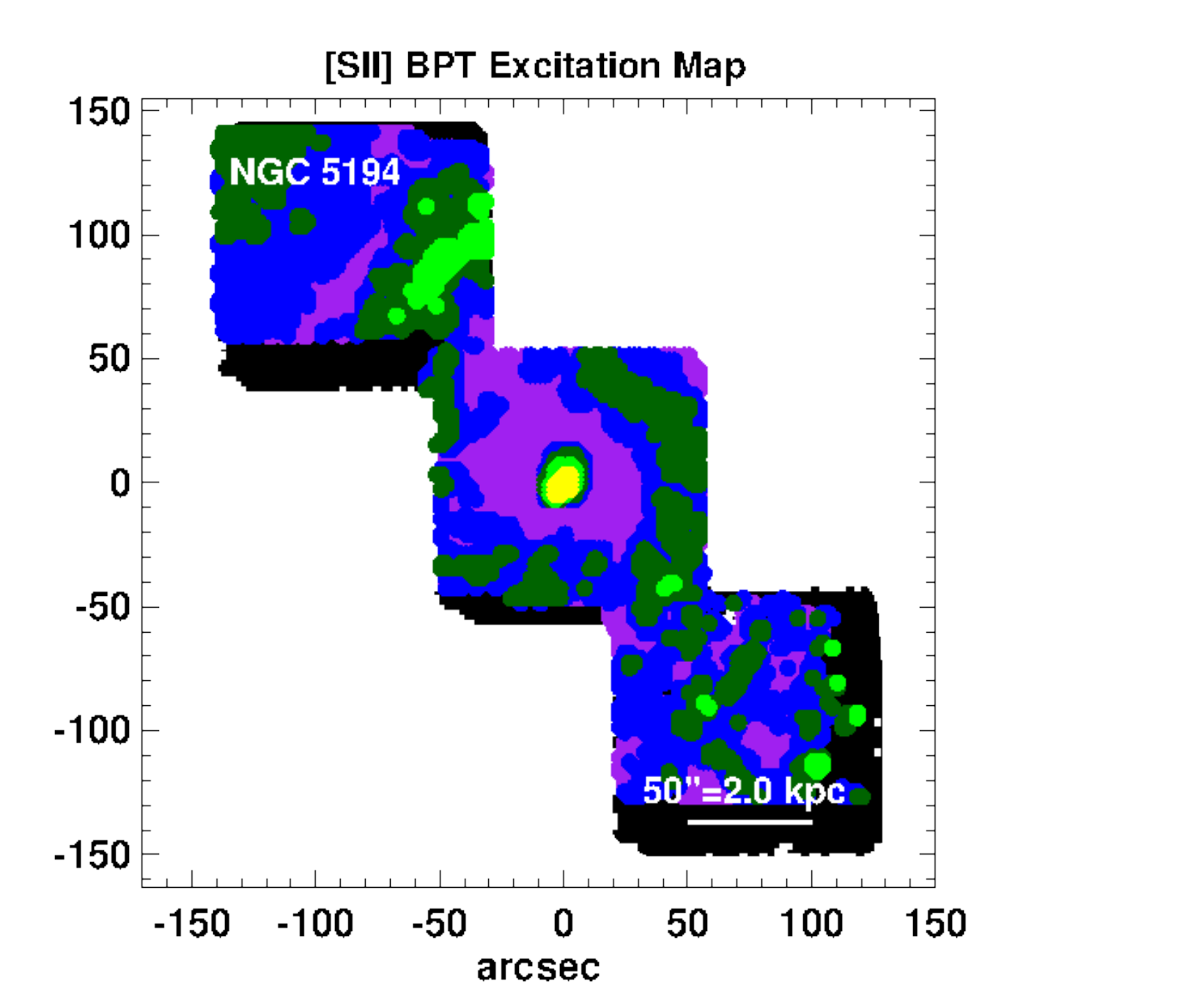} 
\hspace{-0.5 cm}
\includegraphics[width=0.285\textwidth]{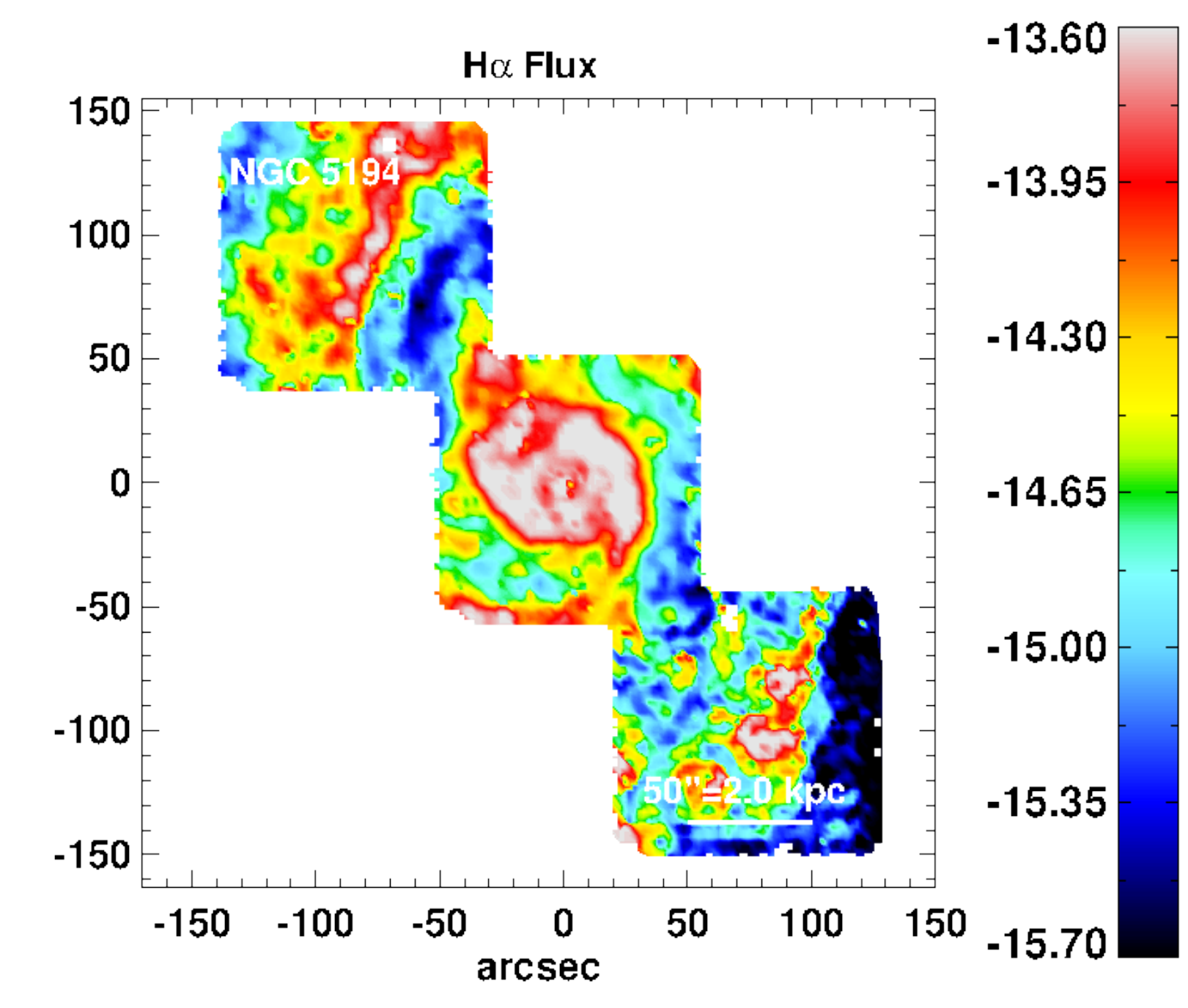} 
\includegraphics[width=0.285\textwidth]{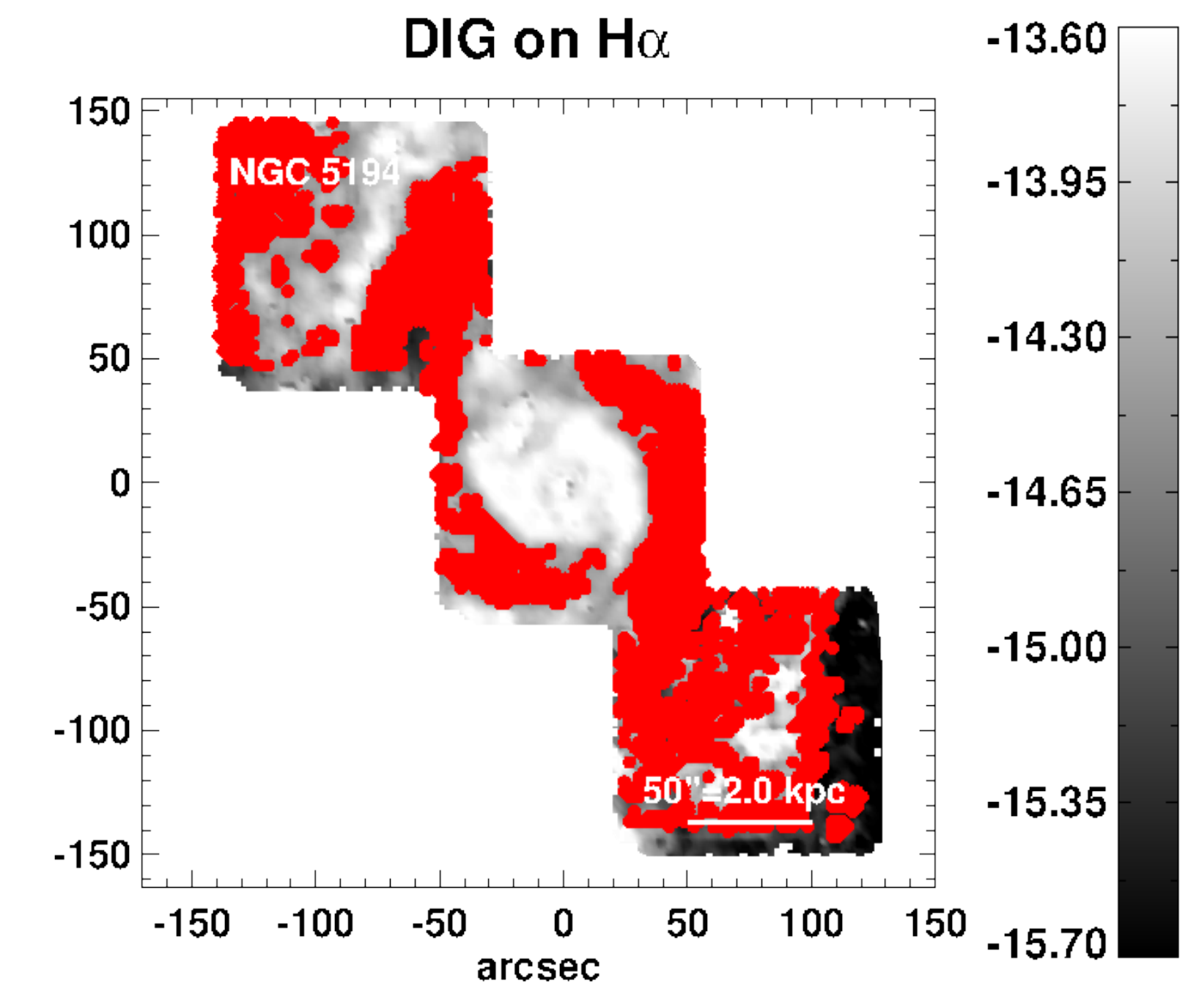}
\includegraphics[width=0.25\textwidth]{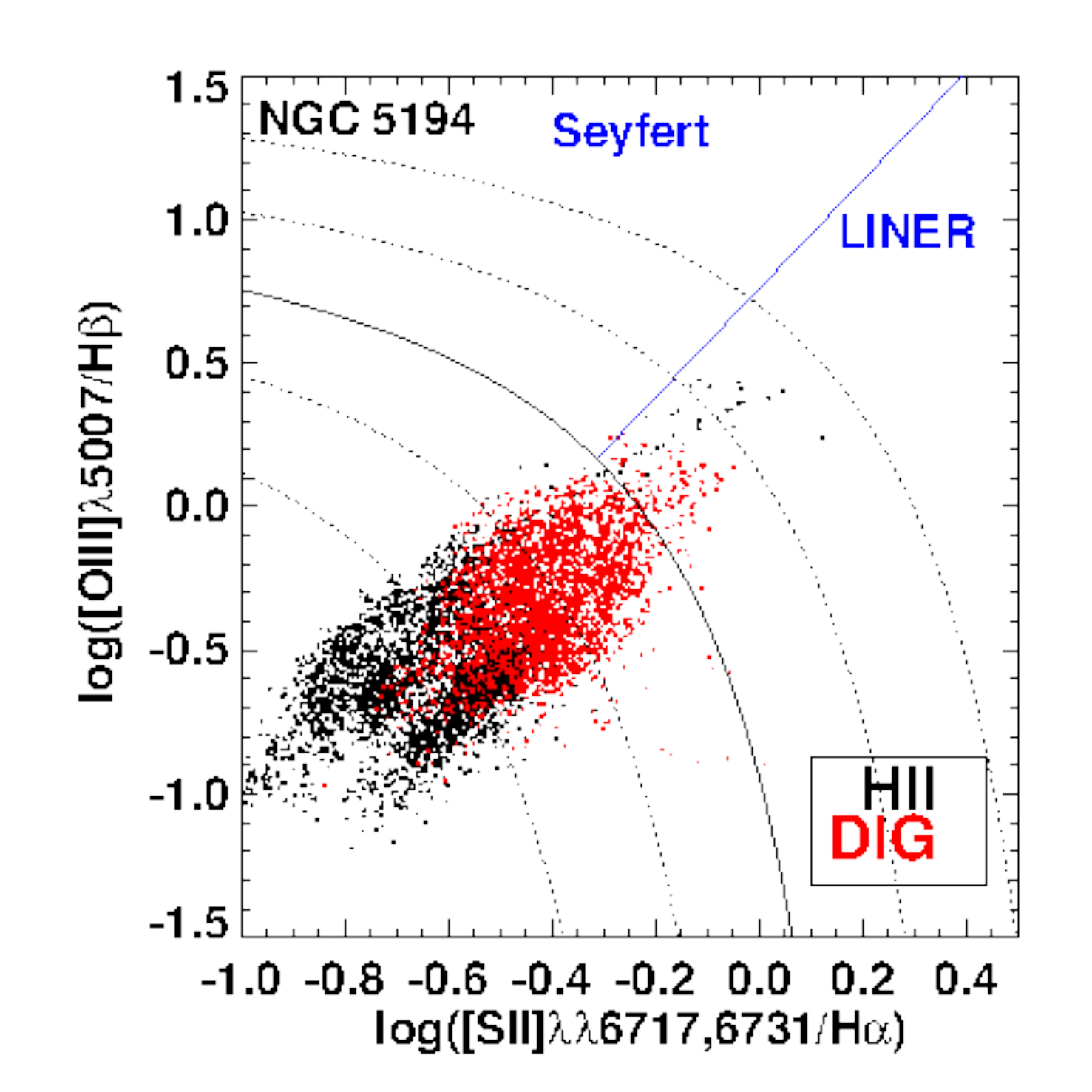}
\caption{Continued:  For NGC 5194}
\end{figure}

\addtocounter{figure}{-1}

\begin{figure}
\includegraphics[width=0.25\textwidth]{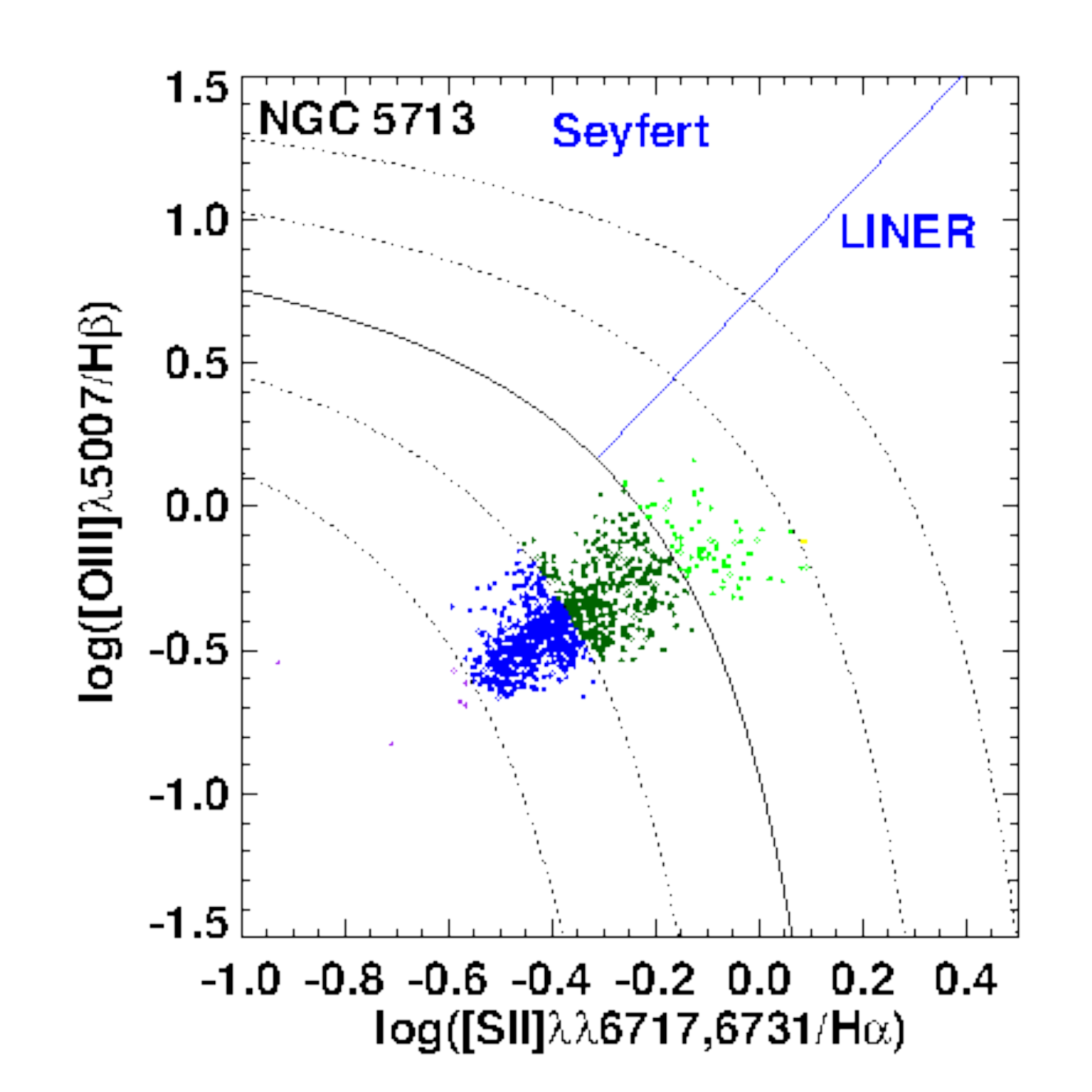}
\includegraphics[width=0.285\textwidth]{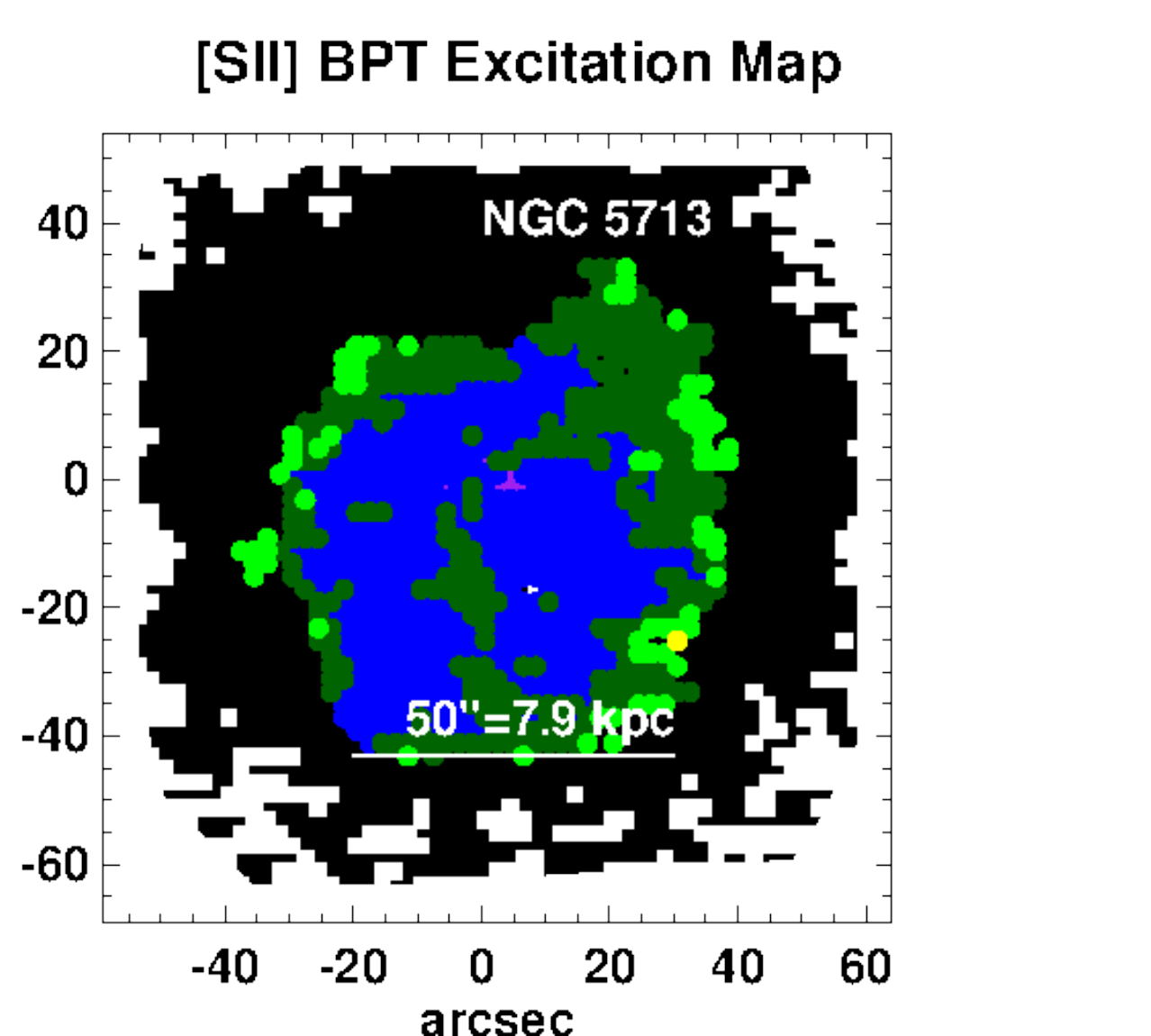}  
\hspace{-0.5 cm}
\includegraphics[width=0.285\textwidth]{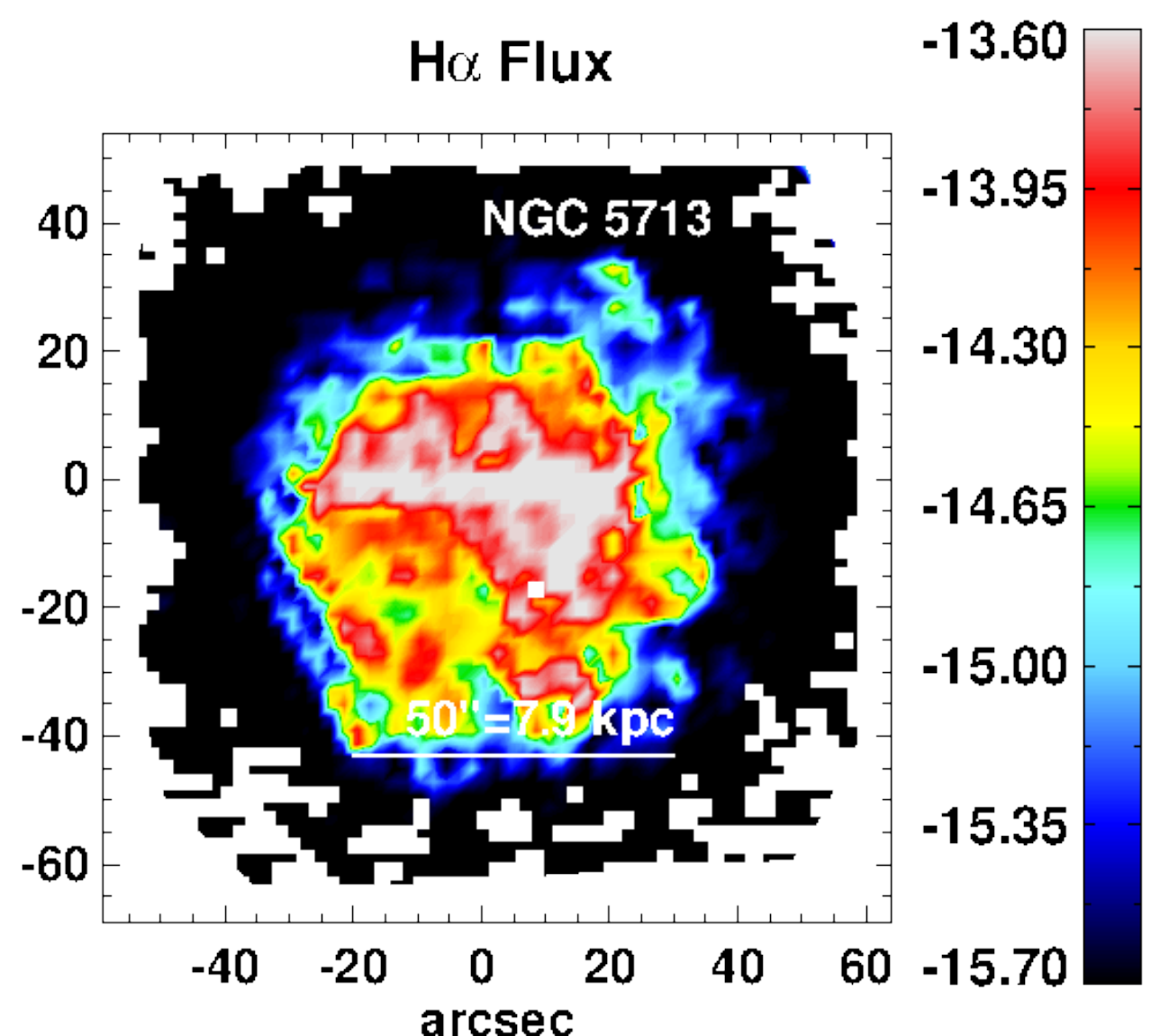} 
\includegraphics[width=0.285\textwidth]{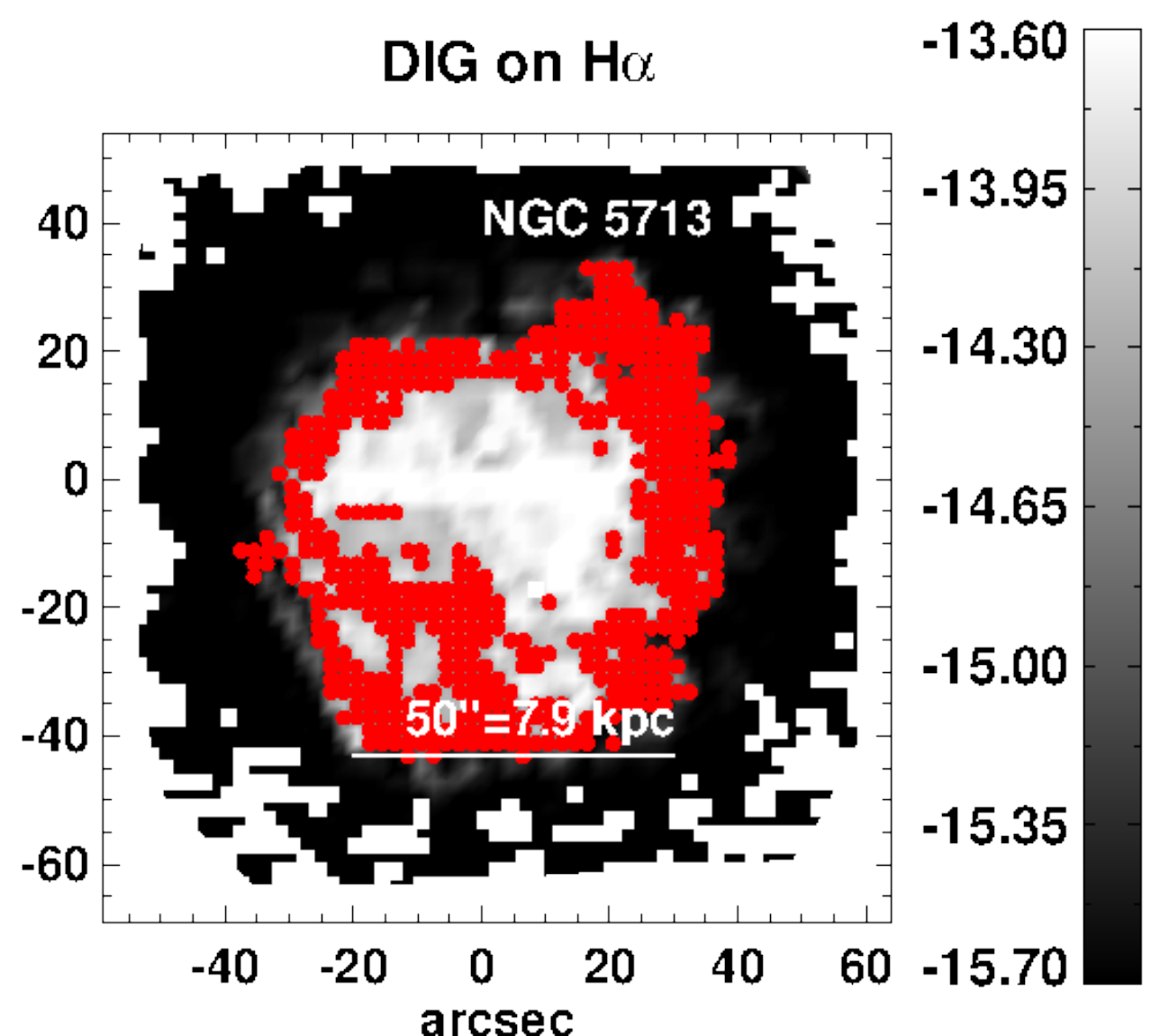}
\includegraphics[width=0.25\textwidth]{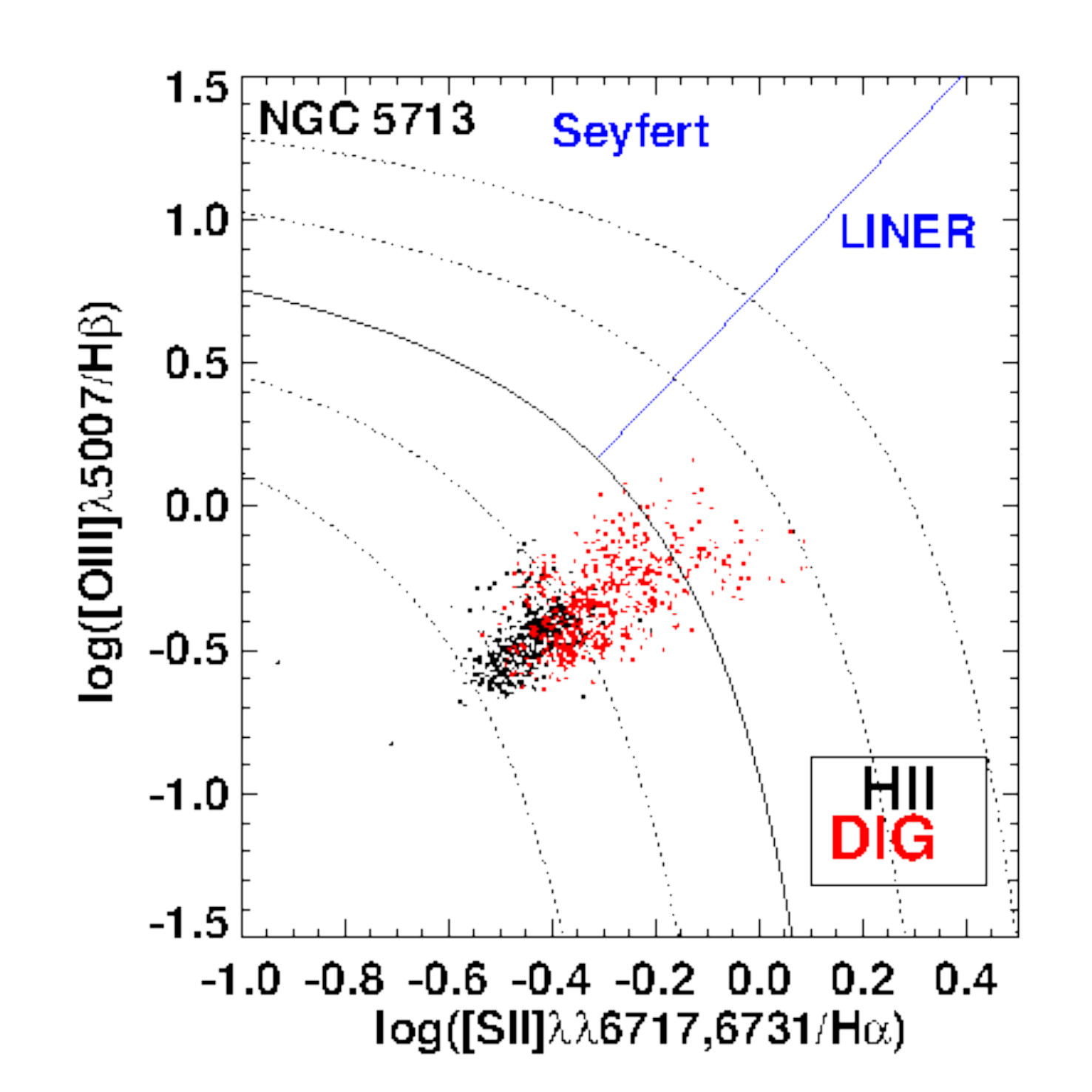}
\caption{Continued:  For NGC 5713.}
\end{figure}

\clearpage

\end{landscape}

\subsection{Identifying and Correcting for Diffuse Ionized Gas} \label{sec:dig}

\begin{figure*}
 \includegraphics[width=0.95\textwidth]{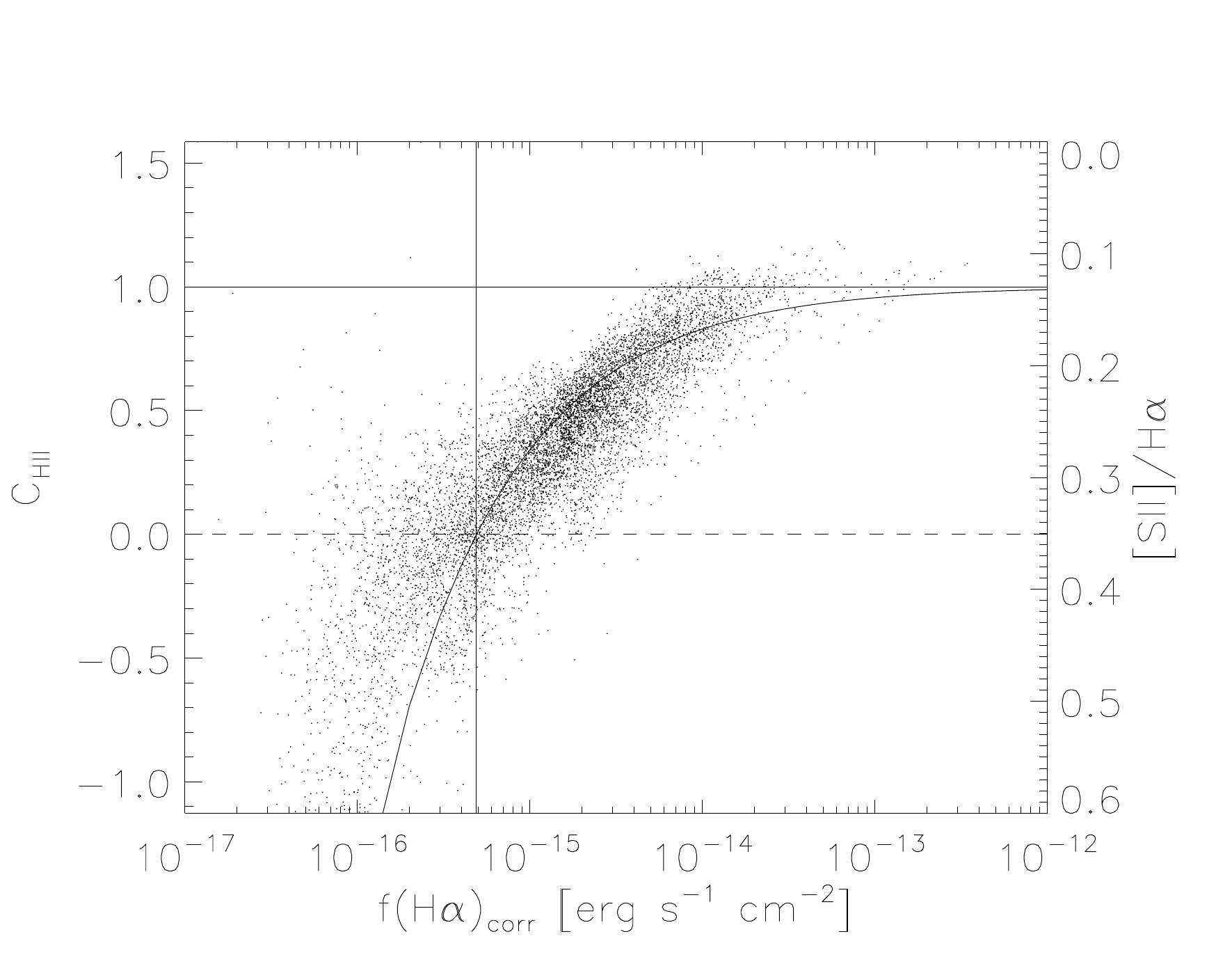}
 \caption{How we identify diffuse ionized gas (DIG) in a galaxy's spaxels
 as described in $\S$ \protect\ref{sec:dig} and building upon the
 methodology and Figure 9 from \protect\cite{blanc2009}. Each data-point
 represents an individual spaxel from NGC 2903. The x-axis is
 $f($H$\alpha)_{corr}$, the extinction corrected flux of H$\alpha$ per
 spaxel.  The right y-axis is the \iontwo{S}{II}/H$\alpha$ ratio per spaxel.  The
 horizontal solid line represents the median \iontwo{S}{II}/H$\alpha$ of the 100
 brightest spaxels $ \left({\mbox{\iontwo{S}{II}}  / \mbox{H}\alpha}
 \right)_{\mbox{\tiny H II}} $, where we assume nearly all the flux is
 coming from gas in \HII{} regions excited by star formation.  The dashed
 horizontal line represents the median \iontwo{S}{II}/H$\alpha$ of the 100 dimmest
 spaxels with sufficient signal-to-noise $  \left( {\mbox{\iontwo{S}{II}}  /
 \mbox{H}\alpha} \right)_{\mbox{\tiny DIG}} $, where we assume 100\% of
 the flux comes from the DIG.
 We make an initial guess for each spaxel of  $C_{\mbox{\tiny H II}}$, the
 fraction of $f($H$\alpha)_{corr}$ coming from star formation, from each
 spaxel's \iontwo{S}{II}/H$\alpha$ value, $ \left({\mbox{\iontwo{S}{II}}  / \mbox{H}\alpha}
 \right)_{\mbox{\tiny H II}} $, \&  $  \left( {\mbox{\iontwo{S}{II}}  /
 \mbox{H}\alpha} \right)_{\mbox{\tiny DIG}} $ as shown on the left y-axis.
   From each spaxel's  $f($H$\alpha)_{corr}$ and the initial guess of
 $C_{\mbox{\tiny H II}}$,
 we fit Equation \ref{eq:chii} (solid curved line) to find the final
 estimate of  $C_{\mbox{\tiny H II}}$ for each spaxel and $f_0$,
    the flux where we assume 100\% of the emission is from the DIG marked
 with the vertical solid line.}
 \label{fig:chii}
 \end{figure*}

Diffuse ionized gas (DIG) consists of warm ($\sim 10^4$ K) ionized gas  that resides up to two kiloparsecs above and below the plane of a spiral galaxy's disc
\citep{mathis2000, madsen2006, Haffner2009}.
This gas has been raised above the galactic disc by superbubbles created from regions in the disc where multiple supernovae have occurred and possibly by Type Ia supernovae at large scale heights \citep{wood2010}.
The DIG consists of the majority of ionized gas in a galaxy \citep{walterbos1998}, and its low level emission is a possible source of contamination superimposed over emission from \HII{} regions ionized by photons from local massive stars in the disc.
The low density, clumpy structure, and sufficient number of photons
from hot OB stars in the galactic disc are enough to keep the DIG
mostly ionized \citep{Haffner2009}, 
although shock ionization might also be a small secondary contributor \citep{martin2000}.
The majority of the energy ionizing the DIG is currently thought to come from massive OB stars in the galactic disc from which ionizing Lyman continuum photons can travel large path lengths, up to several kpc above the  disc, before ionizing the DIG \citep{walterbos1998}.
Therefore the source of ionizing photons for the DIG is non-local.
Current \zgas{} diagnostics are calibrated to \HII{} regions where the ionizing photons originate
from local massive stars, but are not calibrated for DIG where the ionizing photons have a
non-local origin.
In order to avoid contamination by DIG, some  IFU studies of \zgas{}
specifically  target only bright \HII{}  regions (e.g., see
\citealt{sanchez2012}), thereby sacrificing spatial sampling.
In this study, we try to identify and exclude regions dominated 
by DIG emission from our \zgas{} analysis.

The contributions from \HII{} regions and DIG for each individual emission line are hard to disentangle from each other, since we are seeing each galaxy in projection.
The flux from an emission line measured in a single spaxel is integrated along the line of sight through the galaxy and will include some fraction of flux from \HII{} regions and some fraction from DIG.
We explore the contributions to the  H$\alpha$ flux by \HII{} regions and DIG by expanding upon the method developed in \cite{blanc2009}.
For the H$\alpha$ line, we call the fraction of flux coming from \HII{} regions  $C_{\mbox{\tiny H II}}$ and the fraction of flux
coming from DIG $C_{\mbox{\tiny DIG}}$ such that the total H$\alpha$ flux $f$(H$\alpha$) is given by:
\begin{align}
f(\mbox{H}\alpha) &= f(\mbox{H}\alpha)_{\mbox{\tiny H II}} + f(\mbox{H}\alpha)_{\mbox{\tiny DIG}}  \\
				&= C_{\mbox{\tiny HII}} f(\mbox{H}\alpha) + C_{\mbox{\tiny DIG}} f(\mbox{H}\alpha),
\end{align}
where $C_{\mbox{\tiny H II}} = 1 - C_{\mbox{\tiny DIG}}$.

The Wisconsin H$\alpha$ Mapper (WHAM) sky survey \citep{madsen2006} observed DIG and \HII{} regions in the Milky Way and found that
DIG has lower H$\alpha$    surface brightness and higher  \iontwo{S}{II}/H$\alpha$ than \HII{} regions.
The higher  \iontwo{S}{II}/H$\alpha$ observed in DIG appears to mainly be caused by the higher temperatures found in DIG than in \HII{} regions \citep{madsen2006, Haffner2009}, but ionization effects might also play a role.
In our work,
we start by assuming \HII{} regions dominate the observed areas of a galaxy
with the highest H$\alpha$ flux, and DIG dominates the areas with the lowest.
Using all spaxels where $f$(H$\alpha$) $S/N$ $>$ 3, we estimate the characteristic values of \iontwo{S}{II}/H$\alpha$ for \HII{} regions (\iontwo{S}{II}/H$\alpha)_{\mbox{\tiny H II}}$ and DIG (\iontwo{S}{II}/H$\alpha)_{\mbox{\tiny DIG}}$ by finding the median \iontwo{S}{II}/H$\alpha$ in the 100 brightest and dimmest spaxels respectively.
The value of \iontwo{S}{II}/H$\alpha$
in a given galaxy can be affected by several factors including variations in the ionization parameter $q$, the hardness of the local 
radiation field  and \zgas{} \citep{diaz1991, dors2011} so we calculate (\iontwo{S}{II}/H$\alpha)_{\mbox{\tiny H II}}$ \&
(\iontwo{S}{II}/H$\alpha)_{\mbox{\tiny DIG}}$ separately for
each individual galaxy 
following the steps below.

{\it{Step 1: Obtain  initial estimate for each spaxel's  $C_{\mbox{\tiny     H II}}$:}} 
An initial estimate for each spaxel's  $C_{\mbox{\tiny H II}}$ is obtained from each spaxel's \iontwo{S}{II}/H$\alpha$ with the following equation:
\begin{equation}
{\mbox{\iontwo{S}{II}}  \over  \mbox{H}\alpha} = C_{\mbox{\tiny H II}} \left( {\mbox{\iontwo{S}{II}}  \over  \mbox{H}\alpha} \right)_{\mbox{\tiny H II}} +  
								C_{\mbox{\tiny DIG}} \left( {\mbox{\iontwo{S}{II}}  \over  \mbox{H}\alpha} \right)_{\mbox{\tiny DIG}},
\end{equation}
and solving for our initial guess of $C_{\mbox{\tiny H II}}$:
\begin{equation}
C_{\mbox{\tiny HII}} = { {{ \mbox{\iontwo{S}{II}}\over\mbox{H}\alpha} -  \left( {\mbox{\iontwo{S}{II}}  \over  \mbox{H}\alpha} \right)_{\mbox{\tiny DIG}} }   \over
				 { \left( {\mbox{\iontwo{S}{II}}  \over  \mbox{H}\alpha} \right)_{\mbox{\tiny HII}} -  \left( {\mbox{\iontwo{S}{II}}  \over  \mbox{H}\alpha} \right)_{\mbox{\tiny DIG}}    } } .
\end{equation}

{\it{Step 2: Solve for one value of  $f_0$ and $\beta$ for each galaxy:}}
Next we use our  initial estimate for $C_{\mbox{\tiny H II}}$ and
$f$(H$\alpha$) for all the individual spaxels to find one value of
H$\alpha$ flux $f_0$ for a given galaxy where we assume 100\% of the
flux comes from DIG.   
This procedure is illustrated  in  Figure~\ref{fig:chii}.
We solve for $f_0$ by fitting a single curve to the following equation:
\begin{equation} \label{eq:chii}
C_{\mbox{\tiny H II}} = 1.0 - \left( {f_0 \over f(\mbox{H}\alpha) }\right)^\beta ; (\mbox{for}\ f(\mbox{H}\alpha) > f_0),
\end{equation}
weighting the fit by the uncertainty in \iontwo{S}{II}/H$\alpha$.  Variation in
DIG surface brightness is accounted for with the exponent $\beta$,
where the DIG would have a constant surface brightness across the disc
of a galaxy at $\beta = 1$, and tracks the distribution of SF in the
disc at $\beta < 1$.
We report the values found for $f_0$ and $\beta$ for each galaxy in Table \ref{tab:sfr}.
{\it{Step 3: Solve for a final $C_{\mbox{\tiny H II}}$ for each spaxel
    where $f(\mbox{H}\alpha) > f_0$:}}
Once we solve for $f_0$ and $\beta$, we reuse Equation \ref{eq:chii} and each spaxel's $f(\mbox{H}\alpha)$ to solve for a final $C_{\mbox{\tiny H II}}$ for each spaxel where $f(\mbox{H}\alpha) > f_0$.  Any spaxels where  $f(\mbox{H}\alpha) \leq f_0$ are considered to have 100\% contribution from the DIG to  $f(\mbox{H}\alpha)$.
From here, $C_{\mbox{\tiny H II}}$ serves two functions, it acts as a correction for $f(\mbox{H}\alpha)$ such that the H$\alpha$ flux coming from H II regions is $f(\mbox{H}\alpha)_{\mbox{\tiny H II}} =  C_{\mbox{\tiny H II}} f(\mbox{H}\alpha)$, and it allows us to identify and exclude DIG contaminated spaxels from our \zgas{} analysis.  For this study we exclude spaxels where DIG takes up 40\% or more of $f(\mbox{H}\alpha)$, so we only include spaxels where $C_{\mbox{\tiny H II}} >0.6$.

For each galaxy, the far right panel of Figure~\ref{fig:bpt-sliced}
 illustrates the location of the DIG  on the excitation diagnostic diagram, while 
the centre right panel of Figure~\ref{fig:bpt-sliced} shows  
the projected spatial distribution of DIG in the galaxy. 
Most of the DIG-dominated regions occupy the mid-to-upper right side of 
the excitation diagnostic diagram.
This behaviour is expected since high \iontwo{N}{II}/H$\alpha$ and 
\iontwo{S}{II}/H$\alpha$ line ratios are associated with DIG emission.
Within the galaxies,  most of the DIG-dominated  emission  is
located
at large projected  distances from the centre of the
galaxy and is often associated  with  the regions 
between the spiral arms.
A comparison of the centre left and centre right panels of  Figure~\ref{fig:bpt-sliced}
shows that many of the  LINER-type regions, which are located in the 
outer regions  of galaxies and are far from spiral arms or other sites 
of high SFRs,  tend to  be spatially associated with DIG.
This effect has also been observed by \cite{greenawalt1997}, \cite{hoopes2003}, \cite{singh2013}, and \cite{belfiore2015}.

In NGC 2903 we calculate the effect of erroneously 
including DIG when calculating \zgas{}.
We find that including the DIG-dominated regions only changes the average absolute value of \zgas{} for all our diagnostics by at most $\pm 0.1$ dex and the slope of the \zgas{} gradients by $\pm 0.015$ dex kpc$^{-1}$, even in the outer regions of the disc where emission is the most dominated by the DIG.
The uncertainty of 0.1 dex is about the same as the uncertainity
 found when the $T_e$ method is used to estimate
 \zgas{} \citep{kennicutt2003b, hagele2008}.
While this is a small effect in NGC 2903, for accuracy we remove
all  DIG-dominated
regions where $C_{\mbox{\tiny H II}} \leq 0.6$
when calculating \zgas{} in our sub-sample
galaxies.

\begin{table*}
\caption{Star Formation Rates}
\begin{center}
\begin{tabular}{cccccccccc}
\hline \hline\label{tab:sfr} 
{ Galaxy} & $M_\star$ & \frtwofive{} & $R_{\rm vp}$ & $R_{25}$  &  $f_0$ & $\beta$  & VENGA SFR(H$\alpha$) & Global TIR & Global SFR(TIR)     \\
 &  & & (kpc) &  (kpc)  & (erg s$^{-1}$ cm$^{-2}$) & &  ($M_\odot$ yr$^{-1}$)  & Lum. (erg s$^{-1}$) &  ($M_\odot$ yr$^{-1}$) \\
 (1) & (2) & (3) & (4) & (5) & (6) & (7) & (8) & (9) & (10) \\
\hline
NGC 0337 & 1.6E+10 & 0.72 & 5.8   & 8.1  & 5.74E-16 & 0.772  & 1.52  & 4.88E+43 & 1.90\\
NGC 0628 & 2.0E+10 & 0.47 & 6.2   & 13.1  & 2.52E-16 & 0.789  & 0.50  & 3.80E+43 & 1.48  \\
NGC 2903 & 5.0E+10 & 0.38 & 6.2   & 16.3  & 4.90E-16 & 0.587 & 1.48  & 7.73E+43 & 3.01  \\
NGC 3938 & 3.2E+10 & 0.60 & 8.4   & 14.0  & 3.53E-16 &  0.891 & 1.75  & 6.10E+43 & 2.37  \\
NGC 4254 & 4.0E+11 & 0.81 & 9.0   & 11.1  & 5.74E-16 & 0.637 & 4.11 & 1.30E+44 & 5.06  \\
NGC 5194 & 7.9E+10 & 0.56 & 7.7  &  13.8  & 8.68E-17 & 0.804 & 2.27 & 1.05E+44 & 4.08 \\
NGC 5713 & 7.9E+10 & 0.66 & 8.6  & 13.0  & 1.47E-15 & 0.653 & 9.50  & 3.02E+44 & 11.76  \\
\hline\hline
\end{tabular}
\end{center}
Columns: 
(1)~Galaxy NGC number.
(2)~Total stellar mass in units of $M_\odot$ from \cite{blanc2013}.
(3)~Fraction of \rtwofive{} radius covered, as illustrated in Figure \ref{fthumb}. 
(4)~Projected radius of galaxy covered by our  Mitchell Spectrograph  IFU data in kpc.
(5)~The galaxy's projected radius \rtwofive{} from RC3 \citep{devaucouleurs1991}.  
\rtwofive{}  is defined as the radius where the surface brightness
of the outer disc reaches 25 mag arcsec$^{-2}$ in the $B$-band.
(6)~$f_0$ is the H$\alpha$ flux per spaxel in units of erg s$^{-1}$ cm$^{-2}$ where we assume 100\% of the emission is from the DIG.  See Equation \ref{eq:chii} in $\S$ \ref{sec:dig}. 
(7)~$\beta$ quantifies variation in DIG surface brightness across a galaxy's disc.  The DIG would have a constant surface brightness across the disc of a galaxy at $\beta = 1$, and tracks the distribution of SF in the disc at $\beta < 1$.   See Equation \ref{eq:chii} in $\S$ \ref{sec:dig}.
(8)~Extinction corrected H$\alpha$-based SFRs  ($M_\odot$ yr$^{-1}$) measured from integrated H$\alpha$ flux over the regions covered by our pointings of the Mitchell Spectrograph.  See $\S$ \ref{sec:sfr} for details.
(9)~TIR luminosity, in units of solar luminosity, calculated from the IRAS 25, 60 \& 100 $\mu$m bands.  See $\S$ \ref{sec:sfr} for details.
(10)~TIR-based SFR ($M_\odot$ yr$^{-1}$) calculated from the TIR luminosity.  See $\S$ \ref{sec:sfr} for details.
\end{table*}

\subsection{Computing Star Formation Rates} \label{sec:sfr}

We can estimate the current star formation rate (SFR) by measuring the intensity of recombination lines such as H$\alpha$.
UV photons blue-ward of the Lyman break ionize neutral atomic hydrogen and when ionized hydrogen recombines, part of that energy is radiated away as recombination lines such as H$\alpha$ \citep{kennicutt1998}.
Our IFU data include readily available spatially resolved H$\alpha$ emission from which we can compute SFRs.  H$\alpha$ derived SFRs can be used in IFU studies to directly compare against different parameters (e.g. \zgas{}, $q$, or gas excitation) and environments in a galaxy,  be combined with multi-wavelength data to study SF laws and efficiency \citep{blanc2009, blanc2013b}, and be tested against other SFR diagnostics \citep{catalan2015}.
Before any SFR can be measured, we correct the H$\alpha$ flux for extinction
using the
H$\alpha$/H$\beta$ = 2.86
line ratio decrement described in \cite{osterbrock}.
We apply a  reddening correction using a Milky Way like extinction curve, typically  of $E(B-V) \sim 0.5$ although the reddening correction can range anywhere from $E(B-V) = 0$ to 2.
Contaminated regions such as Seyfert \& LINER are masked out, as described in $\S$ \ref{sec:bpt-method}.  
We correct for the DIG contribution to the H$\alpha$ flux
$f(\mbox{H}\alpha)$ by multiplying  it by the  value of $C_{\mbox{\tiny H II}}$  calculated in $\S$~\ref{sec:dig}.
While DIG contamination affects the local SFR measured in a given spaxel, the total SFR across an entire galaxy includes emission from the DIG since the photons ionizing the DIG originated from hot massive young stars.
To account for this, we scale the DIG corrected SFR in each spaxel by dividing by the ratio of the total  $f(\mbox{H}\alpha)$ from \HII{} regions to the total  $f(\mbox{H}\alpha)$ across the galaxy $\left(\sum  C_{\mbox{\tiny H II}}f(\mbox{H}\alpha)\right) / \left(\sum f(\mbox{H}\alpha)\right)$.
The SFR for a given spaxel is calculated from the flux of the H$\alpha$ line 
using the prescription from \cite{kennicutt2012}: 
\begin{equation}
\mbox{SFR}\ (M_\odot\ \mbox{yr}^{-1}) = { 5.37 \times 10^{-42}\   C_{\mbox{\tiny H II}}\ f(\mbox{H}\alpha) \over
		\left(\sum  C_{\mbox{\tiny H II}}f(\mbox{H}\alpha)\right) / \left(\sum f(\mbox{H}\alpha)\right) },
\end{equation}
where $f(\mbox{H}\alpha)$ is in units of erg s$^{-1}$.
We  convert the SFR for each individual spaxel into a SFR density ($\Sigma_{\rm SFR}$ $M_\odot$ yr$^{-1}$ kpc$^{-2}$) by dividing the SFR for each spaxel by
its deprojected area in kpc$^2$.
Table \ref{tab:sfr} lists the total integrated SFR over our spatial coverage for each galaxy.
Maps of  $\Sigma_{\rm SFR}$ can be seen in Figure \ref{fig:2d-maps} and
radial gradients for $\Sigma_{\rm SFR}$ are shown on the right side of Figure \ref{fig:q-sfr-gradients}.

To assess the robustness of our SFR estimates,
we compare
the extinction corrected H$\alpha$ SFRs from our IFU data
to the SFRs inferred from the global 
total infrared (TIR) luminosity between 3-1100 $\mu$m.
Infrared derived SFRs avoid the issues of extinction or contamination (ie. from shocks) that can affect H$\alpha$ derived SFR estimates.
Fluxes from the  25, 60, \&
100 $\mu$m IRAS bands are obtained from the IRAS Revised Bright Galaxy Sample \citep{sanders2003}.
The IRAS fluxes are converted into luminosities given the distance to each galaxy, and then into a global TIR luminosity using Equation 5 from \cite{dale2002}.
TIR SFRs are computed using the TIR SFR calibration from \cite{kennicutt2012}.
The global TIR luminosities and SFRs we calculate for our galaxies can be seen in Table \ref{tab:sfr}.
Although our IFU data only cover a fraction of the disc's \rtwofive{} radius (see Table \ref{tab:venga-subsample}),
our extinction corrected H$\alpha$ based SFR from VENGA is a large fraction (34 to 81\%) of the global SFR$_{\rm TIR}$.
This is reasonable because most of the obscured SF in a galaxy (as traced by TIR light) takes place in the central regions of a galaxy.

\subsection{Computing the Ionization Parameter $q$} \label{sec:how-q}

\begin{table*}
\caption{Summary of the seven \zgas{} diagnostics}
\begin{center}
\begin{tabular}{ccccc}
\hline\hline
{ \zgas{} Diagnostics} & \zgas{}  Indicator & {Reference} & {Calibration Type} & $q$-corrected? \\
 (1)  & (2) & (3) & (4) & (5) \\
\hline
$R_{23}$-M91 & $R_{23}$=(\iontwo{O}{II}$\lambda$3727+\iontwo{O}{III}$\lambda\lambda$4959,5007)/H$\beta$ & \cite{mcgaugh91} & Theoretical & y$^a$ \\
$R_{23}$-KK04 & $R_{23}$=(\iontwo{O}{II}$\lambda$3727+\iontwo{O}{III}$\lambda\lambda$4959,5007)/H$\beta$ & \cite{kobulnicky04} & Theoretical &  y$^b$ \\
$R_{23}$-Z94 &$R_{23}$=(\iontwo{O}{II}$\lambda$3727+\iontwo{O}{III}$\lambda\lambda$4959,5007)/H$\beta$ & \cite{zaritsky1994} & Theoretical &  n \\
\\
N2O2-KD02 & $N202$=\iontwo{N}{II}$\lambda$6584/\iontwo{O}{II}$\lambda$3727 & \cite{kewley2002} & Theoretical & Invariant$^c$  \\
N2-D02 & $N2$=\iontwo{N}{II}$\lambda$6584/H$\alpha$ & \cite{deincolo02} & Empirical &  n\\
N2-PP04 & $N2$=\iontwo{N}{II}$\lambda$6584/H$\alpha$ & \cite{pettini2004} & Empirical & n\\
O3N2-PP04 & $O3N2$=(\iontwo{O}{III}$\lambda\lambda$4959,5007/H$\beta$)/(\iontwo{N}{II}$\lambda$6584/H$\alpha$) & \cite{pettini2004} & Empirical & n \\
\hline\hline
\label{tab:zgas-indicators}
\end{tabular}
\end{center}
\raggedright
Columns:
(1)~Abbreviation for each \zgas{} diagnostic.  Details for each diagnostic can be found in $\S$ \ref{sec:how-zgas}, Appendix \ref{sec:zgas-append}, and in \cite{kewley2008}.
(2)~The \zgas{}  indicator used for each \zgas{} diagnostic, and its
definition in terms of emission line ratios (see Appendix \ref{sec:zgas-append} for details).
(3)~Reference from the literature for each \zgas{} diagnostic.
(4)~Type of calibration  used for each \zgas{} diagnostic.  Theoretical calibrations are based on model \HII{} regions from plasma simulation codes, while empirical calibrations are based on correlations between observed line ratios in \HII{} regions and \zgas{} derived from the direct $T_e$ method.   See $\S$ \ref{sec:how-zgas} for details.
(5)~This column indicates whether this \zgas{} diagnostic is corrected for variations in the ionization parameter $q$.  A `y' indicates `yes' and `n' indicates `no.' \\
$^a$~$q$ corrected for using the \iontwo{O}{III}/\iontwo{O}{II} line ratio.  See Appendix \ref{sec:zgas-append}.\\
$^b$~$q$ and \zgas{} are solved for iteratively as seen in Appendix \ref{sec:q2-append}.\\
$^c$~The \iontwo{N}{II}/\iontwo{O}{II} line ratio is invariant to the value of $q$, as shown in Figure \ref{fig:degeneracy}.
\end{table*}

\cite{kewley2002} define an effective ionization parameter $q$ as the flux of ionizing photons above the Lyman limit (energy $> $13.6 eV) through the surface of a Str\"omgren sphere of radius $R$ 
divided by the electron density $n$:
\begin{equation}
q\ ({\rm cm\ s}^{-1})= {Q \over 4 \uppi R^2 n}
\end{equation}
where $Q$ is the flux of ionizing photons produced at the surface of the central star(s). 
The ionization parameter $q$ for a given spaxel is set by the mass, metallicity, and age of the stellar populations that make up the local ionizing sources and the structure, geometry, metallicity, and dust content of the illuminated gas.
There is a known inverse correlation between $q$ and \zgas{} \citep{freitas-lemes2014, perez-montero2014, rosa2014} which is hypothesized to arise from the \zgas{} tracing the metallicity of the underlying stellar popluations \citep{dors2011, sanchez2015}.
For each spaxel consistent with photo-ionization primarily by massive stars, we calculate $q$ 
with several goals in mind:  (a)~We wish to explore how $q$ varies across regions which have different SFR densities
(see $\S$ \ref{sec:results-components} and Figure \ref{fig:2d-maps});
(b)~We need to know $q$ because several of the \zgas{} diagnostics we use
(listed in
Table \ref{tab:zgas-indicators}) attempt to correct for variations in $q$;
(c)~We wish to compare to \zgas{} diagnostics that 
do not explicitly include $q$ 
(see  Table~\ref{tab:zgas-indicators})
and explore their {\it relative}  radial trends.

In the first method we use to calculate $q$, 
we start by deriving
\zgas{} from the diagnostic based on the  \iontwo{N}{II}/\iontwo{O}{II} ratio \citep{kewley2002}, 
which is fairly invariant to the value of $q$ 
(bottom left panel of Figure \ref{fig:degeneracy})  due to the similar
ionization potentials of N$^+$ and O$^+$.
Once we have \zgas{}, we use the line ratio  \iontwo{O}{III}/\iontwo{O}{II} to
calculate $q$, using the fact that  this ratio  depends heavily on $q$ 
(bottom right panel of Figure \ref{fig:degeneracy}). 
Specifically, we derive $q$ by placing 
 \zgas{} and \iontwo{O}{III}/\iontwo{O}{II} into the following equations from \cite{kobulnicky04}:
\begin{align}
y =& \log\left({\mbox{\iontwo{O}{III}} \over \mbox{\iontwo{O}{II}}}\right) \\ 
Z _{\rm gas} =& \log(\mbox{O/H})+12  \\ 
\log q =& 32.81-1.153y^2  \label{eq:q} \\
 &+ Z_{\rm gas}(-3.396-0.025y+0.1444y^2) / \nonumber \\
 &[4.603-0.3119y-0.163y^2 + \nonumber \\
  &Z_{\rm gas}(-0.48+0.0271y + 0.02037y^2)]  \nonumber
\end{align}

We discuss an alternative method for finding  $q$ in Appendix  \ref{sec:q2-append}. 
That method gets $q$ from the \iontwo{O}{III}/\iontwo{O}{II} line ratio, but it  
solves iteratively for \zgas{}  and $q$,
to ensure they are consistent with multiple line ratios such as 
\iontwo{O}{III}/\iontwo{O}{II}, \iontwo{N}{II}/\iontwo{O}{II}, and $R_{23}$ (Figure \ref{fig:degeneracy}).
Both methods give similar results for $q$.

\subsection{Computation of \zgas{} Using Different Diagnostics} \label{sec:how-zgas}

The most direct method for computing \zgas{} from emission line fluxes
involves measuring  the electron temperature ($T_e$) from 
the flux ratio of the \iontwo{O}{III} lines: \iontwo{O}{III}$\lambda$4363 to
\iontwo{O}{III}$\lambda\lambda$4959,5007.  
By placing the measured value of $T_e$ into a classical \HII{} region model, 
one finds \zgas{} \citep{osterbrock}. 
Our VENGA observations do not go deep enough to detect the emission of
the \iontwo{O}{III}$\lambda$4363 line with high enough $S/N$ to use the direct
$T_e$ method.  
Instead, taking advantage of the broad wavelength coverage (3600-6800 \AA{})
and $\sim 120$ km s$^{-1}$ spectral resolution  of the VENGA IFU  data, 
we use
 several stronger detectable  emission lines  
(\iontwo{O}{II}, \iontwo{O}{III}, H$\beta$, H$\alpha$, \iontwo{N}{II})  to compute 
four different
widely used
\zgas{} indicators    ($R_{23}$, N202, N2, O3N2),
which are defined, respectively,  by  the  line ratios
$R_{23} \equiv$ (\iontwo{O}{II}$\lambda$3727+\iontwo{O}{III}$\lambda\lambda$4959,5007)/H$\beta$,   
N202 $\equiv$ \iontwo{N}{II}$\lambda$6584/\iontwo{O}{II}$\lambda$3727,  
N2 $\equiv$ \iontwo{N}{II}$\lambda$6584/H$\alpha$, and  
 O3N2 $\equiv$ (\iontwo{O}{III}$\lambda\lambda$4959,5007/H$\beta$)/(\iontwo{N}{II}$\lambda$6584/H$\alpha$) (Table \ref{tab:zgas-indicators}).

It is extremely important  to have multiple \zgas{} indicators 
in order to break degeneracies  and  explore the systematics
between  indicators. 
Different methods to determine \zgas{} may differ from each other by up to 
0.7 dex \citep{kewley2008, lopezsanchez2012, blanc2015}.
The variation of $R_{23}$  with \zgas{}  and $q$ is shown in 
the top left of Figure \ref{fig:degeneracy}. 
It can
also be seen  that for moderate to high values of  $R_{23}$,
we have a degeneracy  with two possible ranges of \zgas{} for every 
$R_{23}$  value  at fixed $q$.  We call these ranges the low and high branches of \zgas{} for  $R_{23}$ \citep{kewley2002, kobulnicky04}.
\HII{} regions have temperatures typically around $10^4$ K,
so are hot enough to collisionally excite the forbidden \iontwo{O}{II} and \iontwo{O}{III}
transitions. 
On  the lower branch  of \zgas{} where  $R_{23}$  rises with \zgas{}
at a given $q$,  
the main coolant is oxygen and the strength of the \iontwo{O}{II} \& \iontwo{O}{III} lines increases with
oxygen abundance. 
On the higher branch  of \zgas{},  $R_{23}$ falls  with \zgas{} at a
given $q$ because at high  \zgas{},  cooling is dominated by far-IR
fine structure metal lines and the electron temperature  becomes too
low to collisionally excite the \iontwo{O}{II} \& \iontwo{O}{III} lines 
\citep{pagel1979}.
Adding in other line ratios can help solve for $q$ and break 
\zgas{} degeneracies. 
The \iontwo{O}{III}/\iontwo{O}{II}  line ratio depends heavily on $q$  
(bottom right panel of Figure \ref{fig:degeneracy})
due to the different ionization potentials of  O$^+$ and  O$^{++}$, while
the N202 indicator
(\iontwo{N}{II}/\iontwo{O}{II} line ratio)  does not depend strongly 
on $q$ and maps  monotonically onto \zgas{} (bottom left panel of Figure 
\ref{fig:degeneracy})
due to the similar
ionization potentials of N$^+$ and O$^+$. 

For our study, we do not build our own empirical or theoretical \zgas{} diagnostics,
 but instead use 
 seven of the eight
published
\zgas{} diagnostics detailed in \cite{kewley2008}
based on the four \zgas{} indicators ($R_{23}$, N202, N2, O3N2) to derive \zgas{}
in our targets:
$R_{23}$-M91 \citep{mcgaugh91}, $R_{23}$-KK04 \citep{kobulnicky04}, 
$R_{23}$-Z94 \citep{zaritsky1994}, N2O2-KD02 \citep{kewley2002},  
N2-D02 \citep{deincolo02}, N2-PP04 \citep{pettini2004},
and O3N2-PP04  \citep{pettini2004}
(see list of \zgas{} diagnostics in Table \ref{tab:zgas-indicators}).
We do not use  $R_{23}$-P05 \citep{pilyugin05} due to its large systematic offset from the other diagnostics. 

Different \zgas{}  diagnostics using the same \zgas{} indicator can 
yield different metallicities
due to the different theoretical and empirical calibrations 
 and corrections for variations in the ionization
parameter $q$ (see Appendix 
\ref{sec:zgas-append}).
For instance, $R_{23}$-M91 and $R_{23}$-KK04 both correct for variations in $q$, but correct for $q$ via two 
different methods (outlined in $\S$~\ref{sec:how-q}), while  the 
$R_{23}$-Z94 diagnostic does not correct for $q$.  
Another difference is that  some \zgas{} calibrations we use
are ``empirical''  and others are ``theoretical.''
Empirical calibrations involve calibrating the strong emission line ratios observed in \HII{} regions 
to the \zgas{} derived from the direct $T_e$ method.
Theoretical \zgas{} calibrations base their line ratios on models of stellar
populations, radiative transfer, and nebular photo-ionization \citep{dopita2000, kewley2002}.

We  compute the full suite of  seven \zgas{} diagnostics from our
VENGA data. We start by 
making a signal-to-noise cut of $S/N> 5$ for the Balmer decrement line
 ratio
 (H$\alpha$/H$\beta$ = 2.86)
 to ensure we have a reliable extinction
 correction, and $S/N >3$  for all the emission line fluxes used in each \zgas{} diagnostic.  
We exclude all regions that are LINER or Seyfert on the excitation diagnostic diagrams 
($\S$ \ref{sec:bpt-method}), along with all regions flagged as DIG-dominated ($\S$ \ref{sec:dig}),
and only use regions where the gas is primarily photo-ionized by massive stars from recent SF.  
For each spaxel, we compute the line ratios for the four
indicators ($R_{23}$, N202,N2, O3N2)
and then apply the different calibrations to derive the seven \zgas{}
diagnostics (Table \ref{tab:zgas-indicators}),   following the steps
detailed in \cite{kewley2008} and Appendix \ref{sec:zgas-append}.
This is possible for seven (NGC 0337, 0628, 2903, 3938, 4254, 5194, \& 5713)
of the eight  galaxies in our sample. 
The rejected galaxy, NGC 1068, is classified as a Seyfert 2 AGN and most of the spaxels
in both the nuclear regions and extended disc
show Seyfert type excitation conditions (as seen in Figure \ref{fig:bpt-sliced}) so we remove this galaxy from further \zgas{} analysis.

\section{Results and Discussion} \label{sec:results}

\subsection{2D Maps of SFR, $q$, and \zgas{} Across Different Galactic Components}\label{sec:results-components}

Figure \ref{fig:2d-maps} shows a montage of 2D maps of  $q$ and six \zgas{}
diagnostics ($R_{23}$-KK04, $R_{23}$-M91, N202-KD02, O3N2-PP04, N2-D02, \& N2-PP04), along    
with  the stellar continuum and  extinction-corrected H$\alpha$-based SFRs.
We do not show the seventh \zgas{} diagnostic $R_{23}$-Z94 because it gives very similar results to $R_{23}$-KK04.
Two of these galaxies are barred and isolated (NGC 2907 \& 5713), three are unbarred and isolated (NGC 0628, 3938, \& 4254),
one is barred and weakly interacting (NGC 0337), and one is unbarred
and weakly interacting (NGC 5194).
The high spatial resolution (median FWHM of the PSF (FPSF)  of 387 pc and a median ($R_e$-bulge/FPSF) 
of 2.2)  and IFU data coverage over a large fraction of the galaxy's
disc (median \frtwofive{} of 0.66) in our sub-sample allows us to
resolve individual galaxy components such as the bulge, bar, and outer
disc, and thereby explore how $q$, SFR, and \zgas{} vary across these components for the seven galaxies.

\subsubsection{SFRs} \label{sec:results-sfr}

For the central  10 to 15 kpc mapped by VENGA, our galaxies have an extinction-corrected H$\alpha$-based SFR of 0.50 to 9.50 $\ M_\odot$ yr$^{-1}$.
We report the total measured H$\alpha$ SFRs  for our VENGA subsample of galaxies along with their TIR-based SFRs in Table \ref{tab:sfr}.
Our H$\alpha$-based SFRs for all the galaxies in our sub-sample are
less than their  global TIR-based SFRs (see $\S$ \ref{sec:sfr} and
Table \ref{tab:sfr}), with our H$\alpha$ based SFRs ranging from 
34 to 81\%  of the TIR-based global SFRs.
We show the deprojected radial profiles of the SFR surface density  $\Sigma_{\rm SFR}$ ($M_\odot$ yr$^{-1}$ kpc$^{-2}$) on the right side of  Figure \ref{fig:q-sfr-gradients}.
 The average
 $\Sigma_{\rm SFR}$  falls by over an order of magnitude from the inner 2 kpc to the outer disc.
(e.g. from 0.1 to below 0.01 or from 0.01 to below 0.001 
$M_\odot$ yr$^{-1}$ kpc$^{-2}$).
The 2D maps of the H$\alpha$ based SFR ($\S$~\ref{sec:sfr}) can be seen in the top centre of Figure \ref{fig:2d-maps}, and the maps show that SF tends to be concentrate and  in the nuclear regions, bar, and spiral arms and becomes less prominent in the inter-arm regions of the outer disc.  The SF shows the patchy morphology expected from the distribution of \HII{} regions throughout a spiral galaxy.

\begin{figure*}
\includegraphics[width=0.32\textwidth]{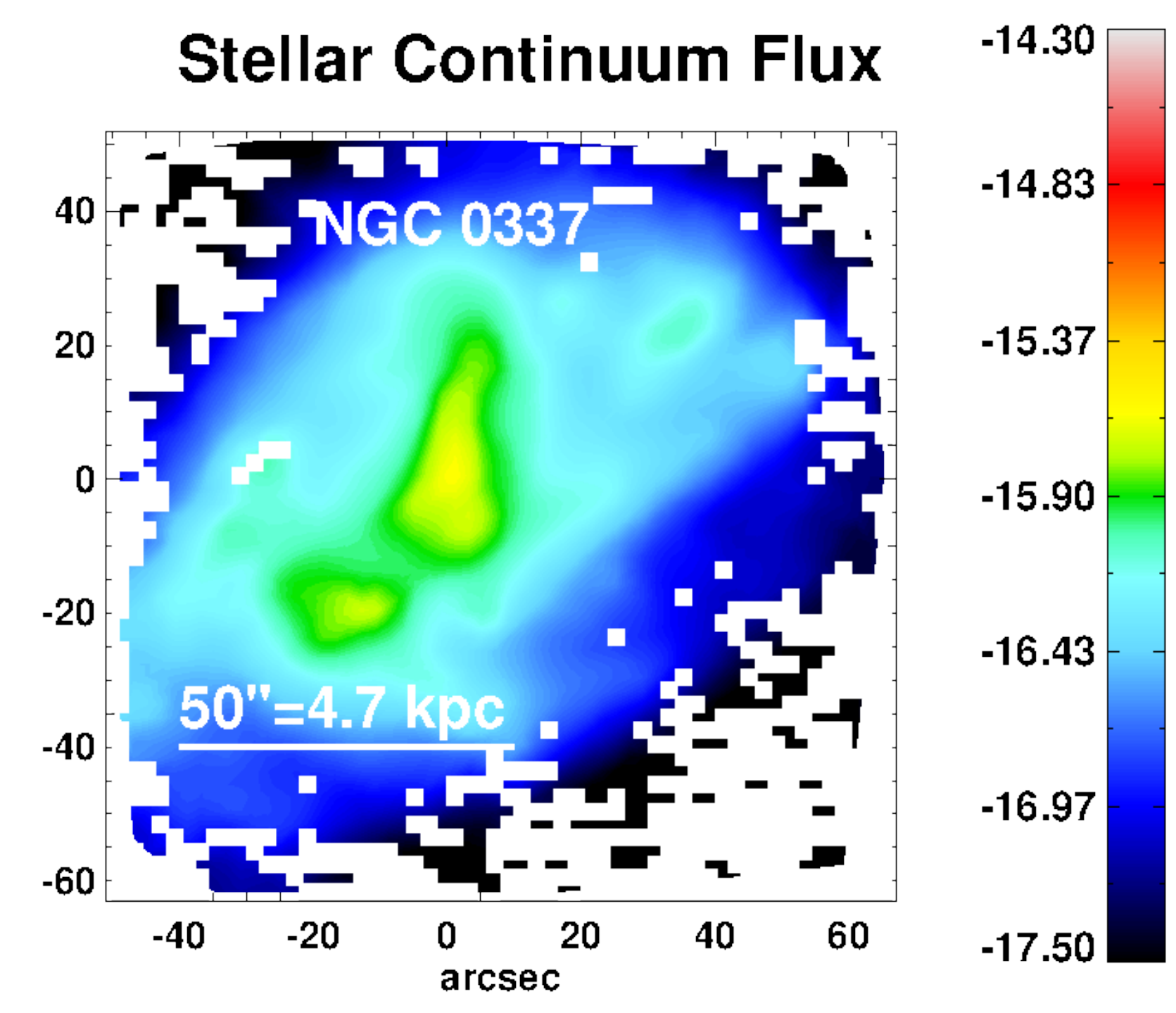}
\includegraphics[width=0.32\textwidth]{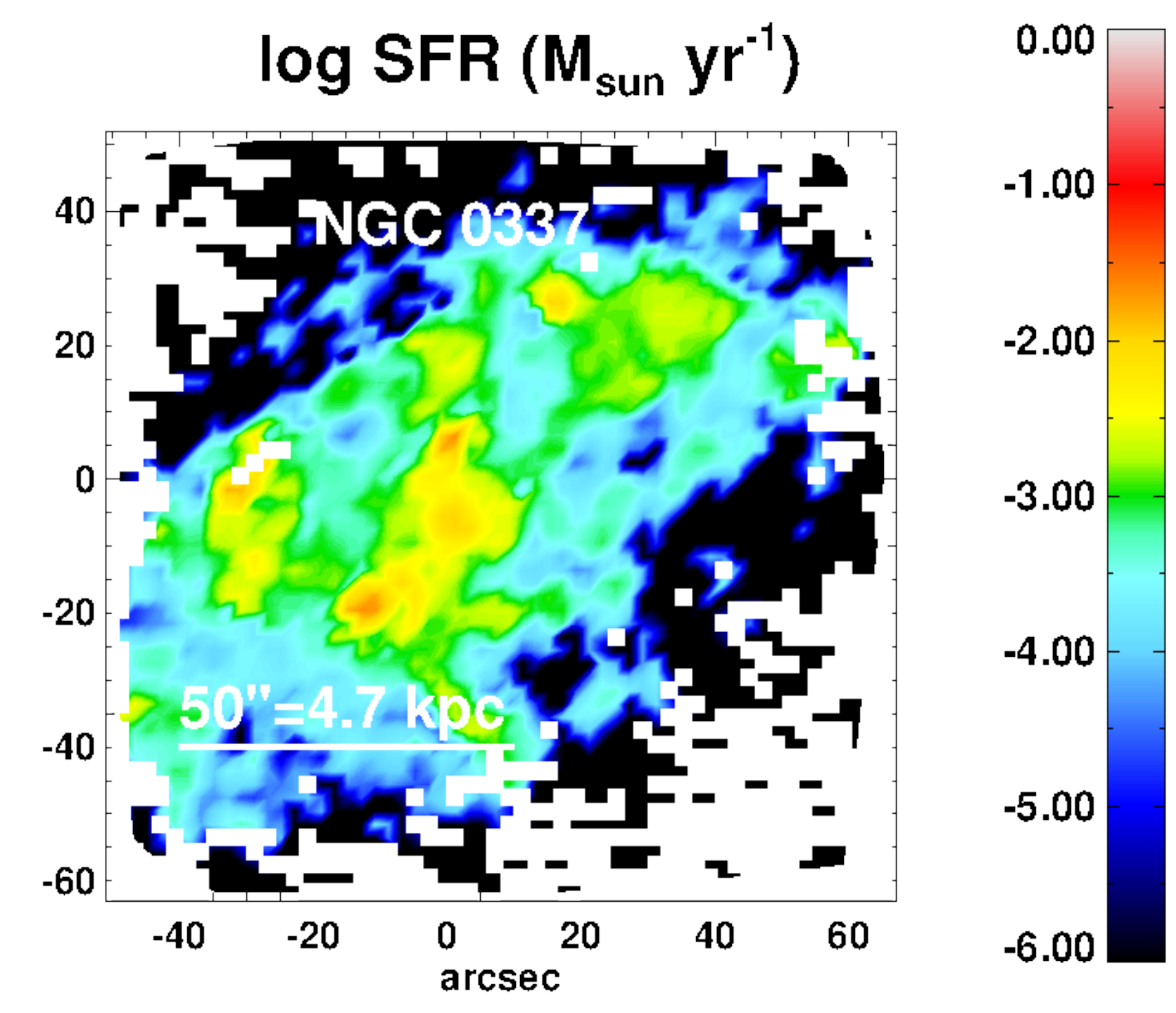}
\includegraphics[width=0.32\textwidth]{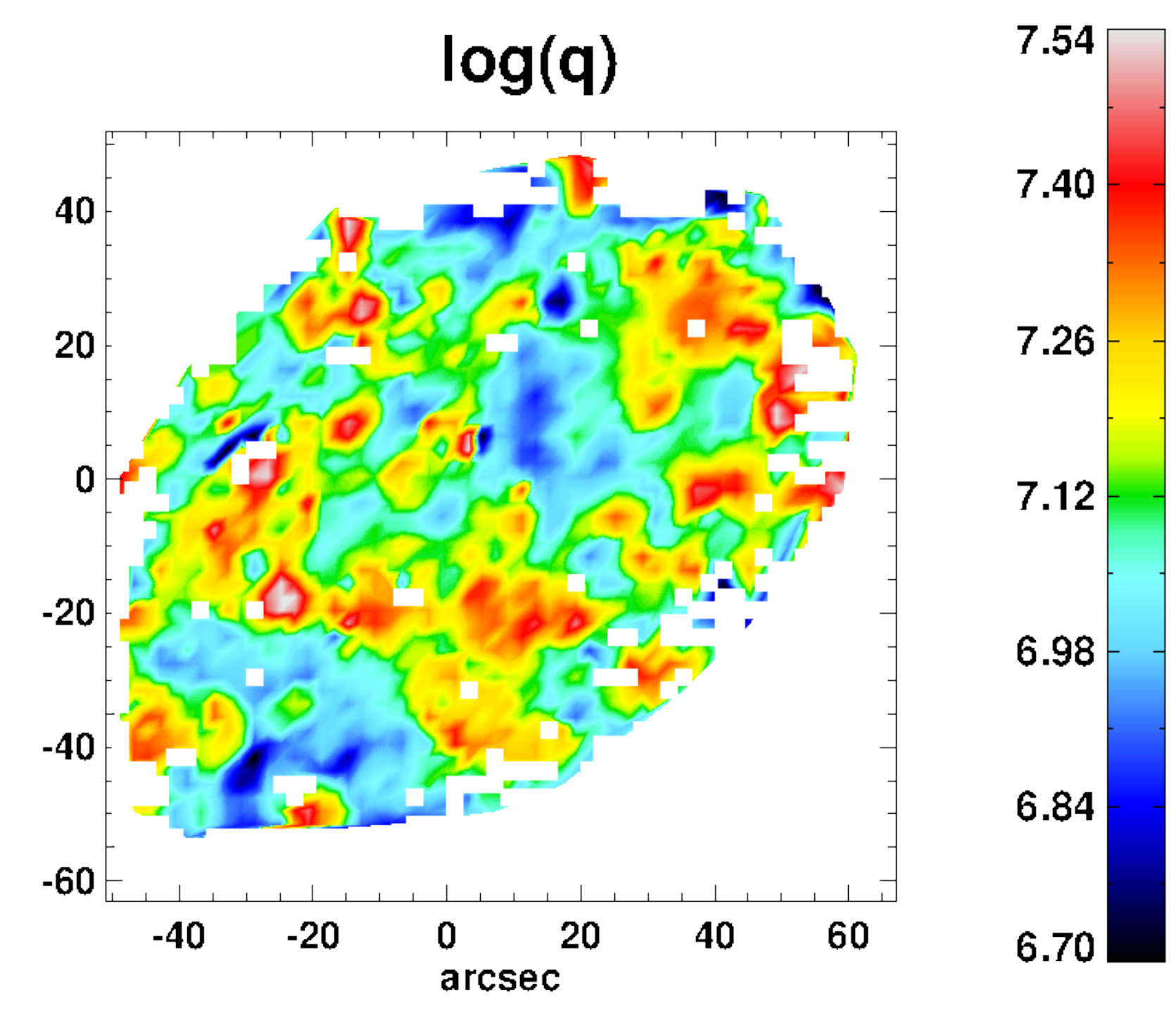}\\
\includegraphics[width=0.32\textwidth]{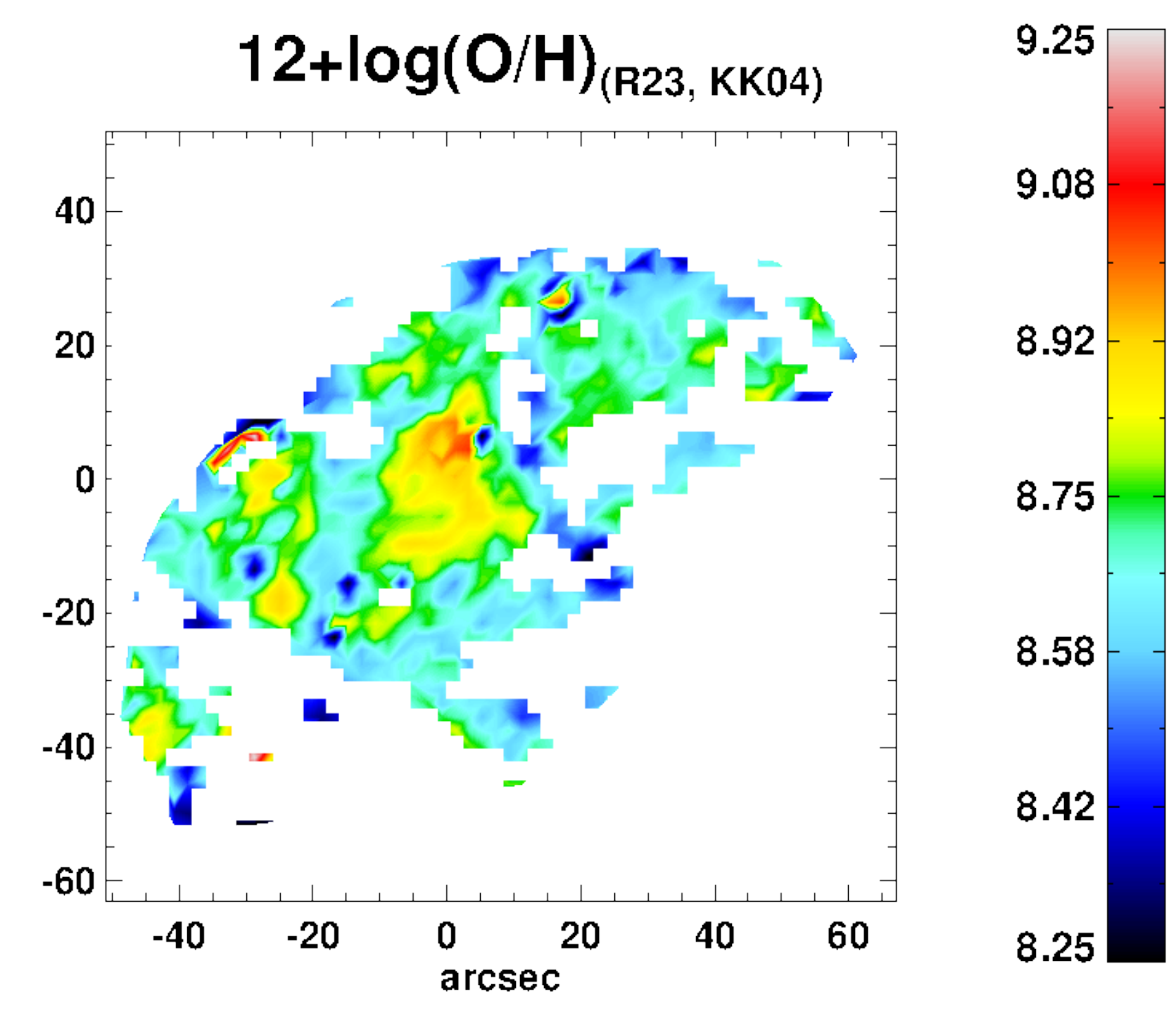}
\includegraphics[width=0.32\textwidth]{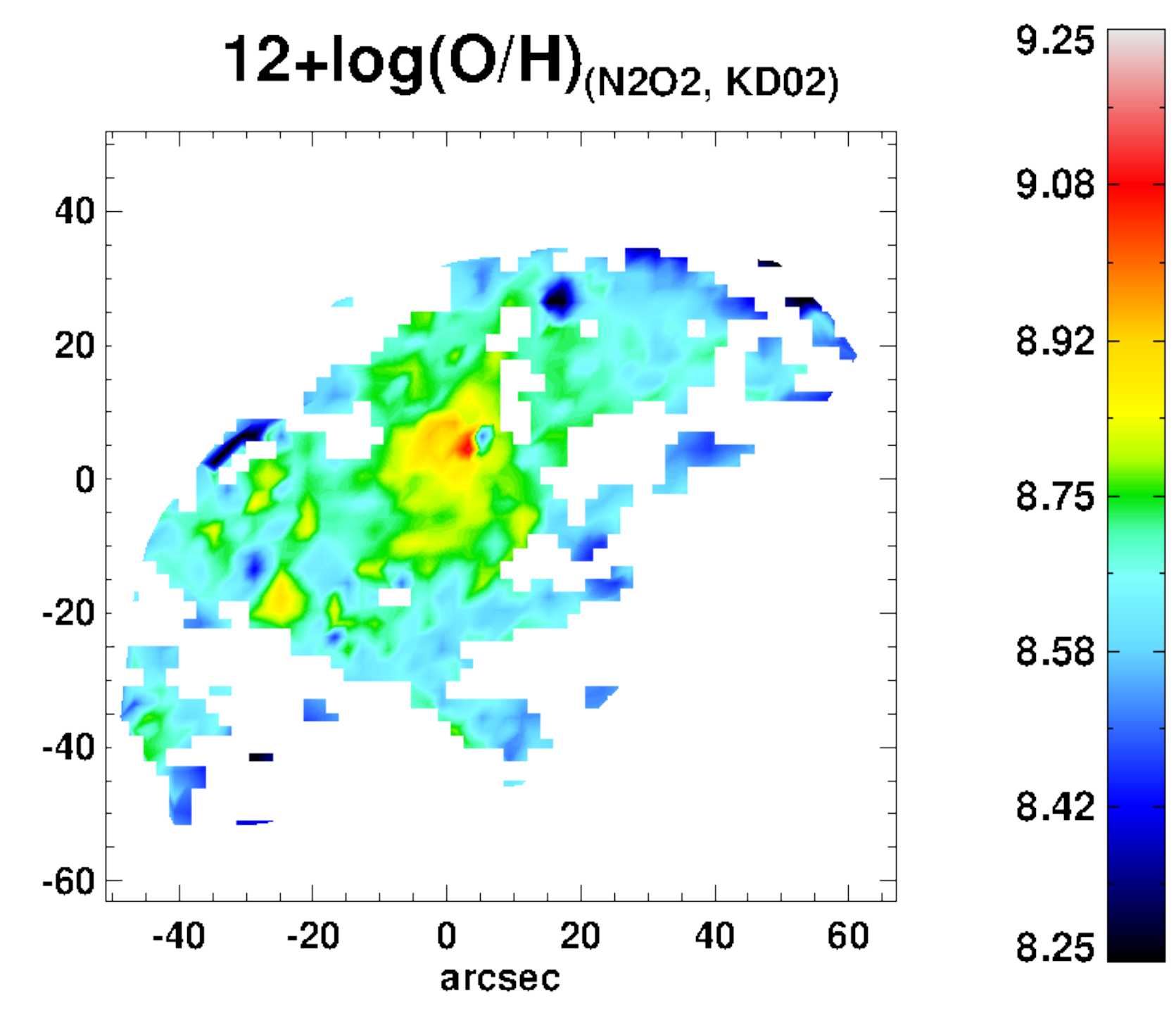}
\includegraphics[width=0.32\textwidth]{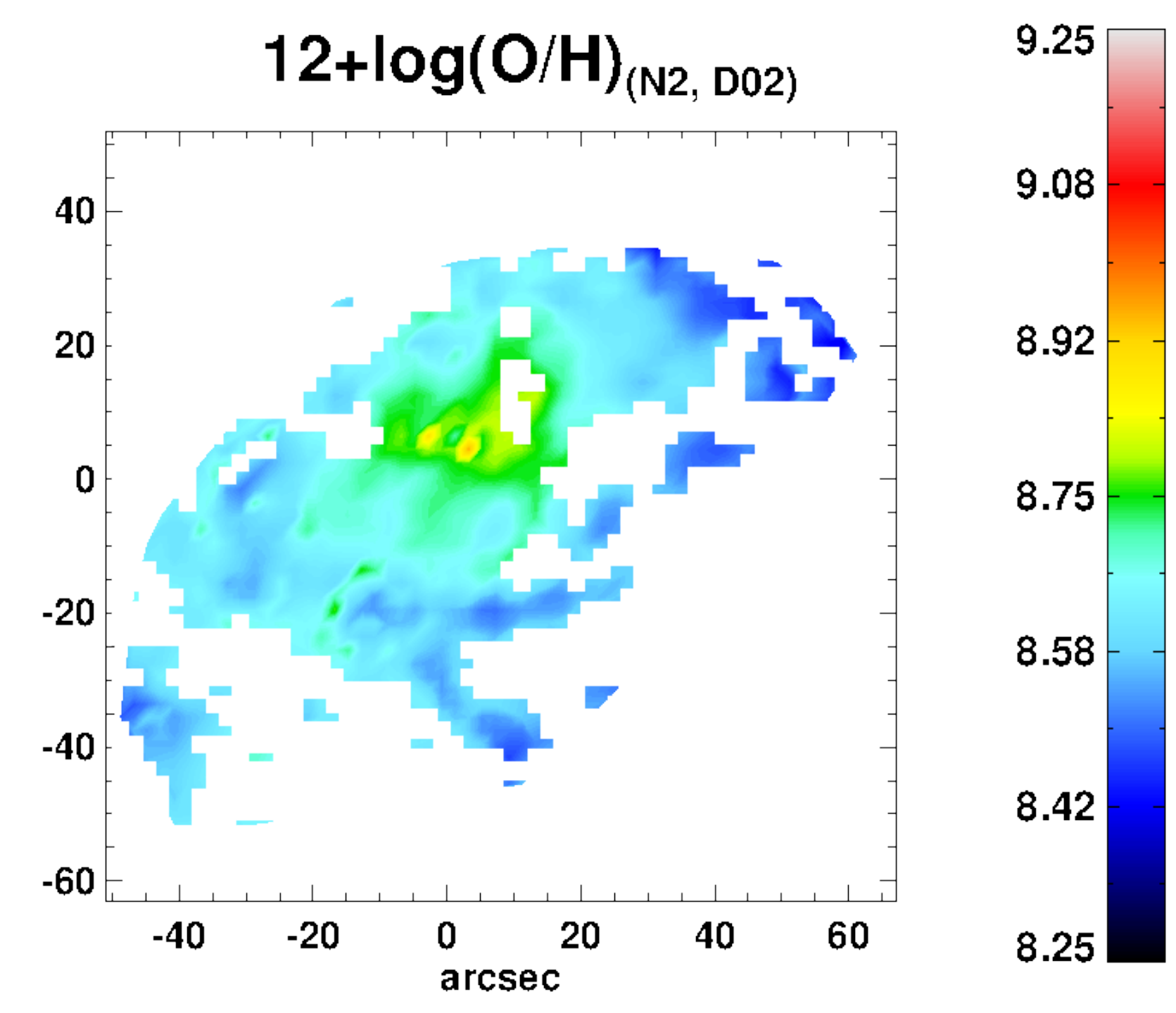} \\
\includegraphics[width=0.32\textwidth]{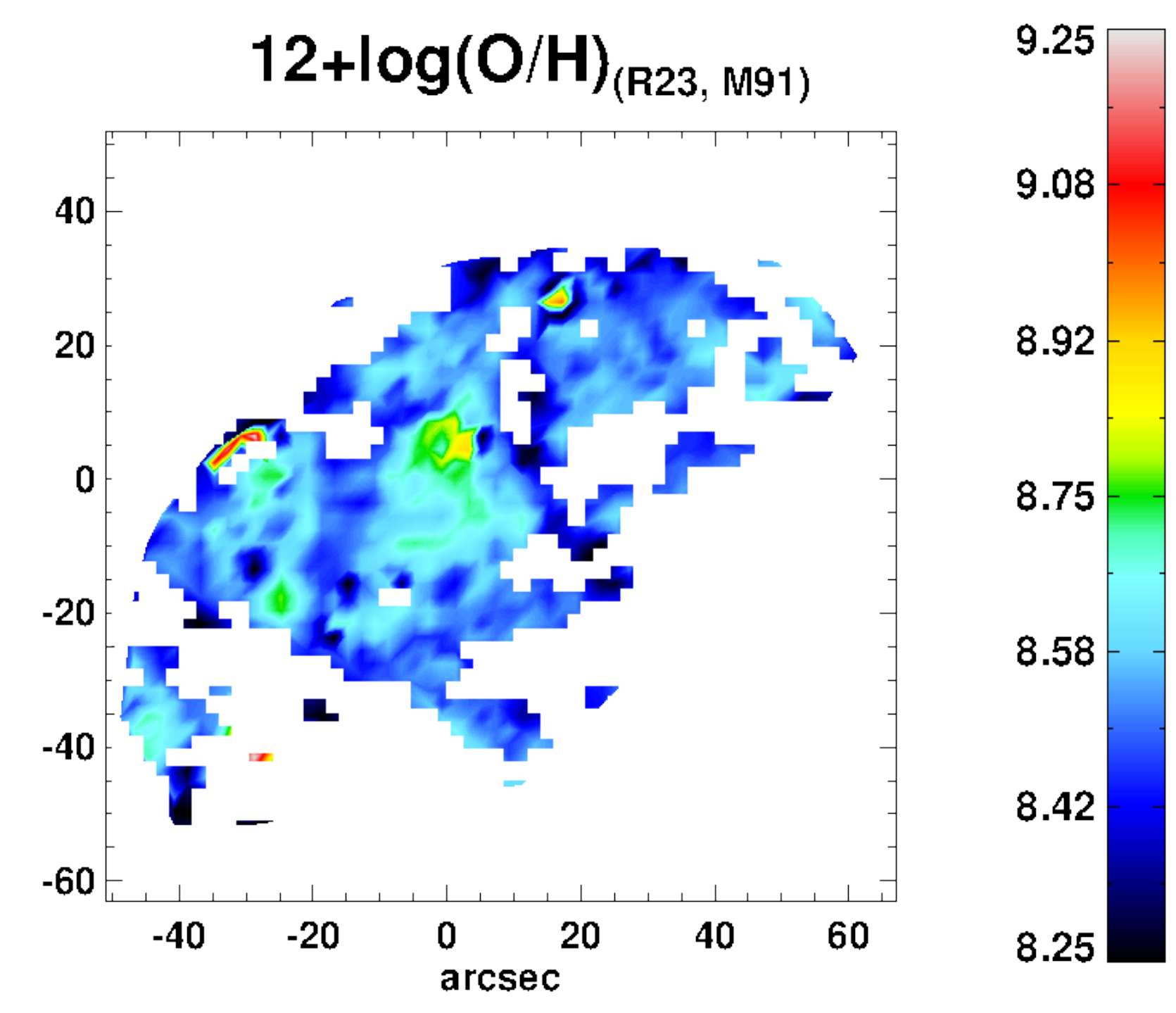}
\includegraphics[width=0.32\textwidth]{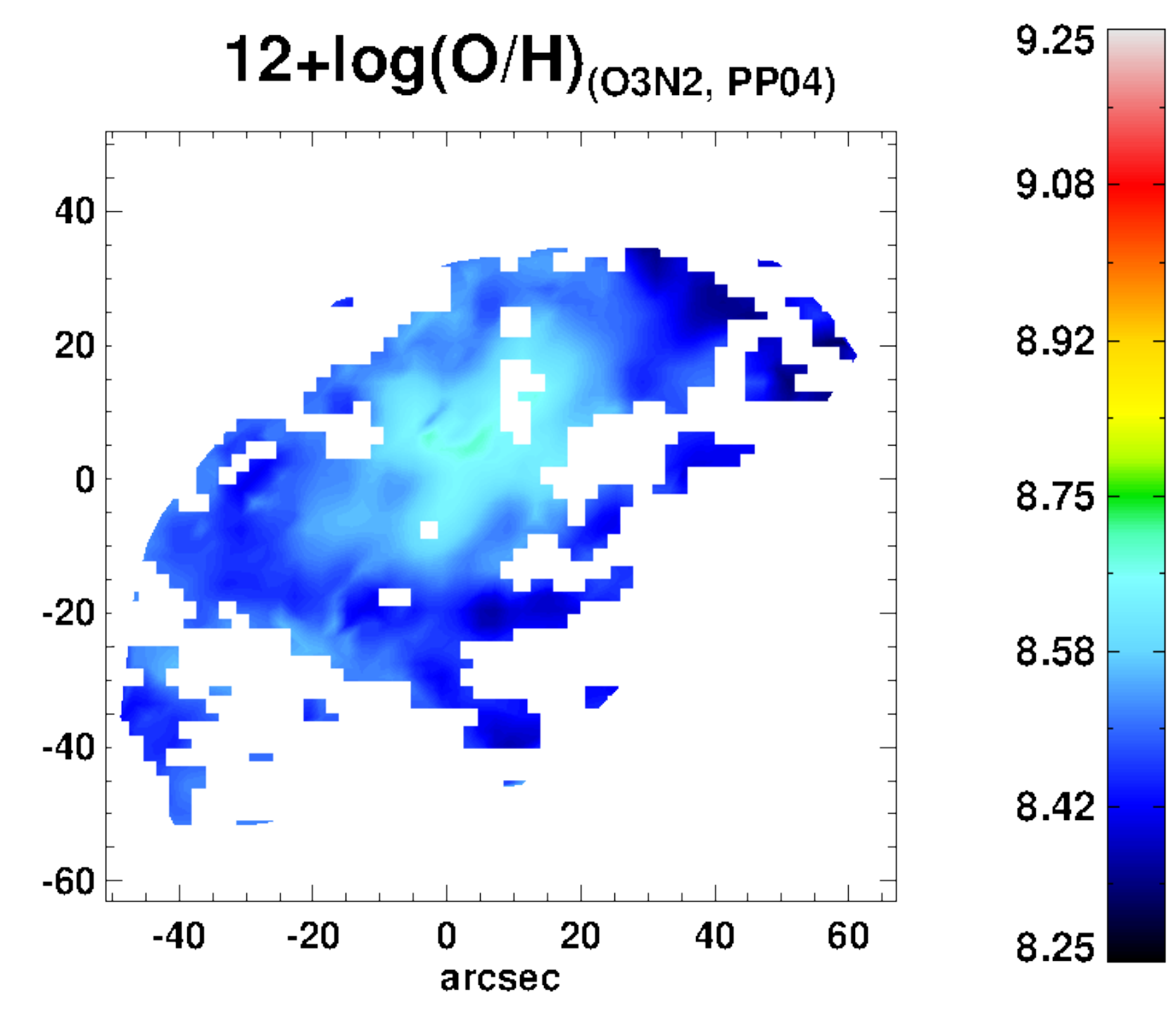}
\includegraphics[width=0.32\textwidth]{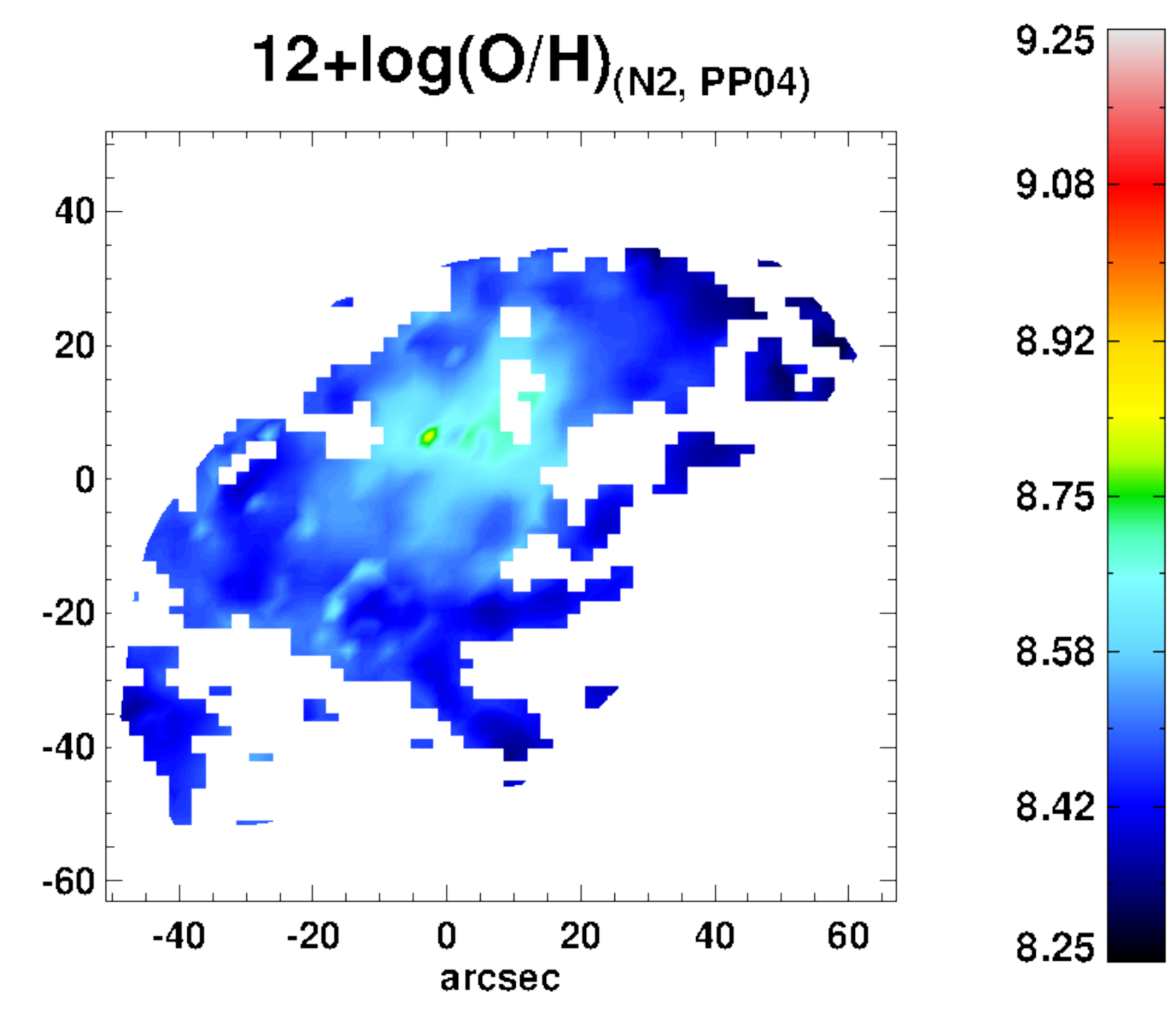}
\caption{
For NGC 0337, we show the following from  left to right:
{\bf Top Row:}  
Bolometric stellar flux in units of erg s$^{-1}$ cm$^{-2}$ \AA{}$^{-1}$; 
SFR  in units of $M_\odot$ yr$^{-1}$ derived from extinction corrected H$\alpha$ flux 
using the calibration from \protect\cite{kennicutt2012}; 
Ionization parameter $q$  in units of the  cm s$^{-1}$ calculated as described 
in $\S$ \ref{sec:how-q}; 
{\bf Middle and Bottom Row:}  
Gas phase metallicity \zgas{}, or log(O/H)+12, based on six 
different \zgas{} diagnostics $R_{23}$-KK04, N2O2-KD02, N2-DO2, 
$R_{23}$-M91, O3N2-PP04, N2-PP04. 
}
\label{fig:2d-maps}
\end{figure*}

\addtocounter{figure}{-1}
\clearpage

\begin{landscape}
\begin{figure}
\includegraphics[width=0.43\textheight]{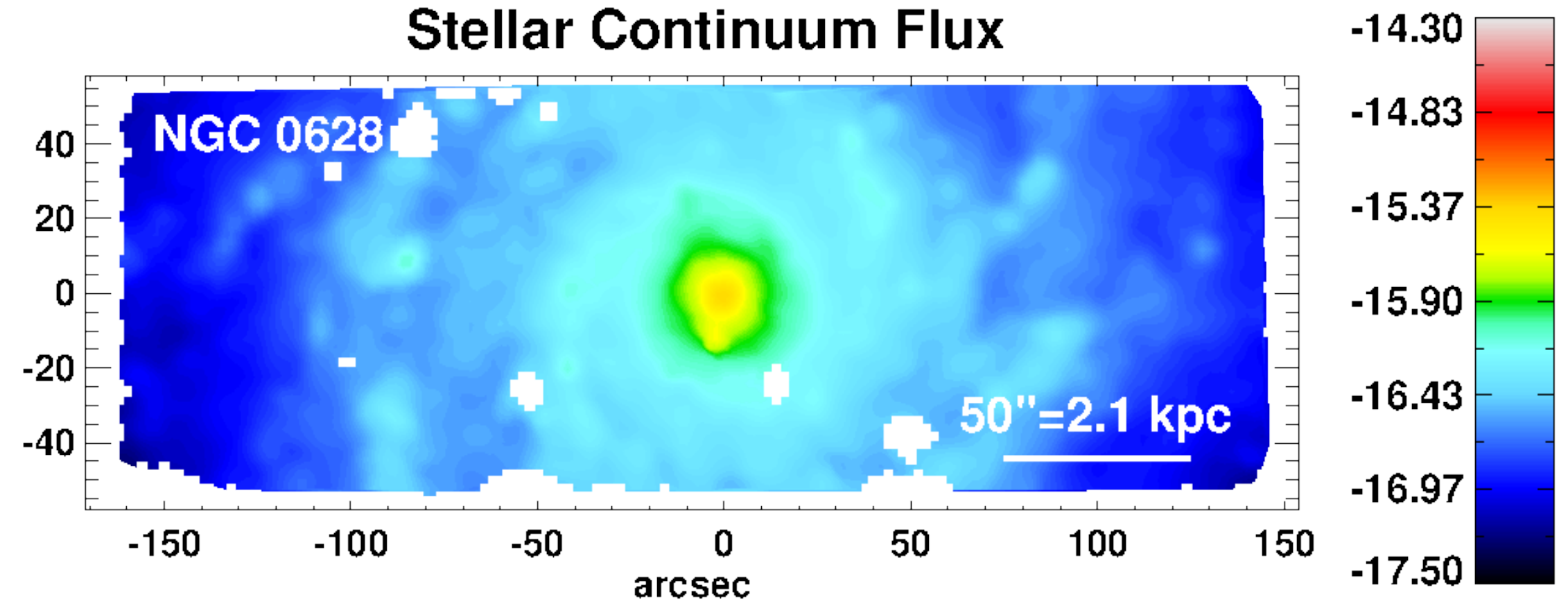}
\includegraphics[width=0.43\textheight]{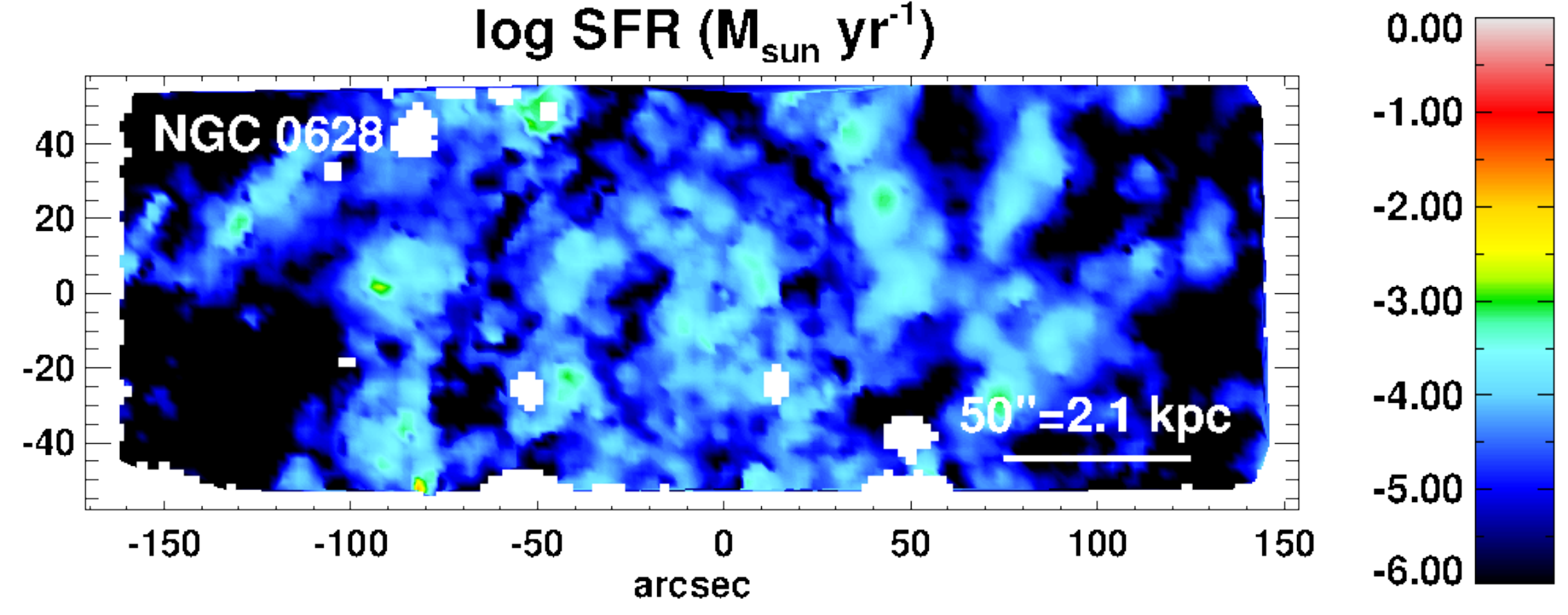}
\includegraphics[width=0.43\textheight]{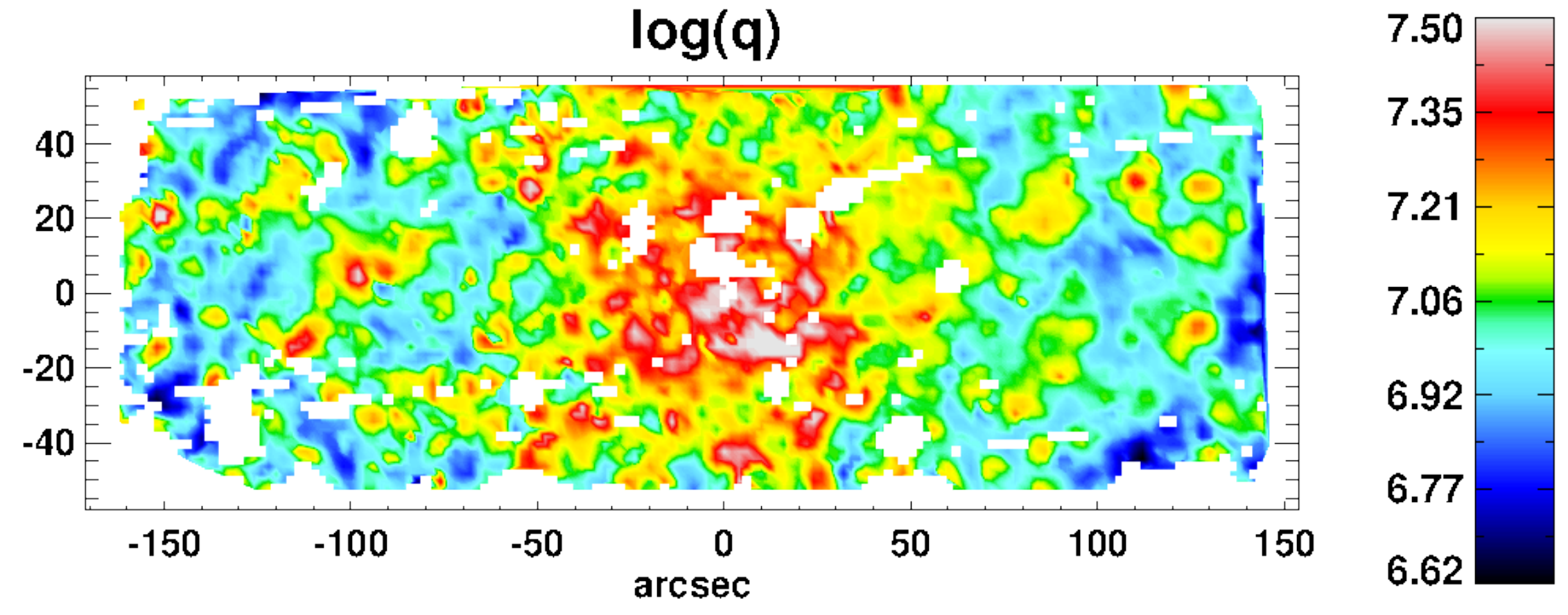}\\
\includegraphics[width=0.43\textheight]{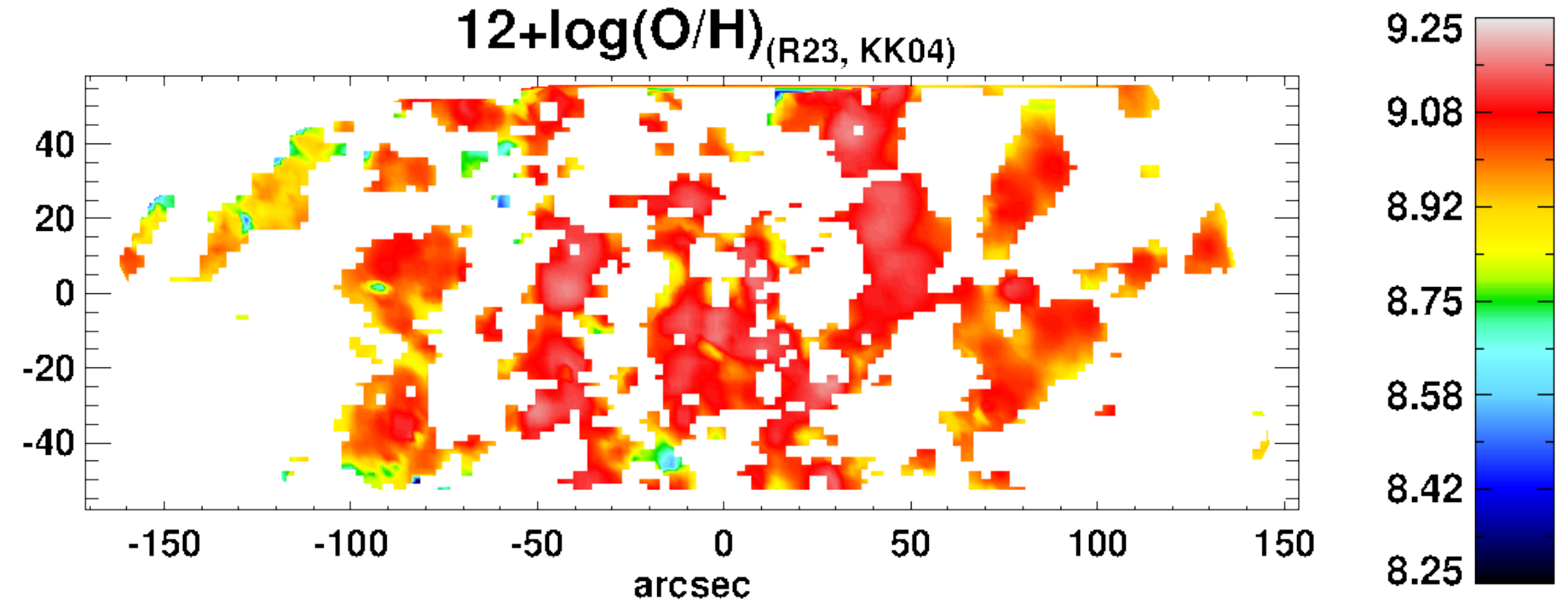}
\includegraphics[width=0.43\textheight]{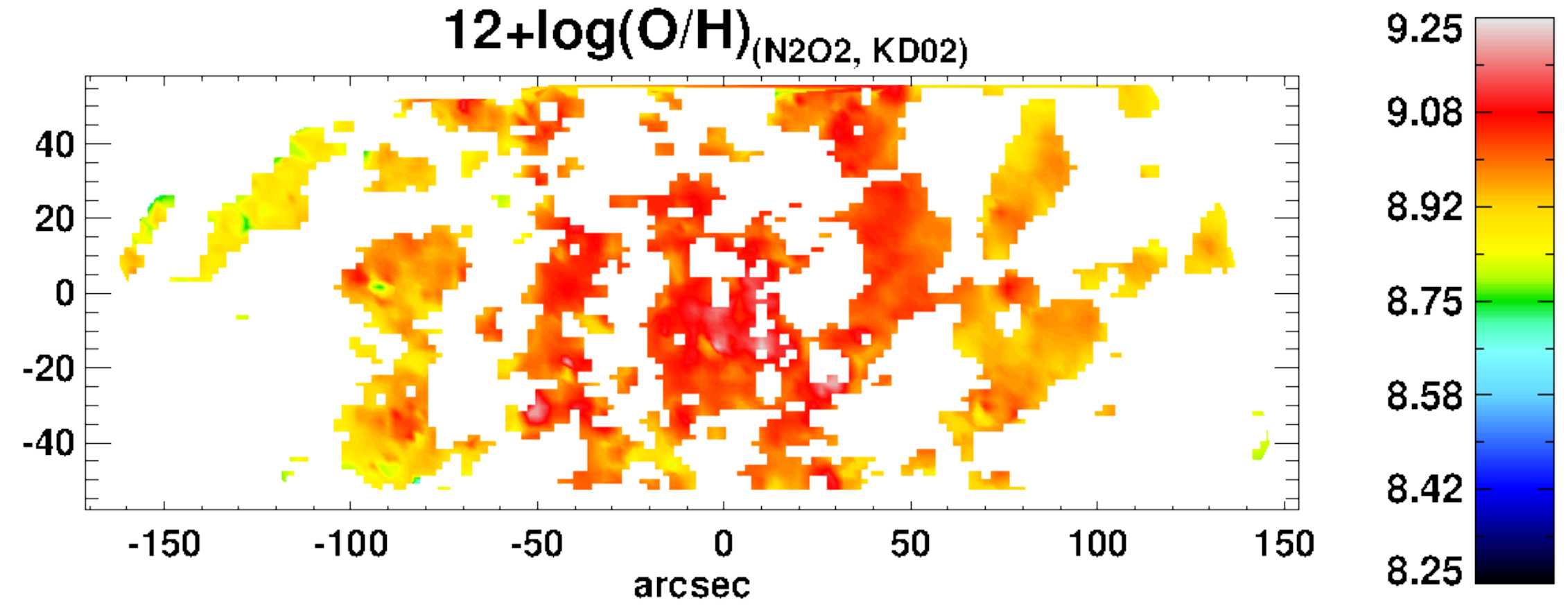}
\includegraphics[width=0.43\textheight]{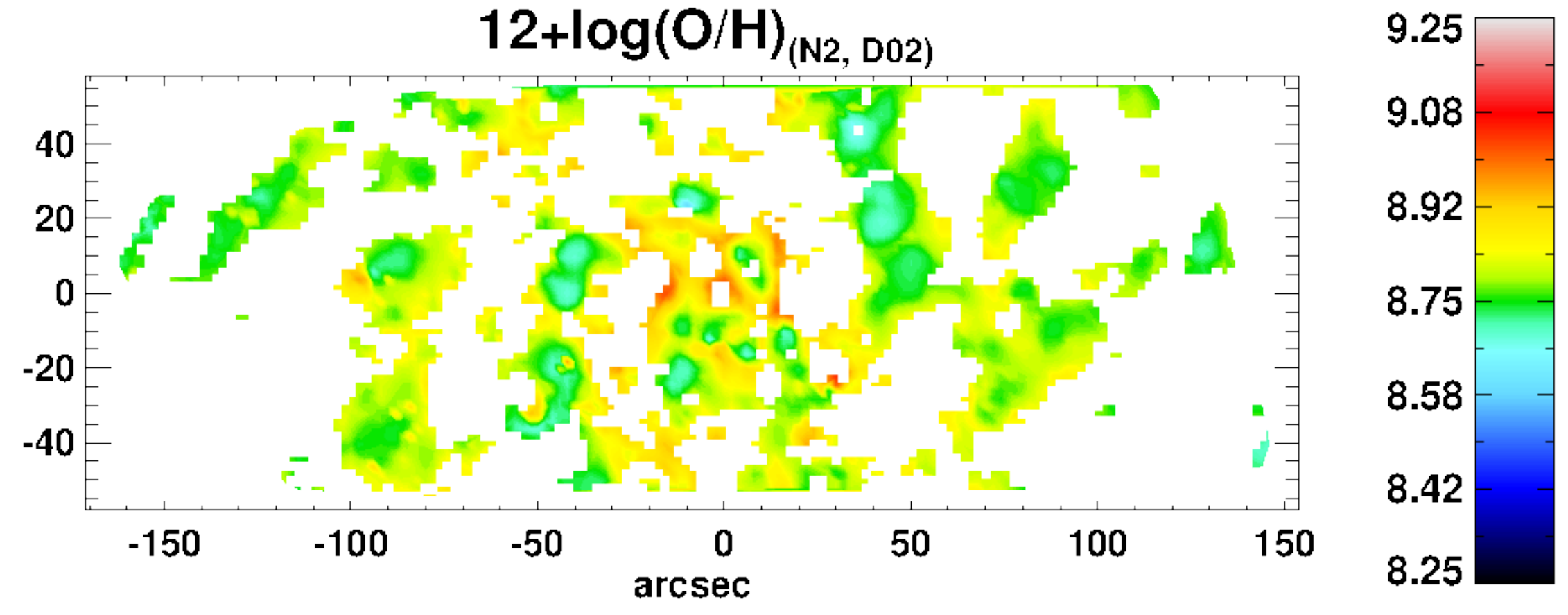} \\
\includegraphics[width=0.43\textheight]{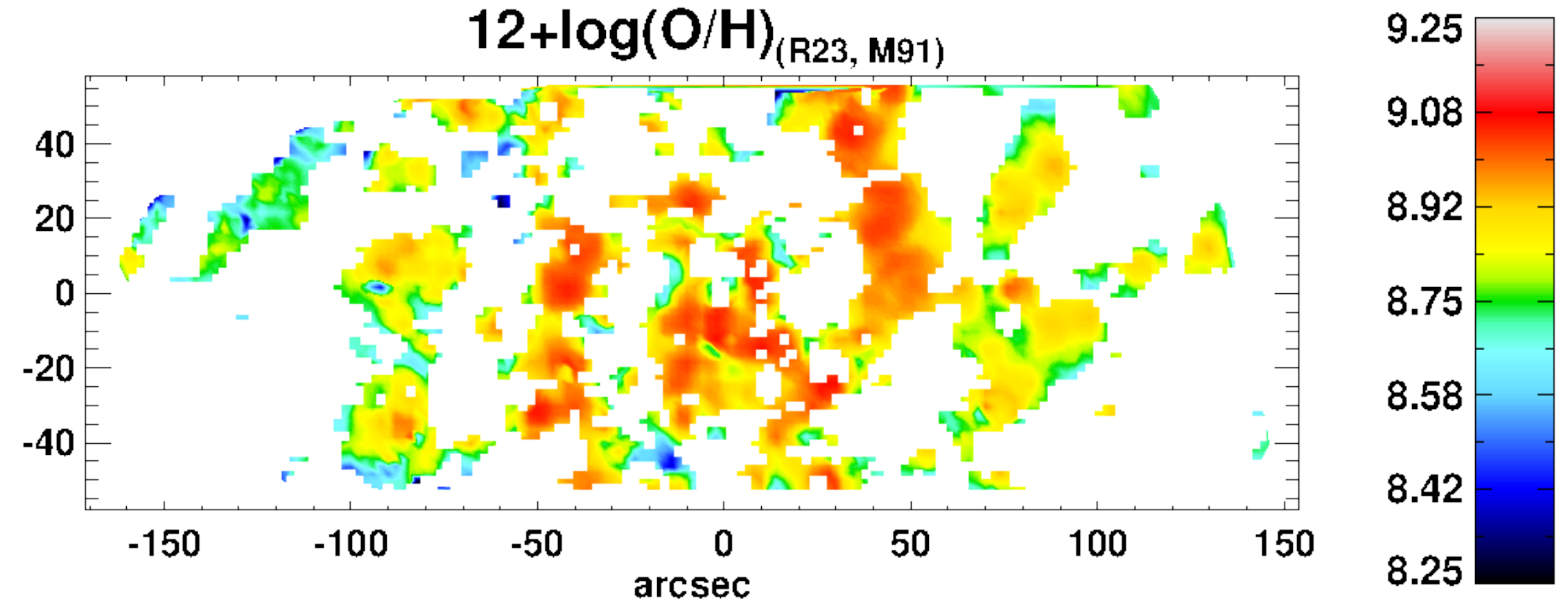}
\includegraphics[width=0.43\textheight]{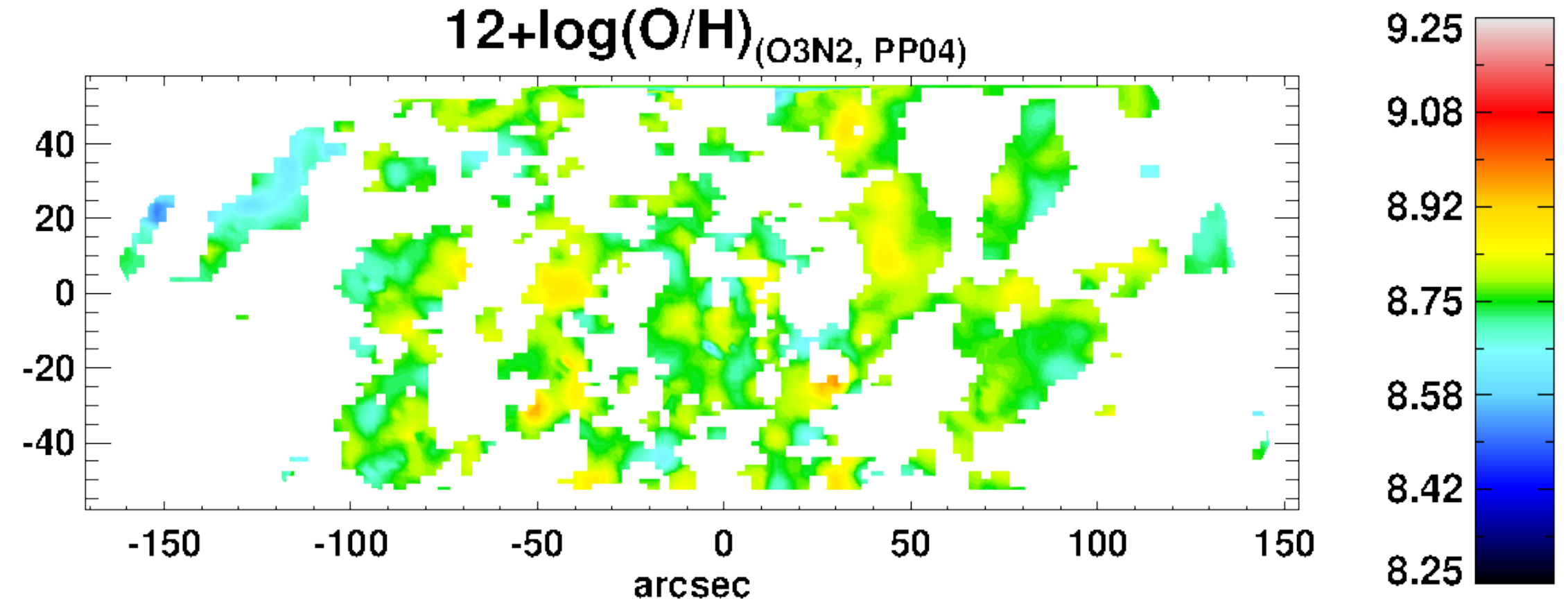}
\includegraphics[width=0.43\textheight]{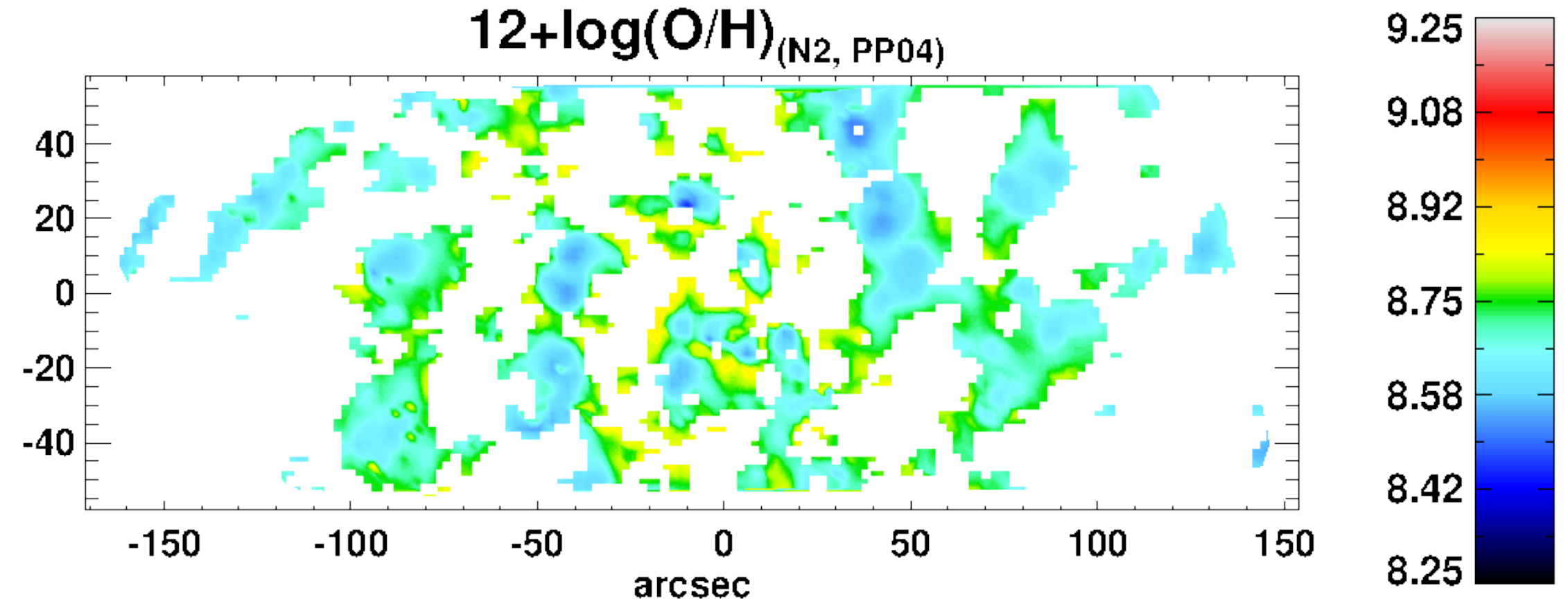}
\caption{Continued: For NGC 0628}
\label{}
\end{figure}
\end{landscape}

\addtocounter{figure}{-1}
\clearpage

\begin{figure*}
\includegraphics[height=0.32\textheight]{figures/spring2013/A3_map_color_stellar_flux.ngc2903.pdf}
\includegraphics[height=0.32\textheight]{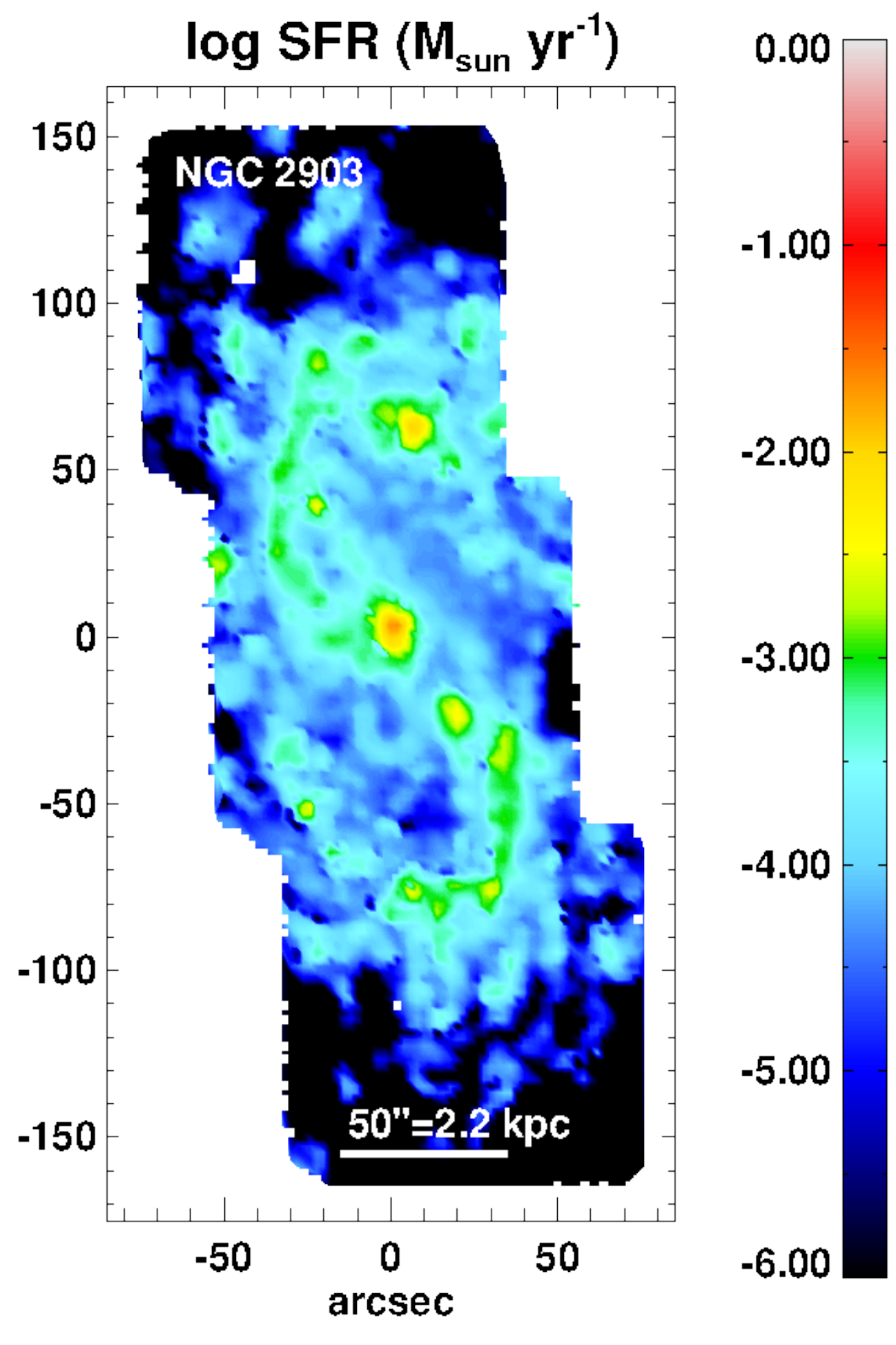}
\includegraphics[height=0.32\textheight]{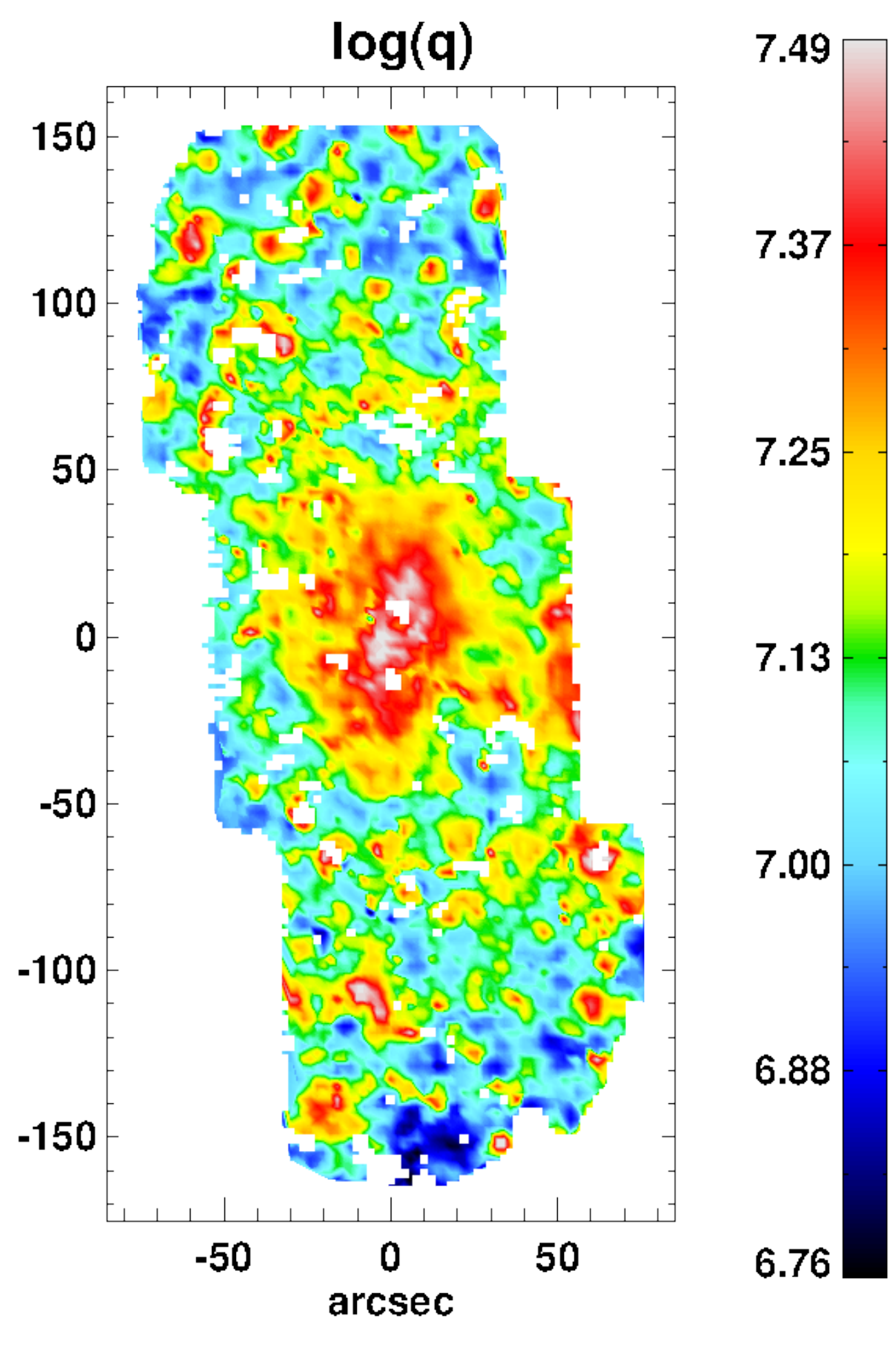}\\\
\includegraphics[height=0.32\textheight]{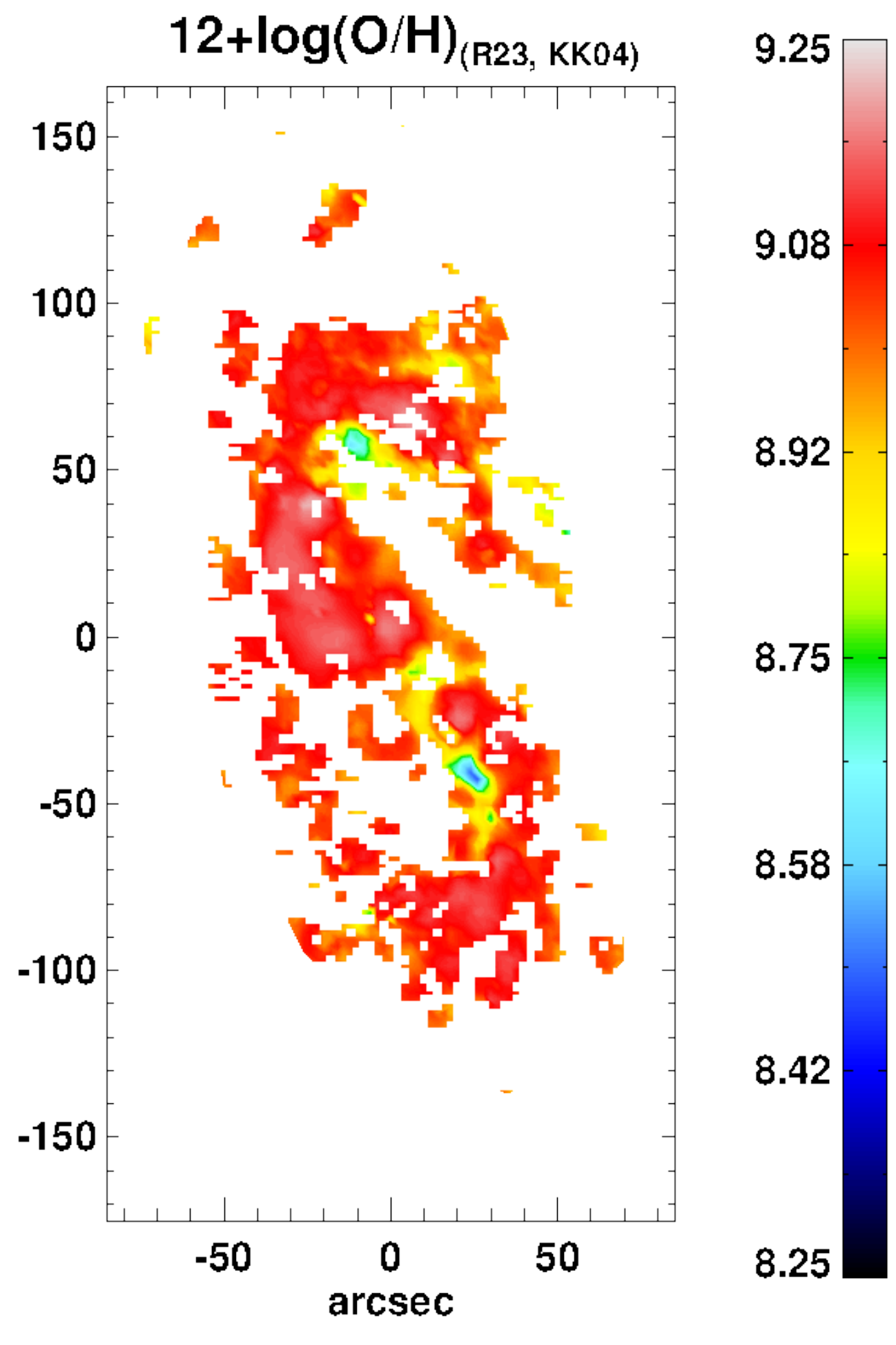}
\includegraphics[height=0.32\textheight]{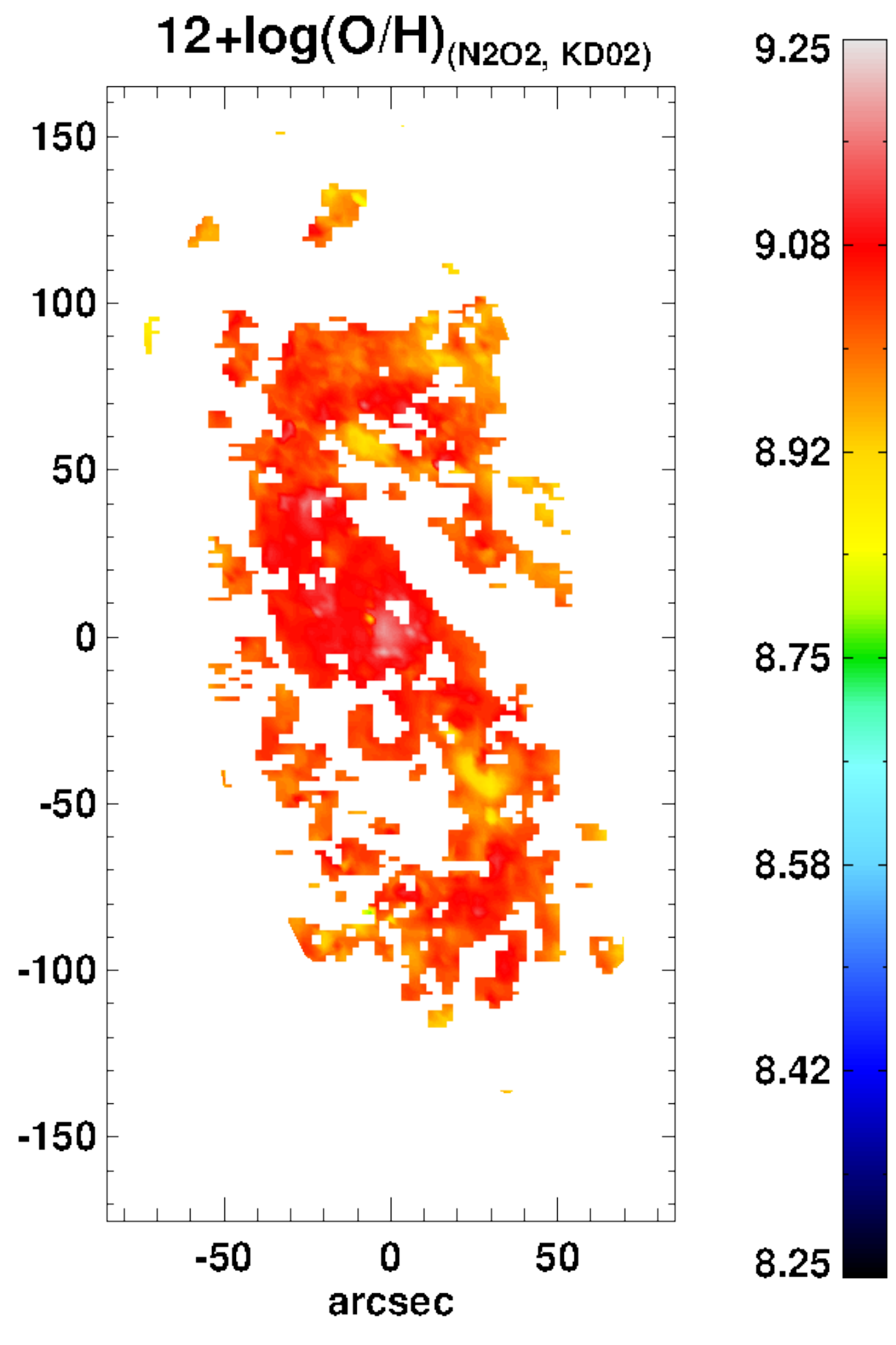}
\includegraphics[height=0.32\textheight]{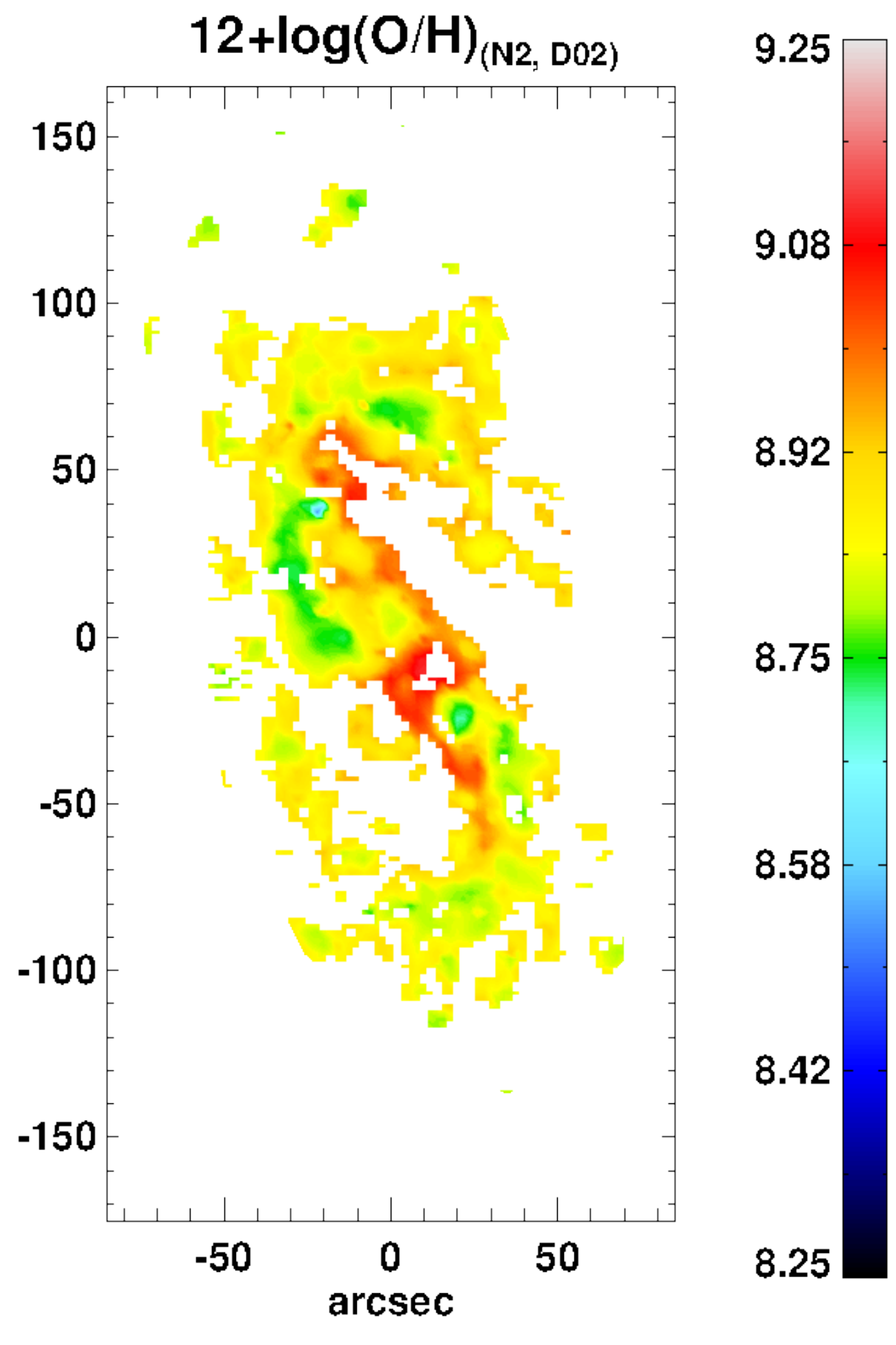} \\
\includegraphics[height=0.32\textheight]{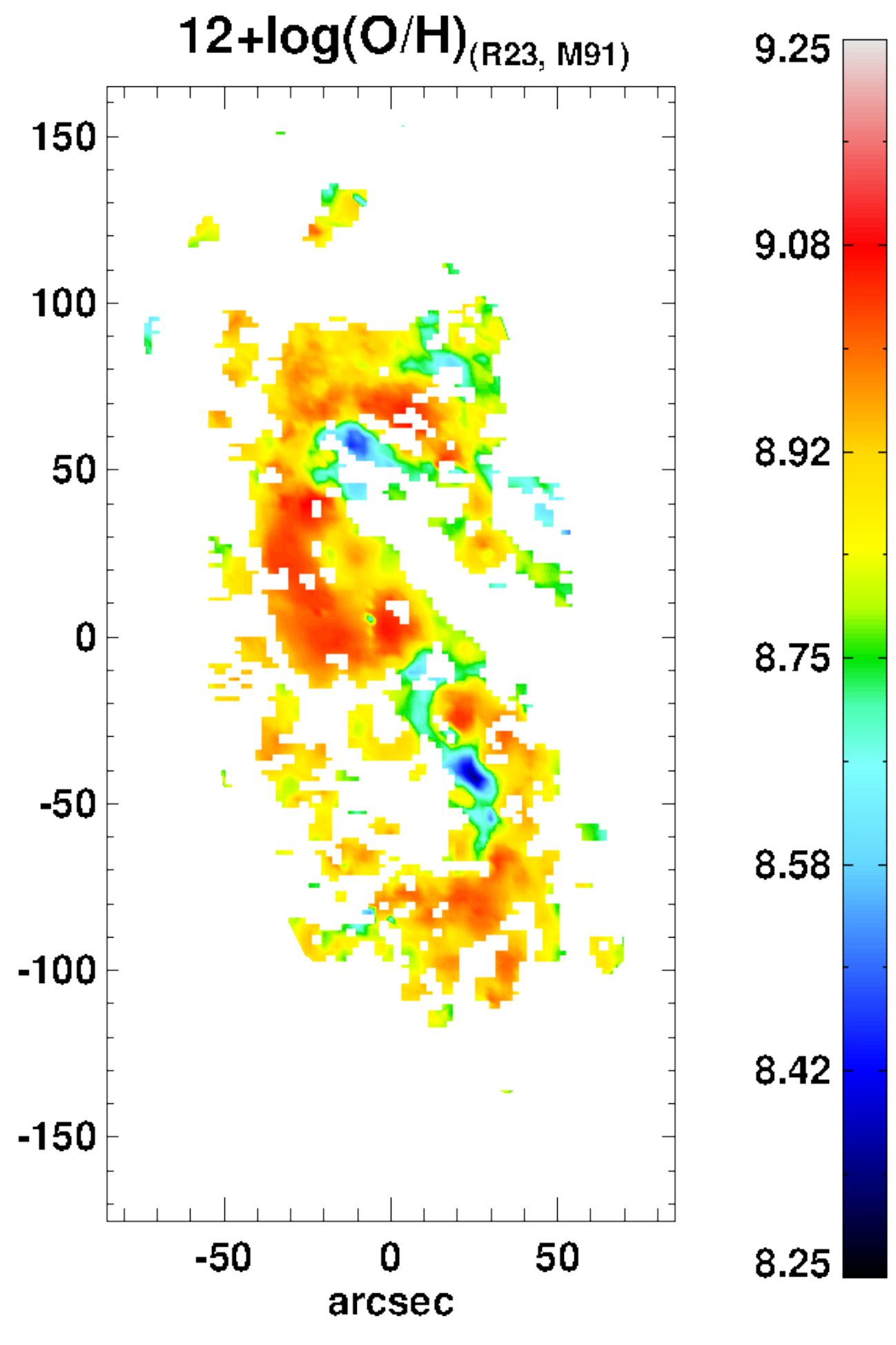}
\includegraphics[height=0.32\textheight]{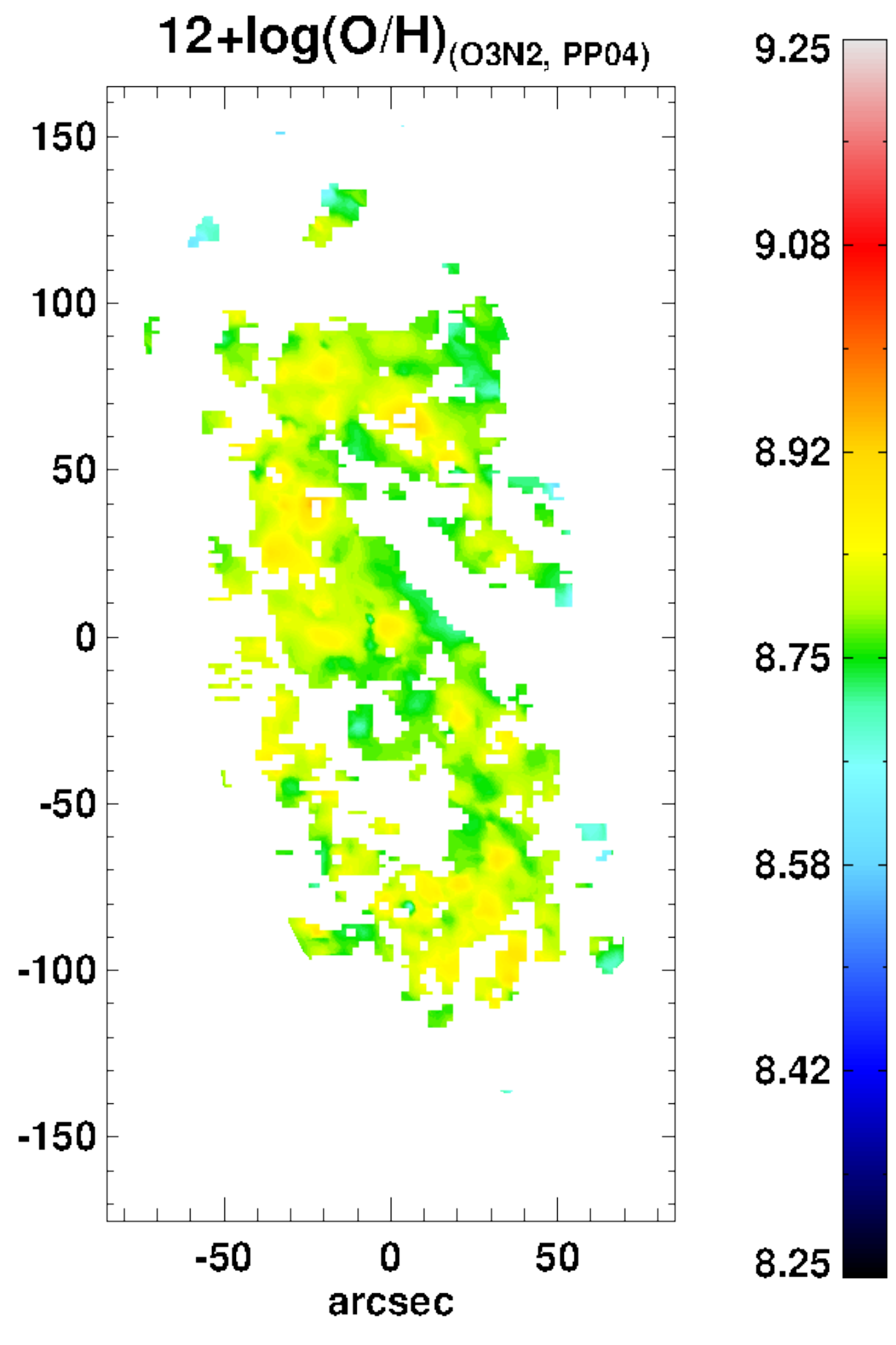}
\includegraphics[height=0.32\textheight]{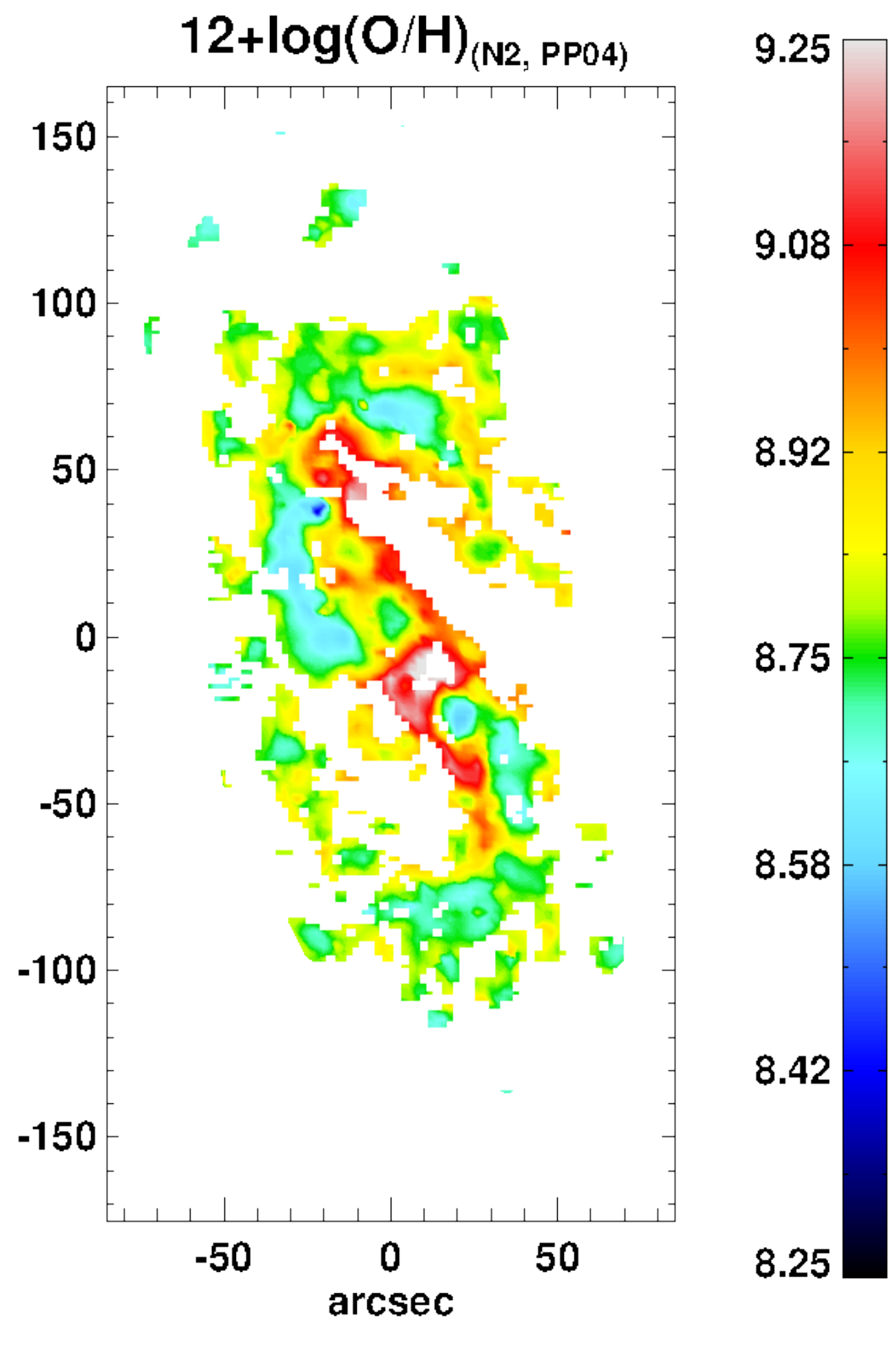}
\caption{Continued: For NGC 2903.}
\label{}
\end{figure*}

\addtocounter{figure}{-1}
\clearpage

\begin{figure*}
\includegraphics[width=0.32\textwidth]{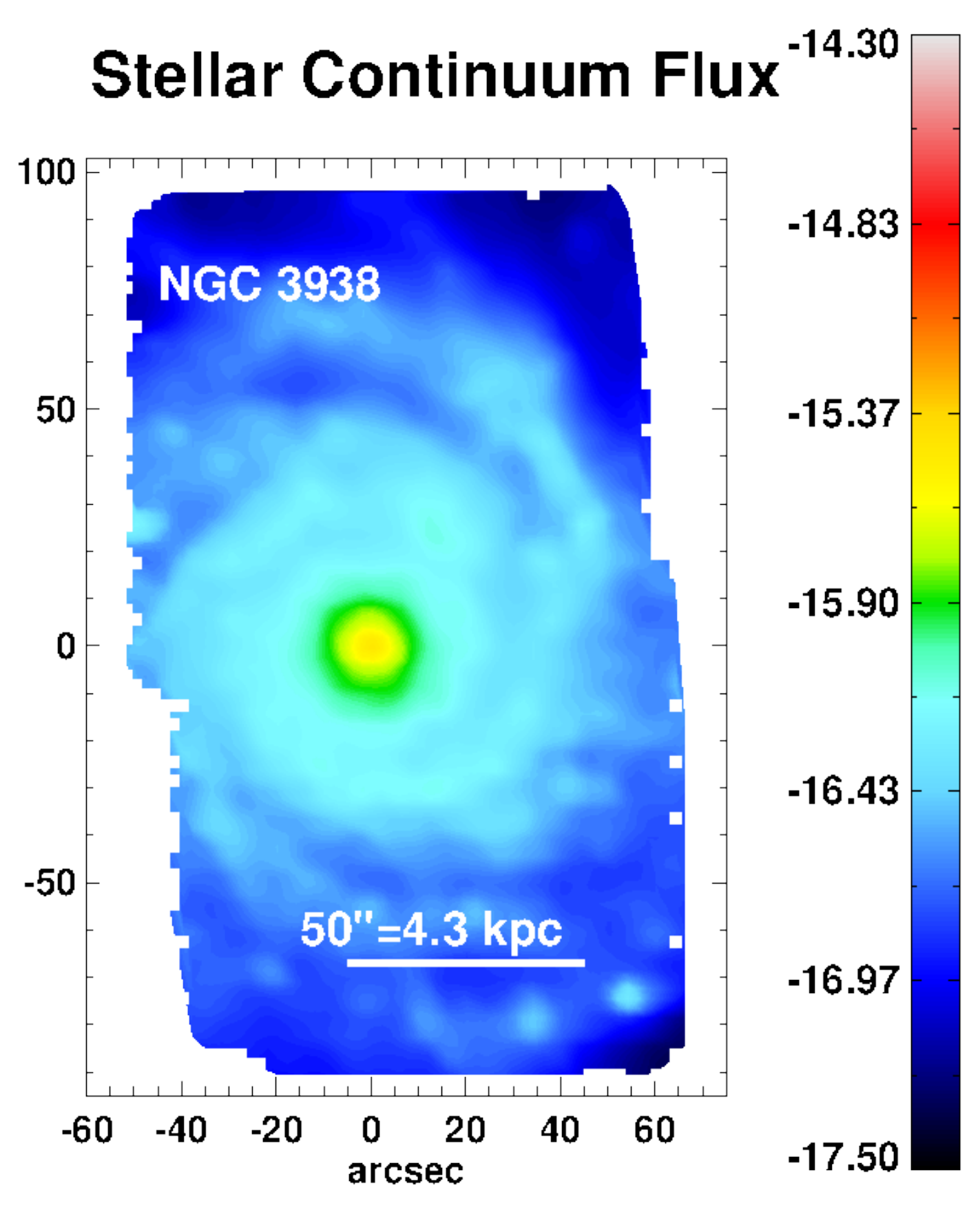}
\includegraphics[width=0.32\textwidth]{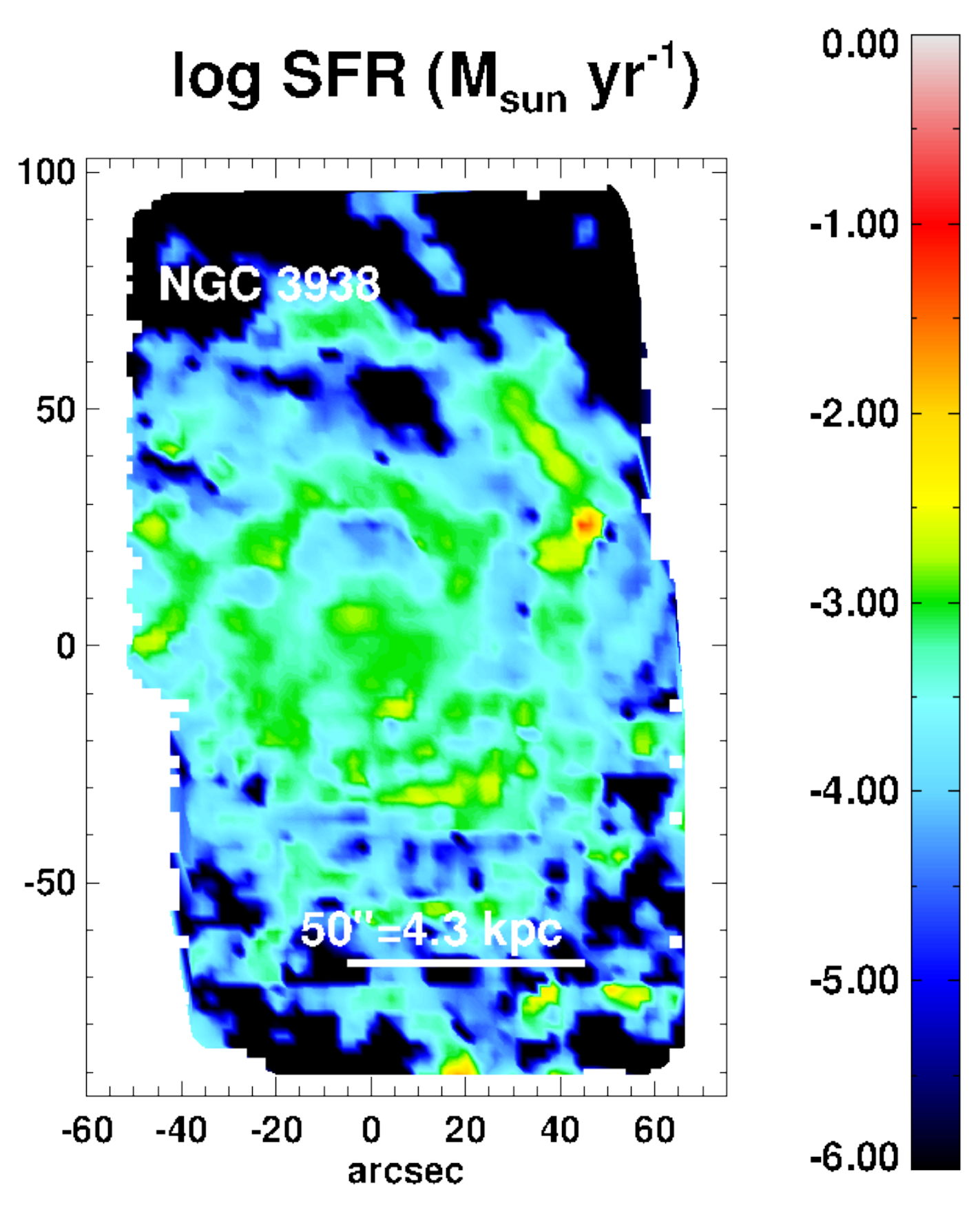}
\includegraphics[width=0.32\textwidth]{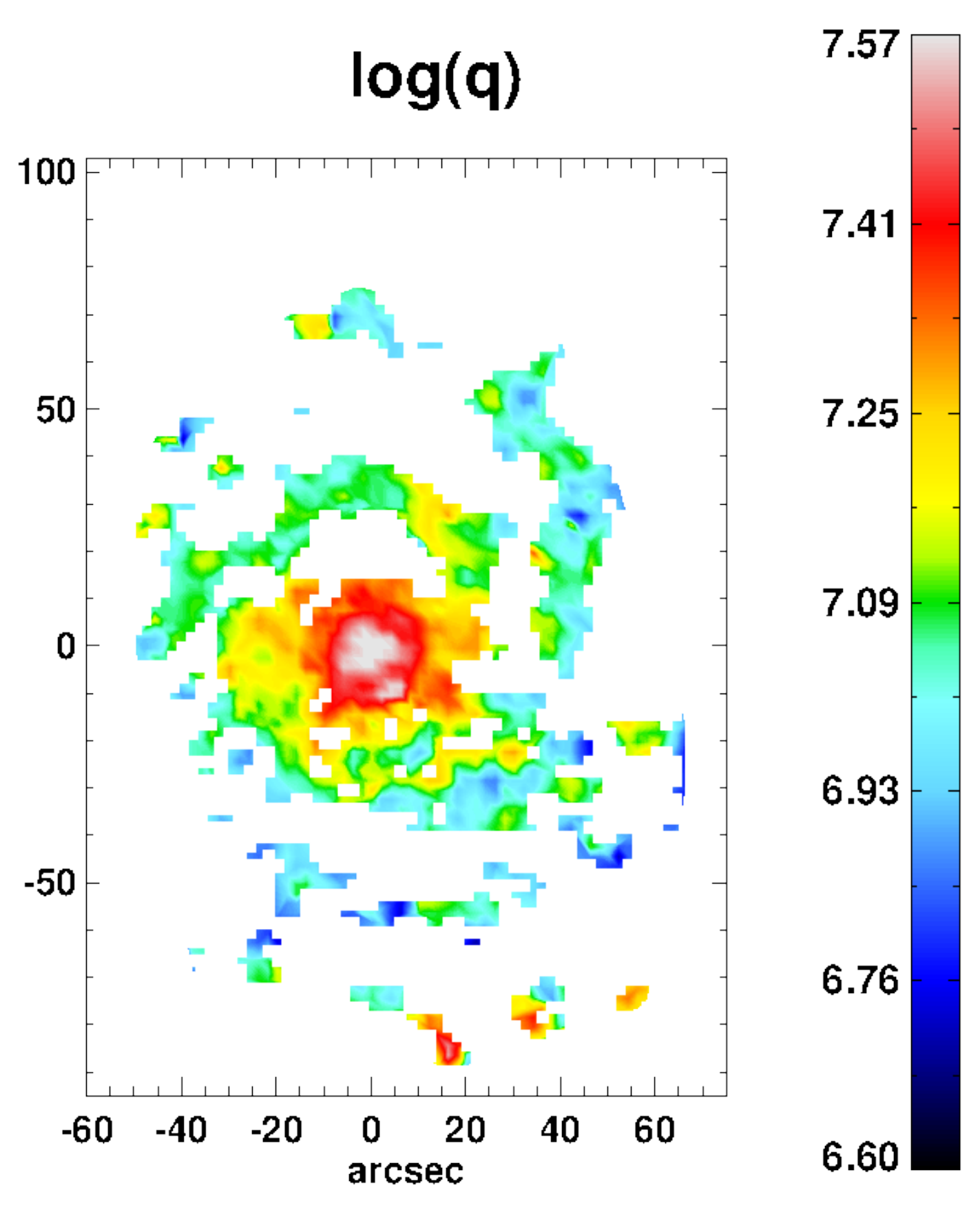}\\
\includegraphics[width=0.32\textwidth]{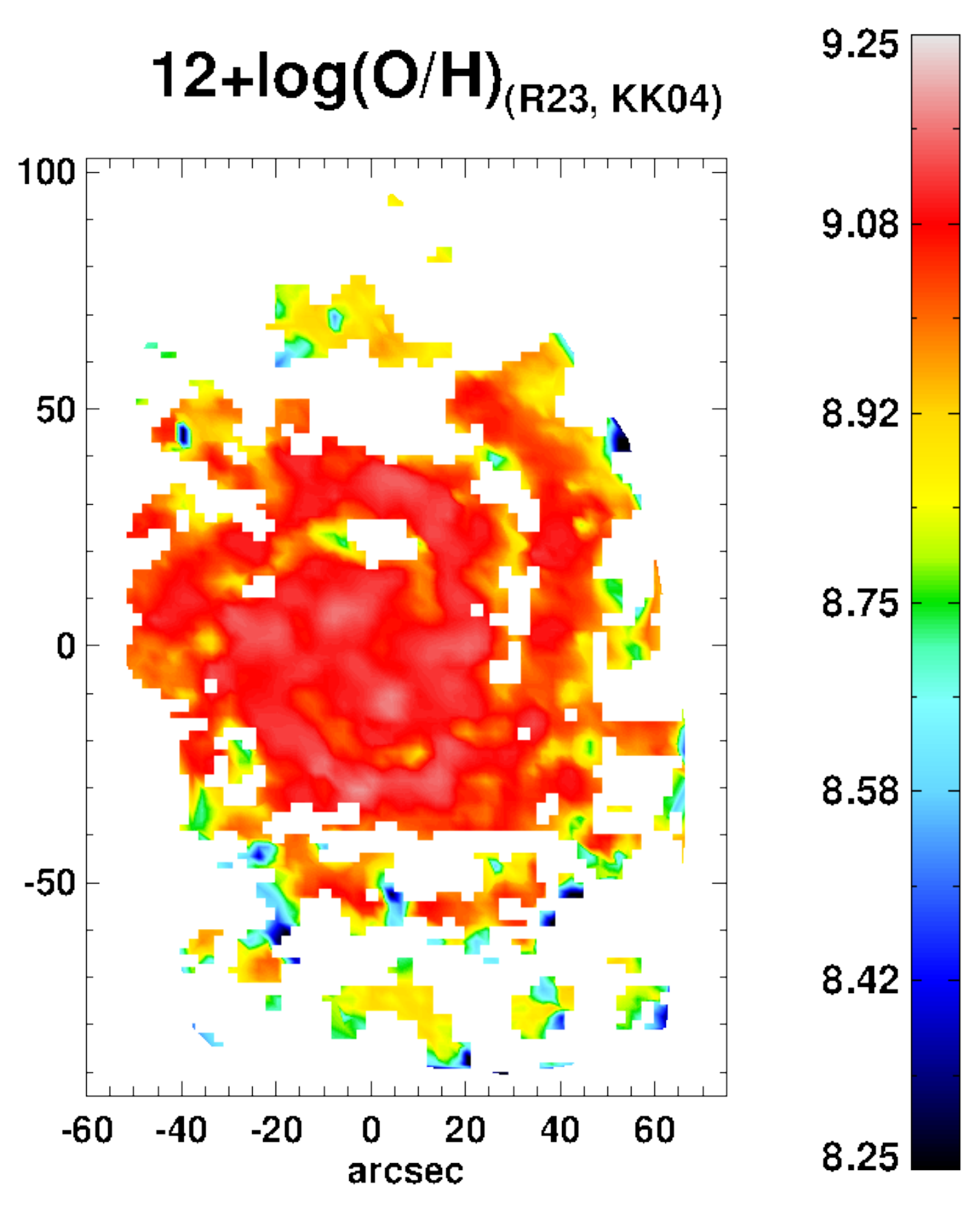}
\includegraphics[width=0.32\textwidth]{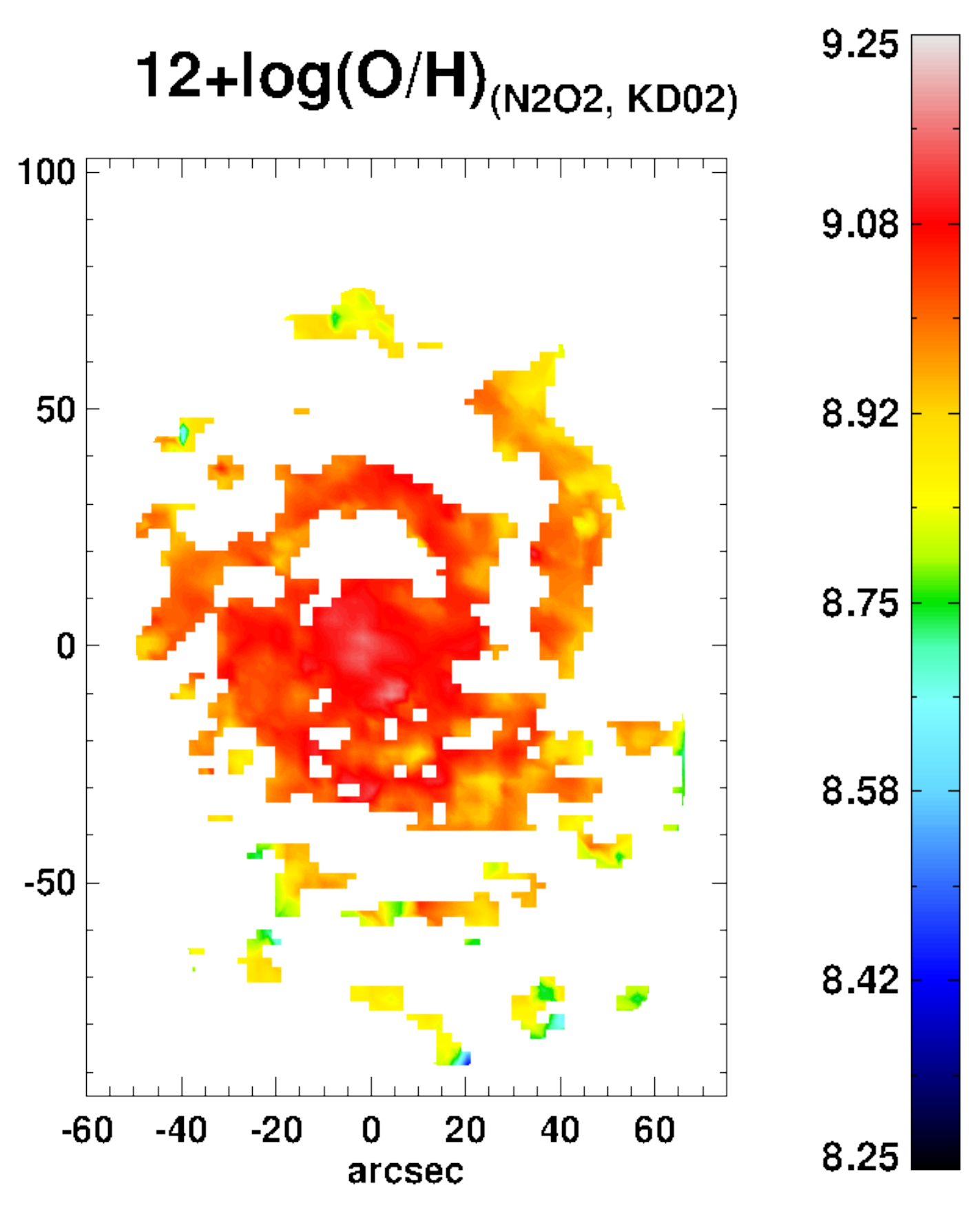}
\includegraphics[width=0.32\textwidth]{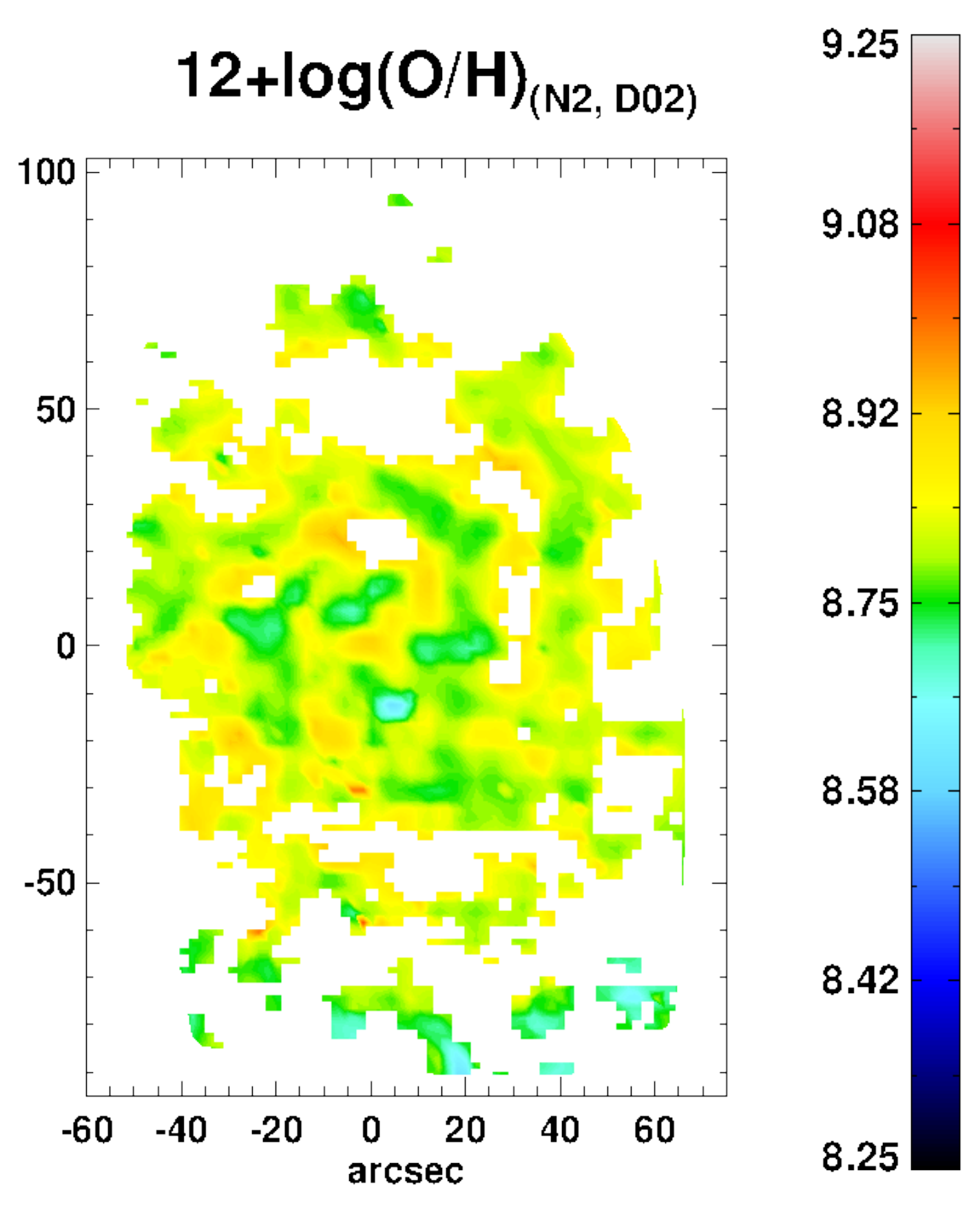} \\
\includegraphics[width=0.32\textwidth]{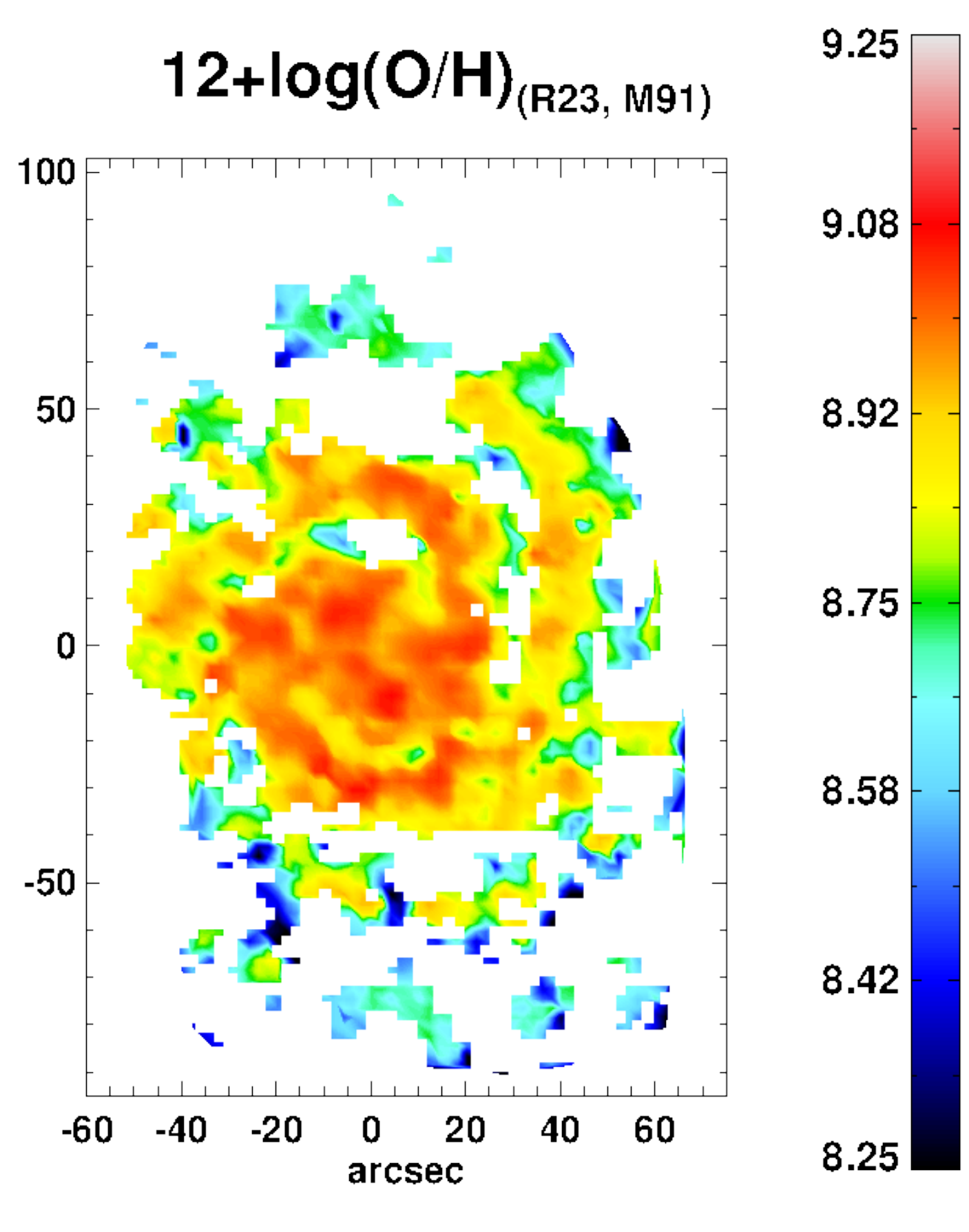}
\includegraphics[width=0.32\textwidth]{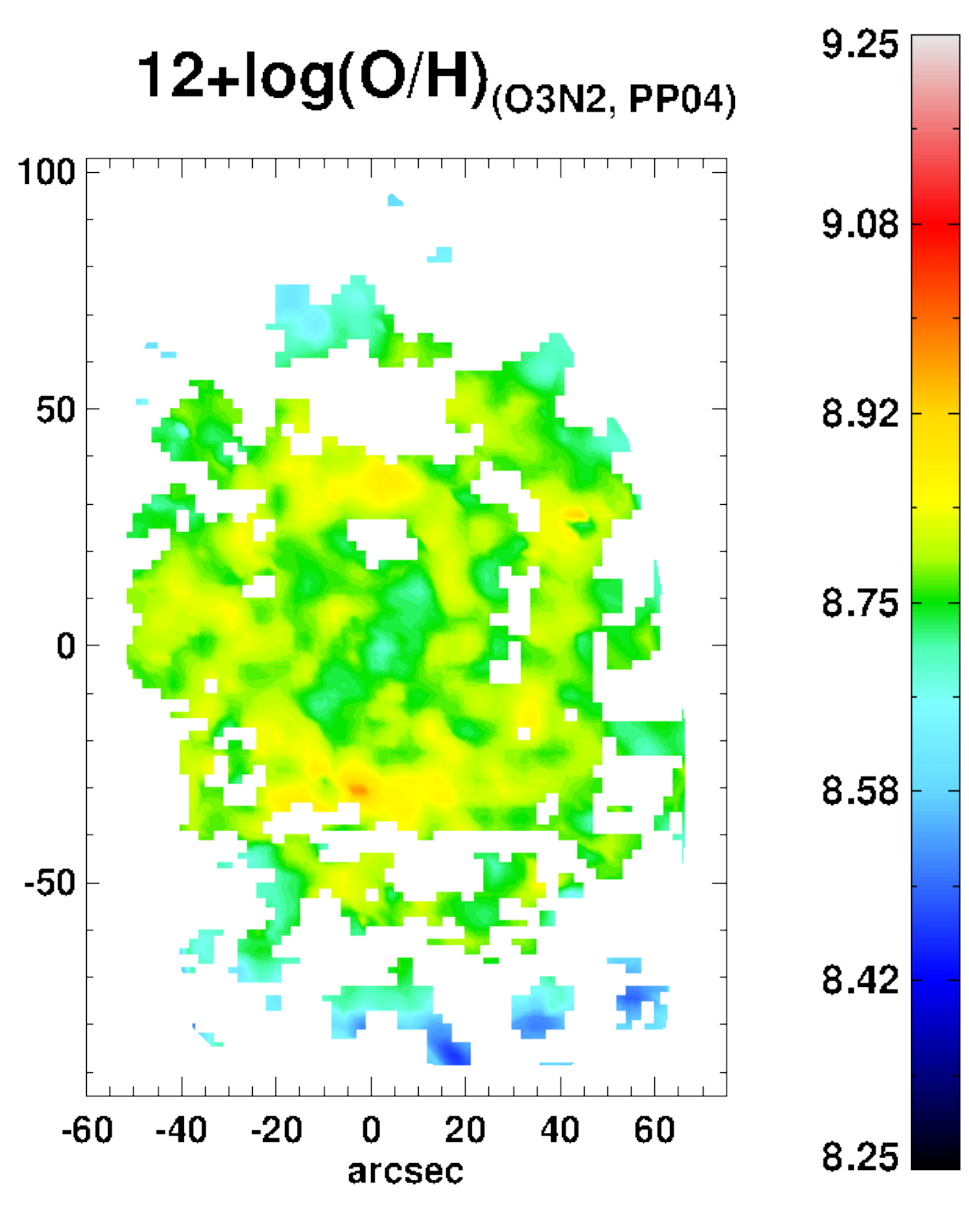}
\includegraphics[width=0.32\textwidth]{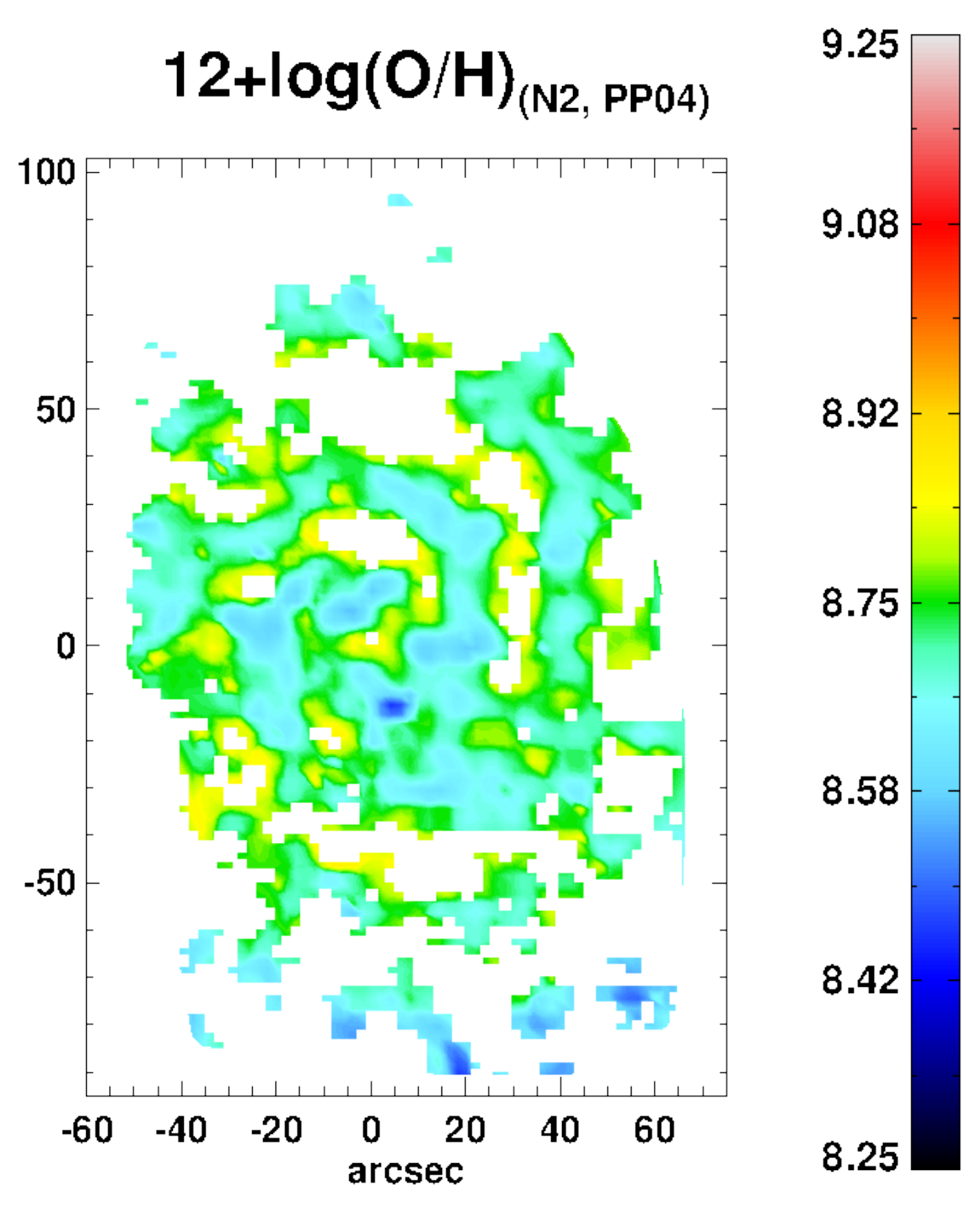}
\caption{Continued: For NGC 3938.}
\label{}
\end{figure*}

\addtocounter{figure}{-1}
\clearpage

\begin{figure*}
\includegraphics[width=0.32\textwidth]{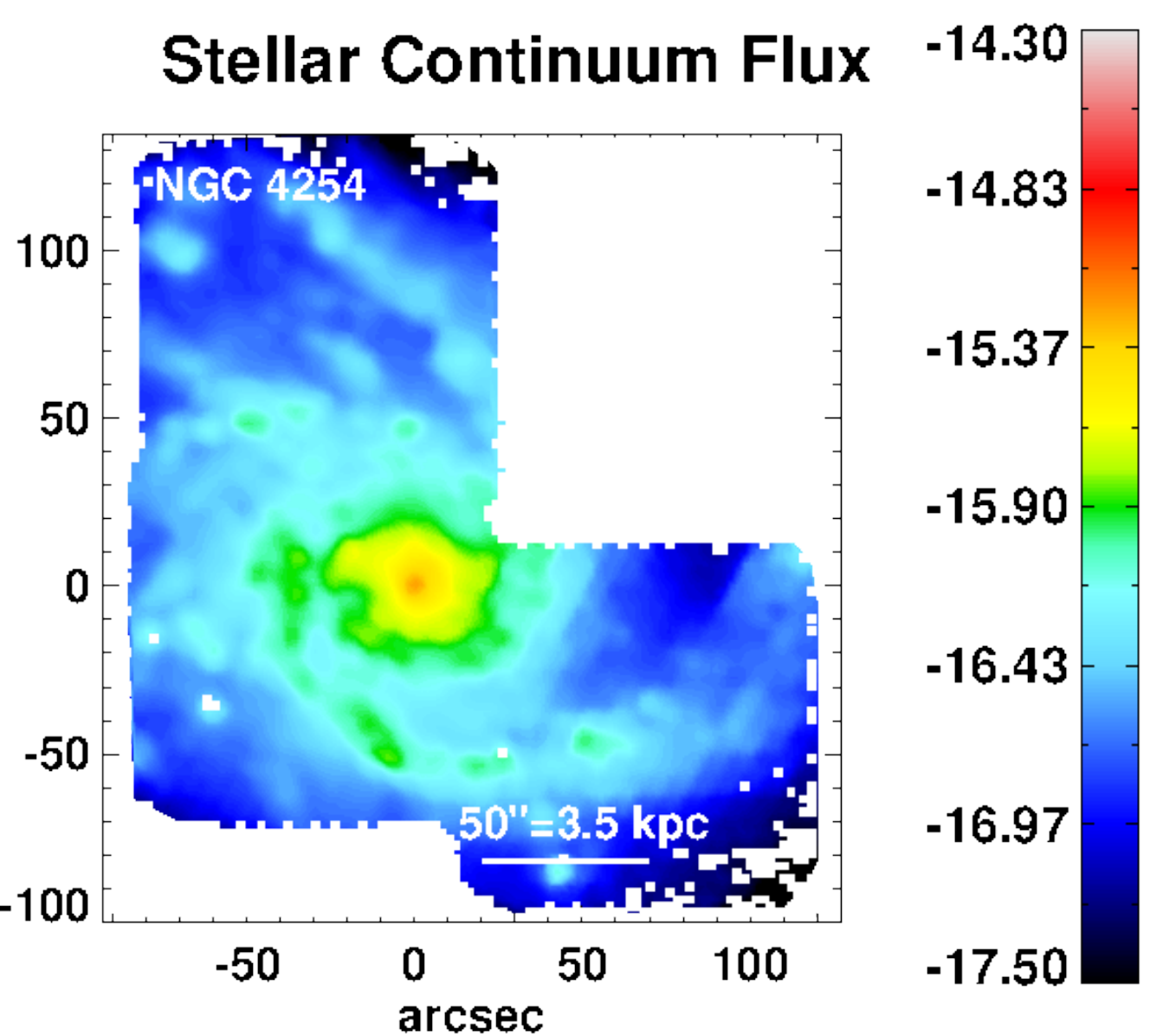}
\includegraphics[width=0.32\textwidth]{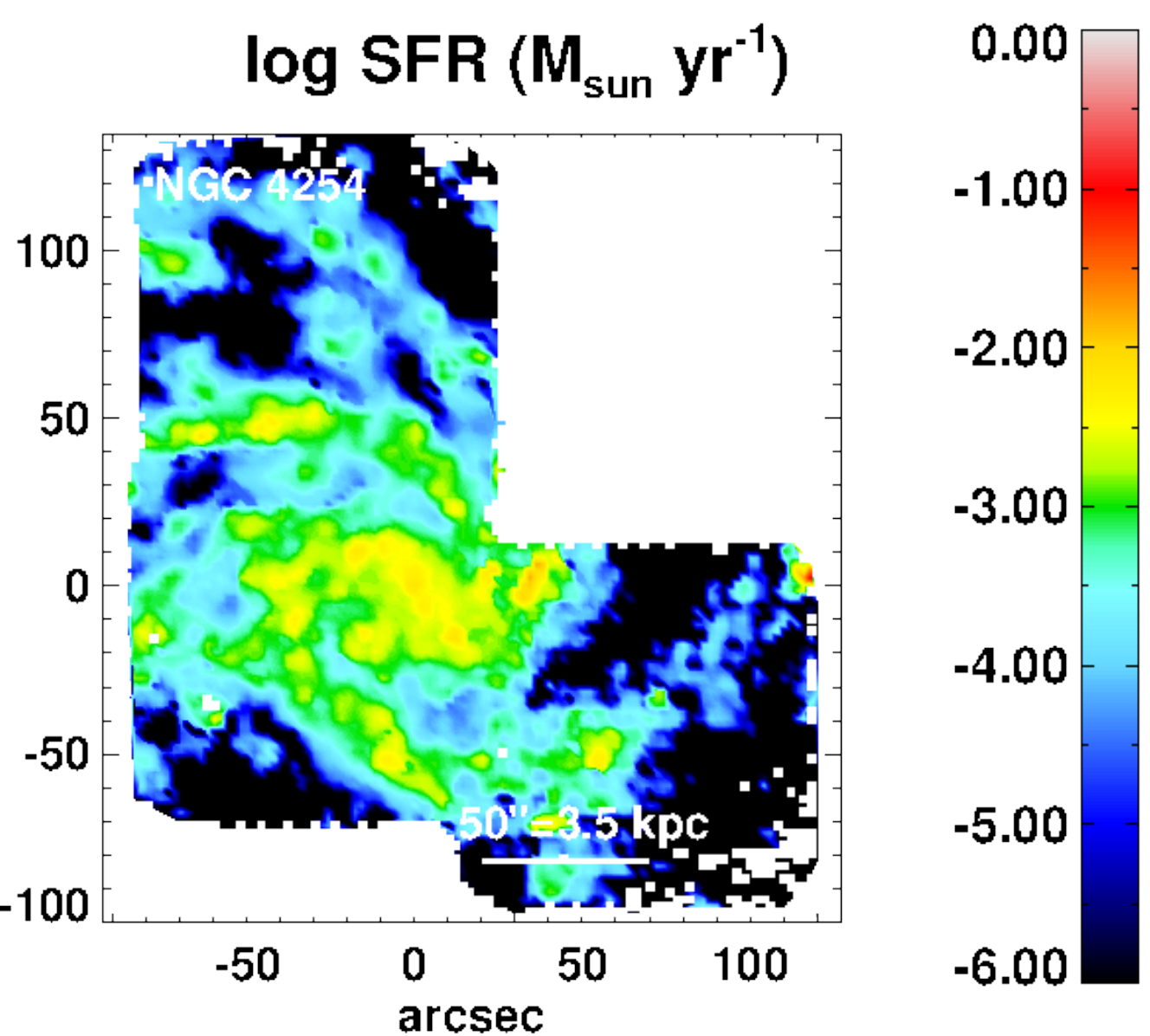}
\includegraphics[width=0.32\textwidth]{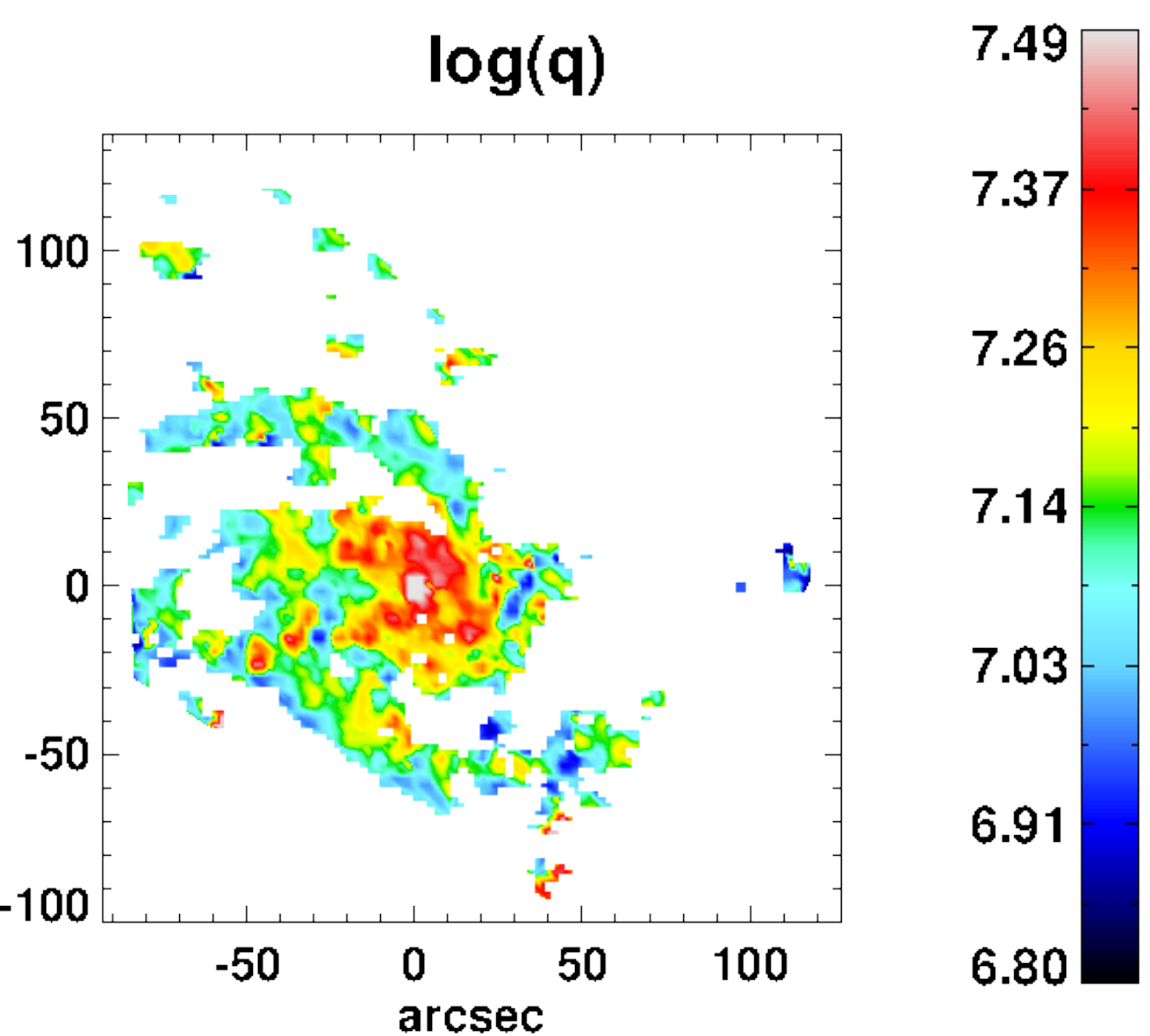}\\
\includegraphics[width=0.32\textwidth]{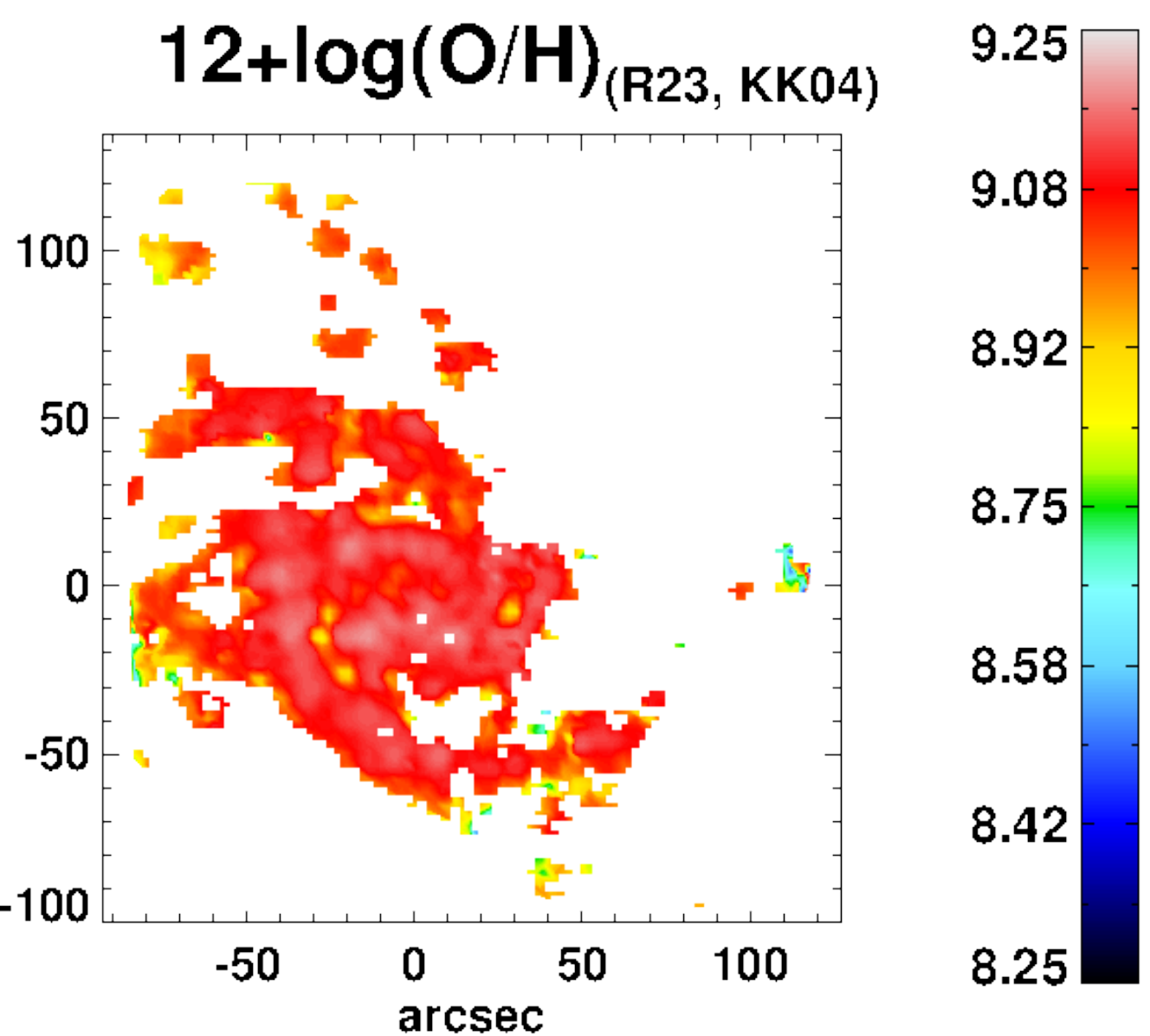}
\includegraphics[width=0.32\textwidth]{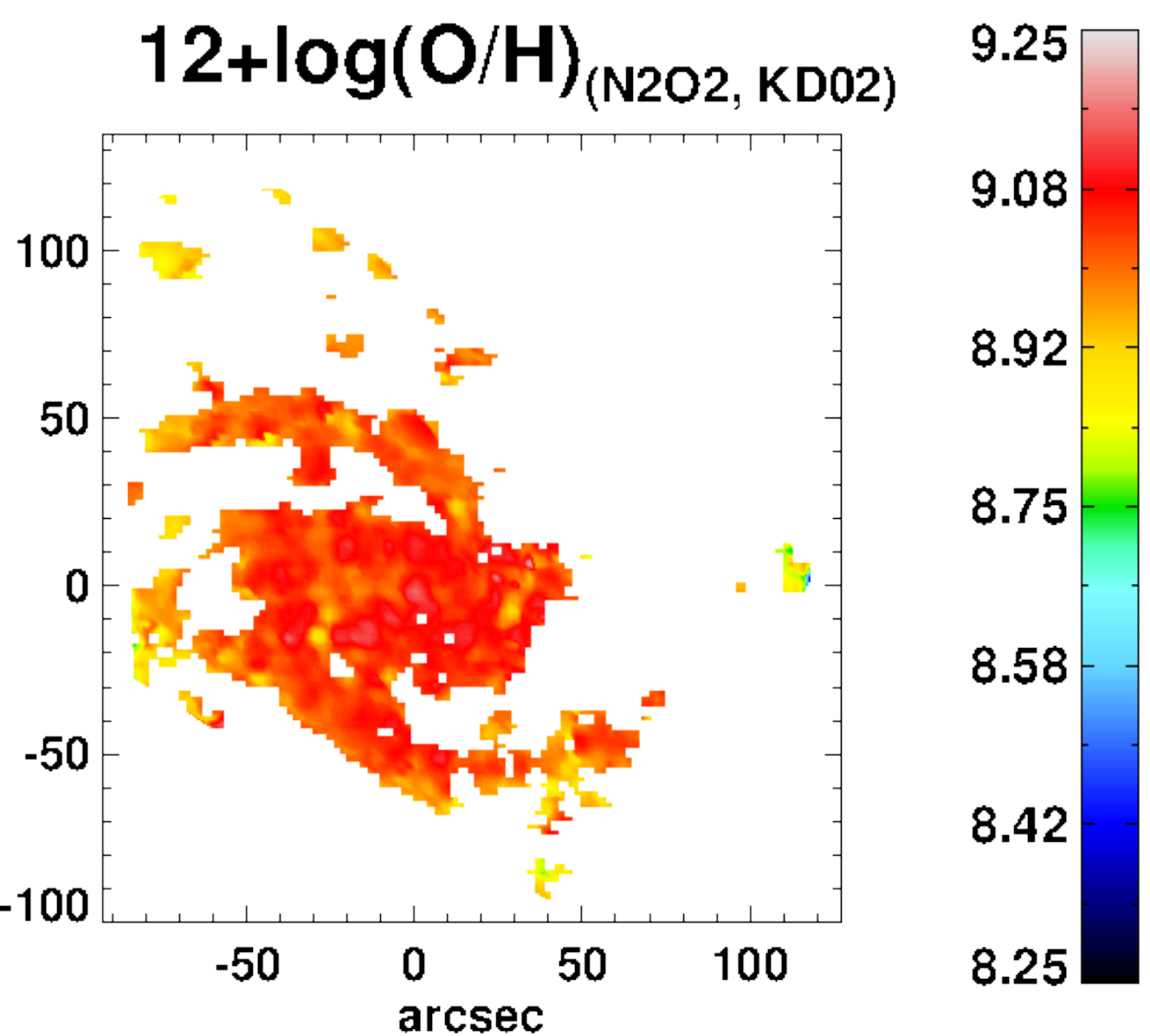}
\includegraphics[width=0.32\textwidth]{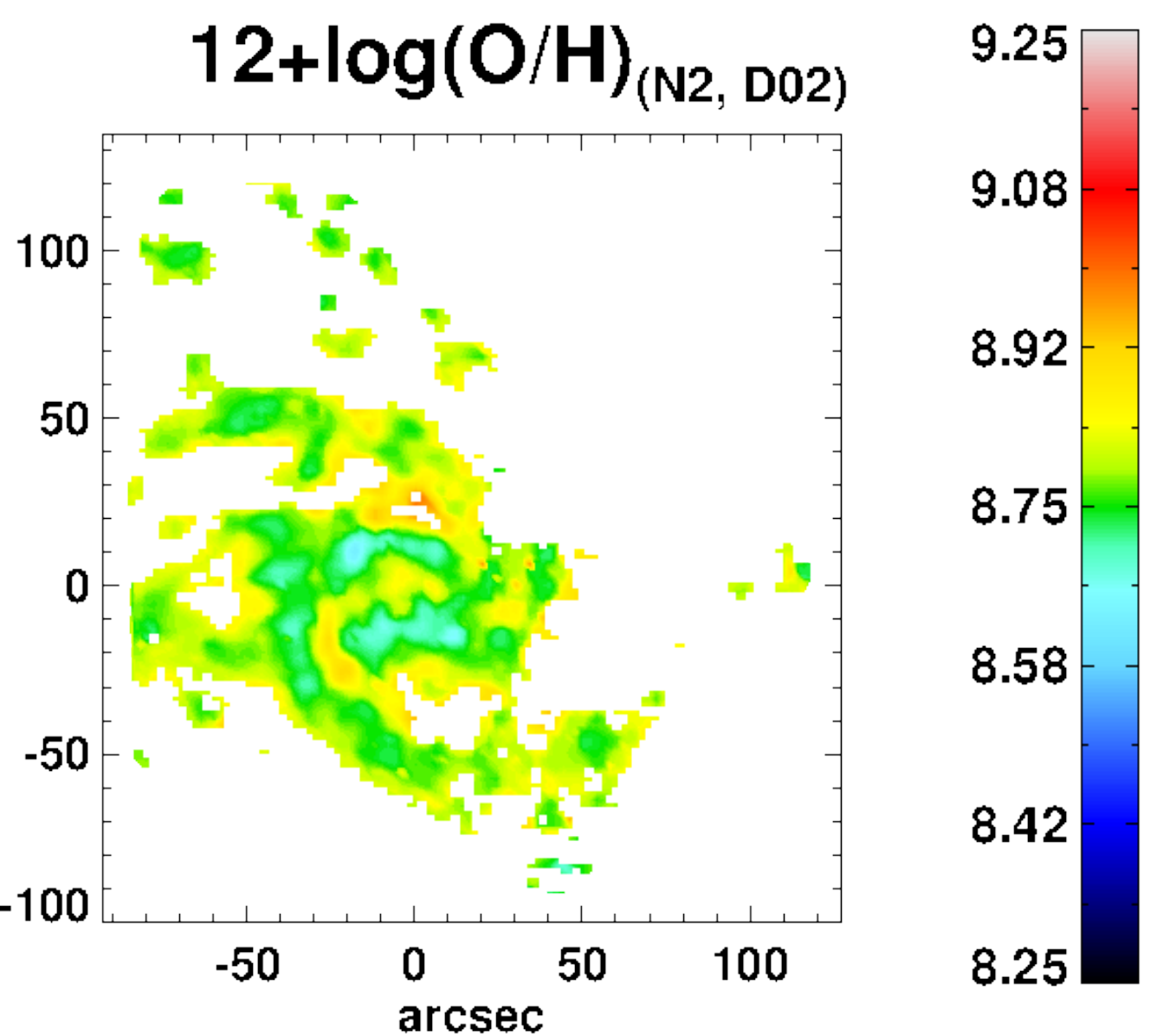} \\
\includegraphics[width=0.32\textwidth]{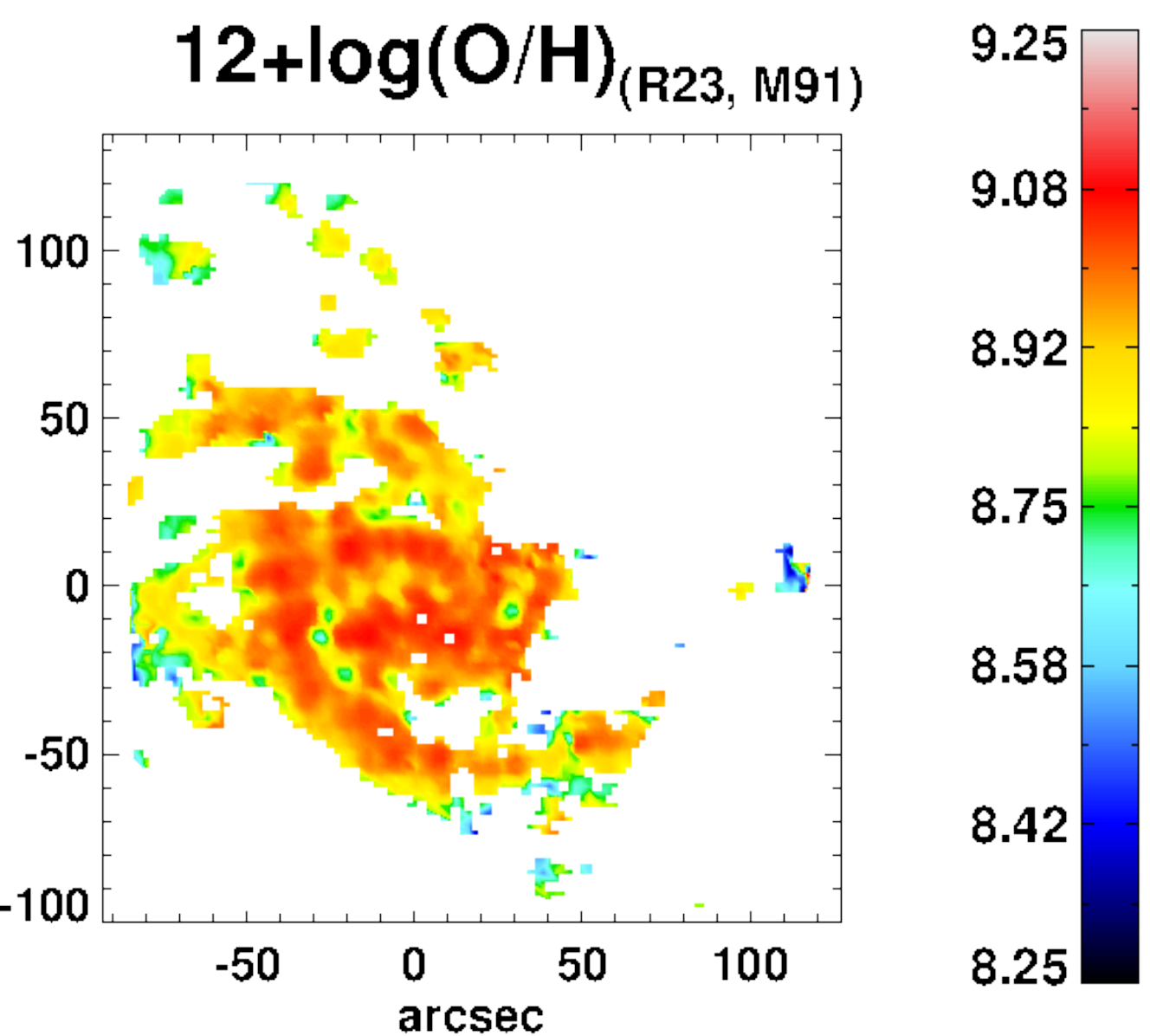}
\includegraphics[width=0.32\textwidth]{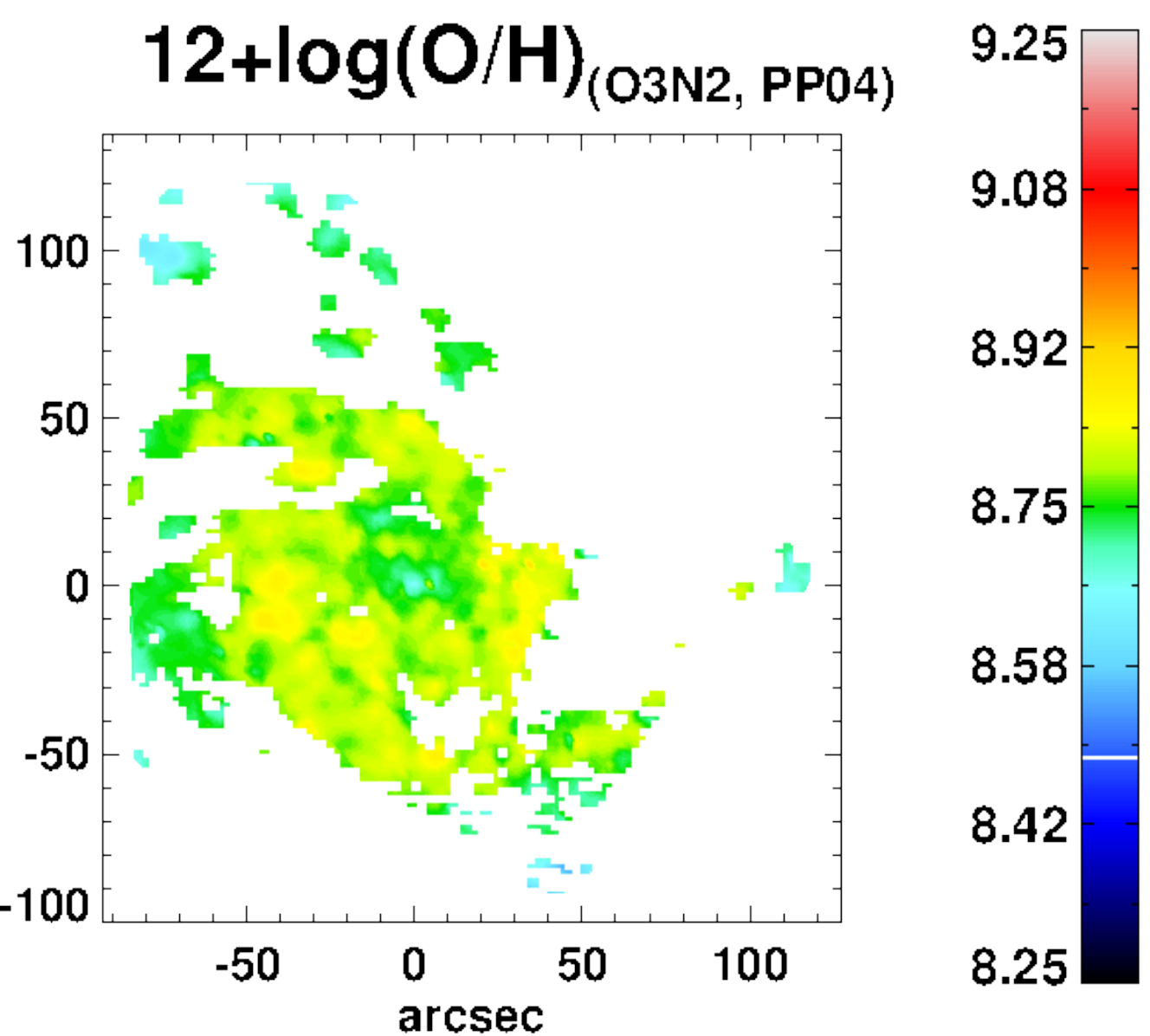}
\includegraphics[width=0.32\textwidth]{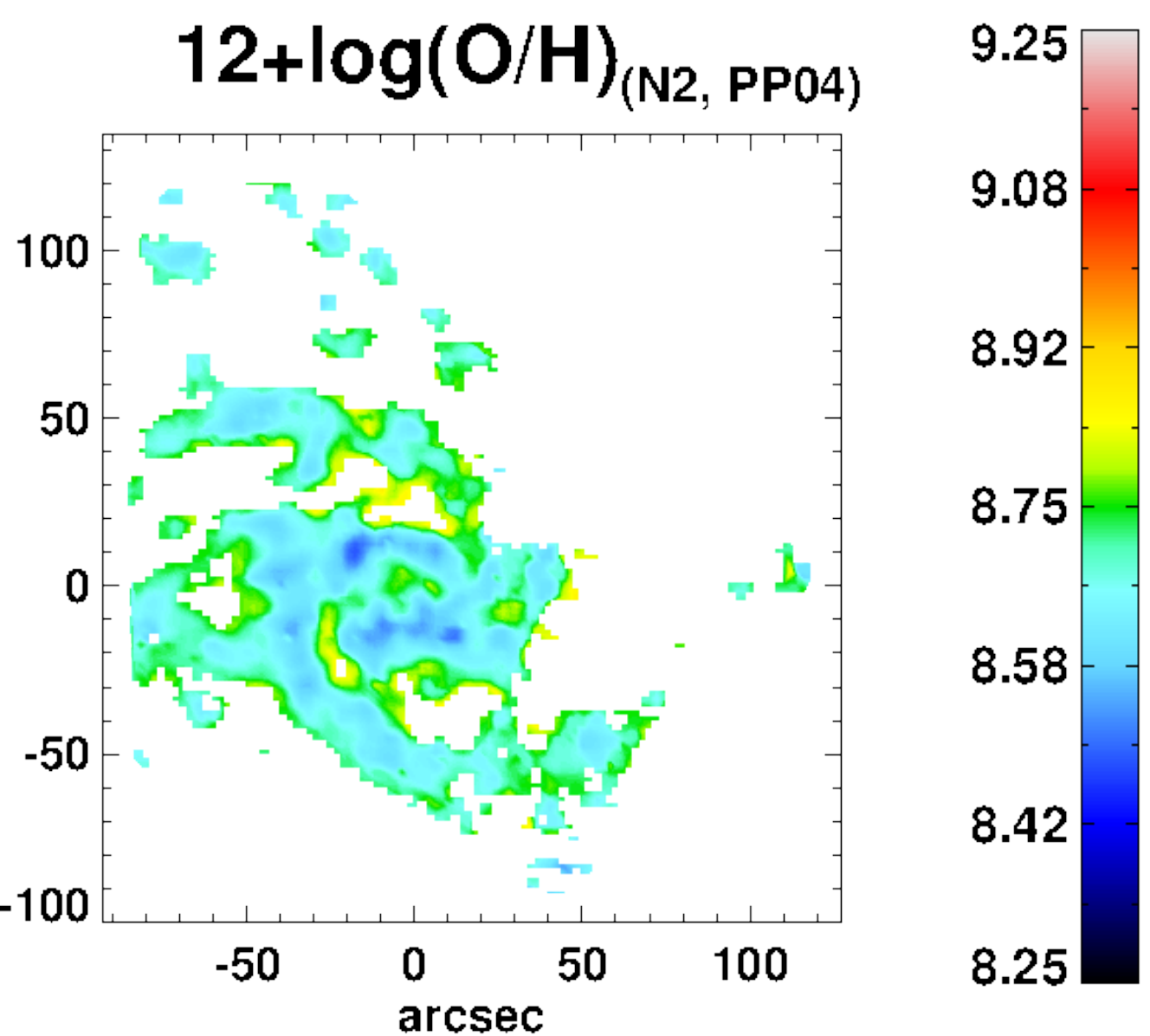}
\caption{Continued: For NGC 4254.}
\label{}
\end{figure*}

\addtocounter{figure}{-1}
\clearpage

\begin{landscape}
\begin{figure}
\includegraphics[width=0.39\textheight]{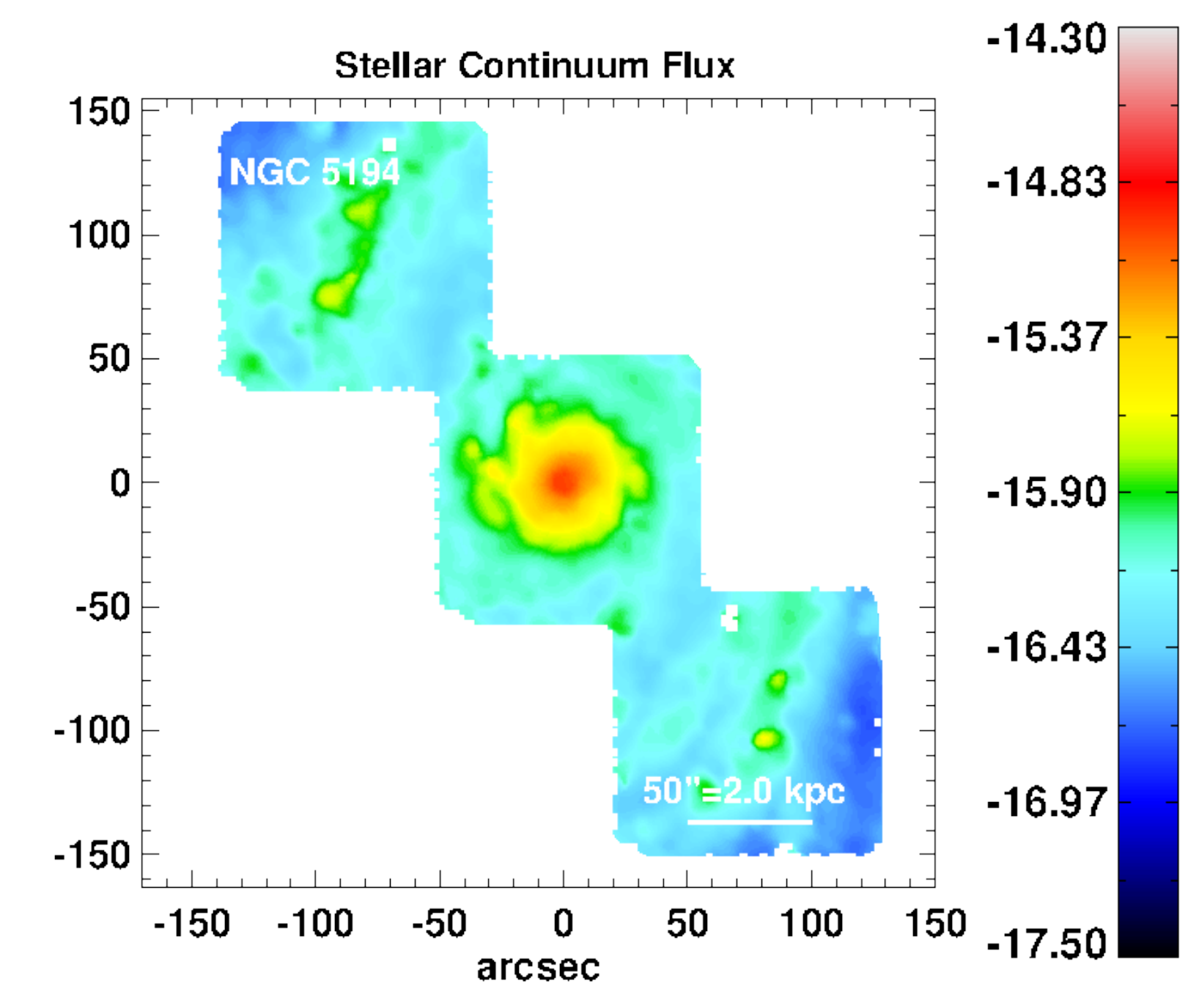}
\includegraphics[width=0.39\textheight]{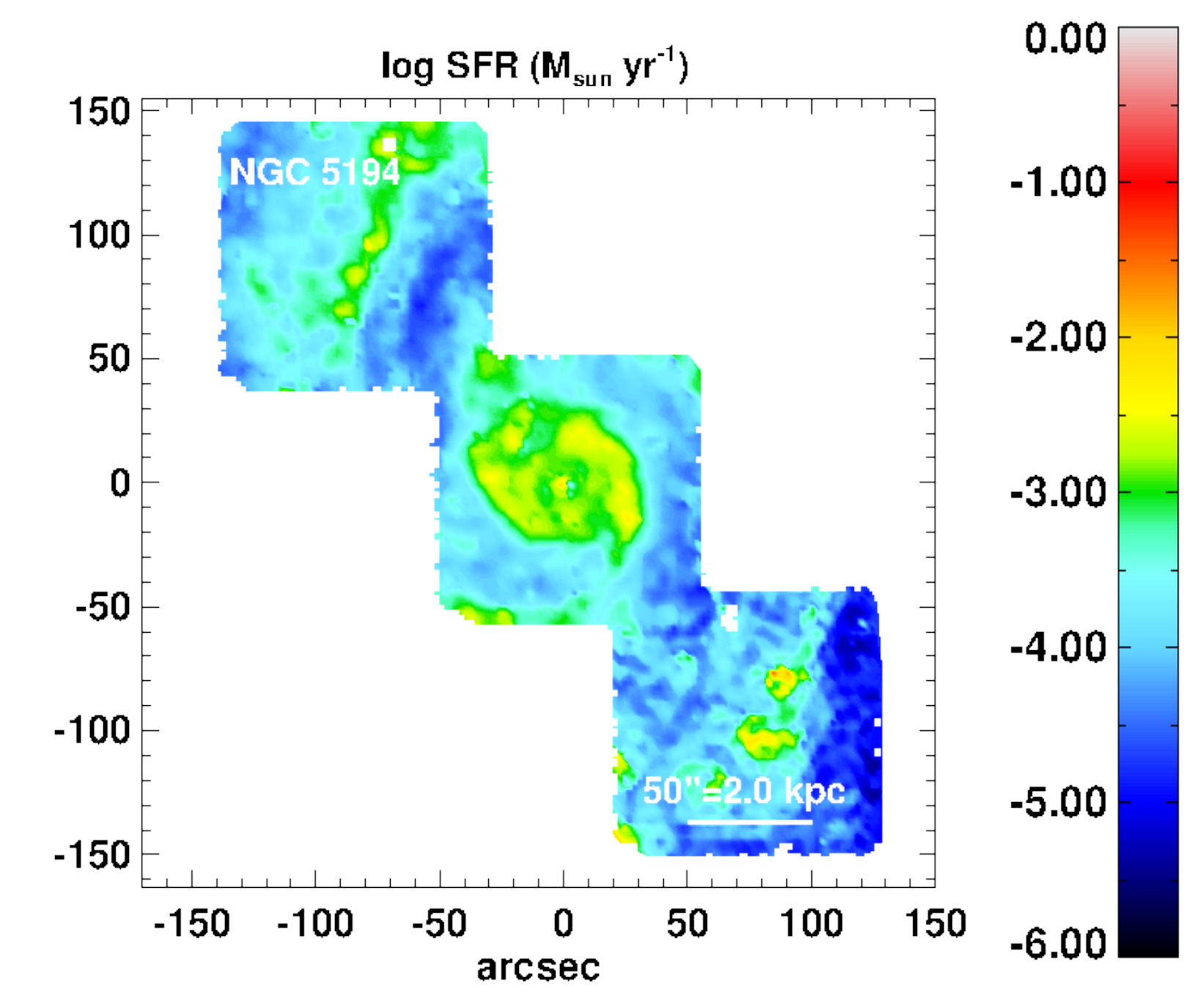}
\includegraphics[width=0.39\textheight]{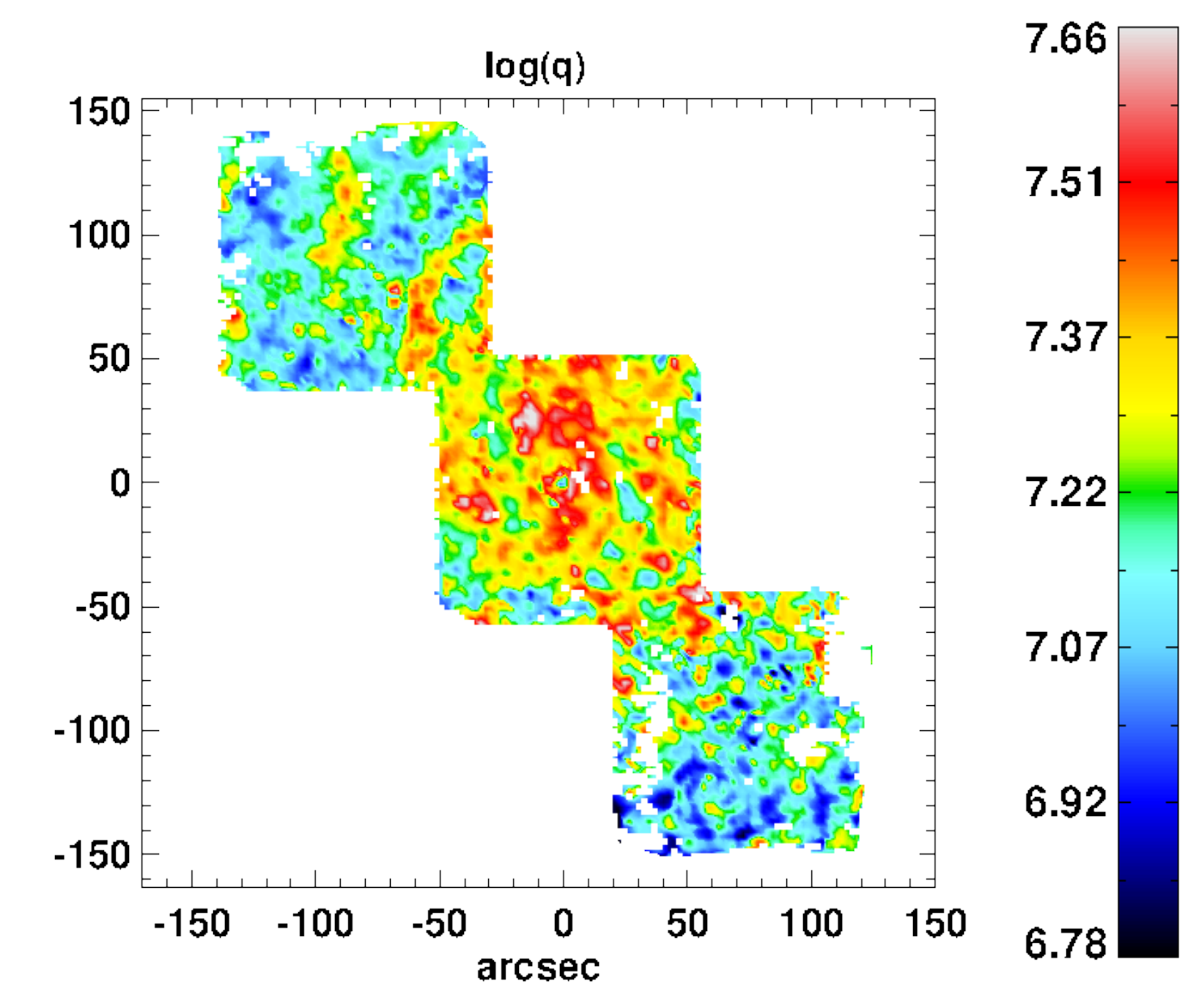}\\
\includegraphics[width=0.39\textheight]{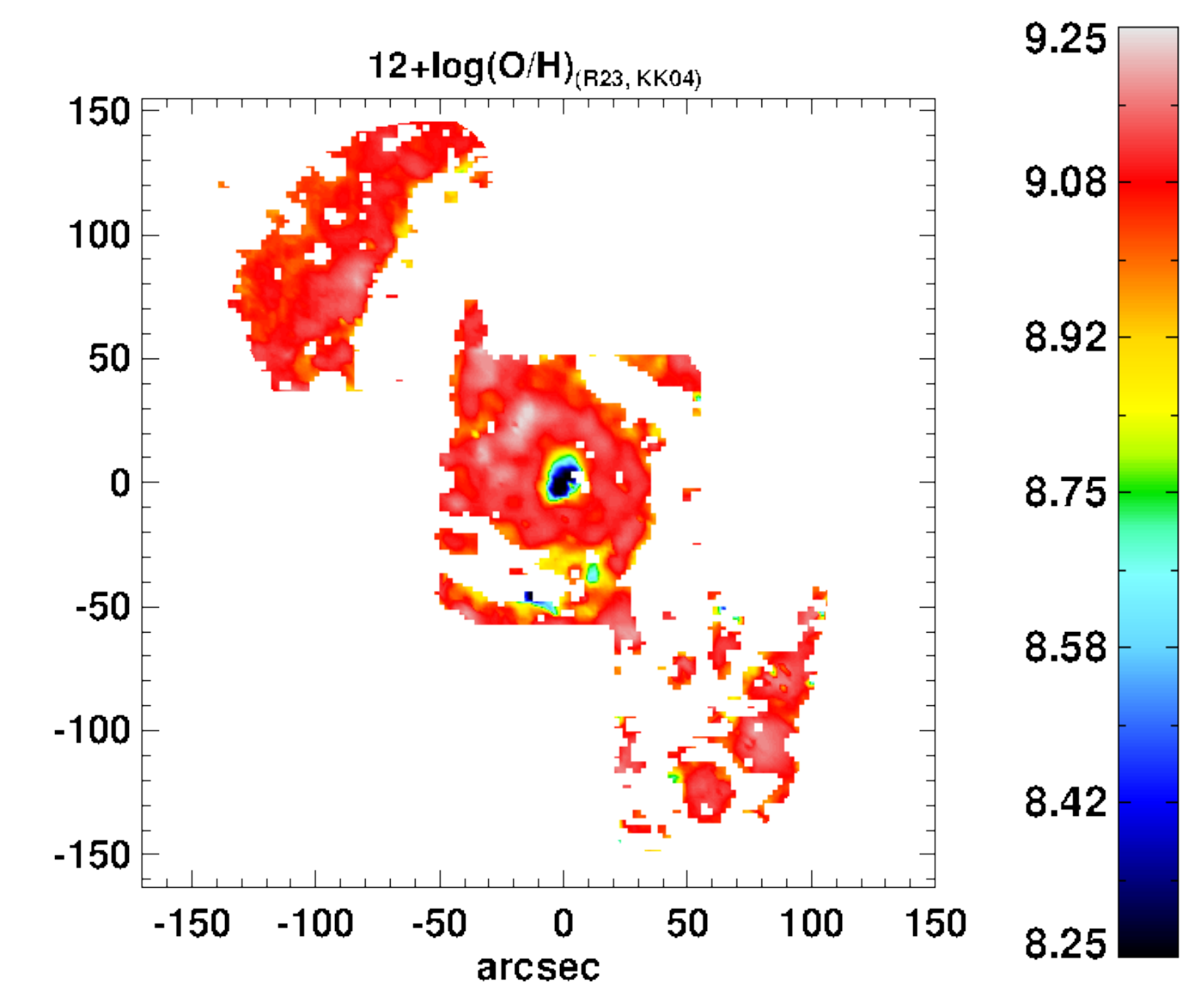}
\includegraphics[width=0.39\textheight]{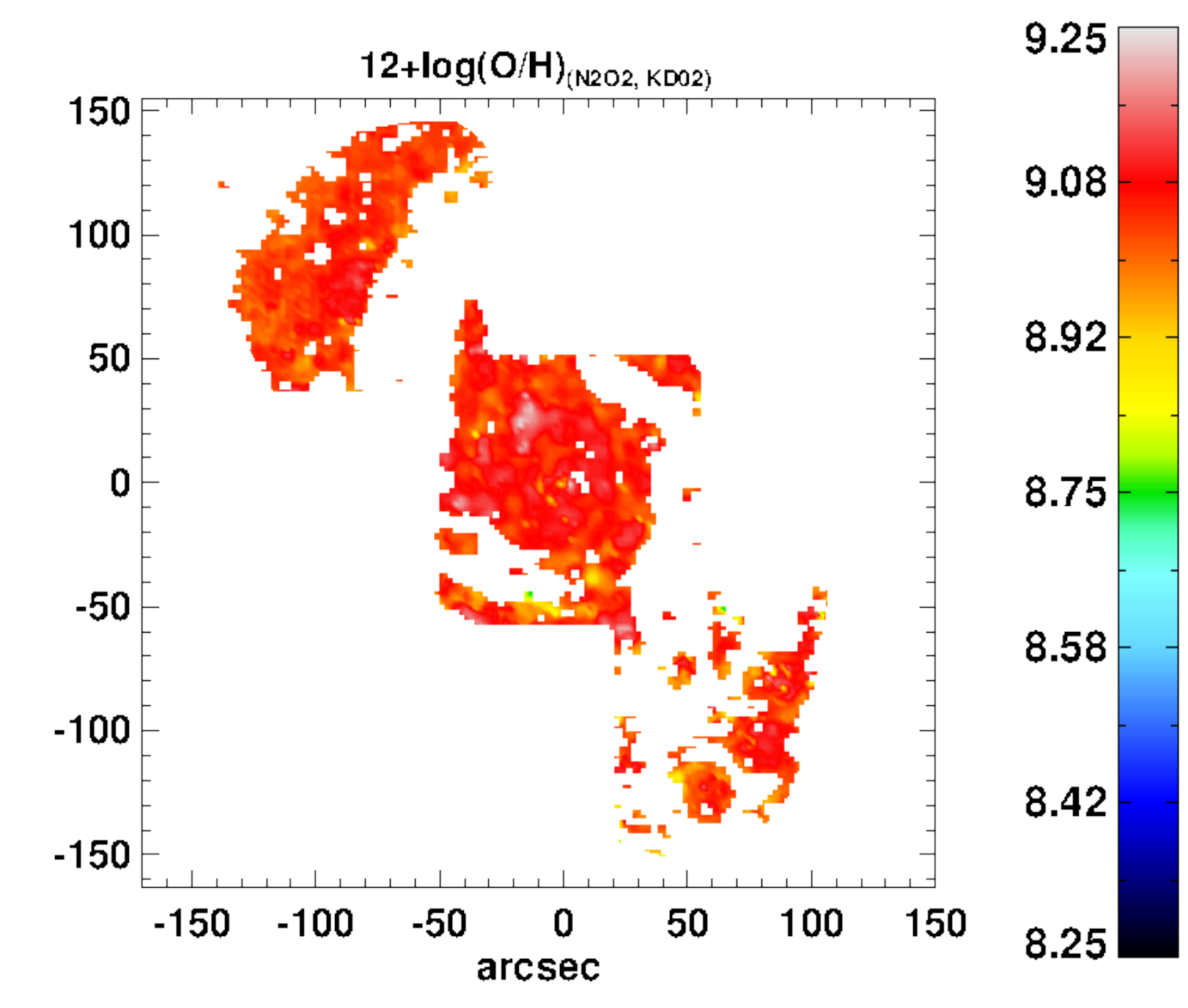}
\includegraphics[width=0.39\textheight]{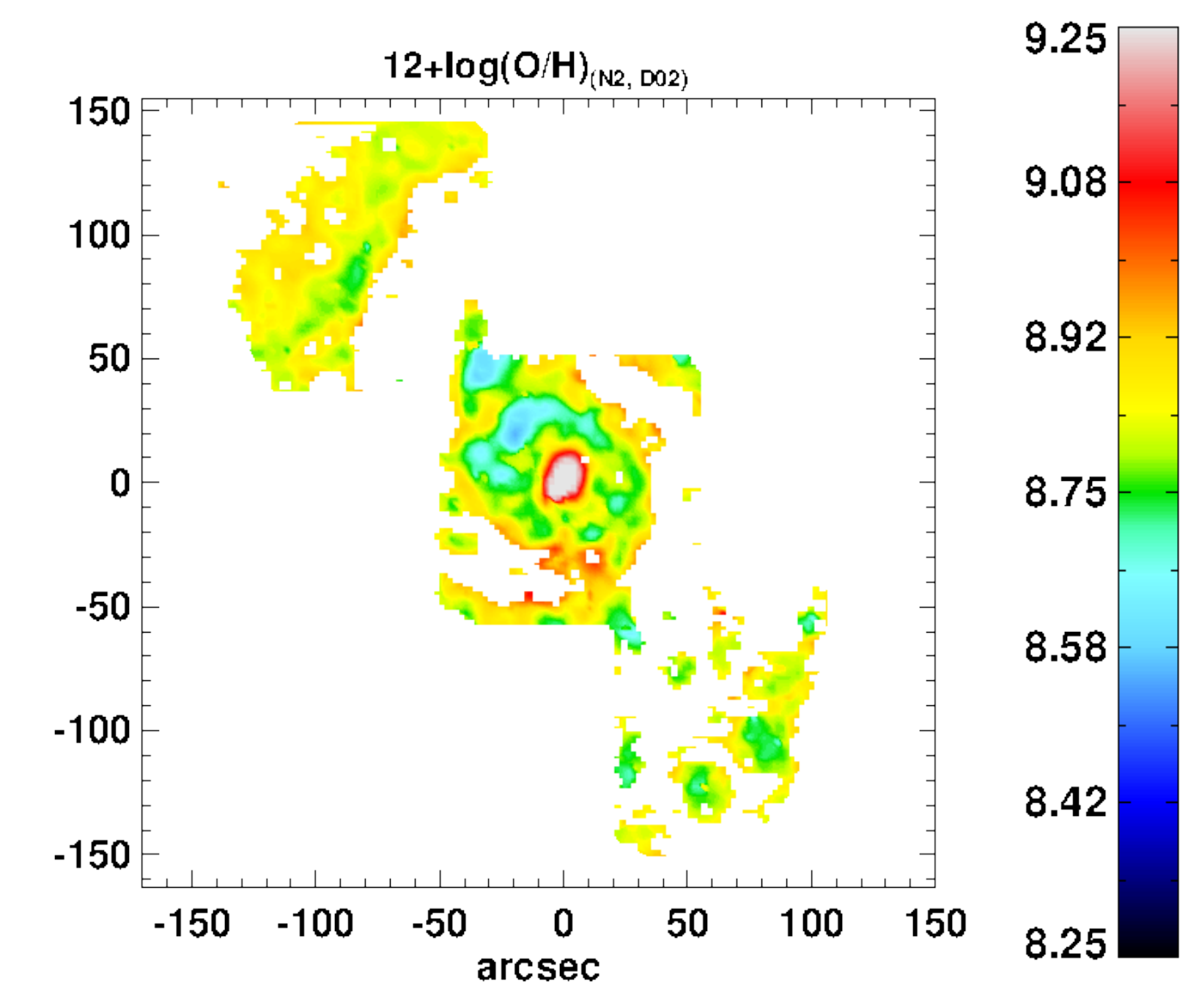} \\
\includegraphics[width=0.39\textheight]{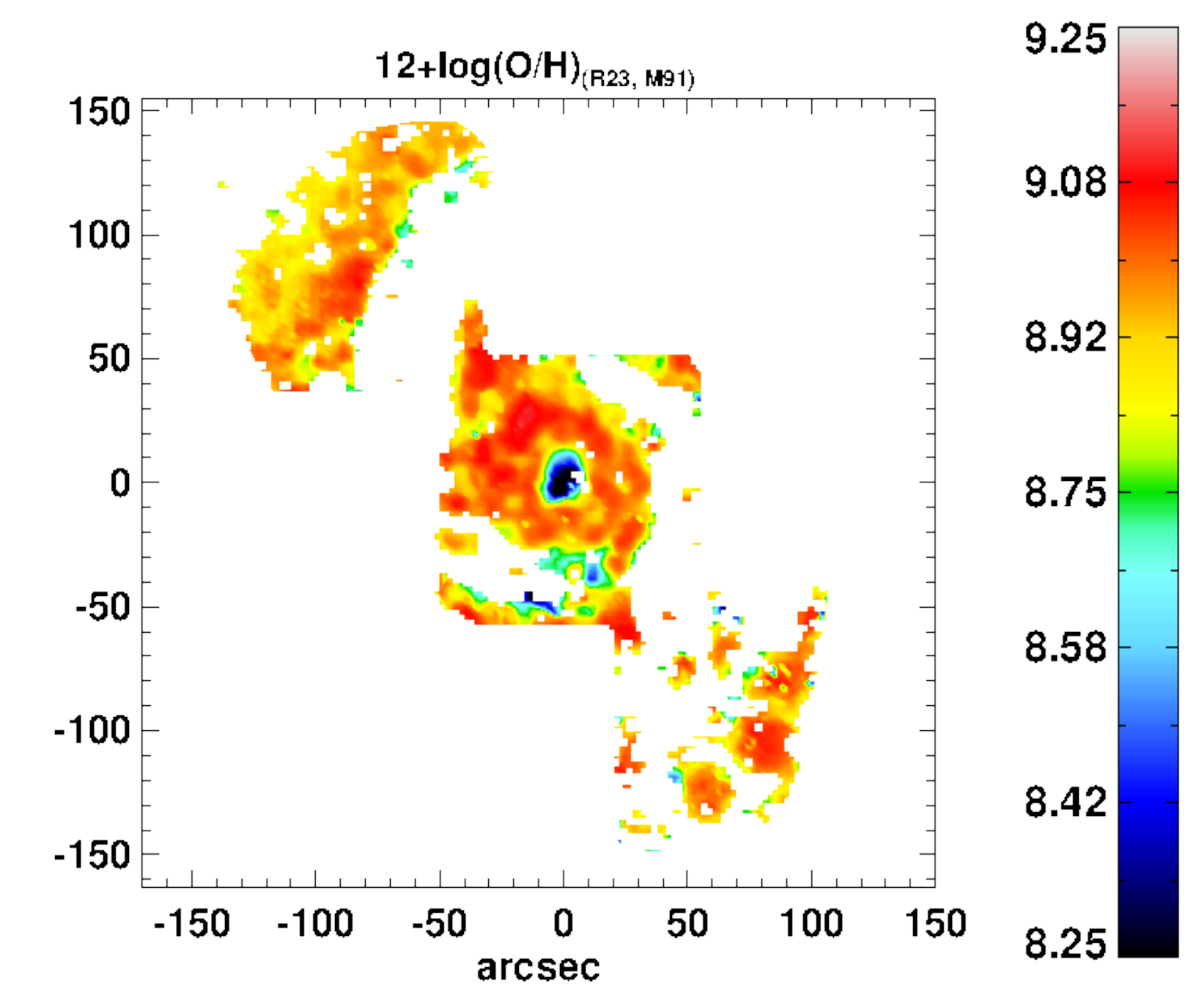}
\includegraphics[width=0.39\textheight]{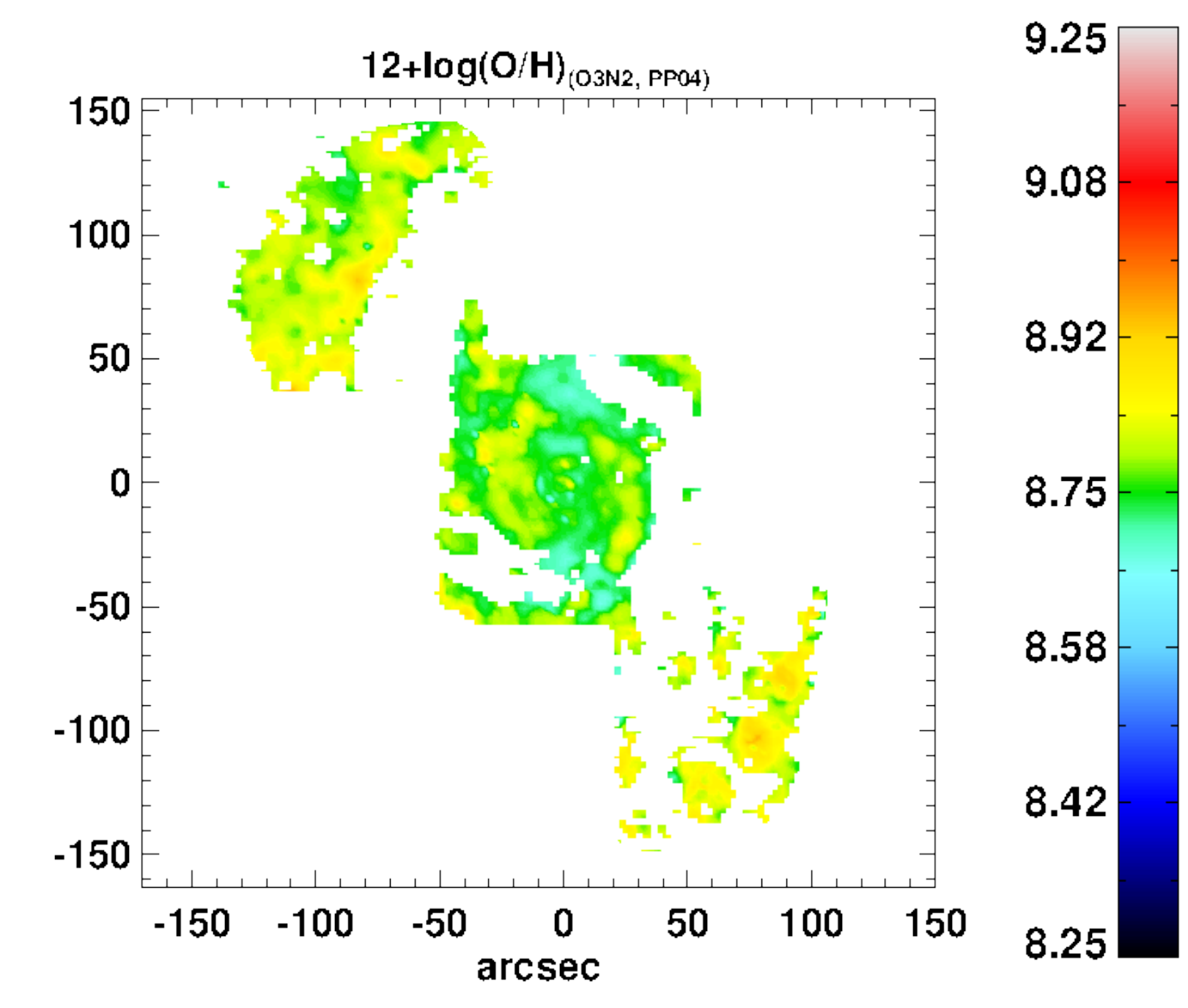}
\includegraphics[width=0.41\textheight]{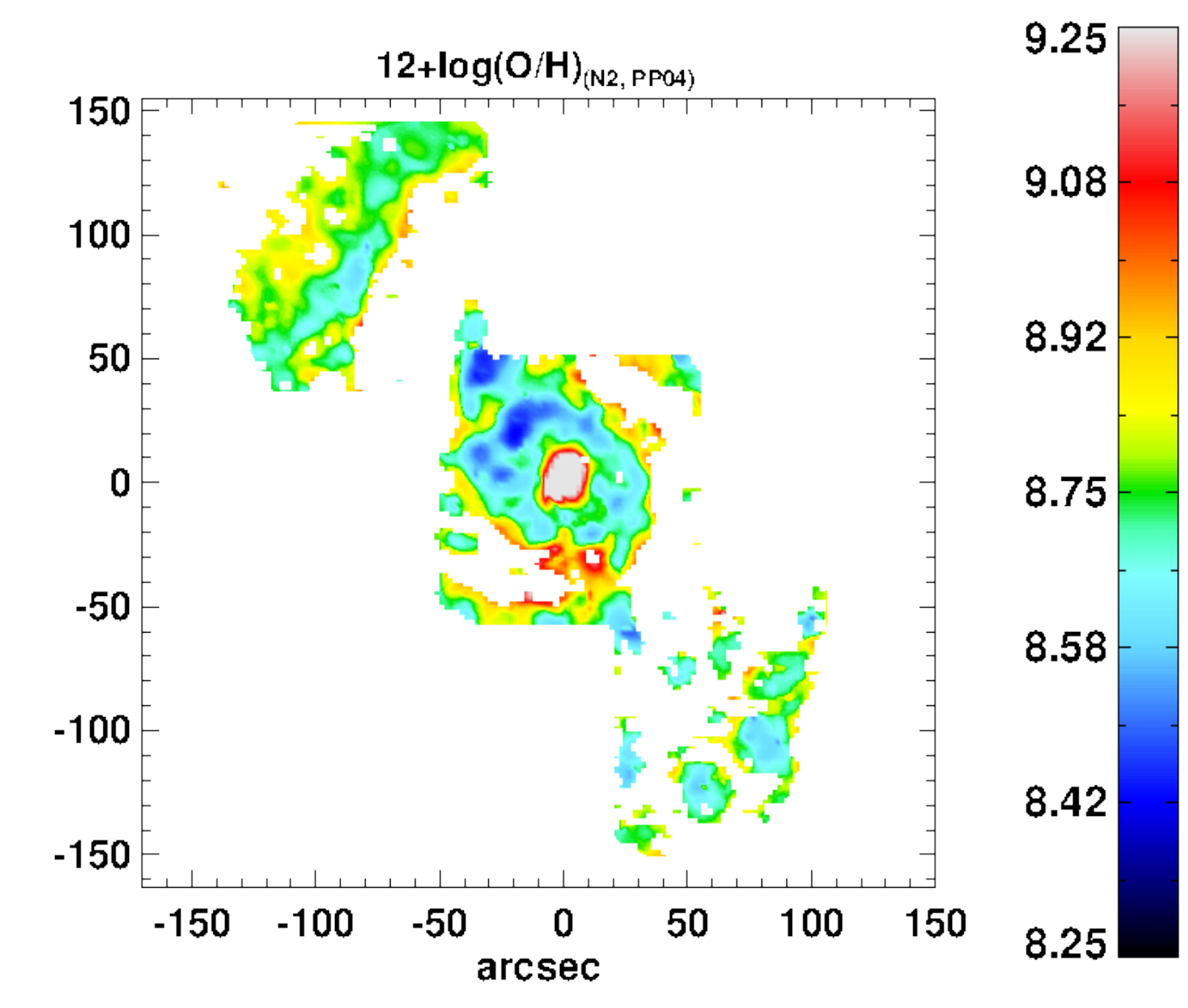}
\caption{Continued: For NGC 5194}
\label{}
\end{figure}
\end{landscape}

\clearpage
\addtocounter{figure}{-1}

\begin{figure*}
\includegraphics[width=0.32\textwidth]{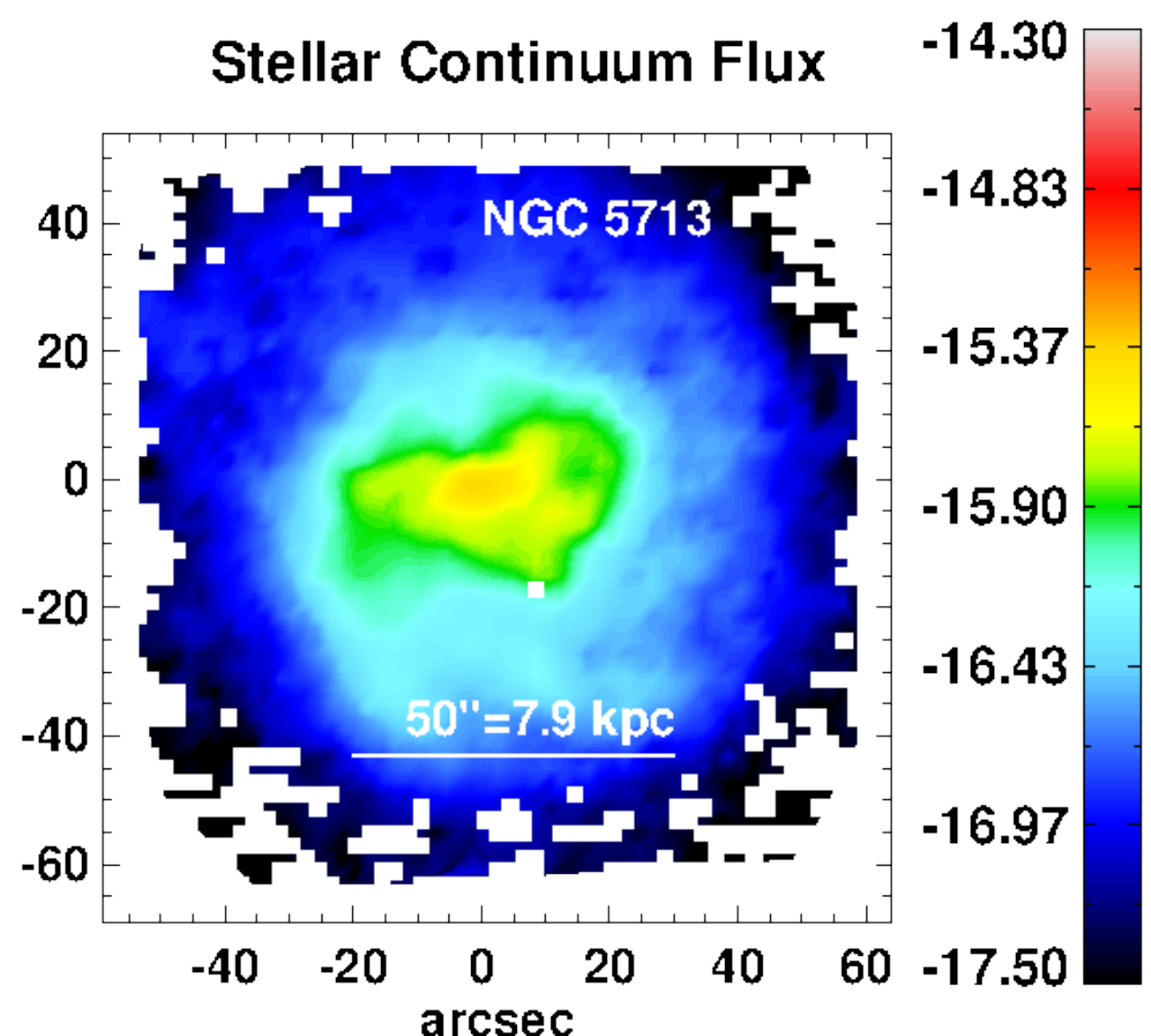}
\includegraphics[width=0.32\textwidth]{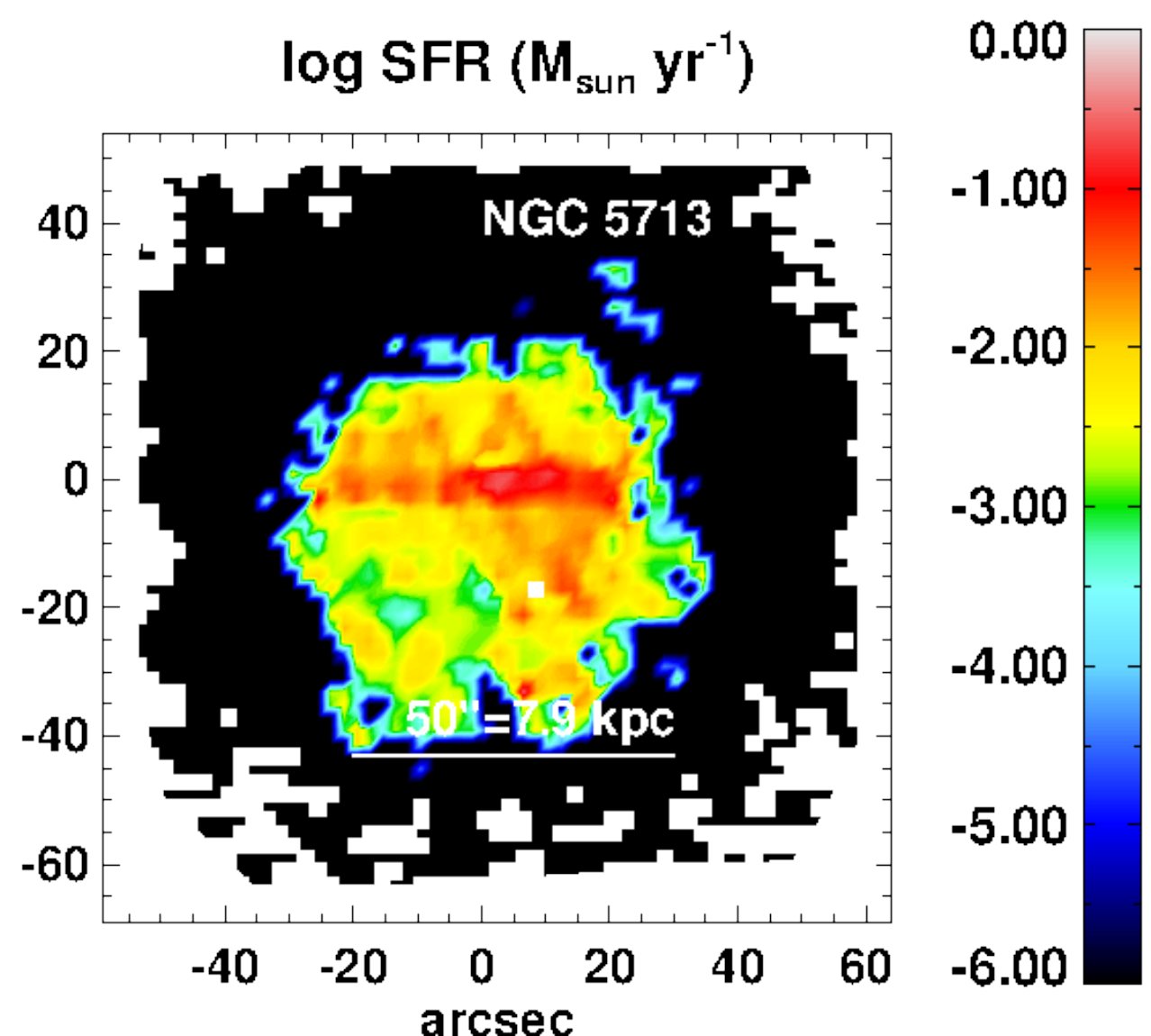}
\includegraphics[width=0.32\textwidth]{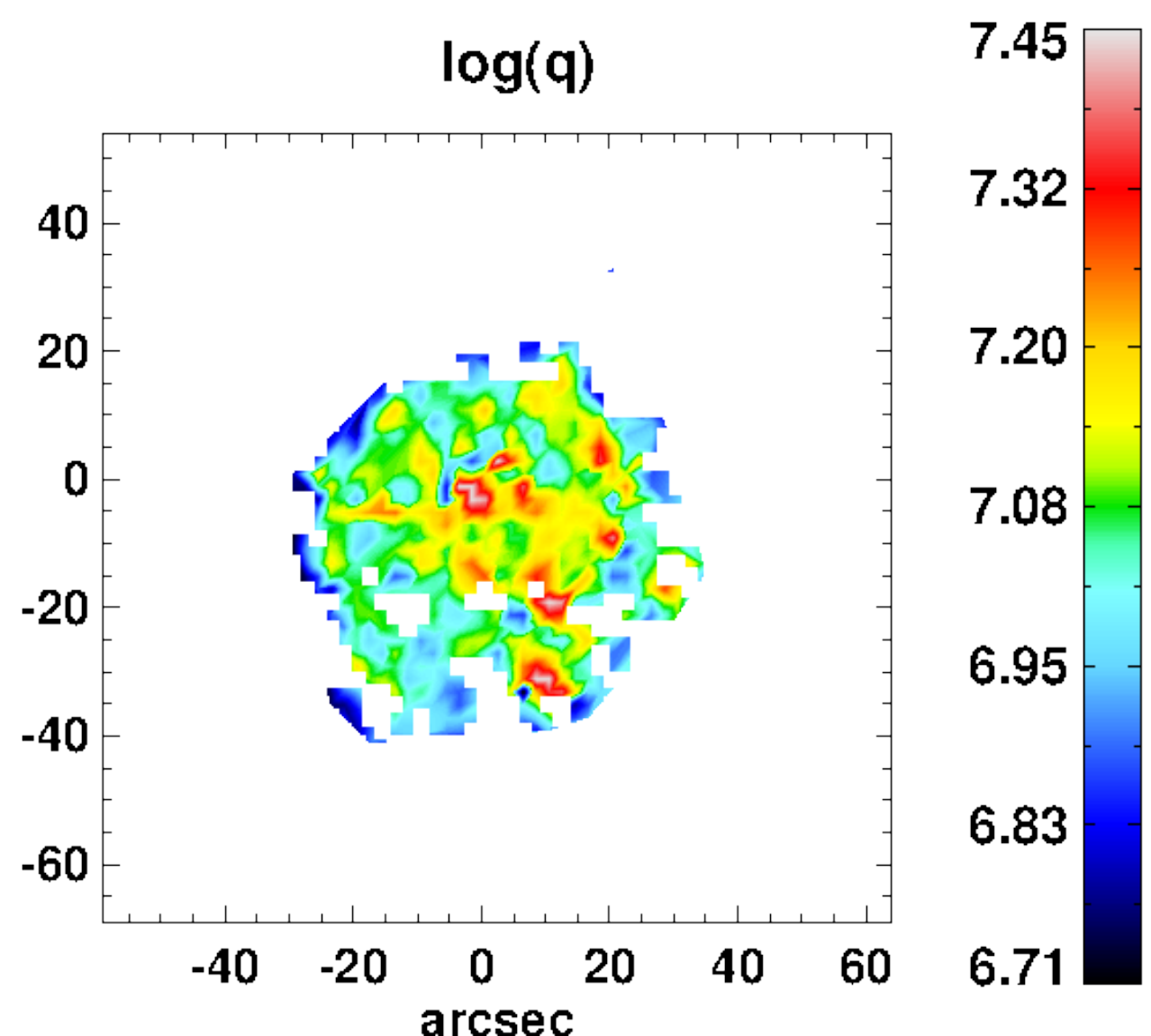}\\\
\includegraphics[width=0.32\textwidth]{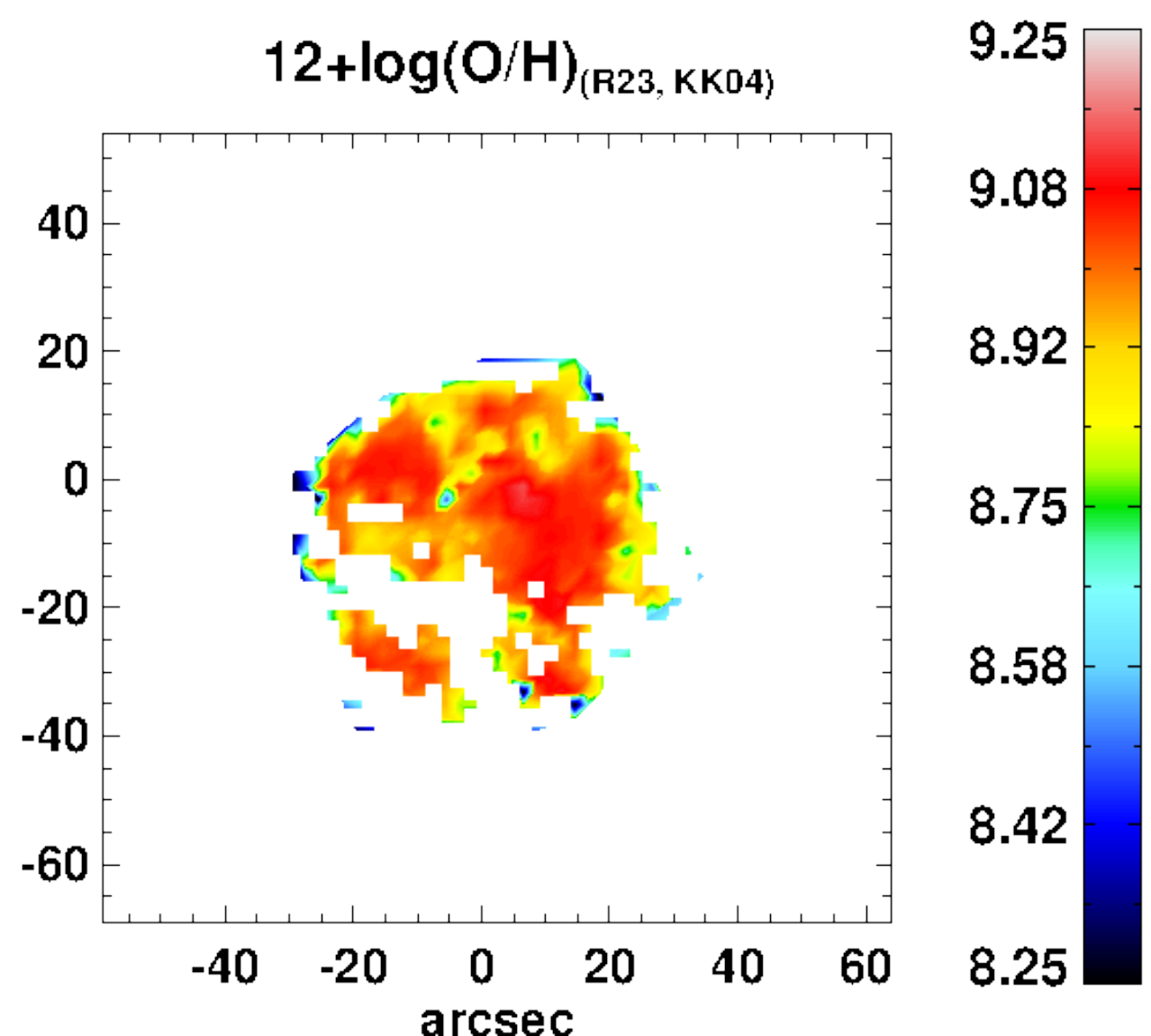}
\includegraphics[width=0.32\textwidth]{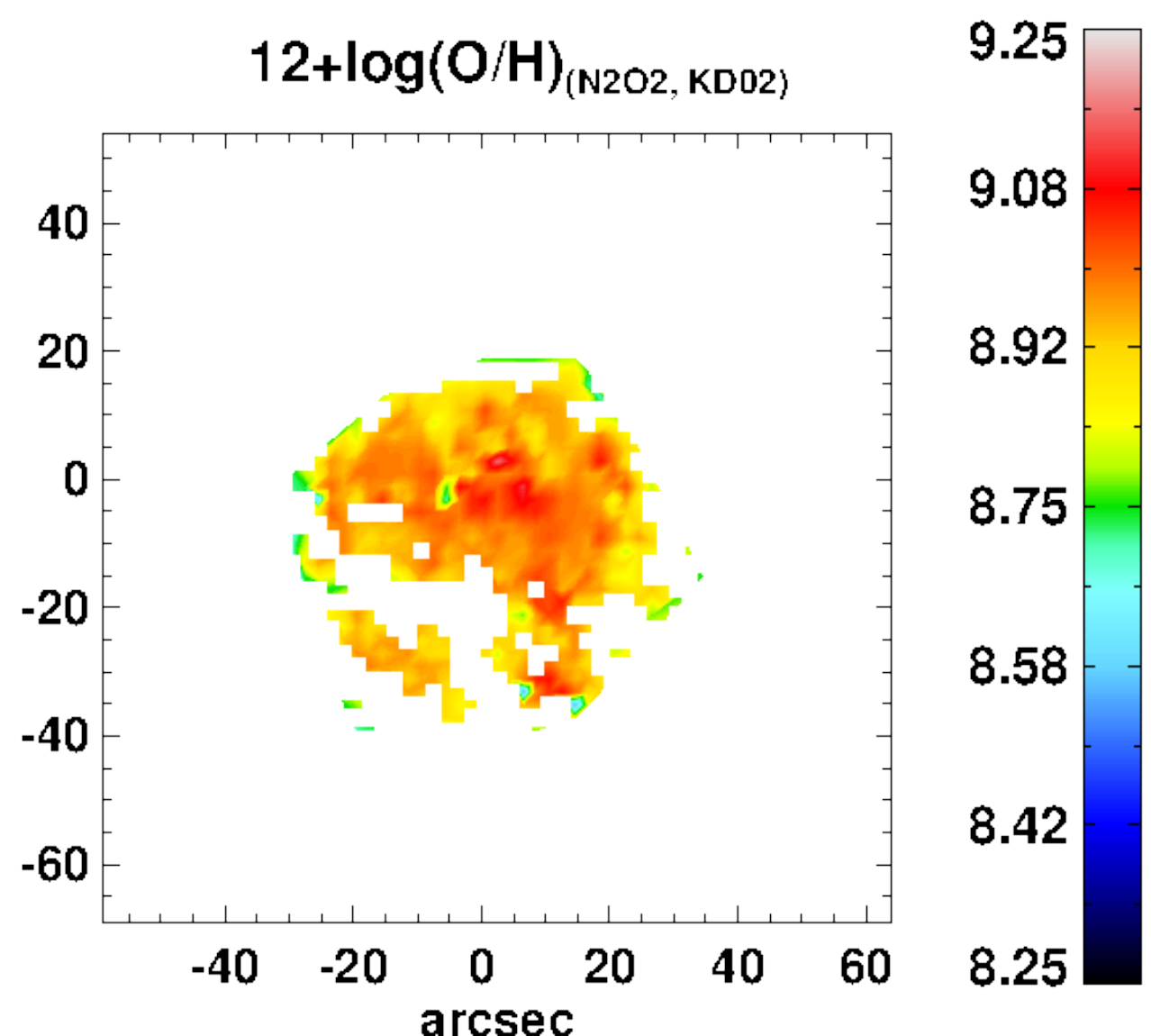}
\includegraphics[width=0.32\textwidth]{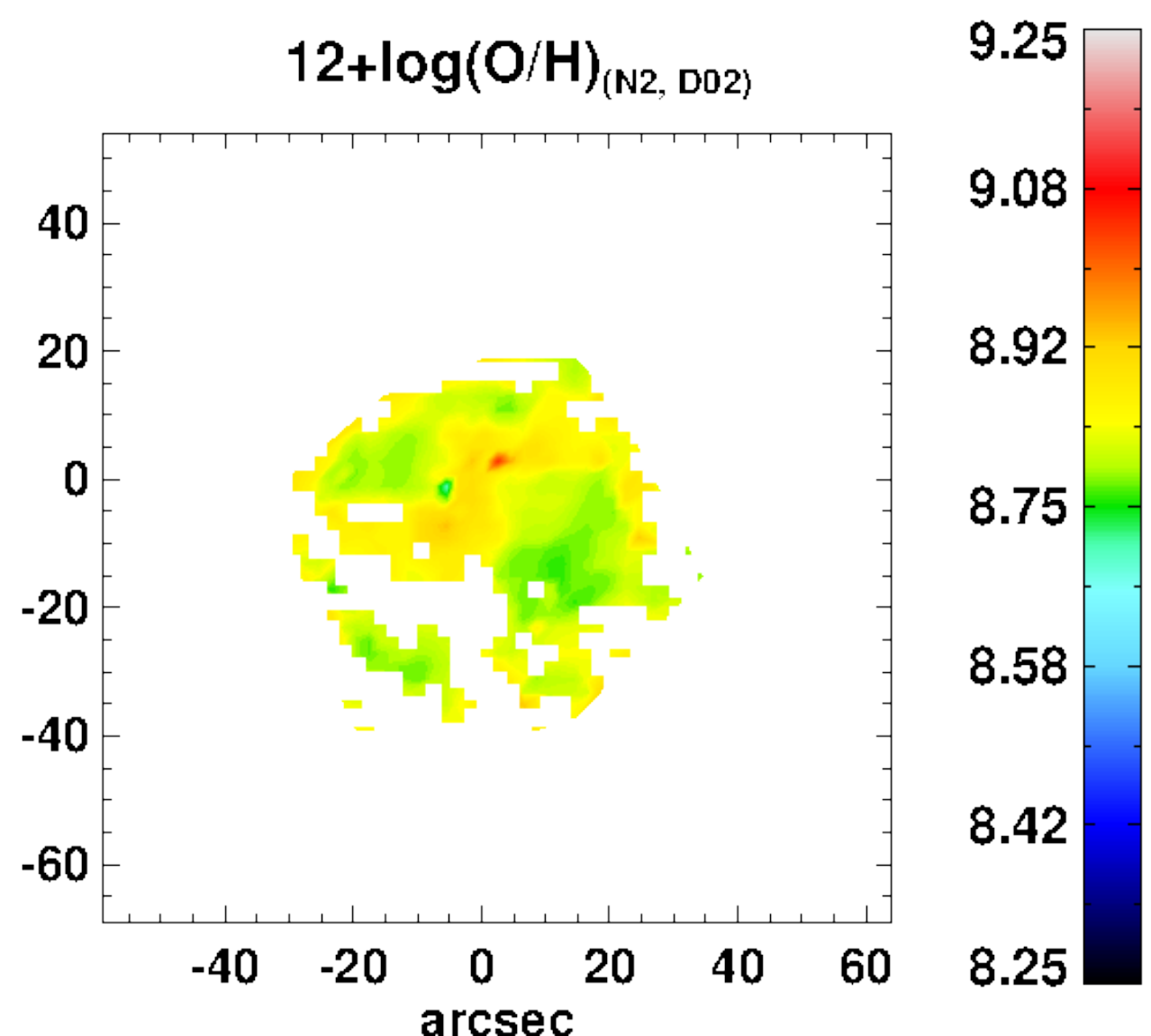} \\
\includegraphics[width=0.32\textwidth]{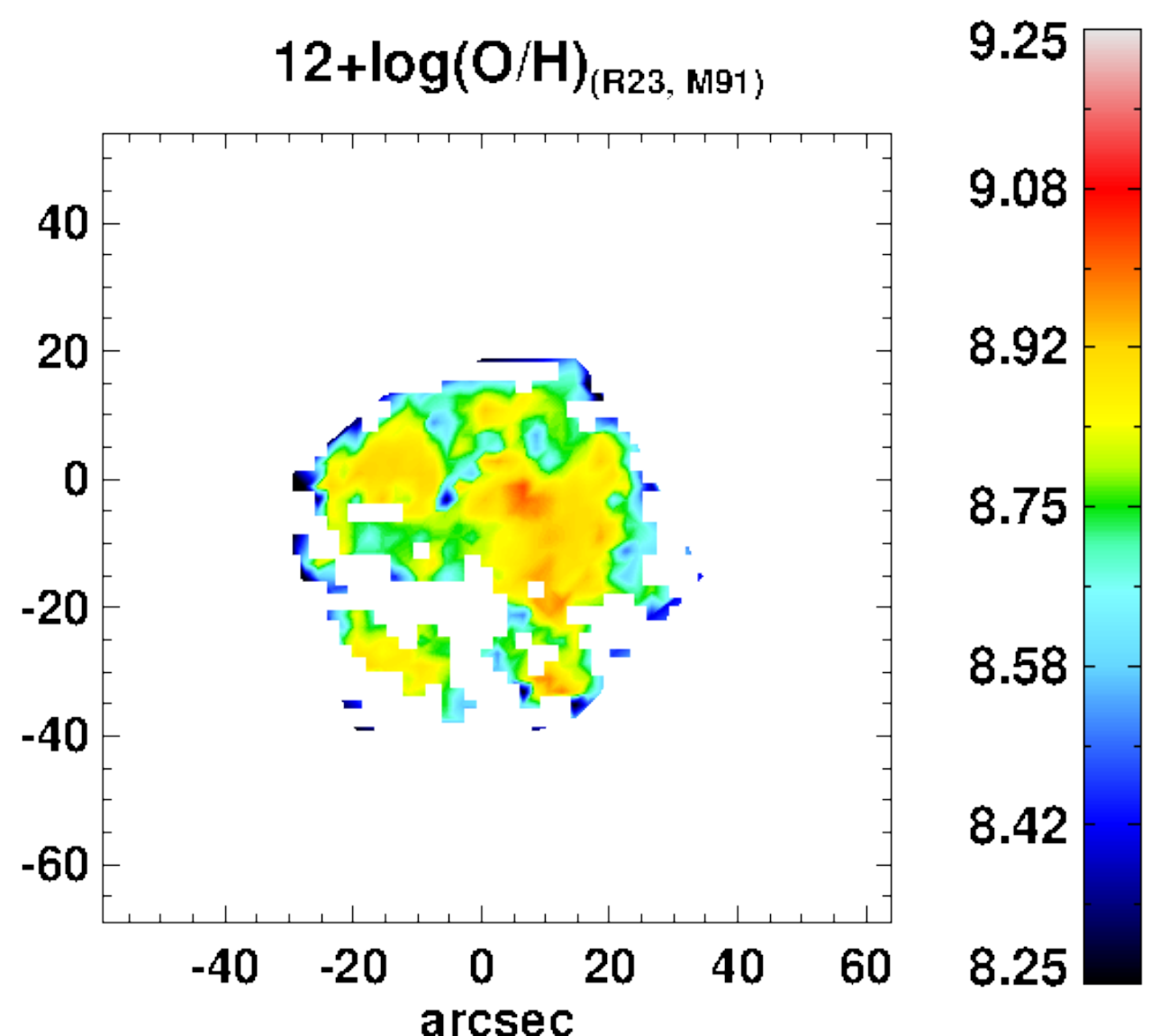}
\includegraphics[width=0.32\textwidth]{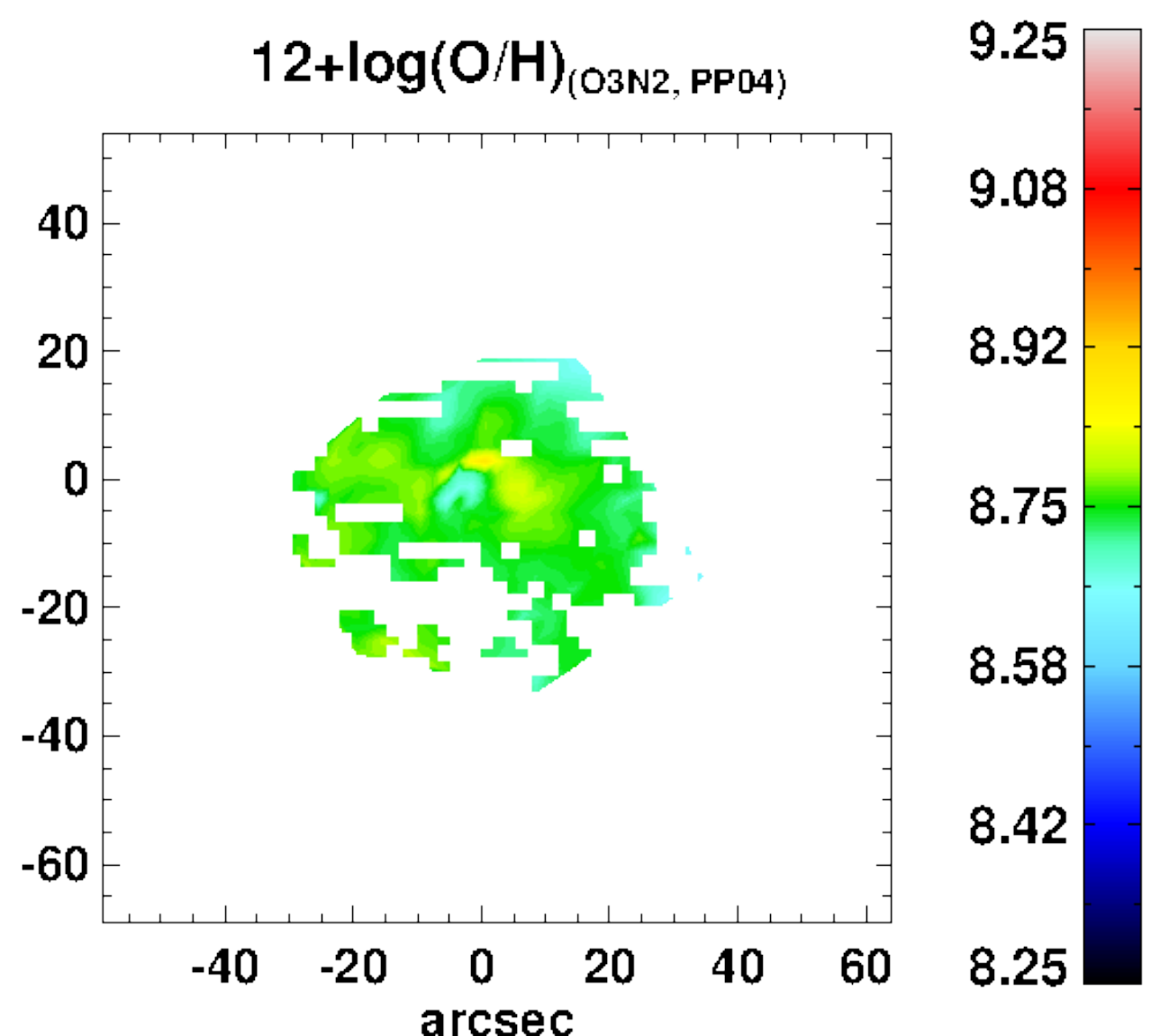}
\includegraphics[width=0.32\textwidth]{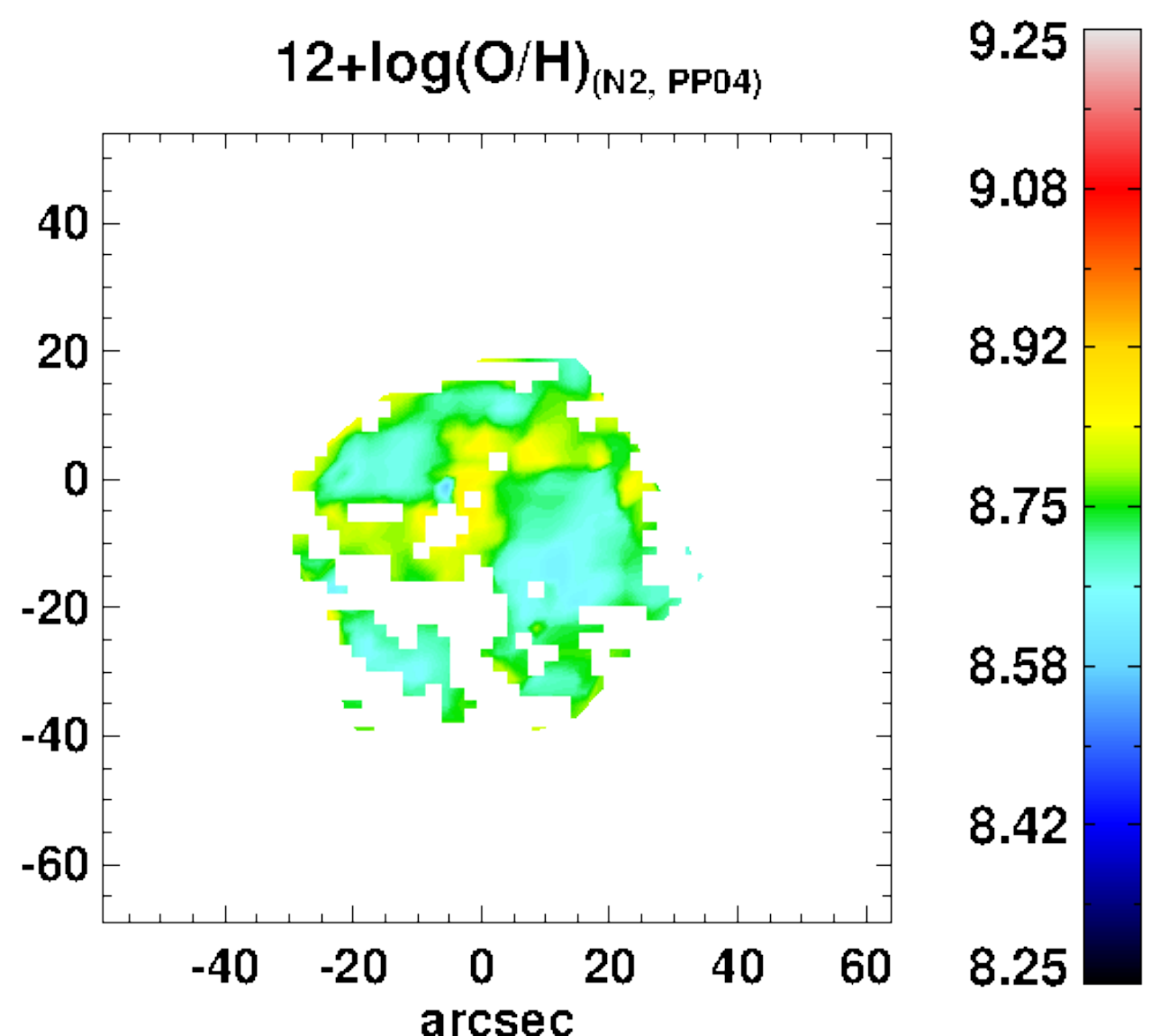}
\caption{Continued: For NGC 5713.}
\label{}
\end{figure*}
\clearpage

\begin{figure*}
\includegraphics[height=0.24\textheight, clip=true, trim=0.1cm 0.00cm 0.8cm 0.8 cm]{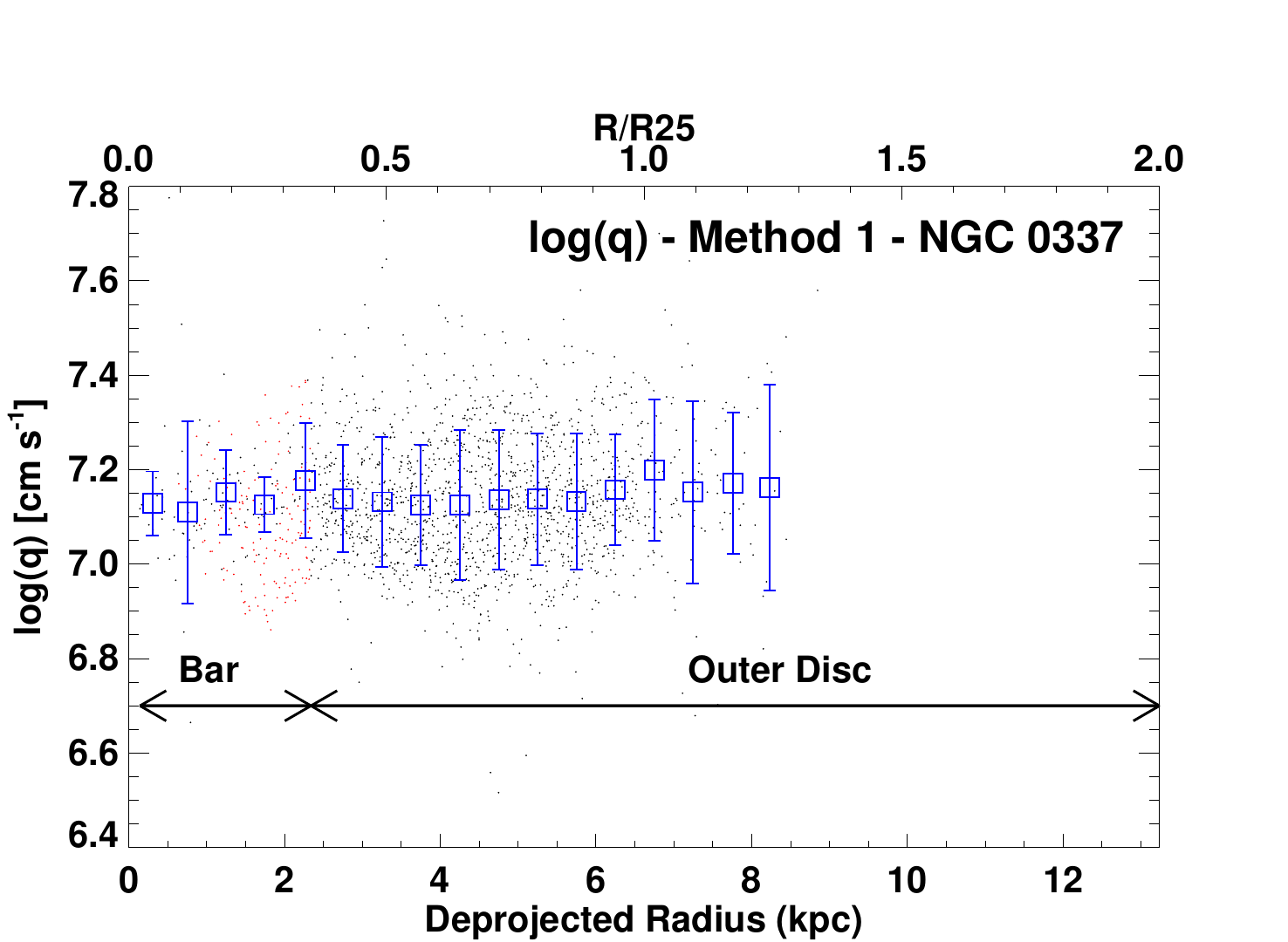}
\includegraphics[height=0.24\textheight, clip=true, trim=0.1cm 0.00cm 0.8cm 0.8 cm]{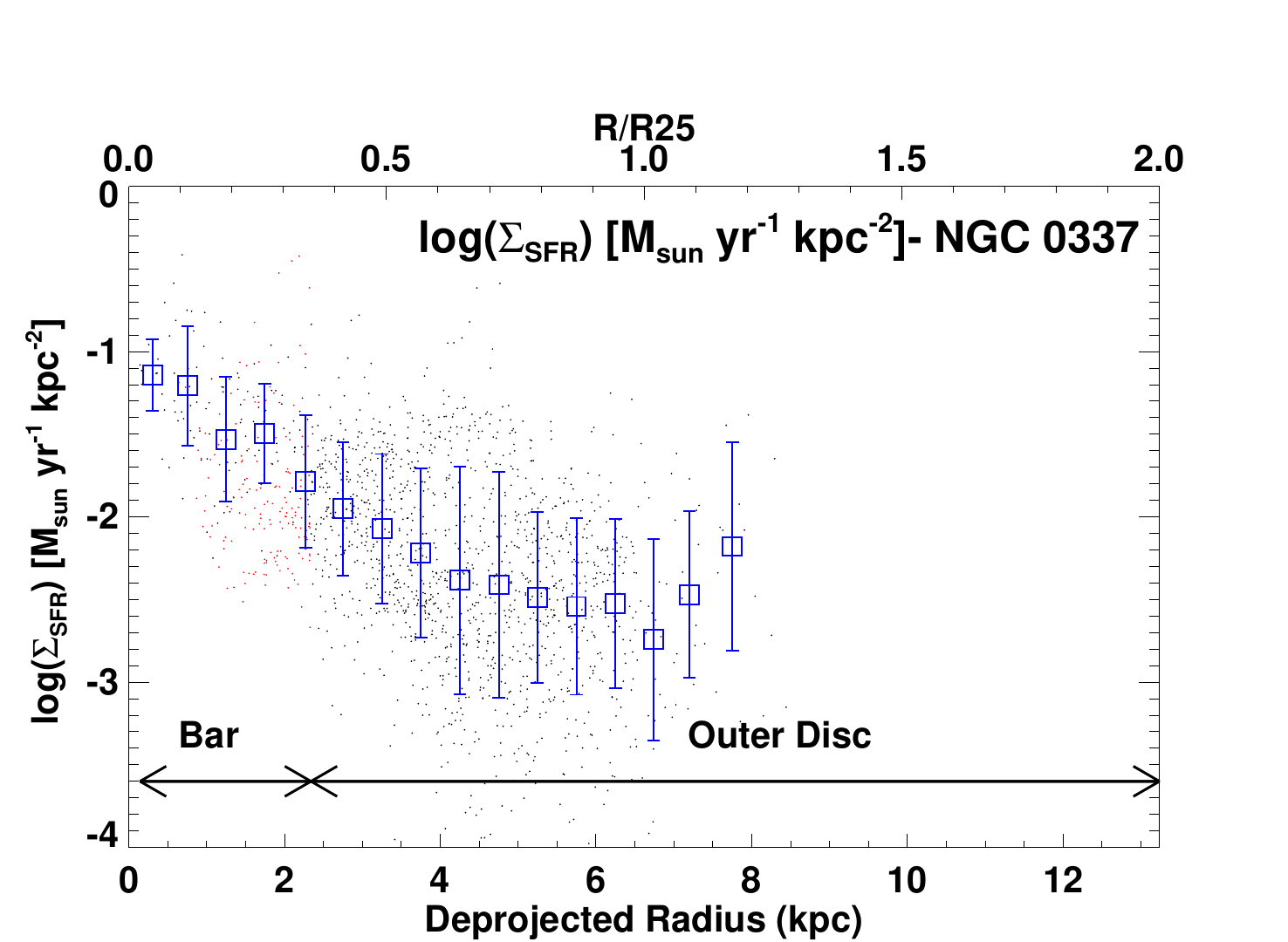}\vspace{-18pt} \\
\includegraphics[height=0.24\textheight, clip=true, trim=0.1cm 0.00cm 0.8cm 0.8 cm]{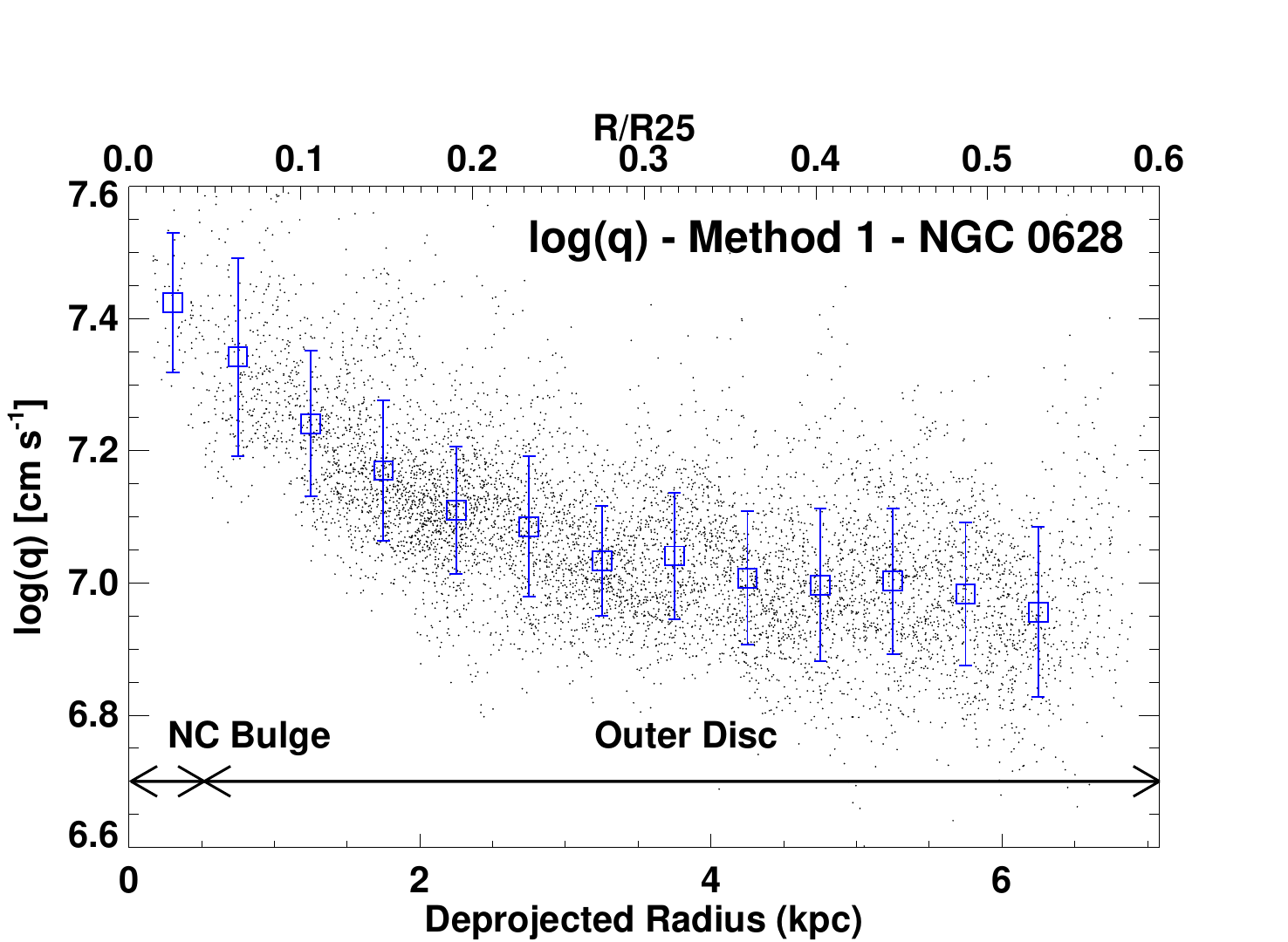}
\includegraphics[height=0.24\textheight, clip=true, trim=0.1cm 0.00cm 0.8cm 0.8 cm]{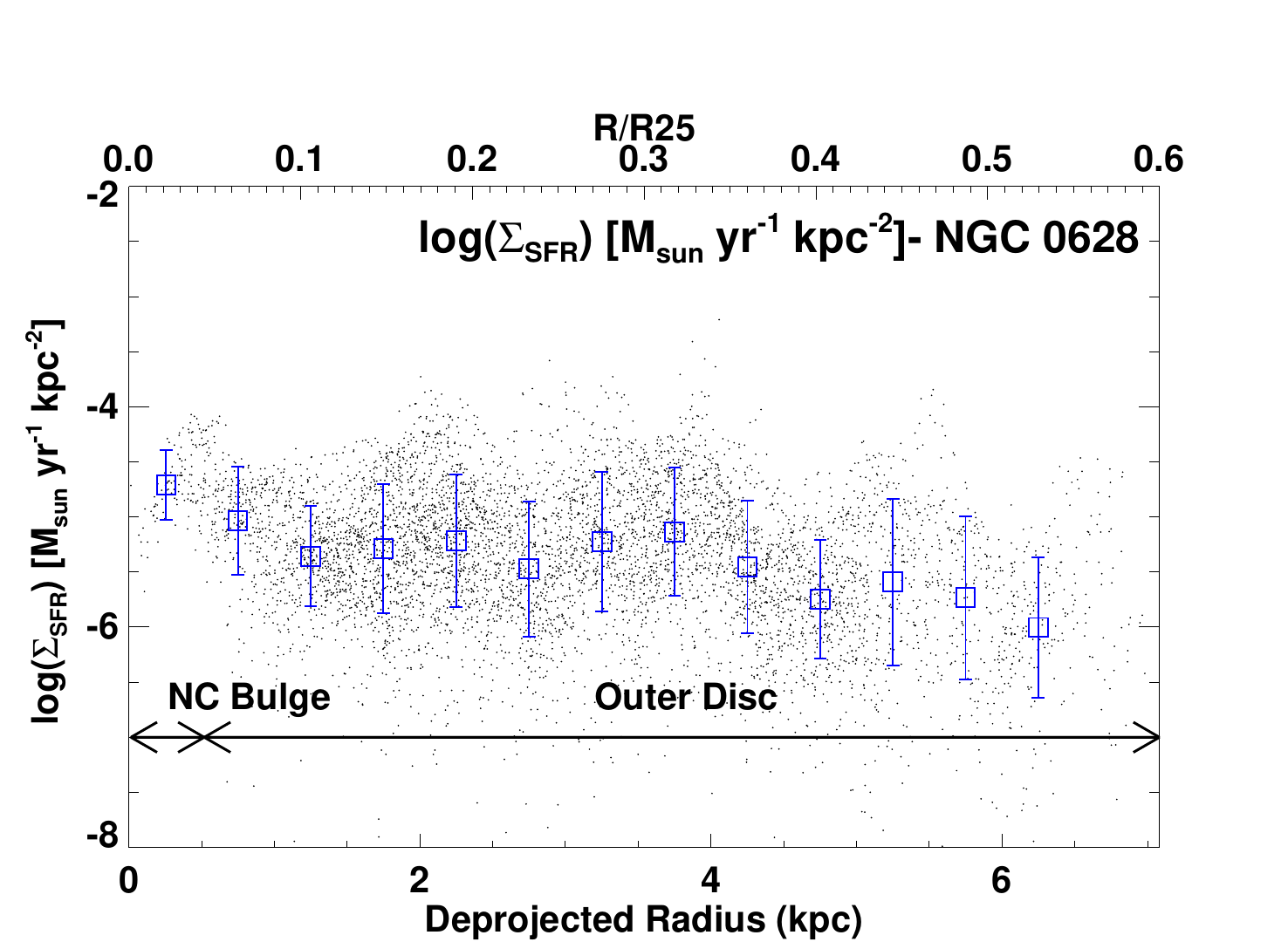}\vspace{-18pt} \\
\includegraphics[height=0.24\textheight, clip=true, trim=0.1cm 0.00cm 0.8cm 0.8 cm]{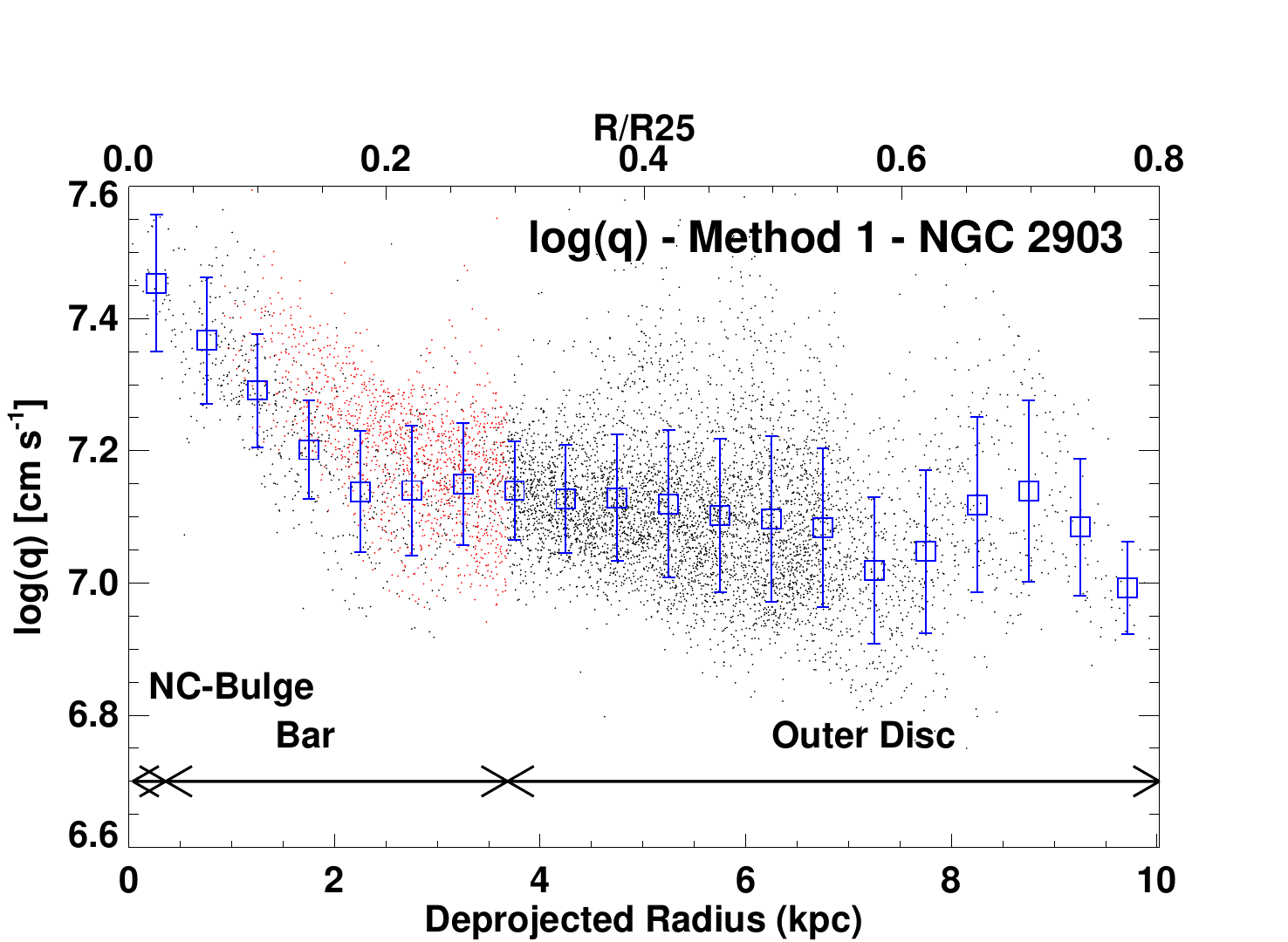}
\includegraphics[height=0.24\textheight, clip=true, trim=0.1cm 0.00cm 0.8cm 0.8 cm]{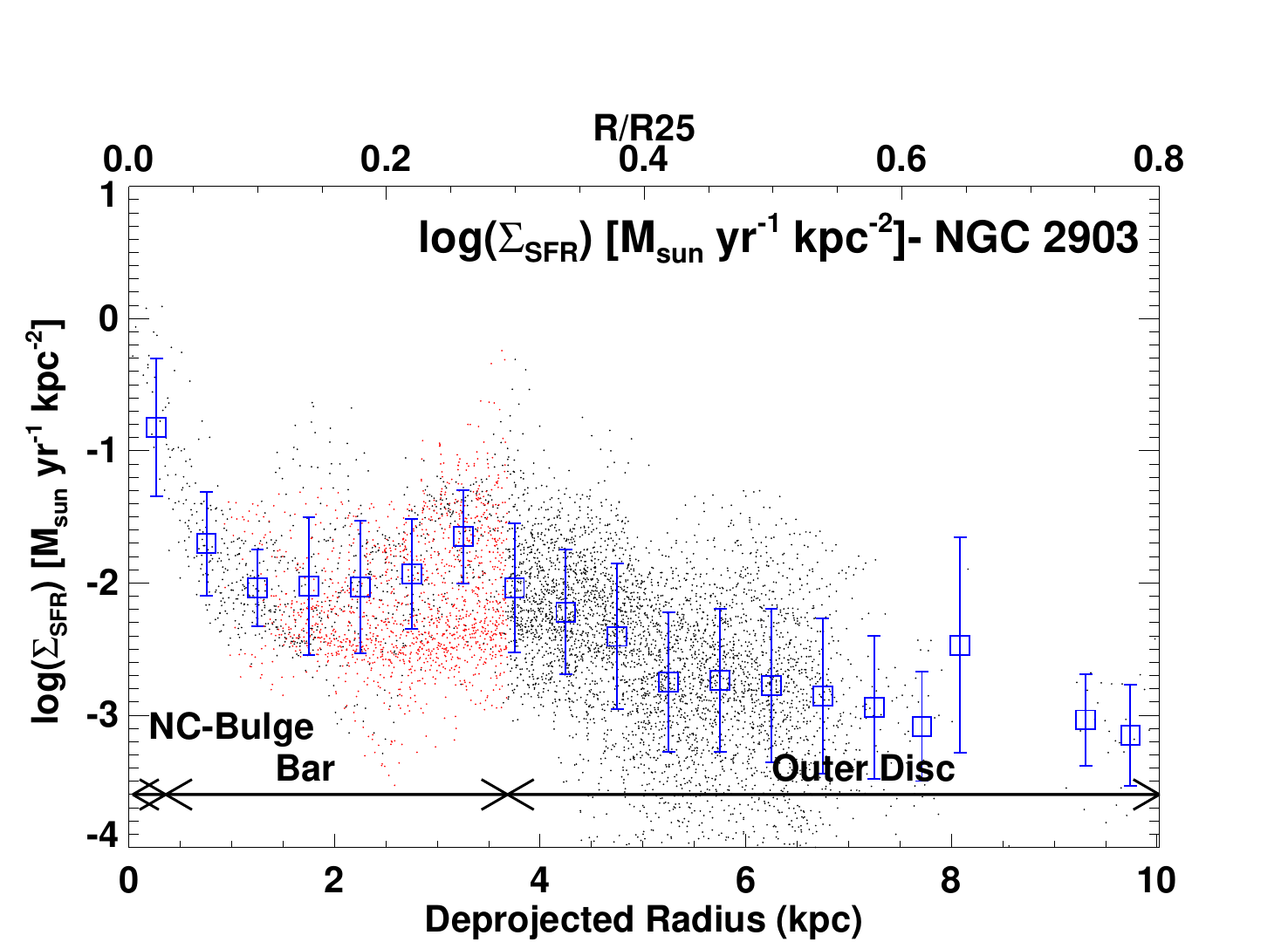}\vspace{-18pt}\\
\includegraphics[height=0.24\textheight, clip=true, trim=0.1cm 0.00cm 0.8cm 0.8 cm]{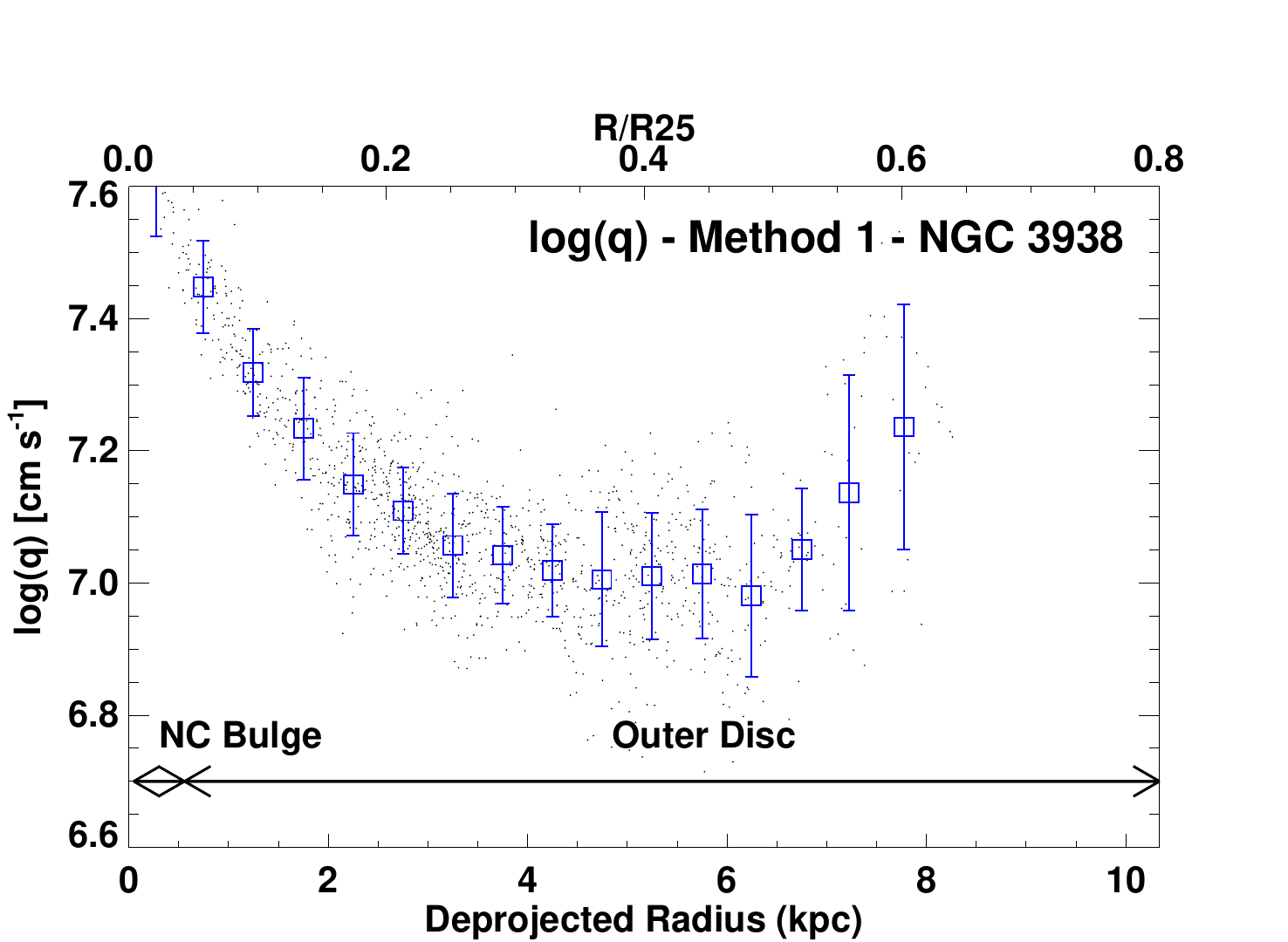}
\includegraphics[height=0.24\textheight, clip=true, trim=0.1cm 0.00cm 0.8cm 0.8 cm]{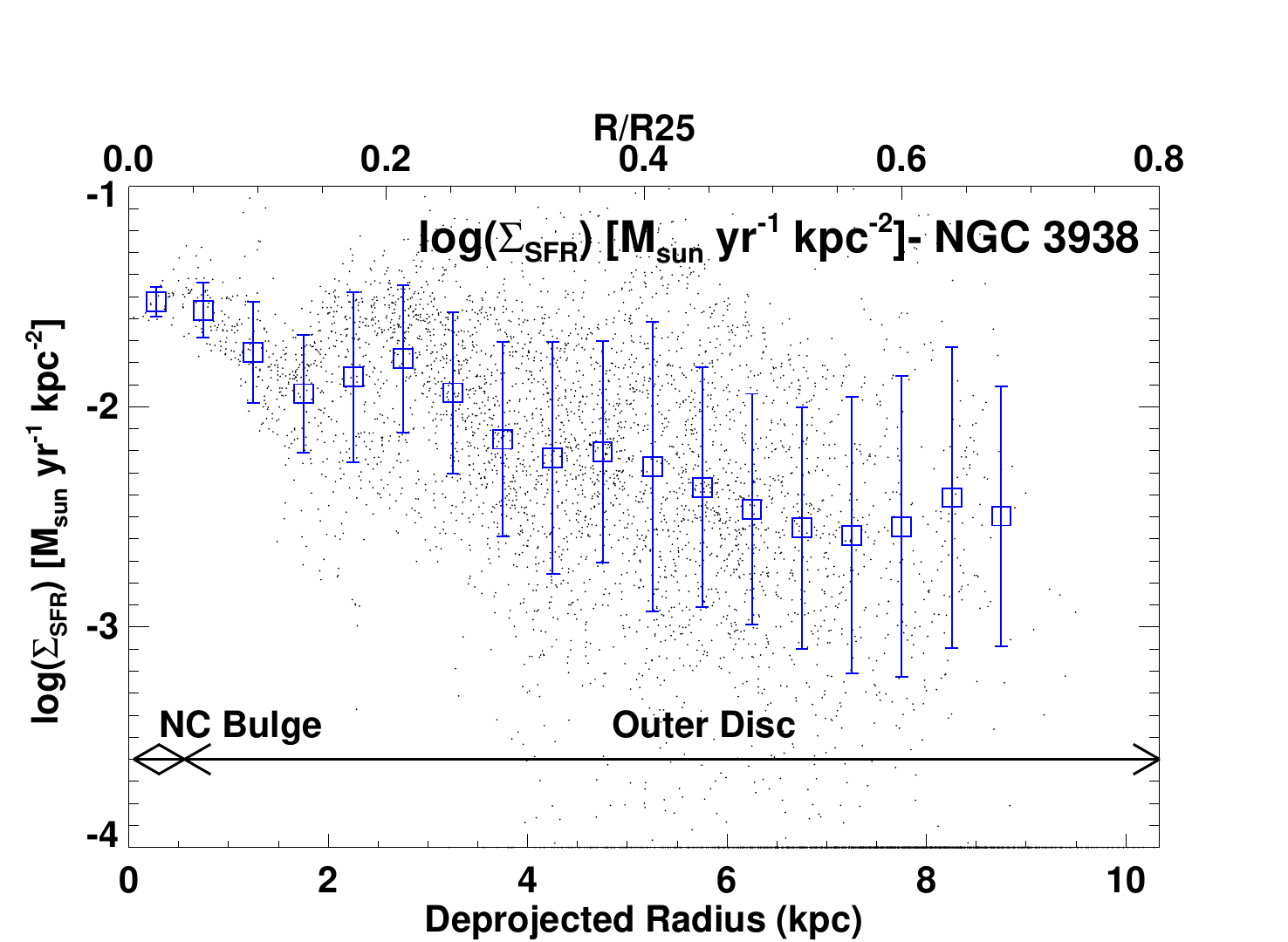}
\caption{Deprojected radial profiles for the ionization parameter $q$ 
({\bf{left}}; $\S$\ref{sec:how-q})  in units of cm s$^{-1}$ and  the SFR surface density $\Sigma_{\rm SFR}$ 
in units of  $M_\odot$ yr$^{-1}$ kpc$^{-2}$ ({\bf{right}}) for NGC 0337, 0628, 2903, \& 3938.
The arrows on the bottom show our decomposition of each galaxy into non-classical (NC)  
bulge (which refers to discy/pseudo bulge), bar, and outer disc.  
The blue squares and error bars are the mean and 1$\sigma$ dispersion of 0.5 kpc bins.
In the case of barred galaxies, the  region labelled `Bar' on the plot refers to 
the radii between the bulge and the end of the bar. In this region, 
we show regions in the bar feature as  black points, and  regions azimuthally offset
from the bar feature as red points. The blue squares denoting the mean only use regions
within the bar.
}
\label{fig:q-sfr-gradients}
\end{figure*}

\addtocounter{figure}{-1}

\begin{figure*}
\includegraphics[height=0.24\textheight, clip=true, trim=0.1cm 0.00cm 0.8cm 0.8 cm]{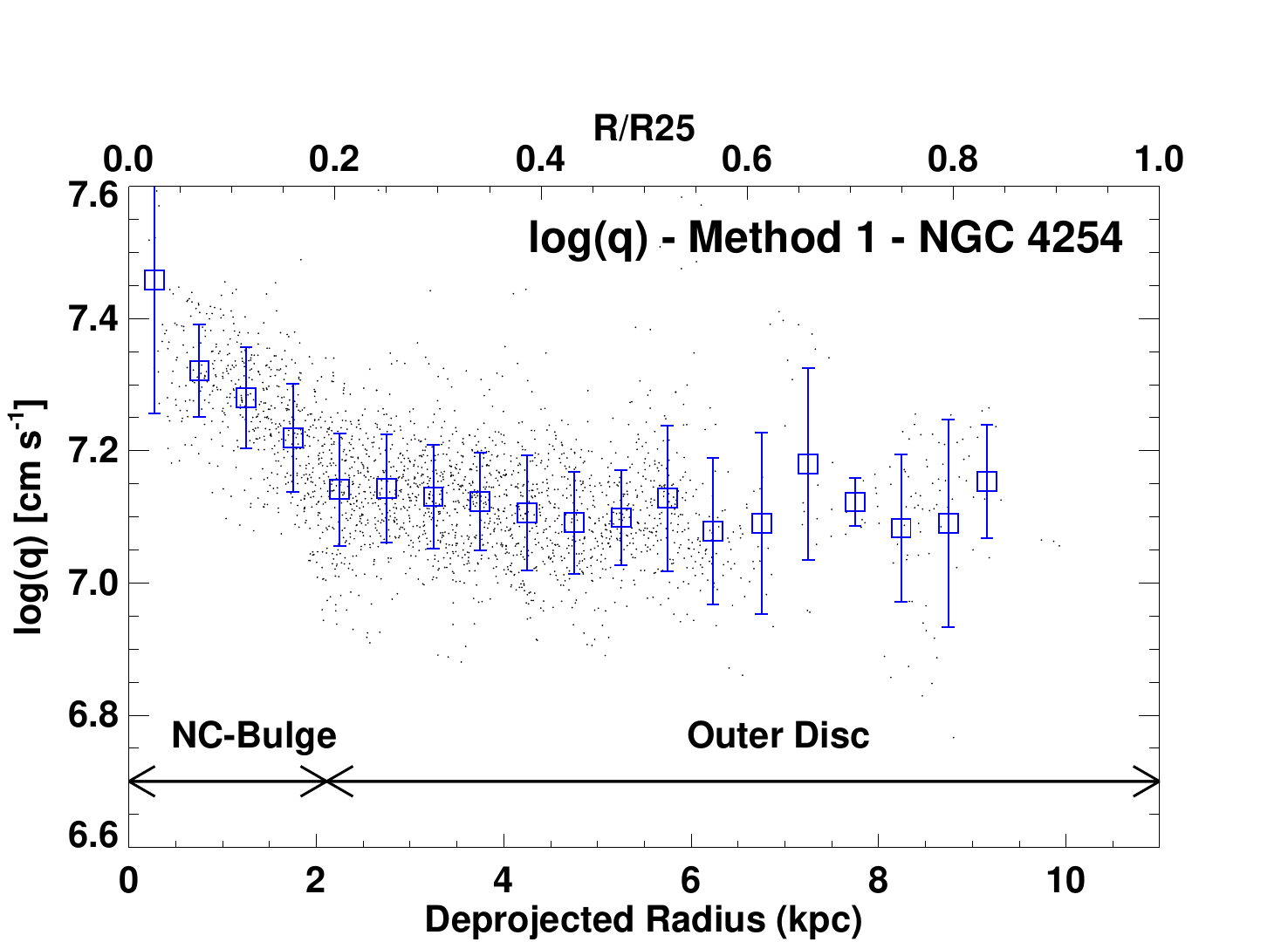}
\includegraphics[height=0.24\textheight, clip=true, trim=0.1cm 0.00cm 0.8cm 0.8 cm]{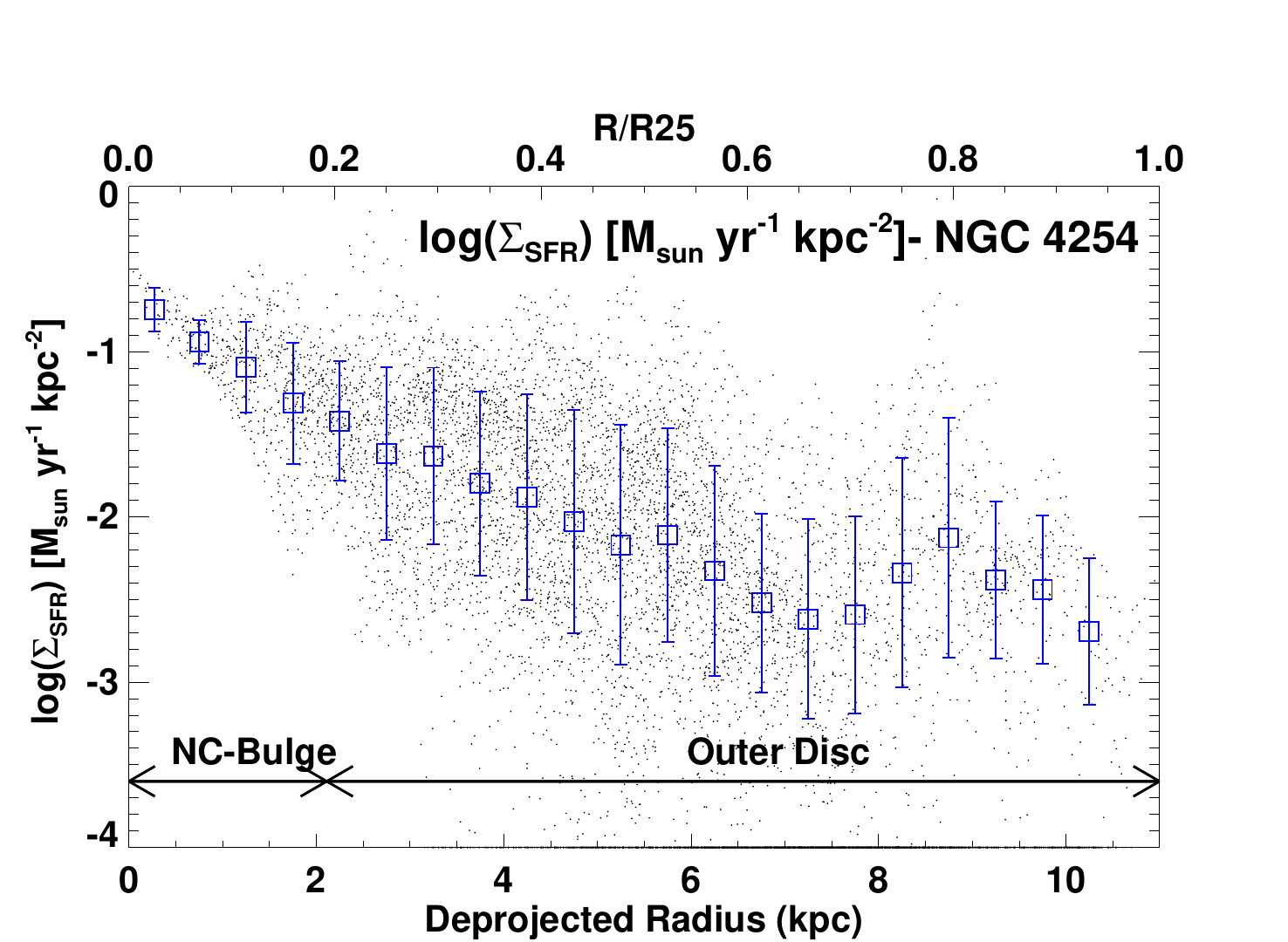}\vspace{-18pt} \\
\includegraphics[height=0.24\textheight, clip=true, trim=0.1cm 0.00cm 0.8cm 0.8 cm]{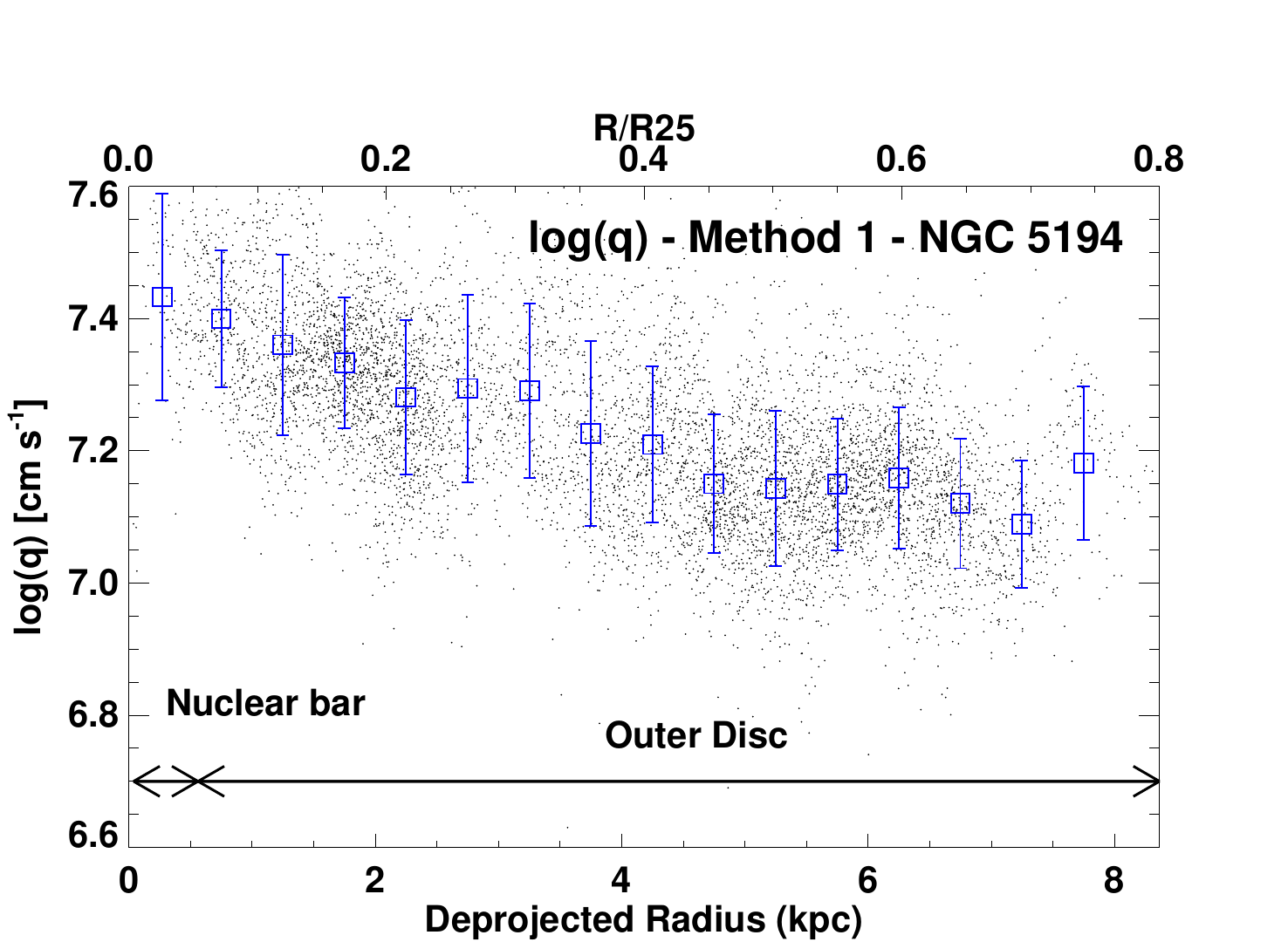}
\includegraphics[height=0.24\textheight, clip=true, trim=0.1cm 0.00cm 0.8cm 0.8 cm]{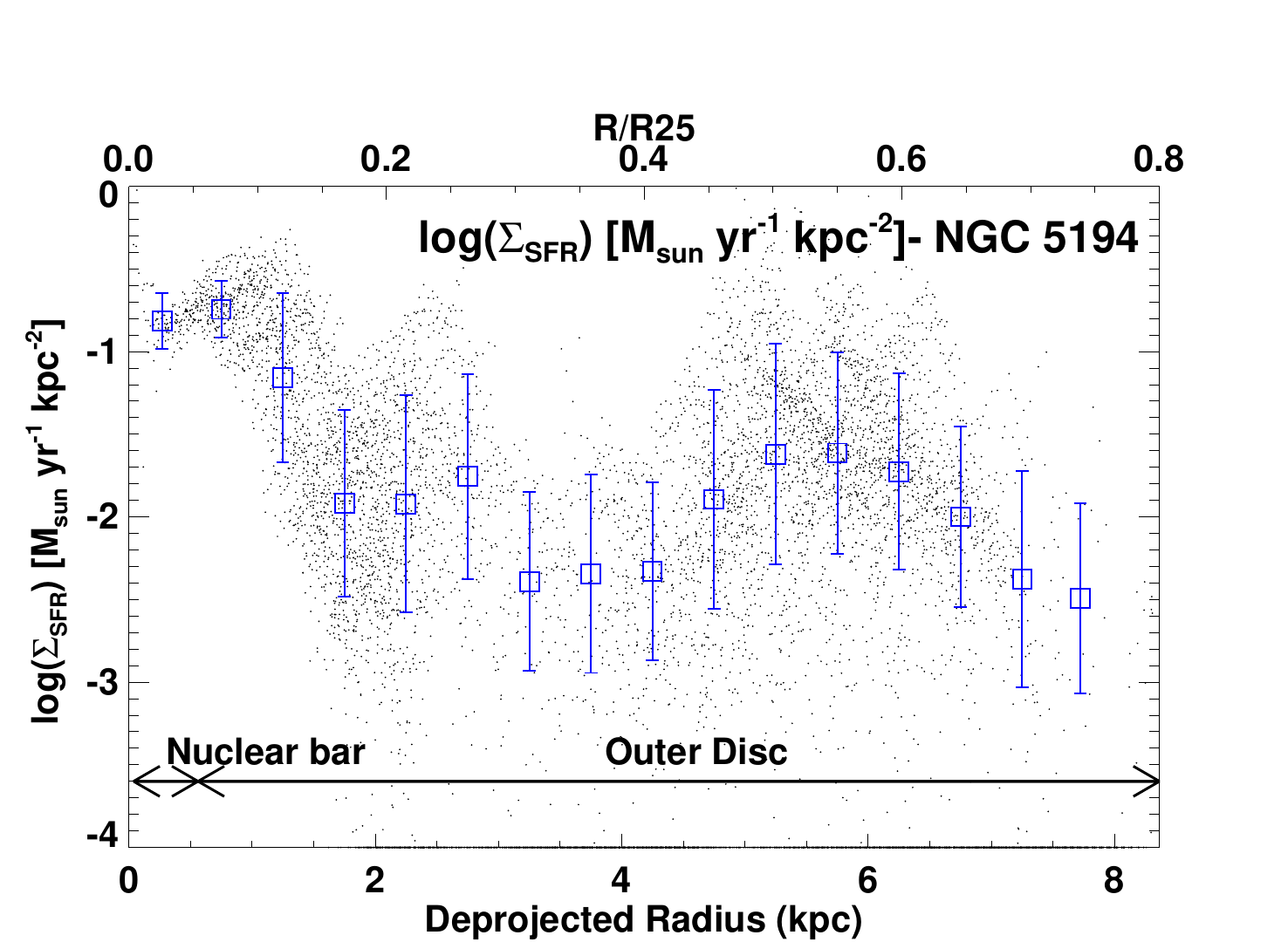}\vspace{-18pt} \\
\includegraphics[height=0.24\textheight, clip=true, trim=0.1cm 0.00cm 0.8cm 0.8 cm]{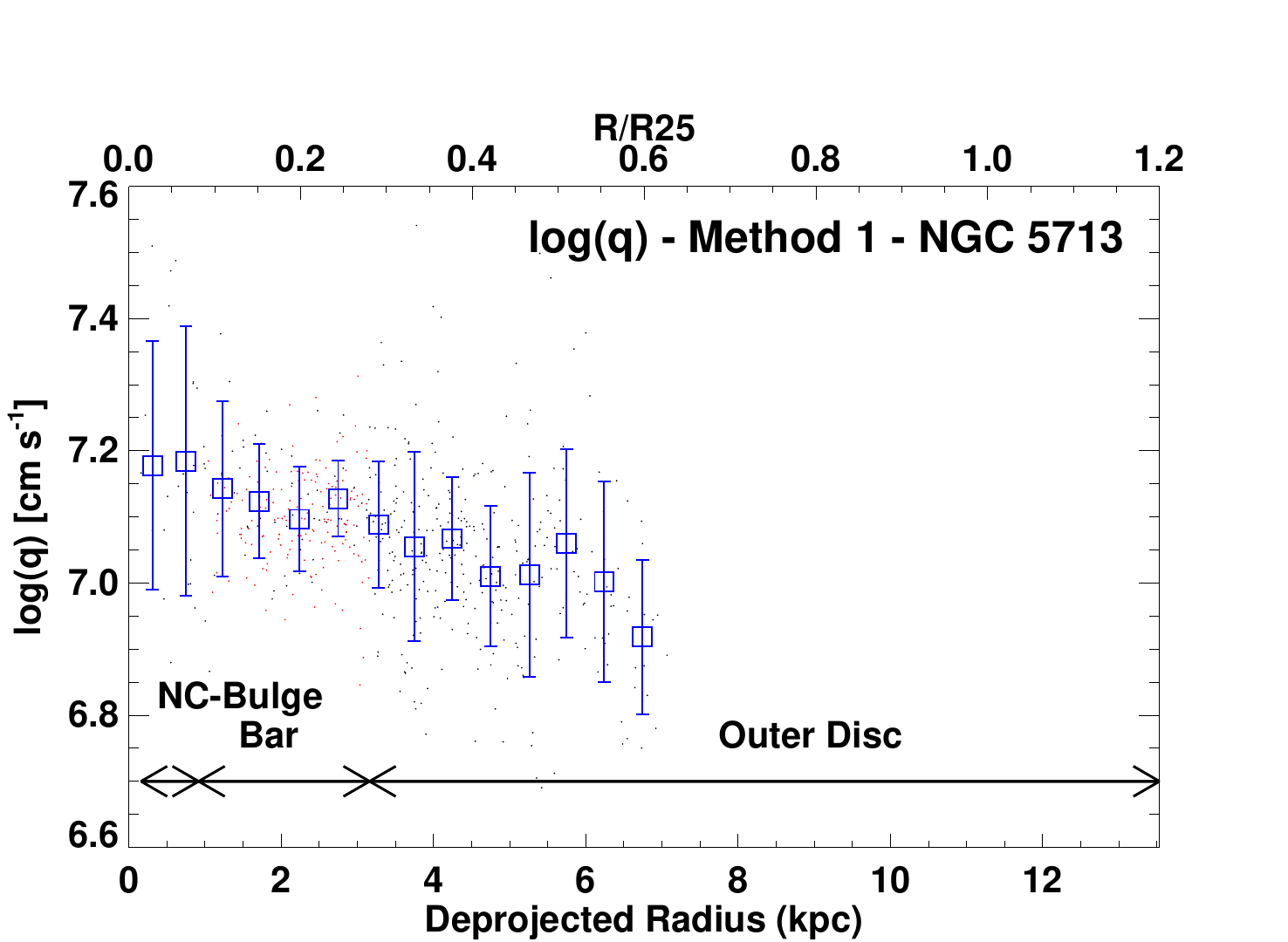}
\includegraphics[height=0.24\textheight, clip=true, trim=0.1cm 0.00cm 0.8cm 0.8 cm]{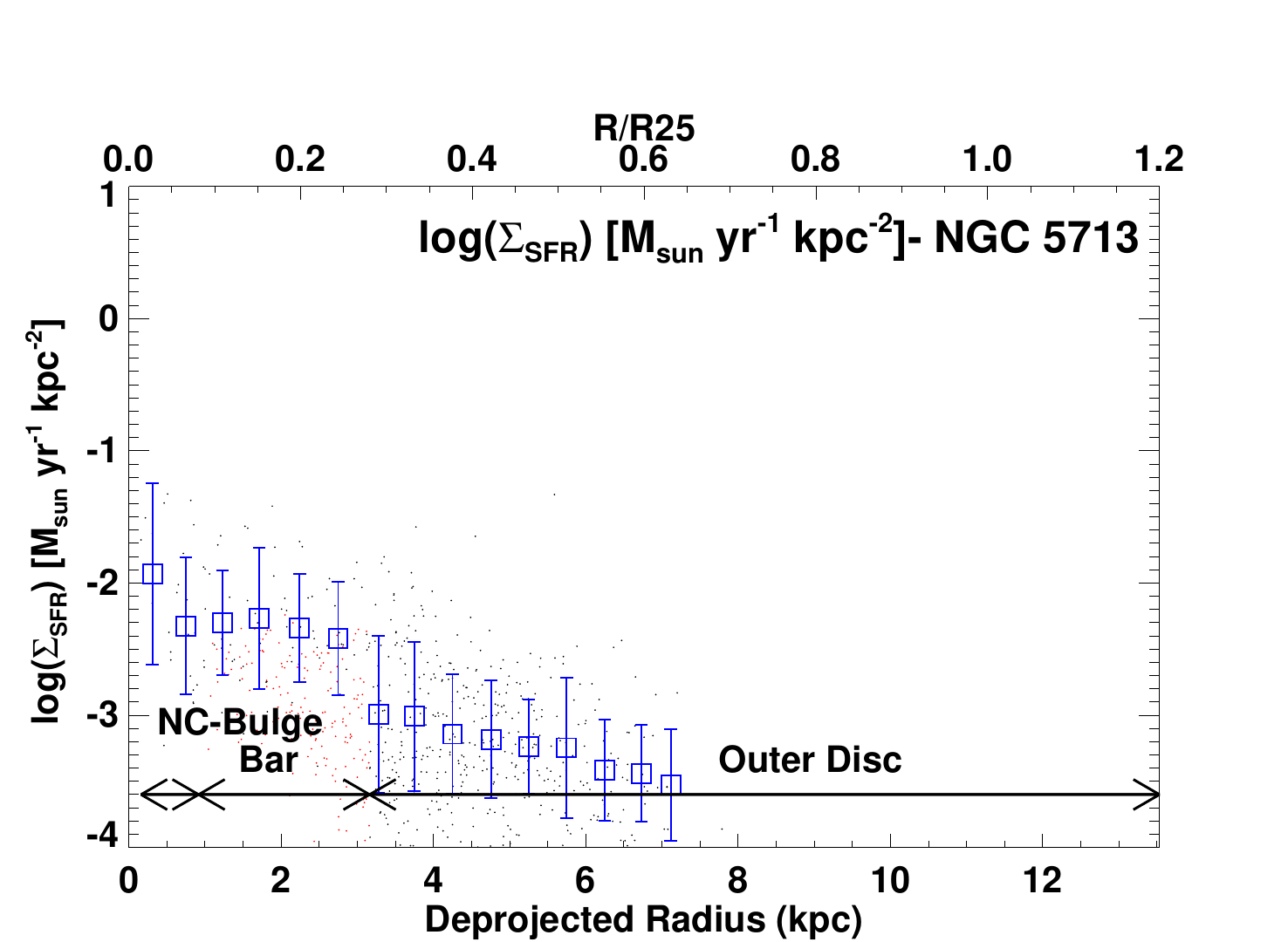}
\caption{Continued: For NGC 4254, 5194, \& 5713.}
\end{figure*}

\clearpage

\begin{figure*}
\includegraphics[width=\textwidth]{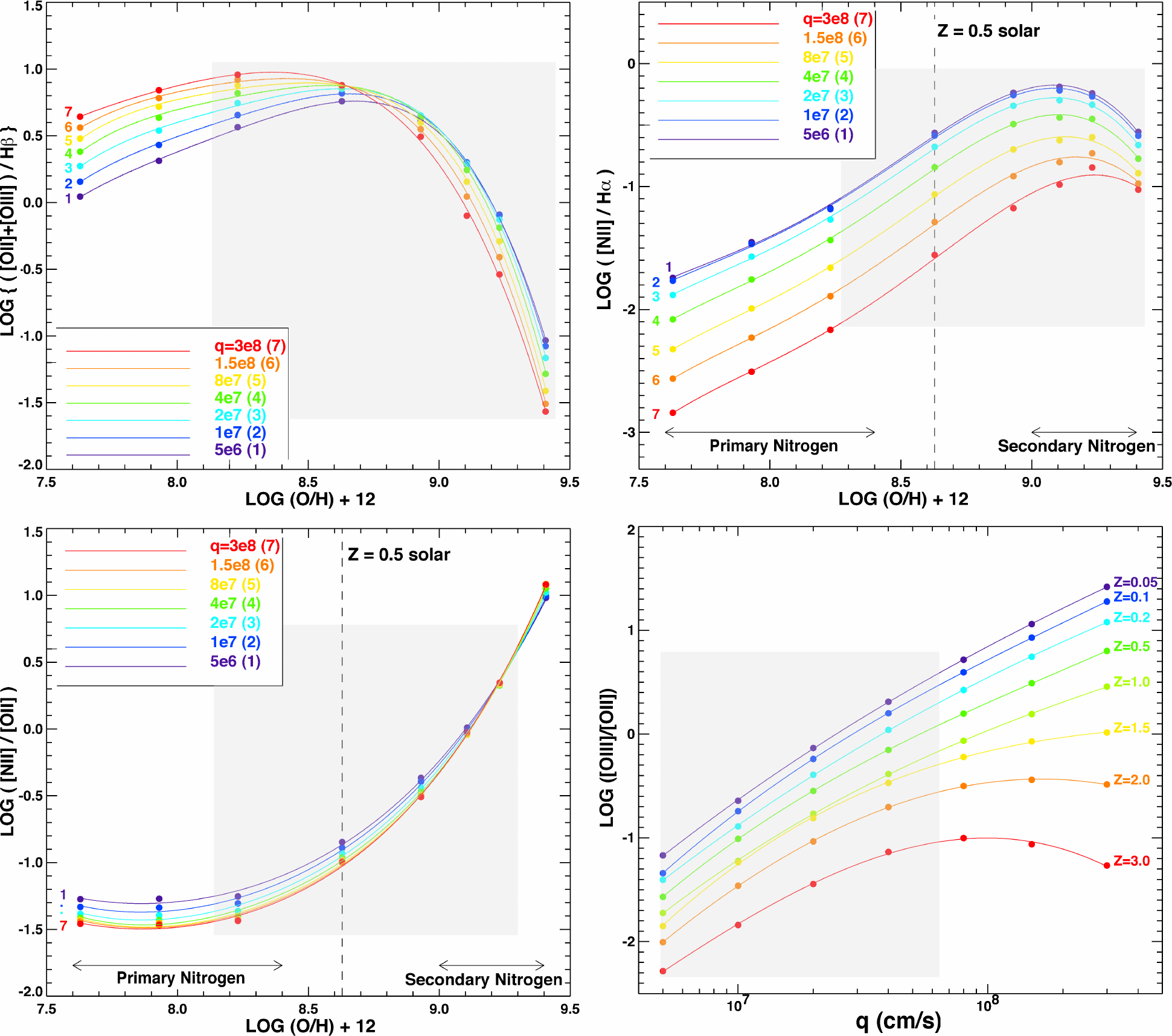} 
\caption{Plots from \protect\cite{kewley2002} showing the theoretical dependence 
of different emission line ratios on the gas phase metallicity \zgas{}  (defined as 
log(O/H)+12)  and ionization parameter $q$ for gas in regions primarily photo-ionized 
by photons from massive stars. 
The shaded rectangles illustrate the range of values for the line ratios, \zgas{}, and $q$ found in our data.
The line ratio  (\iontwo{O}{II}+\iontwo{O}{III})/H$\beta$ defining 
the \zgas{}  indicator $R_{23}$  varies with \zgas{} and $q$.
High values of $R_{23}$ are degenerate with \zgas{}   at a given $q$
({\bf{top left}}).
The \iontwo{N}{II}/\iontwo{O}{II} ratio used to define  the  N202 \zgas{}  indicator is 
not highly sensitive to $q$  ({\bf{bottom left}}).
The \iontwo{N}{II}/H$\alpha$ is  used to define  the  N2 \zgas{}  indicator
({\bf{top right}}).
The  \iontwo{O}{III}/\iontwo{O}{II} line ratio  is highly sensitive to $q$
({\bf{bottom  right}}).
}
\label{fig:degeneracy}
\end{figure*}

\subsubsection{$q$} \label{subsec:q}

The 2D maps of the ionization parameter $q$ ($\S$~\ref{sec:how-q}) 
can be seen in the top right panel in Figure \ref{fig:2d-maps}.  
For most of our galaxies, the maps show a peak in $q$ in the 
central part of the galaxy and regions of localized enhancements 
in the outer disc.
The distribution of $q$  and SFR are qualitatively similar in
many galaxies.

The azimuthally averaged values of $q$ and $\Sigma_{\rm SFR}$ shown 
in Figure \ref{fig:q-sfr-gradients} complement the 2D maps.
For  NGC 0628, 2903, 3938, 4254 \& 5713, 
the azimuthally averaged profiles of $q$ and  $\Sigma_{\rm SFR}$ 
are qualitatively similar, tending to peak in the centre, and
to fall at larger radii. 
 The average
 $\Sigma_{\rm SFR}$  falls by over an order of magnitude
(e.g. from 0.1 to below 0.01 or from 0.01 to below 0.001 
$M_\odot$ yr$^{-1}$ kpc$^{-2}$), the average value of $q$ 
typically varies by no more than a factor of 3 (e.g., 
from 6.95 to 7.40 dex, corresponding to 0.8 to 3 times 
10$^7$  cm s$^{-1}$).

In fact, in these galaxies, $q$ does not fall below $\sim 6.95$ dex 
even at large radii. Similar lower limits in $q$ have been seen by  
\cite{shields1990}, \cite{dopita2000}, \cite{rosa2014}, \cite{perez-montero2014}, and \cite{sanchez2015}.
The  observed lower limit  of $q > 10^7$ cm s$^{-1}$ is not  well understood. 
It might be a limitation of the current photoionization models used in constructing diagnostics for $q$.
Another possible explanation is that
the \iontwo{O}{III} line intensity becomes dimmer and harder to detect
at $q < 10^7$ cm s$^{-1}$ (see evolution of \iontwo{O}{III}/H$\beta$ vs. $q$ in
the figures  found in \citealt{kewley2001} and \citealt{kewley2004}) due to a
combination of a lower number of ionizing photons and a lower
ionization fraction for O$^{++}$, so for any given fibre the \iontwo{O}{III}
emission detected would be weighted towards regions where 
the gas has $q \gtrsim 10^7$ cm s$^{-1}$.
There is evidence that $q$ and \zgas{} are correlated with each other \citep{freitas-lemes2014, perez-montero2014, rosa2014}
since metal poor stars give off more ionizing UV radiation \citep{dors2011, sanchez2015}, but we do not notice any clear trends between $q$ and \zgas{} when comparing them 
across deprojected radii (Figures \ref{fig:q-sfr-gradients} \& \ref{fig:zgas-gradients})
or the 2D maps (Figure \ref{fig:2d-maps}).

We also note that the azimuthally averaged profile of
$q$  in the weakly interacting late-type galaxy
NGC~0337 is markedly different from that of the 
other  six galaxies. 
In NGC~0337, $\Sigma_{\rm SFR}$  rises by over an order of magnitude
from  a  radius of 8 kpc to the centre of the galaxy, but the $q$
profile remains flat  at a value of $\sim 7.1$ dex.
The  value of $q$  in the central 2 kpc radius of  
NGC~0337  does not rise  to the  typical values of 7.4 dex shown
by other galaxies  although  $\Sigma_{\rm SFR}$ in this  region is as high as in 
NGC~2903, NGC~4254, NGC~5194, and  
almost a factor of 10 above that of NGC~3938.
This effect is not due to distance or resolution effects as NGC~0337
is at a similar  distance as NGC~4254 and NGC~3938.
Since stellar metallicity is thought to correlate with $q$ \citep{dors2011, sanchez2015},
the flat radial gradient in $q$ in NGC~0337 might possibly be tied to the galaxy's lower absolute
metallicity when compared to the other galaxies in our subsample.

\subsubsection{\zgas{}}

Our 2D maps (Figure \ref{fig:2d-maps})  of six \zgas{}
diagnostics ($R_{23}$-KK04, $R_{23}$-M91, N202-KD02, O3N2-PP04,
N2-D02, \& N2-PP04) show a range in  \zgas{}  of $\sim$
8.5 to 9.3 dex across our spirals.
For the \zgas{} diagnostics using   the indicators $R_{23}$ 
based on 
(\iontwo{O}{II}+\iontwo{O}{III})/H$\beta$,
N2O2 based on 
\iontwo{N}{II}/\iontwo{O}{II},   and  O3N2 based on (\iontwo{O}{III}/H$\beta$)/(\iontwo{N}{II}/H$\alpha$),
\zgas{} peaks roughly in the same regions as SF, mainly in the 
central  bulge and along the spiral arms.  However,  
the two  \zgas{} diagnostics (N2-D2 \& N2-PP04)   
based on \iontwo{N}{II}/H$\alpha$  often show spatial trends that are
{\it{ opposite to those shown by other  indicators in regions of high SFR}}.  
For example, in NGC 2903, 
the N2-D2 \& N2-PP04 \zgas{} diagnostics appear 
to show lower \zgas{} along  the leading edges of the bar and
along the spiral arms than on the trailing edge, while the opposite
behaviour is seen in the other \zgas{} diagnostics (Figure~\ref{fig:2d-maps}).
Similarly flipped behaviour is clearly seen in almost all 
our galaxies, especially in the spiral arms where there is a 
high SFR. 
The inconsistency between the N2-D2 \& N2-PP04  diagnostics 
and the other \zgas{} diagnostics may be due to several reasons, 
which we  discuss in  $\S$~\ref{sec:results-zgas}, point (iii).

Figure~\ref{fig:zgas-gradients} complements the 2D maps and shows 
deprojected radial \zgas{} gradients 
based on all seven \zgas{} diagnostics for  NGC 0337, 0628, 2903, 3938, 4254, 5194, \& 5713.
In terms of the {\it shape} of the radial \zgas{} profile, 
the seven  \zgas{} diagnostics show fairly good agreement 
beyond the inner 1-2 kpc,  and yield flat to  slightly negative 
gradients (Figure \ref{fig:zgas-gradients}).   
The gradients are shallow
and   in dex kpc$^{-1}$, they  range from
(0.001 to -0.055) in $R_{23}$-KK04,   (-0.007 to -0.041)  in N2O2-KD02
and (0.003 to -0.035)  in N2-PP04,  while in  dex ($R$/\rtwofive{})$^{-1}$ they
range from  (0.01 to -0.77) in $R_{23}$-KK04, (-0.09  to -0.57)  
in N202-KD02, and  (0.03 to -0.57) in N2-PP04. 
For comparison, the \zgas{} gradient for the Milky Way was determined by \cite{shaver1983} using the 
$R_{23}$ ratio and finds the gradient to be -0.07 $\pm$ 0.015 dex kpc$^{-1}$.
Table \ref{tab:zgas-gradients} summarizes the results for our \zgas{} gradients.
It is remarkable that all the galaxies in our sub-sample including unbarred, barred, and interacting
galaxies exhibit such flat \zgas{} gradients across all their galactic components.
We return to this point in $\S$~\ref{sec:results-bars} to \ref{sec:results-redshift}.

\begin{figure*}
\begin{flushleft}
\vspace{-22pt}
\includegraphics[height=0.24\textheight, clip=true, trim=0.1cm 0.00cm 0.8cm 0.8 cm]{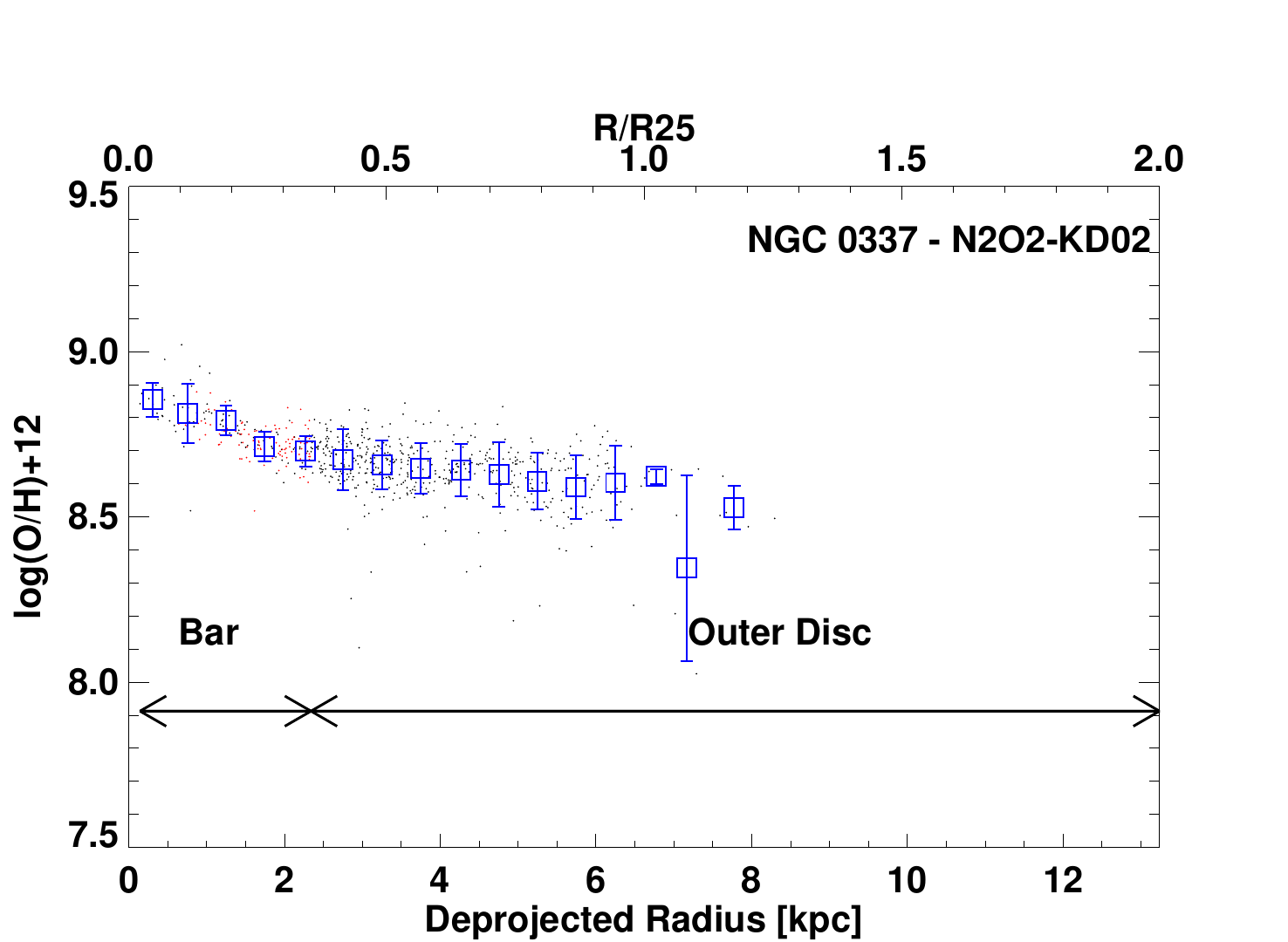}
\includegraphics[height=0.24\textheight, clip=true, trim=0.1cm 0.00cm 0.8cm 0.8 cm]{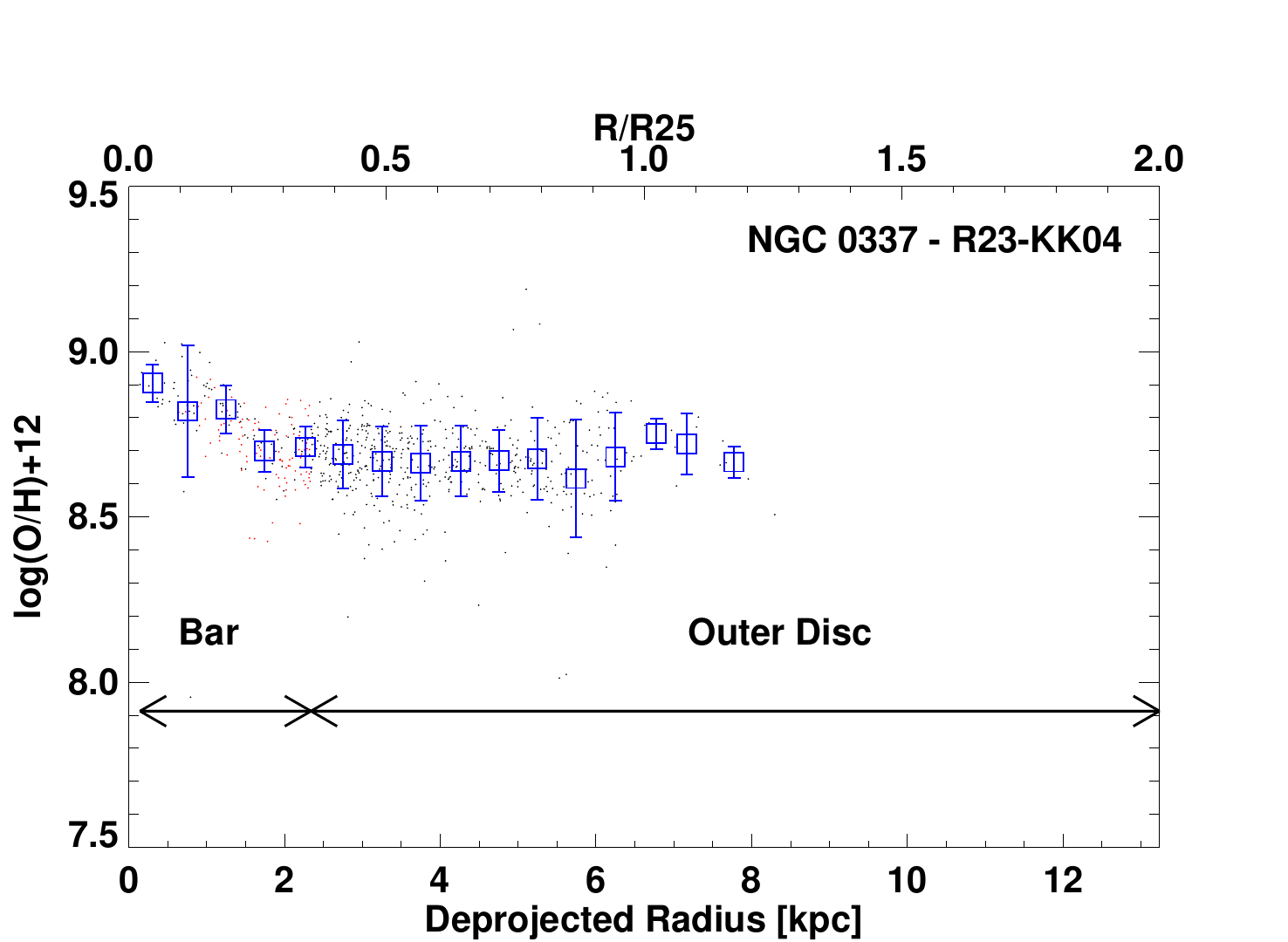}
\vspace{-18pt}\\
\includegraphics[height=0.24\textheight, clip=true, trim=0.1cm 0.00cm 0.8cm 0.8 cm]{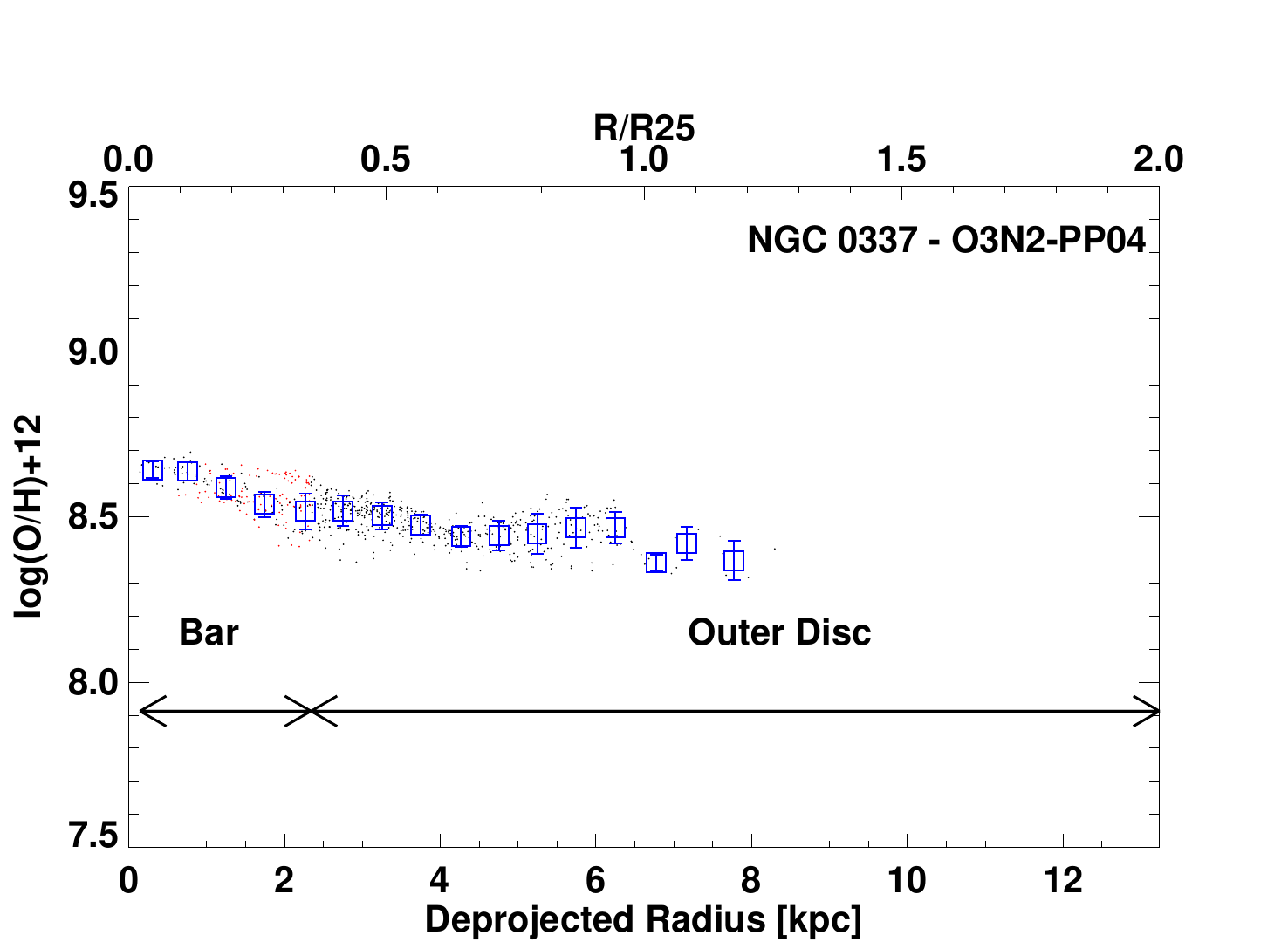}
\includegraphics[height=0.24\textheight, clip=true, trim=0.1cm 0.00cm 0.8cm 0.8 cm]{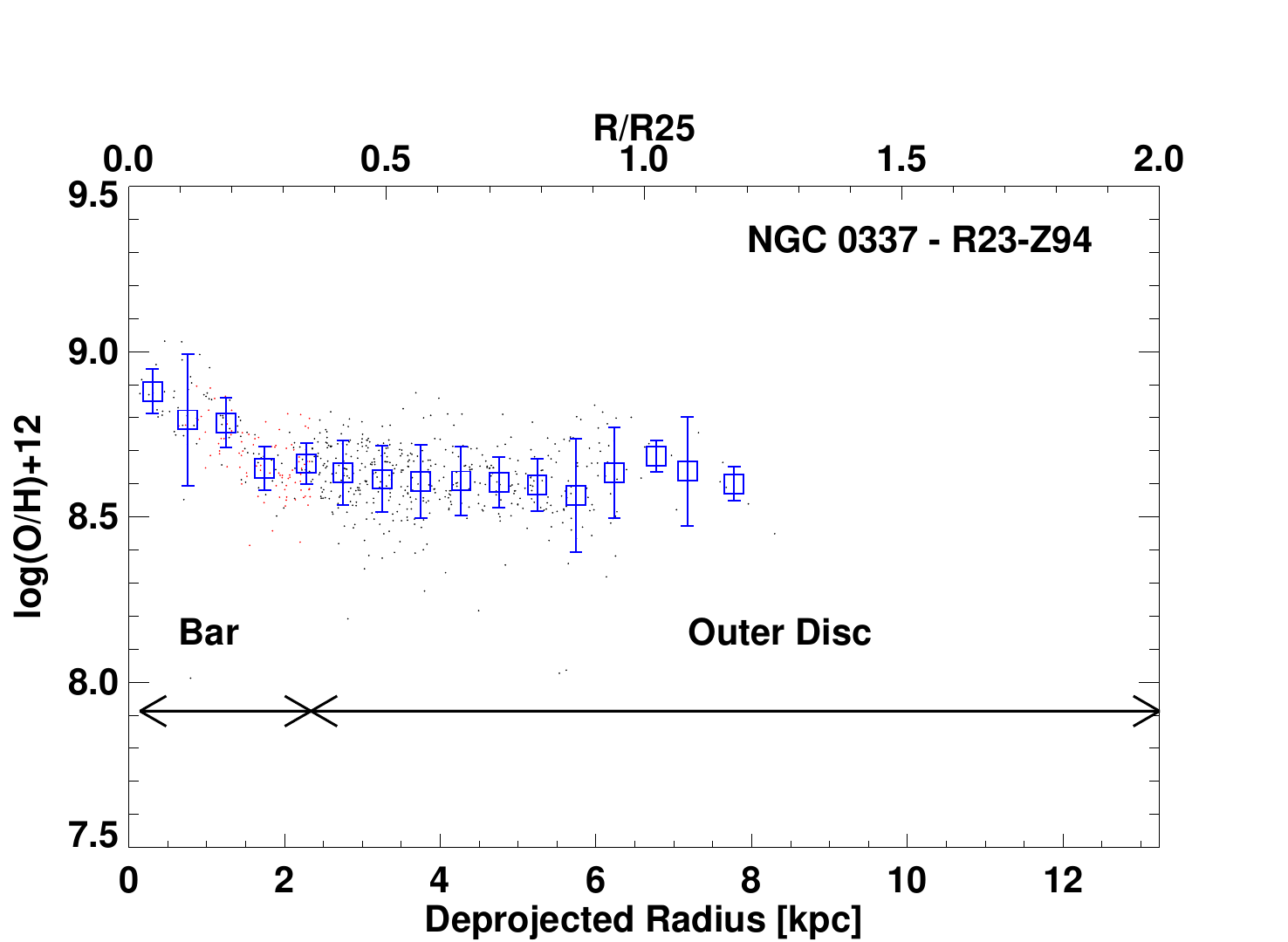}
\vspace{-18pt}\\
\includegraphics[height=0.24\textheight, clip=true, trim=0.1cm 0.00cm 0.8cm 0.8 cm]{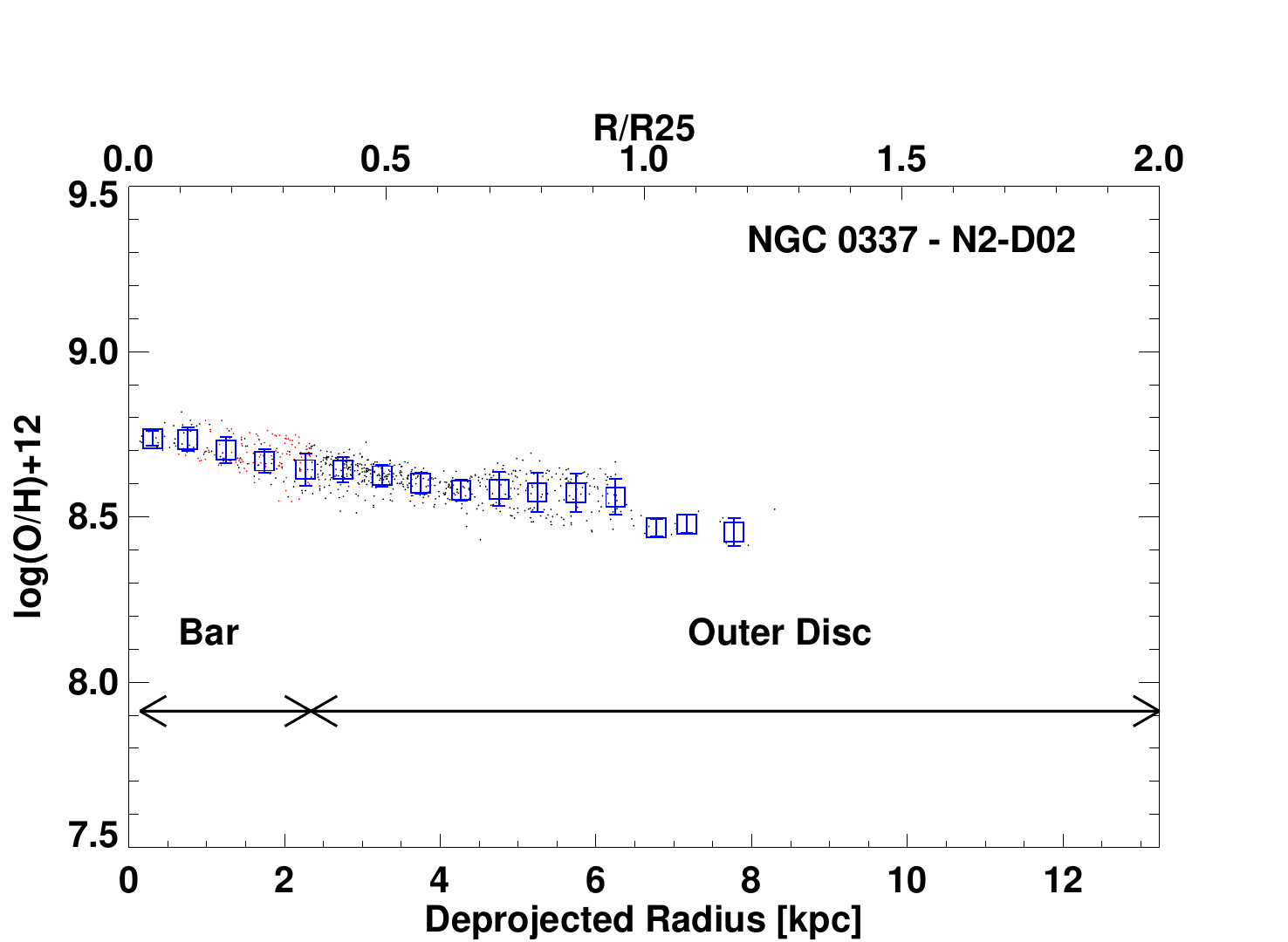}
\includegraphics[height=0.24\textheight, clip=true, trim=0.1cm 0.00cm 0.8cm 0.8 cm]{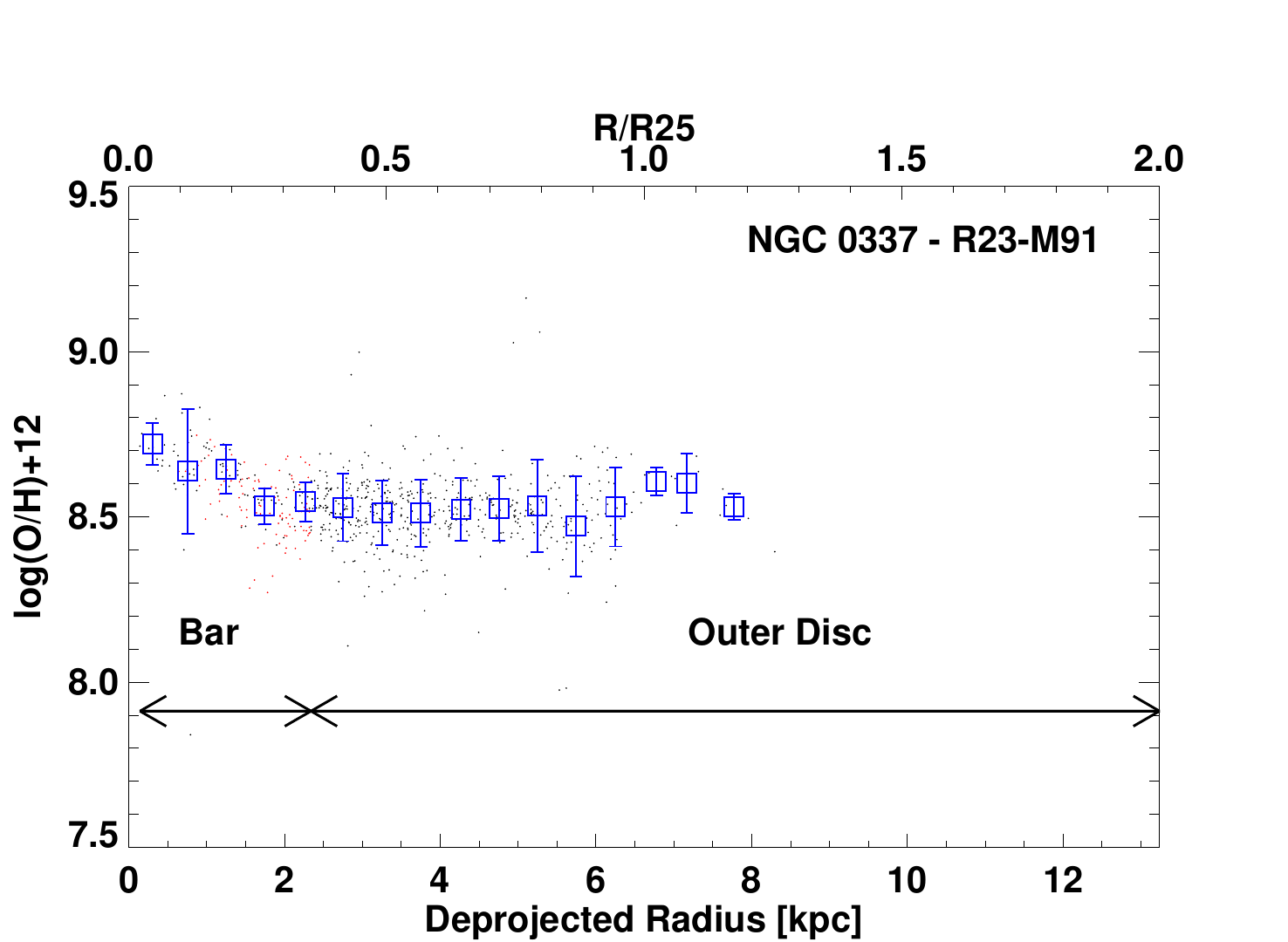}
\vspace{-18pt}\\
\includegraphics[height=0.24\textheight, clip=true, trim=0.1cm 0.00cm 0.8cm 0.8 cm]{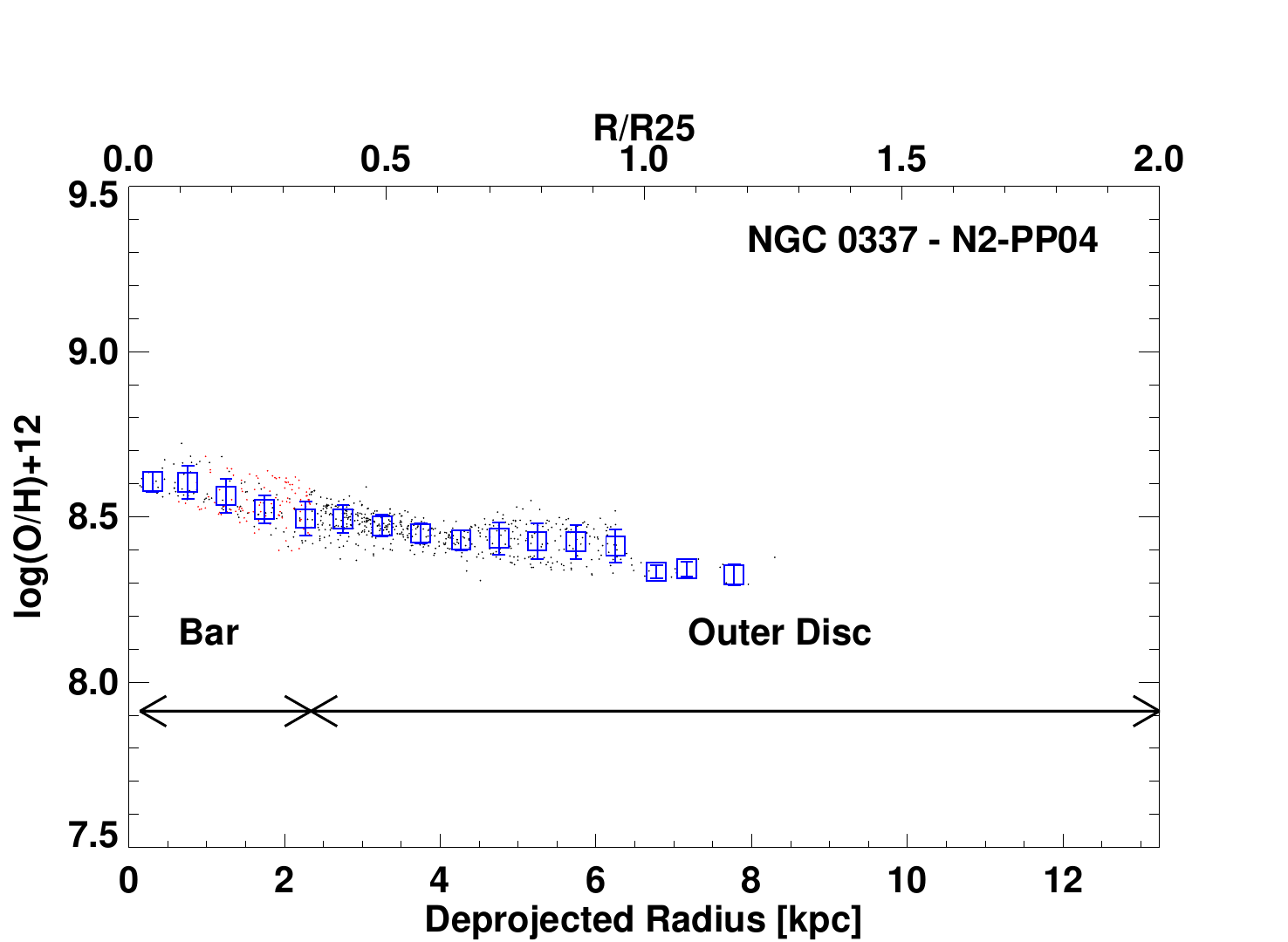}
\end{flushleft}
\caption{
For  NGC 0337, we show the deprojected radial \zgas{} profiles for the seven 
different \zgas{} diagnostics: N202-KD02, $R_{23}$-KK04, O3N2-PP04 $R_{23}$-Z94, 
N2-D02, $R_{23}$-M91, \& N2-PP04. 
The arrows on the bottom show our decomposition of each galaxy into non-classical (NC)  
bulge (which refers to discy/pseudo bulge), bar, and outer disc.  
The blue squares and error bars are the mean and 1$\sigma$ dispersion of 0.5 kpc bins.
In the case of barred galaxies, the  region labelled `Bar' on the plot refers to 
the radii between the bulge and the end of the bar. In this region, 
we show regions in the bar feature as  black points, and  regions azimuthally offset
from the bar feature as red points. The blue squares denoting the mean only use regions
within the bar.
The white masked regions in the maps represent spaxels,
which  were not used in the computation of \zgas{} and $q$
because they are associated with  data of low quality  or dominated
by  DIG or gas with LINER or Seyfert excitation conditions.
}
\label{fig:zgas-gradients}
\end{figure*}

\addtocounter{figure}{-1}
\clearpage

\begin{figure*}
\begin{flushleft}
\vspace{-22pt}
\includegraphics[height=0.24\textheight, clip=true, trim=0.1cm 0.00cm 0.8cm 0.8 cm]{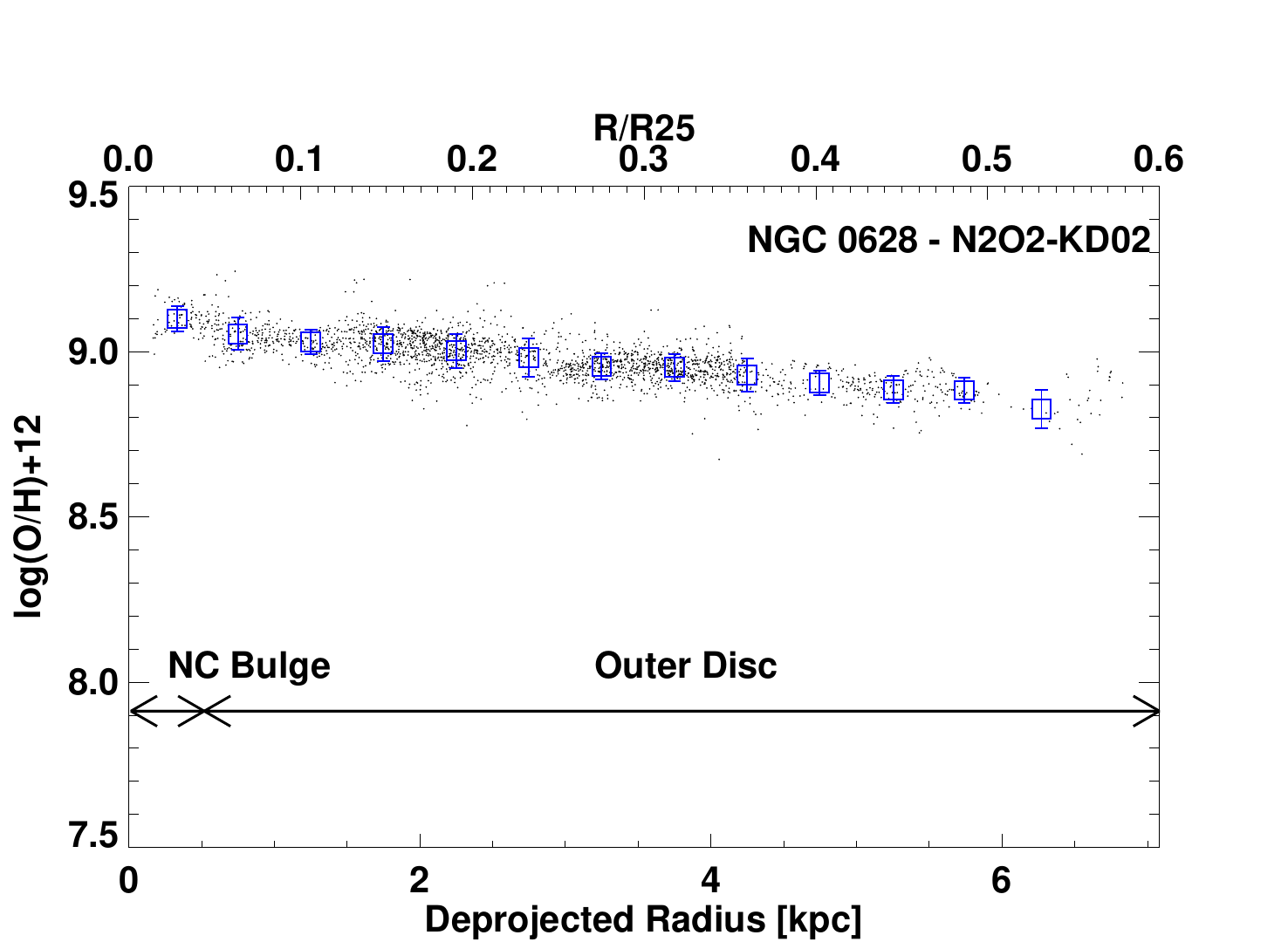}
\includegraphics[height=0.24\textheight, clip=true, trim=0.1cm 0.00cm 0.8cm 0.8 cm]{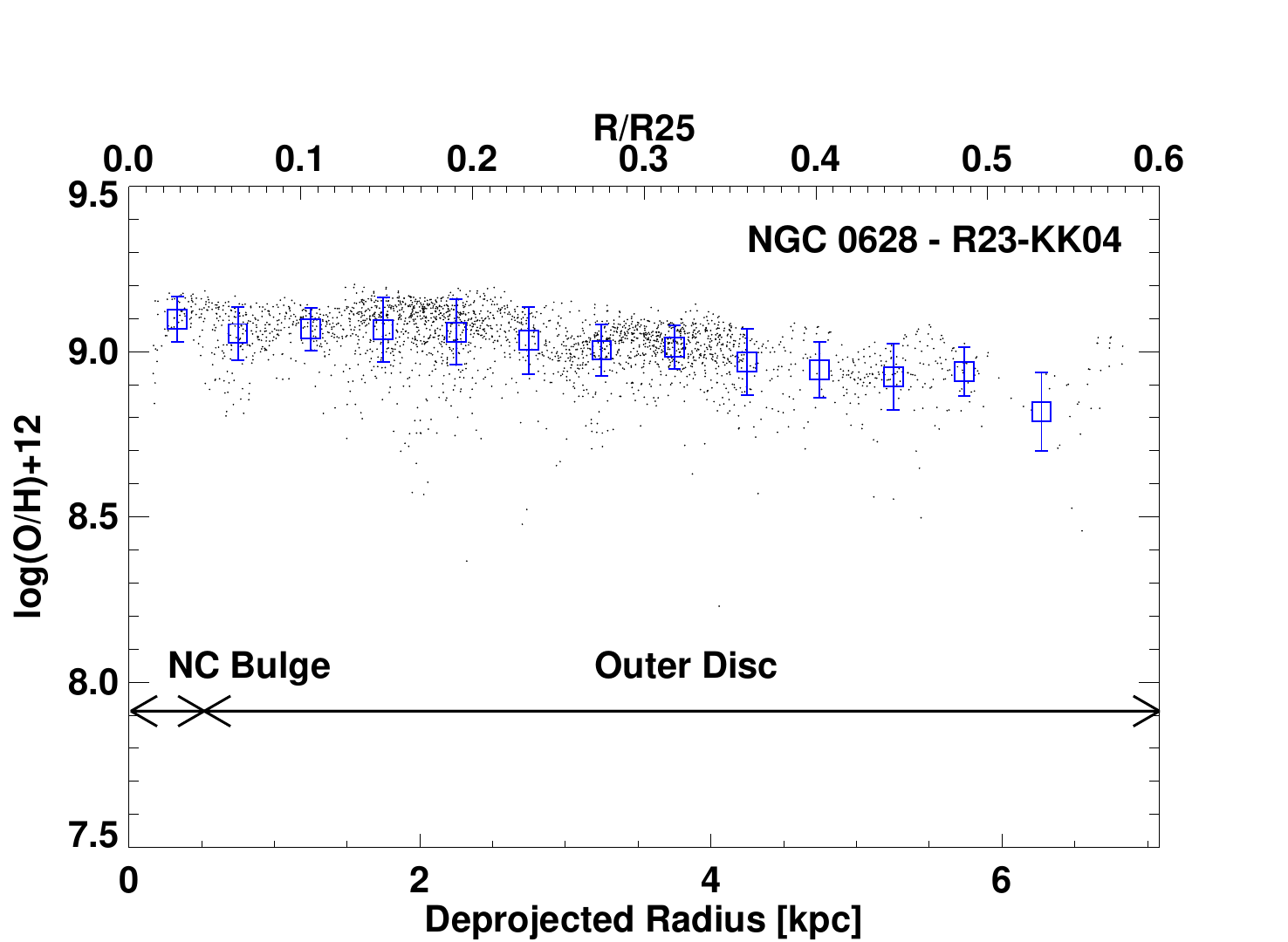}
\vspace{-18pt}\\
\includegraphics[height=0.24\textheight, clip=true, trim=0.1cm 0.00cm 0.8cm 0.8 cm]{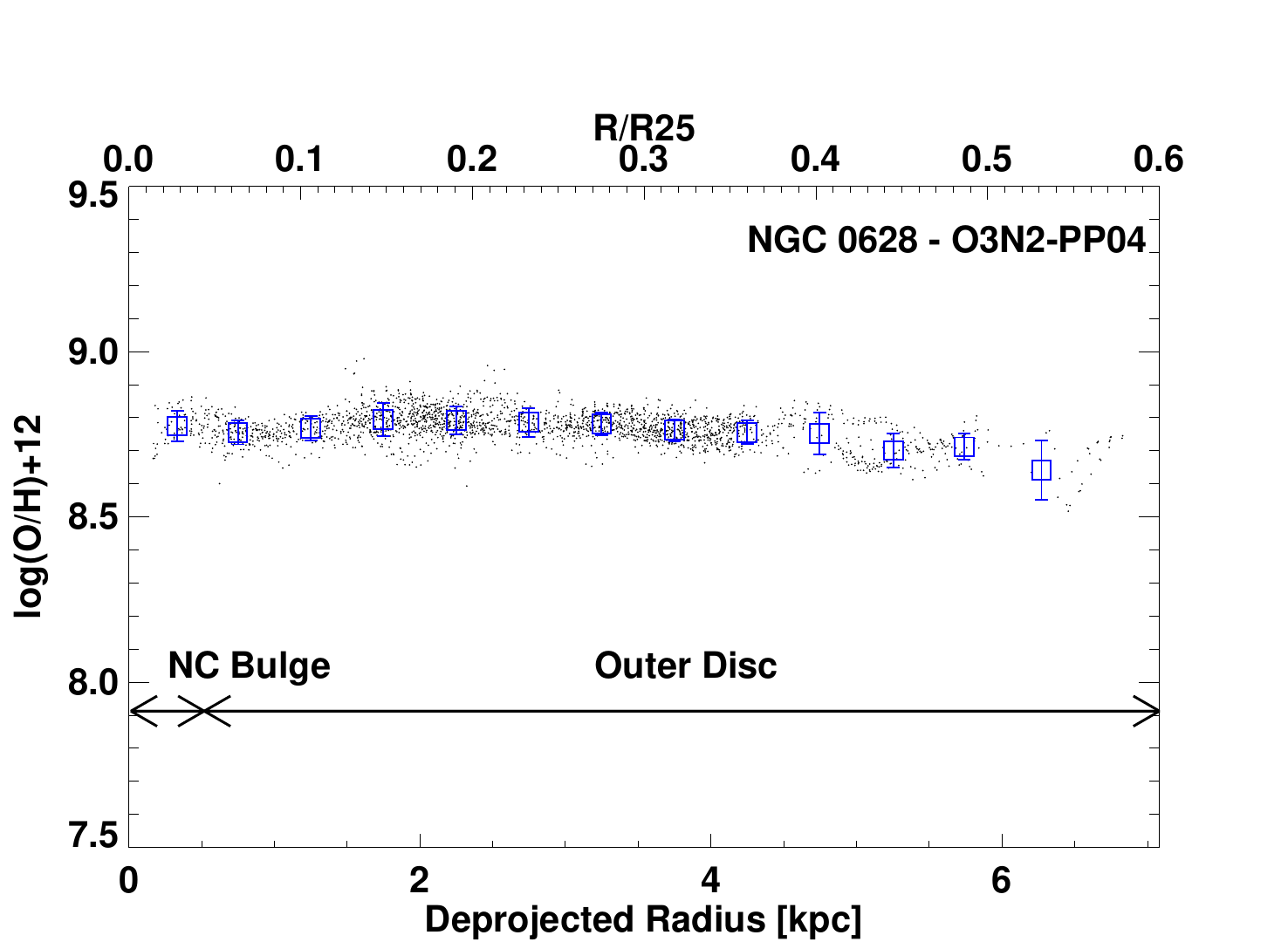}
\includegraphics[height=0.24\textheight, clip=true, trim=0.1cm 0.00cm 0.8cm 0.8 cm]{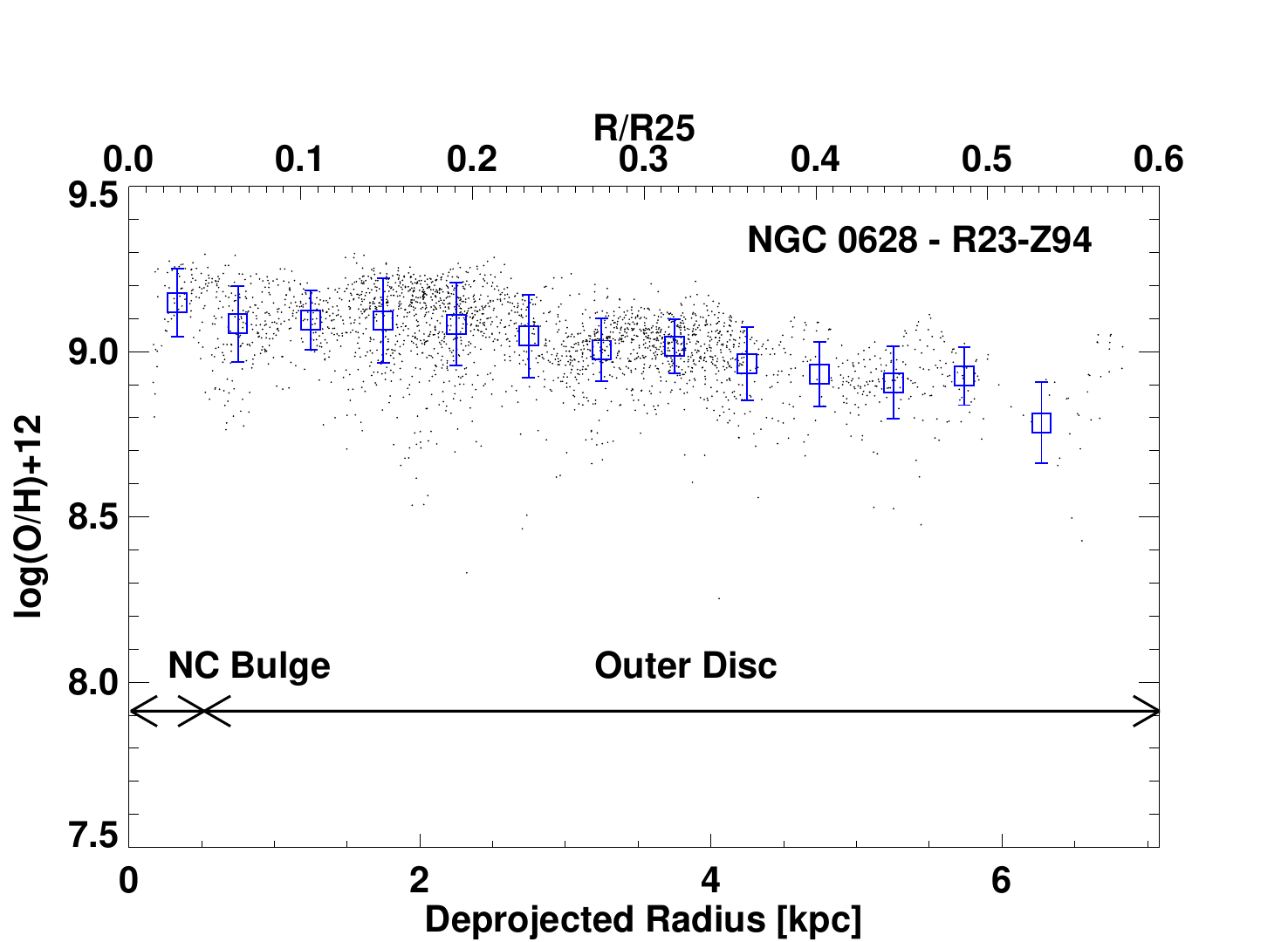}
\vspace{-18pt}\\
\includegraphics[height=0.24\textheight, clip=true, trim=0.1cm 0.00cm 0.8cm 0.8 cm]{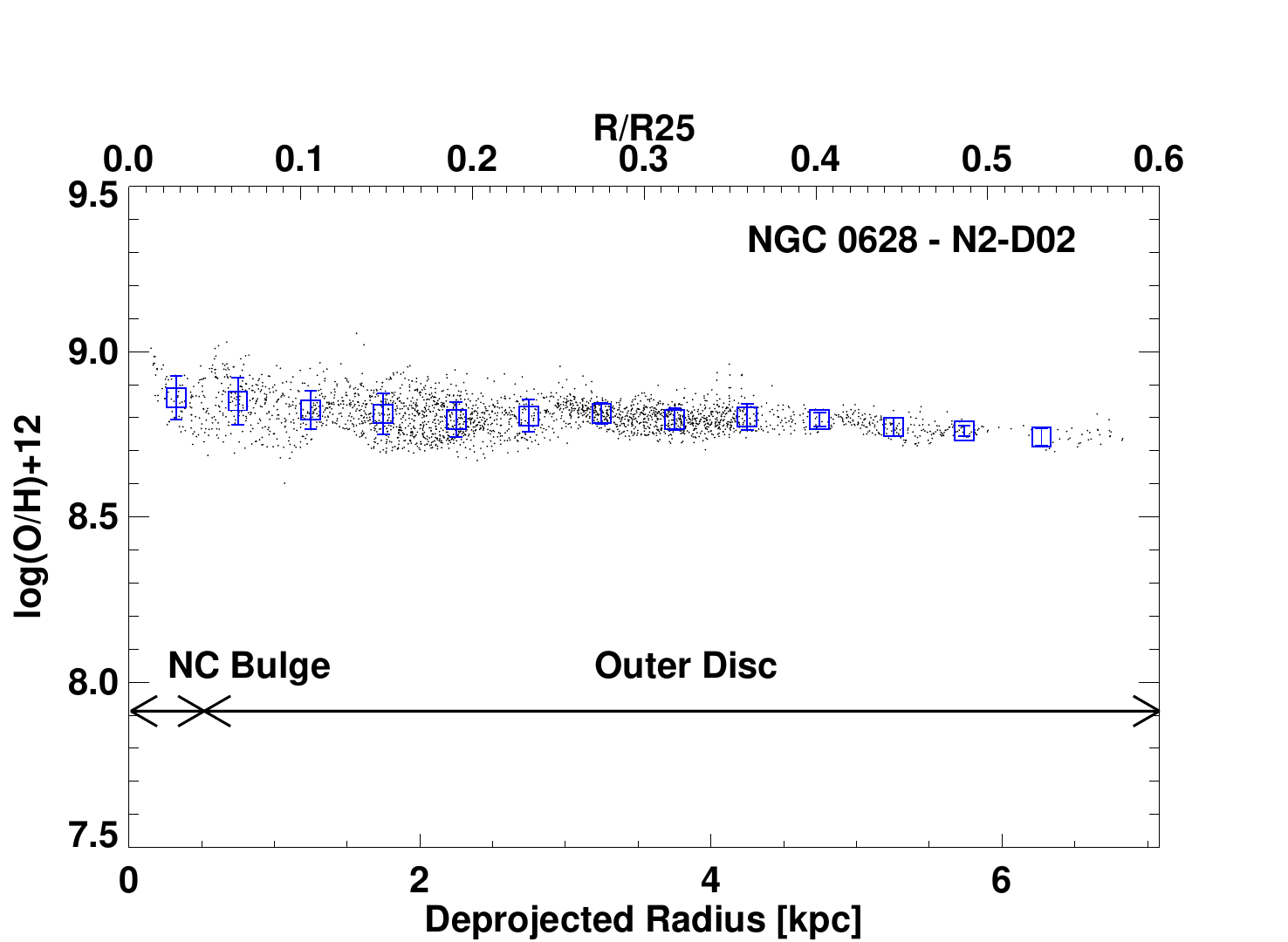}
\includegraphics[height=0.24\textheight, clip=true, trim=0.1cm 0.00cm 0.8cm 0.8 cm]{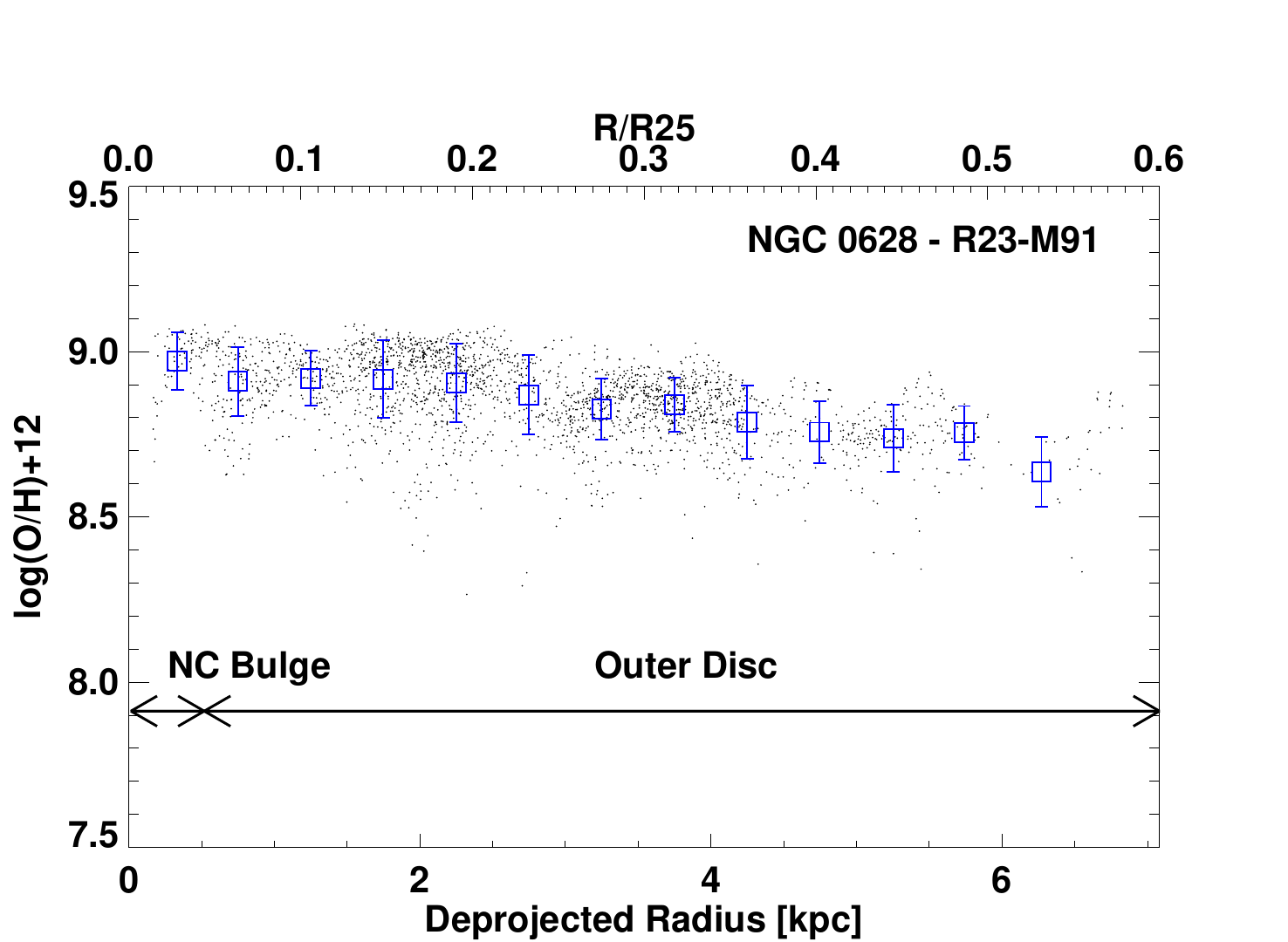}
\vspace{-18pt}\\
\includegraphics[height=0.24\textheight, clip=true, trim=0.1cm 0.00cm 0.8cm 0.8 cm]{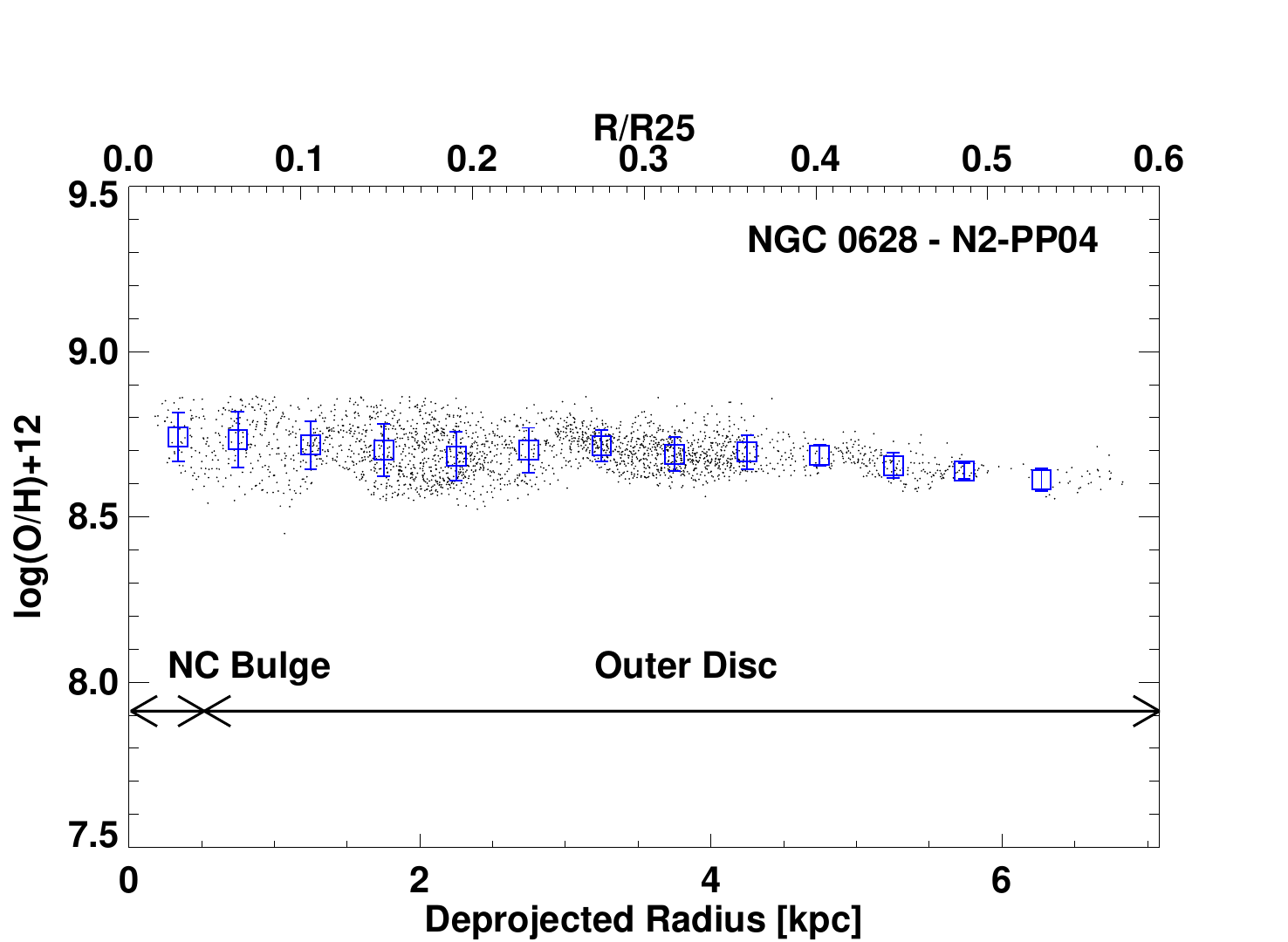}
\end{flushleft}
\caption{Continued: For NGC 0628.}
\label{}
\end{figure*}

\addtocounter{figure}{-1}
\clearpage

\begin{figure*}
\begin{flushleft}
\vspace{-22pt}
\includegraphics[height=0.24\textheight, clip=true, trim=0.1cm 0.00cm 0.8cm 0.8 cm]{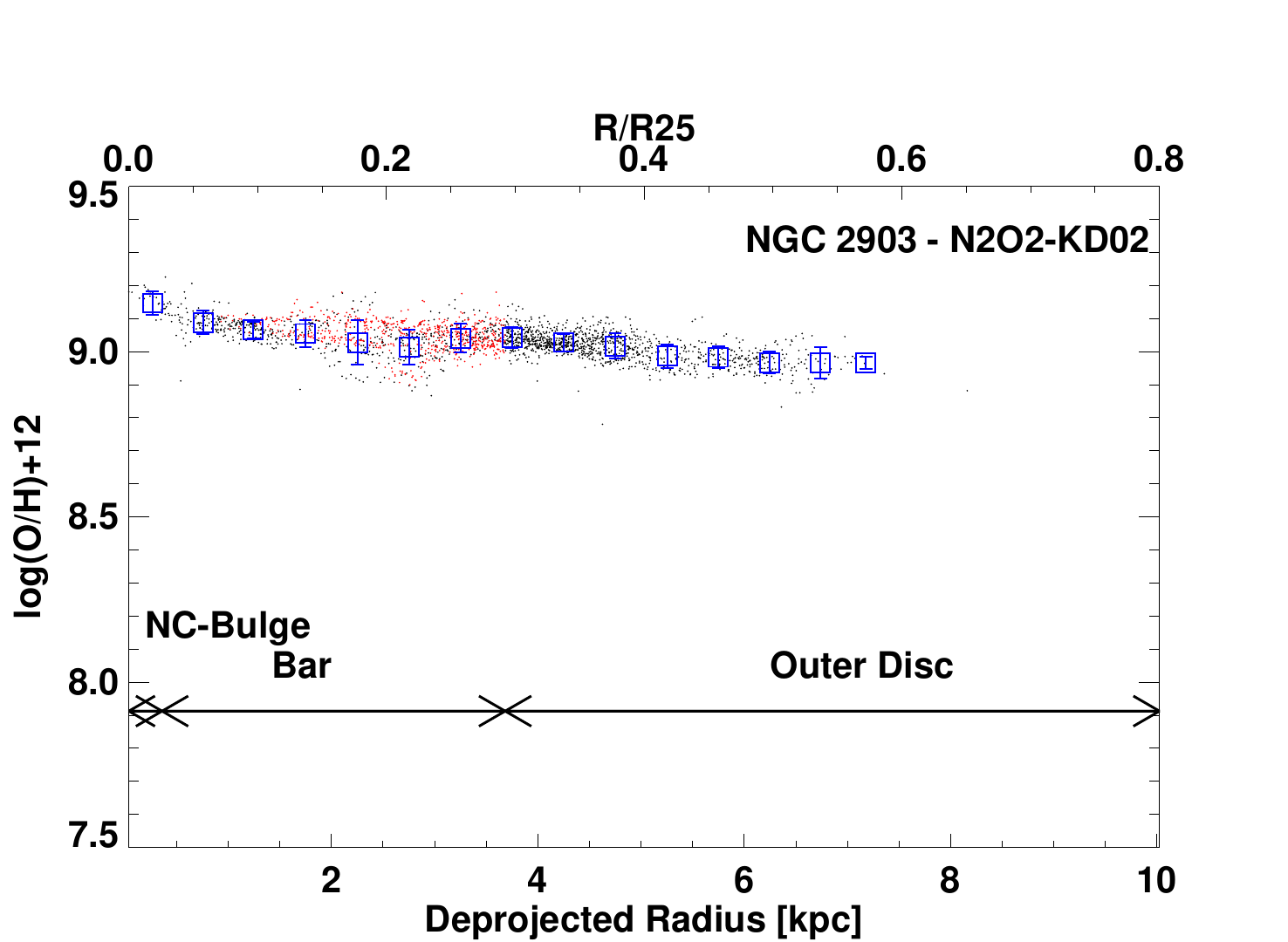}
\includegraphics[height=0.24\textheight, clip=true, trim=0.1cm 0.00cm 0.8cm 0.8 cm]{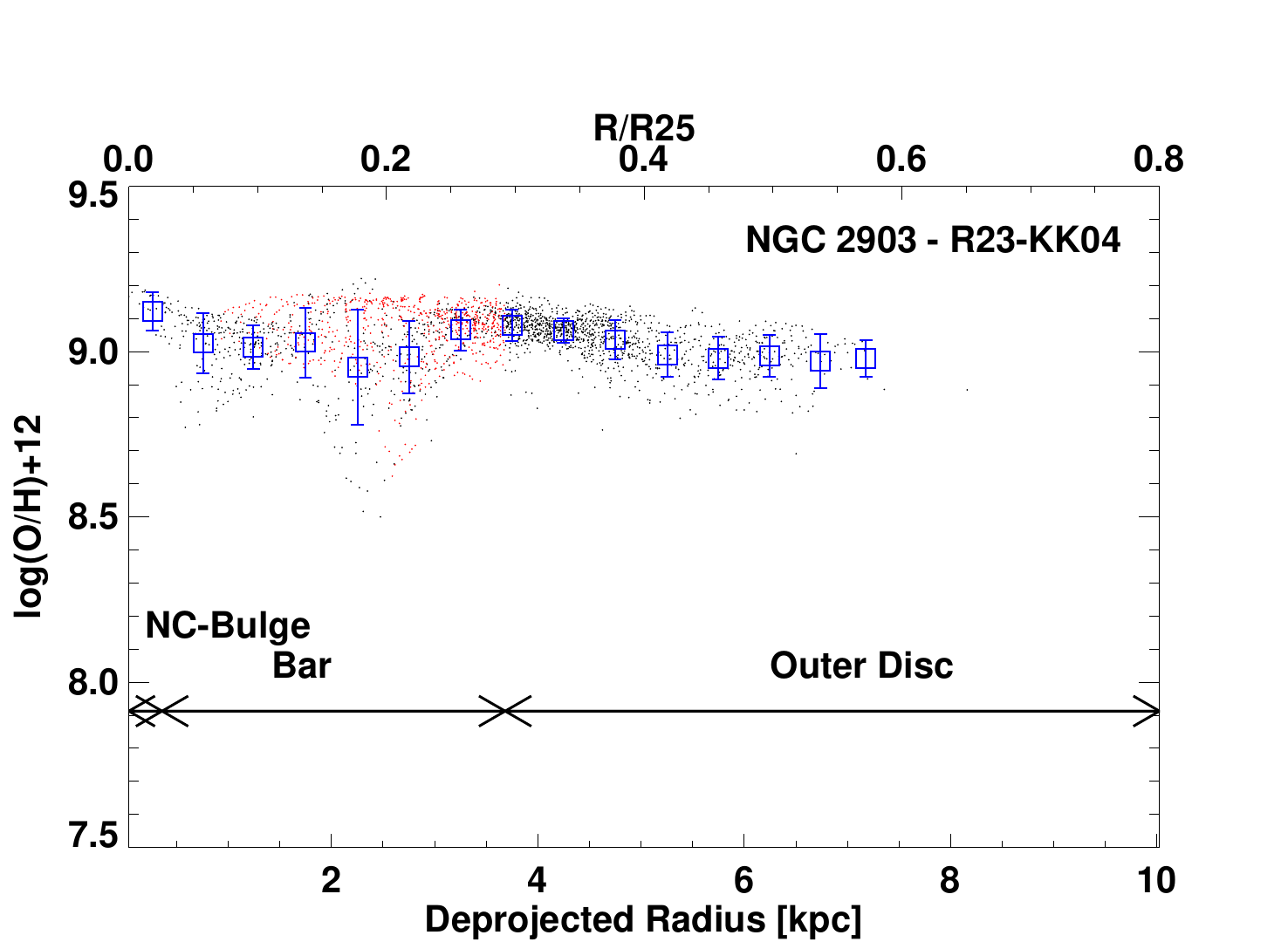}
\vspace{-18pt}\\
\includegraphics[height=0.24\textheight, clip=true, trim=0.1cm 0.00cm 0.8cm 0.8 cm]{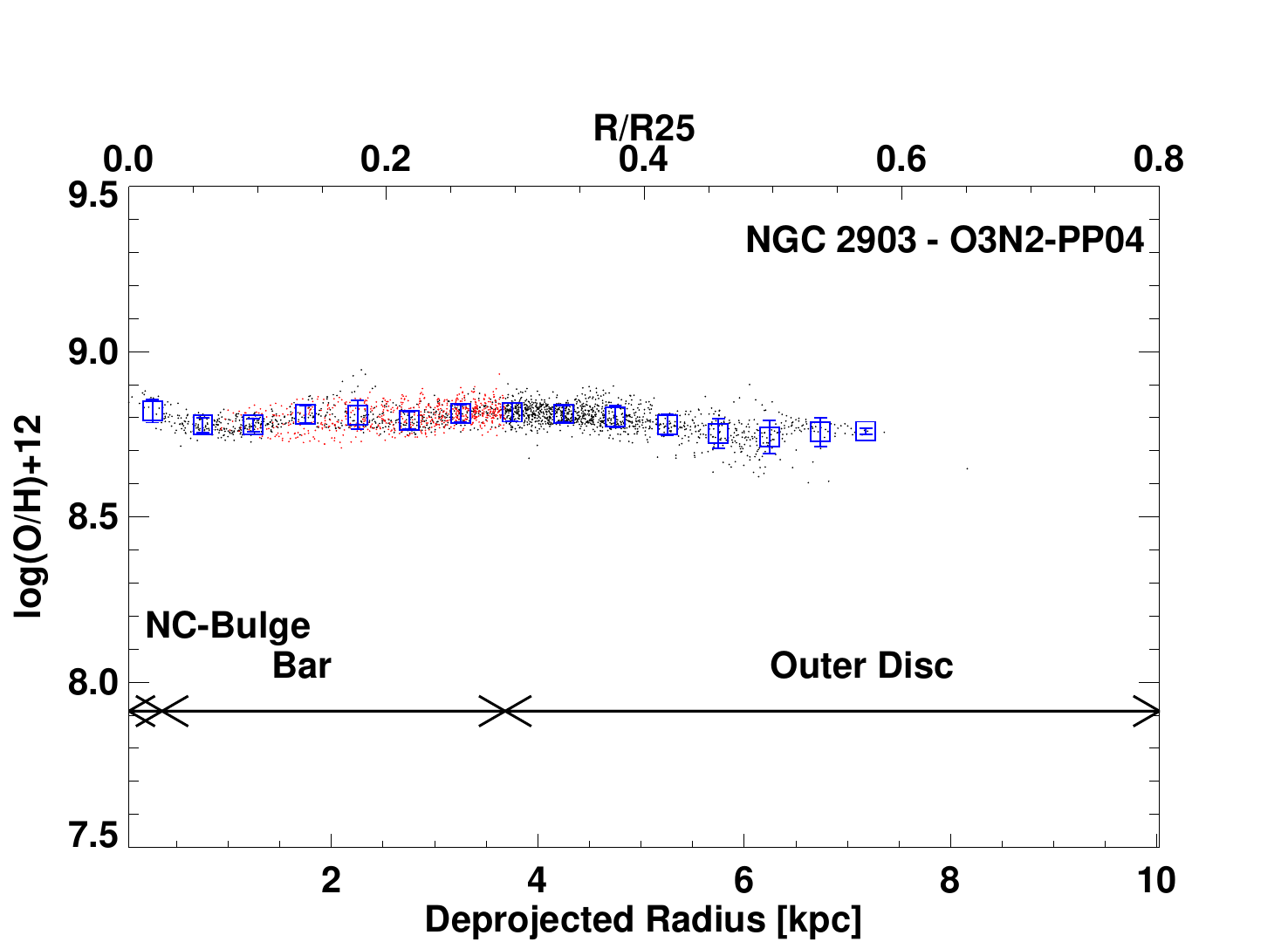}
\includegraphics[height=0.24\textheight, clip=true, trim=0.1cm 0.00cm 0.8cm 0.8 cm]{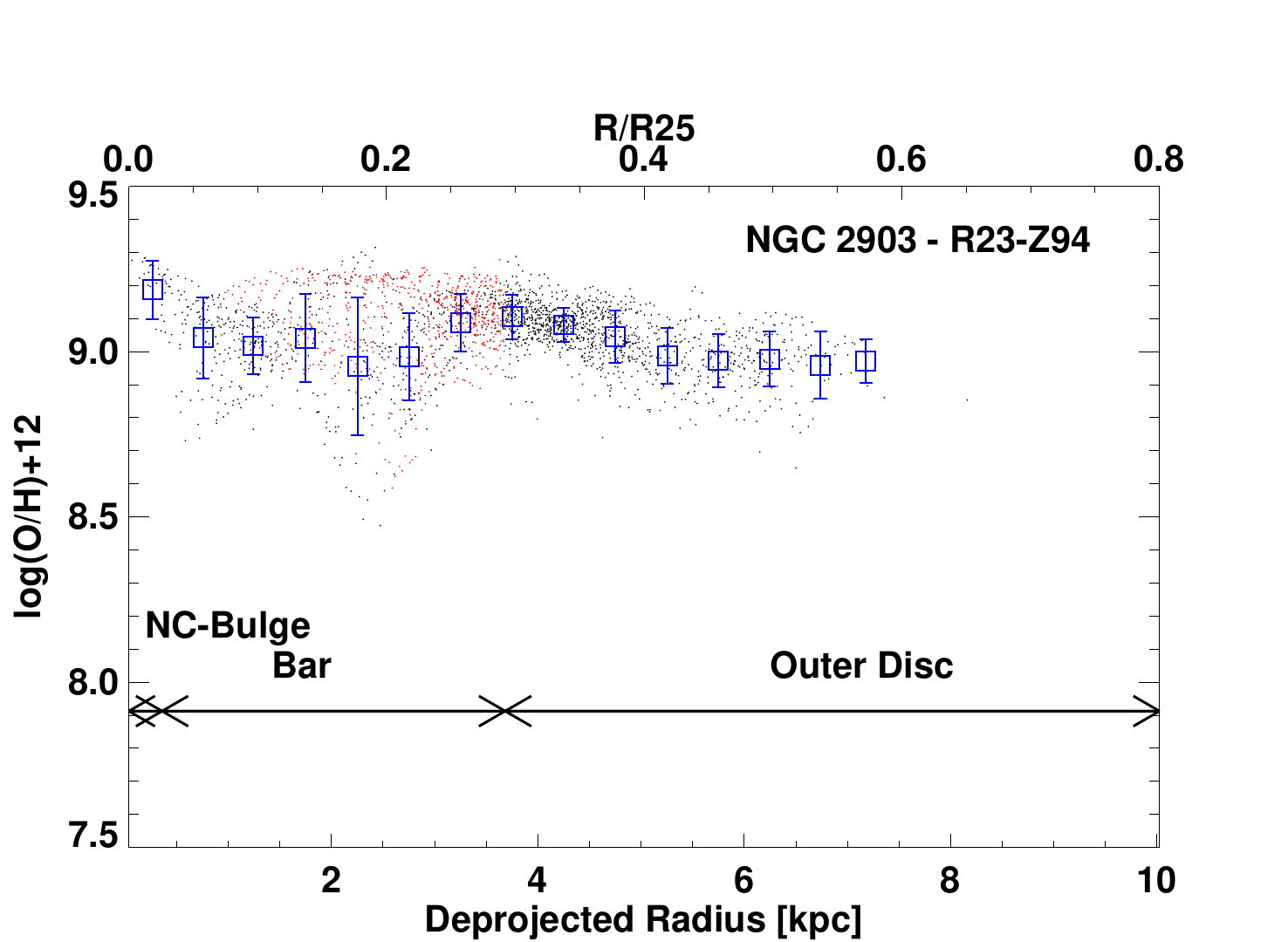}
\vspace{-18pt}\\
\includegraphics[height=0.24\textheight, clip=true, trim=0.1cm 0.00cm 0.8cm 0.8 cm]{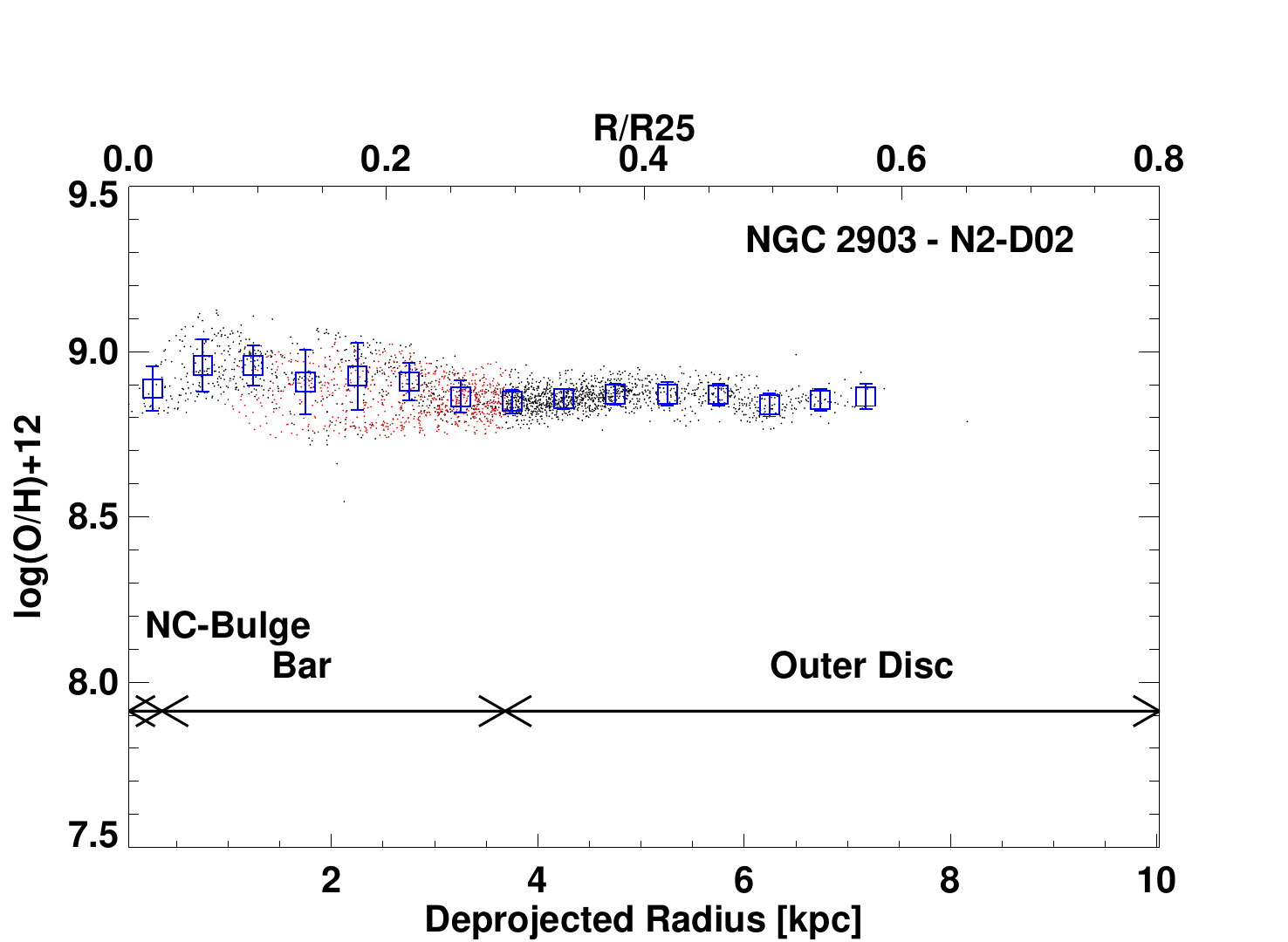}
\includegraphics[height=0.24\textheight, clip=true, trim=0.1cm 0.00cm 0.8cm 0.8 cm]{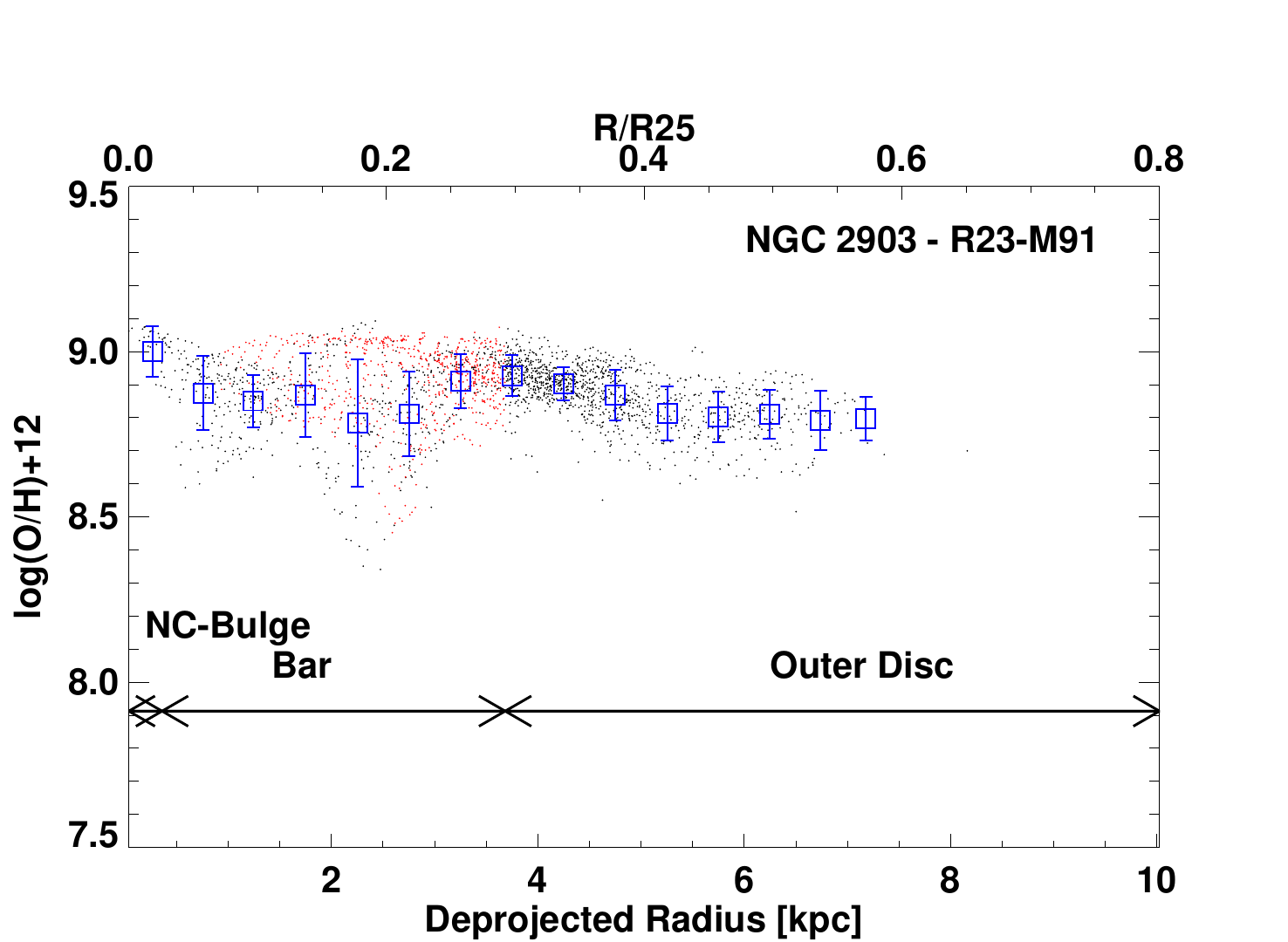}
\vspace{-18pt}\\
\includegraphics[height=0.24\textheight, clip=true, trim=0.1cm 0.00cm 0.8cm 0.8 cm]{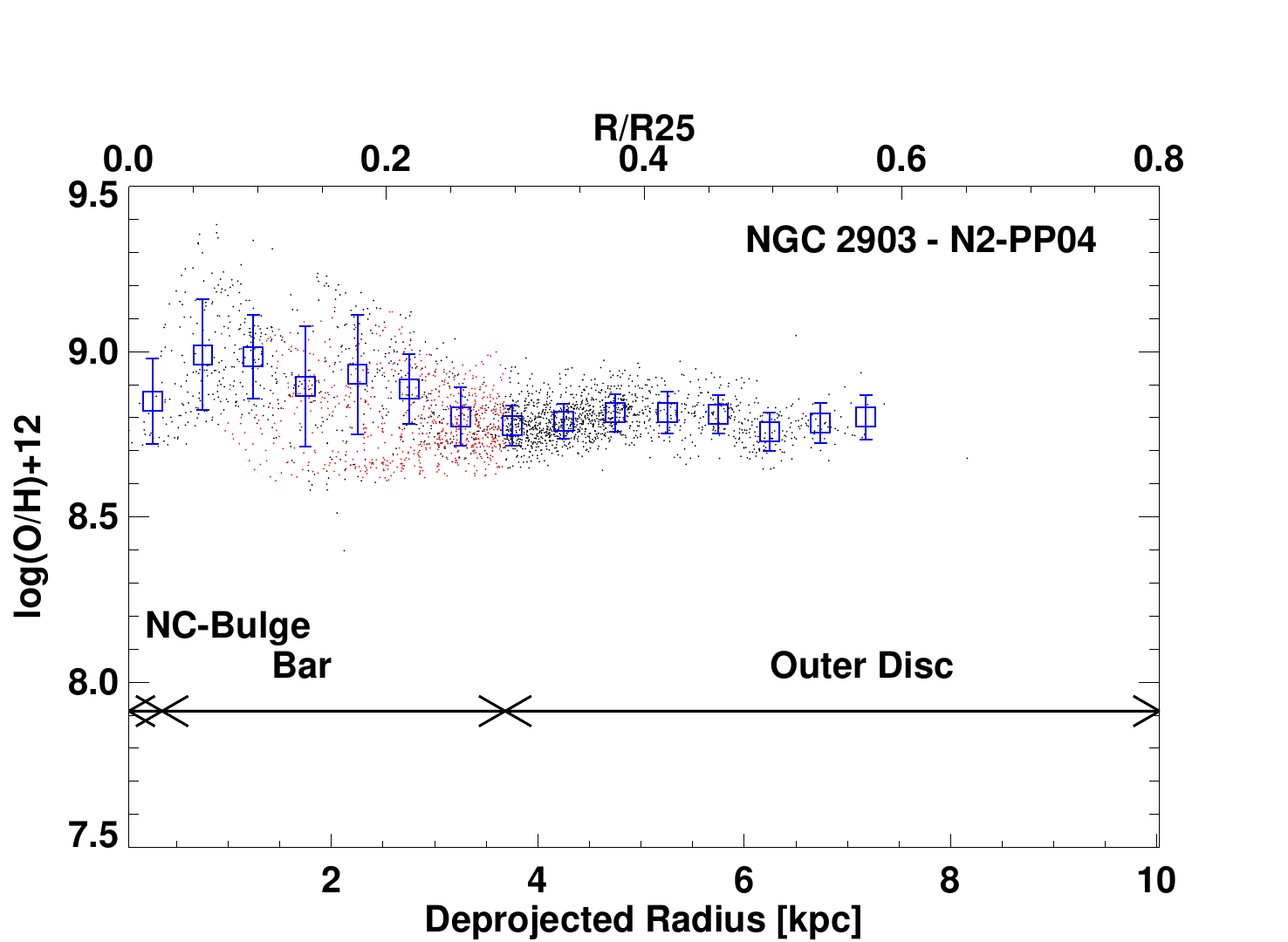}
\end{flushleft}
\caption{Continued: For NGC 2903.}
\label{}
\end{figure*}

\addtocounter{figure}{-1}
\clearpage

\begin{figure*}
\begin{flushleft}
\vspace{-22pt}
\includegraphics[height=0.24\textheight, clip=true, trim=0.1cm 0.00cm 0.8cm 0.8 cm]{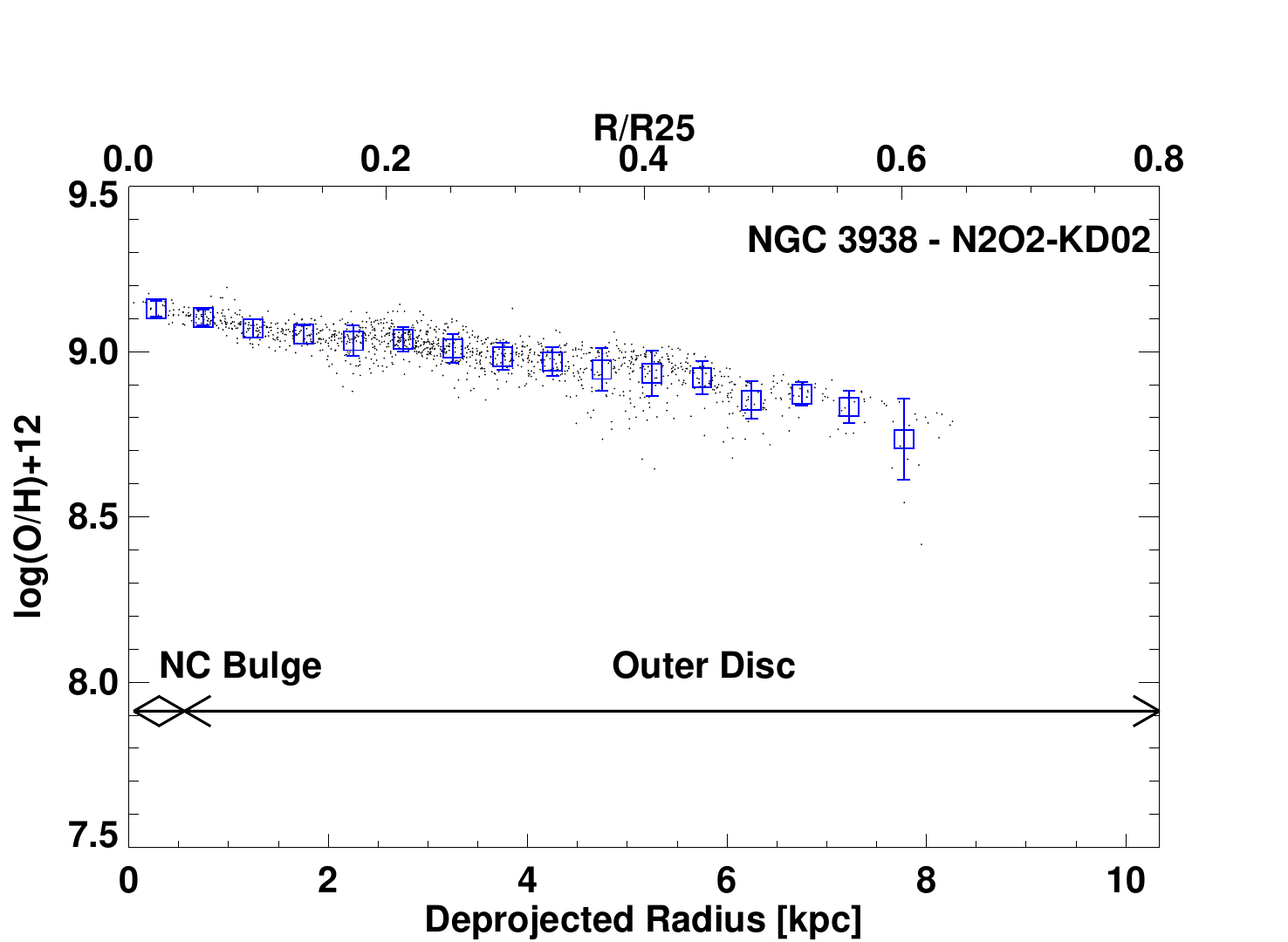}
\includegraphics[height=0.24\textheight, clip=true, trim=0.1cm 0.00cm 0.8cm 0.8 cm]{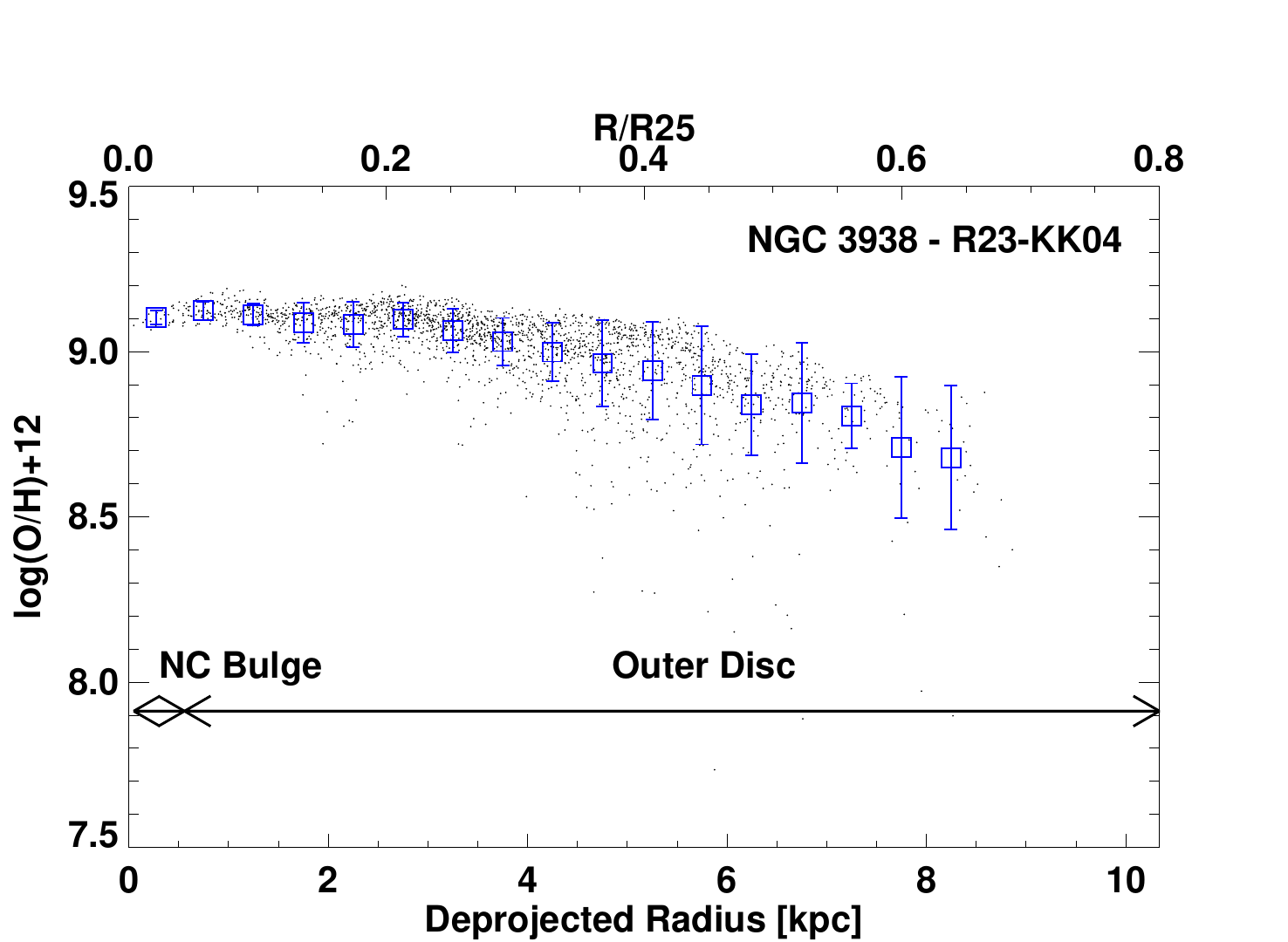}
\vspace{-18pt}\\
\includegraphics[height=0.24\textheight, clip=true, trim=0.1cm 0.00cm 0.8cm 0.8 cm]{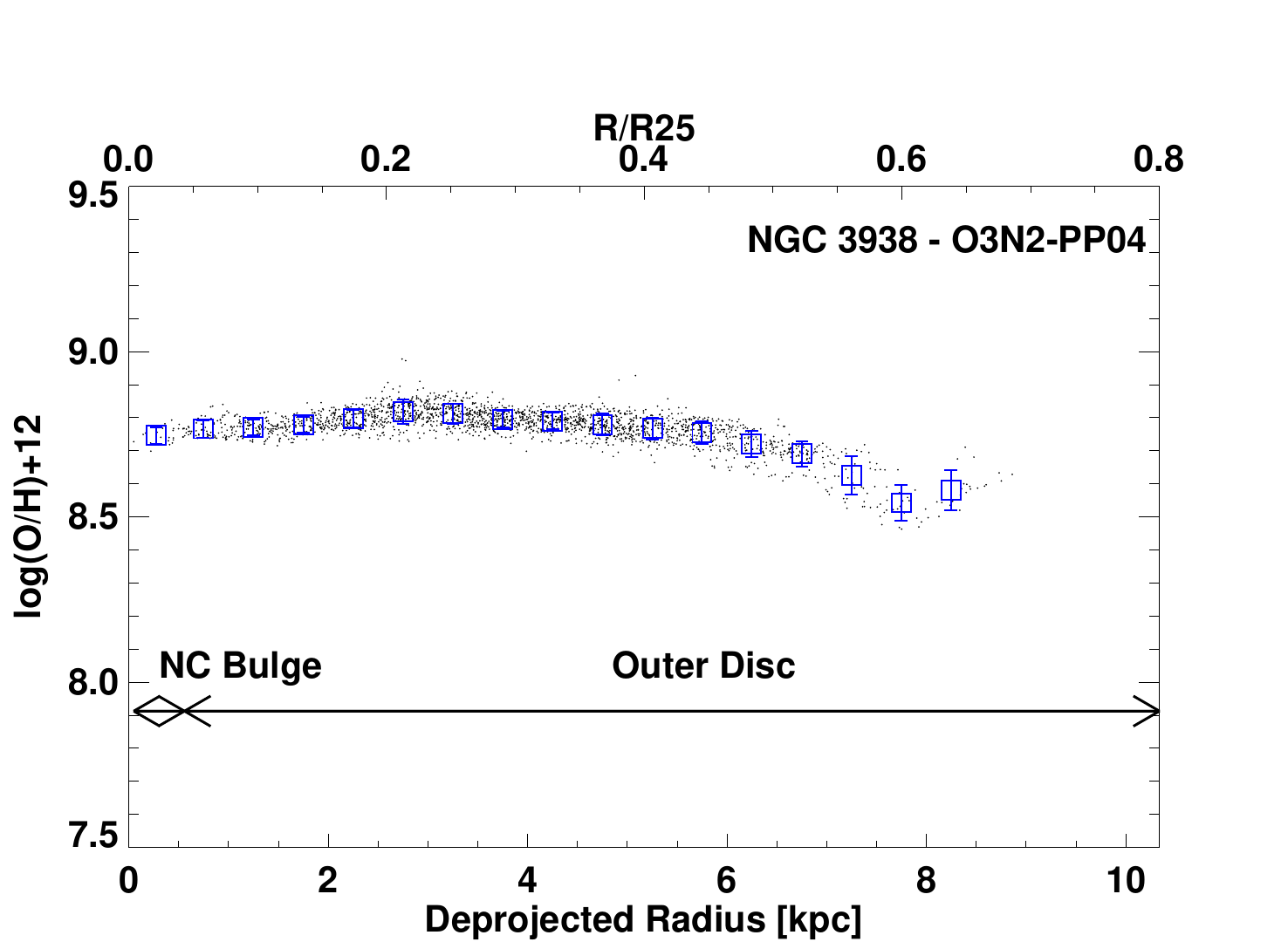}
\includegraphics[height=0.24\textheight, clip=true, trim=0.1cm 0.00cm 0.8cm 0.8 cm]{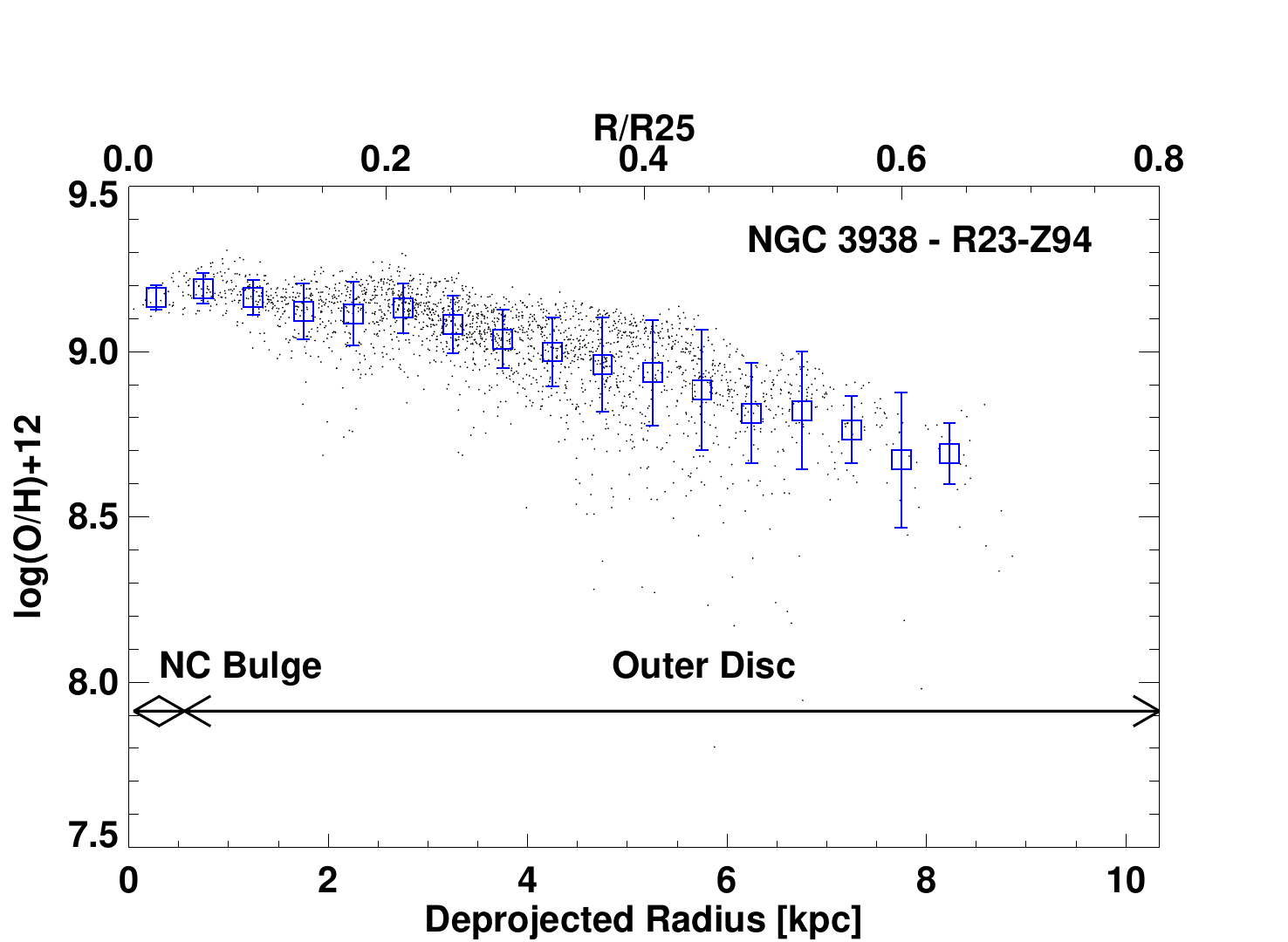}
\vspace{-18pt}\\
\includegraphics[height=0.24\textheight, clip=true, trim=0.1cm 0.00cm 0.8cm 0.8 cm]{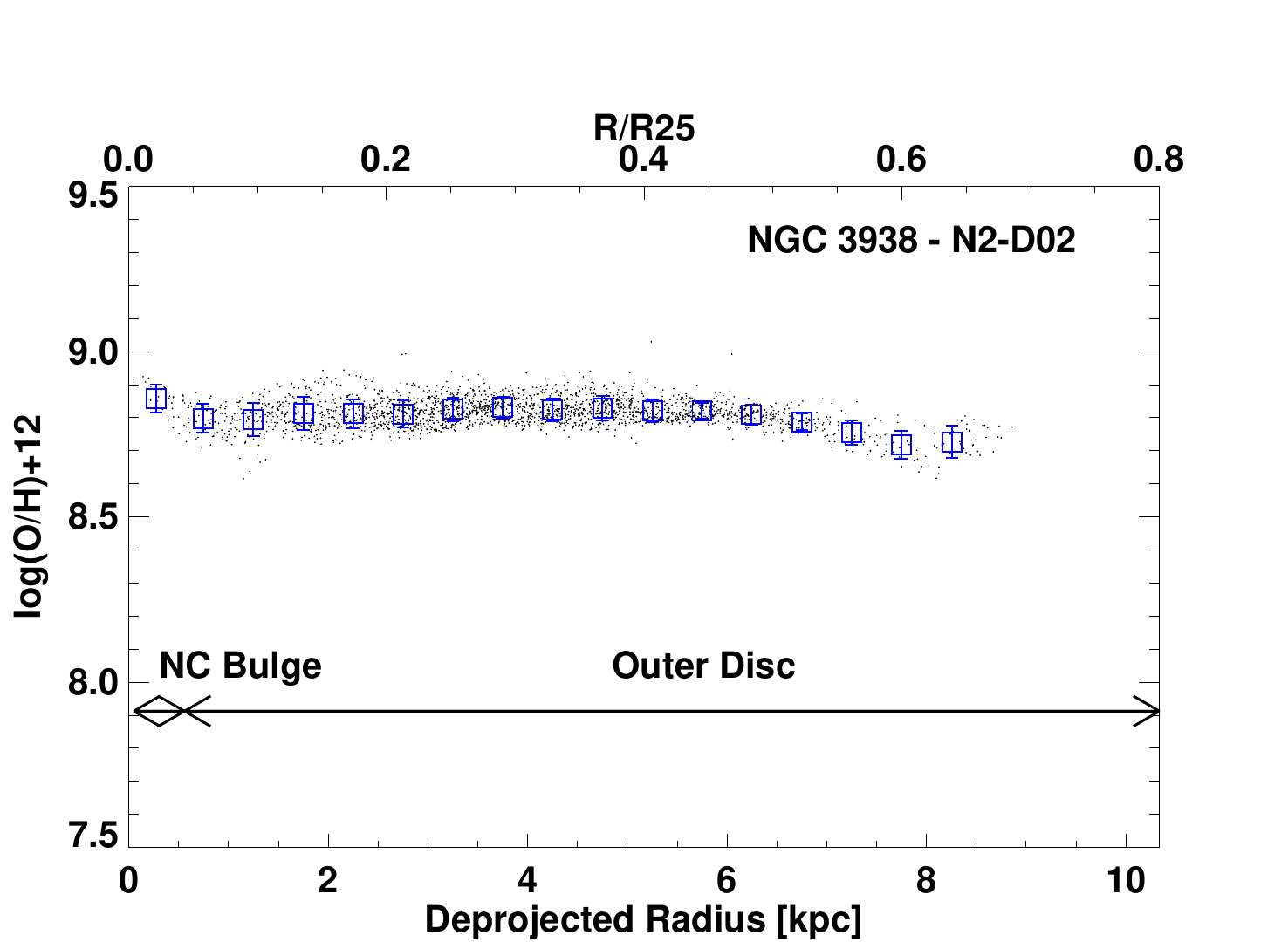}
\includegraphics[height=0.24\textheight, clip=true, trim=0.1cm 0.00cm 0.8cm 0.8 cm]{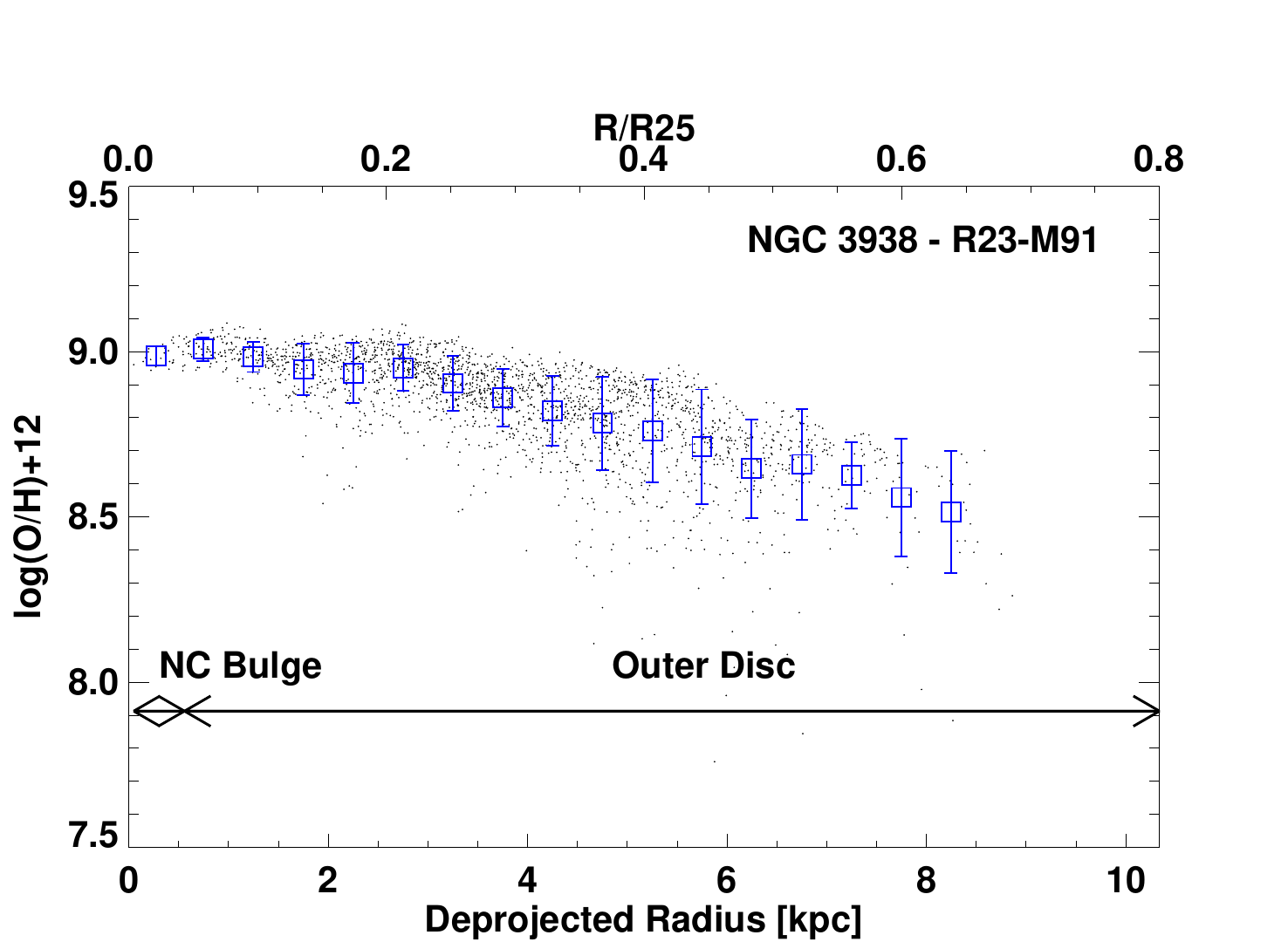}
\vspace{-18pt}\\
\includegraphics[height=0.24\textheight, clip=true, trim=0.1cm 0.00cm 0.8cm 0.8 cm]{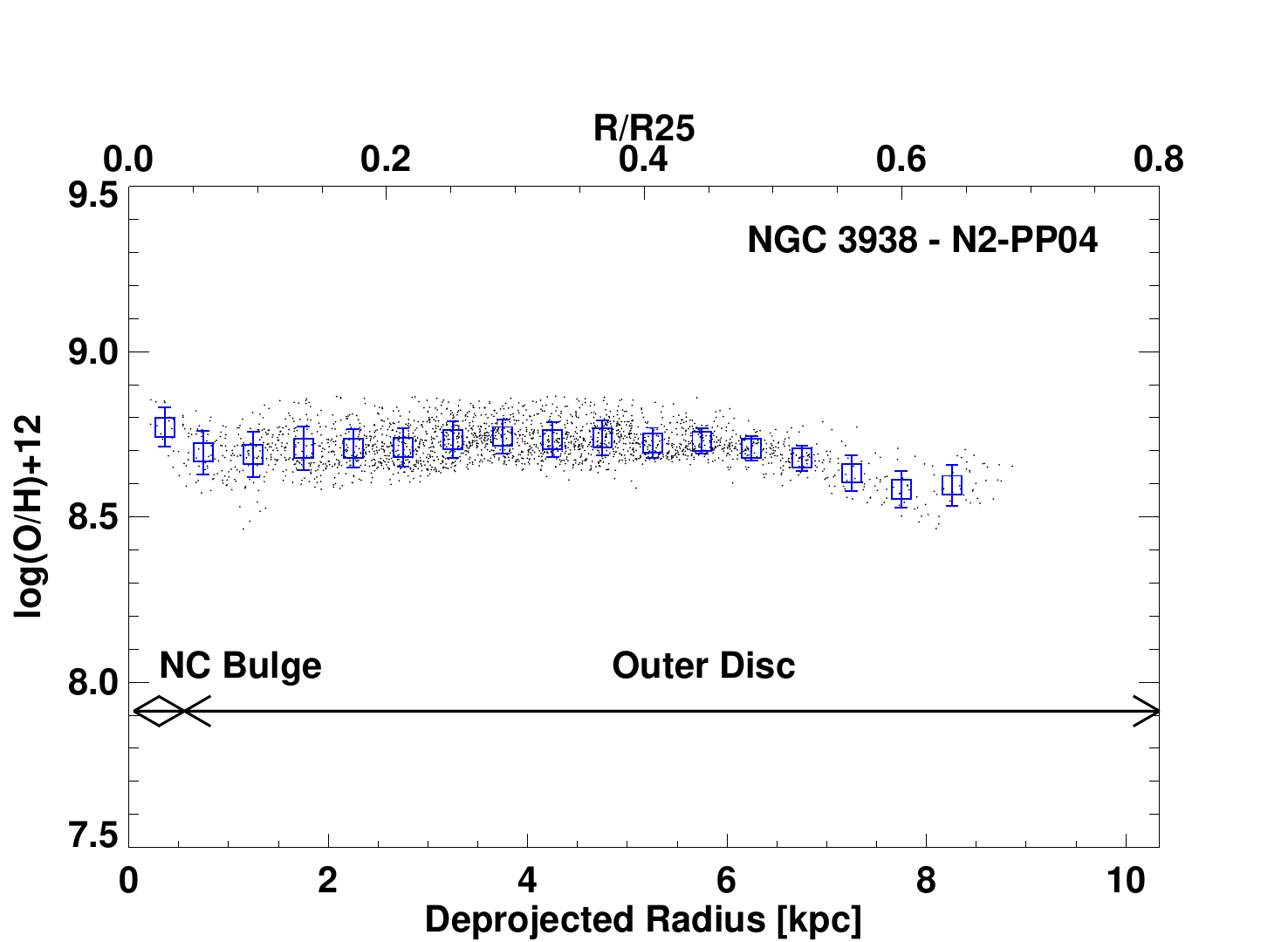}
\end{flushleft}
\caption{Continued: For NGC 3938.}
\label{}
\end{figure*}

\addtocounter{figure}{-1}
\clearpage

\begin{figure*}
\begin{flushleft}
\vspace{-22pt}
\includegraphics[height=0.24\textheight, clip=true, trim=0.1cm 0.00cm 0.8cm 0.8 cm]{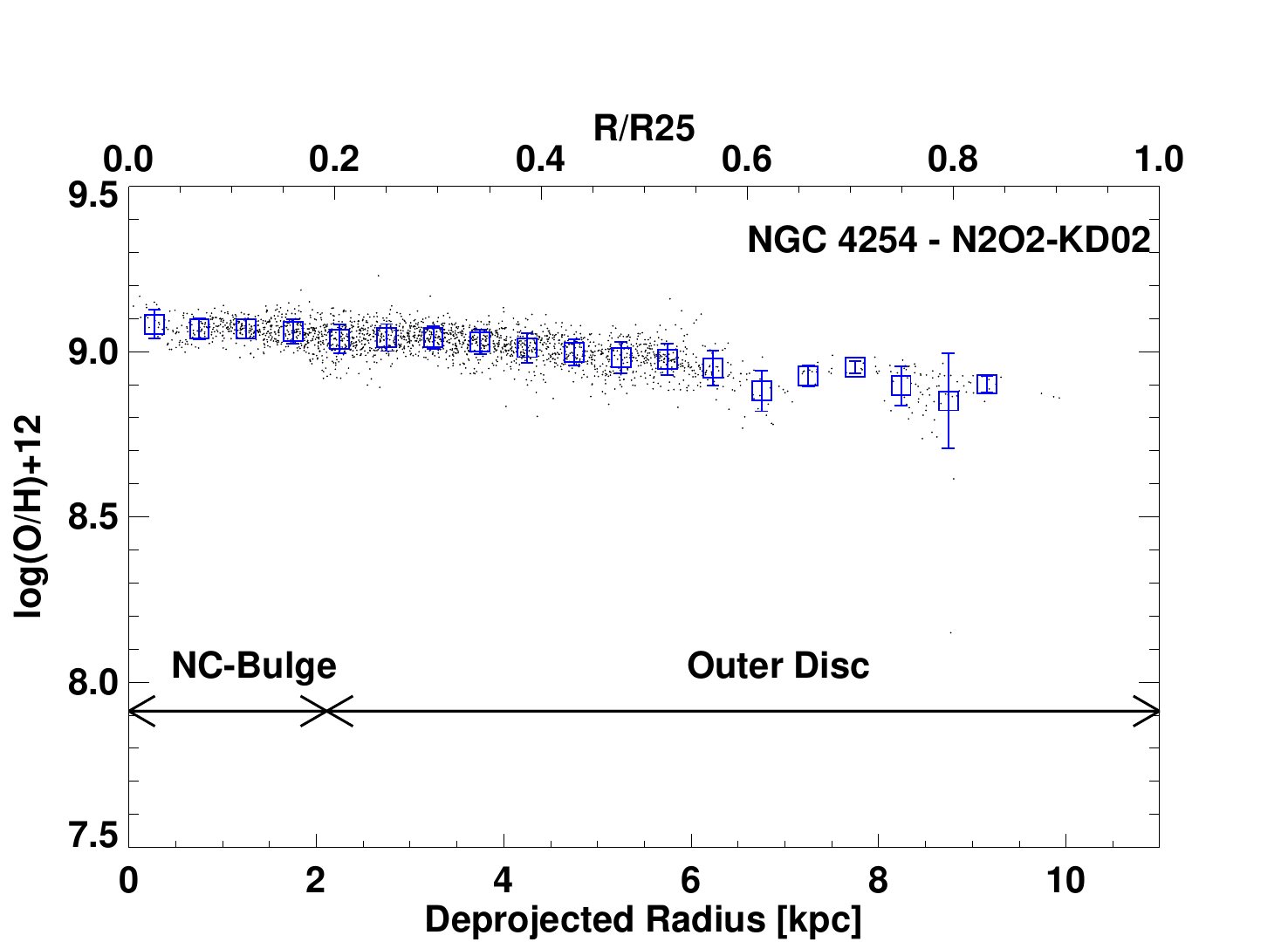}
\includegraphics[height=0.24\textheight, clip=true, trim=0.1cm 0.00cm 0.8cm 0.8 cm]{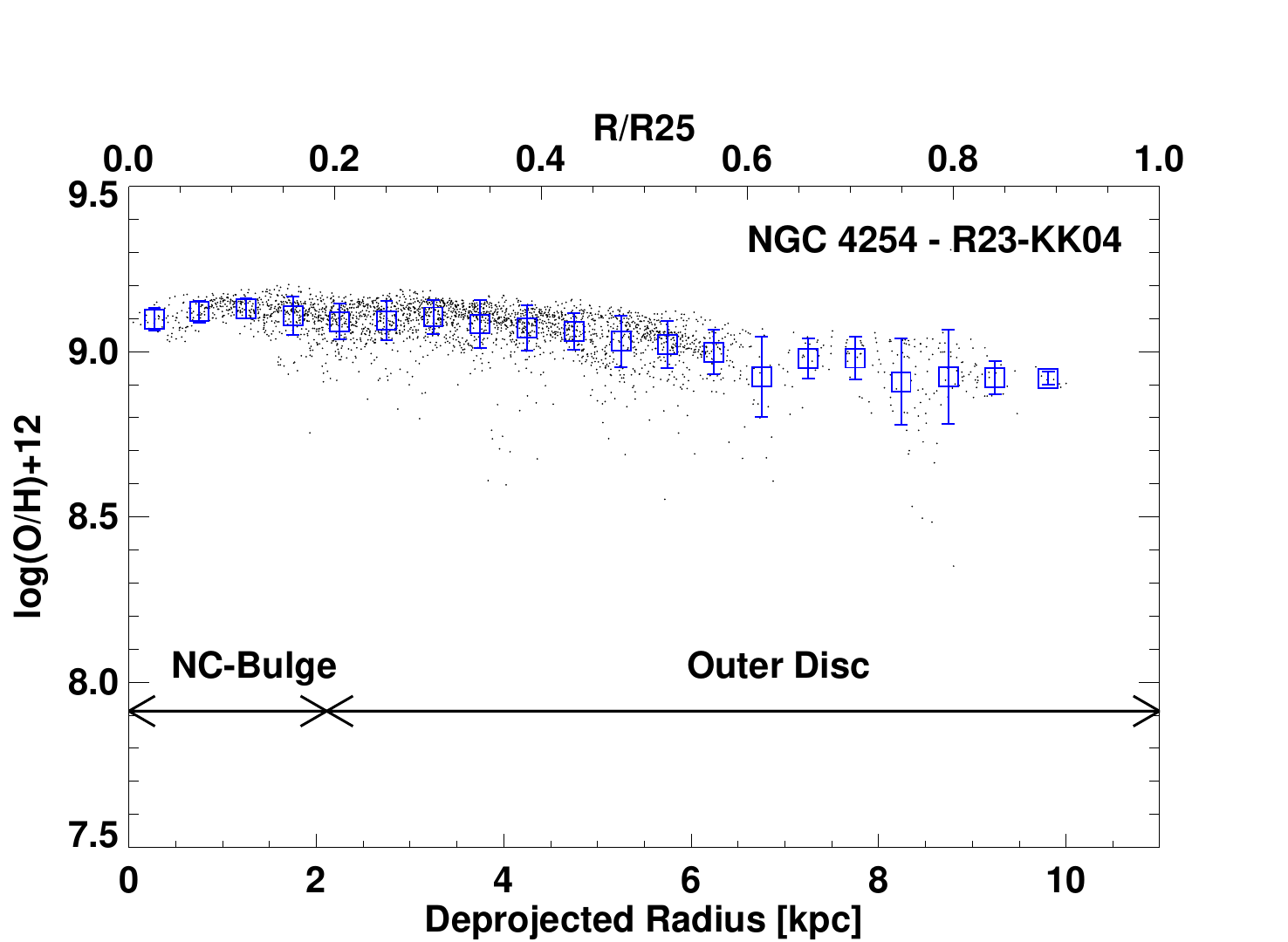}
\vspace{-18pt}\\
\includegraphics[height=0.24\textheight, clip=true, trim=0.1cm 0.00cm 0.8cm 0.8 cm]{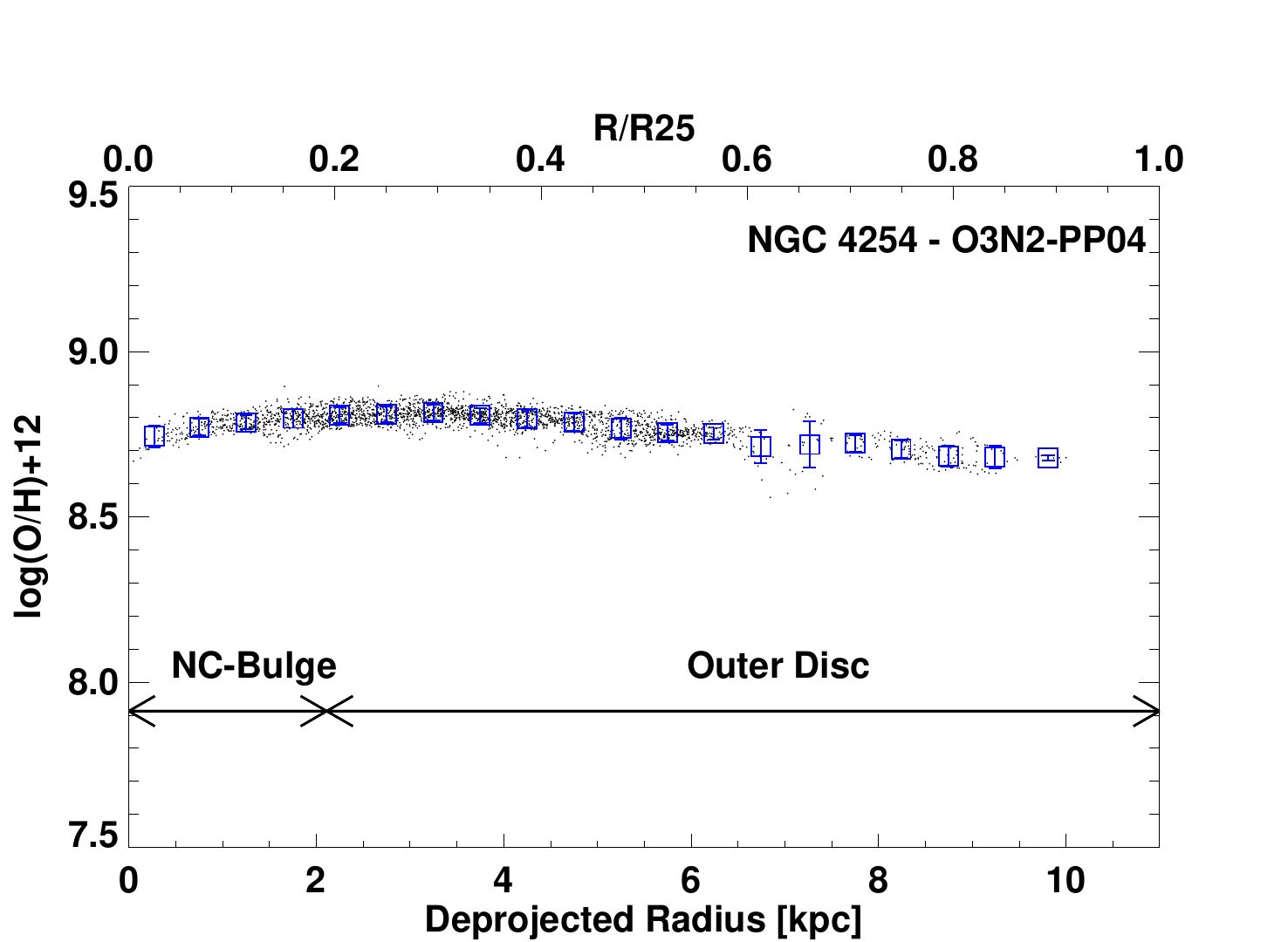}
\includegraphics[height=0.24\textheight, clip=true, trim=0.1cm 0.00cm 0.8cm 0.8 cm]{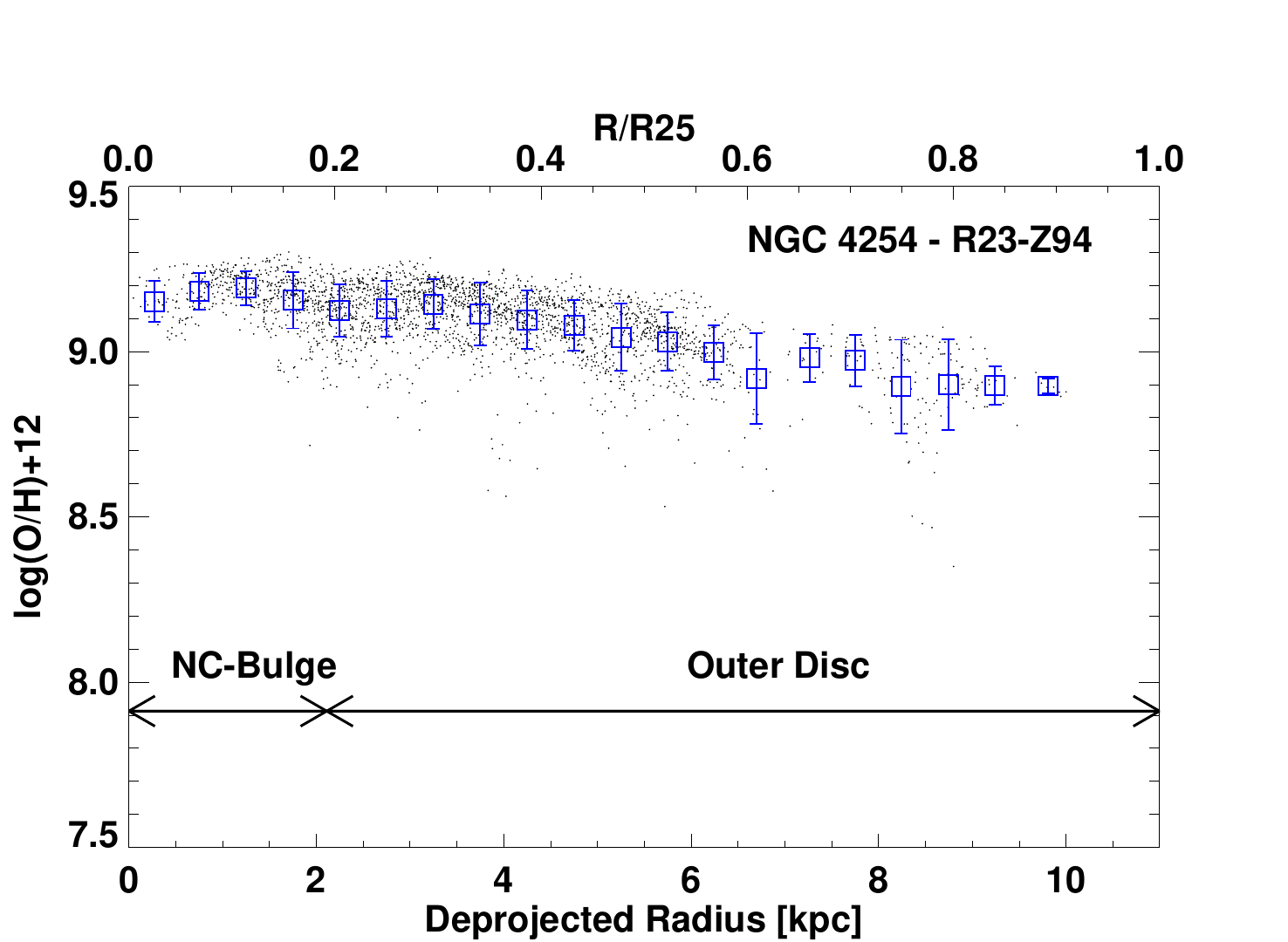}
\vspace{-18pt}\\
\includegraphics[height=0.24\textheight, clip=true, trim=0.1cm 0.00cm 0.8cm 0.8 cm]{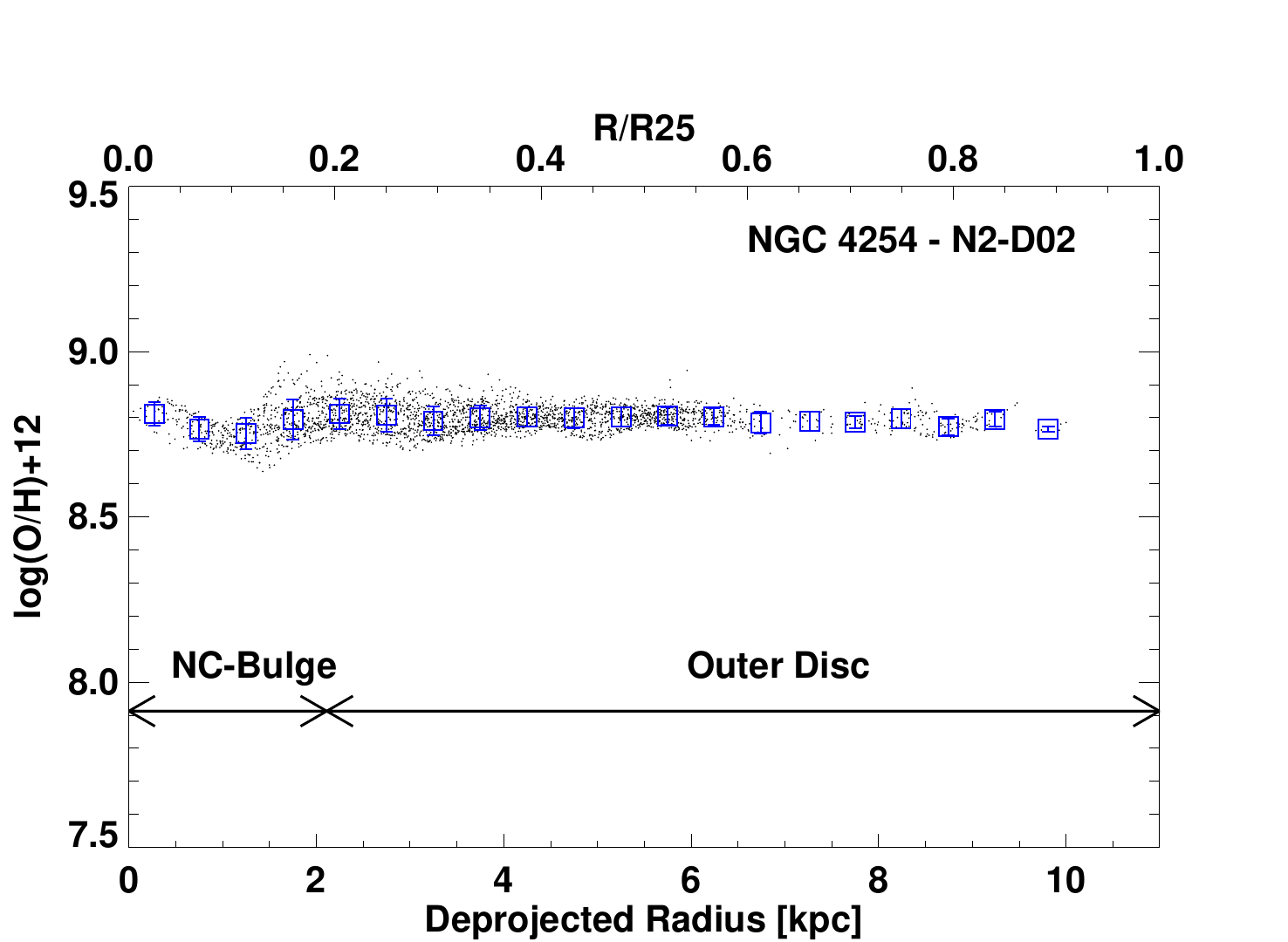}
\includegraphics[height=0.24\textheight, clip=true, trim=0.1cm 0.00cm 0.8cm 0.8 cm]{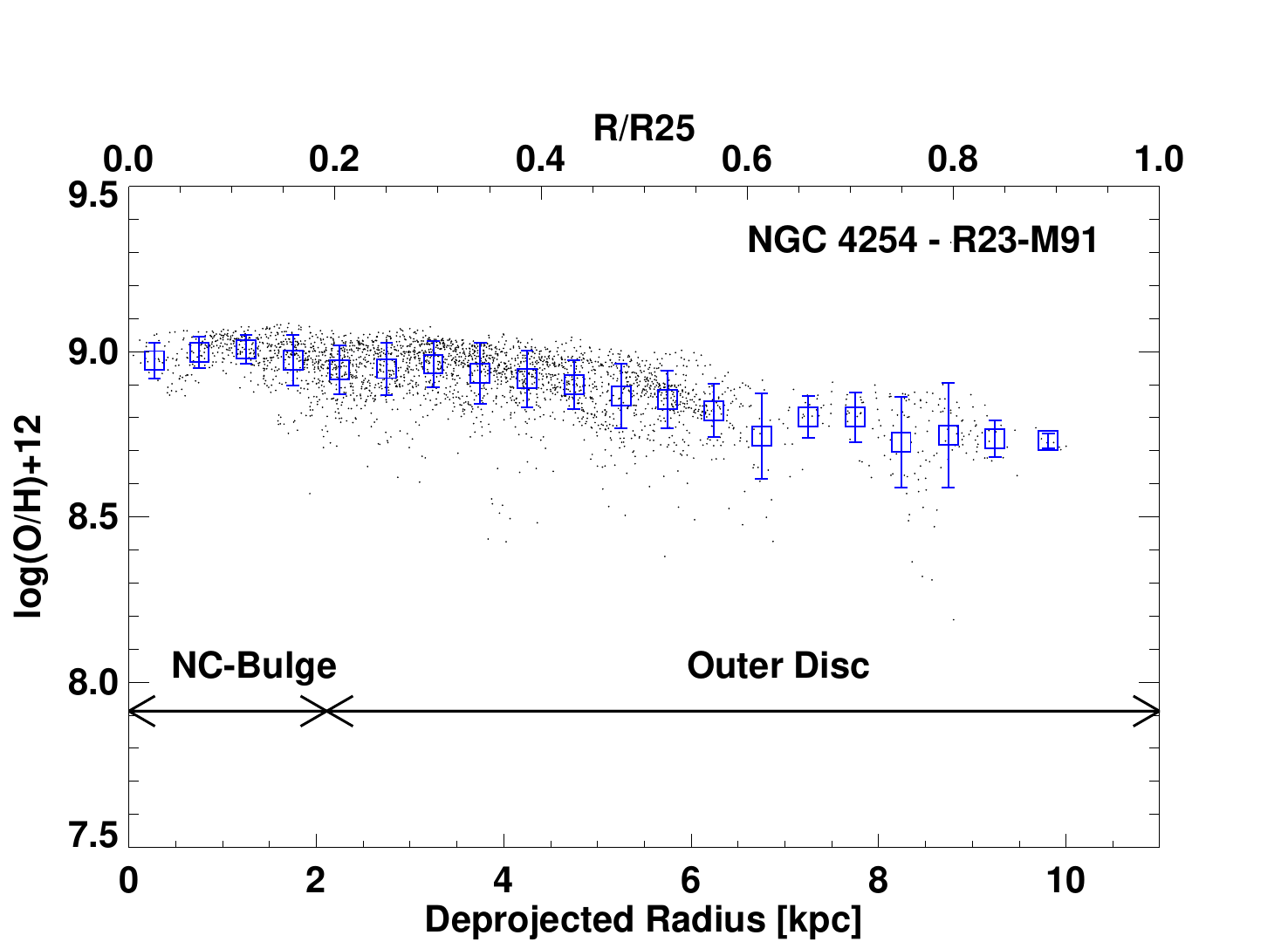}
\vspace{-18pt}\\
\includegraphics[height=0.24\textheight, clip=true, trim=0.1cm 0.00cm 0.8cm 0.8 cm]{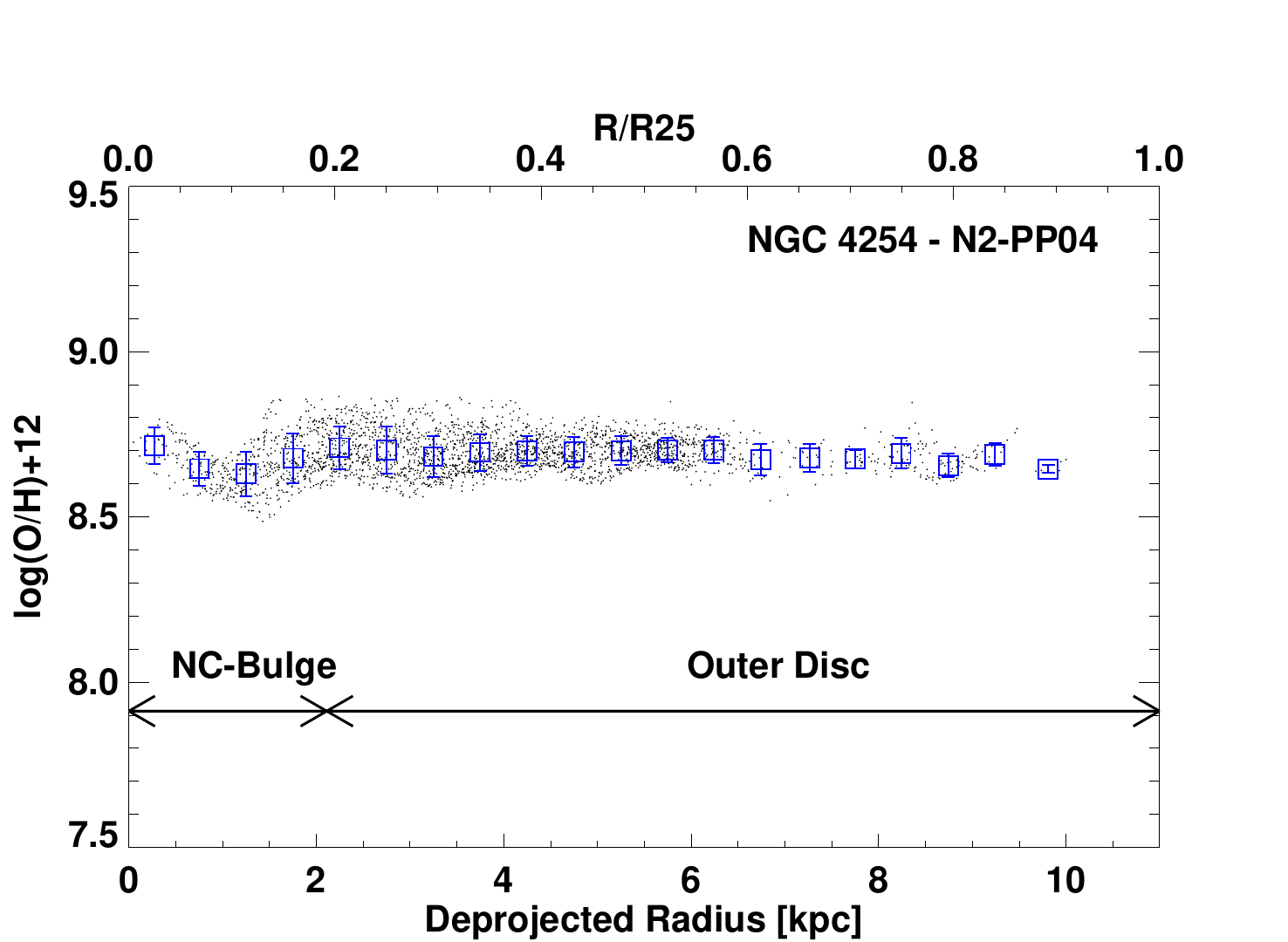}
\end{flushleft}
\caption{Continued: For NGC 4254. 
}
\label{}
\end{figure*}

\addtocounter{figure}{-1}
\clearpage

\begin{figure*}
\begin{flushleft}
\vspace{-22pt}
\includegraphics[height=0.24\textheight, clip=true, trim=0.1cm 0.00cm 0.8cm 0.8 cm]{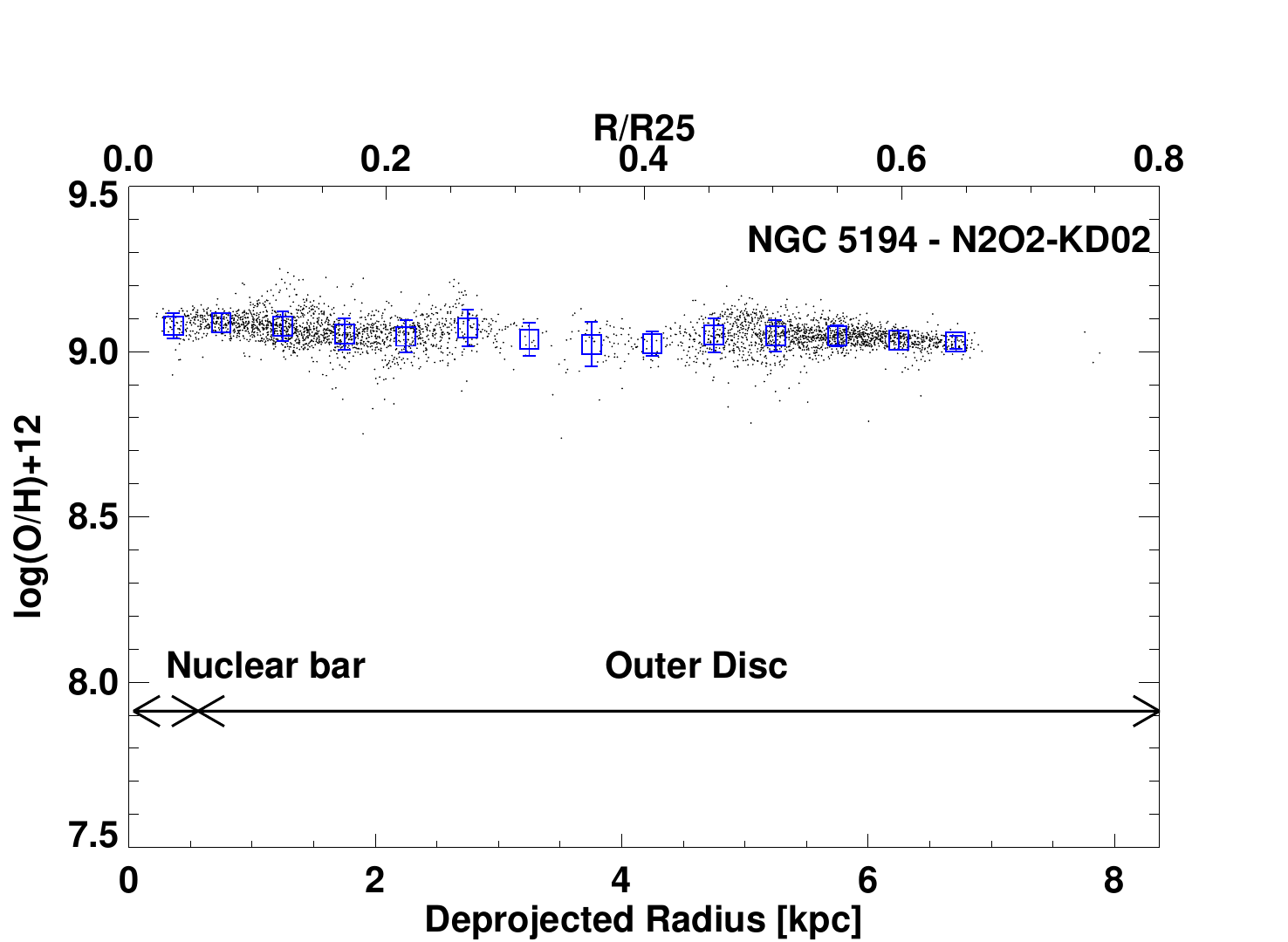}
\includegraphics[height=0.24\textheight, clip=true, trim=0.1cm 0.00cm 0.8cm 0.8 cm]{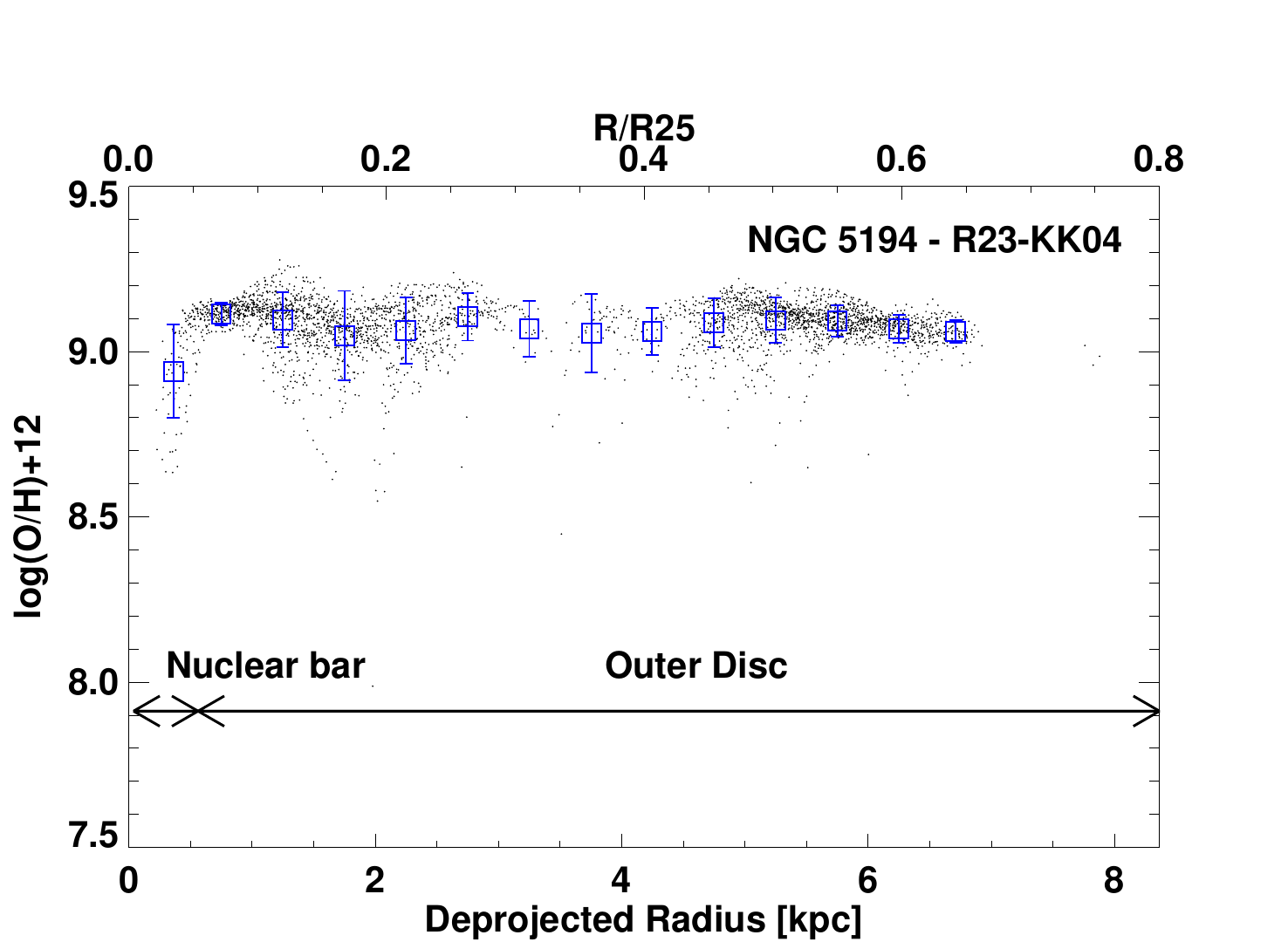}
\vspace{-18pt}\\
\includegraphics[height=0.24\textheight, clip=true, trim=0.1cm 0.00cm 0.8cm 0.8 cm]{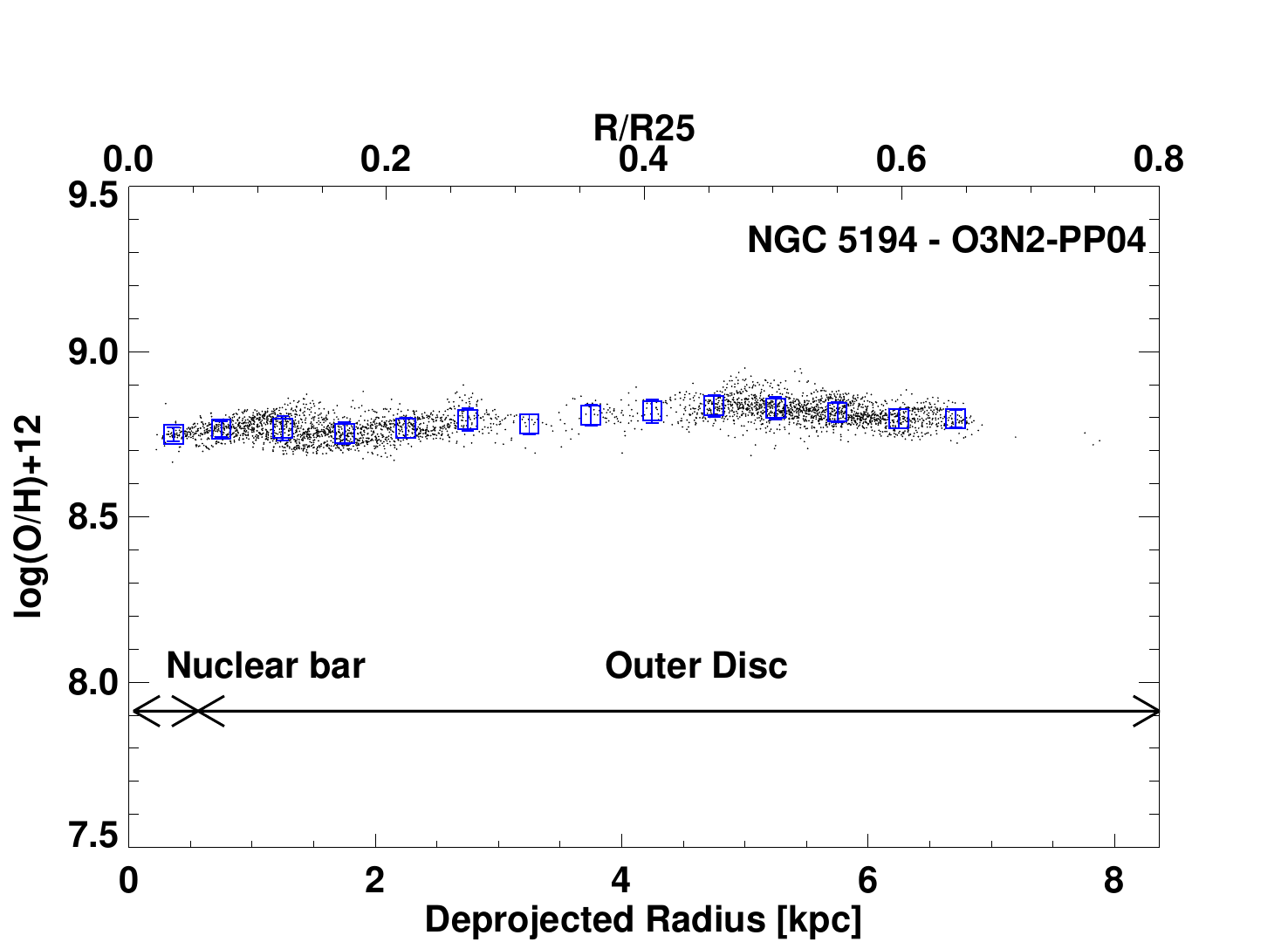}
\includegraphics[height=0.24\textheight, clip=true, trim=0.1cm 0.00cm 0.8cm 0.8 cm]{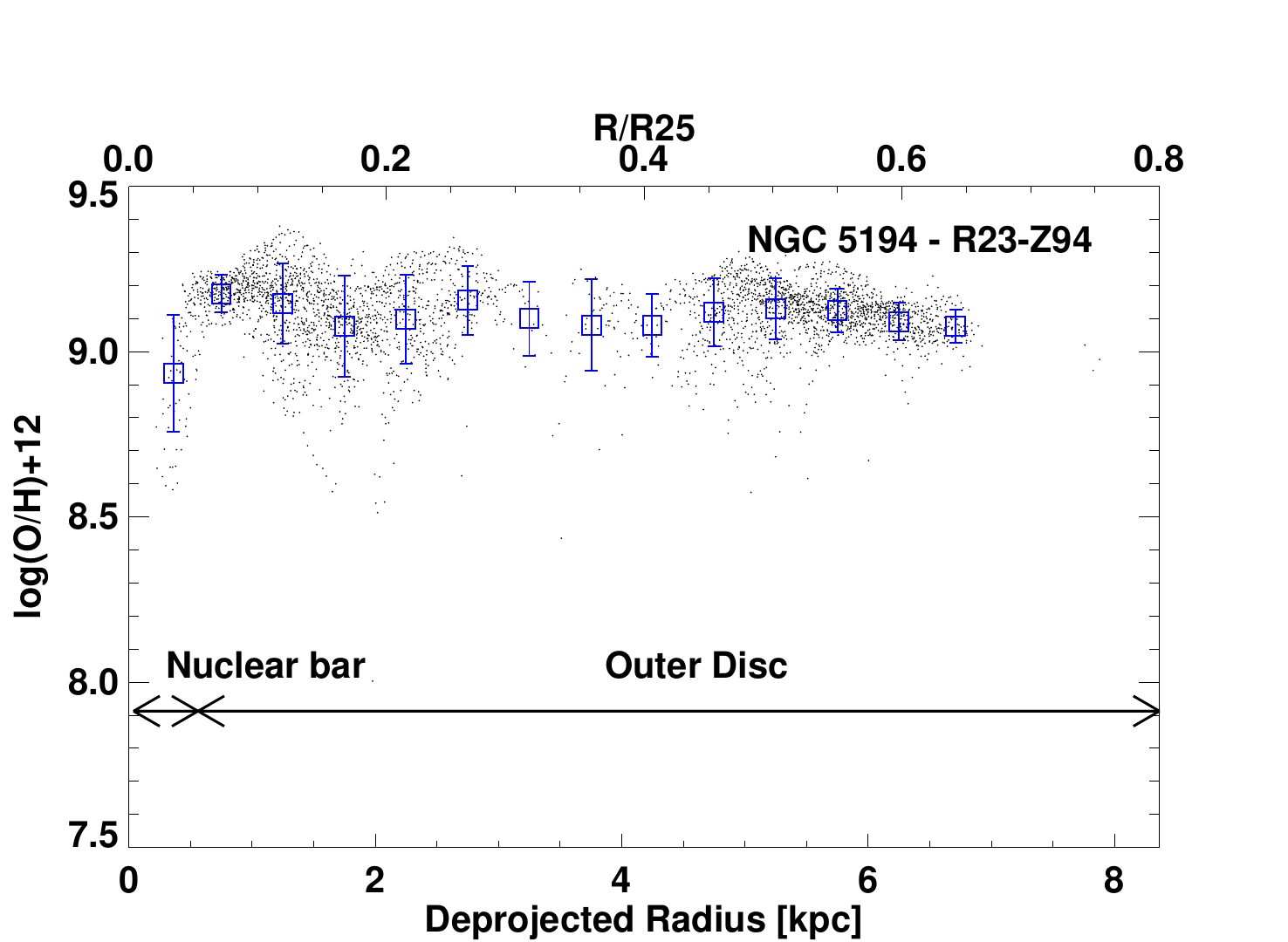}
\vspace{-18pt}\\
\includegraphics[height=0.24\textheight, clip=true, trim=0.1cm 0.00cm 0.8cm 0.8 cm]{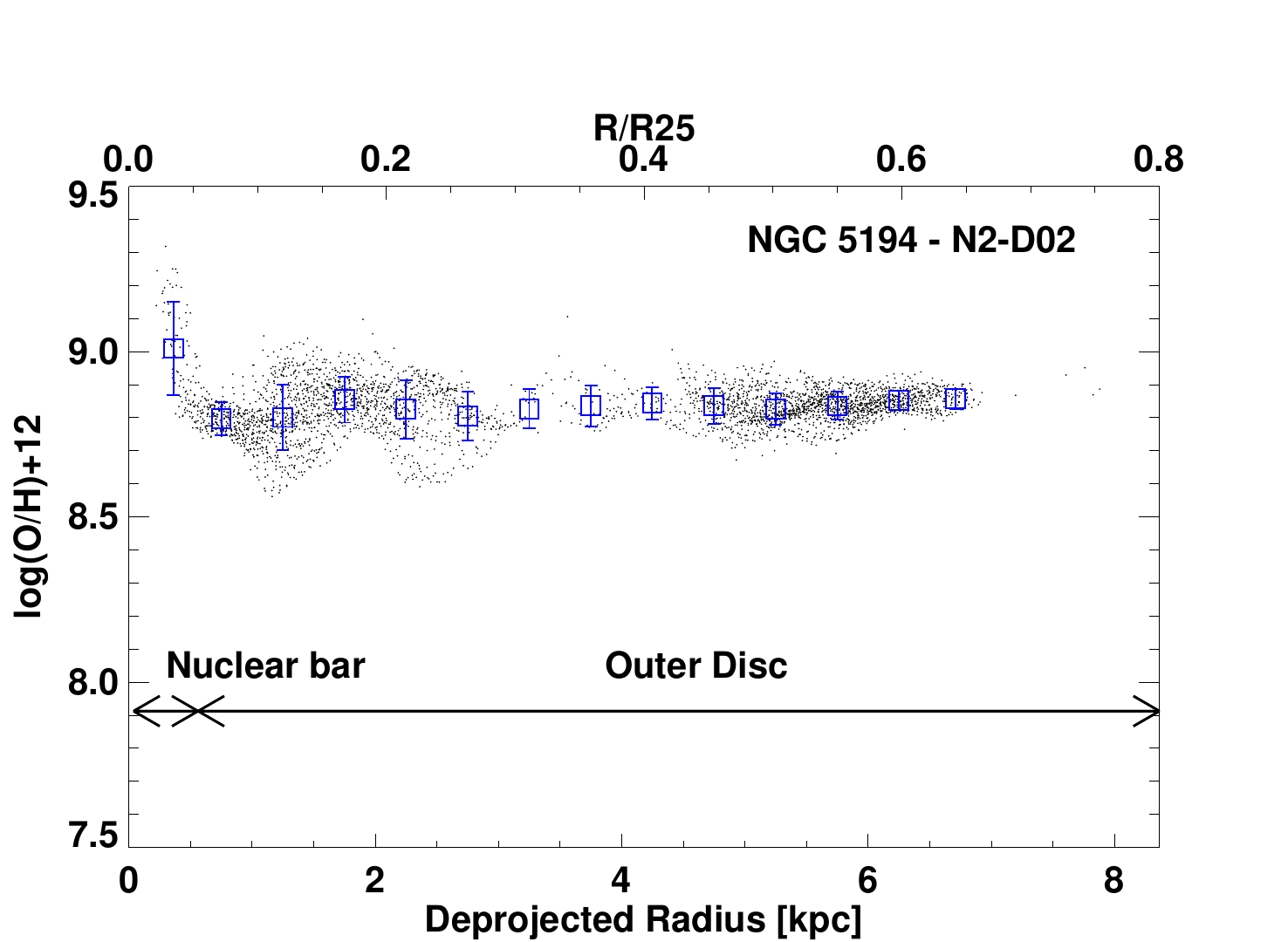}
\includegraphics[height=0.24\textheight, clip=true, trim=0.1cm 0.00cm 0.8cm 0.8 cm]{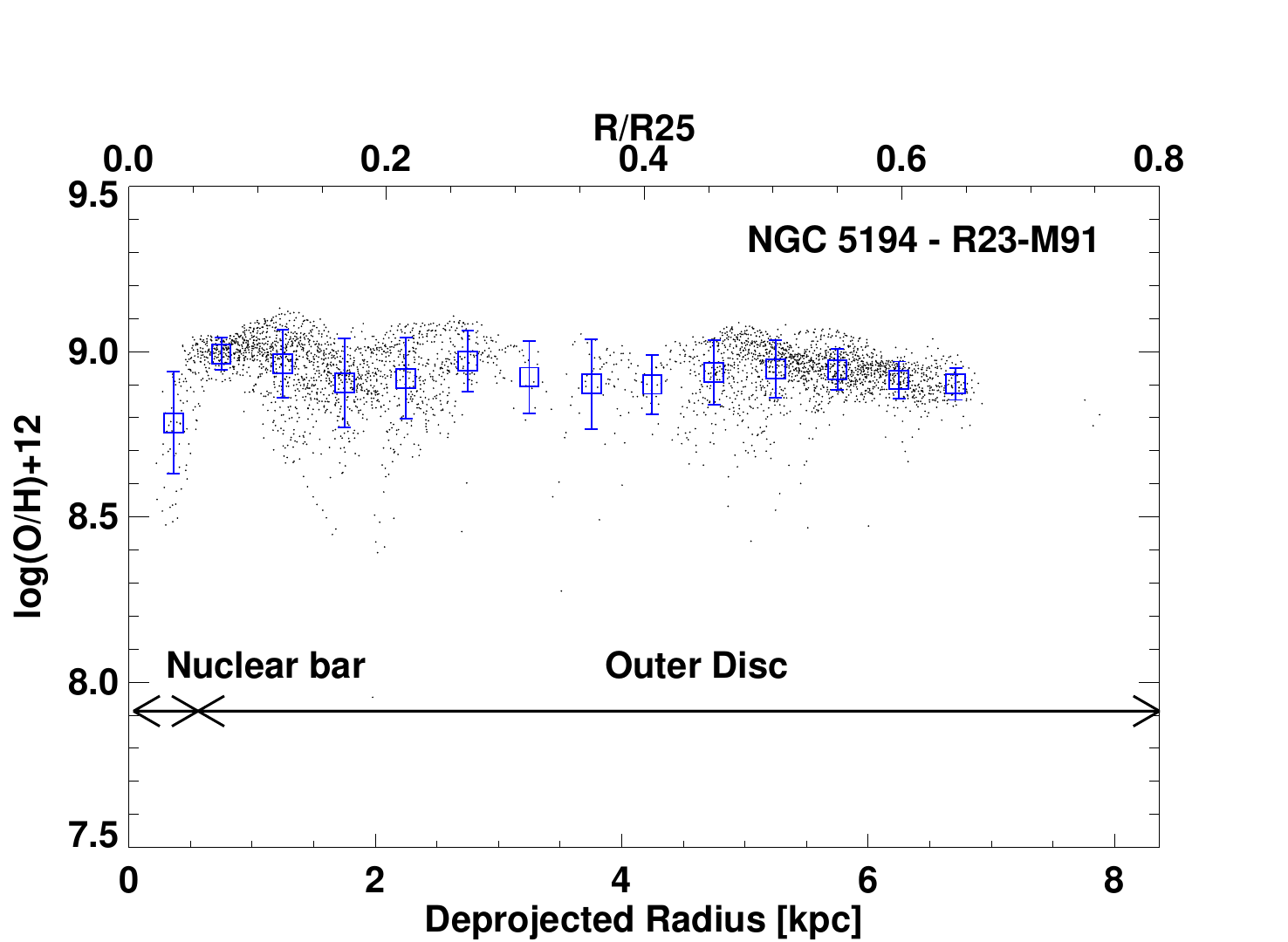}
\vspace{-18pt}\\
\includegraphics[height=0.24\textheight, clip=true, trim=0.1cm 0.00cm 0.8cm 0.8 cm]{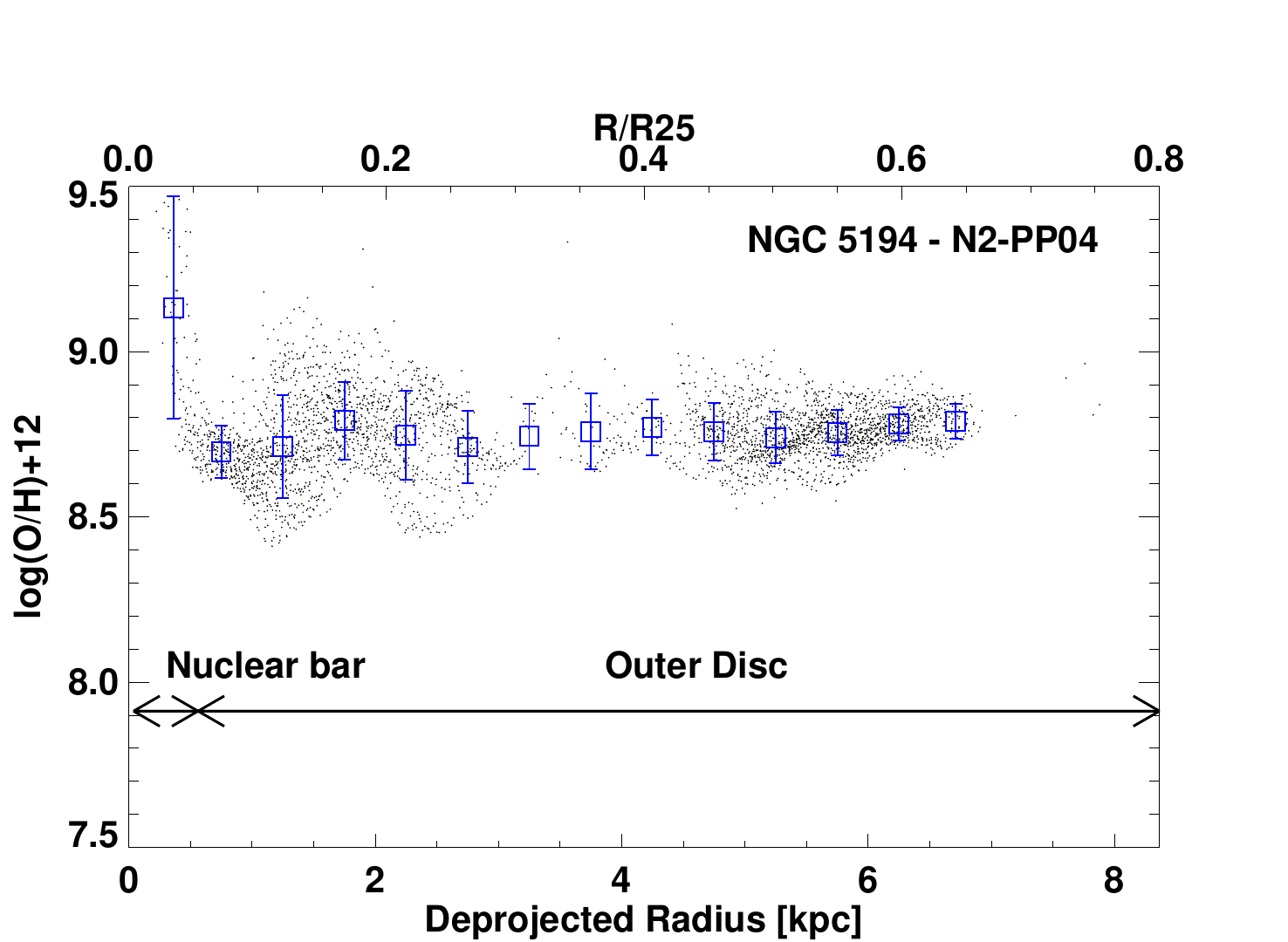}
\end{flushleft}
\caption{Continued: For NGC 5194.}
\label{}
\end{figure*}

\addtocounter{figure}{-1}
\clearpage

\begin{figure*}
\begin{flushleft}
\vspace{-22pt}
\includegraphics[height=0.24\textheight, clip=true, trim=0.1cm 0.00cm 0.8cm 0.8 cm]{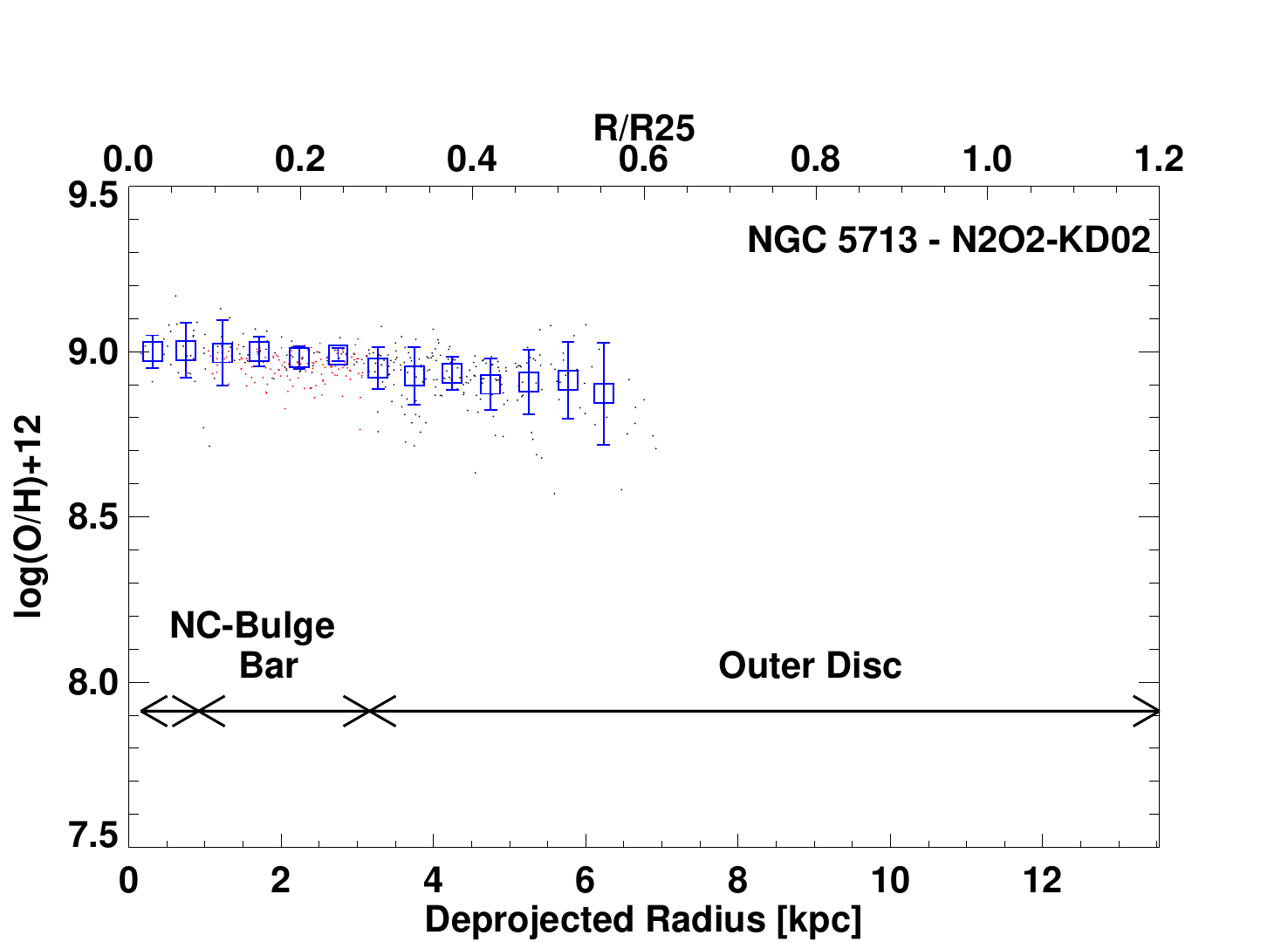}
\includegraphics[height=0.24\textheight, clip=true, trim=0.1cm 0.00cm 0.8cm 0.8 cm]{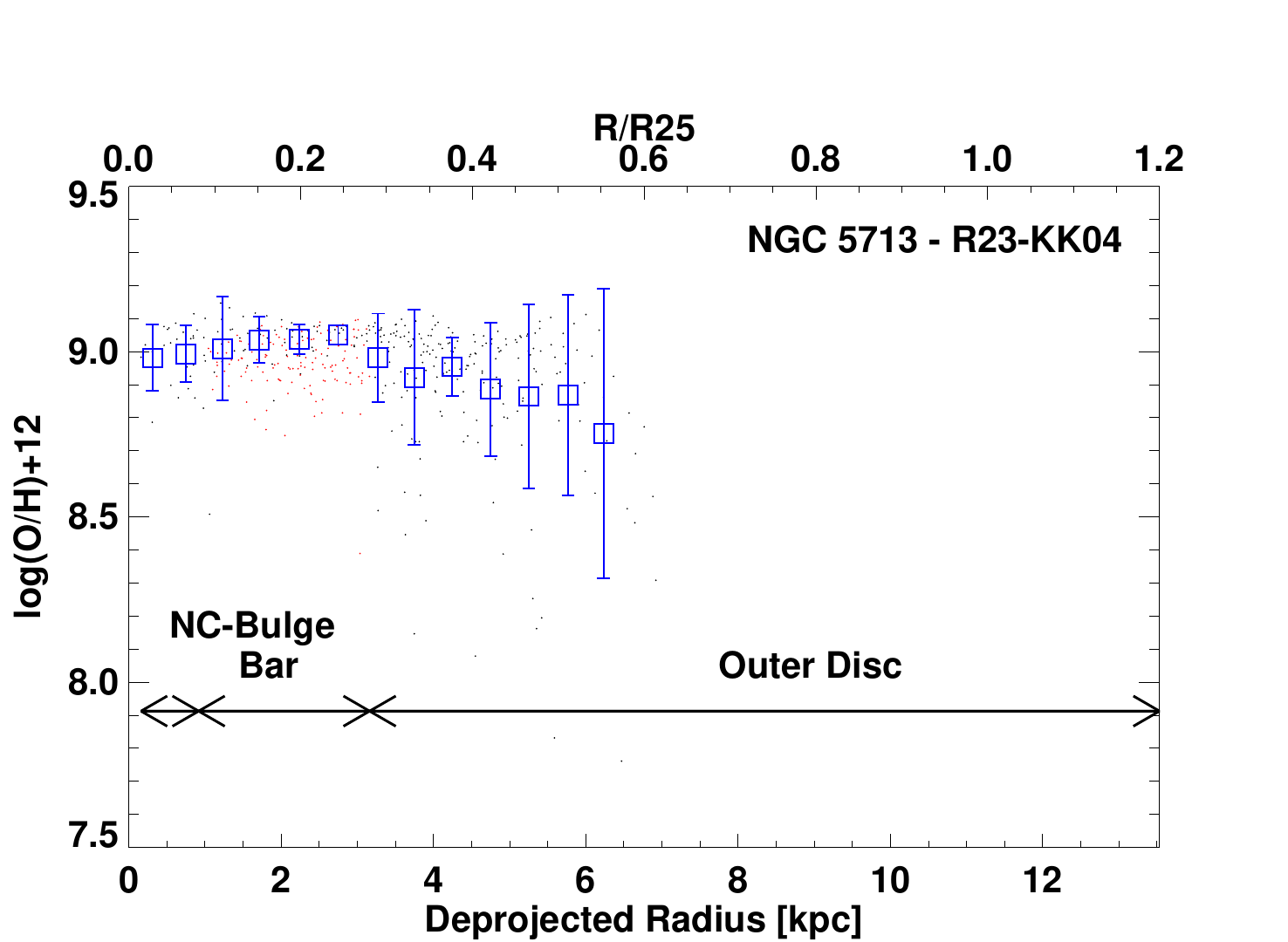}
\vspace{-18pt}\\
\includegraphics[height=0.24\textheight, clip=true, trim=0.1cm 0.00cm 0.8cm 0.8 cm]{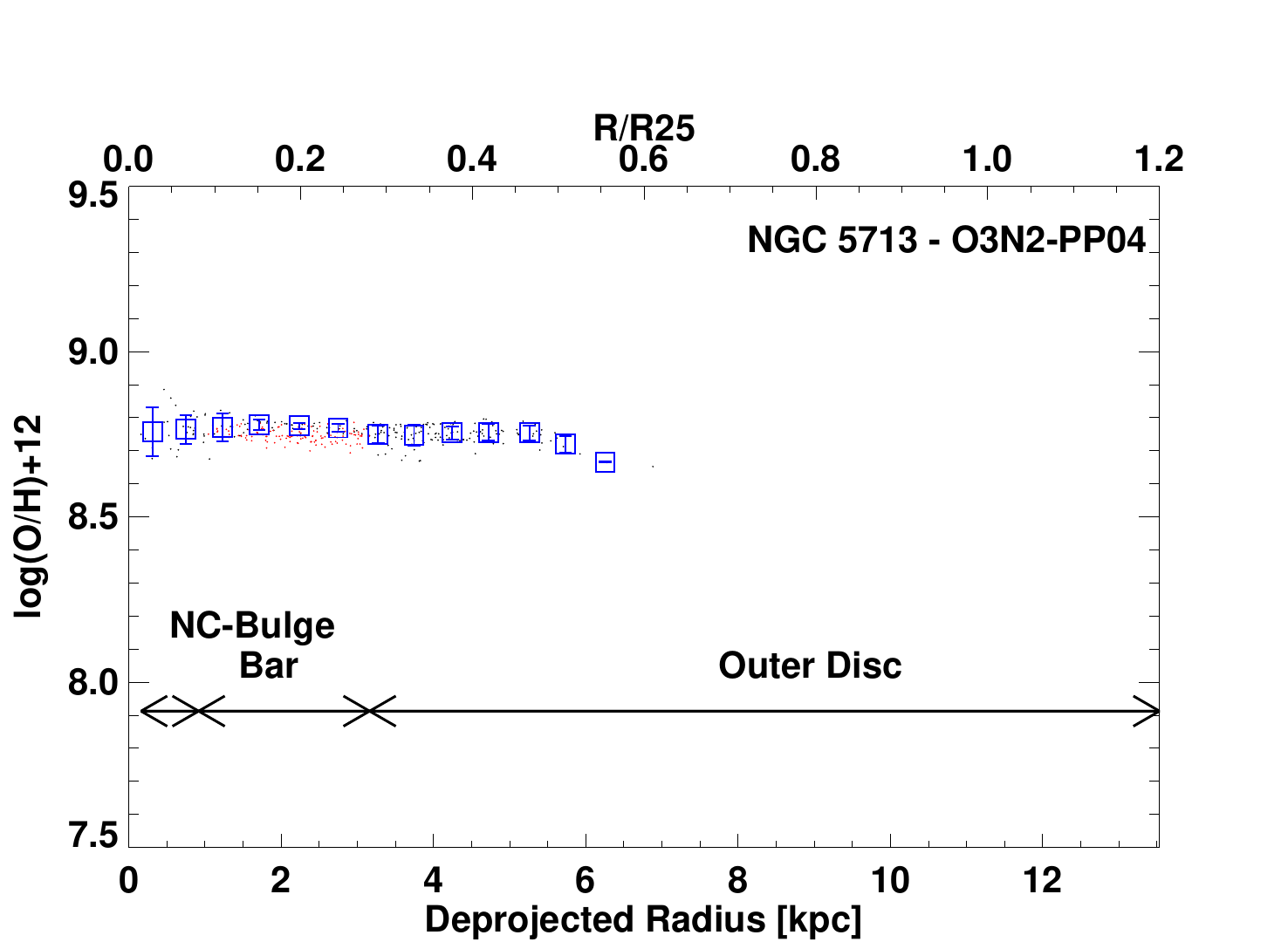}
\includegraphics[height=0.24\textheight, clip=true, trim=0.1cm 0.00cm 0.8cm 0.8 cm]{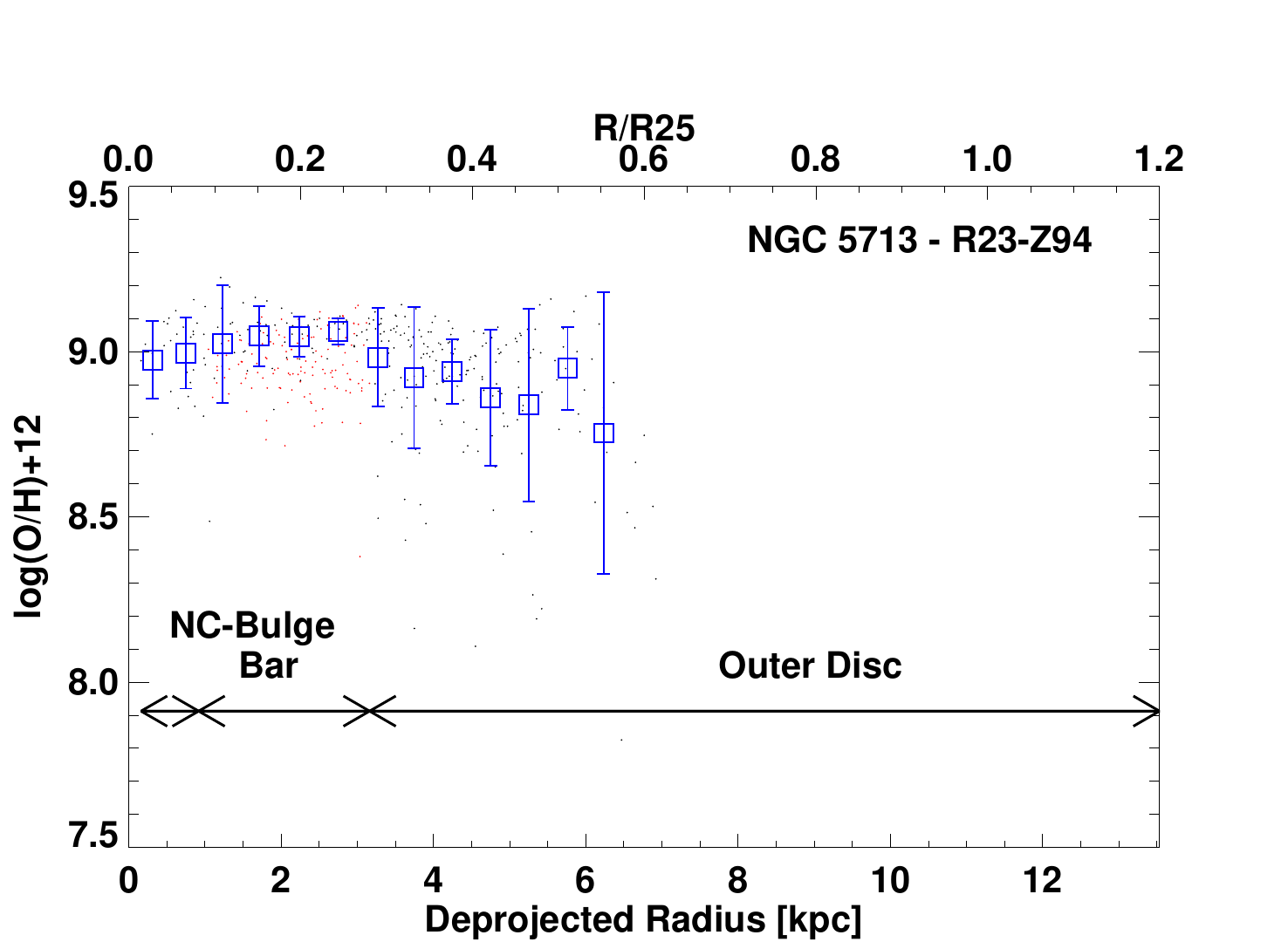}
\vspace{-18pt}\\
\includegraphics[height=0.24\textheight, clip=true, trim=0.1cm 0.00cm 0.8cm 0.8 cm]{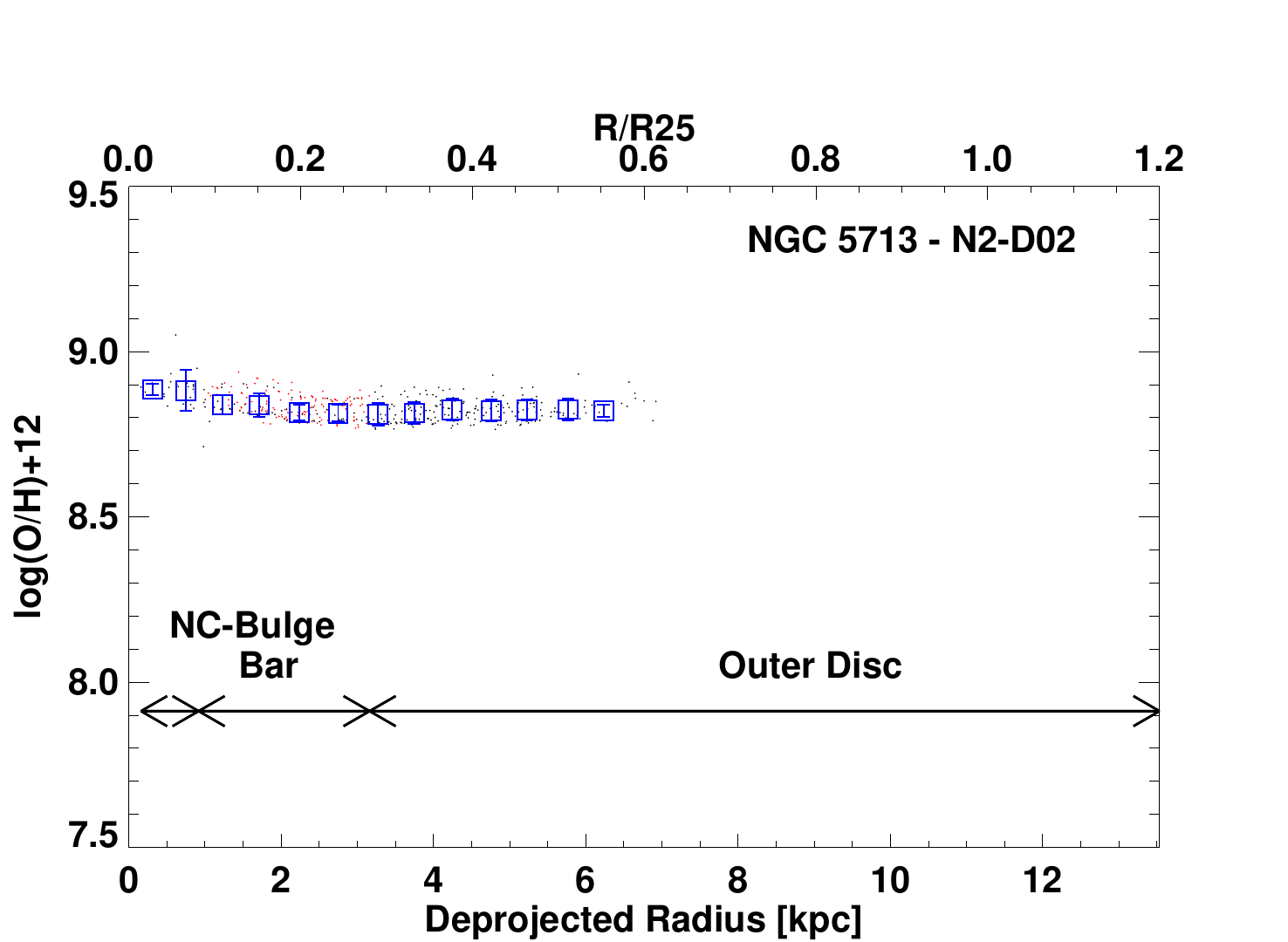}
\includegraphics[height=0.24\textheight, clip=true, trim=0.1cm 0.00cm 0.8cm 0.8 cm]{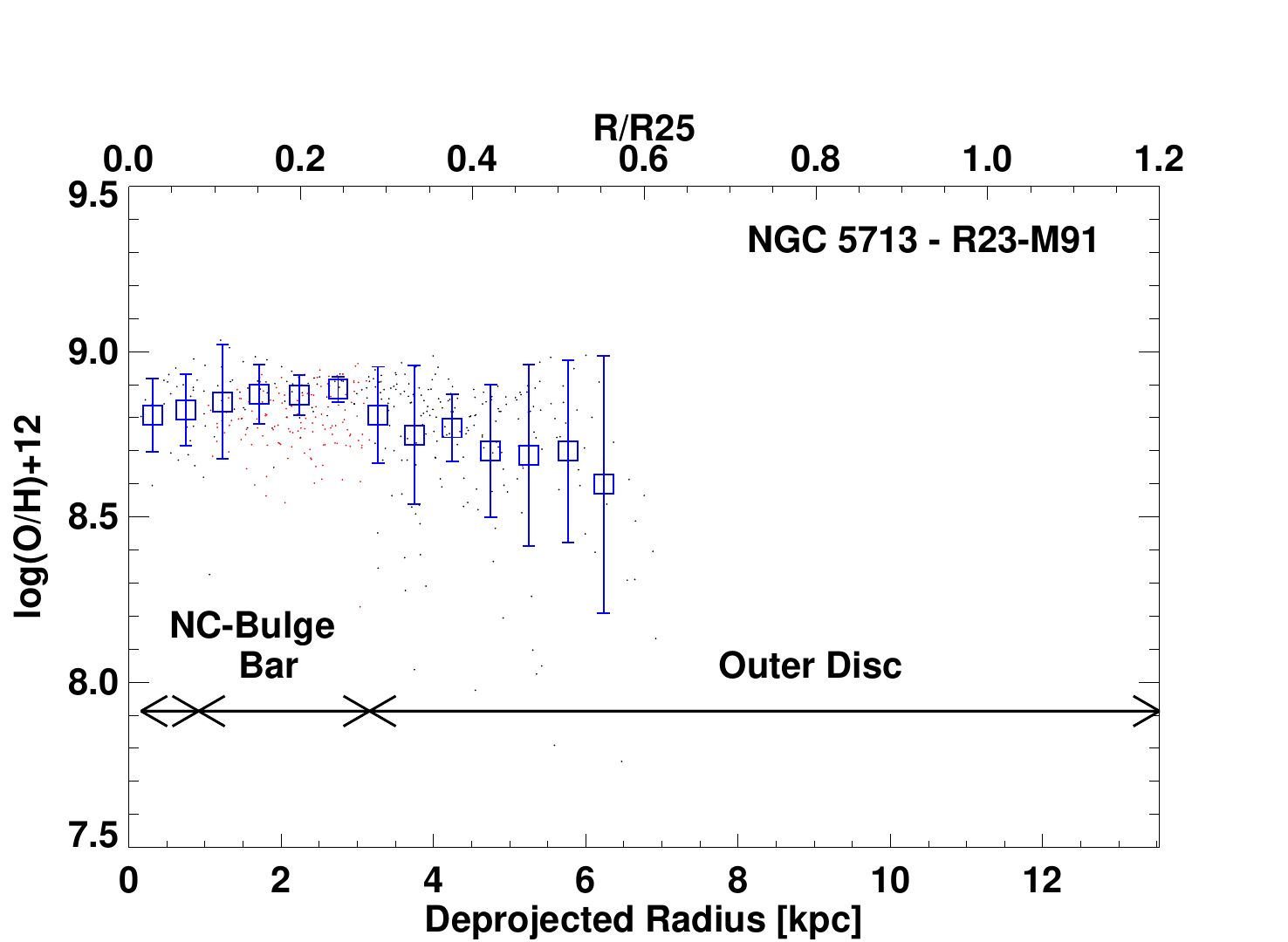}
\vspace{-18pt}\\
\includegraphics[height=0.24\textheight, clip=true, trim=0.1cm 0.00cm 0.8cm 0.8 cm]{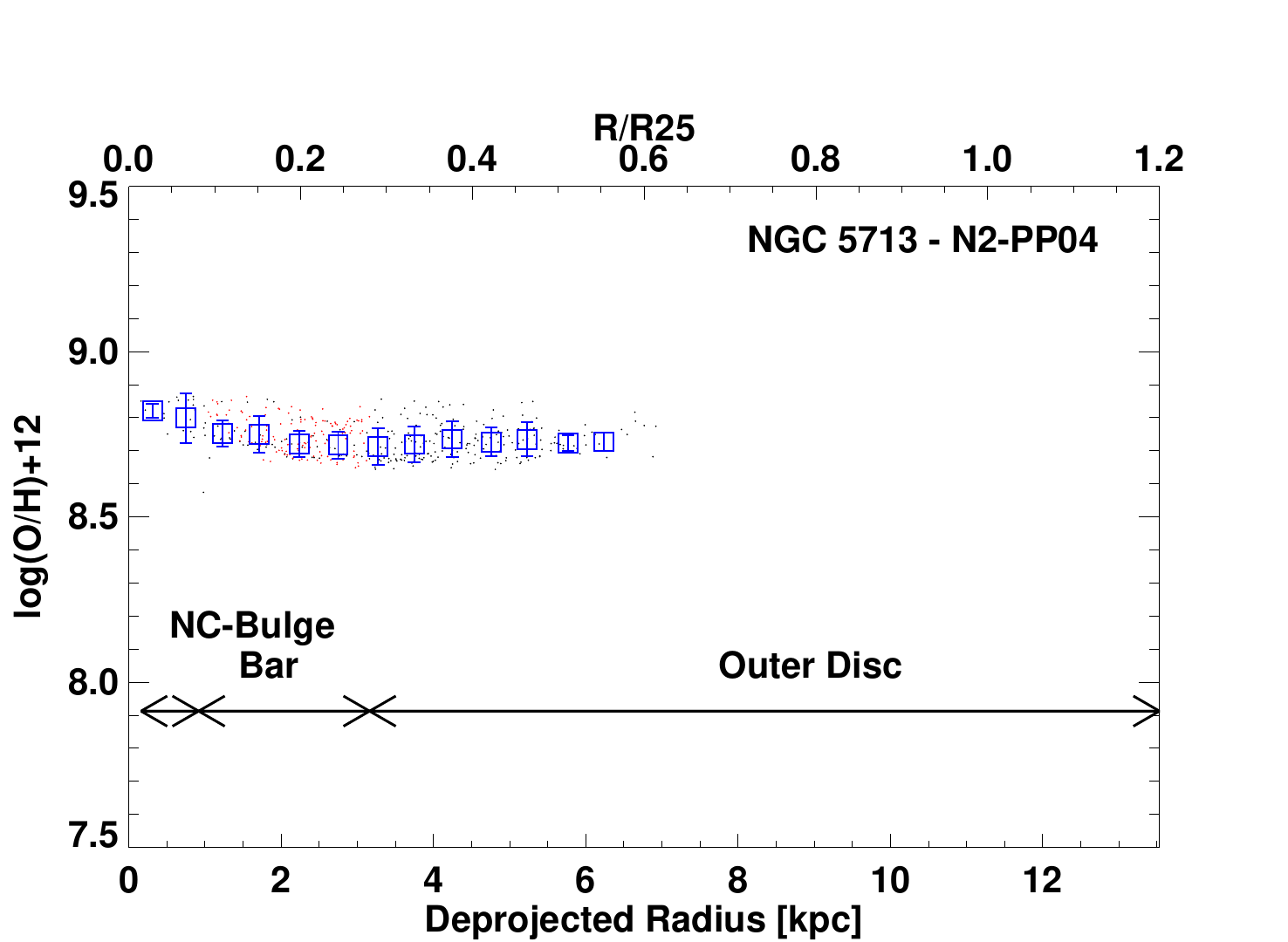}
\end{flushleft}
\caption{Continued: For NGC 5713.}
\label{}
\end{figure*}

\clearpage

\begin{table*}
\caption{Summary of \zgas{} gradients in our study and other published studies}
\hspace{-1cm}
\begin{center}
\begin{tabular}{clccc}
\hline\hline
Reference & Galaxy or Class of Galaxies & \zgas{}  Diagnostic & dex ($R$/\rtwofive{})$^{-1}$ or dex/$(R_e)^{a}$ &  dex   kpc$^{-1}$ \\
 (1) & (2) & (3) & (4) & (5) \\
\hline
\multicolumn{5}{c}{\it Our results for barred and unbarred spirals at $z\sim0^b$ using the N2O2-KD02 \zgas{} diagnostic} \\
This Study & NGC 0337  (Sd, barred/weakly inter.) & N2O2-KD02 &  -0.31 $\pm$ 0.02 & -0.038 $\pm$ 0.002 \\
This Study & NGC 0628  (Sc, unbarred/isolated)  & N2O2-KD02 &  -0.49 $\pm$ 0.01 & -0.037 $\pm$ 0.001 \\
This Study & NGC 2903  (Sbc, barred/isolated)  & N2O2-KD02 &  -0.32 $\pm$ 0.01 & -0.019 $\pm$ 0.001  \\
This Study & NGC 3938  (Sc, unbarred/isolated)  & N2O2-KD02 &  -0.57 $\pm$ 0.01 & -0.041 $\pm$ 0.001\\
This Study & NGC 4254  (Sc, unbarred/isolated)  & N2O2-KD02 &  -0.26 $\pm$ 0.01 & -0.023 $\pm$ 0.001\\
This Study & NGC 5194  (Sbc, unbarred/weakly inter.)  & N2O2-KD02 &  -0.09 $\pm$ 0.01 & -0.007 $\pm$ 0.001 \\
This Study & NGC 5713  (Sbc, barred/isolated)  & N2O2-KD02 & -0.34 $\pm$ 0.04 & -0.026 $\pm$ 0.003 \\
\\
This Study &  All seven galaxies & N2O2-KD02 & -0.34 $\pm$ 0.14 & -0.027 $\pm$ 0.011  \\
This Study &  4 unbarred &  N2O2-KD02 & -0.35 $\pm$ 0.19 & -0.028 $\pm$ 0.013  \\
This Study &  3 barred &  N2O2-KD02 & -0.32 $\pm$ 0.01 & -0.028 $\pm$ 0.008 \\
This Study &  3 unbarred/isolated &  N2O2-KD02 &  -0.44 $\pm$ 0.13 & -0.034 $\pm$ 0.008 \\
This Study &  2 barred/isolated &  N2O2-KD02 & -0.33 $\pm$ 0.01  &  -0.023 $\pm$ 0.003 \\
This Study &  1 unbarred/weakly interacting & N2O2-KD02 & -0.09 $\pm$ 0.01 & -0.007 $\pm$ 0.001 \\
This Study &  1 barred weakly interacting & N2O2-KD02 & -0.31 $\pm$ 0.02 & -0.038 $\pm$ 0.002 \\
\hline
\multicolumn{5}{c}{\it Our results for barred and unbarred spirals at $z\sim0$ using the $R_{23}$-KK04 \zgas{} diagnostic} \\
This Study &  All seven galaxies &  $R_{23}$-KK04 & -0.33 $\pm$ 0.26 & -0.026 $\pm$ 0.018 \\
This Study &  4 unbarred &  $R_{23}$-KK04 & -0.036 $\pm$ 0.28 & -0.027 $\pm$ 0.020  \\
This Study &  3 barred &  $R_{23}$-KK04 & -0.28 $\pm$ 0.21 & -0.024 $\pm$ 0.016 \\
This Study &  3 unbarred/isolated &  $R_{23}$-KK04 & -0.49 $\pm$ 0.21 & -0.037 $\pm$ 0.013 \\
This Study &  2 barred/isolated &  $R_{23}$-KK04 & -0.33 $\pm$ 0.25 & -0.024 $\pm$ 0.019  \\
This Study &  1 unbarred/weakly inter. & $R_{23}$-KK04 & 0.01 $\pm$ 0.01 &  0.001 $\pm$ 0.001 \\
This Study &  1 barred weakly inter. & $R_{23}$-KK04 &  -0.20 $\pm$ 0.03 & -0.025 $\pm$ 0.003 \\
\hline
\multicolumn{5}{c}{\it Our results for barred and unbarred spirals at $z\sim0$ using the O3N2-PP04 \zgas{} diagnostic} \\
This Study &  All seven galaxies & O3N2-PP04 & -0.11 $\pm$ 0.13 & -0.010 $\pm$ 0.012 \\
This Study &  4 unbarred &  O3N2-PP04 & -0.09 $\pm$ 0.15 &  -0.007 $\pm$ 0.011 \\
This Study &  3 barred &  O3N2-PP04 & -0.14 $\pm$ 0.08 & -0.014 $\pm$ 0.012 \\
This Study &  3 unbarred/isolated &  O3N2-PP04 & -0.18 $\pm$ 0.05 & -0.014 $\pm$ 0.003  \\
This Study &  2 barred/isolated &  O3N2-PP04 & -0.09 $\pm$ 0.01 & -0.006 $\pm$ 0.001  \\
This Study &  1 unbarred/weakly interacting & O3N2-PP04 & 0.16 $\pm$ 0.01 & 0.011 $\pm$ 0.001 \\
This Study &  1 barred weakly interacting & O3N2-PP04 & -0.25 $\pm$ 0.01 & -0.031 $\pm$ 0.001  \\
\hline
\multicolumn{5}{c}{\it Our results for barred and unbarred spirals at $z\sim0$ using the N2-PP04$^{c}$ \zgas{} diagnostic} \\
This Study &  All seven galaxies & N2-PP04 & -0.16 $\pm$ 0.19 & -0.012 $\pm$ 0.014  \\
This Study &  4 unbarred &  N2-PP04 & -0.04 $\pm$ 0.08 &  -0.003 $\pm$ 0.006 \\
This Study &  3 barred &  N2-PP04 & -0.31 $\pm$ 0.20 & -0.025 $\pm$ 0.013 \\
This Study &  3 unbarred/isolated &  N2-PP04 & -0.07 $\pm$ 0.08 & -0.005 $\pm$ 0.006  \\
This Study &  2 barred/isolated &  N2-PP04 & -0.33 $\pm$ 0.24 & -0.021 $\pm$ 0.014  \\
This Study &  1 unbarred/weakly interacting &N2-PP04 & 0.02 $\pm$ 0.01  & 0.001 $\pm$ 0.001 \\
This Study &  1 barred weakly interacting & N2-PP04 & -0.26 $\pm$ 0.01 & -0.033 $\pm$ 0.001 \\
\hline
\multicolumn{5}{c}{\it Published IFU studies of spirals at $z \sim 0$ using the O3N2-PP04 \zgas{} diagnostic} \\
\cite{sanchez2014} & 38 barred (RC3 type B) & O3N2-PP04  & -0.09 $\pm$ 0.07$^a$ & - \\
\cite{sanchez2014} & 30 barred (RC3 type AB) & O3N2-PP04  & -0.13 $\pm$ 0.09$^a$ & - \\
\cite{sanchez2014} & 78 unbarred (RC3 type A) & O3N2-PP04 & -0.12 $\pm$ 0.08$^a$ & - \\
\hline
\multicolumn{5}{c}{\it Published studies of isolated and strongly interacting spirals at $z \sim 0$} \\
\cite{rupke2010} & 16 strongly interacting & N2O2-KD02 & -0.25 $\pm$ 0.11 & -0.020 $\pm$ 0.009  \\
\cite{rupke2010} & 11 isolated &   N2O2-KD02 & -0.57 $\pm$ 0.18 & -0.051 $\pm$ 0.030 \\
\cite{kewley2010} & 8 strongly interacting & N2O2-KD02 & -0.25 $\pm$ 0.12 & -  \\
\hline
\multicolumn{5}{c}{\it Published studies at higher $z$ using the N2-PP04 zgas{} diagnostic.} \\
\cite{yuan2011} & Sp1149 ($z \sim 1.5$) & N2-PP04 & -1.92 $\pm$ 0.24 & -0.16 $\pm$ 0.02 \\
\cite{jones2010} & Clone Arc  ($z\sim 2$) & N2-PP04 & - & -0.27 $\pm$ 0.05  \\
\hline
\label{tab:zgas-gradients}
\end{tabular}
\end{center}
\raggedright
Columns:
(1) Reference for \zgas{} study.
(2) Name and/or classification (Hubble class, barred vs. unbarred, isolated vs. weakly or strongly interacting) galaxy or set of galaxies.
(3) \zgas{} diagnostic used for reported \zgas{} gradients.
(4) \zgas{} gradient in units of dex ($R$/\rtwofive{})$^{-1}$ or
dex/($R_e$)$^a$.  Reported gradients for single galaxies are the slope
from linear least squares fit to equal data points, and the reported
uncertainty is the 1-$\sigma$ uncertainty in the slope's fit.
Reported gradients for groups of galaxies are the mean gradient of the group and the uncertainty is the 1-$\sigma$ scatter about the mean.
(5) \zgas{} gradient reported in the same manner as column 4, but in units of dex kpc$^{-1}$. \\
$^{a}$ We give dex ($R$/\rtwofive{})$^{-1}$  for all studies except for \cite{sanchez2014} where we give  dex/($R_e$)$^{-1}$ \\
$^{b}$ Note that NGC 1068 has been excluded from \zgas{} analysis since most of the gas is excited by the central AGN.  See $\S$~\ref{sec:results-components} and Figure \ref{fig:bpt-sliced}.\\
$^{c}$ Note that we consider the N2 \zgas{} diagnostics less reliable than the other \zgas{} diagnostics, as discussed in $\S$ \ref{sec:results-zgas}.  Since N2 diagnostics are frequently used in high redshift \zgas{} studies as discussed in $\S$ \ref{sec:results-redshift}, we report N2-PP04 \zgas{} gradients here for our VENGA sub-sample.
\end{table*}

\subsection{Comparison of Absolute Values of \zgas{} from Different  Diagnostics} 
\label{sec:results-zgas}

Currently, one long-standing major issue in studies of gas-phase  metallicity
is that  there are systematic offsets between  {\it absolute} values of  
\zgas{} given  by  the different diagnostics. 
Attempts have been made to reconcile the differences between the diagnostics
 (for example see 
\citealt{kewley2002, kewley2008, sanchez2010, blanc2015}),
but the issue is far from settled.
Using our  2D maps (Figure \ref{fig:2d-maps})
 and radial gradients (Figure \ref{fig:zgas-gradients})  of the 
the seven \zgas{} diagnostics
 ($R_{23}$-KK04, $R_{23}$-M91, $R_{23}$-Z94, N2O2-KD02,  O3N2-PP04,
 N2-D02, and N2-PP04), 
we explore how
the  absolute value of \zgas{}  varies between
some of the
different \zgas{} diagnostics:

\begin{enumerate}
\item 
{\it {$R_{23}$ \zgas{}  diagnostics:}}
The three $R_{23}$ \zgas{}  diagnostics ($R_{23}$-KK04, 
$R_{23}$-M91, $R_{23}$-Z94)  
show a similar shape in the radial profile of \zgas{} and  yield
absolute values that agree  within 0.1 to 0.2 dex.
The fact that these three \zgas{}  diagnostics agree  within  0.1-0.2 
dex although they do not all  use $q$  to estimate  \zgas{}  
(see Appendices \ref{sec:q2-append} \& \ref{sec:zgas-append}) 
is likely due to the fact that we are on the upper branch of $R_{23}$
where \zgas{}  is high  ($> 8.87$)  and there is only a small dependence on $q$
(see top left of Figure \ref{fig:degeneracy}).

\item 
{\it { The N2O2-KD02 \zgas{} diagnostic:}}
The N2O2-KD02 diagnostic shows a low amount of scatter when
compared to the $R_{23}$ and N2 diagnostics.
N2O2-KD02  gives \zgas{}  values similar to the three main $R_{23}$
diagnostics ($R_{23}$-KK04, $R_{23}$-M91, \& $R_{23}$-Z94).
We note that N2O2-KD02  is a theoretical calibration that uses the
\iontwo{N}{II}/\iontwo{O}{II} line  ratio,  which  has little  dependence on $q$ 
(Figure \ref{fig:degeneracy}).

\item 
{\it {The N2-D02 and N2-PP04 \zgas{} diagnostics:}} 
The  N2-D02 and N2-PP04 \zgas{} diagnostics use the \iontwo{N}{II}/H$\alpha$ line ratio 
and are based on empirical calibrations that do not correct for local variations 
in  $q$.
{\it{N2-PP04  gives  a \zgas{}   that tends to be systematically    lower,  by as 
much as 0.4 dex, compared  to the three main $R_{23}$ 
diagnostics ($R_{23}$-KK04, $R_{23}$-M91,  and $R_{23}$-Z94).}} 
The systematically lower \zgas{}  values given by  the empirically-calibrated
N2-D02 and N2-PP04 diagnostics,  compared to the theoretically-calibrated
$R_{23}$ diagnostics,  is likely related to the temperature gradients and variations
in the nebulae used as empirical calibrators \citep{kewley2008}.
Furthermore, in the inner 2 kpc of some of our galaxies,
the radial gradients in  \zgas{}  given by N2-D02 and N2-PP04 
can {\it{be opposite to  those}} given by other diagnostics.
For instance, in NGC 2903, the  first five \zgas{} diagnostics give 
a slightly negative to flat  \zgas{}  gradient in the inner 1.5 kpc radius,
but N2-KD02 and N2-PP04 give a slightly positive gradient, which then
inverts  (see Figure \ref{fig:zgas-gradients}).  
Similar opposite trends were seen in the  2D maps of \zgas{} 
(see $\S$~\ref {subsec:q}).  
A milder example can be seen in the 
form of a negative central gradient in NGC 4254.

We have explored possible explanations for  the large difference 
between  the N2-D02 and N2-PP04 diagnostics  and the  $R_{23}$ 
diagnostics. 
At typical $q$ values below $10^8$  cm s$^{-1}$,  
high \iontwo{N}{II}/H$\alpha$ ($> 0.1$) values 
allow for two possible values of \zgas{}, which are
typically above and below 9.1 dex (see  upper right panel of 
Figure \ref{fig:degeneracy}).  Therefore, one possible reason for erroneous
\zgas{}  is that the  N2-D02 and N2-PP04 diagnostics are  picking 
the wrong value of \zgas{}. 
A second possibility is that erroneous \zgas{}  values are caused  
by  the  fact that the N2-KD02 and N2-PP04 empirical 
calibrations do not take into account variations in  $q$ when  
computing \zgas{}.

To evaluate the impact of $q$ variations, 
we attempted to calculate  \zgas{}  by interpolating over
the theoretical \iontwo{N}{II}/H$\alpha$ curves from \cite{kewley2002}
which use the  \iontwo{N}{II}/H$\alpha$ ratio and corrects explicitly for
local variations in $q$. For this exercise, we used the $q$ we calculated
based  on the \iontwo{O}{III}/\iontwo{O}{II} line ratio ($\S$ \ref{sec:how-q}). 
We found that the shape and absolute value of the radial profiles did not change significantly
from the empirical N2-D02 and N2-PP04 diagnostics, 
suggesting that variations in $q$ are not at the root of the problem.

\item
{\it { The O3N2-PP04 \zgas{} diagnostic:}}
The O3N2-PP04 diagnostic shows a low amount of scatter  when compared
to the $R_{23}$  diagnostics, and  yields  a \zgas{}  value that is systematically
lower by  0.2  to 0.3 dex compared  to  the three main $R_{23}$ diagnostics.
The fact that the empirically-calibrated diagnostics,  O3N2-PP04, as well as
N2-D02 and N2-PP04 (discussed in point (iii)),  give systematically lower
\zgas{}  than the theoretically-calibrated $R_{23}$ diagnostics,  suggests
that   temperature gradients and variations  in the nebulae used as
empirical calibrators \citep{kewley2008} may be important.
A related factor is that the  O3N2-PP04 diagnostic  does
not correct for local variations in $q$. This is particularly problematic
because our procedure of  using a fixed H$\beta$/H$\alpha$ value
for extinction correction means that  the O3N2-PP04 diagnostic
effectively depends directly on the \iontwo{O}{III}/\iontwo{N}{II} line ratio, which is shown
by \cite{kewley2002} to be strongly  dependent on $q$.

\end{enumerate}

\subsection{Comparison of \zgas{} Between Barred and Unbarred  Galaxies} 
\label{sec:results-bars}

\subsubsection{Our Results}
\label{subsubsec:ourzgas} 

Column 7 of Table
\ref{tab:venga-subsample} indicates  whether our sample galaxies 
host a large-scale stellar bar. In making this assessment, 
we do not rely solely on the RC3 classification as these are based on visual
inspection of optical photographic plates \citep{devaucouleurs1991} and may suffer
from obscuration by dust and SF.
Instead, we also use quantitative analyses of near-infrared images 
(Table~\ref{tab:bars}),  such as
ellipse fits \citep{marinova2007, md07},   
multi-component  bulge-bar-disc decomposition \citep{weinzirl}, 
 and Fourier decomposition \citep{laurikainen02}.
For all sample galaxies, the NIR-based methodologies agree with the RC3
classification on the presence/absence of a large-scale stellar bar.
We classify NGC 1068 \& 5194  as unbarred  because they 
do not host large-scale stellar bars and show evidence of only nuclear
bars with respective $a_{\rm bar} /R_{25} =$  0.07 \& 0.06, 
\citep{md07},   where $a_{\rm bar}$ denotes the bar  semi-major axis. 
 Large-scale stellar bars have  typical  
$a_{\rm bar}/R_{25}$ ranging from   0.1 to 0.5 \citep{erwin2005, marinova2007},   
while nuclear bars have $a_{\rm bar}/R_{25} < 0.1$
\citep{laine2002}, and are thus are too short to significantly impact the outer disc.    
The final division of our  sample  yields
two barred isolated galaxies (NGC~2903 \& 5713), three unbarred
isolated galaxies  (NGC 0628, 3938, \& 4254), 
one barred weakly interacting galaxy (NGC~0337\footnote{There is some
  evidence that NGC 0337  might be undergoing an interaction or minor
  merger.  See notes for column 8 in Table
  \ref{tab:venga-subsample}.}), and one unbarred weakly interacting
galaxy (NGC 5194).

\begin{table*}
\caption{Bar Identification}
\begin{center}
\begin{tabular}{llll}
\hline \hline\label{tab:bars} 
NGC & RC3 Bar Identification & NIR Large-Scale Bar Identification & NIR References \\
(1) & (2) & (3) & (4) \\
\hline
0337              & SB   & Barred      & MD07 \\
0628              & SA   & Unbarred  & MD07 \\
1068              & SA   & Unbarred$^a$      & MD07, LS02 \\
2903              & SAB & Barred       & LS02 \\
3938              & SA    & Unbarred  &  MD07, MJ07, W09 \\
4254              & SA    & Unbarred  &   MD07, MJ07, W09 \\
5194 (M51a) & SA     &Unbarred$^a$  & MD07 \\
5713              & SAB   & Barred     & MD07, MJ07, LS02\\
\hline\hline
\end{tabular}
\end{center}
\raggedright
Columns: (1)~NGC galaxy name. 
(2)~Bar identification in RC3 based on qualitative examination of galaxies imaged with photographic plates at optical wavelengths \citep{devaucouleurs1991}. SA = Unbarred; SAB = Intermediate bar strength; SB = Barred.  See $\S$~\ref{sec:results-bars} for more details.
(3)~Large-scale bar identification based on quantitative analysis of near-infrared (J,H, or K band) images.  We use this classification to definitively identify galaxies with and without large-scale stellar bars, since the qualitative classification from photographic plates at optical wavelengths can be significantly affected by dust obscuration, ongoing star formation, and the subjective judgement of the classifier.   See column (4) for references and $\S$~\ref{sec:results-bars} for more details.
(4)~The abbreviated references for the bar identification in column (3) refer to the following papers and methodologies: MD07 = \citep{md07} ellipse fitting to 2MASS JHK-band images; MJ07 = \citep{marinova2007} ellipse fitting to OSU H-band images taking into account projection effects; W09 = \citep{weinzirl} multi-component bulge-bar-disc decomposition of OSU H-band images; LS02 = \citep{laurikainen02} Fourier decomposition of 2MASS JHK-band images.
$^{a}$~NGC 5194 (M51a) and NGC 1068 do not appear to host large-scale stellar bars, but do host nuclear bars. 
The small size of nuclear bars compared to the total size of the disc in a galaxy limits their ability to radially redistribute gas, so we classify these galaxies as  ``Unbarred.'' 
These nuclear bar structures are confirmed in both galaxies by \cite{md07}, where $a_{\rm bar}/R_{25} < 0.1$.  
\end{table*}

\begin{figure}
\includegraphics[angle=0,width=0.51\textwidth]{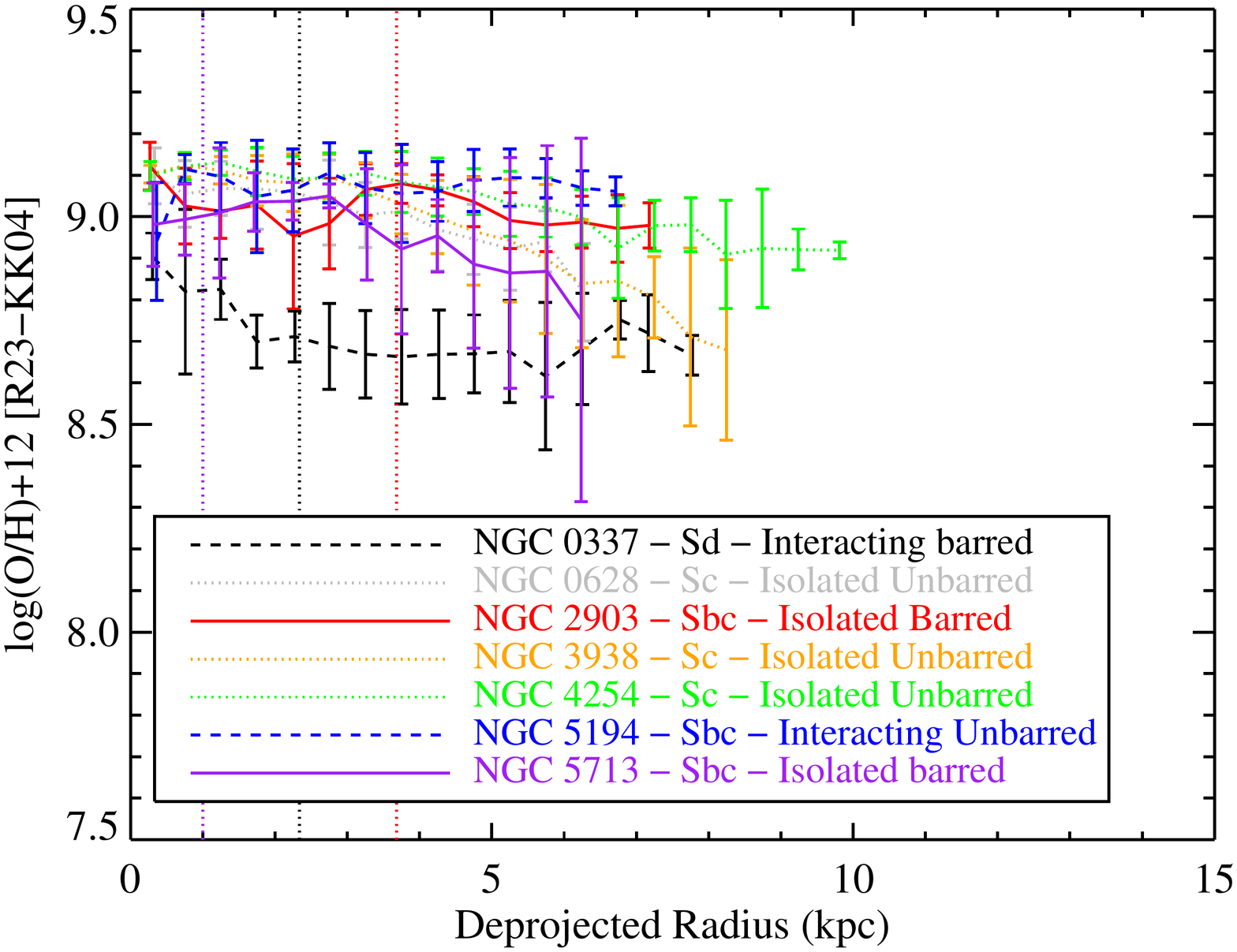}\hspace{-0.8cm}
\includegraphics[angle=0,width=0.51\textwidth]{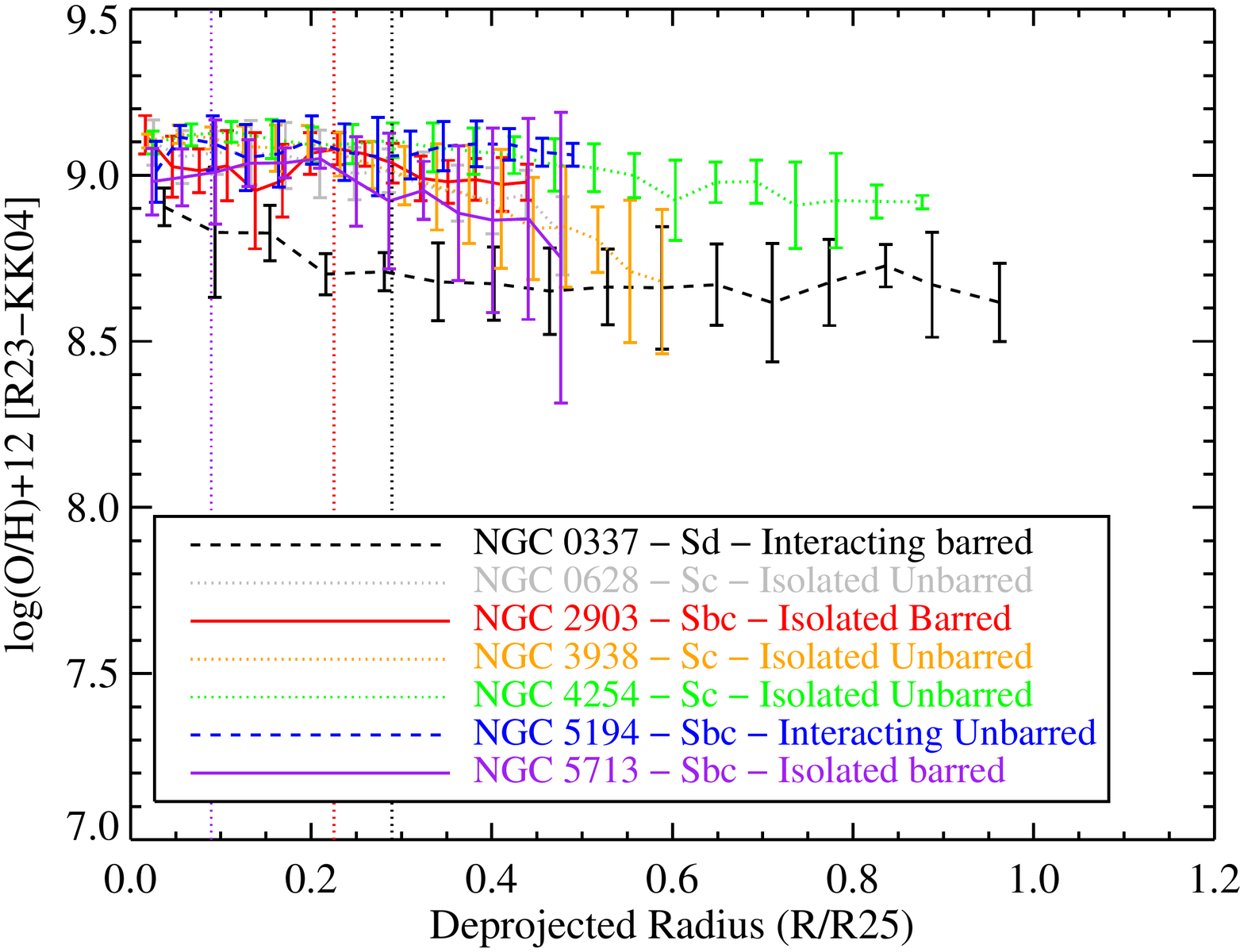}
\caption{
Comparison of  deprojected \zgas{} profiles using the 
$R_{\rm 23}$-KK04 diagnostic for two barred isolated galaxies (NGC~2903 \& NGC~5713; solid line), three unbarred isolated galaxies 
 (NGC 0628, NGC~3938, \& NGC~4254;  dotted lines),  one barred weakly interacting galaxy (NGC 0337; dashed line), and one unbarred weakly interacting galaxy 
(NGC 5194; dashed line).   The horizontal axis shows the deprojected radius in units of kpc {\bf (top plot)}  
and $R/R_{\rm 25}$  {\bf (bottom plot)}. 
The points and error bars are the mean and 1$\sigma$ dispersion of 0.5 kpc 
bins. In the case of barred galaxies, the dotted vertical line indicates the radius 
where the bar ends. 
For barred galaxies, the dotted vertical line indicates the radius 
where the bar ends. 
As expected from the mass-metallicity relation, the  lowest \zgas{}
values are in  NGC~0337, which  has the lowest stellar mass and latest Hubble type.  
The most striking result is that  the barred and unbarred galaxies 
show similarly-shaped \zgas{} profiles,  which are flat  or  show very small negative 
gradients  from the inner kpc out  to large radii (7-10 kpc or 
0.5-1.0 $R_{\rm 25}$) in the outer disc.
} 
\label{fig:zgas-comp-r23kk04}
\end{figure}

Figure~\ref{fig:zgas-comp-r23kk04} over-plots the \zgas{} radial profiles from
the $R_{23}$-KK04 diagnostic.  
In terms of the absolute value of \zgas{},  the lowest mean \zgas{} 
occur in  the galaxy NGC~0337, which has the latest Hubble type (Sd) 
and the lowest stellar mass ($1.6 \times 10^{10}$ M$_\odot$). 
The other six galaxies, which have  higher stellar  masses and later  
Hubble types, show  \zgas{}  values  higher than those  in NGC~0337,  
by $\sim$ 0.1 to 0.4 dex,  depending on the radius being  considered.  
These differences in the value of \zgas{}  are expected from 
the mass-metallicity relation
(e.g. \citealt{tremonti2004, lee2006, zhao2010}).

The most interesting result, however, lies in the similar shape
of the \zgas{} profile shown by barred and unbarred galaxies.
Table \ref{tab:zgas-gradients} shows the \zgas{} gradients for all our
sample galaxies, differentiating between 
isolated barred/unbarred spirals  versus those that may be
weakly interacting barred/unbarred spirals. This is important
in order to try to disentangle the effect of the bar from that of an ongoing
interaction. 
Table \ref{tab:zgas-gradients}
also lists  the gradients found in other \zgas{} studies.
For the individual galaxies, we report the \zgas{} gradients as the
slope of a linear least squares fit to the equally weighted data
points along with the 1-$\sigma$ uncertainty of the slope in the fit. 
For groups of  galaxies,  we
report the mean \zgas{} gradient and 1-$\sigma$ scatter about the mean.  While our high spatial resolution and radial profiles in Figures \ref{fig:zgas-gradients} \& \ref{fig:zgas-comp-r23kk04} shows that the \zgas{} radial profiles in our galaxies have fine structure and are probably not best fit with a single line, we use a linear fit to allow a statistical comparison of the \zgas{} gradients in our galaxy sample to other studies.

{\it{ Isolated barred and unbarred galaxies in our sample exhibit  similarly-shaped 
\zgas{} profiles,  which are flat  or  show very small negative 
gradients  from the inner kpc out  to large radii (7-10 kpc or 
0.5-1.0 $R_{\rm 25}$) in the outer disc.}}
Specifically, the isolated barred galaxies (NGC 2903 \& NGC 5713)  show very shallow 
N202-KD02  \zgas{} gradients
 of $-0.33 \pm 0.01$ in units of  dex ($R$/\rtwofive{})$^{-1}$ and -0.023/ $\pm$ 0.003 in units of dex kpc$^{-1}$.
The three  isolated unbarred galaxies (NGC 0628, 3938, \& 4254) also show 
similarly shallow \zgas{} gradients  of
-0.44 $\pm$ 0.13  in units of  dex ($R$/\rtwofive{})$^{-1}$ and -0.034
$\pm$ 0.008 in units of dex kpc$^{-1}$.
There is no statistically significant difference between the \zgas{}
gradients reported for isolated barred vs. unbarred spirals
in our sample.
The values cited above are for the N2O2-KD02 \zgas{} diagnostic, but 
the other \zgas{} diagnostics give similarly shallow negative
gradients as seen in Table \ref{tab:zgas-gradients} and Figure \ref{fig:zgas-gradients}.

Some studies (e.g., see \citealt{sanchez2012, sanchez2014})  caution 
that differences in \zgas{} profiles may arise due to different units 
used for radial distance  (kpc vs. the disc half-light radius $R_e$). 
It is therefore relevant to note that  our \zgas{} profiles 
show no significant difference between barred and unbarred galaxies 
even when we use different radial units, such as the radius in 
absolute units of kpc  (top panel of Figure~\ref{fig:zgas-comp-r23kk04}) 
or the radius  scaled in terms of half-light radius  $R_e$  or 
\rtwofive{}  (e.g.,  bottom panel of Figure~\ref{fig:zgas-comp-r23kk04})

\begin{landscape}
\begin{figure}
\includegraphics[angle=0,height=0.28\textheight]{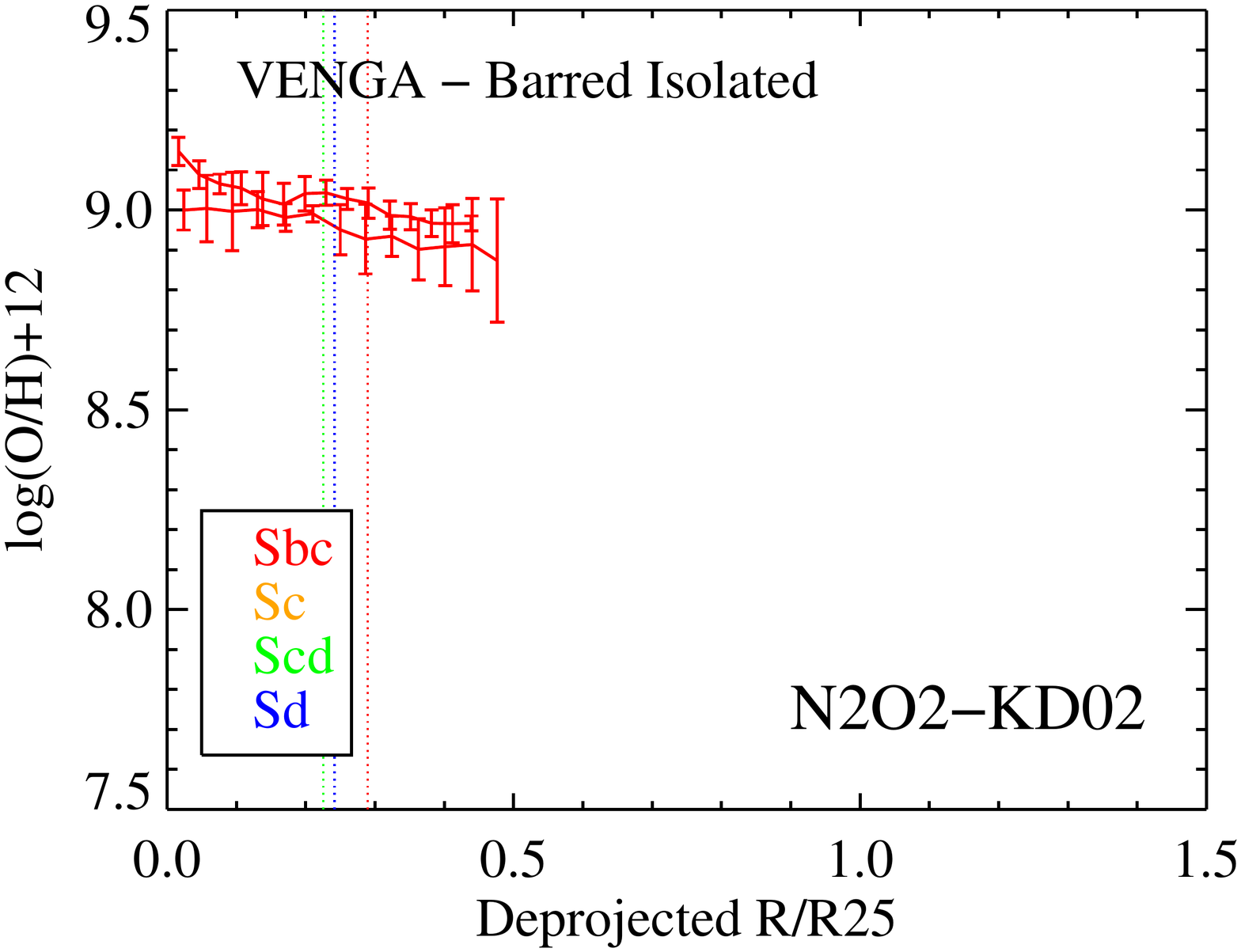}\hspace{-0.8cm}
\includegraphics[angle=0,height=0.28\textheight]{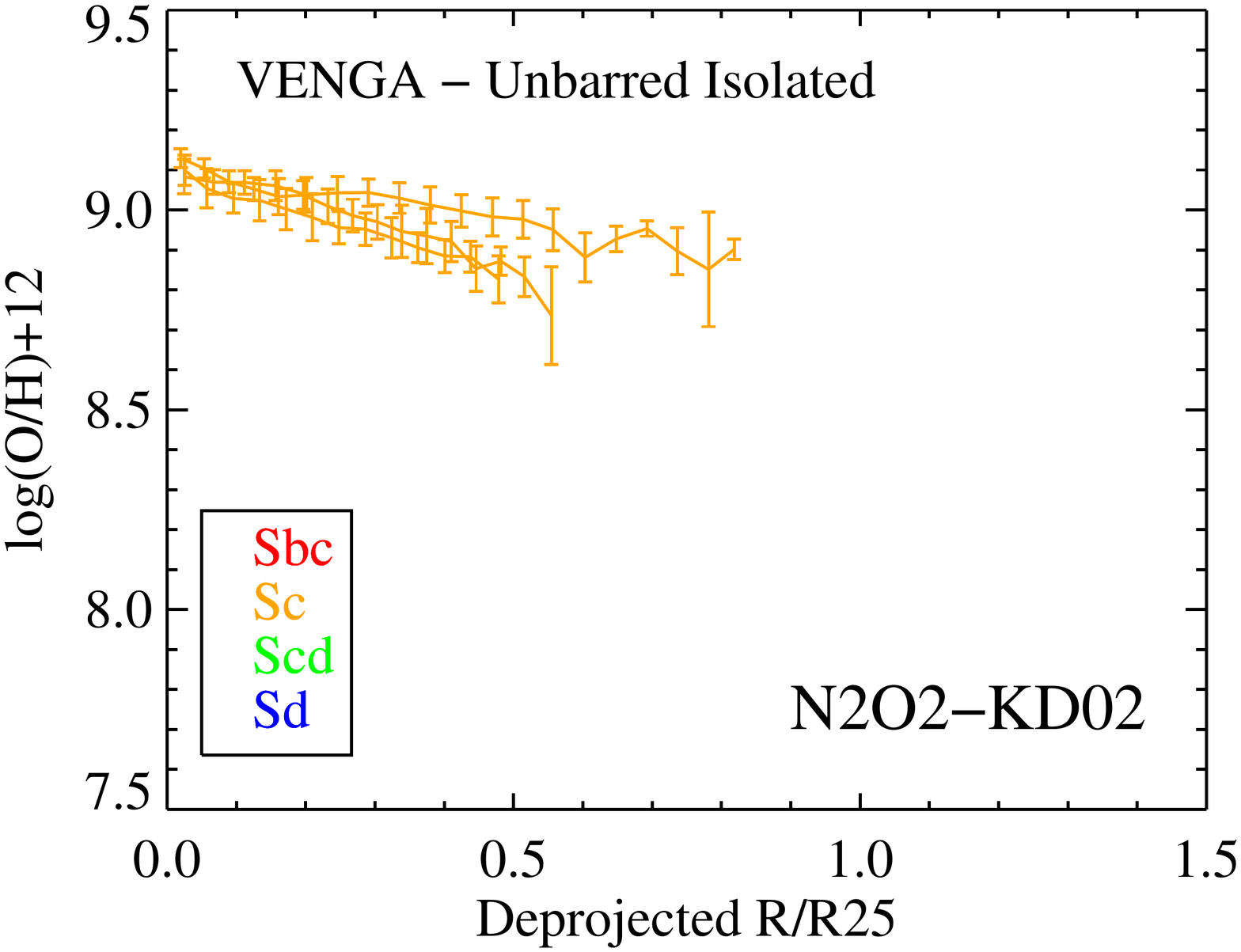}\hspace{-0.8cm}
\includegraphics[angle=0,height=0.28\textheight]{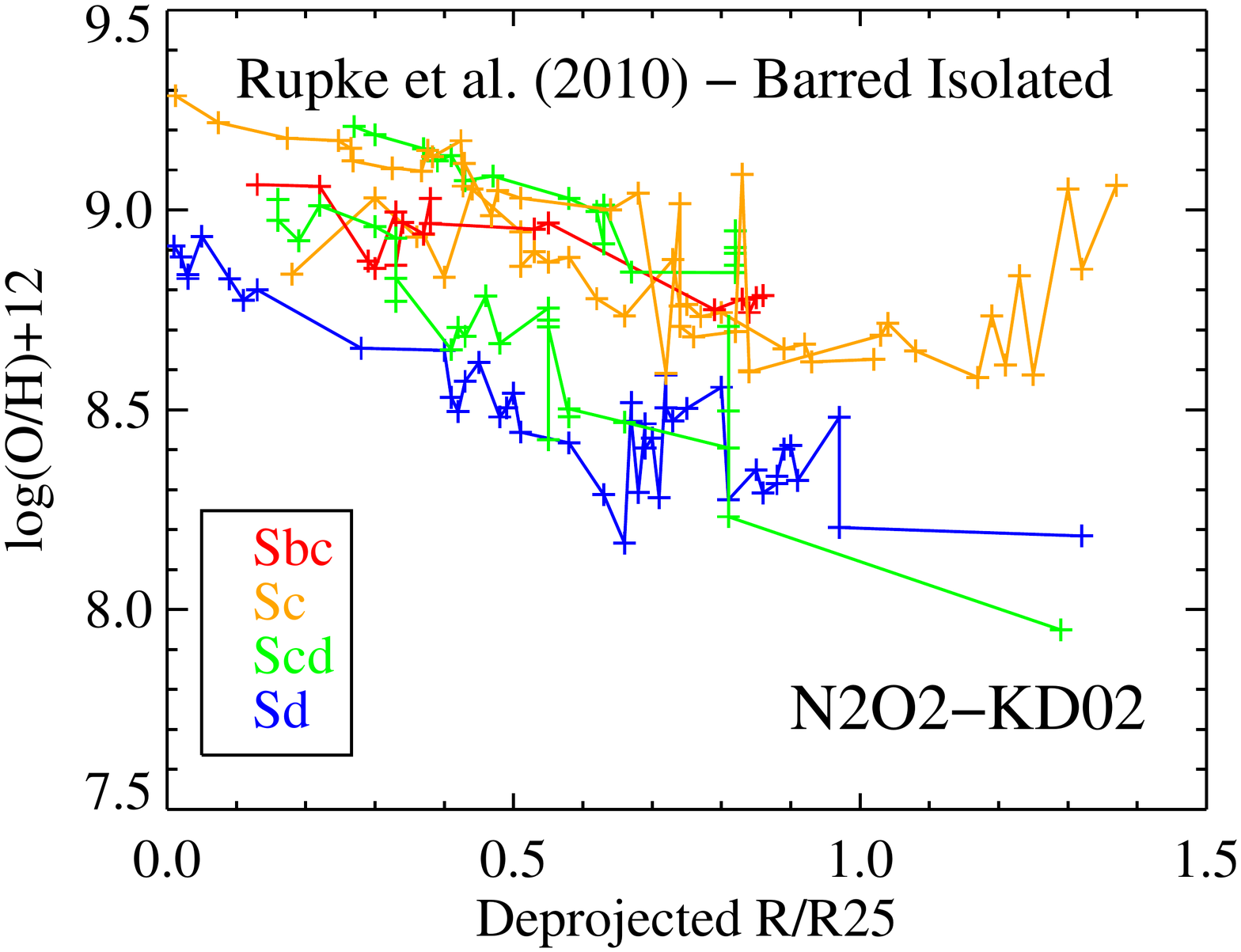} \hspace{-0.8cm}
\includegraphics[angle=0,height=0.28\textheight]{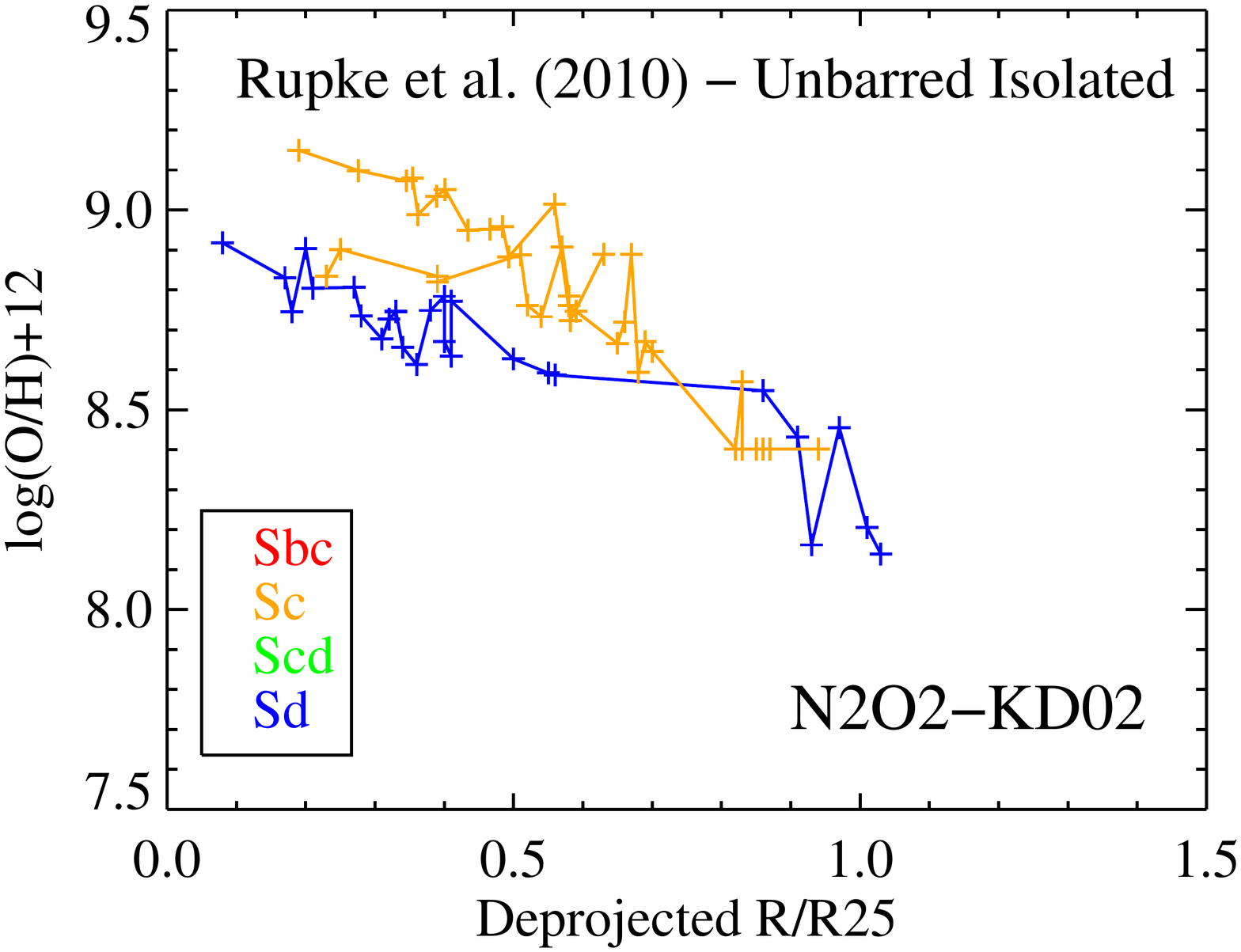}  \hspace{-0.8cm} \\
\includegraphics[angle=0,height=0.28\textheight]{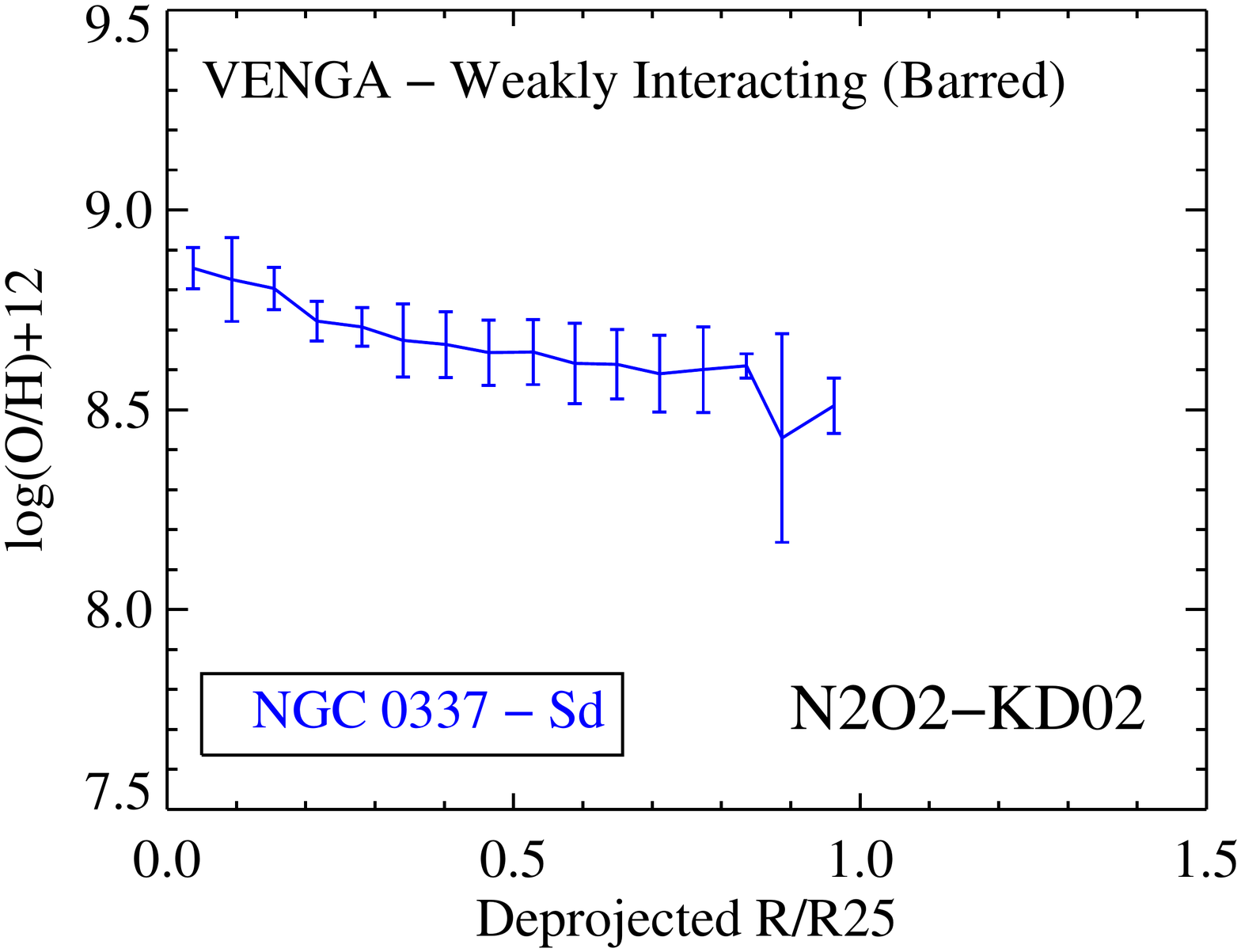} \hspace{-0.8cm}
\includegraphics[angle=0,height=0.28\textheight]{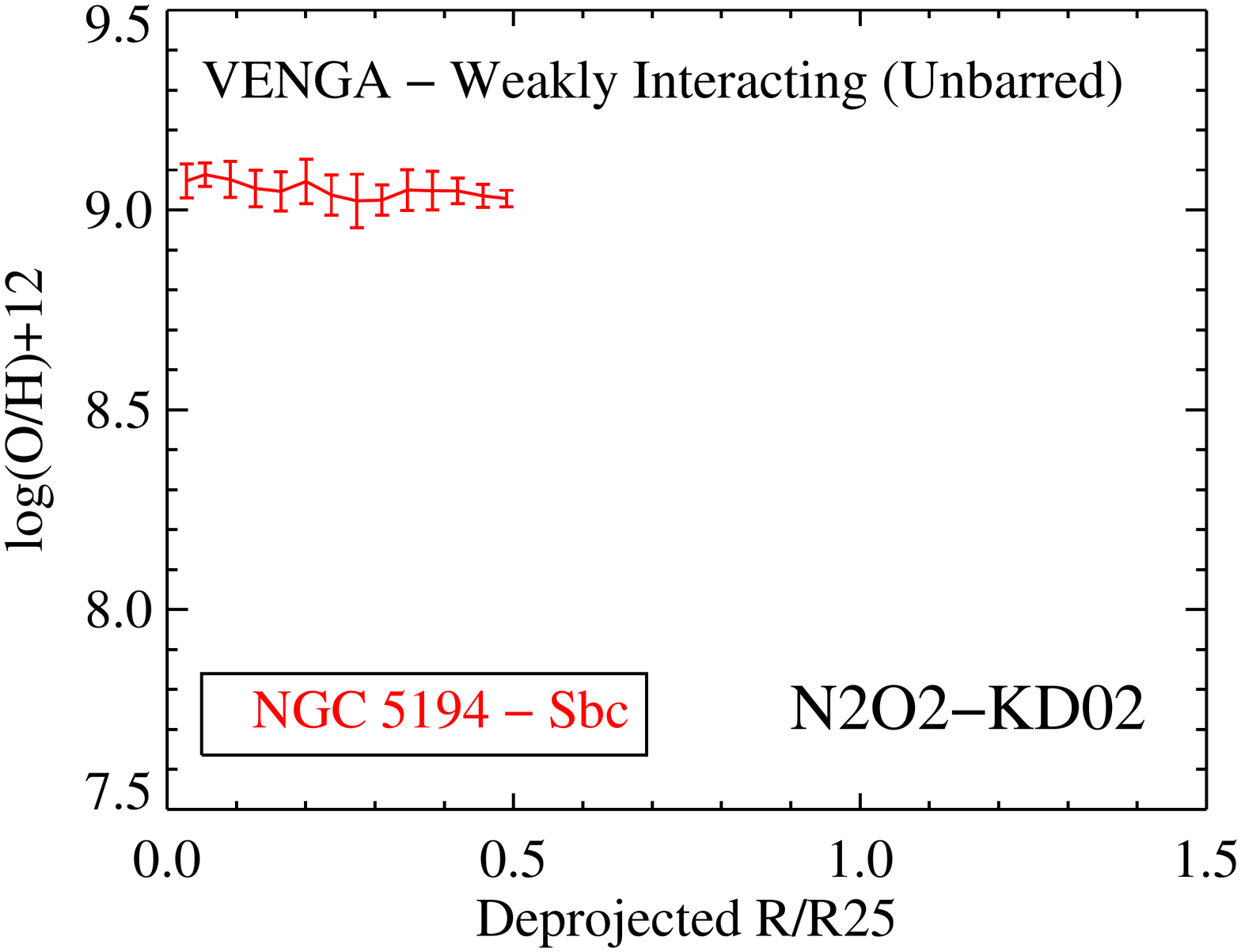} \hspace{-0.8cm}
\includegraphics[angle=0,height=0.28\textheight]{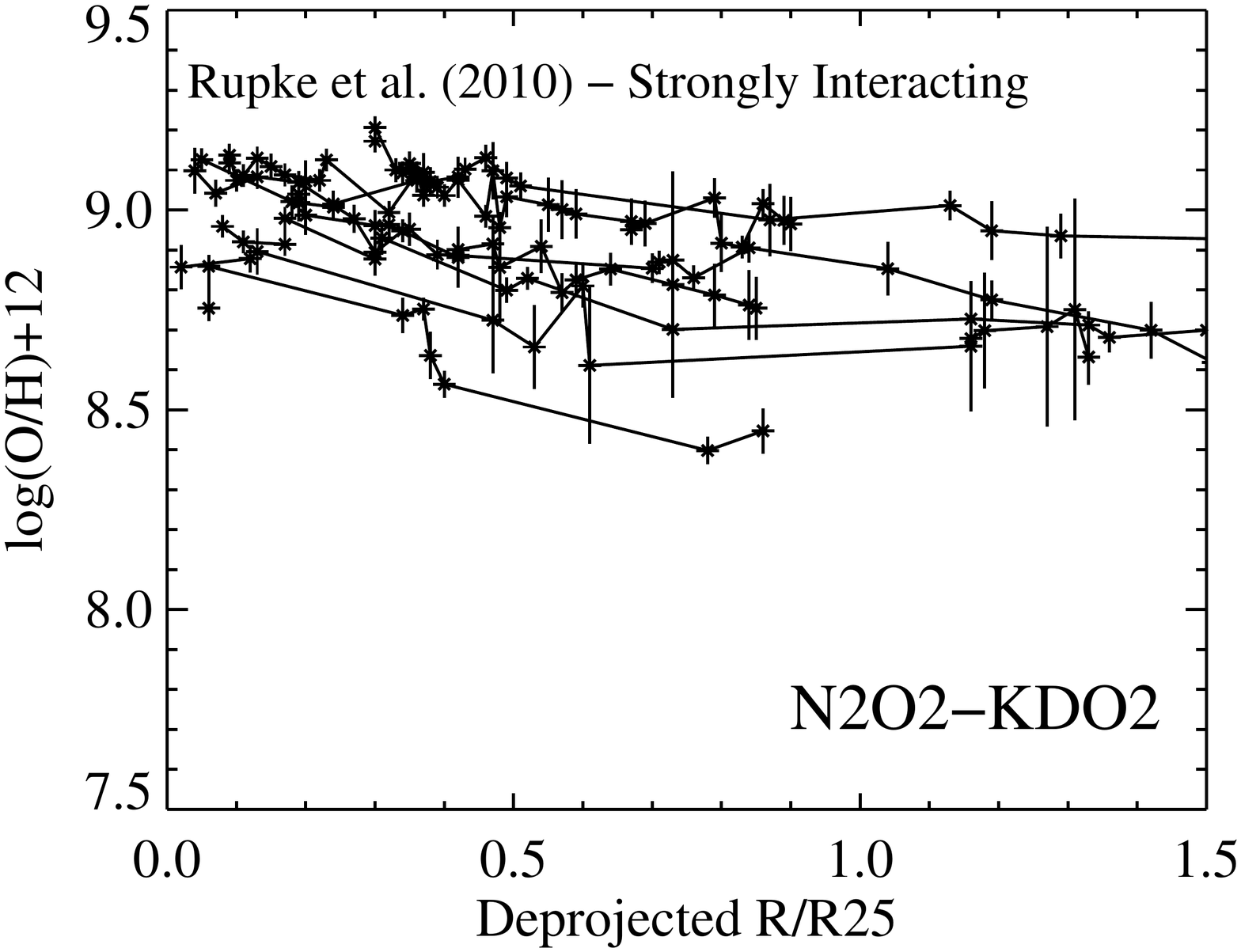}  \hspace{-0.8cm}
\includegraphics[angle=0,height=0.28\textheight]{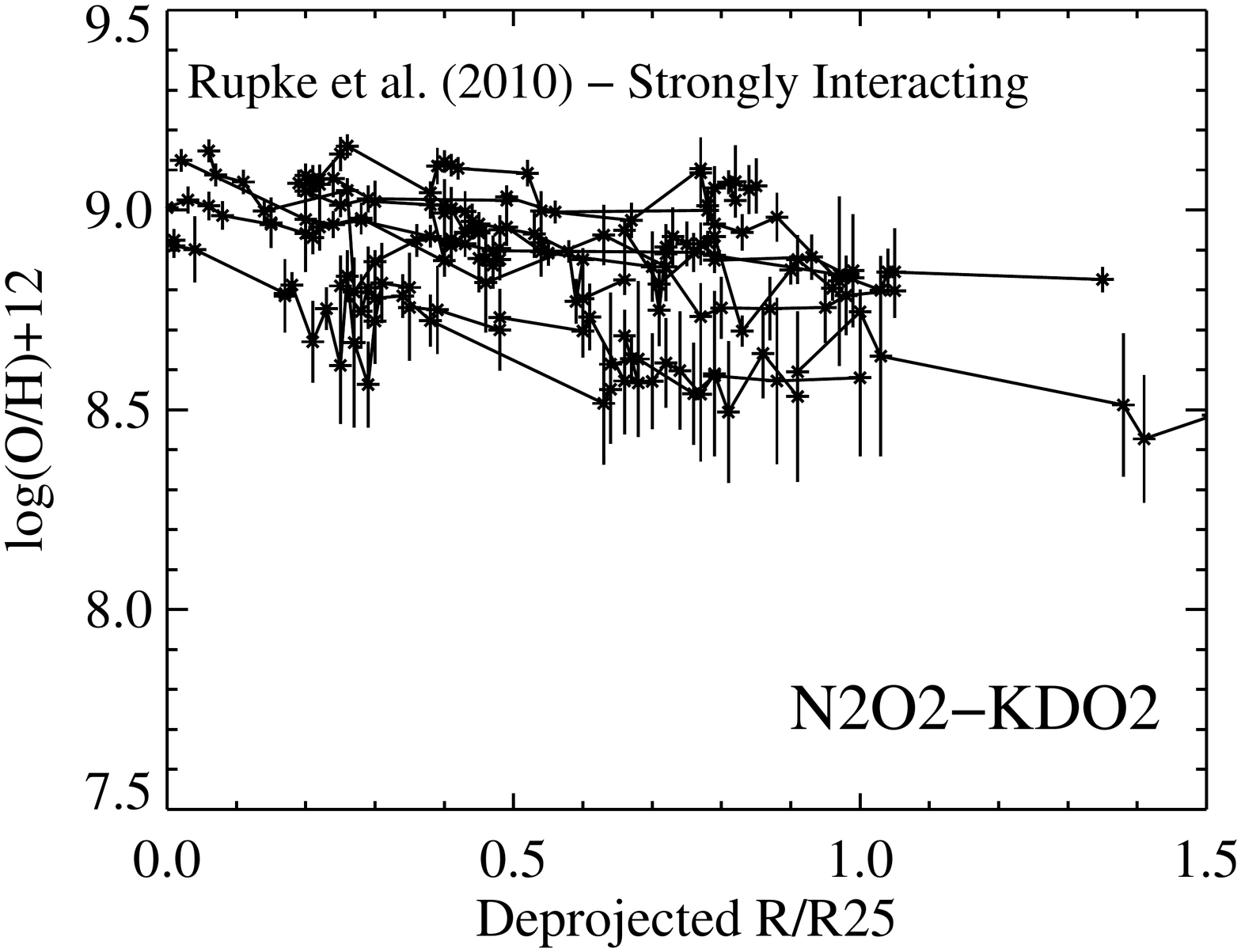}  \hspace{-0.8cm}
\caption{
Radial deprojected \zgas{} profiles based on the  N202-KD02 diagnostic
are shown, with the horizontal axis being the deprojected radius  in units of
$R/R_{\rm 25}$.  
The  {\bf{four left}} panels show  our study's  \zgas{} profiles  for barred and
unbarred galaxies,  divided between isolated  ({\bf{top row}})  and interacting 
 ({\bf{bottom row}}) systems.  The interacting systems in our study are consistent
with minor interactions/mergers with mass ratios below 1:3. 
The  {\bf{four right}} panels show \zgas{} profiles based on 
multi-slit spectroscopic data from  \protect\cite{rupke2010},  
divided between isolated  ({\bf{top row}})  and strongly interacting 
 ({\bf{bottom row}}) systems. 
These  strongly interacting systems  represents major 
interactions/mergers as they have mass ratios close to unity (1:1 to
1:3).
It is striking that the VENGA-based radial \zgas{} 
profiles are  of much superior quality  in terms of spatial 
sampling and scatter than the profiles based on slit-spectroscopy.
}
\label{fig:venga-rupke-n2o2}
\end{figure}
\end{landscape}

\begin{landscape}
\begin{figure}
\includegraphics[angle=0,height=0.28\textheight]{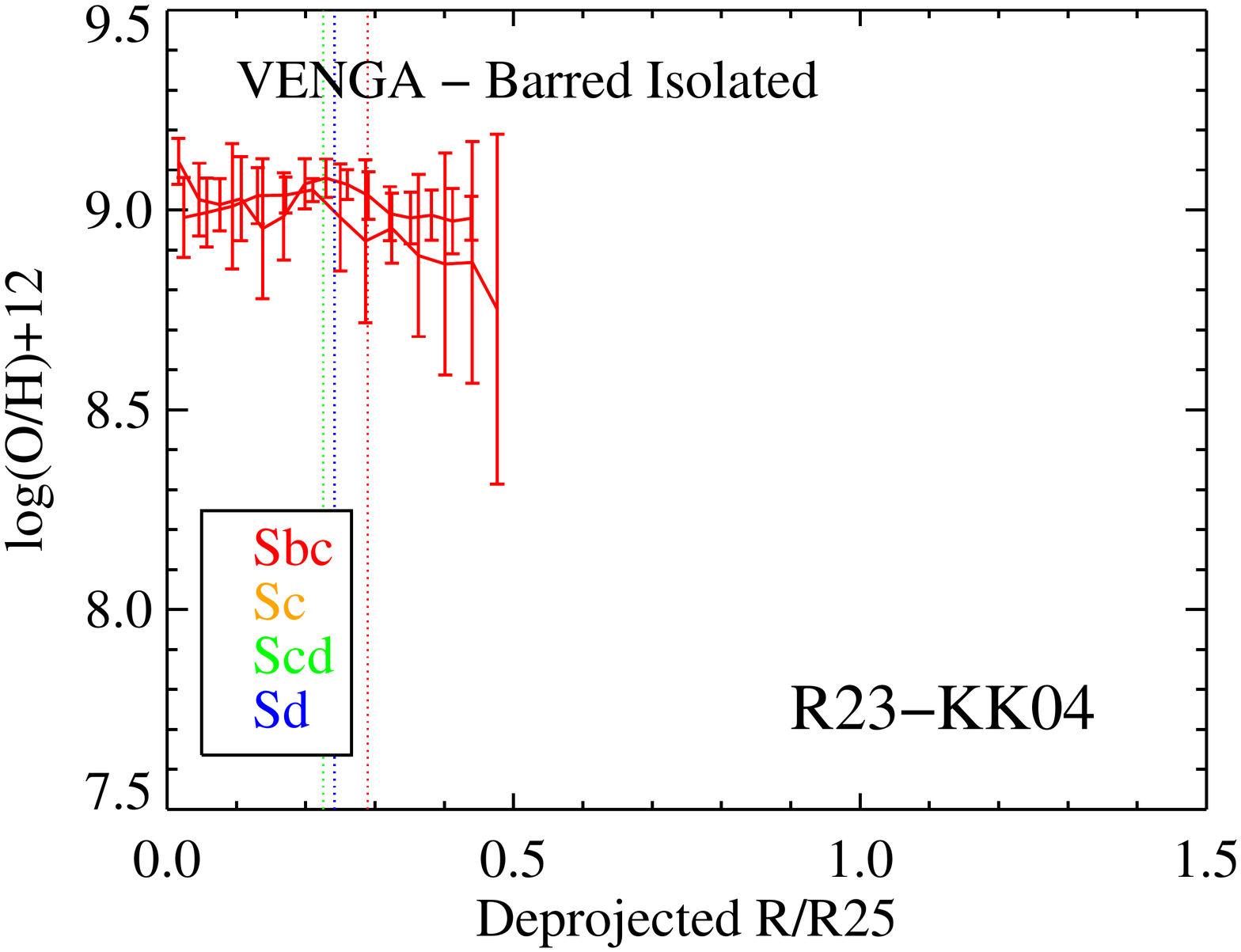}\hspace{-0.8cm}
\includegraphics[angle=0,height=0.28\textheight]{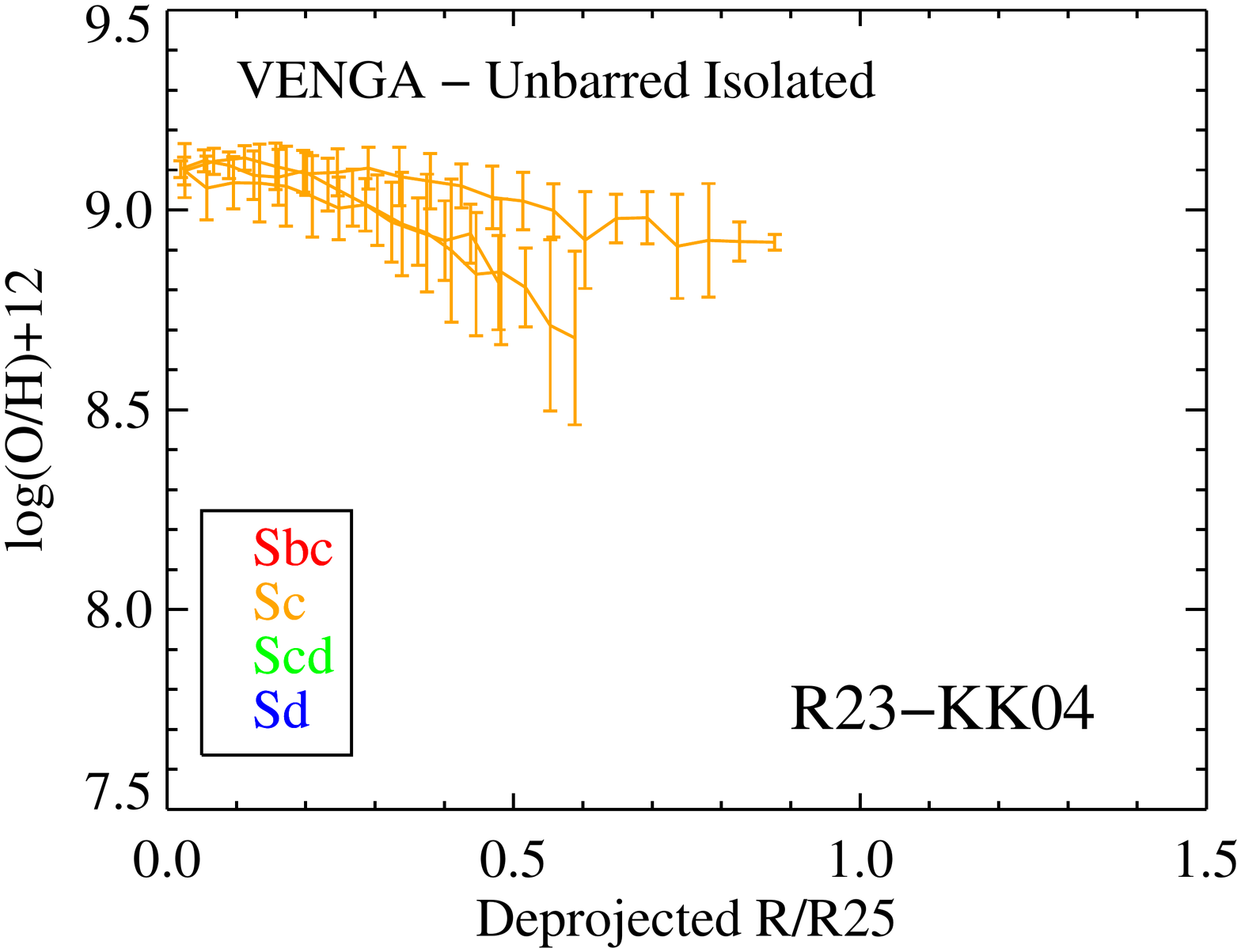}\hspace{-0.8cm}
\includegraphics[angle=0,height=0.28\textheight]{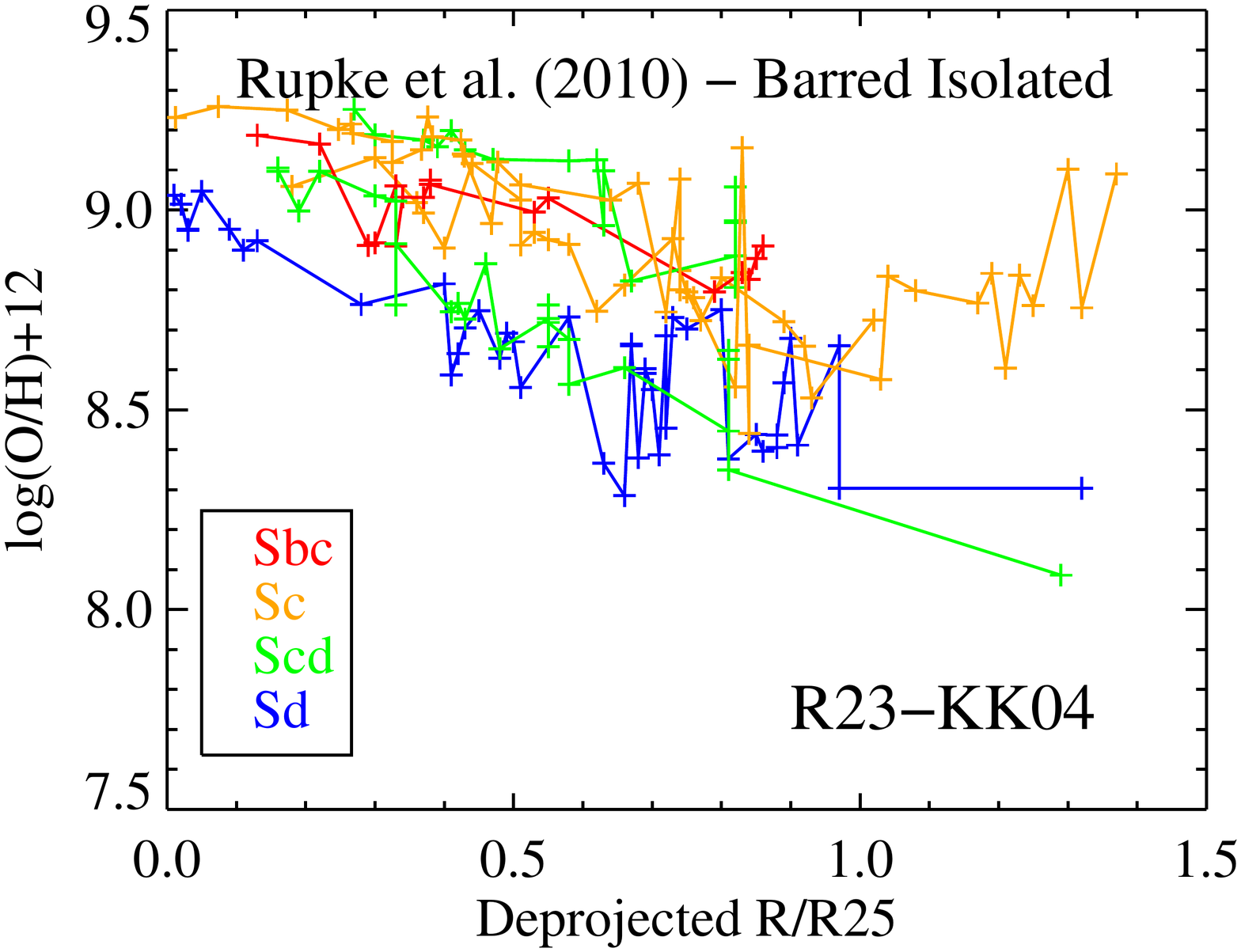}\hspace{-0.8cm}
\includegraphics[angle=0,height=0.28\textheight]{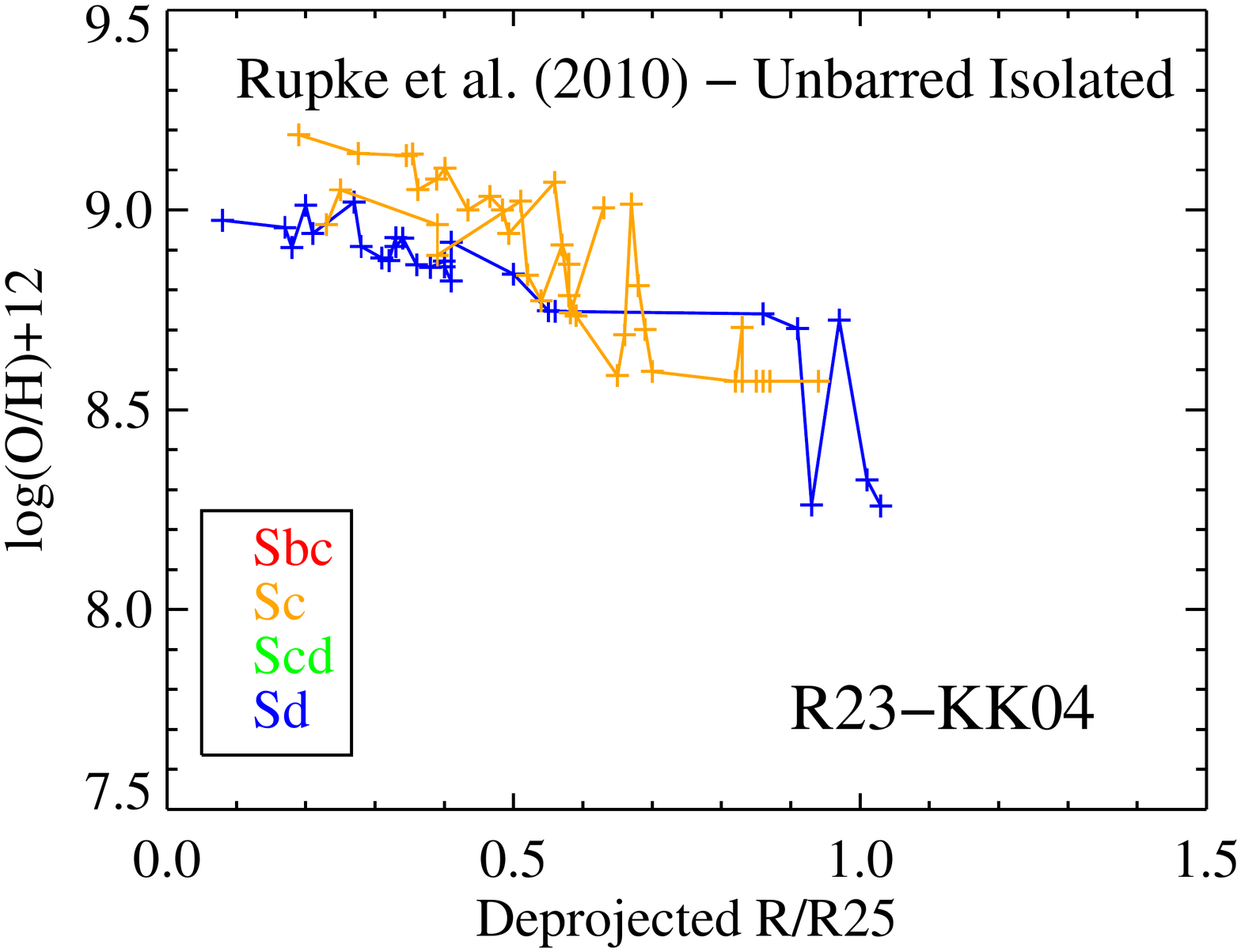} \hspace{-0.8cm} \\
\includegraphics[angle=0,height=0.28\textheight]{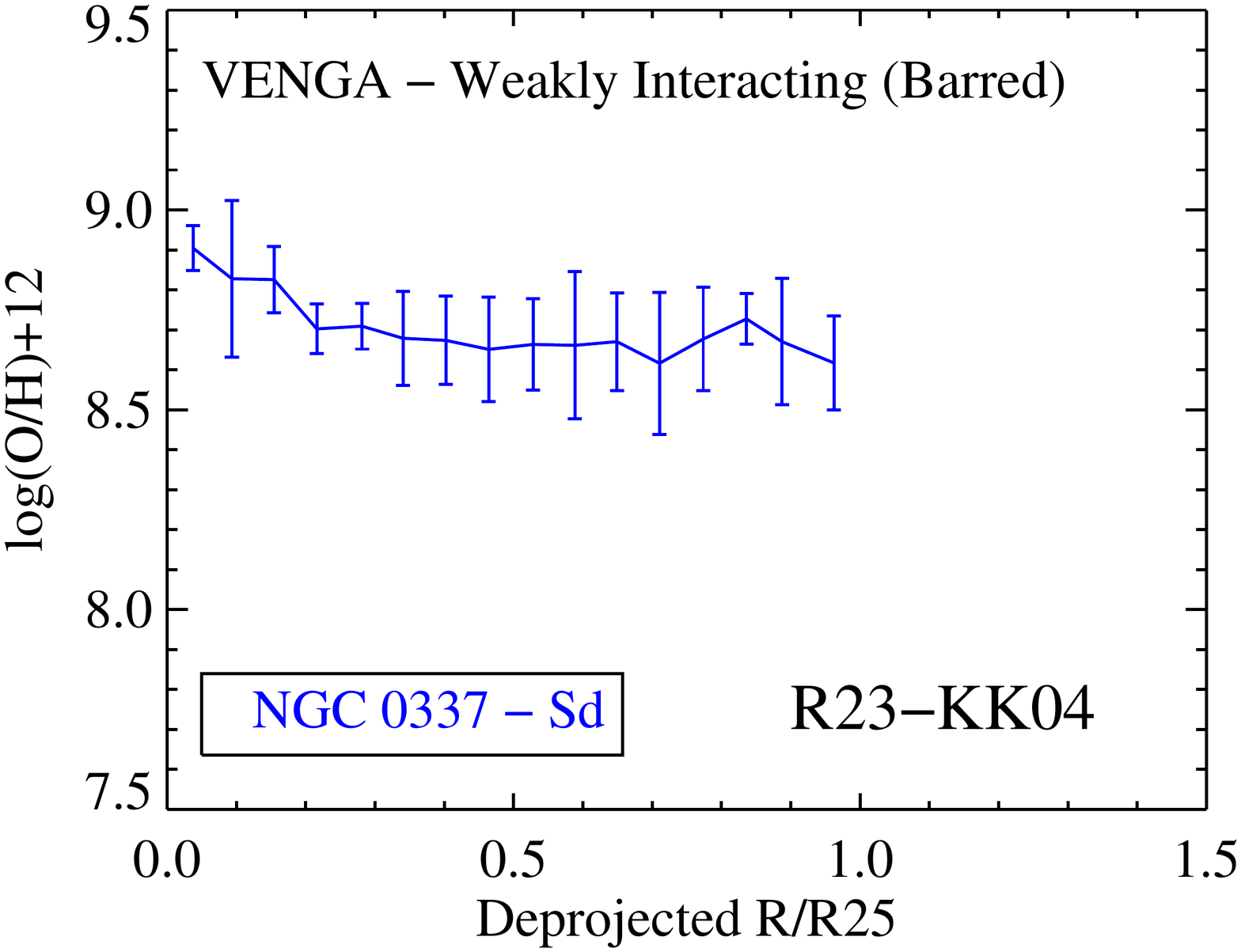} \hspace{-0.8cm}
\includegraphics[angle=0,height=0.28\textheight]{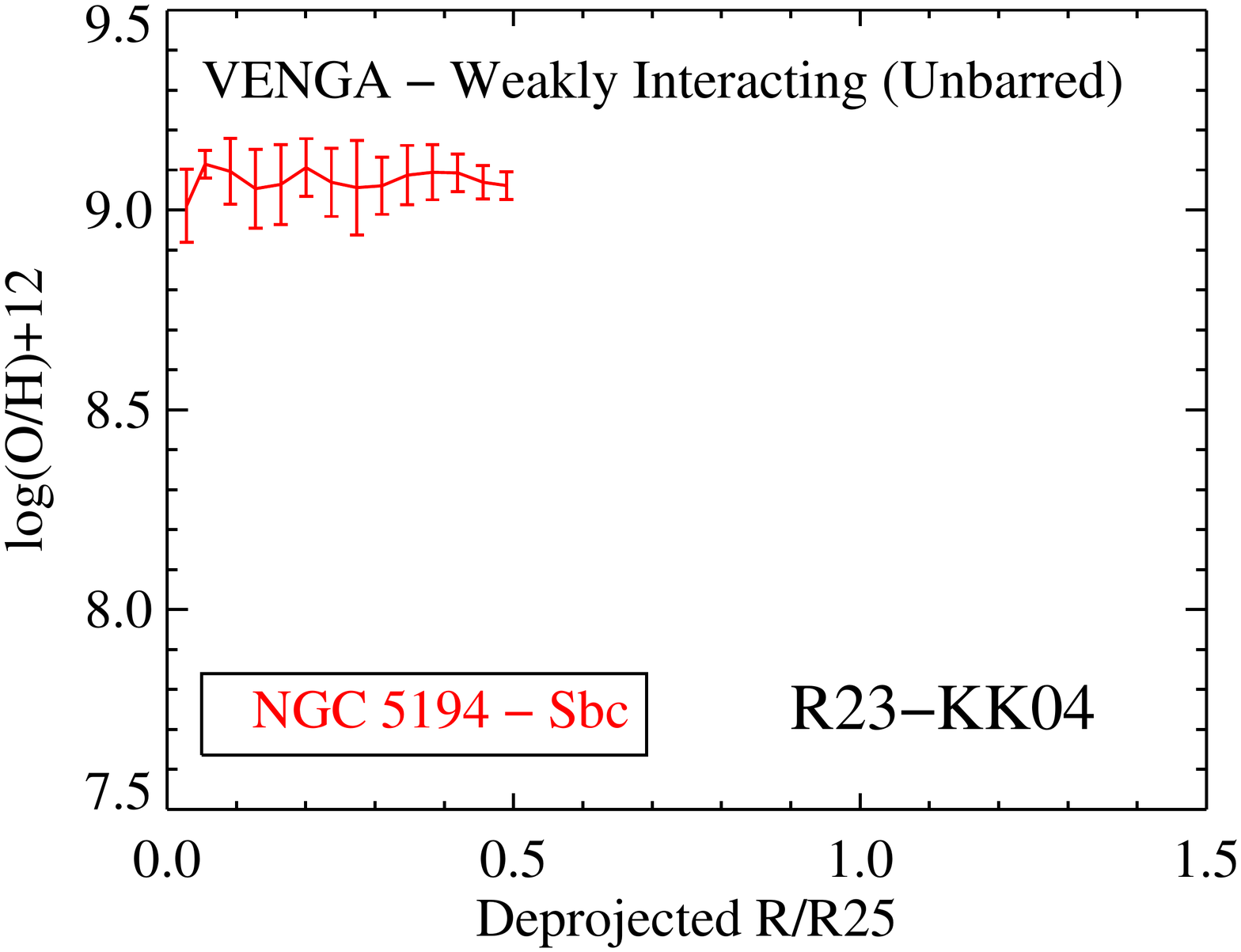} \hspace{-0.8cm}
\includegraphics[angle=0,height=0.28\textheight]{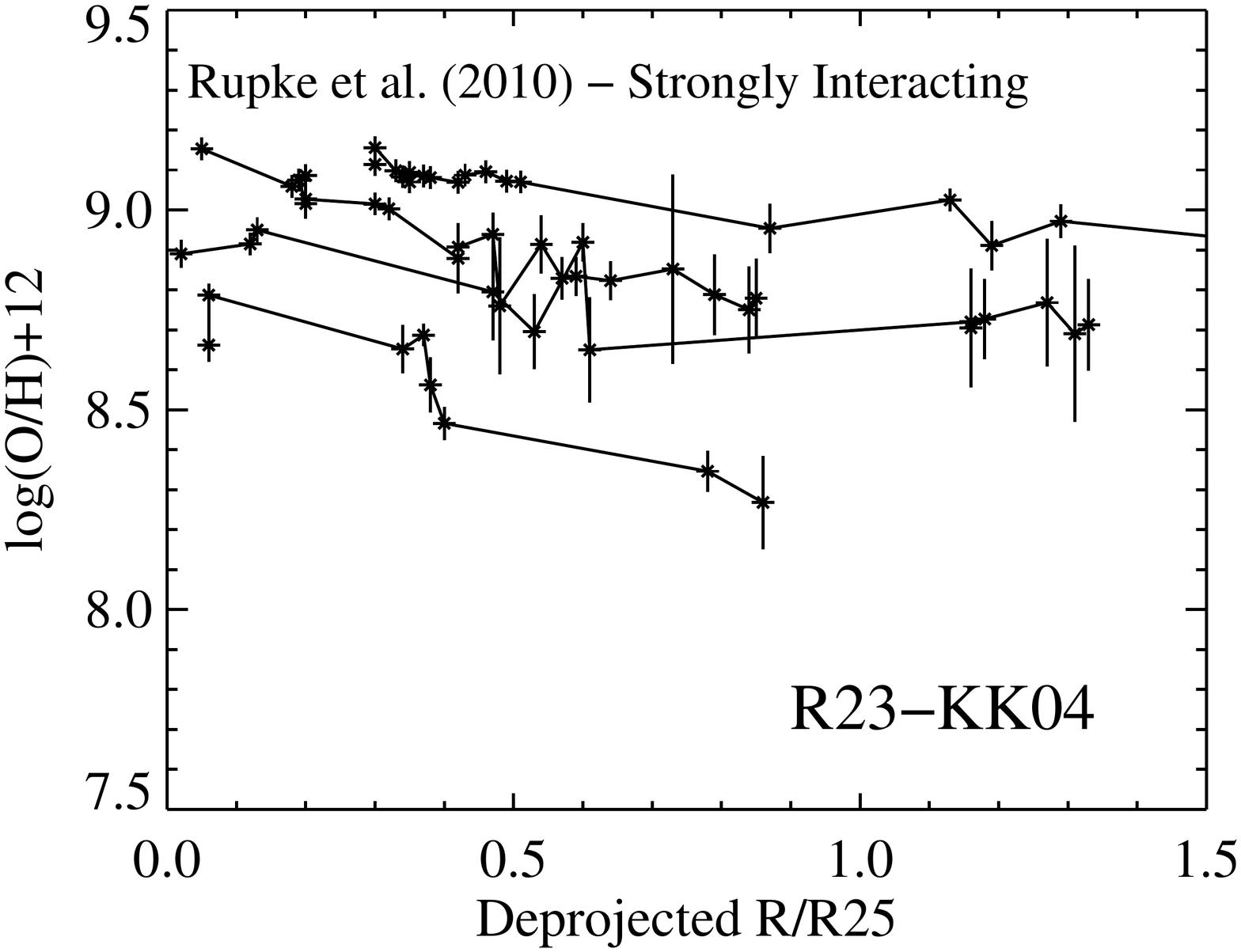}\hspace{-0.8cm}
\includegraphics[angle=0,height=0.28\textheight]{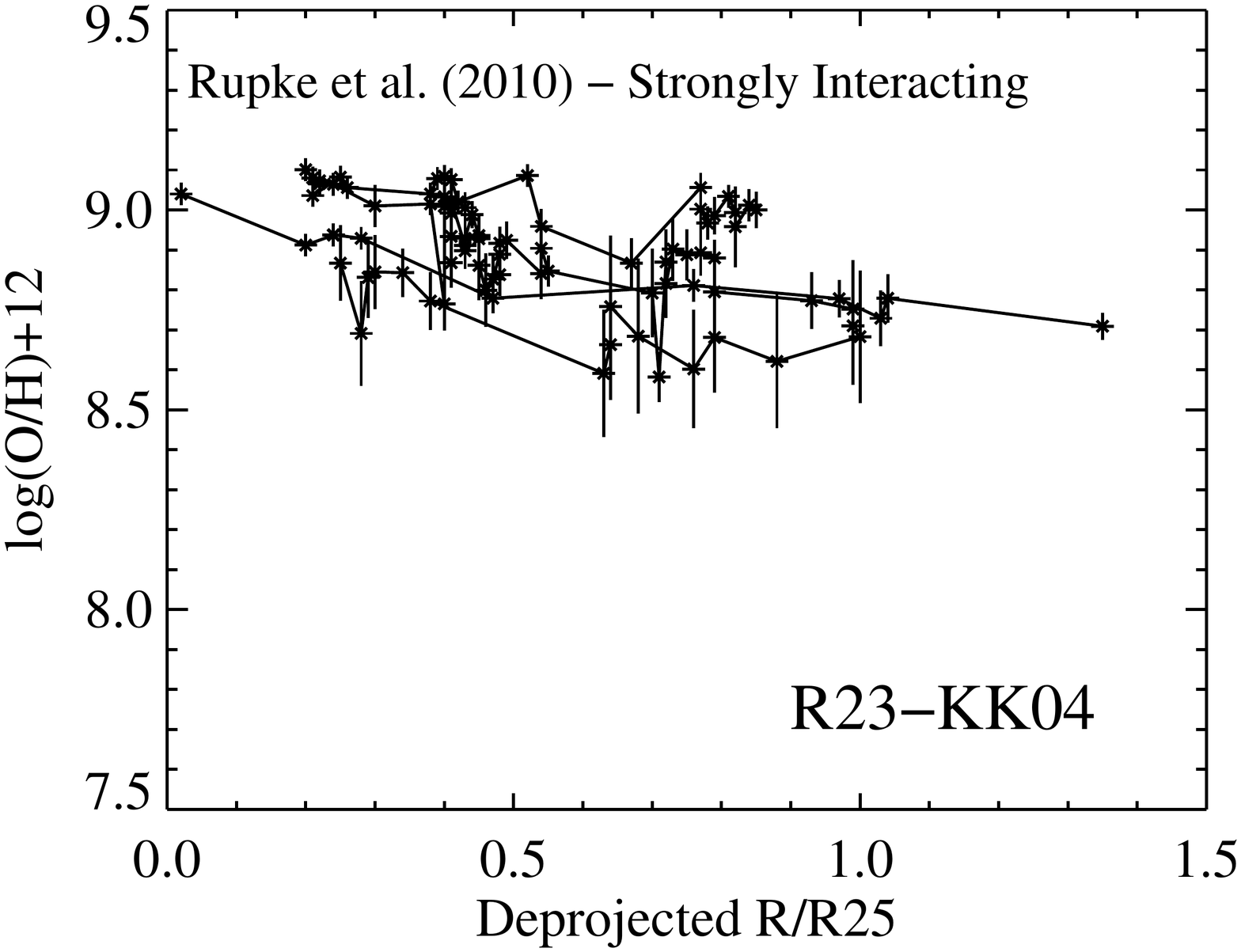}\hspace{-0.8cm}
\caption{
Same as for Figure \ref{fig:venga-rupke-n2o2} but for the $R_{23}$-KK04 \zgas{} diagnostic.
}
\label{fig:venga-rupke-r23}
\end{figure}
\end{landscape}

Disentangling the effect of bars from minor mergers and tidal
interactions  is difficult for several reasons. 
Bars can form spontaneously in an isolated galaxy  or  
can be induced via a tidal interaction or  minor merger 
(e.g., \citealt{hernquist1995}).   
Furthermore, during a tidal interaction or minor merger, gas inflows
are driven by tidal torques  from the smaller  companion, as well as
torques from a bar induced in the disc  of the primary  (e.g., \citealt{hernquist1995}).   
Therefore, we exert extra caution in our analysis by 
separating our {\it{isolated}}   barred/unbarred galaxies 
from {\it{weakly interacting}}  barred/unbarred galaxies.
The two weakly interacting galaxies  (NGC~0337, NGC~5194 or M51a) in our sample
have properties consistent with minor interactions/mergers of mass ratios
below 1:3, and their morphologies are regular enough for a bar type (barred
or unbarred) to be reliably assigned.
The two left panels of  Figures \ref{fig:venga-rupke-n2o2} \& \ref{fig:venga-rupke-r23},
show our data for barred and unbarred galaxies, divided between
isolated (top row)  and interacting  (bottom row).
The N2O2-KD02 \zgas{}  gradient for the interacting barred Sd galaxy NGC 0337 is $-0.31 \pm 0.02$
dex ($R$/\rtwofive{})$^{-1}$,  compared to  -0.33 $\pm$ -0.01 dex ($R$/\rtwofive{})$^{-1}$
for the  isolated barred galaxies (NGC 2903 \& NGC 5713).
Similarly,  the \zgas{} gradient for the  weakly interacting unbarred Sbc galaxy NGC~5194 
is   $-0.09 \pm 0.01$ dex ($R$/\rtwofive{})$^{-1}$, compared to 
-0.44 $\pm$ 0.13 dex ($R$/\rtwofive{})$^{-1}$ 
for the isolated unbarred galaxies (NGC 0628, 3938, \& 4254). 
This comparison hints at the possibility that weakly interacting 
unbarred systems might have a flatter gradient
than  the corresponding isolated unbarred galaxies,  but 
the number statistics
in our subsample of VENGA galaxies
 are too small to draw any general conclusion.
 
\clearpage

All the spiral galaxies in our subsample exhibit remarkably flat \zgas{} gradients, making it is difficult to find distinguishing characteristics in the gradients between the galaxies.  Does the fine structure of the \zgas{} radial profiles seen in Figure \ref{fig:zgas-gradients} provide hints of the underlying enrichment and/or radial mixing processes that can lead to the observed globally flat \zgas{} gradients?
The effects of these processes might be detectable as
small scale breaks or inflection points in the \zgas{} radial profile.
Examples of such breaks have been observed by \cite{roy1997}, \cite{scarano2011}, \cite{scarano2013}, and \cite{rosa2014}. 
\cite{pilyugin2003} cautions that breaks observed with $R_{23}$ diagnostics should be confirmed with other diagnostics due  to the double degeneracy in \zgas{} for a given value of $R_{23}$   (described in $\S$ \ref{sec:how-zgas}).  The \zgas{} in most of our data is high enough that we avoid this issue.
The fine structure of the \zgas{} radial profile for the  isolated barred galaxy NGC 2903 
shows a positive \zgas{} gradient in the nucleus which then flattens then inverts to a negative gradient towards the end
of the bar.  This pattern is seen using all the \zgas{} diagnostics except those based on N2, which shows opposite behavior in the gradient (see discussion in $\S$ \ref{sec:results-zgas}).
The gradient then becomes flatter  and smooth in the outer disc beyond the bar.
The case is less clear for our other isolated barred galaxy NGC 5713 which appears to show a break in the \zgas{} radial profile at the end of its bar using the 
$R_{23}$ based diagnostics but the same break is not seen in the other diagnostics, although the scatter
in $R_{23}$ is large beyond the end of the bar. 
Our one weakly interacting barred spiral NGC 0337 shows a stronger negative gradient over its bar which then becomes flatter in its outer disk.  None of our unbarred isolated galaxies (NGC 0628, 3938, \& 4254) or unbarred interacting galaxy (NGC 5194)
show any clear breaks in their \zgas{} radial profiles beyond the inner few kpc.
This fine structure, especially in NGC 2903, possibly hints at how bars can flatten \zgas{} gradients over time, which we discuss in greater detail in $\S$ \ref{subsubsec:theorybar}

\subsubsection{Comparison of Our Results with Other Studies}
\label{subsubsec:compare1}

Early studies based on  slit spectroscopic or narrow-band imaging data   
(e.g., \citealt{vilacostas1992, zaritsky1994, martin1994,dutil1999, henry1999})
claimed that \zgas{} gradients are flatter in isolated 
barred spiral galaxies than those that are unbarred. However, these studies 
suffer from several caveats.  They use multi-slit spectroscopic 
or narrow band photometric data of targeted bright \HII{} regions and 
consequently, do not fully sample the whole galaxy from the bulge to the outer disc. 
Furthermore, the data from the barred and unbarred galaxies come from
different studies and not observed or analysed in the same consistent
manner.

Next  we compare our  \zgas{} profiles to those in 
\cite{rupke2010},  based on   multi-slit spectroscopy
of \HII{} regions.
The study by \cite{rupke2010} 
builds upon the work done by \cite{kewley2010} by comparing
\zgas{} in 
16 strongly interacting galaxies to a control sample of 11 isolated
galaxies.
The  strongly interacting systems in this study  represent major 
interactions/mergers as they  were selected to have mass ratios close 
to unity (1:1 to 1:3).
\cite{rupke2010}  do not give the bar type of their sample galaxies
as the morphologies of the galaxies are too distorted to reliably
detect a bar and assign a bar type (unbarred or barred).
For their non-interacting sample, we are able to compile the bar type
from the literature, using studies based on  NIR images (e.g.,
\citealt{md07}), CO, HI and  H$\alpha$ kinematics 
\citep{bimasong, garrido2004, hernandez2005, hess2009, muraoka2009}. 
We cannot assign any bar type to the strongly interacting galaxies as 
they have  highly  disturbed morphologies.
In the top row of Figure \ref{fig:venga-rupke-n2o2}, we compare 
N202-KD02 \zgas{} profiles of isolated barred and unbarred galaxies
in the \cite{rupke2010} sample (two right panels) and in our study
(two left panels). 
The top row of Figure~\ref{fig:venga-rupke-r23} repeats this 
comparison using the $R_{23}$-KK04  \zgas{} profiles. 
The \zgas{} profiles in \cite{rupke2010}   
do not appear to be systematically different between  
isolated barred and unbarred galaxies of a given Hubble type,  
in agreement  with our results. 
It appears that some of the \zgas{} profiles in \cite{rupke2010}  
are steeper than the typical \zgas{} profile we see in our data,
but since the datasets have different spatial sampling, it is not possible to do a consistent comparison.

For NGC 0628, this galaxy was observed for the PINGS survey by \cite{pings2011}.
Their radial profiles appear qualitatively similar to our own.
Correcting for the assumed distance to NGC 0628, \cite{pings2011} find a fibre-to-fibre \zgas{} gradient using the $R_{23}$-KK04 diagnostic of $-0.031\pm0.003$ dex kpc$^{-1}$
which is nearly identical to our gradient of  $-0.031\pm0.001$ dex kpc$^{-1}$.
Specifically targeting known \HII{} regions in NGC 0628,  \cite{pings2011}
finds a gradient of $-0.044\pm0.002$ dex kpc$^{-1}$, slightly steeper than our value.

Finally,  we compare our  IFU-based results of \zgas{} to  the recent  IFU studies 
by \cite{sanchez2012, sanchez2014} based on the PINGS
\citep{pings2010}  and CALIFA \citep{sanchez11}  IFU surveys.
Our study
extends upon the work
of \cite{sanchez2012, sanchez2014}  in
several respects.
As shown by  Table \ref{tab:survey-comp}, our study has a smaller  sample but benefits 
from a higher median spatial resolution (387 pc vs 644 pc in \citealt{sanchez2012} and 
985 pc in \citealt{sanchez2014}), 
a higher spectral resolution (120 km s$^{-1}$ at 5000 \AA{} vs 
600  km s$^{-1}$ and  350 km s$^{-1}$), and  the use of  seven 
complementary \zgas{} diagnostics.
\cite{sanchez2012}   base their analyses
on targeting clumpy bright \HII{} regions in their galaxies, 
 while our  
analysis uses all the spaxels in our VENGA IFU data after excluding
those dominated by  Seyfert or LINER  gas excitations 
($\S$ \ref{sec:bpt-method}), and DIG-dominated regions  ($\S$
\ref{sec:dig}).
\cite{sanchez2014} targets clumpy bright \HII{} regions like \cite{sanchez2012}
but then also excludes Seyfert or LINER contaminated regions similar to our methodology.
The \zgas{} gradients reported by \cite{sanchez2012, sanchez2014} are derived from unweighted linear fits, similar to how we derive gradients in our sample, although the fits are limited to radii of $0.3 < r/r_e < 2.1$.
We report the mean \zgas{} gradients and 1-$\sigma$ uncertainties about the mean for barred vs. unbarred galaxies from \cite{sanchez2014} in
Table \ref{tab:zgas-gradients}.  
With our higher spatial and spectral resolution,
 and our suite of seven
\zgas{} diagnostics,
we confirm the results from 
 \cite{sanchez2012, sanchez2014}, 
finding no significant
difference in the \zgas{} gradients  between barred and unbarred spirals.

\subsubsection{Theoretical Implications of Our Results for Bar-Driven Evolution}
\label{subsubsec:theorybar}

Taken at face value, our IFU-based study and the IFU studies of  
\cite{sanchez2014, sanchez2012}   are finding similarly 
shallow \zgas{} profiles in barred and unbarred galaxies out to  a
large fraction  (0.5 to 1.0) of  $R_{\rm 25}$.
Below we discuss the implications of this result and
how bars impact \zgas{}  profiles.

A large-scale stellar bar impacts \zgas{}  by driving gas inflows and
outflows in different parts of the galaxy disc.
Beyond the end of the stellar bar, between the corotation resonance (CR)  and  the outer 
Lindblad resonance (OLR)  of the bar,  gas is driven {\it outward} in the disc
by gravitational  torques  (e.g., reviews by \citealt{buta1996} and by \citealt{jogee2006}).
Conversely, inside the bar,  gas between the CR and the outer 
inner  Lindblad resonances (OILR) of the bar is driven  {\it inward}
until it reaches the inner kpc region where the bulge potential tends
to dominate. 
If a  galaxy with an initially steeply negative \zgas{} profile  (where 
\zgas{} falls with radius)  develops a stellar  bar,  the bar 
 can  flatten  the  \zgas{} profiles by driving   gas inflow  from 
the bar end to the inner kpc region, and gas outflows from the bar's 
end  to the outer disc. 
However, an added layer of complexity is that strong bars often 
induce  high SFRs in the circumnuclear (inner kpc) region of spirals 
(e.g.,  \citealt{kormendy04,  jogee05}),  and this leads to an 
enrichment of  \zgas{}, followed by starburst-driven outflows.

The \zgas{}  profile in a present-day spiral  
depends on the initial \zgas{}  gradient in the early stages of galaxy  
formation, and the subsequent complex cumulative  history of gas
inflow/outflows  and  in-situ SF.  
Therefore, it is relevant to look at cosmological simulations of
barred galaxies.
 Cosmologically motivated simulations of disc galaxies show that bars 
have undergone two distinct phases of development in early and late times
\citep{heller2007, romano2008}.
The first bars that formed in the first few Gyr  (at  
$z>2$)  of a galaxy's life were induced by  asymmetries in the dark matter haloes and they decayed 
and reformed quickly  during that period, while the galaxy is growing
its mass via major and minor mergers.
 These early transient stellar bars exist in almost 
all simulated galaxies, and are  associated with strong radial gas
flows. At later times, when the Universe was around 5-7 Gyr (e.g., $z \sim
1.5$), discs are more massive and a more stable generation of 
bars  start to form,  primarily through tidal interactions.  
These late-time bars are typically long lived ($\sim$ 5-10 Gyr),  
consistent with observations of bars in bright
or massive galaxies  out to $z \sim 1$ \citep{jogee2004, sheth2008, cameron2010}.

Our empirical study finds that  that present-day barred and unbarred massive 
spirals host  similarly shallow \zgas{} profiles.  
These results  imply  that if spirals had steeper metallicity gradient at earlier epochs,
that the flattening of this gradient over time  is not driven 
primarily  by the {\it{present-day}}   large-scale stellar bar.
Instead, it implies that processes not directly related the {\it{present-day}}
large-scale stellar bar  played an important role in shaping the 
\zgas{}  profiles of  massive spirals  over their lifetime.
These processes likely  include 
gas inflows/outflows  driven by  {\it earlier} generations of transient
stellar bars at $z\gg 1$, as well as  gas inflows 
from minor mergers and tidal interactions since $z<2$. 
Another layer of complexity is added by local 
SF activity and outflows driven by starbursts or AGN. 

It is relevant to note here  that observations show that minor mergers 
are very common at $z<1$, being at least three times  more frequent 
than major mergers \citep{jogee09, lotz11}.  While prograde minor 
mergers would induce stellar bars,  retrograde minor mergers are unlikely to
do so, and could therefore help to flatten \zgas{} in an unbarred spiral.
Such scenarios can be tested once we have \zgas{} profiles for large
samples  of barred and unbarred galaxies over $z\sim 0$ to 1.5.

\subsection{Comparison of \zgas{} Gradients Between  Low and High
  Redshift Galaxies}  \label{sec:results-redshift}
\rm 

We explore whether \zgas{} gradients are flattening over cosmic
time by comparing our $z=0$ results to gradients observed at higher
redshifts.
While all our VENGA galaxies at $z \sim 0$ show similarly shallow \zgas{} gradients across both bared and unbarred massive spiral galaxies, our sample size is small.   Do all spirals at $z \sim 0$ show similarly shallow \zgas{} gradients?  \cite{sanchez2014}, with a sample of 193 spirals, finds an average  gradient with the O3N2-PP04 diagnostic of $-0.16\pm0.12$~dex~$(R/R_{\rm 25})^{-1}$, consistent with our value of $-0.11 \pm 0.13$~dex~$(R/R_{\rm 25})^{-1}$.  The sample of 49 galaxies found in \cite{ho2015} finds an average gradient using the N2O2-KD02 diagnostic of  $-0.39\pm0.18$~dex~$(R/R_{\rm 25})^{-1}$, also consistent with our value of $-0.34\pm0.14$~dex~$(R/R_{\rm 25})^{-1}$.
\cite{bresolin2015} finds their sample of 10 low surface brightness spiral galaxies have even shallower average gradients, with  $-0.065\pm0.055$~dex~$(R/R_{\rm 25})^{-1}$ and $-0.133\pm0.072$~dex~$(R/R_{\rm 25})^{-1}$ for O3N2 and N2O2 diagnostics respectively. 
Across all these studies we find that  \zgas{} gradients for these $z\sim0$ spiral galaxies are at least as shallow, on average, as our own spirals.  This suggests that the shallow \zgas{} gradients we observe in our own sample are typical for most massive spirals at $z\sim0$.

\begin{figure}
\hspace{-1cm}
\includegraphics[angle=0,width=0.55\textwidth]{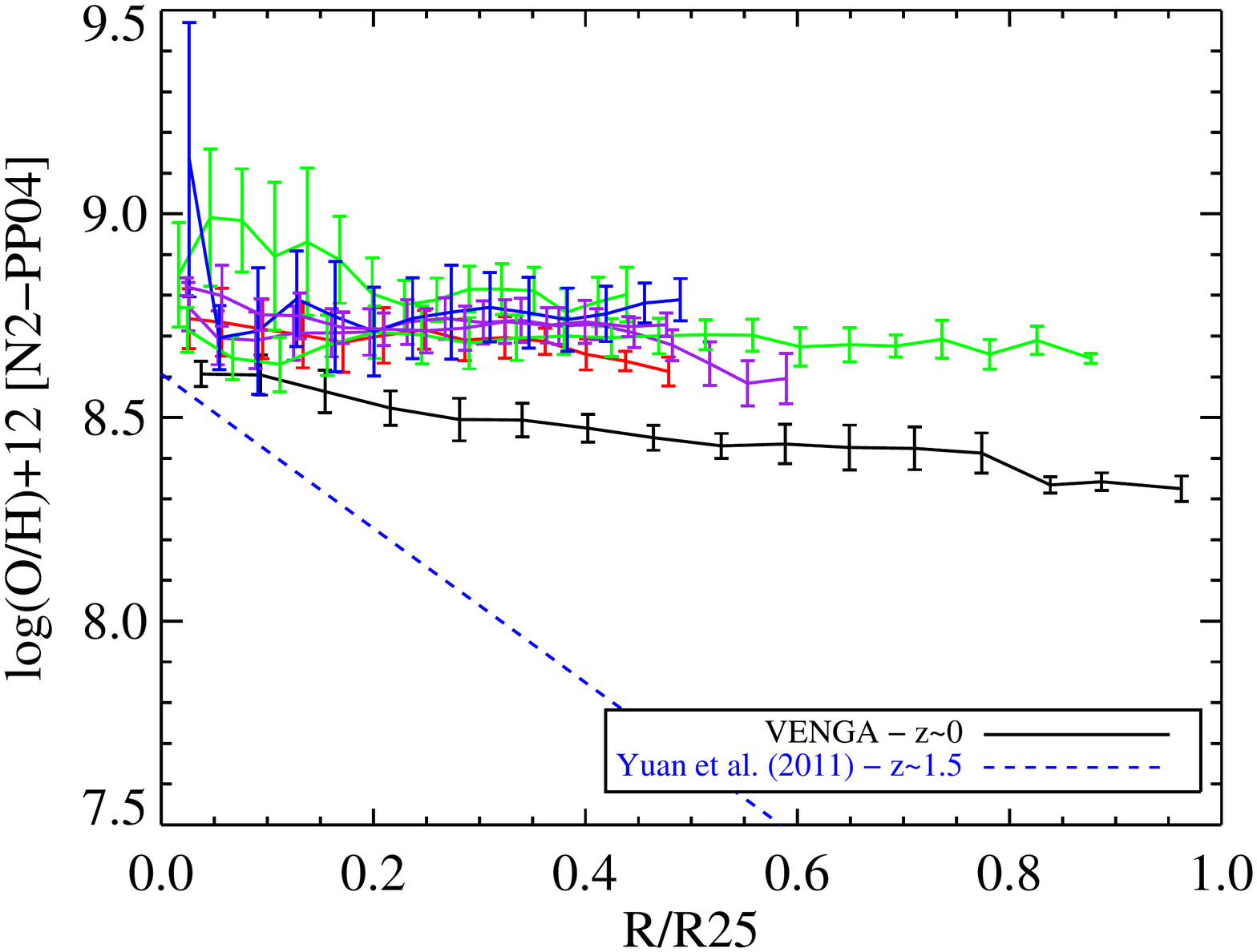}
\caption{
Radial \zgas{} profiles in units of  $R/R_{\rm 25}$ are shown for 
our local ($z \sim 0$) VENGA sub-sample of spirals  (solid lines), and  a 
higher redshift ($z \sim 1.5$) gravitationally lensed disc galaxy Sp1149 
\protect\citep{yuan2011}.  For comparison in radial units of kpc, see Figure \ref{fig:boissier}.
Sp1149 shows a significantly steeper gradient
than our $z \sim 0$ VENGA sub-sample.  These observations are consistent with the
idea that \zgas{} gradients flatten over cosmic time.
}
\label{fig:yuan2011}
\end{figure}

\begin{figure}
\hspace{-1cm}
\includegraphics[angle=0,width=0.55\textwidth]{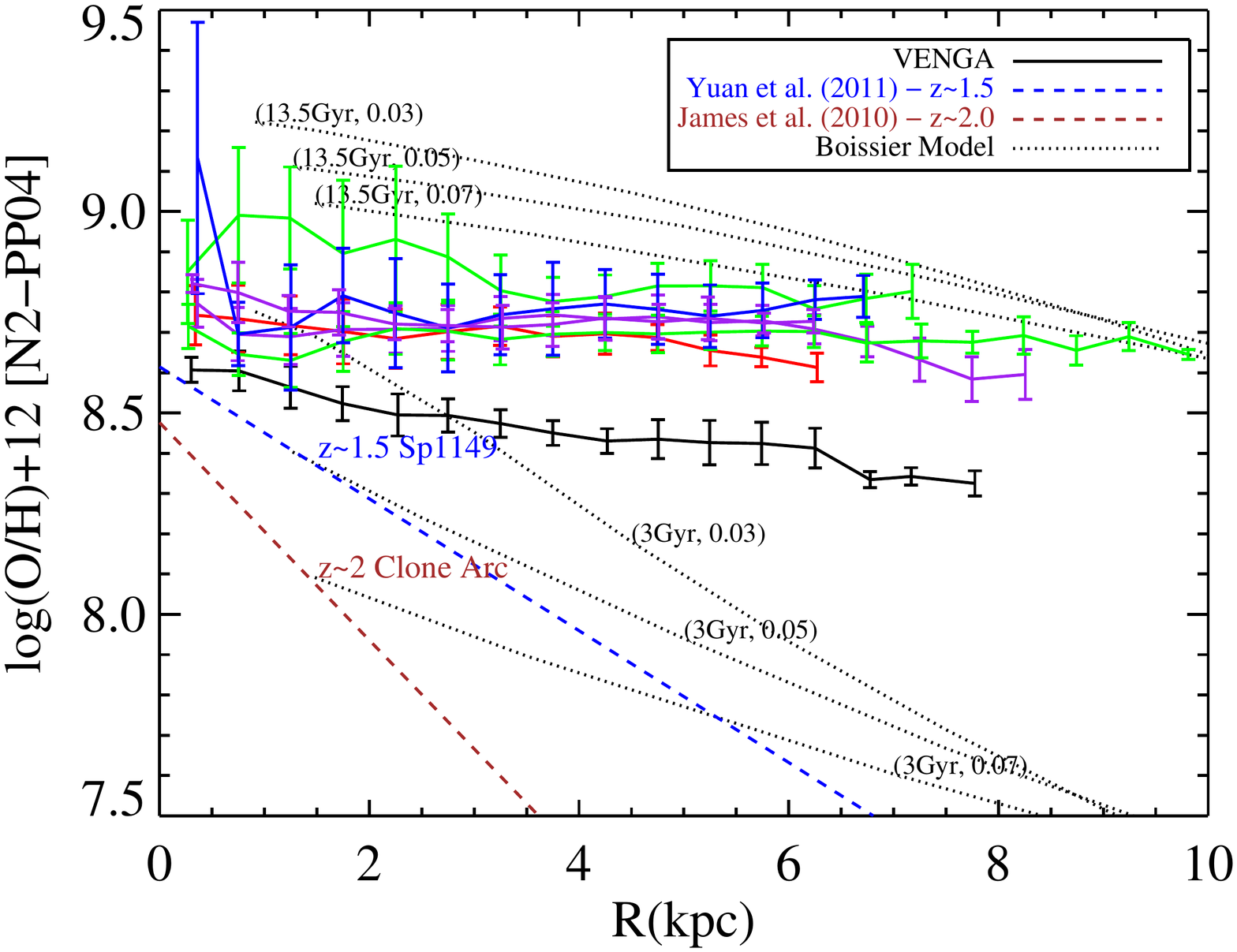}
\caption{
Comparison of the radial \zgas{} profiles in units of kpc
between  VENGA  galaxies at  $z \sim 0$, 
a  gravitationally lensed disc galaxy Sp1149 at $z \sim 1.5$ 
\protect\citep{yuan2011} and the  clone arc system at $z\sim 2$  \protect\citep{jones2010}.
Both Sp1149  and the ``clone arc'' shows a significantly steeper gradient
than our $z \sim 0$ VENGA sub-sample.  These observations are consistent with the
idea that \zgas{} gradients flatten over cosmic time.
Also over-plotted are predictions from the theoretical models of 
\protect\cite{boissier1999}.
The models at $z \sim1.5$  (age $\sim 3$ Gyr) and at $z\sim0$ (age $\sim13.5$ Gyr)
assume parameters to match Sp1149, with  a circular velocity of 200 km
s$^{-1}$ and dimensionless spin parameters of 0.03, 0.05, \& 0.07.
The age and dimensionless spin parameters are labelled in parenthesis
on the model plots.
}
\label{fig:boissier}
\end{figure}

High redshift studies typically lack the spatial resolution to
separate out different 
galactic components and
excitation regions, and the effects of low spatial resolution can
artificially flatten inferred \zgas{} gradients as shown by \cite{yuan2013a} and \cite{mast2014}.
To overcome these limitations, recent work using gravitational lensing
\citep{yuan2011} and adaptive optics (AO) \citep{swinbank2012}
has begun to study radial metallicity gradients at higher redshifts.
Figures \ref{fig:yuan2011} \& \ref{fig:boissier}
compare  the \zgas{} gradients of our sub-sample of $z \sim 0$ VENGA 
spirals to two gravitationally lensed galaxies observed with AO: Sp1149  at $z\sim1.5$ \citep{yuan2011} 
and the ``clone arc''  at $z\sim2$ \citep{jones2010}.
We use the  N2-PP04 \zgas{} diagnostic as it is common to both low and  
high redshift systems.  
The radial extent of the galactic disc changes with redshift and 
stellar mass so we compare the \zgas{} gradients 
using the $R$/\rtwofive{} scale in Figure \ref{fig:yuan2011}.   
Since some systems do not have \rtwofive{} reported, we also compared  
using radius in kpc in Figure \ref{fig:boissier}.  

Both higher redshift galaxies have measured dynamical masses of 
$\sim 2.5\times10^{10} M_\odot$ \citep{jones2010, yuan2011}.  
If  at most $\sim 20\%$ of their mass is composed of stars, the stellar
mass for both galaxies would be  $\le 2\times10^9 M_\odot$.
Therefore, the stellar masses  of these  two  $z \sim 1.5 - 2.0$ galaxies 
are one to two orders of magnitude  smaller than those  
of the $z\sim0$  spirals in  our sample 
($1.6 \times 10^{10}$ M$_\odot$ to  $4 \times 11^{11}$ M$_\odot$).
Sp1149  at $z\sim1.5$ \citep{yuan2011} has  an N2-PP04 \zgas{} gradient
of $-0.16 \pm 0.02$ dex kpc$^{-1}$ and 
$-1.92 \pm  0.24$  dex $(R/R_{25})^{-1}$ (Table \ref{tab:zgas-gradients}).
The ``clone arc''  at $z\sim2$ \citep{jones2010} has a gradient of 
$-0.27 \pm 0.05$ dex kpc$^{-1}$. 
In contrast,  the  more massive $z \sim 0$  galaxies in our 
sample show much flatter N2-PP04 \zgas{} gradients, spanning values from 
-0.034 to  0.003 dex kpc$^{-1}$
and -0.57 to 0.03 dex $(R/R_{25})^{-1}$,  
with a mean of  -0.012 $\pm$ 0.014  dex kpc$^{-1}$  and -0.16 $\pm$ 0.19 dex
$(R/R_{25})^{-1}$  (Table \ref{tab:zgas-gradients}).
We therefore conclude that these two lower mass galaxies 
(Sp1149 and the Clone Arc)   at $z \sim 1.5 - 2.0$ 
have \zgas{}  profiles that are much steeper than the shallow 
\zgas{}  profiles exhibited by  our $z\sim 0$ massive 
spirals. 
If these systems are representative of progenitor 
and descendant populations, then our comparisons suggest that
gas phase metallicity gradients flatten over cosmic time as a galaxy 
grows in  stellar mass.
Similar results are reported by \cite{yuan2011} who compare their
data to slit based $z\sim 0$ spectroscopic data.

A few recent observations by \cite{stott2014} and references therein 
report that some systems have flat \zgas{} radial profiles at $z \sim 0.8$.
If these results are representative, then they allow for the
possibility  that  there are highly efficient processes, which can
rapidly convert steep  \zgas{} profiles  at $z\ge 2$  into  flat
profiles by  $z \sim 0.8$.
However,  the observations by \cite{stott2014} do not use AO,  and it is
possible that the flatness of the  \zgas{}  profiles in their $z\sim
0.8$ systems is due to  the lower spatial 
resolution \citep{yuan2013a, mast2014},  despite their attempts
to correct for resolution effects.

More rigorous studies of the  chemical evolution 
of galaxies since $z\sim 2$  will  require  high resolution
observations of \zgas{} profiles in large  samples of galaxies of 
different stellar masses over a range of different redshifts between $z\sim 2$ to 0.
It is unclear at this stage whether the handful of existing observations 
over this redshift range are 
representative of the general population.

\subsection{Comparison of \zgas{} Gradients to Theoretical Models} \label{sec:results-models}

In order to explore some of the baryonic physics
that shapes the metallicity profiles of galaxies, 
we compare  different theoretical galaxy  chemical evolution models to the 
empirical \zgas{} profiles ($\S$\ref{sec:results-redshift})  
of our $z\sim 0$ sample of spirals,  
and published  \zgas{} profiles  in higher redshift  $z\sim 1.5-2.0$ 
systems.

We start by comparing our \zgas{} gradients to those derived by the
updated Boissier model.
The Boissier model was first employed by \cite{boissier1999} to
simulate the chemical evolution of the Milky Way's disc, and was then
generalized to other discs in \cite{prantzos2000} by allowing the
rotation curve and dimensionless spin parameter to scale.  The scaling
laws for the rotation curve and dimensionless spin parameter are
deduced from $\Lambda$CDM simulations of disc formation by
\cite{mo1998}.  In the Boissier model, a single disc is modeled as a
set of independent concentric rings with no radial inflows or
outflows.  The stellar distribution assumes a Kroupa IMF
\citep{kroupa2001}.  For the model we compare to, the SFR law has been
updated to match the empirical SFR laws found in nearby spirals
\citep{mateos2011}.   
Primordial gas infall decreases exponentially with time, with larger
gas infall timescales at greater radii.  Gas accretes for a shorter
amount of time in the centre than at the edge of the disc.  
Since this model  does not include radial mixing of gas through gas  
inflows and outflows,  it should only be considered as a 
model of the ``inside-out'' disc formation scenario. 

Figure~\ref{fig:boissier}  compares the Boissier model at different
ages  to  radial \zgas{}  profiles of  our  
sample of $z \sim 0$ massive spirals, and the lower mass 
gravitationally lensed galaxies
Sp1149  at z $\sim$ 1.5 \citep{yuan2011}, and the ``clone arc" at 
$z\sim2$ \citep{jones2010}  (see $\S$\ref{sec:results-redshift}).
We use models with a circular velocity of 200
km s$^{-1}$ and dimensionless spin parameters of 0.03, 0.05, \& 0.07. 
The Boissier models exhibit a somewhat  shallower \zgas{} gradient at 
$z \sim 1.5$  than Sp1149. They subsequently flatten with time from 
$z \sim 1.5$  to 0, but do not flatten enough to  match the 
shallow \zgas{} gradients in 
our sample of  $z \sim 0$ massive spirals.
This is perhaps not surprising since the Boissier models are missing
important aspects of galaxy evolution and baryonic physics.  They do not  
include the  radial gas mixing via  inflows and outflows, 
driven by  bars or interactions. Furthermore, since they are not 
cosmological hydrodynamic simulations, they do not include galaxy mergers, 
cold mode gas accretion, or feedback.

Next we compare
the same observations
to  more realistic 
simulations of the assembly and chemical evolution of galaxies,
which  are  cosmologically-motivated,  include more baryonic physics, 
and allow for radial inflow/outflow of gas.  
We consider simulated disc galaxies in  two different set of models: the MUGS models 
from \cite{pilkington2012}, and 
the MaGICC  models  \citep{gibson2013}.
The first set of models from \cite{pilkington2012}, called MUGS, uses
the gravitational N-body and SPH code Gasoline \citep{wadsley2004} to
simulate 16 isolated disc galaxies that are randomly drawn from a 50
h$^{-1}$ Mpc $\Lambda$CDM
simulation
with WMAP3 cosmology.  
Each galaxy is re-simulated at much higher resolution \citep{klypin2001}.  
\cite{pilkington2012} selects four galaxies with the most prominent
discs, including g1536 and g15784.  The MUGS model includes star
formation and SNe feedback implemented with the 
``blast wave formalism'' described in \citep{stinson2006}, heating by a
background UV field, gas cooling derived using Cloudy
\citep{ferland1998}, and gas  enrichment via Type II \& Ia SNe assuming
a Kroupa IMF.  

A similar Gasoline code called MaGICC \citep{brook2012}, is used to also
simulate g1536 and g15784 \citep{gibson2013} using ``enhanced''
feedback with about double the energy per SN in MaGICC ($10^{51}$
ergs) heating the ISM, compared to  MUGS  ($4\times10^{50}$ ergs).  
MaGICC also includes radiative energy feedback from massive stars shortly before they go SN, 
and a \cite{chabrier2001} IMF, forming larger numbers of high mass stars than the Kroupa IMF in MUGS.
Both simulated disc galaxies (g1536 \& g15784) are selected to be isolated and have somewhat quiescent assembly histories \citep{gibson2013} which are identical in both MUGS and MaGICC.  

\begin{figure}
\vspace{-0.9cm}
\includegraphics[angle=0,width=0.50\textwidth]{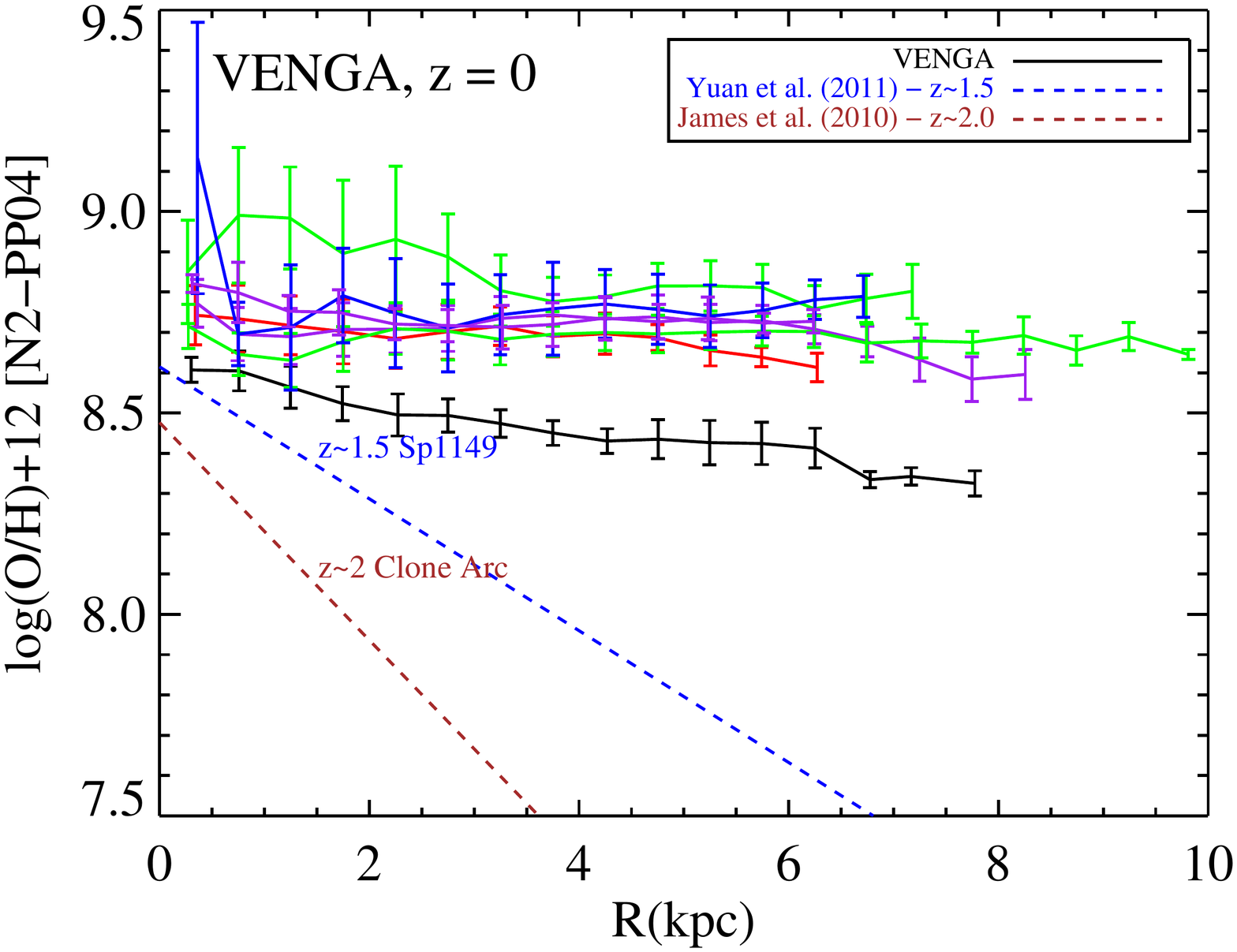}\vspace{-0.8cm}\\
\includegraphics[angle=0,width=0.50\textwidth]{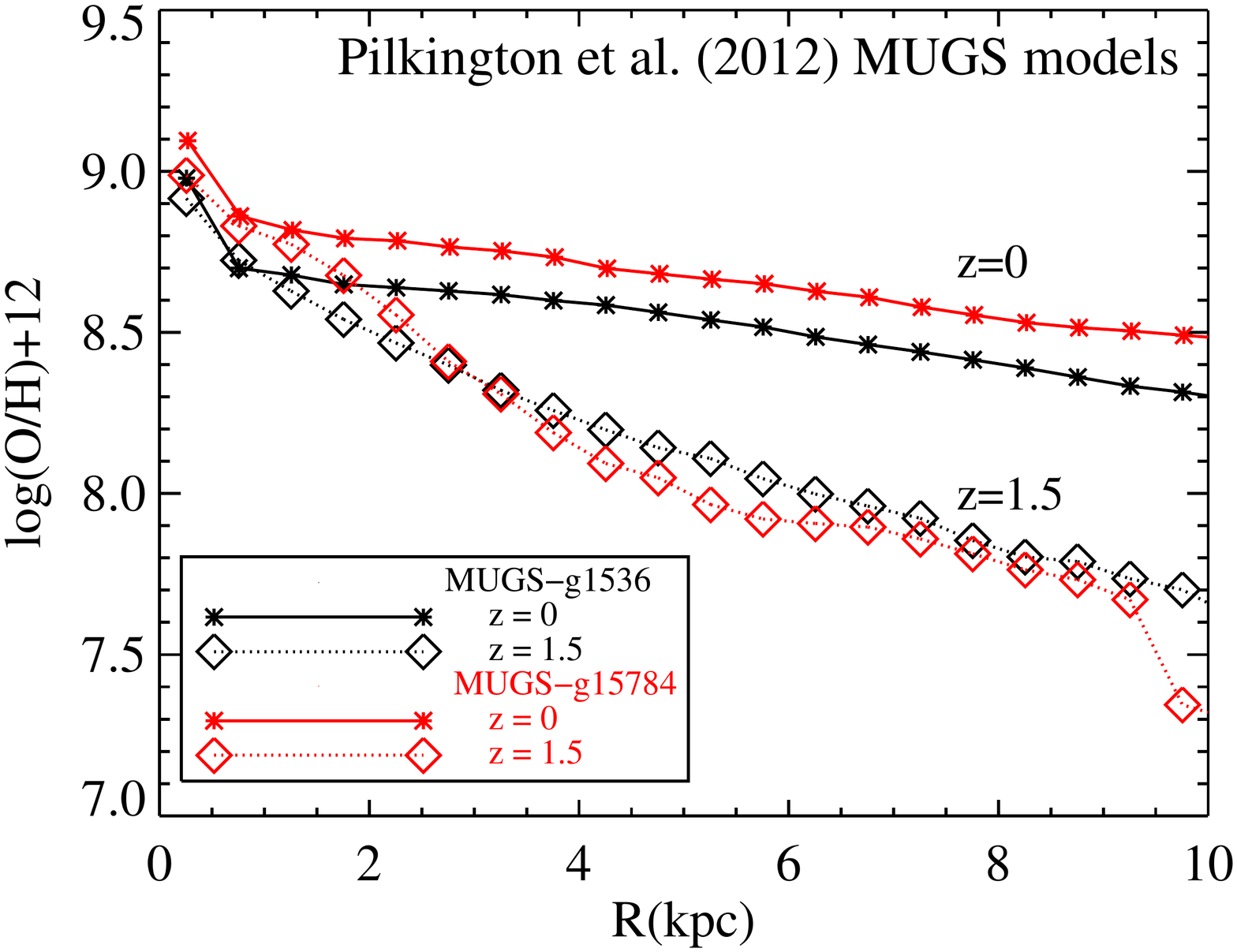}\vspace{-0.8cm}\\
 \includegraphics[angle=0,width=0.50\textwidth]{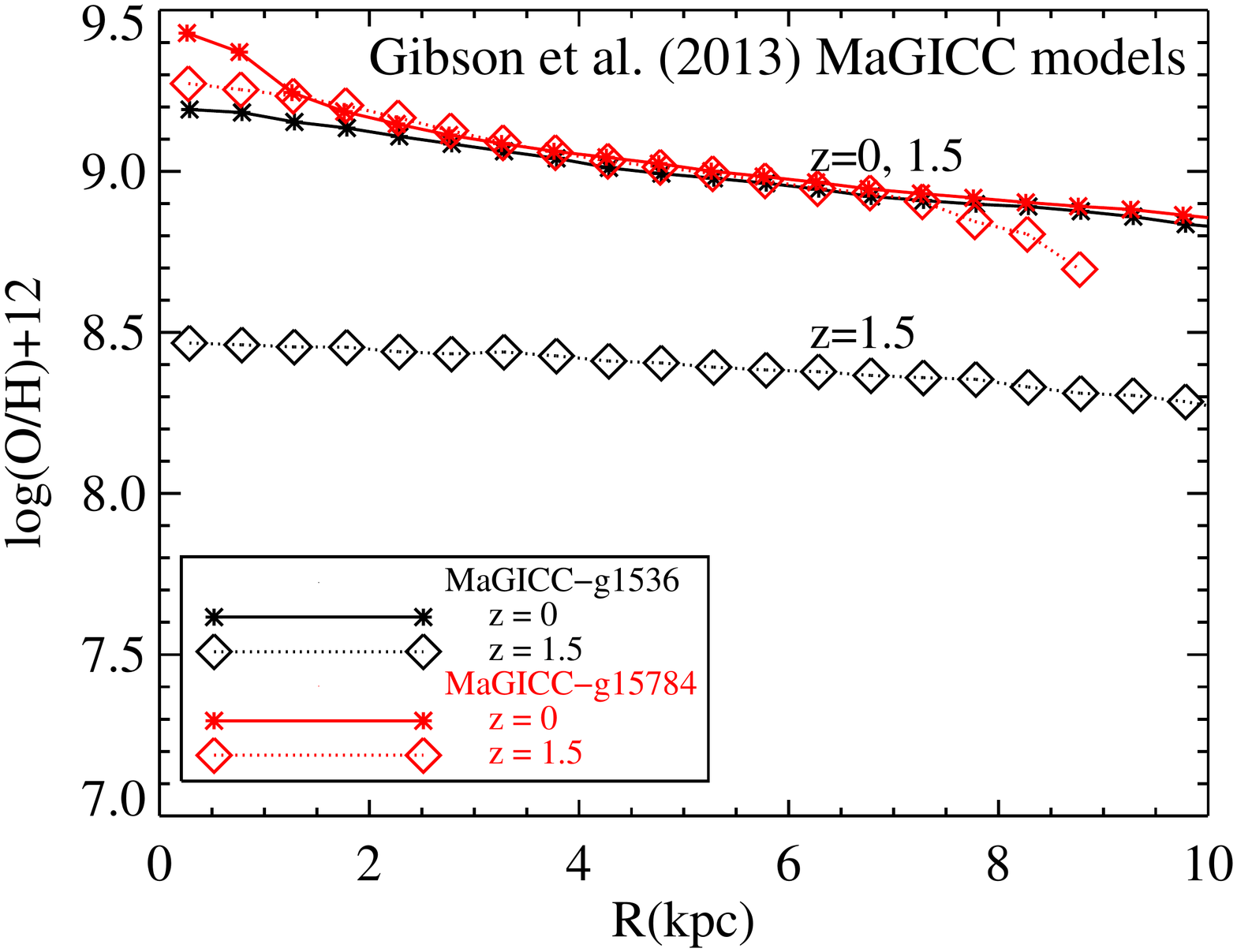}\vspace{-0.7cm}
\caption{The radial \zgas{} profiles in units of kpc are shown for 
our local $z \sim 0$ VENGA sub-sample, the higher redshift $z \sim 1.5$
gravitationally lensed disc galaxy Sp1149 \protect\citep{yuan2011}, and 
the $z \sim 2$ clone arc \citep{jones2010} {\bf{(top)}}; 
and for the  $z=0$  and $z=1.5$  MUGS {\bf{(middle)}} and MaGICC {\bf{(bottom)}} models 
for the simulated disc galaxies g1536 \& g15784 \protect\citep{gibson2013, pilkington2012}.
For the simulations, the MUGS results are in much better agreement than the MaGICC results with the evolution from high to low redshift in both the absolute value and gradients of \zgas{}, with what we see when comparing the flat gradients in our VENGA sub-sample to the strongly negative gradients seen at high redshift.  MaGICC includes ``enhanced'' feedback which drives stronger radial mixing and more numerous outflows to flatten \zgas{} gradients by $z \sim 1.5$ with only a slight flattening from $z \sim 1.5 $ to 0, while MUGS has more ``conventional'' feedback prescriptions that
appear to better match the physics driving the observed evolution of \zgas{} gradients  from $z \sim 1.5$ to 0 in spiral galaxies.
}
\label{fig:pilkington}
\end{figure}

The stellar mass of the galaxies in MaGICC grow $\sim 4 \times$ in
stellar mass from $z = 1.5$ to 0  ($\sim10^8$ to $4.5 \times 10^8
M_\odot$ for g1536 and $\sim10^9$ to $4.2 \times 10^9 M_\odot$ for
g15784; \citealt{obreja2014})  
while the final stellar masses of the MUGS galaxies are up to two orders of magnitude larger ($6.0\times10^{10} M_\odot$ for g1536 and $1.1\times10^{11} M_\odot$ for g15784; \citealt{stinson2010}).
Figure \ref{fig:pilkington} shows how the \zgas{} gradients for g1536
\& g15784 simulated in both MUGS and MaGICC evolve from redshift 1.5
to 0, and compares them 
to our sample of $z \sim 0$ massive spirals, and the lower mass 
galaxies Sp1149  at z $\sim$ 1.5 \citep{yuan2011}, and the ``clone arc" at 
$z\sim2$ \citep{jones2010}  (see $\S$\ref{sec:results-redshift}).
It is clear from Figure \ref{fig:pilkington}, that the primary difference in \zgas{} gradients 
for these simulations depends on the type of feedback being used, and
less on which specific disc galaxy (g1536 or g15784) is being
simulated. 
MaGICC, with its ``enhanced'' feedback, drives stronger radial mixing
and more numerous outflows,  so that at redshift $z \sim 1.5$ 
MaGICC has higher \zgas{} and flatter gradients than MUGS. 
The \zgas{}  gradients in MaGICC 
are already so flat by  $z \sim 1.5$  that they do not show much
evolution or extra flattening  from redshift 1.5 to 0
\citep{gibson2013}.
The galaxies in the MUGS simulation have more ``conventional''
feedback and their \zgas{} gradients start out strongly negative
at $z\sim 1.5$ and flatten significantly between redshift
1.5 to 0 \citep{pilkington2012}. 

The simulations MUGS and MaGICC imply that the strength of the feedback can significantly affect the evolution of \zgas{} gradients \citep{gibson2013}.
Although these comparisons between simulations and data are very
preliminary, ``conventional'' feedback used in MUGS results are in much
better agreement 
with our  sample of $z\sim 0$ massive spirals, 
Sp1149  at $z \sim 1.5$ \citep{yuan2011}, and the ``clone arc" at 
$z\sim2$ \citep{jones2010}   
in terms of 
the absolute value of \zgas{},  the shape of the \zgas{} profiles, 
and the evolution in stellar masses.
However, we note that the interval $z\sim 1.5$ to 0  cover a large 
cosmic interval of  $\sim$ 10 Gyr  and we need  a finer grid of
theoretical and empirical profiles at different redshifts over this 
interval to further  constrain the models and the rate at which
\zgas{} gradients are flattened.

\section{Summary and Conclusions}
\label{sec:summary}

We have presented a high resolution study of the excitation conditions and  
metallicity (\zgas{})  of  ionized gas  in a sample of eight nearby  
barred and unbarred spiral galaxies drawn from the VENGA  IFU survey.
Compared to  other integral field spectroscopy studies of 
spirals, our  study benefits  from a high spatial resolution (median $\sim$~387
pc),  a large spatial coverage of the galaxy (from the bulge to
the outer disc),   broad wavelength range (3600-6800 \AA{}),  
high spectral resolution ($\sim$~120 km s$^{-1}$ at 5000 \AA{}), and the
use of a full suite seven  \zgas{} diagnostics (Table \ref{tab:survey-comp}).
The combination of spatial coverage and  resolution allows us to resolve
individual galactic components, such as the bulge, primary stellar bar,
outer disc, and separate regions of widely
different
excitation
(e.g., \HII{} regions, spiral arms, starburst or AGN driven outflows,
diffuse ionized gas, etc.). 
Our results are summarized below:

\begin{enumerate}[(1)]

\vspace{1mm}
\item
{\it{Distribution of gas with  excitation conditions typical of Seyfert, LINER, 
and  Star-Forming Regions: }}
We use excitation diagnostic diagrams to  separate spaxels  hosting
gas whose excitations conditions are indicative of 
a hard UV radiation field or  
shocks (Seyfert or LINERS),
and photoionization by massive stars.
We also develop a procedure to identify spaxels dominated
by diffuse ionized gas (DIG),  
and to correct for the contribution  of DIG to the 
H$\alpha$ flux in  other spaxels.
While it is  conventionally assumed that LINERs are caused by
central starburst-driven outflows or low luminosity AGN in
the nuclear regions,  we  find that LINER-type  excitations are
present at large  projected  distances
(3-10 kpc) from the centre,  and tend to be associated 
with DIG (Figure~\ref{fig:bpt-sliced}).

\vspace{1mm}
\item
{\it{Spatially resolved maps of $q$, SFR, and \zgas{} :}}
After  excluding spaxels dominated by Seyfert, LINER, and
DIG, we produce accurate high resolution maps of the  ionization parameter,  
$q$, the star formation rate (SFR), and   seven  \zgas{}  diagnostics based
on the R23, N202, O3N2, and N2  metallicity indicators (Figure \ref{fig:2d-maps}). 
The distribution of $q$  and SFR are qualitatively similar in
many galaxies. However, while 
the average  $\Sigma_{\rm SFR}$  falls by over an order of magnitude
(e.g. from 0.1 to below 0.01 or from 0.01 to below 0.001 
$M_\odot$ yr$^{-1}$ kpc$^{-2}$),  the average value of $q$ 
typically varies by no more than a factor of  three  (e.g., 
from 6.95 to 7.40 dex, corresponding to 0.8 to 3 times 
10$^7$  cm s$^{-1}$)  (Figure \ref{fig:q-sfr-gradients}).

\vspace{1mm}
\item 
{\it{Comparison  of  Different \zgas{}   Diagnostics:}}
We explored how the  absolute values of \zgas{}  vary between
the seven  \zgas{} diagnostics (Table \ref{tab:zgas-indicators}):
(i)~The three $R_{23}$ \zgas{}  diagnostics ($R_{23}$-KK04, 
$R_{23}$-M91, $R_{23}$-Z94)  based on  
the  (\iontwo{O}{III}/H$\beta$)/(\iontwo{N}{II}/H$\alpha$) ratios 
show similar \zgas{}  profiles and  yield
absolute values that agree  within 0.1 to 0.2 dex; 
(ii)~ N2O2-KD02  (based on the \iontwo{N}{II}/\iontwo{O}{II} line  ratio) 
gives \zgas{}  values similar to the $R_{23}$ diagnostics, 
but  shows a lower scatter.  This may be related to the
relative insensitivity of  \iontwo{N}{II}/\iontwo{O}{II}  to variations in $q$.
(iii)~O3N2-PP04 (based on  the \iontwo{N}{II}/[H$\alpha$] and  
\iontwo{O}{III}/[H$\beta$] line ratios) gives  a \zgas{}   that is
systematically lower 
by  0.2  to 0.3 dex compared  to  the $R_{23}$ diagnostics; 
(iv)~N2-D02 and N2-PP04  (based on the
\iontwo{N}{II}/H$\alpha$ ratio)  yield  a \zgas{}   that is
systematically  
lower,  by as much as 0.4 dex, compared  to the  $R_{23}$ 
diagnostics. Furthermore, in  the inner 2 kpc of some of our galaxies,
the radial gradients in  \zgas{}  given by N2-D02 and N2-PP04 
can {\it{be opposite to  those}} given by other diagnostics.
Caution should therefore be exerted in using the N2-D02 and N2-PP04 \zgas{}
diagnostics.

\vspace{1mm}
\item 
{\it{Comparison of \zgas{}  profiles of isolated barred and unbarred galaxies:}}
The maps  (Figure \ref{fig:2d-maps}) and radial profiles  (Figure \ref{fig:zgas-gradients}) 
of  $q$ and \zgas{} are  of much higher quality 
in terms of spatial resolution and sampling than most other maps 
published to date. 
We find that isolated  barred and unbarred spirals galaxies exhibit similarly  
shallow \zgas{} profiles from the inner kpc out  to large radii (7-10 kpc or 
0.5-1.0 $R_{\rm 25}$) in the outer disc.
Specifically, with the  N2O2-KD02 \zgas{} diagnostic, the gradients are  $-0.44 \pm 0.13$ 
and  $-0.33 \pm 0.01$ dex ($R$/\rtwofive{})$^{-1}$, respectively,  
 for  isolated  unbarred and isolated barred spirals (Table~\ref{tab:zgas-gradients}).
Our result on the similar \zgas{}  gradients in isolated barred and unbarred 
galaxies contradicts earlier claims based on slit spectroscopy, but agrees with
recent IFU studies by \cite{sanchez2012, sanchez2014}.
If spirals had steeper metallicity gradient at earlier epochs,
then our result  implies that the flattening of this gradient over
time  is not driven primarily  by the present-day bar. Instead, the
evolution in the \zgas{} profile  is likely shaped  by the cumulative history of  
gas inflows/outflows  driven by  {\it{earlier}} generations of transient
stellar bars at $z\gg 1$, as well as  gas inflows 
from minor mergers and tidal interactions since $z<2$. 
Another layer of complexity is added by local 
SF activity and outflows driven by starbursts or AGN. 
We do not have a large enough sample to do a full comparison 
of isolated and interacting galaxies, but we note that
 the only  weakly interacting unbarred galaxy in our 
sample
shows a  tantalizing hint of having a shallower gradient  
($-0.09 \pm 0.01$ dex ($R$/\rtwofive{})$^{-1}$;  Table~\ref{tab:zgas-gradients})
than unbarred isolated spirals.

\vspace{1mm}
\item{\it{Comparison of \zgas{} Gradients At  Low and High
  Redshift in Data and Models:}}
The \zgas{} gradients  in our  $z\sim 0$  massive spirals are
markedly shallower, by $\sim 0.2$ dex kpc$^{-1}$,  
compared to published gradients for lensed lower mass 
galaxies at  redshifts $z\sim 1.5-2.0$ 
(Figures \ref{fig:yuan2011} \& \ref{fig:boissier}) 
If these systems are representative of progenitor 
and descendant populations, then our comparisons suggest that
gas phase metallicity gradients flatten over cosmic time as a galaxy 
grows in  stellar mass.
Simple non-cosmological theoretical models  (e.g.,
\citealt{boissier1999}) that  mimic an inside-out disc 
formation scenario through gas accretion, but do not include radial mixing, 
fail to provide a good match to   the low or high redshift data (Figure~\ref{fig:boissier}).
More realistic cosmologically-motivated  
hydrodynamical simulations  (e.g., \citealt{gibson2013,  pilkington2012})  
that  include more baryonic physics, and allow for radial inflow/outflow 
of gas appear to better match the data, but are highly sensitive to the 
adopted stellar feedback prescription (Figure \ref{fig:pilkington}).
It appears that the MUGS simulations
with ``conventional'' feedback  are in better agreement than the MaGICC 
simulations with  ``enhanced'' feedback  with  the data 
in terms of the  evolution of  \zgas{} values, \zgas{}  gradients, and 
stellar masses.
However, we caution that general conclusions on the evolution of 
\zgas{} profiles since $z<2$ will  require  high resolution
observations of \zgas{} profiles in large  samples of galaxies of 
different stellar masses at different redshifts between $z\sim 2$ to 0,  as well  as  finer grids of theoretical predictions. 
It is unclear at this stage whether the handful of existing observations 
at $z>1$ are  representative of the general population at these epochs.

\end{enumerate}

\section*{Acknowledgements}
G.B. is supported by CONICYT/FONDECYT, Programa de Iniciacion, Folio 11150220.  SJ and KK acknowledge support from NSF grant NSF AST-1413652 and  the National Aeronautics and Space Administration (NASA) JPL SURP Program.  This paper includes data taken at The McDonald Observatory of The University of Texas at Austin.

\footnotesize{
\bibliographystyle{mnras}
\bibliography{final_to_arxiv}
}

\appendix

\section{How Ionization Parameter $q$ is determined iteratively for the $R_{23}$-KK04 diagnostic - Method 2}\label{sec:q2-append}

The iterative determination of $q$ is based off of the $R_{23}$ \zgas{} diagnostic in \cite{kobulnicky04}, and this procedure is detailed in Appendix A2.1 of \cite{kewley2008}. 
$R_{23}$ is dependent on $q$ (see Figure \ref{fig:degeneracy} on the top left) so we iterate to converge on a \zgas{} and $q$ for this \zgas{} diagnostic. 
First, an initial \zgas{} is assumed by breaking the double valued degeneracy of $R_{23}$ using \iontwo{N}{II}/\iontwo{O}{II}.
Breaking the double valued degeneracy  is described in detail in Appendix A1 of \cite{kewley2008}, in summary:
\begin{align}
\mbox{upper branch}:& \ \ \ \log\left({\mbox{\iontwo{N}{II}} / \mbox{\iontwo{O}{II}}} \right) \ge -1.2 \label{eq:upperbranch} \\
\mbox{lower branch}:& \ \ \ \log\left({\mbox{\iontwo{N}{II}} / \mbox{\iontwo{O}{II}}} \right) < -1.2 
\end{align}
For each branch, the code assumes an initial \zgas{} of:
\begin{align}
\mbox{upper branch} : \ \ \ \mbox{initial } \log(\mbox{O/H})+12 =  8.7 \\
\mbox{lower branch} : \ \ \ \mbox{initial } \log(\mbox{O/H})+12 =  8.2
\end{align}
This first initial guess for \zgas{} and \iontwo{O}{III}/\iontwo{O}{II} is used to make an initial guess at $q$ using the following (See Eq. 13 in  \citealt{kobulnicky04}):
\begin{align}
y &= \log\left({\mbox{\iontwo{O}{III}} / \mbox{\iontwo{O}{II}}}\right) \\ 
Z  &= \mbox{initial } \log(\mbox{O/H})+12 \label{eq:logz} \\ 
\log q &= {32.81-1.153y^2 + Z(-3.396-0.025y+0.1444y^2)  \over 4.603-0.3119y-0.163y^2 + Z(-0.48+0.0271y + 0.02037y^2) }  \label{eq:q}
\end{align}
With the initial guess for \zgas{} and $q$ for each fibre, the we iterate up to 10 times to try to constrain \zgas{} and $q$.  First \zgas{} is calculated from $R_{23}$ and the initial guess for $q$, then we recalculates $q$ using the \zgas{} we just calculated and \iontwo{O}{III}/\iontwo{O}{II}.  For the next iteration, the calculated $q$ is then used to recalculate \zgas{}.  If the \zgas{} difference between two iterations is $\Delta$\zgas{} $< 0.01$ then the iterations stop and \zgas{} and $q$ are set for that fibre.  If, after 10 iterations, there is no convergence, that fibre is flagged as a bad fit.  This iteration makes use of $R_{23}$:
\begin{equation}
R_{23} = ({\mbox{\iontwo{O}{II}} + \mbox{\iontwo{O}{III}}) / H\beta}
\end{equation}
For each iteration, \zgas{} is calculated for the upper and lower branches as:
\begin{align}
 \log(\mbox{O/H})_{upper}+12 =& 9.72 -0.777 R_{23} - 0.951 R_{23}^2 \\
 	& - 0.072 R_{23}^3 - 0.811 R_{23}^4 \nonumber  \\
 	&- \log(q)(0.0737 - 0.0713 R_{23}) \nonumber   \\
	& -0.141 R_{23}^2 + 0.0373 R_{23}^3 - 0.058 R_{23}^4 \nonumber  \\
 \log(\mbox{O/H})_{lower}+12 =& 9.40 + 4.65 R_{23} - 3.17 R_{23}^2 \\
 	&  - \log(q)(0.272+0.547 R_{23} - 0.513 R_{23}^2) \nonumber 
\end{align}
The ionization parameter $q$ is then calculated from \zgas{} using Equation \ref{eq:q}, which is then used to recalculate \zgas{} for the next iteration.  This continues until convergence ($\Delta$\zgas{} $< 0.01$), or up to 10 iterations, after which the fibre is flagged as a bad fit.

\section{How \zgas{} is determined for our seven different diagnostics}\label{sec:zgas-append}
{\bf Note:}  These techniques are nearly identical to those described in Appendix A2 of \cite{kewley2008}.
\begin{enumerate}
\item {\bf $R_{23}$ - \cite{mcgaugh91}} \\
{\bf Assumed {\it q}:} Not fixed, models attempt to correct for variation in $q$ using the \iontwo{O}{III}/\iontwo{O}{II} line ratio defined in $y$ below.  \\
{\bf Method:} This method is doubly degenerate so which branch we are in is determined using \iontwo{N}{II}/\iontwo{O}{II} as described in Equation \ref{eq:upperbranch}.  McGaugh and Koblunicky performed detailed simulations of \HII{} regions and have come up with the following analytical fits to the results from their models.  These fits for the upper and lower branches are given in \cite{kobulnicky1999}:
\begin{align}
x =& \log(R_{23}) = \log \left( ({\mbox{\iontwo{O}{II}}+\mbox{\iontwo{O}{III}}) / H\beta}\right)  \\ 
y =& \log \left( {\mbox{\iontwo{O}{III}} \over \mbox{\iontwo{O}{II}} }  \right) \\
Z_{upper} =& 12 - 2.939 - 0.2x - 0.237x^2 - 0.305 x^3 \\
	& - 0.0283x^4 - y (0.0047 - 0.0221 x -0.102 x^2   \nonumber \\
	&- 0.0817 x^3 - 0.00717 x^4 ) \nonumber  \\
Z_{lower} =& 12 - 4.994  +0.767x+0.602x^2 \\
	& -y(0.29+0.332x-0.331x^2) \nonumber 
\end{align}\\
\item {\bf $R_{23}$ - \cite{kobulnicky04}} \\
{\bf Assumed {\it q}:} Iterated upon. \\
{\bf Method: } First the \zgas{} and $q$ are guessed at depending on if you are in the upper or lower $R_{23}$ branch, then the code iterates to converge on a \zgas{} and $q$.
The equations behind this method are described in detail above in Appendix \ref{sec:q2-append}.\\
\item {\bf $R_{23}$ - \cite{zaritsky1994}} \\
{\bf Assumed {\it q}:} No solution included.  \\
{\bf Method: } 
This method is derived by averaging the three previous calibrations for $R_{23}$ done by \cite{edmunds1984}, \cite{dopita1986}, and  \cite{mccall85}.
This method only works for the upper branch of the $R_{23}$ double degeneracy.  Which branch we are in is determined using \iontwo{N}{II}/\iontwo{O}{II} as described in Equation \ref{eq:upperbranch}.  If we are in the lower branch, the fibre is disregarded.  If we are in the upper branch we use the following equations to calculate \zgas{}:
\begin{equation}
x = \log(R_{23}) = \log \left( ({\mbox{\iontwo{O}{II}}+\mbox{\iontwo{O}{III}}) / H\beta}\right)
\end{equation}
\begin{equation}
Z_{upper} = 9.265 - 0.33x - 0.202x^2 - 0.207x^3-0.333x^4
\end{equation}\\
\item {\bf \iontwo{N}{II}/\iontwo{O}{II} - \cite{kewley2002} \label{sec:n2o2}} \\
{\bf Assumed {\it q}:} $2 \times 10^7$, although $q$ is nearly invariant for this \zgas{} diagnostic.\\
{\bf Method: } For the above assumed $q$, we set the line ratio equal to the following fourth degree polynomial for \zgas{} (see Table 3 from \citealt{kewley2002}):
\begin{align}
&\log\left({\mbox{\iontwo{N}{II}} / \mbox{\iontwo{O}{II}}} \right) = \\
& \ \ \ 1106.87 - 532.154 Z + 96.3733 Z^2 - 7.81061 Z^3 + 0.239282 Z^4  \nonumber 
\end{align}
The \textsc{IDL} function {\it fz\_roots.pro} then solves for the roots of this polynomial to find \zgas{}.\\
\item {\bf (\iontwo{O}{III}/H$\beta$)/(\iontwo{N}{II}/H$\alpha$) - \cite{pettini2004}} \\
{\bf Assumed {\it q}:} None, fit is empirical  \\
{\bf Method:}
Empirical fit of line ratios to 137 \HII{} regions, 131 have metallicities measured used the direct $T_e$ method, while 6 are derived using strong line methods.
Note that we use the H$\beta$/H$\alpha$ line ratio to correct for extinction, so 
we fix H$\beta$/H$\alpha$ to the same constant
in all our data so this diagnostic is actually just based on the
\iontwo{O}{III}/\iontwo{N}{II} line ratio. 
While this fit is empirical and does not correct for $q$, \cite{kewley2002} shows that the \iontwo{O}{III}/\iontwo{N}{II} line ratio is strongly dependent on $q$.
\begin{equation}
x = \log \left( {\mbox{\iontwo{O}{III}} / H\beta \over \mbox{\iontwo{N}{II}} / \mbox{H}\alpha}\right)
\end{equation}
\begin{equation}
Z = 8.73 -0.32x
\end{equation}\\
\item {\bf \iontwo{N}{II}/H$\alpha$ - \cite{deincolo02}} \\
{\bf Assumed {\it q}:} None, fit is empirical. \\
{\bf Method:}
Empirical fit to 236 galaxies.    Approximately half are  metal poor and half are metal rich to cover a wide range in \zgas{}.  A linear least squares fit to the data gives:
\begin{align}
x =& \log \left(  {\mbox{\iontwo{N}{II}} / \mbox{H}\alpha} \right) \\
Z =& 9.12 + 0.73 x
\end{align}
This empirical fit covers a range of \HII{} regions with -2.5 $<$ log(\iontwo{N}{II}/H$\alpha$) $<$ -0.3  which corresponds to a range in \zgas{} of 7.30 $<$ 12+log(O/H) $<$ 8.90 in this fit. 
We consider values in our data of log(\iontwo{N}{II}/H$\alpha$) $>$ -0.3 corresponding to 12+log(O/H) $>$ 8.90 to be an extrapolation of the fit.
Figure \ref{fig:n2_fits} in Appendix \ref{fig:n2_fits} shows this fit as a dashed line.
\\
\item {\bf \iontwo{N}{II}/H$\alpha$ - \cite{pettini2004}} \\
{\bf Assumed {\it q}:} None, fit is empirical \\
{\bf Method:} Empirical fit of line ratios to 137 \HII{} regions, 131 have metallicities measured used the direct $T_e$ method, while 6 are derived using strong line methods.
\begin{equation}
x = \log \left( {\mbox{\iontwo{N}{II}}  / \mbox{H}\alpha}\right)
\end{equation}
\begin{equation}
Z = 9.37 + 2.03 x +1.26 x^2 + 0.32 x^3
\end{equation}
The \HII{} regions observed cover a range of  $-2.5 < \log(\mbox{\iontwo{N}{II}}/\mbox{H}\alpha) < -0.3$ which corresponds to a range in \zgas{} of 7.17$<$ 12+log(O/H) $<$ 8.87 in this fit.
We consider values in our data of log(\iontwo{N}{II}/H$\alpha$) $>$ -0.3 corresponding to 12+log(O/H) $>$ 8.87 to be an extrapolation of the fit.
Figure \ref{fig:n2_fits}  in Appendix \ref{fig:n2_fits} shows this fit as a solid line.

\end{enumerate}

\section{Comparison of \iontwo{N}{II}/H$\alpha$ \zgas{} diagnostics}
\label{sec:n2-append}

\begin{figure}
\includegraphics[width=0.51\textwidth]{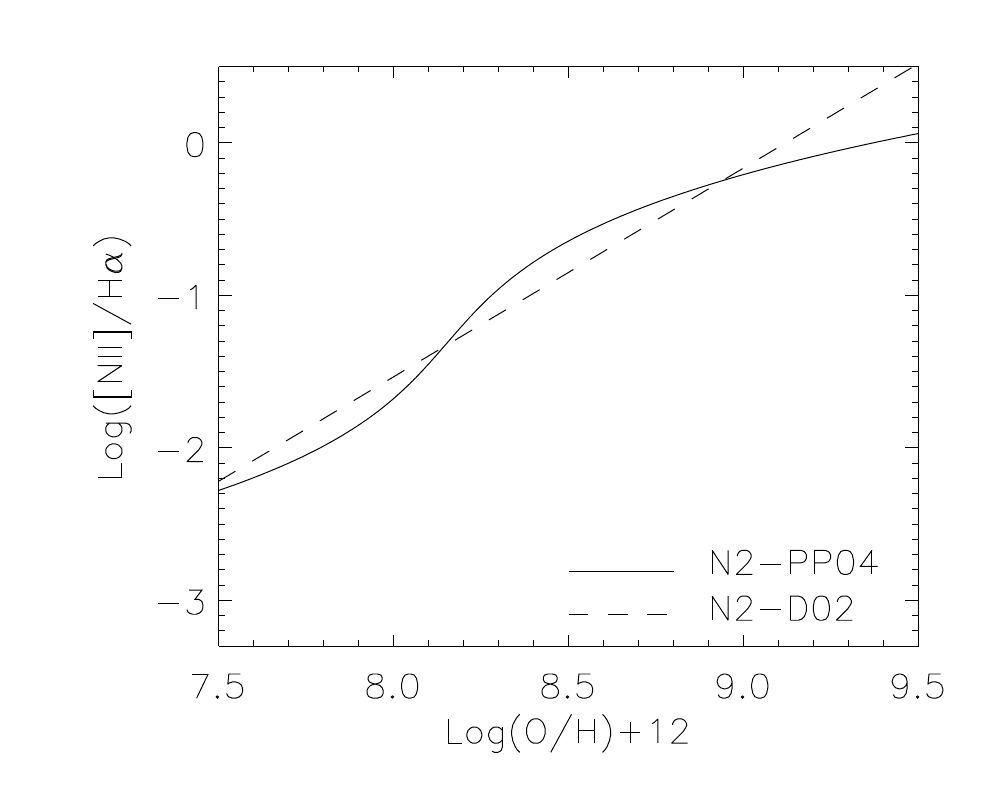}
\caption{Plot showing the two empirical fits for the \zgas{} calibrations based off the \iontwo{N}{II}/H$\alpha$ line ratio.  The solid line denotes the empirical fit of a 3rd order polynomial for N2-PP04 \protect\citep{pettini2004} and the dashed line denotes the linear fit for N2-D02 \protect\citep{deincolo02}.  The axes are shown the with the same range as the top right plot in Figure \protect\ref{fig:degeneracy} for comparison to the theoretical fits from \protect\cite{kewley2002}.
For the \zgas{} range covered in this study (8.3 $<$ log(O/H)+12 $<$ 9.4), a change in \iontwo{N}{II}/H$\alpha$ leads to a larger change in \zgas{} inferred from the N2-PP04 diagnostic than from the N2-D02 diagnostic.
Details for these \iontwo{N}{II}/H$\alpha$ \zgas{} calibrations can be found in $\S$ \ref{sec:results-zgas} and Appendix \ref{sec:zgas-append}.
}
\label{fig:n2_fits}
\end{figure}

\bsp	
\label{lastpage}
\end{document}